\documentclass{pasj00}

\usepackage[OT2,T1]{fontenc}

\def\commenta{$^*$}
\def\commentb{$^\dagger$}
\def\commentc{$^\ddagger$}
\def\commentd{$^\S$}
\def\commente{$^\|$}

\newcounter{author}
\def\authorcount#1#2{\refstepcounter{author}\label{#1}
                     \altaffiltext{\ref{#1}}{#2}}

\def\Ohtprep{T. Ohshima et al. in preparation}
\def\Nakataprep{C. Nakata et al. in preparation}
\def\Isogaiprep{K. Isogai et al. in preparation}

\begin{document}
\SetRunningHead{T. Kato et al.}{Period Variations in SU UMa-Type Dwarf Novae VII}

\Received{201X/XX/XX}
\Accepted{201X/XX/XX}

\title{Survey of Period Variations of Superhumps in SU UMa-Type Dwarf Novae.
    VII: The Seventh Year (2014--2015)}

\author{Taichi~\textsc{Kato},\altaffilmark{\ref{affil:Kyoto}*}
        Franz-Josef~\textsc{Hambsch},\altaffilmark{\ref{affil:GEOS}}$^,$\altaffilmark{\ref{affil:BAV}}$^,$\altaffilmark{\ref{affil:Hambsch}}
        Pavol~A.~\textsc{Dubovsky},\altaffilmark{\ref{affil:Dubovsky}}
        Igor~\textsc{Kudzej},\altaffilmark{\ref{affil:Dubovsky}}
        Berto~\textsc{Monard},\altaffilmark{\ref{affil:Monard}}$^,$\altaffilmark{\ref{affil:Monard2}}
        Ian~\textsc{Miller},\altaffilmark{\ref{affil:Miller}}
        Hiroshi~\textsc{Itoh},\altaffilmark{\ref{affil:Ioh}}
        Seiichiro~\textsc{Kiyota},\altaffilmark{\ref{affil:Kis}}
        Kazunari~\textsc{Masumoto},\altaffilmark{\ref{affil:OKU}}
        Daiki~\textsc{Fukushima},\altaffilmark{\ref{affil:OKU}}
        Hiroki~\textsc{Kinoshita},\altaffilmark{\ref{affil:OKU}}
        Kazuki~\textsc{Maeda},\altaffilmark{\ref{affil:OKU}}
        Jyunya~\textsc{Mikami},\altaffilmark{\ref{affil:OKU}}
        Risa~\textsc{Matsuda},\altaffilmark{\ref{affil:OKU}}
        Naoto~\textsc{Kojiguchi},\altaffilmark{\ref{affil:OKU}}
        Miho~\textsc{Kawabata},\altaffilmark{\ref{affil:OKU}}
        Megumi~\textsc{Takenaka},\altaffilmark{\ref{affil:OKU}}
        Katsura~\textsc{Matsumoto},\altaffilmark{\ref{affil:OKU}}
        Enrique~de~\textsc{Miguel},\altaffilmark{\ref{affil:Miguel}}$^,$\altaffilmark{\ref{affil:Miguel2}}
        Yutaka~\textsc{Maeda},\altaffilmark{\ref{affil:Mdy}}
        Tomohito~\textsc{Ohshima},\altaffilmark{\ref{affil:Kyoto}}
        Keisuke~\textsc{Isogai},\altaffilmark{\ref{affil:Kyoto}}
        Roger~D.~\textsc{Pickard},\altaffilmark{\ref{affil:BAAVSS}}$^,$\altaffilmark{\ref{affil:Pickard}}
        Arne~\textsc{Henden},\altaffilmark{\ref{affil:AAVSO}}
        Stella~\textsc{Kafka},\altaffilmark{\ref{affil:AAVSO}}
        Hidehiko~\textsc{Akazawa},\altaffilmark{\ref{affil:OUS}}
        Noritoshi~\textsc{Otani},\altaffilmark{\ref{affil:OUS}}
        Sakiko~\textsc{Ishibashi},\altaffilmark{\ref{affil:OUS}}
        Minako~\textsc{Ogi},\altaffilmark{\ref{affil:OUS}}
        Kenji~\textsc{Tanabe},\altaffilmark{\ref{affil:OUS}}
        Kazuyoshi~\textsc{Imamura},\altaffilmark{\ref{affil:OUS}}
        William~\textsc{Stein},\altaffilmark{\ref{affil:Stein}}
        Kiyoshi~\textsc{Kasai},\altaffilmark{\ref{affil:Kai}}
        Tonny~\textsc{Vanmunster},\altaffilmark{\ref{affil:Vanmunster}}
        Peter~\textsc{Starr},\altaffilmark{\ref{affil:Starr}}
        Arto~\textsc{Oksanen},\altaffilmark{\ref{affil:Nyrola}}
        Elena~P.~\textsc{Pavlenko},\altaffilmark{\ref{affil:CrAO}}
        Oksana~I.~\textsc{Antonyuk},\altaffilmark{\ref{affil:CrAO}}
        Kirill~A.~\textsc{Antonyuk},\altaffilmark{\ref{affil:CrAO}}
        Aleksei~A.~\textsc{Sosnovskij},\altaffilmark{\ref{affil:CrAO}}
        Nikolaj~V.~\textsc{Pit},\altaffilmark{\ref{affil:CrAO}}
        Julia~V.~\textsc{Babina},\altaffilmark{\ref{affil:CrAO}}
        Aleksandr~\textsc{Sklyanov},\altaffilmark{\ref{affil:Kazan}} 
        Rudolf~\textsc{Nov\'ak},\altaffilmark{\ref{affil:Novak}}
        Shawn~\textsc{Dvorak},\altaffilmark{\ref{affil:Dvorak}}
        Ra\'ul~\textsc{Michel},\altaffilmark{\ref{affil:UNAM}}
        Gianluca~\textsc{Masi},\altaffilmark{\ref{affil:Masi}}
        Colin~\textsc{Littlefield},\altaffilmark{\ref{affil:LCO}}
        Joseph~\textsc{Ulowetz},\altaffilmark{\ref{affil:Ulowetz}}
        Sergey~Yu.~\textsc{Shugarov},\altaffilmark{\ref{affil:Sternberg}}$^,$\altaffilmark{\ref{affil:Slovak}}
        Polina~Yu.~\textsc{Golysheva},\altaffilmark{\ref{affil:Sternberg}}
        Drahomir~\textsc{Chochol},\altaffilmark{\ref{affil:Slovak}}
        Viktoriia~\textsc{Krushevska},\altaffilmark{\ref{affil:MainUkraine}}
        Javier~\textsc{Ruiz},\altaffilmark{\ref{affil:Ruiz1}}$^,$\altaffilmark{\ref{affil:Ruiz2}}$^,$\altaffilmark{\ref{affil:Ruiz3}}
        Tam\'as~\textsc{Tordai},\altaffilmark{\ref{affil:Polaris}}
        Etienne~\textsc{Morelle},\altaffilmark{\ref{affil:Morelle}}
        Richard~\textsc{Sabo},\altaffilmark{\ref{affil:Sabo}}
        Hiroyuki~\textsc{Maehara},\altaffilmark{\ref{affil:OAO}}
        Michael~\textsc{Richmond},\altaffilmark{\ref{affil:RIT}}
        Natalia~\textsc{Katysheva},\altaffilmark{\ref{affil:Sternberg}}
        Kenji~\textsc{Hirosawa},\altaffilmark{\ref{affil:Hsk}}
        William~N.~\textsc{Goff},\altaffilmark{\ref{affil:Goff}}
        Franky~\textsc{Dubois},\altaffilmark{\ref{affil:Dubois}}
        Ludwig~\textsc{Logie},\altaffilmark{\ref{affil:Logie}}
        Steve~\textsc{Rau},\altaffilmark{\ref{affil:Rau}}
        Irina~B.~\textsc{Voloshina},\altaffilmark{\ref{affil:Sternberg}}
        Maksim~V.~\textsc{Andreev},\altaffilmark{\ref{affil:Terskol}}$^,$\altaffilmark{\ref{affil:ICUkraine}}
        Kazuhiko~\textsc{Shiokawa},\altaffilmark{\ref{affil:Siz}} 
        Vitaly~V.~\textsc{Neustroev},\altaffilmark{\ref{affil:Neustroev}}
        George~\textsc{Sjoberg},\altaffilmark{\ref{affil:Sjoberg}}$^,$\altaffilmark{\ref{affil:AAVSO}}
        Sergey~\textsc{Zharikov},\altaffilmark{\ref{affil:UNAM}}
        Nick~\textsc{James},\altaffilmark{\ref{affil:James}}
        Greg~\textsc{Bolt},\altaffilmark{\ref{affil:Bolt}}
        Tim~\textsc{Crawford},\altaffilmark{\ref{affil:Crawford}}
        Denis~\textsc{Buczynski},\altaffilmark{\ref{affil:Buczynski}}
        Lewis~M.~\textsc{Cook},\altaffilmark{\ref{affil:LewCook}}
        Christopher~S.~\textsc{Kochanek},\altaffilmark{\ref{affil:Ohio}}
        Benjamin~\textsc{Shappee},\altaffilmark{\ref{affil:Ohio}}
        Krzysztof~Z.~\textsc{Stanek},\altaffilmark{\ref{affil:Ohio}}
        Jos\'e~L.~\textsc{Prieto},\altaffilmark{\ref{affil:DiegoPortales}}$^,$\altaffilmark{\ref{affil:Princeton}}
        Denis~\textsc{Denisenko},\altaffilmark{\ref{affil:Denisenko}}
        Hideo~\textsc{Nishimura},\altaffilmark{\ref{affil:Nmh}}
        Masaru~\textsc{Mukai},\altaffilmark{\ref{affil:Mukai}}
        Shizuo~\textsc{Kaneko},\altaffilmark{\ref{affil:Kaneko}}
        Seiji~\textsc{Ueda},\altaffilmark{\ref{affil:SeijiUeda}}
        Rod~\textsc{Stubbings},\altaffilmark{\ref{affil:Stubbings}}
        Masayuki~\textsc{Moriyama},\altaffilmark{\ref{affil:Myy}}
        Patrick~\textsc{Schmeer},\altaffilmark{\ref{affil:Schmeer}}
        Eddy~\textsc{Muyllaert},\altaffilmark{\ref{affil:VVSBelgium}}
        Jeremy~\textsc{Shears},\altaffilmark{\ref{affil:Shears}}$,$\altaffilmark{\ref{affil:BAAVSS}}
        Robert~J.~\textsc{Modic},\altaffilmark{\ref{affil:MRV}}
        Kevin~B.~\textsc{Paxson},\altaffilmark{\ref{affil:Paxson}}
}

\authorcount{affil:Kyoto}{
     Department of Astronomy, Kyoto University, Kyoto 606-8502, Japan}
\email{$^*$tkato@kusastro.kyoto-u.ac.jp}

\authorcount{affil:GEOS}{
     Groupe Europ\'een d'Observations Stellaires (GEOS),
     23 Parc de Levesville, 28300 Bailleau l'Ev\^eque, France}

\authorcount{affil:BAV}{
     Bundesdeutsche Arbeitsgemeinschaft f\"ur Ver\"anderliche Sterne
     (BAV), Munsterdamm 90, 12169 Berlin, Germany}

\authorcount{affil:Hambsch}{
     Vereniging Voor Sterrenkunde (VVS), Oude Bleken 12, 2400 Mol, Belgium}

\authorcount{affil:Dubovsky}{
     Vihorlat Observatory, Mierova 4, Humenne, Slovakia}

\authorcount{affil:Monard}{
     Bronberg Observatory, Center for Backyard Astrophysics Pretoria,
     PO Box 11426, Tiegerpoort 0056, South Africa}

\authorcount{affil:Monard2}{
     Kleinkaroo Observatory, Center for Backyard Astrophysics Kleinkaroo,
     Sint Helena 1B, PO Box 281, Calitzdorp 6660, South Africa}

\authorcount{affil:Miller}{
     Furzehill House, Ilston, Swansea, SA2 7LE, UK}

\authorcount{affil:Ioh}{
     Variable Star Observers League in Japan (VSOLJ),
     1001-105 Nishiterakata, Hachioji, Tokyo 192-0153, Japan}

\authorcount{affil:Kis}{
     VSOLJ, 7-1 Kitahatsutomi, Kamagaya, Chiba 273-0126, Japan}

\authorcount{affil:OKU}{
     Osaka Kyoiku University, 4-698-1 Asahigaoka, Osaka 582-8582, Japan}

\authorcount{affil:Miguel}{
     Departamento de F\'isica Aplicada, Facultad de Ciencias
     Experimentales, Universidad de Huelva,
     21071 Huelva, Spain}

\authorcount{affil:Miguel2}{
     Center for Backyard Astrophysics, Observatorio del CIECEM,
     Parque Dunar, Matalasca\~nas, 21760 Almonte, Huelva, Spain}

\authorcount{affil:Mdy}{
     Kaminishiyamamachi 12-14, Nagasaki, Nagasaki 850-0006, Japan}

\authorcount{affil:BAAVSS}{
     The British Astronomical Association, Variable Star Section (BAA VSS),
     Burlington House, Piccadilly, London, W1J 0DU, UK}

\authorcount{affil:Pickard}{
     3 The Birches, Shobdon, Leominster, Herefordshire, HR6 9NG, UK}

\authorcount{affil:AAVSO}{
     American Association of Variable Star Observers, 49 Bay State Rd.,
     Cambridge, MA 02138, USA}

\authorcount{affil:OUS}{
     Department of Biosphere-Geosphere System Science, Faculty of Informatics,
     Okayama University of Science, 1-1 Ridai-cho, Okayama,
     Okayama 700-0005, Japan}

\authorcount{affil:Stein}{
     6025 Calle Paraiso, Las Cruces, New Mexico 88012, USA}

\authorcount{affil:Kai}{
     Baselstrasse 133D, CH-4132 Muttenz, Switzerland}

\authorcount{affil:Vanmunster}{
     Center for Backyard Astrophysics Belgium, Walhostraat 1A,
     B-3401 Landen, Belgium}

\authorcount{affil:Starr}{
     Warrumbungle Observatory, Tenby, 841 Timor Rd,
     Coonabarabran NSW 2357, Australia}

\authorcount{affil:Nyrola}{
     Hankasalmi observatory, Jyvaskylan Sirius ry, Vertaalantie
     419, FI-40270 Palokka, Finland}

\authorcount{affil:CrAO}{
     Crimean Astrophysical Observatory, p/o Naychny, 298409,
     Republic of Crimea}

\authorcount{affil:Novak}{
     Research Centre for Toxic Compounds in the Environment, Faculty of 
     Science, Masaryk University, Kamenice 3, 625 00 Brno, Czech Republic}

\authorcount{affil:Kazan}{
     Kazan Federal University, Kremlevskaya str., 18, Kazan, 420008, Russia}

\authorcount{affil:Dvorak}{
     Rolling Hills Observatory, 1643 Nightfall Drive,
     Clermont, Florida 34711, USA}

\authorcount{affil:UNAM}{
     Instituto de Astronom\'{\i}a UNAM, Apartado Postal 877, 22800 Ensenada
     B.C., M\'{e}xico}

\authorcount{affil:Masi}{
     The Virtual Telescope Project, Via Madonna del Loco 47, 03023
     Ceccano (FR), Italy}

\authorcount{affil:LCO}{
     Department of Physics, University of Notre Dame, Notre Dame,
     Indiana 46556, USA}

\authorcount{affil:Ulowetz}{
     Center for Backyard Astrophysics Illinois,
     Northbrook Meadow Observatory, 855 Fair Ln, Northbrook,
     Illinois 60062, USA}

\authorcount{affil:Sternberg}{
     Sternberg Astronomical Institute, Lomonosov Moscow State University, 
     Universitetsky Ave., 13, Moscow 119992, Russia}

\authorcount{affil:Slovak}{
     Astronomical Institute of the Slovak Academy of Sciences, 05960,
     Tatranska Lomnica, the Slovak Republic}

\authorcount{affil:MainUkraine}{
     Main astronomical observatory of the National Academy of Sciences of
     Ukraine, 27 Akademika Zabolotnoho ave., 03680 Kyiv, Ukraine}

\authorcount{affil:Ruiz1}{
     Observatorio de C\'antabria, Ctra. de Rocamundo s/n, Valderredible, 
     Cantabria, Spain}

\authorcount{affil:Ruiz2}{
     Instituto de F\'{\i}sica de Cantabria (CSIC-UC), Avenida Los Castros s/n, 
     E-39005 Santander, Cantabria, Spain}

\authorcount{affil:Ruiz3}{
     Agrupaci\'on Astron\'omica C\'antabria, Apartado 573,
     39080, Santander, Spain}

\authorcount{affil:Polaris}{
     Polaris Observatory, Hungarian Astronomical Association,
     Laborc utca 2/c, 1037 Budapest, Hungary}

\authorcount{affil:Morelle}{
     9 rue Vasco de GAMA, 59553 Lauwin Planque, France}

\authorcount{affil:Sabo}{
     2336 Trailcrest Dr., Bozeman, Montana 59718, USA}

\authorcount{affil:OAO}{
     Okayama Astrophysical Observatory, National Astronomical Observatory 
     of Japan, Asakuchi, Okayama 719-0232, Japan}

\authorcount{affil:RIT}{
     Physics Department, Rochester Institute of Technology, Rochester,
     New York 14623, USA}

\authorcount{affil:Hsk}{
     216-4 Maeda, Inazawa-cho, Inazawa-shi, Aichi 492-8217, Japan}

\authorcount{affil:Goff}{
     13508 Monitor Ln., Sutter Creek, California 95685, USA}

\authorcount{affil:Dubois}{
     Astrolab team, Poelkapellestraat 57 Langemark, Belgium}

\authorcount{affil:Logie}{
     Astrolab team, Gezellestraat 9, 8908 Vlamertinge, Belgium}

\authorcount{affil:Rau}{
     Astrolab team, Veldstraat 6, 8400 Oostende, Belgium}

\authorcount{affil:Terskol}{
     Institute of Astronomy, Russian Academy of Sciences, 361605 Peak Terskol,
     Kabardino-Balkaria, Russia}

\authorcount{affil:ICUkraine}{
     International Center for Astronomical, Medical and Ecological Research
     of NASU, Ukraine 27 Akademika Zabolotnoho Str. 03680 Kyiv,
     Ukraine}

\authorcount{affil:Siz}{
     Moriyama 810, Komoro, Nagano 384-0085, Japan}

\authorcount{affil:Neustroev}{
     Astronomy and Space Physics, PO Box 3000,
     FIN-90014 University of Oulu, Finland}

\authorcount{affil:Sjoberg}{
     The George-Elma Observatory, 9 Contentment Crest, \#182,
     Mayhill, New Mexico 88339, USA}

\authorcount{affil:James}{
     11 Tavistock Road, Chelmsford, Essex CM1 6JL, UK}

\authorcount{affil:Bolt}{
     Camberwarra Drive, Craigie, Western Australia 6025, Australia}

\authorcount{affil:Crawford}{
     Arch Cape Observatory, 79916 W. Beach Road, Arch Cape, Oregon 97102, USA}

\authorcount{affil:Buczynski}{
     Conder Brow Observatory, Fell Acre, Conder Brow, Little Fell Lane,
     Scotforth, Lancs LA2 0RQ, England}

\authorcount{affil:LewCook}{
     Center for Backyard Astrophysics Concord, 1730 Helix Ct. Concord,
     California 94518, USA}

\authorcount{affil:Ohio}{
     Department of Astronomy, the Ohio State University, Columbia,
     OH 43210, USA}

\authorcount{affil:DiegoPortales}{
     N\'ucleo de Astronom\'ia de la Facultad de Ingenier\'ia, Universidad
     Diego Portales, Av. Ej\'ercito 441, Santiago, Chile}

\authorcount{affil:Princeton}{
     Department of Astrophysical Sciences, Princeton University,
     NJ 08544, USA}

\authorcount{affil:Denisenko}{
     Space Research Institute (IKI), Russian Academy of Sciences, Moscow,
     Russia}

\authorcount{affil:Nmh}{
     Miyawaki 302-6, Kakegawa, Shizuoka 436-0086, Japan}

\authorcount{affil:Mukai}{
     JCPM Kagoshima Station, Kagoshima City, Kagoshima 892-0871, Japan}

\authorcount{affil:Kaneko}{
     14-7 Kami-Yashiki, Kakegawa, Shizuoka 436-0049, Japan}

\authorcount{affil:SeijiUeda}{
     6-23-9 Syowa-Minami, Kushiro City, Hokkaido 084-0909, Japan
     }

\authorcount{affil:Stubbings}{
     Tetoora Observatory, Tetoora Road, Victoria, Australia}

\authorcount{affil:Myy}{
     290-383, Ogata-cho, Sasebo, Nagasaki 858-0926, Japan}

\authorcount{affil:Schmeer}{
     Bischmisheim, Am Probstbaum 10, 66132 Saarbr\"{u}cken, Germany}

\authorcount{affil:VVSBelgium}{
     Vereniging Voor Sterrenkunde (VVS), Moffelstraat 13 3370
     Boutersem, Belgium}

\authorcount{affil:Shears}{
     ``Pemberton'', School Lane, Bunbury, Tarporley, Cheshire, CW6 9NR, UK}

\authorcount{affil:MRV}{
     351 Fairlawn Dr., Richmond Heights, Ohio 44143, USA 
}

\authorcount{affil:Paxson}{
     20219 Eden Pines, Spring, Texas 77379, USA}



\maketitle

\begin{abstract}
Continuing the project described by \citet{Pdot}, we collected
times of superhump maxima for 102 SU UMa-type dwarf novae 
observed mainly during the 2014--2015 season and characterized
these objects.  Our project has greatly improved
the statistics of the distribution of orbital periods,
which is a good approximation of the distribution of
cataclysmic variables at the terminal evolutionary stage,
and confirmed the presence of a period minimum at a period of
0.053~d and a period spike just above this period.
The number density monotonically decreased toward the longer
period and there was no strong indication of a period gap.
We detected possible negative superhumps in Z Cha.
It is possible that normal outbursts are also suppressed
by the presence of a disk tilt in this system.
There was no indication of enhanced orbital humps
just preceding the superoutburst, and this result favors
the thermal-tidal disk instability as the origin of
superoutbursts.  We detected superhumpsin three 
AM CVn-type dwarf novae.
Our observations and recent other detections suggest that
8\% of objects showing dwarf nova-type outbursts
are AM CVn-type objects.  AM CVn-type objects and
EI Psc-type object may be more abundant than previously
recognized.  OT J213806, a WZ Sge-type object, exhibited
a remarkably different feature between the 2010 and 2014
superoutbursts.  Although the 2014 superoutburst was
much fainter the plateau phase was shorter than the 2010 one,
the course of the rebrightening phase was similar.
This object indicates that the $O-C$ diagrams of superhumps
can be indeed variable at least in WZ Sge-type objects.
Four deeply eclipsing SU UMa-type dwarf novae
(ASASSN-13cx, ASASSN-14ag, ASASSN-15bu, NSV 4618) were identified. 
We studied long-term trends in supercycles
in MM Hya and CY UMa and found systematic variations
of supercycles of $\sim$20\%.
\end{abstract}

\section{Introduction}

   Cataclysmic variables (CVs) are close binary systems
transferring matter from a low-mass dwarf secondary to
a white dwarf.  The transferred matter forms an accretion
disk.  Thermal instability od the disk caused by partial ionization
of hydrogen results outbursts in dwarf novae (DNe),
a subclass of CVs.
Tidal instability of the disk caused by the 3:1 resonance 
with the orbiting secondary is considered to develop an eccentric
(or flexing) disk in SU UMa-type dwarf novae, a subclass of DNe.
This eccentric disk is responsible for
superhumps, which have periods a few percent longer
than the orbital period [see e.g. \citet{whi88tidal};
\citet{hir90SHexcess}; \citet{lub91SHa}].
The enhanced mass accretion by tidal instability
causes long-lasting superoutbursts
[thermal tidal instability (TTI) model: \citet{osa89suuma};
\citet{osa96review}].  Although there had been intensive
discussions whether superoutbursts are a result of
tidal instability or an enhanced mass-transfer from
the secondary [e.g. \citet{sma91suumamodel}, \citet{sma04EMT},
\citet{sma08zcha}], recent detailed analyses of
high-precision Kepler observations have favored
the TTI model as the only viable model for ordinary
SU UMa-type dwarf novae (\cite{osa13v1504cygKepler};
\cite{osa13v344lyrv1504cyg}; \cite{osa14v1504cygv344lyrpaper3}).
[For general information of CVs, DNe, SU UMa-type 
dwarf novae and superhumps, see e.g. \citet{war95book}].

   This paper is one of series of papers \citet{Pdot},
\citet{Pdot2}, \citet{Pdot3}, \citet{Pdot4}, \citet{Pdot5}
and \citet{Pdot6}.  These papers originally intended to
clarify the period variations of superhumps in SU UMa-type
dwarf novae and first succeeded in identifying superhump
stages (stages A, B and C) in \citet{Pdot}.
Among them, stage A superhumps have recently been identified
to be reflect the precession of the eccentric disk 
at the radius of the 3:1 resonance, and have been one of 
the most promising tools in determining the mass-ratios in
SU UMa-type dwarf novae and in following the terminal
evolution of CVs \citep{kat13qfromstageA}.

   Continuing this project, we report observations of
superhumps and associated phenomena in SU UMa-type
dwarf novae whose superoutbursts were observed in 2014--2015.
In this paper, we report basic observational materials
and discussions in relation to individual objects.
General discussion related to WZ Sge-type dwarf novae,
which are a subclass of SU UMa-type dwarf novae
with infrequent, large amplitude superoutbursts,
will be given as in a planned summary paper by
\citet{kat15wzsge}.

   Starting from \citet{Pdot6}, we have been intending
these series of papers to be also a source of compiled
information, including historical, of individual dwarf novae
since there have been no compiled publication
since \citet{GlasbyDNbook}.

   The material and methods of analysis are given in
section \ref{sec:obs}, observations and analysis of
individual objects are given in section \ref{sec:individual},
including some discussions particular to the objects,
the general discussion is given in section
\ref{sec:discuss} and the summary is given in section
\ref{sec:summary}.

\section{Observation and Analysis}\label{sec:obs}

\subsection{General Procedure}

   The data were obtained under campaigns led by 
the VSNET Collaboration \citep{VSNET}.
For some objects, we used the public data from 
the AAVSO International Database\footnote{
   $<$http://www.aavso.org/data-download$>$.
}.

   The majority of the data were acquired
by time-resolved CCD photometry by using 30cm-class telescopes
located world-wide, whose observational details will be
presented in future papers dealing with analysis and discussion
on individual objects of interest.
The list of outbursts and observers is summarized in 
table \ref{tab:outobs}.
The data analysis was performed just in the same way described
in \citet{Pdot} and \citet{Pdot6} and we mainly used
R software\footnote{
   The R Foundation for Statistical Computing:\\
   $<$http://cran.r-project.org/$>$.
} for data analysis.
In de-trending the data, we used both lower (1--5th order)
polynomial fitting and locally-weighted polynomial regression 
(LOWESS: \cite{LOWESS}).
The times of superhumps maxima were determined by
the template fitting method as described in \citet{Pdot}.
The times of all observations are expressed in 
barycentric Julian Days (BJD).

   The abbreviations used in this paper are the same
as in \citet{Pdot6}: $P_{\rm orb}$ means
the orbital period and $\varepsilon \equiv P_{\rm SH}/P_{\rm orb}-1$ for 
the fractional superhump excess.   Following \citet{osa13v1504cygKepler},
the alternative fractional superhump excess in the frequency unit
$\varepsilon^* \equiv 1-P_{\rm orb}/P_{\rm SH}-1 = \varepsilon/(1+\varepsilon)$
has been introduced because this fractional superhump excess
can be directly compared to the precession rate.  We therefore
used $\varepsilon^*$ in referring the precession rate.

   We used phase dispersion minimization (PDM; \cite{PDM})
for period analysis and 1$\sigma$ errors for the PDM analysis
was estimated by the methods of \citet{fer89error} and \citet{Pdot2}.
We present evidence for an SU UMa-type dwarf nova
by presenting period analysis and averaged superhump profile
if the paper provides the first solid presentation of
individual objects as such.

   The resultant $P_{\rm SH}$, $P_{\rm dot}$ and other parameters
are listed in table \ref{tab:perlist} in same format as in
\citet{Pdot}.  The definitions of parameters $P_1, P_2, E_1, E_2$
and $P_{\rm dot}$ are the same as in \citet{Pdot}.\footnote{
   The intervals ($E_1$ and $E_2$) for the stages B and C given in the table
   sometimes overlap because of occasional observational ambiguity
   in determining the stages.
}
Comparisons of $O-C$ diagrams between different
superoutbursts are also presented whenever available,
since this comparison was one of the main motivations
in of these series papers (cf. \cite{uem05tvcrv}).
Combined $O-C$ diagrams also help identifying
superhump stages particularly when observations are
insufficient.
In drawing combined $O-C$ diagrams, we usually used
$E=0$ for the start of the superoutburst, which usually
refers to the first positive detection of the outburst.
This epoch usually has an accuracy of $\sim$1 d for
well-observed objects, and if the outburst was not sufficiently
observed, we mentioned in the figure caption how to estimate
$E$ in such an outburst.
We also present $O-C$ diagrams and 
light curves especially for WZ Sge-type dwarf novae, 
which are not expected to undergo outbursts in the near future.
In all figures, the binned magnitudes and $O-C$ values
are accompanied by 1$\sigma$ error bars, which are omitted
when the error is smaller than the plot mark.

   We used the same terminology of superhumps summarized in
\citet{Pdot3}.  We especially call attention to
the term ``late superhumps''.  Although this term has been
used to refer to various phenomena, we only used
the concept of ``traditional'' late superhumps when
an $\sim$0.5 phase shift is detected
[\citet{vog83lateSH}; see also table 1 in \citet{Pdot3} 
for various types of superhumps], 
since we suspect that many of the past
claims of detections of ``late superhumps'' were likely
stage C superhumps [cf. \citet{Pdot}; note that the Kepler 
observation of V585 Lyr also demonstrated this persistent stage C
superhumps without a phase shift \citep{kat13j1939v585lyrv516lyr},
and most recently it is confirmed in another Kepler CV
by \citet{bro15j1915}].

   Early superhumps are double-wave humps seen during the early stages
of WZ Sge-type dwarf novae, and have period close to the orbital
periods (\cite{kat96alcom}; \cite{kat02wzsgeESH}; 
\cite{osa02wzsgehump}).  We are going to discuss
this phenomenon in the planned paper \citep{kat15wzsge}.
We used the period of early superhumps as approximate
orbital period \citep{Pdot6}.

   As in \citet{Pdot}, we have used coordinate-based 
optical transient (OT) designations for some objects, such as
apparent dwarf nova candidates reported in
the Transient Objects Confirmation Page of
the Central Bureau for Astronomical Telegrams\footnote{
   $<$http://www.cbat.eps.harvard.edu/unconf/tocp.html$>$.
}
and listed the original identifiers in table \ref{tab:outobs}.
For objects detected in the Catalina Real-time Transient Survey
(CRTS; \cite{CRTS})\footnote{
   $<$http://nesssi.cacr.caltech.edu/catalina/$>$.
   For the information of the individual Catalina CVs, see
   $<$http://nesssi.cacr.caltech.edu/catalina/AllCV.html$>$.
} transients, we preferably used the names provided
in \citet{dra14CRTSCVs}.
If these names are not yet available,
we used the International Astronomical
Union (IAU)-format names provided by the CRTS team 
in the public data release\footnote{
  $<$http://nesssi.cacr.caltech.edu/DataRelease/$>$.
}

   Since ASAS-SN detectors have relatively poor angular
resolutions (7.5 arcsec/pixel), we provided
coordinates from our own astrometry and astrometric catalogs
for ASAS-SN CVs.
We used SDSS, the Initial Gaia Source List (IGSL,
\cite{IGSL}) and Guide Star Catalog 2.3.2.
The coordinates used in this paper are J2000.0.

\subsection{Period Selection}

   Questions have been raised to our surveys how to select
the periods among the aliases and what is the uncertainty.
Such question are natural if one only sees PDM diagrams.
We should note that PDM (and most of other period finding
algorithms) assumes ``uncorrelated'' (in time) observations
and defines the statistics.  The actual data have more
information, such as the superhump timing data.
Even if the PDM result shows strong aliases, we can resolve
the alias problem if we have sufficiently long continuous
observations (continuous data produce no aliases).
Our period selection is mostly based on this principle,
when alias selection is inconclusive by the PDM analysis only.
The second approach is to examine the trends of $O-C$ values
against the trial periods.  A period longer than the actual
one produces systematically decreasing $O-C$ values within
each night, and our measurements of the superhump maxima
have typical errors of 0.001~d, which is usually sufficient
to select one-day aliases when multiple superhumps were
detected each night.  If the case is not, we describe
the uncertainty of the selection.

\begin{figure}
  \begin{center}
    \FigureFile(88mm,70mm){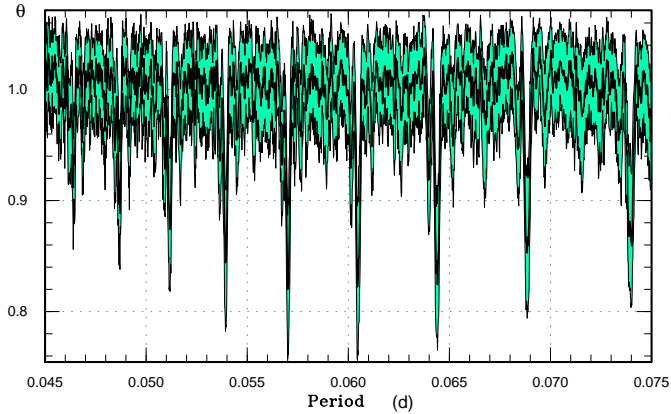}
  \end{center}
  \caption{PDM analysis of FI Cet (see subsection \ref{obj:ficet})
  in a wider period range.}
  \label{fig:ficetshpdm2}
\end{figure}

\begin{figure}
  \begin{center}
    \FigureFile(88mm,70mm){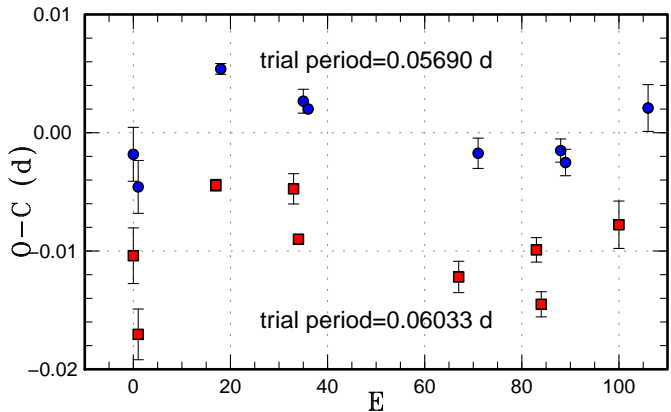}
  \end{center}
  \caption{$O-C$ diagrams assuming two trial periods
  (two aliases in figure \ref{fig:ficetshpdm2}).
  Scatter in the $O-C$ values and systematic trends within 
  each night are apparent for trial period 0.06033~d,
  which is rejected.}
  \label{fig:ficetoccomp}
\end{figure}

   We show an example how our method works by using the
data of FI Cet (subsection \ref{obj:ficet}).
A PDM analysis over a wider range of periods
(figure \ref{fig:ficetshpdm2}) gives an impression that
there are many period candidates and a period of 0.057~d
and 0.060~d are equally acceptable.  If we rely on statistics
assuming temporarily ``uncorrelated'' observations, this would give
equal significance to these periods.
An example for FI Cet is shown in figure \ref{fig:ficetoccomp}.
Scatter in the $O-C$ values and systematic trends within 
each night are apparent for trial period 0.06033~d,
which is rejected.  Although we do not show similar figures
for other objects due to the limitation of space,
we made similar analysis for objects when ambiguities
in period selection remained.

\begin{figure}
  \begin{center}
    \FigureFile(88mm,70mm){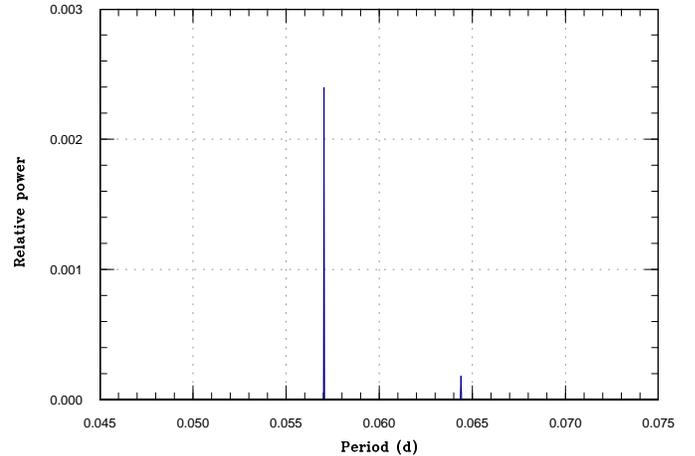}
  \end{center}
  \caption{Lasso period analysis of FI Cet.
  The period of 0.057~d is strongly selected.}
  \label{fig:ficetlasso}
\end{figure}

   Readers may be interested in the result of
least absolute shrinkage and selection operator (Lasso)
(\cite{lasso}; \cite{kat12perlasso}) analysis,
which has been proven
to be very effective in detecting rapidly varying periods
in unevenly sampled data (e.g. \cite{kat13j1924};
\cite{osa13v344lyrv1504cyg}; \cite{kat13j1939v585lyrv516lyr};
\cite{ohs14eruma}).  The result is shown in figure
\ref{fig:ficetlasso}, and the impression is so different
from the PDM result (figure \ref{fig:ficetshpdm2}).
However, we didn't widely used this method in selecting
the aliases since our model used in Lasso analysis also
assumed temporarily uncorrelated observations, and 
the suppression of the aliases is simply a result of
highly non-linear characteristics of compressed sensing.
This figure (if compared to classical figures) would
give misleading impression that there is no
possibility for aliases and we have not used this
type of figure in this paper.

\begin{table*}
\caption{List of Superoutbursts.}\label{tab:outobs}
\begin{center}
\begin{tabular}{ccccl}
\hline
Subsection & Object & Year & Observers or references\commenta & ID\commentb \\
\hline
\ref{obj:kxaql}   & KX Aql          & 2014 & RPc & \\
\ref{obj:nncam}   & NN Cam          & 2014 & DPV & \\
\ref{obj:v342cam} & V342 Cam        & 2014 & OKU, Kis, Ioh & \\
\ref{obj:oycar}   & OY Car          & 2014 & SPE & \\
                  &                 & 2015 & SPE & \\
\ref{obj:ficet}   & FI Cet          & 2014 & HaC & \\
\ref{obj:zcha}    & Z Cha           & 2014 & HaC & \\
\ref{obj:yzcnc}   & YZ Cnc          & 2014b & AAVSO & \\
\ref{obj:v337cyg} & V337 Cyg        & 2014 & IMi & \\
\ref{obj:v503cyg} & V503 Cyg        & 2014 & Kis, DPV & \\
                  &                 & 2014b & RPc & \\
\ref{obj:bcdor}   & BC Dor          & 2015 & HaC, COO & \\
\ref{obj:v660her} & V660 Her        & 2014 & Ioh & \\
\ref{obj:cthya}   & CT Hya          & 2015 & SPE, Kis & \\
\ref{obj:lyhya}   & LY Hya          & 2014 & HaC & \\
\ref{obj:mmhya}   & MM Hya          & 2014 & Mdy, Ioh & \\
\ref{obj:rzlmi}   & RZ LMi          & 2014 & Mic & \\
\ref{obj:aylyr}   & AY Lyr          & 2014 & OUS, Aka & \\
\ref{obj:v453nor} & V453 Nor        & 2014 & HaC & \\
\ref{obj:dtoct}   & DT Oct          & 2014 & HaC & \\
\ref{obj:uvper}   & UV Per          & 2014 & Ioh, DPV, IMi, Mdy, Kis, & \\
                  &                 &      & Nov, RPc & \\
\ref{obj:hypsc}   & HY Psc          & 2014 & OKU, Ioh, IMi, Kis & \\
\hline
  \multicolumn{5}{l}{\parbox{500pt}{\commenta Key to observers:
Aka (H. Akazawa, OUS),
Buc (D. Buczynski),
COO (L. Cook),
CRI (Crimean Astrophys. Obs.),
CTX\commentb (T. Crawford), 
deM (E. de Miguel),
DKS\commentc (S. Dvorak),
DPV (P. Dubovsky),
Dub (F. Dubois team),
GBo (G. Bolt),
GFB\commentc (W. Goff),
HaC (F.-J. Hambsch, remote obs. in Chile), 
Ham (F.-J. Hambsch),
Han (Hankasalmi Obs., by A. Oksanen),
Hsk (K. Hirosawa),
IMi\commentc (I. Miller),
Ioh (H. Itoh),
Kai (K. Kasai),
Kis (S. Kiyota),
KU (Kyoto U., campus obs.),
LCO (C. Littlefield),
MEV\commentc (E. Morelle),
MLF (B. Monard),
Mas (G. Masi),
Mdy (Y. Maeda),
Mhh (H. Maehara),
Mic (R. Michel-Murillo team),
NDJ (N. James),
Neu (V. Neustroev team),
NKa (N. Katysheva and S. Shugarov),
Nov (R. Nov\'ak),
OKU (Osaya Kyoiku U.),
OUS (Okayama U. of Science),
OkC\commentc (A. Oksanen, remote obs. in Chile),
RIT (M. Richmond),
RPc\commentc (R. Pickard),
Rui (J. Ruiz),
SPE\commentc (P. Starr),
SRI\commentc (R. Sabo),
SWI\commentc (W. Stein),
Shu (S. Shugarov team),
Siz (K. Shiokawa),
Ter (Terskol Obs.),
Trt (T. Tordai),
UJH\commentc (J. Ulowetz),
Van (T. Vanmunster),
Vol (I. Voloshina),
AAVSO (AAVSO database)
}} \\
  \multicolumn{5}{l}{\commentb Original identifications, discoverers or data source.} \\
  \multicolumn{5}{l}{\commentc Inclusive of observations from the AAVSO database.} \\
\end{tabular}
\end{center}
\end{table*}

\addtocounter{table}{-1}
\begin{table*}
\caption{List of Superoutbursts (continued).}
\begin{center}
\begin{tabular}{ccccl}
\hline
Subsection & Object & Year & Observers or references\commenta & ID\commentb \\
\hline
--                & CC Scl          & 2014 & \citet{kat15ccscl} & \\
\ref{obj:qwser}   & QW Ser          & 2014 & DPV & \\
\ref{obj:v418ser} & V418 Ser        & 2014 & CRI, AAVSO, LCO, SWI, KU, GFB, & \\
                  &                 & 2014 & Mdy, deM, UJH, Ioh, SRI, DPV & \\
\ref{obj:v701tau} & V701 Tau        & 2015 & RPc, Kai, Trt & \\
\ref{obj:suuma}   & SU UMa          & 2014 & Kis, Nov, Dub, AAVSO & \\
\ref{obj:cyuma}   & CY UMa          & 2014 & Nov, Aka, Kis, DPV, Mdy, Ham, Han & \\
\ref{obj:nsv1436} & NSV 1436        & 2014 & IMi, DPV, Aka, NKa, KU, Hsk & \\
\ref{obj:nsv4618} & NSV 4618        & 2015 & Ioh, HaC, SWI, Ter & \\
\ref{obj:j1853}   & 1RXS J185310    & 2014 & IMi, LCO, Mdy, Ioh & 1RXS J185310.0$+$594509 \\
\ref{obj:j2319}   & 1RXS J231935    & 2014 & Kai, OKU, CRI, Mdy, IMi, DPV & 1RXS J231935.0$+$364705 \\
\ref{obj:j1304}   & 2QZ J130441     & 2014 & HaC, Mdy & 2QZ J130441.7$+$010330 \\
\ref{obj:asassn13cx} & ASASSN-13cx & 2014 & deM, IMi, Han, OkC, Rui, & \\
                     &             &      & RPc, SWI, Mdy, Ham & \\
\ref{obj:asassn14ag} & ASASSN-14ag & 2014 & Mas, Kai, HaC, deM, Kis & \\
\ref{obj:asassn14aj} & ASASSN-14aj & 2014 & HaC & \\
\ref{obj:asassn14au} & ASASSN-14au & 2014 & deM & \\
\ref{obj:asassn14aw} & ASASSN-14aw & 2014 & deM & \\
\ref{obj:asassn14bh} & ASASSN-14bh & 2014 & MLF & \\
\ref{obj:asassn14cl} & ASASSN-14cl & 2014 & OkC, CRI, SWI, AAVSO, IMi, HaC, & \\
                     &             &      & OKU, DPV, RIT, LCO, NDJ, Kai, & \\
                     &             &      & DKS, SPE, CTX, UJH, deM & \\
\ref{obj:asassn14cq} & ASASSN-14cq & 2014 & MLF, HaC & \\
---                  & ASASSN-14cv & 2014 & \Nakataprep & \\ 
\ref{obj:asassn14dm} & ASASSN-14dm & 2014 & MLF & \\
\ref{obj:asassn14do} & ASASSN-14do & 2014 & MLF, HaC & \\
\ref{obj:asassn14dw} & ASASSN-14dw & 2014 & HaC & \\
\ref{obj:asassn14eh} & ASASSN-14eh & 2014 & HaC, Ioh & \\
\ref{obj:asassn14eq} & ASASSN-14eq & 2014 & MLF & \\
\ref{obj:asassn14fr} & ASASSN-14fr & 2014 & MLF & \\
\ref{obj:asassn14gd} & ASASSN-14gd & 2014 & MLF & \\
\ref{obj:asassn14gq} & ASASSN-14gq & 2014 & MLF & \\
\ref{obj:asassn14gx} & ASASSN-14gx & 2014 & MLF, AAVSO, HaC, OKU & \\
\ref{obj:asassn14hk} & ASASSN-14hk & 2014 & Ioh, OKU & \\
\ref{obj:asassn14hl} & ASASSN-14hl & 2014 & MLF & \\
\ref{obj:asassn14hs} & ASASSN-14hs & 2014 & MLF, HaC, Ioh & \\
\hline
\end{tabular}
\end{center}
\end{table*}

\addtocounter{table}{-1}
\begin{table*}
\caption{List of Superoutbursts (continued).}
\begin{center}
\begin{tabular}{ccccl}
\hline
Subsection & Object & Year & Observers or references\commenta & ID\commentb \\
\hline
\ref{obj:asassn14ia} & ASASSN-14ia & 2014 & OKU & \\
\ref{obj:asassn14id} & ASASSN-14id & 2014 & IMi, OKU, KU & \\
\ref{obj:asassn14it} & ASASSN-14it & 2014 & KU & \\
\ref{obj:asassn14iv} & ASASSN-14iv & 2014 & Mdy, HaC & \\
\ref{obj:asassn14je} & ASASSN-14je & 2014 & MLF & \\
\ref{obj:asassn14jf} & ASASSN-14jf & 2014 & HaC & \\
\ref{obj:asassn14jq} & ASASSN-14jq & 2014 & LCO, DPV, KU, Kis, IMi, & \\
                     &             &      & deM, Kai, Ioh & \\
\ref{obj:asassn14jv} & ASASSN-14jv & 2014 & SWI, OUS, Van, DPV, AAVSO, & \\
                     &             &      & CRI, OKU, Dub, Mdy, Buc, & \\
                     &             &      & Aka, deM, Kis, DKS & \\
\ref{obj:asassn14kf} & ASASSN-14kf & 2014 & MLF & \\
\ref{obj:asassn14kk} & ASASSN-14kk & 2014 & MLF & \\
\ref{obj:asassn14ku} & ASASSN-14ku & 2014 & MLF, HaC & \\
\ref{obj:asassn14lk} & ASASSN-14lk & 2014 & MLF & \\
\ref{obj:asassn14mc} & ASASSN-14mc & 2014 & MLF, HaC & \\
\ref{obj:asassn14md} & ASASSN-14md & 2014 & MLF, HaC & \\
\ref{obj:asassn14mh} & ASASSN-14mh & 2014 & HaC, Ioh & \\
\ref{obj:asassn14mj} & ASASSN-14mj & 2014 & Kis, Shu, Ioh, IMi, DPV & \\
\ref{obj:asassn15ah} & ASASSN-15ah & 2015 & MLF, HaC, Ioh & \\
\ref{obj:asassn15ap} & ASASSN-15ap & 2015 & MLF & \\
\ref{obj:asassn15aq} & ASASSN-15aq & 2015 & MLF, HaC & \\
\ref{obj:asassn15aw} & ASASSN-15aw & 2015 & Van & \\
\ref{obj:asassn15bg} & ASASSN-15bg & 2015 & MLF, HaC & \\
\ref{obj:asassn15bp} & ASASSN-15bp & 2015 & Kis, deM, SPE, DKS, Ioh, & \\
                     &             &      & OKU, HaC, AAVSO, DPV, UJH, & \\
                     &             &      & OUS, SWI, IMi, Van, Neu, & \\
                     &             &      & Trt, CRI, Aka & \\   
\ref{obj:asassn15bu} & ASASSN-15bu & 2015 & deM, SWI, SRI, DPV, UJH & \\
\ref{obj:asassn15bv} & ASASSN-15bv & 2015 & MLF & \\
\ref{obj:asassn15dq} & ASASSN-15dq & 2015 & Van, HaC, Kai & \\
\ref{obj:j0819}   & CRTS J081936   & 2013 & Han & CRTS J081936.1$+$191540 \\
\ref{obj:j1720}   & CRTS J172038   & 2014 & Mic & CRTS J172038.7$+$183802 \\
\ref{obj:j2027}   & CRTS J202731   & 2014 & HaC, OKU & CRTS J202731.2$-$224002 \\
\ref{obj:j2147}   & CRTS J214738   & 2014 & Kai, KU, Kis, DPV & CRTS J214738.4$+$244554 \\
\ref{obj:j1740}   & CSS J174033    & 2014 & AAVSO, GFB, DPV, SRI, Mas, & CSS J174033.5$+$414756 \\
                  &                &      & IMi, Nov, RPc, Ioh, RIT & \\
\hline
\end{tabular}
\end{center}
\end{table*}

\addtocounter{table}{-1}
\begin{table*}
\caption{List of Superoutbursts (continued).}
\begin{center}
\begin{tabular}{ccccl}
\hline
Subsection & Object & Year & Observers or references\commenta & ID\commentb \\
\hline
\ref{obj:j0316}   & MASTER J031600  & 2014 & Van & MASTER OT J031600.08$+$175824.4 \\
\ref{obj:j0439}   & MASTER J043915  & 2014 & OKU & MASTER OT J043915.60$+$424232.3 \\
\ref{obj:j0558}   & MASTER J055845  & 2014 & IMi & MASTER OT J055845.55$+$391533.4 \\
\ref{obj:j0858}   & MASTER J085854  & 2015 & SPE, Kis, GBo, MLF & MASTER OT J085854.16$-$274030.7 \\
\ref{obj:j1055}   & MASTER J105545  & 2014 & DPV, IMi & MASTER OT J105545.20$+$573109.7 \\
\ref{obj:j0309}   & OT J030929      & 2014 & DPV, OKU, Mdy, deM, Kis, & PNV J03093063$+$2638031 \\
                  &                 &      & Ioh, Mas, Mic, Siz, Kai, & \\
                  &                 &      & MEV, Van, SPE & \\
---               & OT J060009      & 2014 & \Nakataprep & PNV J06000985$+$1426152 \\
\ref{obj:j0648}   & OT J064833      & 2014 & OKU, HaC, Mdy, DPV, IMi, & PNV J06483343$+$0656236 \\
                  &                 &      & Ioh, Van, Kis, Kai, Aka, NKa & \\
\ref{obj:j2138}   & OT J213806      & 2014 & OKU, DPV, Trt, Aka, Vol, & OT J213806.6$+$261957 \\
                  &                 &      & RPc, DKS, HaC, Kis, AAVSO, & \\
                  &                 &      & MEV, Han, Hsk & \\
\ref{obj:j2305}   & OT J230523      & 2014 & OKU, Kis, Mas & PNV J23052314$-$0225455 \\
---               & PNV J171442     & 2014 & \Nakataprep & PNV J17144255$-$2943481 \\
\ref{obj:j1729}   & PNV J172929     & 2014 & Kis, OKU, Mdy, HaC, Shu, & PNV J17292916$+$0054043 \\
                  &                 &      & Ioh, DPV, Mhh, deM, Aka, Rui & \\
\ref{obj:j0719}   & PTF1 J071912    & 2014 & IMi, DPV & PTF1 J071912.13$+$485834.0 \\
\ref{obj:j0334}   & SDSS J033449    & 2014 & Mdy & SDSS J033449.86$-$071047.8 \\
\ref{obj:j0814}   & SDSS J081408    & 2014 & Mic, IMi, Rui, RPc & SDSS J081408.42$+$090759.1 \\
\ref{obj:j0902}   & SDSS J090221    & 2014 & SWI, deM, DPV, AAVSO, IMi, & SDSS J090221.35$+$381941.9 \\
                  &                 &      & KU, CRI, OKU, LCO, Han, & \\
                  &                 &      & Ioh, Mas, RPc, DKS, Kis, & \\
                  &                 &      & UJH, Mdy, Mic, Shu, Nov & \\
                  &                 &      & \citep{kat14j0902} & \\
\ref{obj:j1202}   & SDSS J120231    & 2014 & MEV, deM, IMi & SDSS J120231.01$+$450349.1 \\
\ref{obj:j1400}   & SDSS J140037    & 2015 & IMi, SWI, DPV, RPc & SDSS J140037.99$+$572341.3 \\
\ref{obj:j1723}   & SDSS J172325    & 2014 & OKU, CRI, RPc, IMi, DPV & SDSS J172325.99$+$330414.1 \\
                  &                 &      & & =ASASSN-14gz \\
\ref{obj:j1730}   & SDSS J173047    & 2014 & deM, SWI, DKS, Nov, RPc, Mhh, DPV, Ham & SDSS J173047.59$+$554518.5 \\
\ref{obj:j1605}   & TCP J160548     & 2014 & DPV, OKU, Van, AAVSO & TCP J16054809$+$2405338 \\
\hline
\end{tabular}
\end{center}
\end{table*}

\begin{table*}
\caption{Superhump Periods and Period Derivatives}\label{tab:perlist}
\begin{center}
\begin{tabular}{c@{\hspace{7pt}}c@{\hspace{7pt}}c@{\hspace{7pt}}c@{\hspace{7pt}}c@{\hspace{7pt}}c@{\hspace{7pt}}c@{\hspace{7pt}}c@{\hspace{7pt}}c@{\hspace{7pt}}c@{\hspace{7pt}}c@{\hspace{7pt}}c@{\hspace{7pt}}c@{\hspace{7pt}}c}
\hline
Object & Year & $P_1$ (d)\commenta & err & \multicolumn{2}{c}{$E_1$\commentb} & $P_{\rm dot}$\commentc & err\commentc & $P_2$ (d)\commenta & err & \multicolumn{2}{c}{$E_2$\commentb} & $P_{\rm orb}$ (d)\commentd & Q\commente \\
\hline
NN Cam & 2014 & -- & -- & -- & -- & -- & -- & 0.073916 & 0.000019 & 0 & 70 & 0.0717 & C \\
V342 Cam & 2014 & -- & -- & -- & -- & -- & -- & 0.078063 & 0.000047 & 0 & 90 & 0.07531 & C \\
SY Cap & 2014 & -- & -- & -- & -- & -- & -- & 0.063414 & 0.000029 & 24 & 89 & -- & C \\
OY Car & 2014 & 0.064595 & 0.000042 & 0 & 140 & 6.9 & 5.5 & -- & -- & -- & -- & 0.063121 & CM \\
OY Car & 2015 & 0.064464 & 0.000063 & 137 & 153 & -- & -- & -- & -- & -- & -- & 0.063121 & C \\
FI Cet & 2014 & 0.056911 & 0.000028 & 18 & 106 & 9.7 & 2.1 & -- & -- & -- & -- & 0.05594 & BE \\
Z Cha & 2014 & 0.077360 & 0.000082 & 25 & 65 & -- & -- & 0.076948 & 0.000023 & 64 & 143 & 0.074499 & B \\
YZ Cnc & 2014b & 0.090707 & 0.000127 & 0 & 21 & -- & -- & -- & -- & -- & -- & 0.0868 & C \\
V337 Cyg & 2014 & 0.070190 & 0.000034 & 0 & 14 & -- & -- & -- & -- & -- & -- & -- & C \\
V503 Cyg & 2014 & 0.081215 & 0.000209 & 0 & 17 & -- & -- & -- & -- & -- & -- & 0.077759 & C \\
BC Dor & 2003 & 0.068048 & 0.000036 & 44 & 145 & -- & -- & -- & -- & -- & -- & -- & C \\
BC Dor & 2015 & 0.068026 & 0.000024 & 53 & 141 & -- & -- & -- & -- & -- & -- & -- & C \\
LY Hya & 2014 & -- & -- & -- & -- & -- & -- & 0.076973 & 0.000025 & -- & -- & 0.0748 & C \\
MM Hya & 2014 & 0.058851 & 0.000030 & 0 & 119 & $-$1.0 & 4.6 & -- & -- & -- & -- & 0.057590 & CG \\
BR Lup & 2014 & 0.082241 & 0.000037 & 23 & 97 & 1.1 & 4.5 & 0.081816 & 0.000090 & 96 & 145 & 0.07948 & B \\
V453 Nor & 2014 & 0.064977 & 0.000045 & 16 & 94 & 16.4 & 3.5 & 0.064590 & 0.000017 & 108 & 232 & 0.063381 & B \\
DT Oct & 2014b & 0.074667 & 0.000043 & 0 & 67 & $-$7.1 & 5.5 & -- & -- & -- & -- & 0.072707 & CG \\
UV Per & 2014 & 0.066750 & 0.000039 & 46 & 112 & 6.1 & 5.3 & 0.066124 & 0.000054 & 135 & 198 & 0.06489 & C \\
HY Psc & 2014 & 0.079942 & 0.000021 & 0 & 74 & -- & -- & -- & -- & -- & -- & 0.0767 & C \\
\hline
  \multicolumn{13}{l}{\commenta $P_1$ and $P_2$ are mean periods of stage B and C superhumps, respectively.} \\
  \multicolumn{13}{l}{\parbox{500pt}{\commentb Interval used for calculating the period (corresponding to $E$ in the individual tables in section \ref{sec:individual}).}} \\
  \multicolumn{13}{l}{\commentc $P_{\rm dot} = \dot{P}/P$ for stage B superhumps, unit $10^{-5}$.} \\
  \multicolumn{13}{l}{\parbox{500pt}{\commentd References: NN Cam (Denisenko, D. 2007, vsnet-alert 9557),
V342 Cam \citep{she11j0423},
Z Cha \citep{dai09zcha},
YZ Cnc \citep{sha88yzcnc},
V503 Cyg \citep{Pdot5},
LY Hya \citep{sti94lyhya},
MM Hya \citep{pat03suumas},
BR Lup (this work based in \cite{men98brlup}, see text),
V453 Nor \citep{ima06asas1600},
DT Oct \citep{Pdot6},
UV Per \citep{tho97uvpervyaqrv1504cyg},
HY Psc Dillon et al. cited in \citet{gan09SDSSCVs},
CC Scl \citep{kat15ccscl},
SU UMa \citep{tho86suuma},
CY UMa \citep{tho96Porb},
QZ Vir \citep{sha84tleo},
CRTS J214738 \citep{Pdot4},
CSS J174033 (\Ohtprep),
OT J213806 \citep{Pdot2},
ASASSN-14cv, PNV J171442 (\Nakataprep),
PTF1 J071912 \citep{lev11j0719},
SDSS J090221 \citep{rau10HeDN},
FI Cet, ASASSN-13cx, ASASSN-14ag, ASASSN-14cl,
ASASSN-14cq, ASASSN-14gx, ASASSN-14id, ASASSN-14jf,
ASASSN-14jv, ASASSN-15bp, OT J030929, OT J230523, PNV J172929 (this work)
}}\\
  \multicolumn{13}{l}{\parbox{500pt}{\commente Data quality and comments. A: excellent, B: partial coverage or slightly low quality, C: insufficient coverage or observations with large scatter, G: $P_{\rm dot}$ denotes global $P_{\rm dot}$, M: observational gap in middle stage, 2: late-stage coverage, the listed period may refer to $P_2$, E: $P_{\rm orb}$ refers to the period of early superhumps, P: $P_{\rm orb}$ refers to a shorter stable periodicity recorded in outburst.}} \\
\end{tabular}
\end{center}
\end{table*}

\addtocounter{table}{-1}
\begin{table*}
\caption{Superhump Periods and Period Derivatives (continued)}
\begin{center}
\begin{tabular}{c@{\hspace{7pt}}c@{\hspace{7pt}}c@{\hspace{7pt}}c@{\hspace{7pt}}c@{\hspace{7pt}}c@{\hspace{7pt}}c@{\hspace{7pt}}c@{\hspace{7pt}}c@{\hspace{7pt}}c@{\hspace{7pt}}c@{\hspace{7pt}}c@{\hspace{7pt}}c@{\hspace{7pt}}c}
\hline
Object & Year & $P_1$ & err & \multicolumn{2}{c}{$E_1$} & $P_{\rm dot}$ & err & $P_2$ & err & \multicolumn{2}{c}{$E_2$} & $P_{\rm orb}$ & Q \\
\hline
CC Scl & 2014 & 0.05998 & 0.00002 & -- & -- & -- & -- & 0.059523 & 0.000006 & -- & -- & 0.058567 & C \\
V418 Ser & 2014 & 0.044669 & 0.000013 & 0 & 83 & 6.1 & 2.6 & 0.044408 & 0.000007 & 106 & 311 & -- & B \\
V701 Tau & 2015 & 0.068838 & 0.000141 & 13 & 72 & -- & -- & -- & -- & -- & -- & -- & C \\
SU UMa & 2014 & 0.079252 & 0.000037 & 0 & 25 & -- & -- & -- & -- & -- & -- & 0.07635 & CG \\
CY UMa & 2014 & 0.072202 & 0.000018 & 5 & 35 & $-$17.8 & 5.1 & 0.072017 & 0.000022 & 43 & 103 & 0.06957 & B \\
QZ Vir & 2014 & 0.060368 & 0.000015 & 20 & 126 & 8.9 & 1.1 & 0.059980 & 0.000086 & 125 & 170 & 0.05882 & B \\
NSV 1436 & 2014 & 0.072843 & 0.000014 & 12 & 84 & 3.7 & 2.0 & 0.072403 & 0.000048 & 109 & 150 & -- & B \\
NSV 4618 & 2015 & -- & -- & -- & -- & -- & -- & 0.067506 & 0.000242 & 0 & 31 & 0.065769 & C \\
1RXS J185310 & 2014 & 0.059521 & 0.000032 & 0 & 65 & -- & -- & -- & -- & -- & -- & -- & C \\
1RXS J231935 & 2014 & 0.066105 & 0.000059 & 0 & 51 & -- & -- & 0.065387 & 0.000059 & 81 & 140 & -- & C \\
2QZ J130441 & 2014 & -- & -- & -- & -- & -- & -- & 0.058069 & 0.000061 & 0 & 72 & -- & C \\
ASASSN-13cx & 2014 & 0.083098 & 0.000042 & 7 & 52 & -- & -- & 0.082647 & 0.000035 & 48 & 133 & 0.079650 & B \\
ASASSN-14ag & 2014 & 0.062059 & 0.000055 & 0 & 38 & -- & -- & -- & -- & -- & -- & 0.060311 & C \\
ASASSN-14aj & 2014 & 0.082028 & 0.000041 & 0 & 90 & $-$3.0 & 5.2 & -- & -- & -- & -- & -- & C2 \\
ASASSN-14au & 2014 & 0.082486 & 0.000084 & 0 & 37 & -- & -- & -- & -- & -- & -- & -- & C \\
ASASSN-14aw & 2014 & 0.097586 & 0.000154 & 0 & 31 & -- & -- & -- & -- & -- & -- & -- & C \\
ASASSN-14bh & 2014 & -- & -- & -- & -- & -- & -- & 0.061368 & 0.000065 & 0 & 35 & -- & C \\
ASASSN-14cl & 2014 & 0.060008 & 0.000013 & 27 & 174 & 8.5 & 0.4 & 0.059738 & 0.000014 & 174 & 249 & 0.05838 & AE \\
ASASSN-14cq & 2014 & 0.057354 & 0.000011 & 34 & 194 & 4.6 & 0.4 & -- & -- & -- & -- & 0.05660 & BE \\
ASASSN-14cv & 2014 & 0.060413 & 0.000007 & 109 & 227 & 0.9 & 0.9 & -- & -- & -- & -- & 0.059917 & AE \\
ASASSN-14dm & 2014 & 0.068335 & 0.000042 & 0 & 76 & -- & -- & 0.068160 & 0.000130 & 73 & 104 & -- & C \\
ASASSN-14do & 2014 & 0.056528 & 0.000032 & 14 & 105 & 4.6 & 3.2 & -- & -- & -- & -- & -- & C \\
ASASSN-14dw & 2014 & 0.075630 & 0.000045 & 0 & 54 & -- & -- & 0.075195 & 0.000029 & 53 & 94 & -- & C \\
ASASSN-14eh & 2014 & 0.062907 & 0.000027 & 0 & 96 & 6.0 & 3.0 & -- & -- & -- & -- & -- & C \\
ASASSN-14eq & 2014 & 0.079467 & 0.000069 & 0 & 52 & $-$40.2 & 10.0 & -- & -- & -- & -- & -- & CG \\
ASASSN-14gd & 2014 & 0.078957 & 0.000054 & 0 & 14 & -- & -- & -- & -- & -- & -- & -- & C \\
ASASSN-14gx & 2014 & 0.056088 & 0.000016 & 18 & 194 & 5.1 & 0.6 & -- & -- & -- & -- & 0.05488 & BE \\
ASASSN-14hk & 2014 & 0.060001 & 0.000019 & 0 & 135 & 2.8 & 2.1 & -- & -- & -- & -- & -- & CG \\
ASASSN-14hs & 2014 & 0.093660 & 0.000059 & 10 & 102 & $-$3.4 & 4.9 & -- & -- & -- & -- & -- & C2 \\
ASASSN-14id & 2014 & 0.079366 & 0.000028 & 0 & 137 & $-$2.2 & 1.7 & -- & -- & -- & -- & 0.076857 & CG \\
ASASSN-14iv & 2014 & 0.069192 & 0.000051 & 0 & 22 & -- & -- & -- & -- & -- & -- & -- & C \\
ASASSN-14je & 2014 & -- & -- & -- & -- & -- & -- & 0.069070 & 0.000054 & 0 & 74 & -- & C \\
\hline
\end{tabular}
\end{center}
\end{table*}

\addtocounter{table}{-1}
\begin{table*}
\caption{Superhump Periods and Period Derivatives (continued)}
\begin{center}
\begin{tabular}{c@{\hspace{7pt}}c@{\hspace{7pt}}c@{\hspace{7pt}}c@{\hspace{7pt}}c@{\hspace{7pt}}c@{\hspace{7pt}}c@{\hspace{7pt}}c@{\hspace{7pt}}c@{\hspace{7pt}}c@{\hspace{7pt}}c@{\hspace{7pt}}c@{\hspace{7pt}}c@{\hspace{7pt}}c}
\hline
Object & Year & $P_1$ & err & \multicolumn{2}{c}{$E_1$} & $P_{\rm dot}$ & err & $P_2$ & err & \multicolumn{2}{c}{$E_2$} & $P_{\rm orb}$ & Q \\
\hline
ASASSN-14jf & 2014 & 0.055949 & 0.000005 & 54 & 341 & 1.1 & 0.2 & -- & -- & -- & -- & 0.05539 & BE \\
ASASSN-14jq & 2014 & 0.055178 & 0.000013 & 0 & 142 & 4.3 & 1.2 & -- & -- & -- & -- & -- & C \\
ASASSN-14jv & 2014 & 0.055102 & 0.000013 & 59 & 210 & 4.9 & 0.7 & -- & -- & -- & -- & 0.05442 & BE \\
ASASSN-14kf & 2014 & 0.072095 & 0.000038 & 13 & 70 & -- & -- & -- & -- & -- & -- & -- & C \\
ASASSN-14kk & 2014 & 0.056361 & 0.000060 & 0 & 89 & -- & -- & -- & -- & -- & -- & -- & C \\
ASASSN-14ku & 2014 & 0.079066 & 0.000079 & 0 & 54 & -- & -- & -- & -- & -- & -- & -- & C2 \\
ASASSN-14lk & 2014 & 0.061432 & 0.000030 & 0 & 130 & $-$2.7 & 3.5 & -- & -- & -- & -- & -- & CG \\
ASASSN-14mc & 2014 & 0.055463 & 0.000017 & 18 & 127 & 1.7 & 2.1 & -- & -- & -- & -- & -- & C \\
ASASSN-14md & 2014 & 0.066878 & 0.000074 & 28 & 74 & -- & -- & -- & -- & -- & -- & -- & C \\
ASASSN-14mh & 2014 & 0.062754 & 0.000018 & 16 & 144 & 1.2 & 1.6 & -- & -- & -- & -- & -- & C \\
ASASSN-14mj & 2014 & 0.060262 & 0.000032 & 18 & 63 & -- & -- & -- & -- & -- & -- & -- & C \\
ASASSN-15ah & 2015 & 0.055469 & 0.000032 & 35 & 145 & 6.2 & 3.2 & -- & -- & -- & -- & -- & C \\
ASASSN-15ap & 2015 & 0.091340 & 0.000042 & 0 & 67 & $-$5.5 & 6.0 & -- & -- & -- & -- & -- & C \\
ASASSN-15aq & 2015 & 0.072297 & 0.000064 & 0 & 80 & 6.6 & 7.3 & 0.071842 & 0.000035 & 80 & 139 & -- & C \\
ASASSN-15aw & 2015 & 0.0615 & 0.0003 & 0 & 3 & -- & -- & -- & -- & -- & -- & -- & C \\
ASASSN-15bg & 2015 & 0.065669 & 0.000071 & 0 & 32 & -- & -- & -- & -- & -- & -- & -- & C \\
ASASSN-15bp & 2015 & 0.056702 & 0.000009 & 35 & 256 & 4.5 & 0.3 & 0.056656 & 0.000013 & 269 & 396 & 0.05563 & AE \\
ASASSN-15bu & 2015 & 0.080049 & 0.000039 & 0 & 80 & $-$8.3 & 4.8 & -- & -- & -- & -- & 0.076819 & CG \\
ASASSN-15dq & 2015 & 0.082062 & 0.000037 & 0 & 47 & 2.3 & 7.4 & -- & -- & -- & -- & -- & C \\
CRTS J214738 & 2014 & 0.096770 & 0.000137 & 41 & 80 & -- & -- & -- & -- & -- & -- & 0.09273 & C \\
CRTS J202731 & 2014 & 0.071499 & 0.000064 & 0 & 78 & -- & -- & 0.071237 & 0.000050 & 78 & 164 & -- & C \\
CSS J174033 & 2014 & 0.045591 & 0.000003 & 0 & 166 & 2.0 & 0.3 & 0.045526 & 0.000008 & 59 & 332 & 0.045048 & A \\
MASTER J043915 & 2014 & 0.062452 & 0.000047 & 0 & 48 & -- & -- & -- & -- & -- & -- & -- & C \\
MASTER J055845 & 2014 & -- & -- & -- & -- & -- & -- & 0.056300 & 0.000400 & 0 & 3 & -- & C \\
MASTER J085854 & 2015 & 0.055560 & 0.000019 & 0 & 124 & 8.1 & 1.0 & -- & -- & -- & -- & -- & B \\
MASTER J105545 & 2014 & 0.066937 & 0.000052 & 0 & 46 & $-$10.8 & 10.3 & -- & -- & -- & -- & -- & CG \\
OT J030929 & 2014 & 0.057437 & 0.000015 & 35 & 199 & 6.8 & 0.5 & 0.057076 & 0.000022 & 190 & 264 & 0.05615 & BE \\
OT J060009 & 2014 & 0.063311 & 0.000007 & 116 & 275 & $-$1.2 & 0.6 & -- & -- & -- & -- & -- & A \\
OT J064833 & 2014 & 0.100326 & 0.000056 & 38 & 75 & 33.4 & 7.5 & -- & -- & -- & -- & -- & B \\
OT J213806 & 2014 & 0.055046 & 0.000011 & 16 & 174 & 6.5 & 0.5 & 0.054905 & 0.000040 & 190 & 228 & 0.054523 & AP \\
OT J230523 & 2014 & 0.055595 & 0.000023 & 34 & 125 & 8.2 & 1.3 & -- & -- & -- & -- & 0.05456 & CE \\
PNV J171442 & 2014 & 0.060092 & 0.000009 & 67 & 188 & 4.4 & 0.7 & -- & -- & -- & -- & 0.059558 & AE \\
\hline
\end{tabular}
\end{center}
\end{table*}

\addtocounter{table}{-1}
\begin{table*}
\caption{Superhump Periods and Period Derivatives (continued)}
\begin{center}
\begin{tabular}{c@{\hspace{7pt}}c@{\hspace{7pt}}c@{\hspace{7pt}}c@{\hspace{7pt}}c@{\hspace{7pt}}c@{\hspace{7pt}}c@{\hspace{7pt}}c@{\hspace{7pt}}c@{\hspace{7pt}}c@{\hspace{7pt}}c@{\hspace{7pt}}c@{\hspace{7pt}}c@{\hspace{7pt}}c}
\hline
Object & Year & $P_1$ & err & \multicolumn{2}{c}{$E_1$} & $P_{\rm dot}$ & err & $P_2$ & err & \multicolumn{2}{c}{$E_2$} & $P_{\rm orb}$ & Q \\
\hline
PNV J172929 & 2014 & 0.060282 & 0.000015 & 32 & 173 & 2.6 & 1.2 & -- & -- & -- & -- & 0.05973 & BE \\
PTF1 J071912 & 2014 & 0.018808 & 0.000011 & 0 & 159 & -- & -- & -- & -- & -- & -- & 0.01859 & C \\
SDSS J081408 & 2014 & 0.100929 & 0.000112 & 0 & 27 & $-$40.8 & 10.9 & -- & -- & -- & -- & -- & CG \\
SDSS J090221 & 2014 & 0.033714 & 0.000005 & 0 & 140 & 1.5 & 0.8 & 0.033593 & 0.000009 & 150 & 246 & 0.03355 & A \\
SDSS J120231 & 2014 & 0.059801 & 0.000042 & 0 & 134 & 9.0 & 1.3 & 0.059577 & 0.000024 & 134 & 202 & -- & B \\
SDSS J140037 & 2015 & 0.063954 & 0.000021 & 0 & 125 & -- & -- & -- & -- & -- & -- & -- & C \\
SDSS J172325 & 2014 & 0.059200 & 0.000021 & 6 & 110 & 5.3 & 2.6 & -- & -- & -- & -- & -- & C \\
SDSS J173047 & 2014 & 0.024609 & 0.000006 & 0 & 166 & $-$0.7 & 0.9 & -- & -- & -- & -- & -- & BG \\
TCP J160548 & 2014 & 0.054989 & 0.000017 & 0 & 316 & 1.6 & 0.2 & -- & -- & -- & -- & -- & C \\
\hline
\end{tabular}
\end{center}
\end{table*}

\section{Individual Objects}\label{sec:individual}

\subsection{KX Aquilae}\label{obj:kxaql}

   For the history of KX Aql, see \citet{Pdot2} and
\citet{Pdot5}.  The 2014 outburst was detected by
C. Chiselbrook at a visual magnitude of 12.8
on December 15 (cf. BAAVSS alert 3897).
The outburst was confirmed to be a superoutburst
by the presence of a superhump and long outburst
duration (vsnet-alert 18083, 18090).\footnote{
   VSNET alert message can be accessed at
   $<$http://ooruri.kusastro.kyoto-u.ac.jp/\\
pipermail/vsnet-alert/$>$.
}
Only one superhump maximum was obtained:
BJD 2457011.2633(15) ($N$=38).

\subsection{NN Camelopardalis}\label{obj:nncam}

   NN Cam = NSV 1485 was recognized as a dwarf nova
by \citet{khr05nsv1485}.  The object was given a GCVS
designation in \citet{NameList79}.\footnote{
  The object is unofficially called ``Samus's star'',
  since the name can be read N. N. Samus, the editor
  in chief of the GCVS, in Russian spelling.
$<$http://www.ka-dar.ru/forum/index.php/topic,1040.15.html$>$.
}
The first superoutburst was
detected in 2007 \citep{Pdot}.  The orbital period
was reported to be 0.0717~d (vsnet-alert 9557).

\begin{figure}
  \begin{center}
    \FigureFile(88mm,70mm){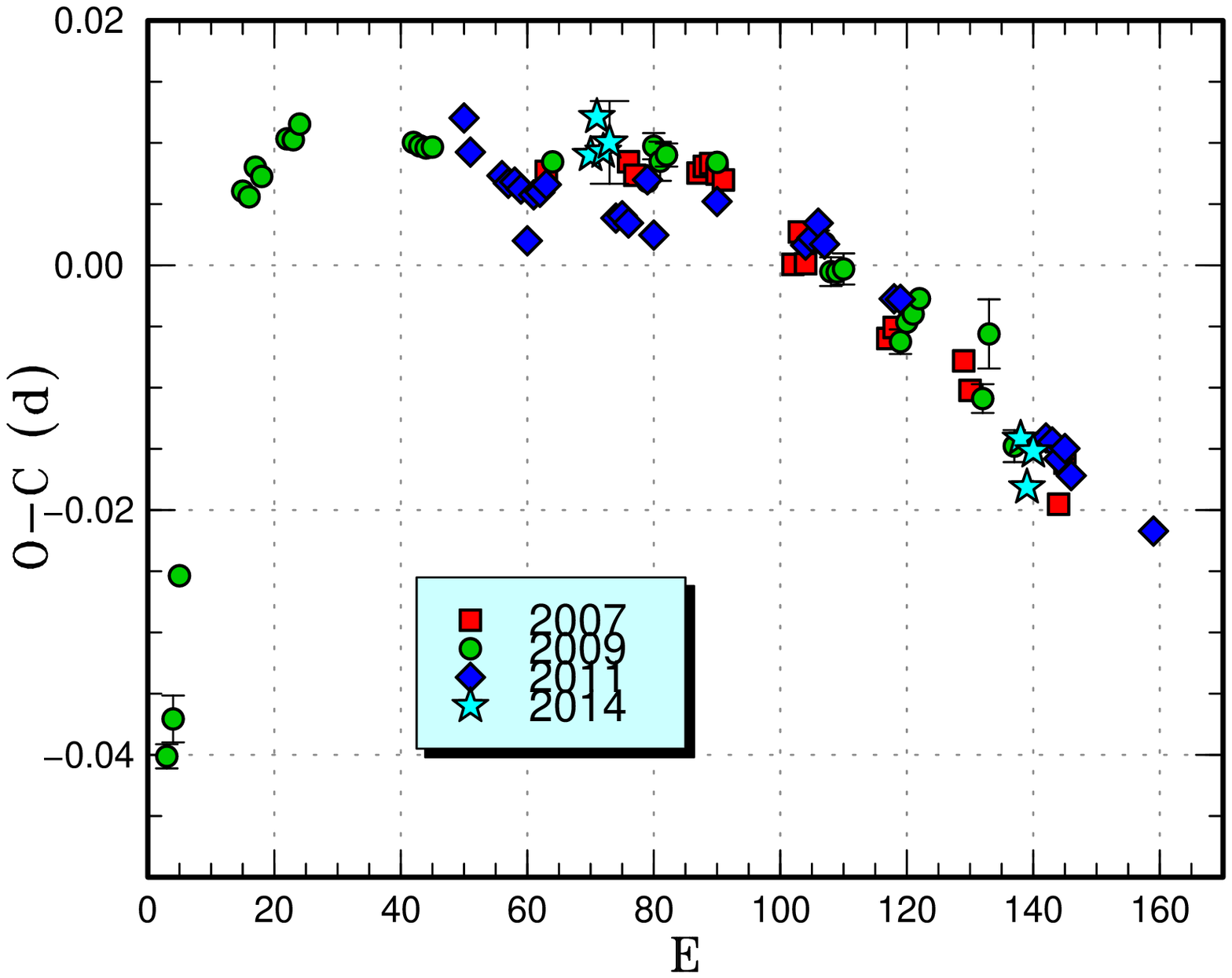}
  \end{center}
  \caption{Comparison of $O-C$ diagrams of NN Cam between different
  superoutbursts.  A period of 0.07430~d was used to draw this figure.
  Approximate cycle counts ($E$) after the start of the superoutburst
  were used.}
  \label{fig:nncamcomp2}
\end{figure}

   The 2014 superoutburst was detected by M. Moriyama
on February 19 (vsnet-alert 16934).
The times of superhump maxima are listed in table
\ref{tab:nncam2014}.  The observation covered
the later stage of the superoutburst.
A comparison of the $O-C$ diagrams (figure \ref{fig:nncamcomp2})
suggests that we mainly observed stage C superhumps.

   A list of recent outbursts is given in table
\ref{tab:nncamout}.  A linear regression to the epochs of
superoutbursts yielded a mean supercycle of 386(6)~d.
The value of supercycle may be halved since this object
is not well observed around solar conjunctions.

\begin{table}
\caption{Superhump maxima of NN Cam (2014)}\label{tab:nncam2014}
\begin{center}
\begin{tabular}{rp{55pt}p{40pt}r@{.}lr}
\hline
\multicolumn{1}{c}{$E$} & \multicolumn{1}{c}{max\commenta} & \multicolumn{1}{c}{error} & \multicolumn{2}{c}{$O-C$\commentb} & \multicolumn{1}{c}{$N$\commentc} \\
\hline
0 & 56713.2745 & 0.0005 & $-$0&0017 & 76 \\
1 & 56713.3519 & 0.0005 & 0&0018 & 75 \\
2 & 56713.4233 & 0.0005 & $-$0&0006 & 76 \\
3 & 56713.4984 & 0.0034 & 0&0005 & 9 \\
68 & 56718.3037 & 0.0005 & 0&0013 & 47 \\
69 & 56718.3740 & 0.0004 & $-$0&0023 & 72 \\
70 & 56718.4512 & 0.0005 & 0&0010 & 58 \\
\hline
  \multicolumn{6}{l}{\commenta BJD$-$2400000.} \\
  \multicolumn{6}{l}{\commentb Against max $= 2456713.2761 + 0.073916 E$.} \\
  \multicolumn{6}{l}{\commentc Number of points used to determine the maximum.} \\
\end{tabular}
\end{center}
\end{table}

\begin{table*}
\caption{List of recent outbursts of NN Cam.}\label{tab:nncamout}
\begin{center}
\begin{tabular}{cccccl}
\hline
Year & Month & max\commenta & magnitude & type & source \\
\hline
2007 & 9 & 54354 & 12.6 & precursor + super & vsnet-alert 9557; \citet{Pdot}; \citet{she11nncam} \\
2008 & 3 & 54535 & 12.5 & ? & AAVSO \\
2008 & 10 & 54758 & 13.2 & super & vsnet-alert 10588; \citet{she11nncam} \\
2009 & 11 & 55137 & 13.0 & super & BAAVSS alert 2130; \citet{Pdot2}; \citet{she11nncam} \\
2011 & 12 & 55905 & 12.4 & super & vsnet-alert 13937; \citet{Pdot4} \\
2012 & 11 & 56241 & 12.7 & super & cvnet-outburst 5039; \citet{Pdot5} \\
2013 & 9 & 56563 & 13.1 & normal & AAVSO \\
2014 & 2 & 56708 & 12.8 & super & this paper \\
\hline
  \multicolumn{5}{l}{\commenta JD$-$2400000.} \\
\end{tabular}
\end{center}
\end{table*}

\subsection{V342 Camelopardalis}\label{obj:v342cam}

\begin{figure}
  \begin{center}
    \FigureFile(88mm,70mm){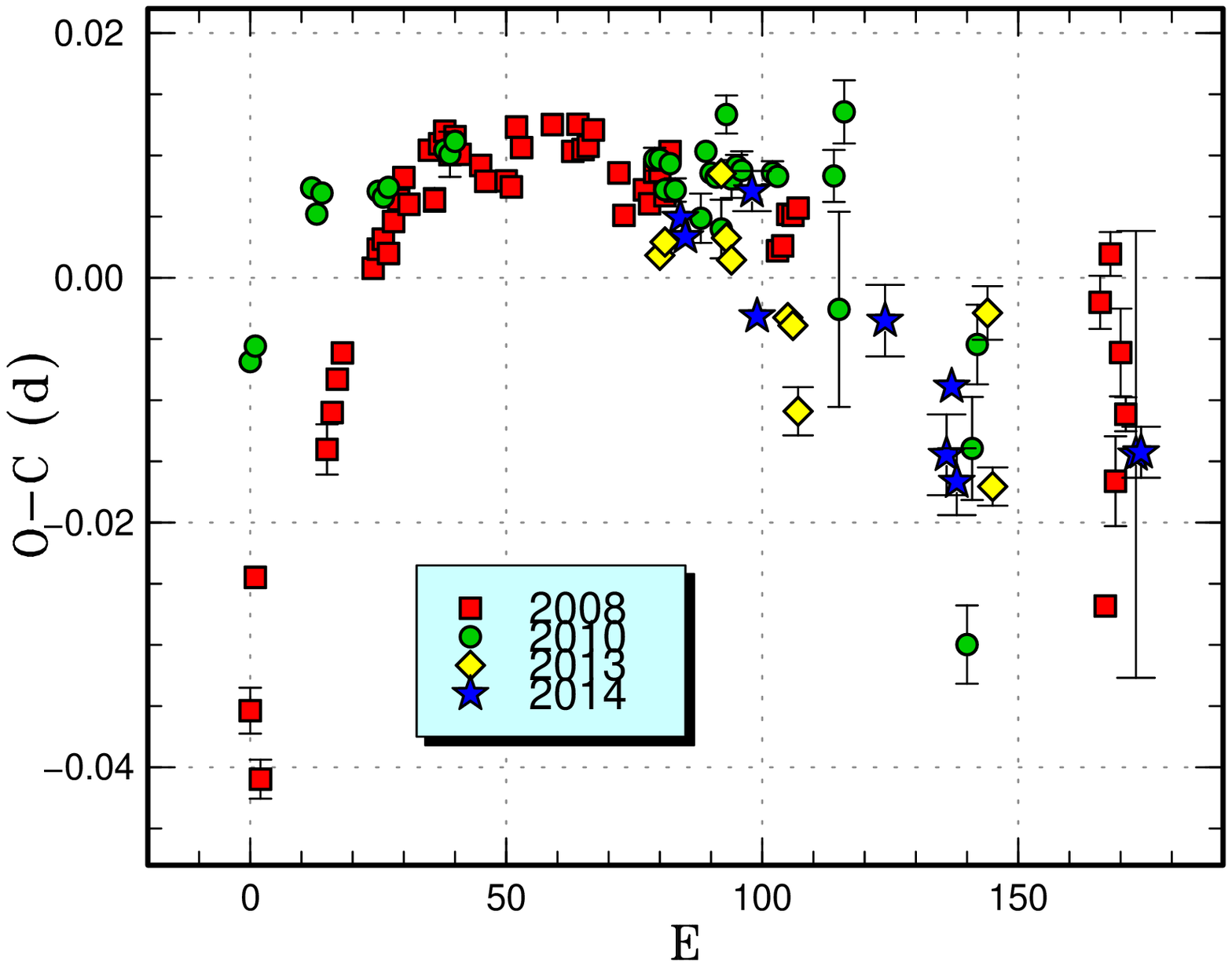}
  \end{center}
  \caption{Comparison of $O-C$ diagrams of V342 Cam between different
  superoutbursts.  A period of 0.07830~d was used to draw this figure.
  Approximate cycle counts ($E$) after the start of the superoutburst
  were used.  Since the start of the 2013 superoutburst
  was not well constrained, we shifted the $O-C$ diagram
  to best fit the others.
  }
  \label{fig:v342camcomp3}
\end{figure}

   For the history of this object (=1RXS J042332$+$745300
=HS 0417$+$7445), see \citet{Pdot6}.
The 2014 superoutburst was detected on December 28
by E. Muyllaert (vsnet-alert 18128).
Time-series observations started on 2015 January 3
and detected superhumps (vsnet-alert 18149,
18157, 18171).
The times of superhump maxima are listed in table
\ref{tab:v342camoc2014}.
Since the observation covered the final part of
the superoutburst, we likely observed only stage C
superhumps, which is also supported by a comparison
of the $O-C$ diagrams (figure \ref{fig:v342camcomp3}).

\begin{table}
\caption{Superhump maxima of V342 Cam (2014)}\label{tab:v342camoc2014}
\begin{center}
\begin{tabular}{rp{55pt}p{40pt}r@{.}lr}
\hline
\multicolumn{1}{c}{$E$} & \multicolumn{1}{c}{max\commenta} & \multicolumn{1}{c}{error} & \multicolumn{2}{c}{$O-C$\commentb} & \multicolumn{1}{c}{$N$\commentc} \\
\hline
0 & 57026.0470 & 0.0003 & 0&0012 & 121 \\
1 & 57026.1237 & 0.0003 & $-$0&0001 & 168 \\
14 & 57027.1454 & 0.0016 & 0&0067 & 85 \\
15 & 57027.2134 & 0.0003 & $-$0&0032 & 162 \\
40 & 57029.1706 & 0.0029 & 0&0023 & 30 \\
52 & 57030.0992 & 0.0033 & $-$0&0058 & 162 \\
53 & 57030.1831 & 0.0009 & 0&0000 & 162 \\
54 & 57030.2536 & 0.0027 & $-$0&0075 & 137 \\
89 & 57032.9963 & 0.0183 & 0&0030 & 68 \\
90 & 57033.0748 & 0.0021 & 0&0034 & 112 \\
\hline
  \multicolumn{6}{l}{\commenta BJD$-$2400000.} \\
  \multicolumn{6}{l}{\commentb Against max $= 2457026.0457 + 0.078063 E$.} \\
  \multicolumn{6}{l}{\commentc Number of points used to determine the maximum.} \\
\end{tabular}
\end{center}
\end{table}

\subsection{SY Capriconi}\label{obj:sycap}

   SY Cap was originally discovered as a long-period
variable star with a photographic range of 12.6 to
fainter than 14.5 (=56.1927, \cite{bel27sycap}).
The variable was independently discovered
at a photographic magnitude of 11 on 1915 September 9
by \citet{ros28sycap} (Ross 328).  \citet{GCVS} listed
this star as a possible Mira-type variable. 
Noting that there is no bright 2MASS counterpart,
T. Kato identified this object as a dwarf nova
(likely SU UMa-type one) using ASAS-3 \citep{ASAS3}
observations (vsnet-alert 10025).
D. Denisenko confirmed this identification (vsnet-alert 10027).

   Soon after this re-classification, an outburst was
detected by ASAS-3 on 2008 August 18 (vsnet-outburst 9336).
The outburst turned out to be a superoutburst
by the detection of superhumps \citep{Pdot}.
\citet{tho12CRTSCVs} made spectroscopic confirmation
as a CV.\footnote{
   Although the object is given CRTS designation in
   \citet{tho12CRTSCVs}, we should note that the object
   was already recognized as a known dwarf nova at the time
   of CRTS detection.
}
Another superoutburst was observed in 2011 August--September
\citep{Pdot4}.

\begin{figure}
  \begin{center}
    \FigureFile(88mm,70mm){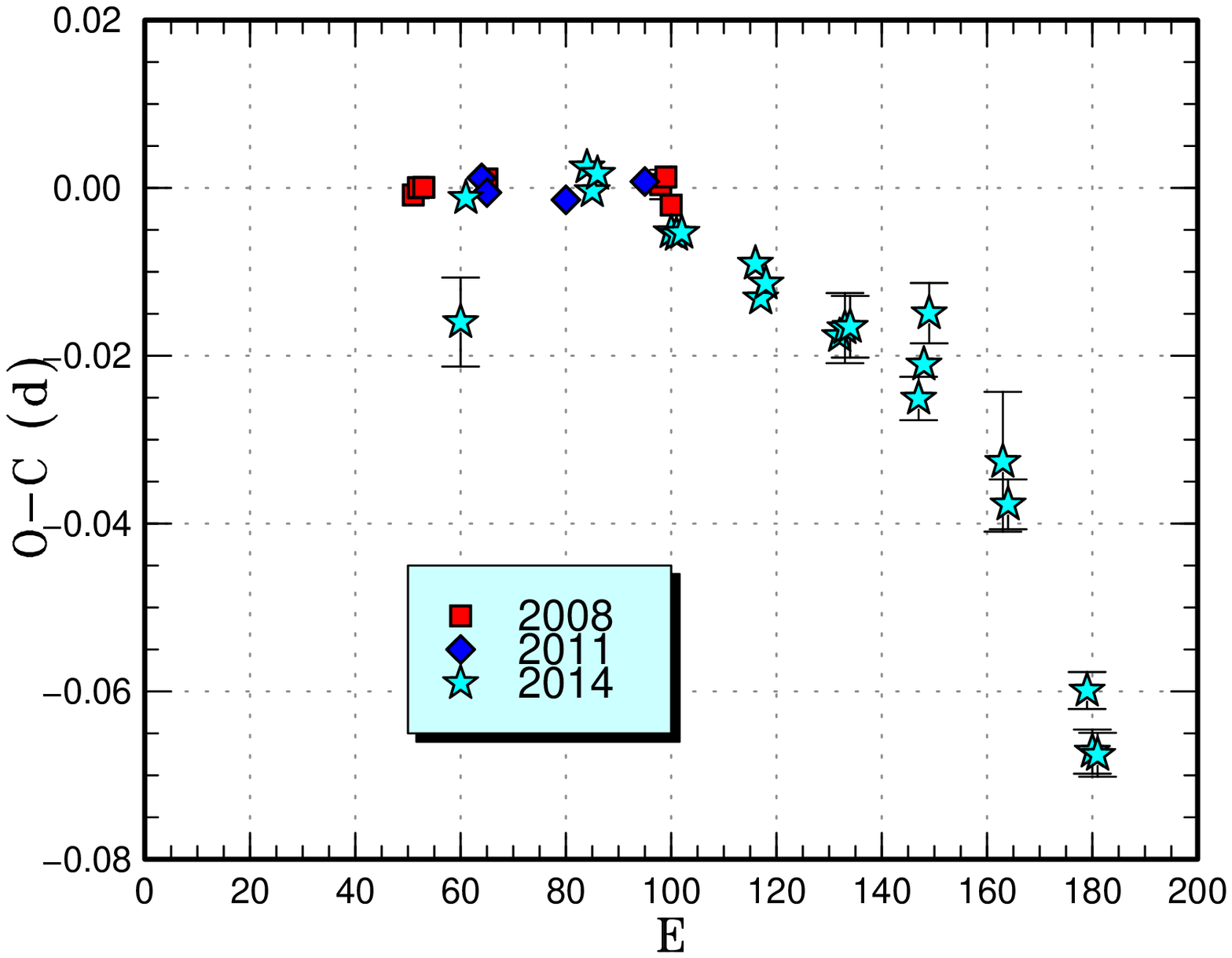}
  \end{center}
  \caption{Comparison of $O-C$ diagrams of SY Cap between different
  superoutbursts.  A period of 0.06376~d was used to draw this figure.
  Approximate cycle counts ($E$) after the start of the superoutburst
  were used.  The 2014 superoutburst was shifted by 60 cycles
  to best match the others.}
  \label{fig:sycapcomp}
\end{figure}

   The 2014 superoutburst was detected by R. Stubbings
at a visual magnitude of 12.8 on September 18
(vsnet-alert 17747).  Subsequent observations detected
superhumps (vsnet-alert 17750, 17764, 17769).
The times of superhump maxima are listed in table
\ref{tab:sycapoc2014}.  We most likely observed
stage C superhump between $E$=24 and $E$=89 and
the value in table \ref{tab:perlist} is based on
this identification (figure \ref{fig:sycapcomp}).
The maxima for $E \ge 119$ were post-superoutburst superhumps.

   Recent outbursts of SY Cap are listed in table
\ref{tab:sycapout}.  The typical cycle length of
normal outbursts is estimated to be $\sim$30~d.
The three shortest intervals between superoutbursts
were 187, 220 and 235~d.  The intervals between
superoutburst clustered around 317--353~d
(six occasions).  The cycle lengths of superoutbursts
(supercycles) may be either bimodally distributed
(187--235~d and 317--353~d) or the recorded longer
supercycles represent two supercycles.  If the latter
is the case, the supercycle may be around 180~d.

\begin{table}
\caption{Superhump maxima of SY Cap (2014)}\label{tab:sycapoc2014}
\begin{center}
\begin{tabular}{rp{55pt}p{40pt}r@{.}lr}
\hline
\multicolumn{1}{c}{$E$} & \multicolumn{1}{c}{max\commenta} & \multicolumn{1}{c}{error} & \multicolumn{2}{c}{$O-C$\commentb} & \multicolumn{1}{c}{$N$\commentc} \\
\hline
0 & 56920.9768 & 0.0053 & $-$0&0268 & 27 \\
1 & 56921.0554 & 0.0005 & $-$0&0116 & 73 \\
24 & 56922.5255 & 0.0007 & 0&0031 & 20 \\
25 & 56922.5864 & 0.0008 & 0&0007 & 18 \\
26 & 56922.6523 & 0.0015 & 0&0033 & 17 \\
40 & 56923.5379 & 0.0012 & 0&0030 & 16 \\
41 & 56923.6014 & 0.0008 & 0&0032 & 18 \\
42 & 56923.6654 & 0.0007 & 0&0039 & 19 \\
56 & 56924.5544 & 0.0009 & 0&0069 & 15 \\
57 & 56924.6140 & 0.0007 & 0&0032 & 16 \\
58 & 56924.6795 & 0.0010 & 0&0055 & 13 \\
72 & 56925.5659 & 0.0021 & 0&0059 & 16 \\
73 & 56925.6306 & 0.0042 & 0&0073 & 16 \\
74 & 56925.6945 & 0.0037 & 0&0080 & 7 \\
87 & 56926.5148 & 0.0026 & 0&0056 & 12 \\
88 & 56926.5827 & 0.0019 & 0&0102 & 16 \\
89 & 56926.6525 & 0.0036 & 0&0168 & 17 \\
103 & 56927.5275 & 0.0083 & 0&0058 & 10 \\
104 & 56927.5861 & 0.0030 & 0&0012 & 16 \\
119 & 56928.5204 & 0.0022 & $-$0&0138 & 14 \\
120 & 56928.5768 & 0.0026 & $-$0&0207 & 16 \\
121 & 56928.6402 & 0.0026 & $-$0&0206 & 16 \\
\hline
  \multicolumn{6}{l}{\commenta BJD$-$2400000.} \\
  \multicolumn{6}{l}{\commentb Against max $= 2456921.0037 + 0.063282 E$.} \\
  \multicolumn{6}{l}{\commentc Number of points used to determine the maximum.} \\
\end{tabular}
\end{center}
\end{table}

\begin{table*}
\caption{List of recent outbursts of SY Cap.}\label{tab:sycapout}
\begin{center}
\begin{tabular}{cccccl}
\hline
Year & Month & max\commenta & magnitude & type & source \\
\hline
2001 & 5 & 52040 & 13.2 & super & ASAS-3 \\
2002 & 4 & 52390 & 13.1 & super & ASAS-3 \\
2002 & 8 & 52502 & 13.7 & normal?\commentb & ASAS-3 \\
2003 & 4 & 52741 & 13.5 & normal?\commentb & ASAS-3 \\
2003 & 4 & 52756 & 14.3 & normal?\commentb & ASAS-3 \\
2003 & 11 & 52945 & 13.6 & super & ASAS-3 \\
2004 & 6 & 53165 & 13.1 & super & ASAS-3 \\
2005 & 4 & 53467 & 13.0 & super & ASAS-3 \\
2005 & 7 & 53568 & 14.1 & normal?\commentb & ASAS-3 \\
2005 & 9 & 53644 & 13.0 & super & ASAS-3 \\
2006 & 6 & 53914 & 14.7 & normal? & ASAS-3 \\
2007 & 4 & 54194 & 13.3 & super?\commentb & ASAS-3 \\
2007 & 9 & 54344 & 13.2 & super & ASAS-3 \\
2008 & 4 & 54560 & 13.6 & ? & ASAS-3 \\
2008 & 8 & 54697 & 13.1 & super & ASAS-3, \citet{Pdot} \\
2008 & 9 & 54725 & 13.9 & normal\commentb & ASAS-3 \\
2008 & 10 & 54760 & 13.6 & normal & ASAS-3, vsnet-outburst 9567 \\
2009 & 4 & 54951 & 13.0 & normal? & vsnet-outburst 10221, ASAS-3 \\
2009 & 7 & 55014 & 12.9 & super & ASAS-3 \\
2009 & 7 & 55038 & 13.4 & normal & vsnet-outburst 10408 \\
2009 & 8 & 55060 & 14.1 & normal & vsnet-outburst 10464 \\
2009 & 10 & 56113 & 13.1 & normal & vsnet-outburst 10611 \\
2010 & 5 & 55340 & 14.1 & normal & vsnet-outburst 11235 \\
2010 & 7 & 55390 & 13.2 & normal? & vsnet-outburst 11379 \\
2010 & 10 & 55497 & 15.4 & normal & vsnet-outburst 11720 \\
2011 & 6 & 55742 & 13.1 & normal & vsnet-outburst 12988 \\
2011 & 8 & 55776 & 13.4 & normal & cvnet-outburst 4252 \\
2011 & 8 & 55799 & 13.0 & super & \citet{Pdot4} \\
2011 & 9 & 55827 & 14.1 & normal & vsnet-outburst 13235 \\
2011 & 10 & 55861 & 15.1 & normal & vsnet-outburst 13363 \\
2011 & 11 & 55890 & 15.0 & normal & vsnet-outburst 13478 \\
2012 & 7 & 56127 & 13.0 & super & vsnet-outburst 14481 \\
2013 & 5 & 56434 & 13.2 & normal? & vsnet-outburst 15436 \\
\hline
  \multicolumn{5}{l}{\commenta JD$-$2400000.} \\
  \multicolumn{5}{l}{\commentb Single detection.} \\
\end{tabular}
\end{center}
\end{table*}

\addtocounter{table}{-1}
\begin{table*}
\caption{List of recent outbursts of SY Cap (continued).}
\begin{center}
\begin{tabular}{cccccl}
\hline
Year & Month & max\commenta & magnitude & type & source \\
\hline
2013 & 7 & 56500 & 13.2 & normal? & vsnet-outburst 15721 \\
2013 & 8 & 56532 & 13.6 & normal & vsnet-alert 16296 \\
2013 & 9 & 56560 & 13.0 & normal & vsnet-outburst 16027 \\
2013 & 10 & 56584 & 13.1 & super & vsnet-alert 16550 \\
2014 & 4 & 56753 & 13.0 & normal? & vsnet-outburst 16749 \\
2014 & 6 & 56817 & 13.4 & normal? & vsnet-outburst 16970 \\
2014 & 9 & 56819 & 12.8 & super & this paper \\
2014 & 10 & 56847 & 13.2 & normal & this paper \\
2014 & 11 & 56982 & 13.5 & normal & this paper \\
\hline
  \multicolumn{5}{l}{\commenta JD$-$2400000.} \\
  \multicolumn{5}{l}{\commentb Single detection.} \\
\end{tabular}
\end{center}
\end{table*}

\subsection{OY Carinae}\label{obj:oycar}

   OY Car was discovered as a dwarf nova by \citet{hof63VSS61}.
The object has been monitored by amateur astronomers
(notably the Variable Star Section of the Royal Astronomical
Society of New Zealand, VSS RASNZ) since 1963 and many outbursts were
recorded.  \citet{war76CVmultipleperiod} listed the object
as an SU UMa-type dwarf nova with a supercycle
(called as super-period then) of $\sim$300~d based on
observations by the VSS RASNZ.
The object received special attention after the discovery of
the eclipsing nature (\cite{vog79ektraiauc}; \cite{vog81oycar}).
The object has been intensively studied since then
(e.g. \cite{rit80oycar}; \cite{bai81oycar}; \cite{she82oycarIR};
\cite{sch83oycar}; \cite{vog83oycar}).  
Superhumps were first detected during the 1980 outburst
\citep{krz85oycarsuper}.  Despite that the object
is one of the best known SU UMa-type dwarf novae, no
systematic study of superhumps has been reported.
Since \citet{Pdot}, it has been established that objects with
short superhumps periods almost always show positive
$P_{\rm dot}$.  Although \citet{Pdot} suggested a positive
$P_{\rm dot}$ for the 1980 data by \citet{krz85oycarsuper},
a historical analysis by \citet{pat93vyaqr} suggested
a negative $P_{\rm dot}$.  New observations of OY Car
have therefore been desired.

   A Markov-Chain Monte Carlo (MCMC) analysis \citep{Pdot4}
of the eclipse observations of in the
AAVSO database (2009--2015) yielded the following
orbital ephemeris:
\begin{equation}
{\rm Min(BJD)} = 2456502.09846(1) + 0.0631209050(6) E .
\label{equ:oycarecl}
\end{equation}
The epoch of this ephemeris corresponds to the mean epoch
of the observations.  The decrease from the period in
\citet{gre06oycar} has been confirmed.  The rate of secular
decrease was about a half ($-0.7 \times 10^{-13}$) of that
reported in \citet{gre06oycar}.  This value agrees with the one
reported by \citet{han15oycar} within a factor of two.
The value corresponds to the time-scale of
$P_{\rm orb}/\dot{P}_{\rm orb}$ of $\sim 2\times 10^9$ yr.
This times scale is similar to what is expected
for angular momentum loss only by the gravitational
wave radiation
[$1.1 \times 10^9$ yr, from equations 9.5b and 9.20 in
\citet{war95book} assuming conservative mass-transfer;
the binary parameters are from RKcat 7.21 \citep{RKCat}]
One should be careful in interpreting secular period
variations in CVs, however, since the same method
is known to give much different times scales of variation,
the well-known case being Z Cha \citep{coo81zchapdot}.
We used this modern ephemeris in analyzing the data.

   We observed two superoutbursts in 2014 and 2015.
Although the initiation of the 2014 superoutburst was
not recorded, the time-resolved observations started
within two days of the start of the superoutburst.
This superoutburst was followed by a post-superoutburst
rebrightening.  The 2015 superoutburst started with
a precursor outburst (vsnet-alert 18313, 18325) and
the early stage of the superoutburst was observed.
This superoutburst was also followed by a post-superoutburst
rebrightening (vsnet-alert 18408).
The times of superhump maxima are listed in tables
\ref{tab:oycaroc2014} and \ref{tab:oycaroc2015}, respectively.
Although the 2015 observations recorded stage A superhumps,
the times of superhump maxima in this stage were not
well determined because they happened to be close
to eclipses.  The cycle count between $E=18$ and $E=137$
for the 2015 data is uncertain.  Note that the superhump period
for the 2014 superoutburst listed in table \ref{tab:perlist}
only refers to the initial part of stage B and it is
expected to be significantly shorter than the mean period of
stage B superhumps.

\begin{figure}
  \begin{center}
    \FigureFile(88mm,70mm){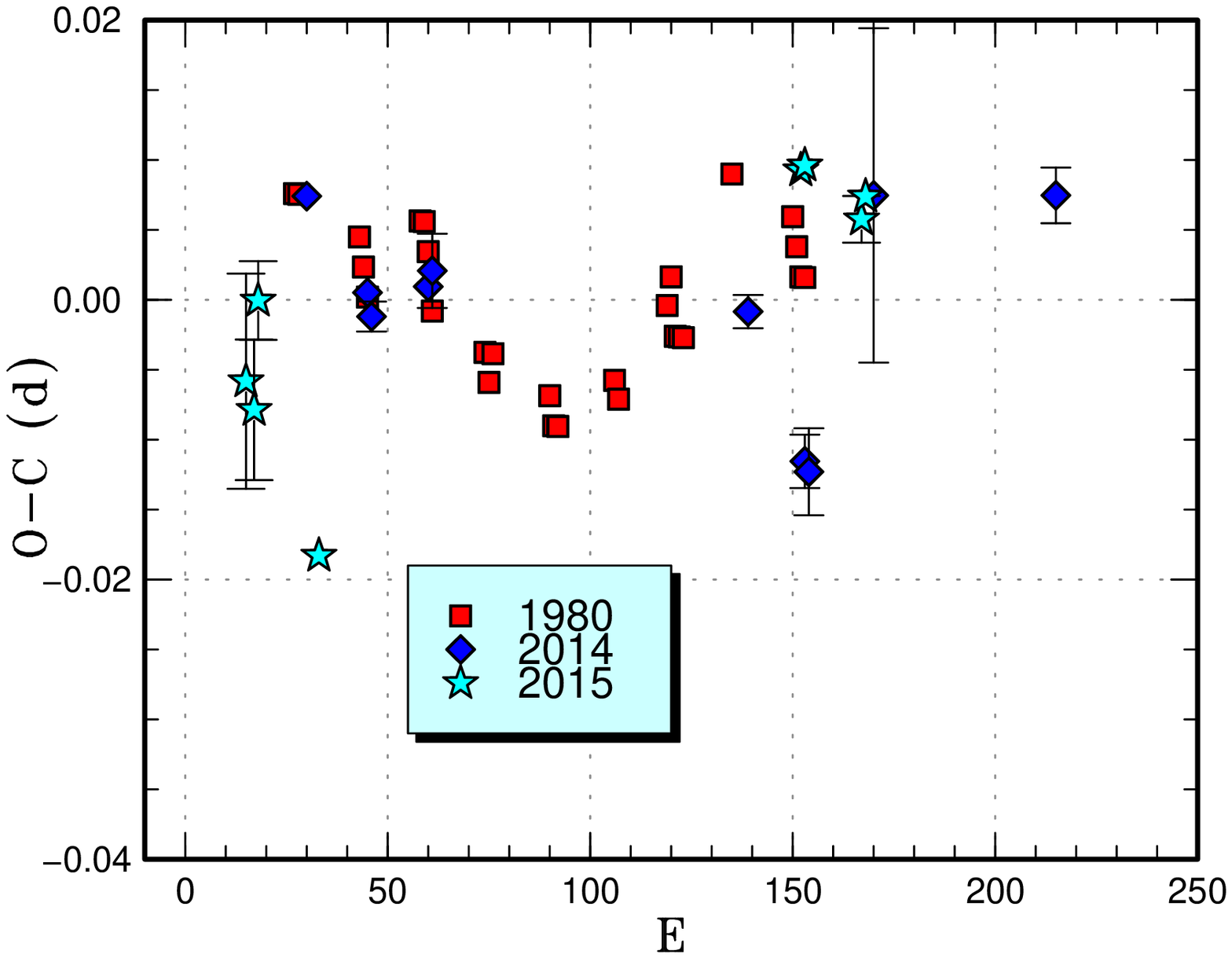}
  \end{center}
  \caption{Comparison of $O-C$ diagrams of OY Car between different
  superoutbursts.  A period of 0.06465~d was used to draw this figure.
  Approximate cycle counts ($E$) after the starts of outbursts
  were used.  The 2015 superoutburst with a separate precursor outburst
  was shifted by 15 cycles to best match the others.}
  \label{fig:oycarcomp}
\end{figure}

   As shown in the comparison of $O-C$ diagrams
(figure \ref{fig:oycarcomp}), the present observations
were limited in coverage than the 1980 one and the result
was not conclusive.  We hope we have a better luck next
time to record the full evolution of superhumps.

\begin{table}
\caption{Superhump maxima of OY Car (2014)}\label{tab:oycaroc2014}
\begin{center}
\begin{tabular}{rp{50pt}p{30pt}r@{.}lcr}
\hline
$E$ & max\commenta & error & \multicolumn{2}{c}{$O-C$\commentb} & phase\commentc & $N$\commentd \\
\hline
0 & 56818.0448 & 0.0009 & 0&0064 & 0.41 & 46 \\
15 & 56819.0076 & 0.0008 & $-$0&0003 & 0.67 & 37 \\
16 & 56819.0706 & 0.0011 & $-$0&0020 & 0.67 & 36 \\
30 & 56819.9778 & 0.0010 & 0&0003 & 0.04 & 36 \\
31 & 56820.0436 & 0.0027 & 0&0014 & 0.08 & 34 \\
109 & 56825.0834 & 0.0012 & $-$0&0004 & 0.92 & 34 \\
123 & 56825.9778 & 0.0019 & $-$0&0109 & 0.09 & 50 \\
124 & 56826.0417 & 0.0031 & $-$0&0117 & 0.11 & 54 \\
140 & 56827.0958 & 0.0120 & 0&0083 & 0.81 & 28 \\
185 & 56830.0051 & 0.0020 & 0&0089 & 0.90 & 28 \\
\hline
  \multicolumn{7}{l}{\commenta BJD$-$2400000.} \\
  \multicolumn{7}{l}{\commentb Against max $= 2456818.0384 + 0.064636 E$.} \\
  \multicolumn{7}{l}{\commentc Orbital phase.} \\
  \multicolumn{7}{l}{\commentd Number of points used to determine the maximum.} \\
\end{tabular}
\end{center}
\end{table}

\begin{table}
\caption{Superhump maxima of OY Car (2015)}\label{tab:oycaroc2015}
\begin{center}
\begin{tabular}{rp{50pt}p{30pt}r@{.}lcr}
\hline
$E$ & max\commenta & error & \multicolumn{2}{c}{$O-C$\commentb} & phase\commentc & $N$\commentd \\
\hline
0 & 57067.0848 & 0.0077 & 0&0024 & 0.86 & 53 \\
2 & 57067.2121 & 0.0050 & 0&0001 & 0.88 & 53 \\
3 & 57067.2846 & 0.0028 & 0&0078 & 0.02 & 24 \\
18 & 57068.2361 & 0.0009 & $-$0&0121 & 0.10 & 43 \\
137 & 57075.9570 & 0.0010 & 0&0026 & 0.42 & 51 \\
138 & 57076.0220 & 0.0013 & 0&0028 & 0.45 & 63 \\
152 & 57076.9232 & 0.0017 & $-$0&0026 & 0.73 & 23 \\
153 & 57076.9895 & 0.0006 & $-$0&0010 & 0.78 & 45 \\
\hline
  \multicolumn{7}{l}{\commenta BJD$-$2400000.} \\
  \multicolumn{7}{l}{\commentb Against max $= 2457067.0825 + 0.064759 E$.} \\
  \multicolumn{7}{l}{\commentc Orbital phase.} \\
  \multicolumn{7}{l}{\commentd Number of points used to determine the maximum.} \\
\end{tabular}
\end{center}
\end{table}

\subsection{FI Ceti}\label{obj:ficet}

   FI Cet was originally discovered as a transient
ROTSE3 J015118.59$-$022300.1 on 2001 October 13 by ROTSE-IIIa
telescope.  The object was detected at 14.71 mag and
faded by two magnitudes over 13 days.  The object was
originally suspected to be a nova based on the large
amplitude, but it was also suspected to be either
a recurrent nova or a WZ Sge-type dwarf nova \citep{smi02ficet}.
The object was named FI Cet in \citet{NameList78}, which
classified it as a possible dwarf nova.

   Although amateur observers recognized that the object is
most likely a larger-amplitude dwarf nova and started monitoring
since 2002, no secure outburst had been detected until
the detection on 2014 June 27 by the ASAS-SN team
\citep{dav15ASASSNCVAAS} at $V$=14.4 (vsnet-alert 17423).

\begin{figure}
  \begin{center}
    \FigureFile(88mm,110mm){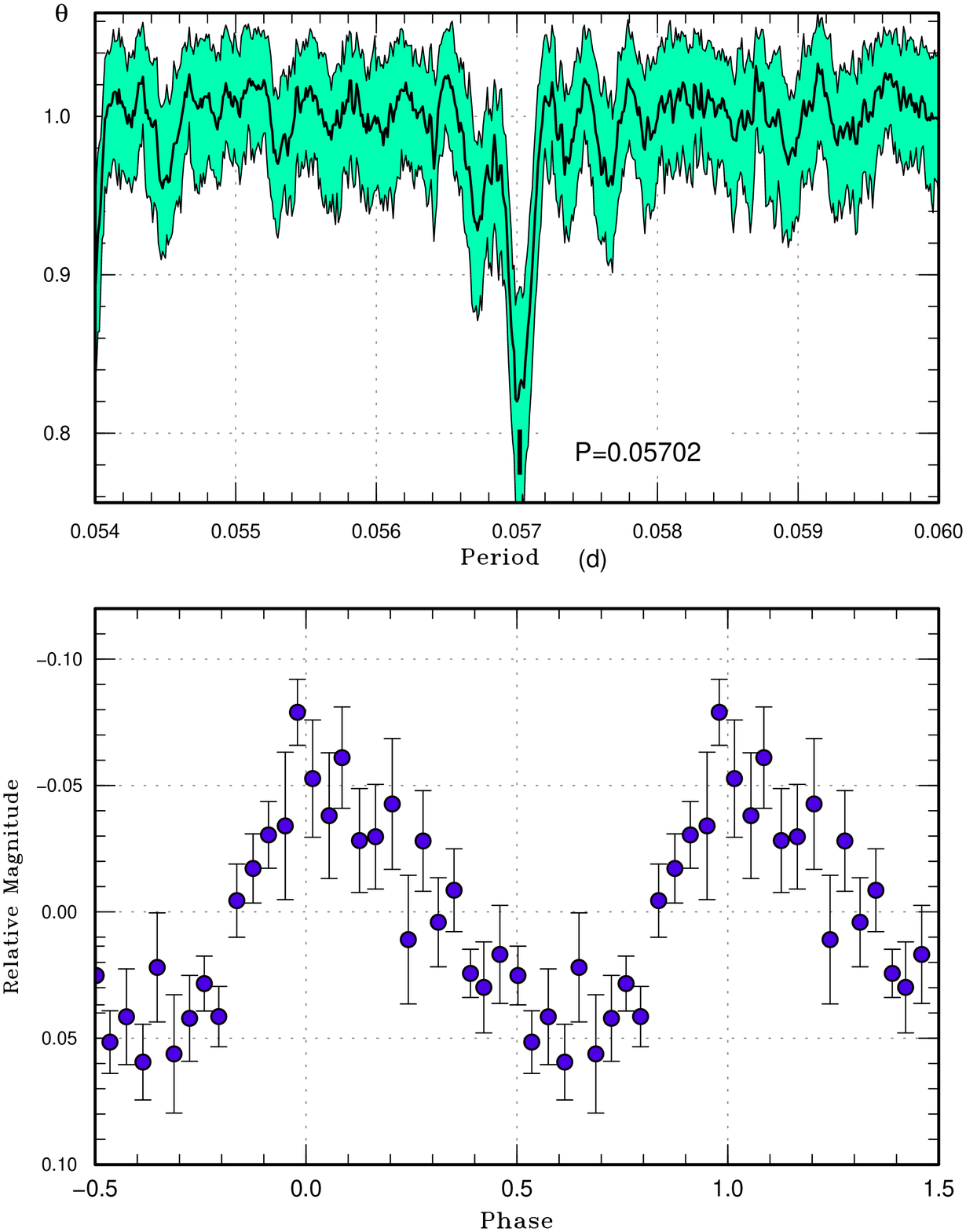}
  \end{center}
  \caption{Superhumps in FI Cet (2014).  (Upper): PDM analysis.
     (Lower): Phase-averaged profile.}
  \label{fig:ficetshpdm}
\end{figure}

\begin{figure}
  \begin{center}
    \FigureFile(88mm,110mm){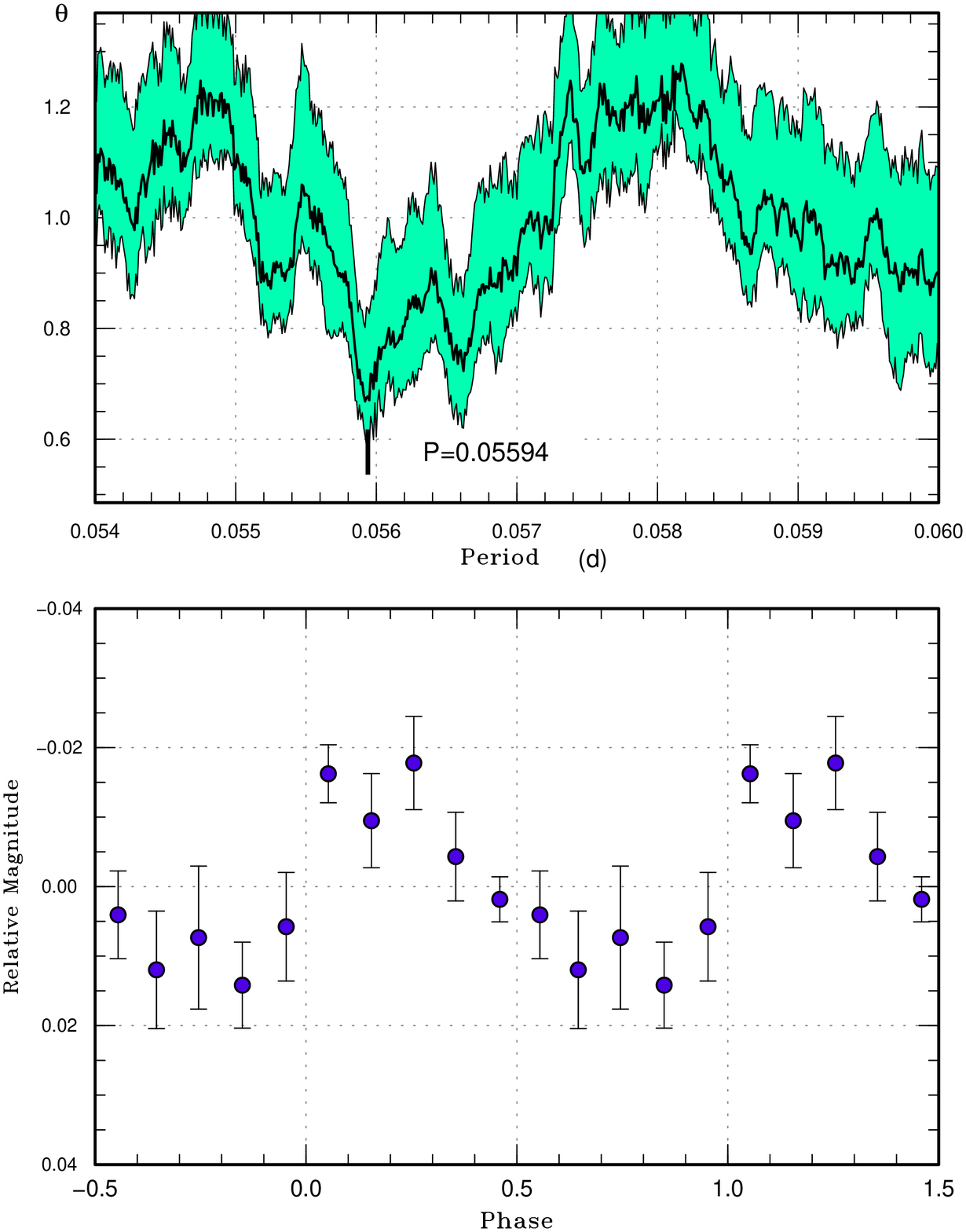}
  \end{center}
  \caption{Possible early superhumps in FI Cet (2014).  (Upper): PDM analysis.
     (Lower): Phase-averaged profile.}
  \label{fig:ficeteshpdm}
\end{figure}

   Subsequent observations initially did not detect superhumps
(vsnet-alert 17428, 17436).  The object started showing
superhumps on July 2 and the amplitudes further grew
over two days (vsnet-alert 17440, 17453; figure \ref{fig:ficetshpdm}).
The times of superhump maxima are listed in table
\ref{tab:ficetoc2014}.  The period identification is
based on the $O-C$ analysis on individual nights.
The adopted period among the candidates only give acceptably
small $O-C$ variations within the same nights.
A stage A-B transition and stage B
with a positive $P_{\rm dot}$ can be well recognized. 
Due to the faintness of the object,
the times of superhump maxima could not be measured in the
later part of the superoutburst.  The light curve,
however, showed a brightening trend around July 13.
Since such a trend in the light curve is usually associated
with development of stage C superhumps, superhumps after
this epoch were likely stage C superhumps.
Although we were not able to measure the period of stage A
superhumps, an analysis of the first two nights yielded
a period of 0.05594(3)~d, which we tentatively identified
to be the period of early superhumps (figure \ref{fig:ficeteshpdm}).
Note that this identification of the period is not
conclusive and awaits determination of the orbital period
in quiescence.
All the pieces of evidence support the WZ Sge-type 
classification as originally proposed.  The large positive
$P_{\rm dot}$ suggests that this object is not an extreme
WZ Sge-type dwarf nova.

\begin{table}
\caption{Superhump maxima of FI Cet (2014)}\label{tab:ficetoc2014}
\begin{center}
\begin{tabular}{rp{55pt}p{40pt}r@{.}lr}
\hline
\multicolumn{1}{c}{$E$} & \multicolumn{1}{c}{max\commenta} & \multicolumn{1}{c}{error} & \multicolumn{2}{c}{$O-C$\commentb} & \multicolumn{1}{c}{$N$\commentc} \\
\hline
0 & 56840.8397 & 0.0023 & $-$0&0018 & 15 \\
1 & 56840.8939 & 0.0022 & $-$0&0046 & 19 \\
18 & 56841.8724 & 0.0005 & 0&0054 & 19 \\
35 & 56842.8382 & 0.0010 & 0&0027 & 16 \\
36 & 56842.8945 & 0.0004 & 0&0020 & 19 \\
71 & 56844.8847 & 0.0013 & $-$0&0017 & 19 \\
88 & 56845.8534 & 0.0010 & $-$0&0015 & 17 \\
89 & 56845.9094 & 0.0011 & $-$0&0025 & 17 \\
106 & 56846.8825 & 0.0020 & 0&0021 & 19 \\
\hline
  \multicolumn{6}{l}{\commenta BJD$-$2400000.} \\
  \multicolumn{6}{l}{\commentb Against max $= 2456840.8415 + 0.056971 E$.} \\
  \multicolumn{6}{l}{\commentc Number of points used to determine the maximum.} \\
\end{tabular}
\end{center}
\end{table}

\subsection{Z Chameleontis}\label{obj:zcha}

\subsubsection{Superhumps during the 2014 superoutburst}

   Z Cha is one of the best known SU UMa-type dwarf novae
since the early history of research of SU UMa-type
dwarf novae (e.g. \cite{war74zcha}; \cite{bai79zcha};
\cite{vog80suumastars}).
Its deep eclipses has provided us wealth
of information about the structure of the accretion disk
and its variation over the course of outburst and superoutburst
(e.g. \cite{vog82zcha}; \cite{hor84superhump}).
Despite its importance, the object was mostly observed
in the era of photoelectric photometry
and no publicly available data for superoutbursts
are published.  This has been an obstacle to compare
the classical knowledge in SU UMa-type dwarf novae
with the one with the modern CCD observations.
In order to improve the situation, we undertook
a campaign in 2013--2014 to cover a full supercycle.
Since such a long-term campaign requires enormous
effort, we did not attempt to record high time-resolution
observations to resolve eclipses, instead we focused
on longer-term (orbital modulations and superhumps)
variations.  The entire data are now publicly available
in the AAVSO database.

\begin{figure}
  \begin{center}
    \FigureFile(88mm,110mm){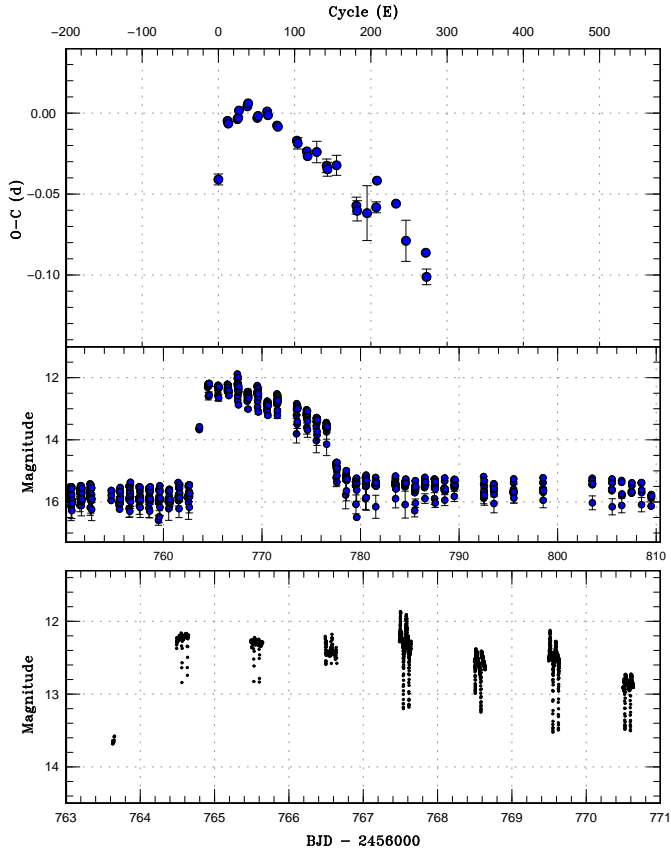}
  \end{center}
  \caption{$O-C$ diagram of superhumps in Z Cha (2014).
     (Upper): $O-C$.
     We used a period of 0.07736~d for calculating the $O-C$ residuals.
     (Middle): Light curve.  The data were binned to 0.0077~d.
     (Lower): Enlarged light curve of showing the precursor and
     evolution of superhumps.  After the full growth of superhumps,
     strong beat phenomena were present.
  }
  \label{fig:zchahumpall}
\end{figure}

   As in the rest of this paper, we first deal with
the superoutburst which occurred in 2014 April.
The rise to the outburst was detected on April 15,
and the object stayed the precursor part for three days,
during which superhumps evolved (see bottom panel of figure
\ref{fig:zchahumpall}).  On April 20, fully
grown superhumps were recorded associated with
an increase of the brightness by $\sim$0.2 mag.
The object entered the post-superoutburst stage
on April 30.  The general behavior is in good agreement
with the modern knowledge: a superoutburst is
triggered by a normal outburst, which appears as
a precursor, followed by development of superhumps
(cf. \cite{osa13v1504cygKepler}).

   The times of superhump maxima were determined during
the superoutburst after subtracting the mean orbital
variation (mostly eclipses) and template fitting
as in other systems.  This simple method assumes 
the constancy of the orbital light curve, which is
obviously wrong because the orbital and superhump
variations interact each other to produce a beat
phenomenon.  We used, however, this method since 
there is no other suitable method to determine times of
superhumps.  Readers should be careful in interpreting
the resultant values (they should contain systematic
errors dur to this simple treatment other than the
nominal errors given in the tables) in such
a high-inclination system.

   The result clearly indicates the presence of stage A
($E \le 14$) and stage B to C transition around $E=64$.
The $O-C$ diagram for stage B superhumps is not as smooth
as other non-eclipsing objects due to the strong beat
phenomenon.  The measured periods of stage B and C
superhumps are in good agreement with the previous
values from re-identification of the published
epochs in the literature \citep{Pdot} and analysis of
recent limited data \citep{Pdot2}.

   The three growing superhumps ($E \le 14$) during 
the precursor phase (stage A superhumps) unfortunately
happened around the phase of eclipses, and the times
of maxima could not be determined around their peaks.
The times of these superhump maxima should therefore
contain considerable errors.  Although the period of
stage A superhumps can be measured as 0.08017(3)~d,
this value should be treated with caution.
The fractional superhump excess corresponds to
$q=0.22(1)$.  We hope observations in the future,
when times of stage A superhumps do not overlap
with eclipses, could provide a more reliable $q$ value
directly comparable to the one from eclipse observations.

   The superhumps persisted in the post-superoutburst phase
(table \ref{tab:zchaoc2014post}), whose times were also
determined after subtracting the mean orbital modulation.
The period was similar to that of stage C superhumps.
This superhump signal could be traced to BJD 2456790
with the PDM method, but became unclear after this.

\begin{figure}
  \begin{center}
    \FigureFile(88mm,70mm){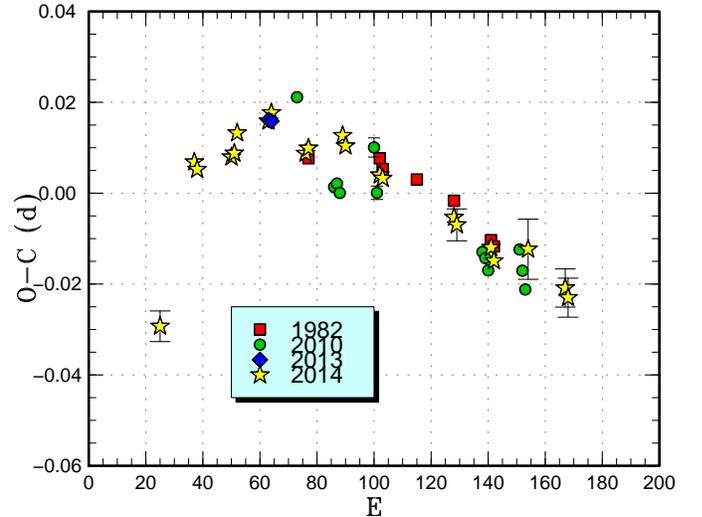}
  \end{center}
  \caption{Comparison of $O-C$ diagrams of Z Cha between different
  superoutbursts.  A period of 0.07736~d was used to draw this figure.
  Approximate cycle counts ($E$) after the starts of outbursts
  were used.  The 1982 superoutburst had a separate precursor outburst
  and the start of the outburst was defined as the start
  of the main outburst.
  The 1982 superoutburst was shifted by 50 cycles
  to best match the others.}
  \label{fig:zchacomp}
\end{figure}

   A comparison of the $O-C$ diagrams between different
superoutbursts is shown in figure \ref{fig:zchacomp}.
The 1982 superoutburst had a separate precursor outburst
and we needed to shift 50 cycles for this superoutburst
to make a match with the others.  This indicates that
superhumps started to grow 50 cycles (about 4~d) before
the start of the main outburst.  This value implies
that superhumps started to develop 1~d after the
precursor outburst, since it took five days before
the main outburst occurs.

\begin{table}
\caption{Superhump maxima of Z Cha (2014)}\label{tab:zchaoc2014}
\begin{center}
\begin{tabular}{rp{50pt}p{30pt}r@{.}lcr}
\hline
$E$ & max\commenta & error & \multicolumn{2}{c}{$O-C$\commentb} & phase\commentc & $N$\commentd \\
\hline
0 & 56765.5347 & 0.0034 & $-$0&0410 & 0.10 & 26 \\
12 & 56766.4992 & 0.0018 & $-$0&0027 & 0.04 & 28 \\
13 & 56766.5749 & 0.0016 & $-$0&0042 & 0.06 & 33 \\
25 & 56767.5060 & 0.0004 & 0&0006 & 0.56 & 91 \\
26 & 56767.5842 & 0.0005 & 0&0016 & 0.61 & 147 \\
27 & 56767.6660 & 0.0016 & 0&0062 & 0.71 & 38 \\
38 & 56768.5195 & 0.0010 & 0&0107 & 0.16 & 133 \\
39 & 56768.5987 & 0.0008 & 0&0128 & 0.23 & 147 \\
51 & 56769.5181 & 0.0006 & 0&0059 & 0.57 & 122 \\
52 & 56769.5966 & 0.0008 & 0&0072 & 0.62 & 147 \\
64 & 56770.5277 & 0.0008 & 0&0121 & 0.12 & 148 \\
65 & 56770.6028 & 0.0006 & 0&0099 & 0.13 & 140 \\
77 & 56771.5247 & 0.0007 & 0&0056 & 0.50 & 142 \\
78 & 56771.6014 & 0.0008 & 0&0050 & 0.53 & 148 \\
103 & 56773.5268 & 0.0017 & 0&0008 & 0.38 & 41 \\
104 & 56773.6025 & 0.0035 & $-$0&0007 & 0.39 & 33 \\
116 & 56774.5258 & 0.0013 & $-$0&0036 & 0.79 & 40 \\
117 & 56774.6002 & 0.0018 & $-$0&0064 & 0.78 & 32 \\
129 & 56775.5311 & 0.0066 & $-$0&0018 & 0.28 & 34 \\
142 & 56776.5283 & 0.0042 & $-$0&0080 & 0.66 & 33 \\
143 & 56776.6035 & 0.0043 & $-$0&0100 & 0.67 & 21 \\
\hline
  \multicolumn{7}{l}{\commenta BJD$-$2400000.} \\
  \multicolumn{7}{l}{\commentb Against max $= 2456765.5757 + 0.077187 E$.} \\
  \multicolumn{7}{l}{\commentc Orbital phase.} \\
  \multicolumn{7}{l}{\commentd Number of points used to determine the maximum.} \\
\end{tabular}
\end{center}
\end{table}

\begin{table}
\caption{Superhump maxima of Z Cha (2014) (post-superoutburst)}\label{tab:zchaoc2014post}
\begin{center}
\begin{tabular}{rp{50pt}p{30pt}r@{.}lcr}
\hline
$E$ & max\commenta & error & \multicolumn{2}{c}{$O-C$\commentb} & phase\commentc & $N$\commentd \\
\hline
0 & 56777.5343 & 0.0062 & 0&0047 & 0.17 & 32 \\
26 & 56779.5207 & 0.0052 & $-$0&0088 & 0.83 & 30 \\
27 & 56779.5949 & 0.0062 & $-$0&0115 & 0.83 & 24 \\
40 & 56780.5991 & 0.0169 & $-$0&0073 & 0.31 & 17 \\
52 & 56781.5310 & 0.0034 & 0&0016 & 0.82 & 28 \\
53 & 56781.6249 & 0.0011 & 0&0185 & 0.08 & 9 \\
78 & 56783.5447 & 0.0025 & 0&0152 & 0.85 & 28 \\
91 & 56784.5274 & 0.0127 & $-$0&0020 & 0.04 & 28 \\
117 & 56786.5314 & 0.0018 & 0&0020 & 0.94 & 80 \\
118 & 56786.5938 & 0.0048 & $-$0&0125 & 0.77 & 27 \\
\hline
  \multicolumn{7}{l}{\commenta BJD$-$2400000.} \\
  \multicolumn{7}{l}{\commentb Against max $= 2456777.5295 + 0.076922 E$.} \\
  \multicolumn{7}{l}{\commentc Orbital phase.} \\
  \multicolumn{7}{l}{\commentd Number of points used to determine the maximum.} \\
\end{tabular}
\end{center}
\end{table}

\subsubsection{Search for negative superhumps}

   Our entire observation covered one complete supercycle
consisting of the 2013 August superoutburst (BJD 2456524), 
three normal outbursts (BJD 2456572, 2456624, 2456676)
and the 2014 April superoutburst (BJD 2456764).

   We searched for possible negative superhumps.  After
dividing the quiescent data into 10--20~d bins
(depending on the gaps in observations and
the goodness of coverage), we subtracted the mean
orbital variation and applied the PDM analysis.
We detected possible signal of negative superhumps
in the following three bins: BJD 2456610--2456623
(before the second normal outburst), period 0.07342(4)~d
and amplitude 0.05 mag; BJD 2456651--2456662 (before
the third normal outburst), period of 0.07352(8)~d
and amplitude 0.05 mag; BJD 2456663--2456675 (after
the third normal outburst), period of 0.07314(4)~d
and amplitude 0.04 mag.  Although these possible
detections were close to the detection limit and
may have been affected by (probably variable)
strong orbital modulations and uneven sampling of
observations, these negative superhumps may have
indeed transiently existed.

   It may be noteworthy that the interval of the third 
normal outburst and the second superoutburst was relatively
long (88~d).  In VW Hyi, there exists ``L'' (long intervals
of normal outbursts) and ``S'' (short intervals)
type supercycles \citep{sma85vwhyi}, and it has been
recently established by Kepler observations of
that L-type and S-type supercycles are related to
negative superhumps, which are considered to be
a result of the tilted disk \citep{osa13v1504cygKepler}.
A similar suggestion was reported in VW Hyi \citep{Pdot6}.
The relatively long interval of outbursts after
the third normal outburst in Z Cha might be also
related to the existence of transient negative superhumps.

\subsubsection{Secular brightness variation}

   \citet{vaname87vwhyi} and \citet{vaname90zcha} studied
secular variation of the brightness of VW Hyi and Z Cha,
respectively.  Although \citet{vaname90zcha} could not
yield convincing results, our much improved and homogeneous
data are expected to provide a better clue to understanding
the quiescent disk.  As in \citet{vaname87vwhyi}, we used
phase 0.20 segment to represent the mean magnitude.
Since Z Cha has an maximum of the quiescent orbital humps
at around the orbital phase 0.86, we used phase 0.36$\pm$0.10
for this purpose.

\begin{figure}
  \begin{center}
    \FigureFile(88mm,70mm){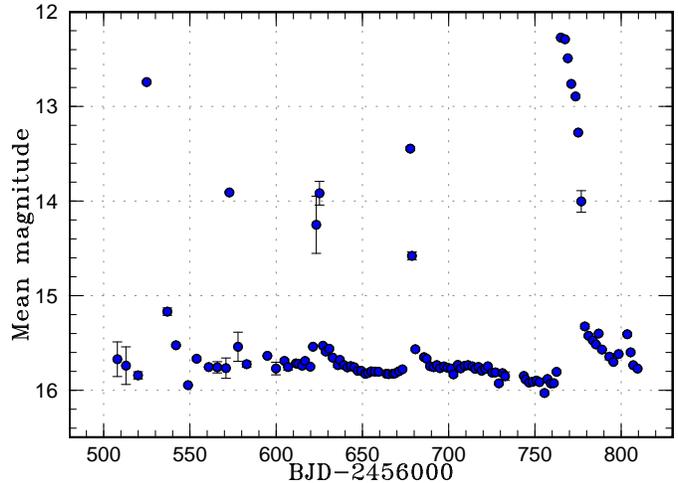}
  \end{center}
  \caption{Brightness variation of Z Cha outside orbital humps
  (orbital phase 0.26--0.46).  Each dot represent 2-d average.}
  \label{fig:zchamean}
\end{figure}

   The result clearly shows that the mean
magnitude is variable in quiescence (figure \ref{fig:zchamean}).
The mean magnitude in quiescence clearly monotonically faded
between the third normal outburst and the second superoutburst.
The situation is somewhat different in the quiescence
between the second and third normal outburst, and there
was a secular fading trend followed by a rising trend
before the third normal outburst.  It appears that
the mean magnitude in quiescence shows a secular fading
trend in a supercycle.  The mean magnitude after the
second superoutburst is much brighter than before
the superoutburst, but this trend was not so evident
in the first superoutburst.  In any case, an explanation
for such a trend is a future task for the disk instability model.

\begin{figure}
  \begin{center}
    \FigureFile(88mm,70mm){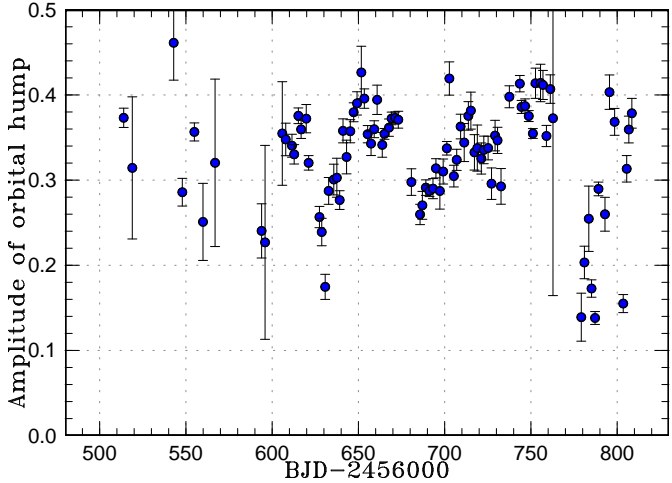}
  \end{center}
  \caption{Amplitude of orbital humps in Z Cha in quiescence
  (difference between orbital phase 0.76--0.96 and
  orbital phase 0.26--0.46).  Each dot represent 2-d average.}
  \label{fig:zchahump}
\end{figure}

   The variation of the orbital humps in quiescence was
also examined (figure \ref{fig:zchahump}).  The amplitude
is defined here between magnitudes of orbital phase 0.76--0.96
and orbital phase 0.26--0.46.  The amplitude of orbital humps
increases during quiescence between normal outbursts.
This most likely reflect the increased release of
the gravitational potential energy as the disk shrinks
in quiescence.  There was no indication of enhanced
orbital humps before any outburst, which severely constrain
the mass-transfer instability model.

\subsection{YZ Cancri}\label{obj:yzcnc}

\begin{figure}
  \begin{center}
    \FigureFile(88mm,70mm){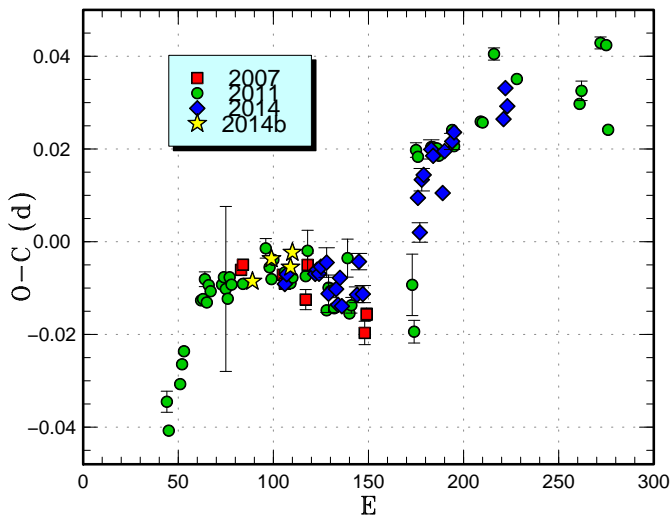}
  \end{center}
  \caption{Comparison of $O-C$ diagrams of YZ Cnc between different
  superoutbursts.  A period of 0.09050~d was used to draw this figure.
  Approximate cycle counts ($E$) after the start of the superoutburst
  were used (in the case of YZ Cnc,
  this refers to the precursor outburst).
  Since the start of the 2014 superoutburst
  was not well constrained, we shifted the $O-C$ diagram
  to best fit the others.  The figure was drawn based on
  an assumption that the preceding outburst before
  the 2014b superoutburst was a precursor outburst.
  The superhumps apparently appeared immediately
  following this precursor outburst.}
  \label{fig:yzcnccomp2}
\end{figure}

   YZ Cnc is a well-known active SU UMa-type dwarf nova
(e.g. \cite{szk84AAVSO}).  See \citet{Pdot6} for more
history.  In 2014, another superoutburst in
November--December was observed.  The times of superhump
maxima are listed in table \ref{tab:yzcncoc2014b}.
According to the AAVSO observations, there was
an outburst (BJD 2456977.7, November 16)
5~d before the start of the superoutburst.
Since the object did not reach the ordinary quiescent
level after this outburst, it was likely that this
outburst was a separate precursor outburst.
Although the superhump observations started only
$\sim$24 cycles after the final rise to the superoutburst,
there was no hint of stage A superhumps.
It was most likely that stage A superhumps appeared
following the separate precursor outburst.
This suggestion was supported by a comparison of
$O-C$ diagrams (figure \ref{fig:yzcnccomp2}).

\begin{table}
\caption{Superhump maxima of YZ Cnc (2014b)}\label{tab:yzcncoc2014b}
\begin{center}
\begin{tabular}{rp{55pt}p{40pt}r@{.}lr}
\hline
\multicolumn{1}{c}{$E$} & \multicolumn{1}{c}{max\commenta} & \multicolumn{1}{c}{error} & \multicolumn{2}{c}{$O-C$\commentb} & \multicolumn{1}{c}{$N$\commentc} \\
\hline
0 & 56985.9582 & 0.0002 & $-$0&0009 & 72 \\
10 & 56986.8680 & 0.0005 & 0&0019 & 169 \\
20 & 56987.7712 & 0.0003 & $-$0&0020 & 184 \\
21 & 56987.8649 & 0.0004 & 0&0010 & 179 \\
\hline
  \multicolumn{6}{l}{\commenta BJD$-$2400000.} \\
  \multicolumn{6}{l}{\commentb Against max $= 2456985.9591 + 0.090707 E$.} \\
  \multicolumn{6}{l}{\commentc Number of points used to determine the maximum.} \\
\end{tabular}
\end{center}
\end{table}

\subsection{V337 Cygni}\label{obj:v337cyg}

   V337 Cyg was discovered as a long-period variable
(AN 101.1928) with a range of 14.2 to fainter than 16.4
\citep{baa28VS}.  \citet{gut33VSnamelist} suggested
that this object is likely a dwarf nova.
The object had been long lost since then, and
both \citet{BruchCVatlas} and \citet{DownesCVatlas1}
could not identify the object.  In 1996, J. Manek
identified three outbursts of this object in the
Sonneberg archive (vsnet 775).  J. Manek further supplied
the dates of six outbursts (vsnet 782).
In 2006 May, a new outburst was detected and subsequent
observations confirmed the SU UMa-type nature of
this object (\cite{boy07v337cyg}; \cite{Pdot}).
\citet{Pdot2} further studied the 2010 superoutburst.

   The 2014 superoutburst was detected by J. Shears
on June 29 (BAAVSS alert 3728; vsnet-alert 17427).
Time-series observations were undertaken on two
nights and the three times of superhump maxima
were measured (table \ref{tab:v337cygoc2014}).
A PDM analysis yielded a period of 0.07019(3)~d.
We probably observed stage B superhumps as judged
from a comparison with the 2007 and 2010 data.

\begin{table}
\caption{Superhump maxima of V337 Cyg (2014)}\label{tab:v337cygoc2014}
\begin{center}
\begin{tabular}{rp{55pt}p{40pt}r@{.}lr}
\hline
\multicolumn{1}{c}{$E$} & \multicolumn{1}{c}{max\commenta} & \multicolumn{1}{c}{error} & \multicolumn{2}{c}{$O-C$\commentb} & \multicolumn{1}{c}{$N$\commentc} \\
\hline
0 & 56840.5108 & 0.0003 & $-$0&0003 & 75 \\
1 & 56840.5817 & 0.0006 & 0&0003 & 43 \\
14 & 56841.4932 & 0.0005 & $-$0&0000 & 67 \\
\hline
  \multicolumn{6}{l}{\commenta BJD$-$2400000.} \\
  \multicolumn{6}{l}{\commentb Against max $= 2456840.5112 + 0.070150 E$.} \\
  \multicolumn{6}{l}{\commentc Number of points used to determine the maximum.} \\
\end{tabular}
\end{center}
\end{table}

\subsection{V503 Cygni}\label{obj:v503cyg}

   For this famous SU UMa-type dwarf nova, see \citet{Pdot6}
for a description.  The 2014 July superoutburst was
observed.  On the initial two nights (BJD 2456864--2456865),
typical superhumps were recorded.  After BJD 2456867,
the superhumps became double-humped.  Since we could not
distinguish which maxima correspond to the smooth continuation
of the earlier superhump maxima, we listed both hump maxima
in table \ref{tab:v503cygoc2014}.  The value given in table
\ref{tab:perlist} refers to the initial part only.
Similar variation in the superhump profile can be
found in \citet{har95v503cyg}.

\begin{table}
\caption{Superhump maxima of V503 Cyg (2014)}\label{tab:v503cygoc2014}
\begin{center}
\begin{tabular}{rp{55pt}p{40pt}r@{.}lr}
\hline
\multicolumn{1}{c}{$E$} & \multicolumn{1}{c}{max\commenta} & \multicolumn{1}{c}{error} & \multicolumn{2}{c}{$O-C$\commentb} & \multicolumn{1}{c}{$N$\commentc} \\
\hline
0 & 56864.0992 & 0.0006 & 0&0031 & 133 \\
1 & 56864.1776 & 0.0007 & 0&0000 & 142 \\
12 & 56865.0712 & 0.0010 & $-$0&0020 & 140 \\
13 & 56865.1530 & 0.0008 & $-$0&0017 & 142 \\
14 & 56865.2391 & 0.0018 & 0&0030 & 45 \\
16 & 56865.3917 & 0.0012 & $-$0&0072 & 64 \\
17 & 56865.4815 & 0.0007 & 0&0011 & 81 \\
40 & 56867.3698 & 0.0017 & 0&0166 & 79 \\
41 & 56867.4216 & 0.0009 & $-$0&0129 & 74 \\
65 & 56869.4034 & 0.0015 & 0&0146 & 74 \\
66 & 56869.4532 & 0.0029 & $-$0&0170 & 69 \\
66 & 56869.4928 & 0.0018 & 0&0226 & 72 \\
67 & 56869.5314 & 0.0036 & $-$0&0202 & 65 \\
\hline
  \multicolumn{6}{l}{\commenta BJD$-$2400000.} \\
  \multicolumn{6}{l}{\commentb Against max $= 2456864.0961 + 0.081425 E$.} \\
  \multicolumn{6}{l}{\commentc Number of points used to determine the maximum.} \\
\end{tabular}
\end{center}
\end{table}

   Another superoutburst occurred in 2014 October
(vsnet-alert 17827; 2014b outburst in table \ref{tab:outobs}).
Time-series observations were
obtained only on two nights and two superhumps maxima
were obtained: BJD 2456942.4363(7) ($N$=68)
and 2456952.4321(10) ($N$=90).

\subsection{BC Doradus}\label{obj:bcdor}

   This object was originally designated as CAL 86
in the direction of the Large Megellanic Cloud (LMC).
This star was originally selected as an Einstein X-ray source.
\citet{cow84lmcXBID} provided an optical identification.
\citet{sch02cal86} reported the detection of a short
(0.066~d) orbital period and at least five outbursts in
the MACHO observations.  Some of these outbursts reached
$V$=14.  \citet{kat04nsv10934mmscoabnorcal86} confirmed
the SU UMa-type nature by the detection of
superhumps during the 2003 outburst.

\begin{figure}
  \begin{center}
    \FigureFile(88mm,70mm){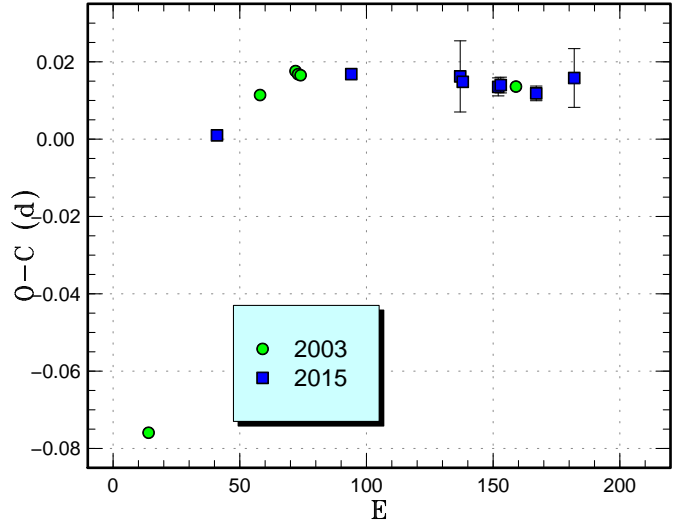}
  \end{center}
  \caption{Comparison of $O-C$ diagrams of BC Dor between different
  superoutbursts.  A period of 0.06806~d was used to draw this figure.
  Approximate cycle counts ($E$) after the maximum of the superoutburst
  were used.
  }
  \label{fig:bcdorcomp}
\end{figure}

   The 2015 superoutburst was visually detected by
R. Stubbings on February 20 (vsnet-alert 18315).
The outburst was also detected by ASAS-SN
(vsnet-alert 18316).
There was a gap in the observation in the early part
of the superoutburst.
The times of superhump maxima are listed in table
\ref{tab:bcdoroc2015}.  After a comparison with the 2003
result (figure \ref{fig:bcdorcomp}), we noticed that
the stage identifications for the 2003 superhumps were incorrect.
We provide an updated result in table \ref{tab:perlist}.

\begin{table}
\caption{Superhump maxima of BC Dor (2015)}\label{tab:bcdoroc2015}
\begin{center}
\begin{tabular}{rp{55pt}p{40pt}r@{.}lr}
\hline
\multicolumn{1}{c}{$E$} & \multicolumn{1}{c}{max\commenta} & \multicolumn{1}{c}{error} & \multicolumn{2}{c}{$O-C$\commentb} & \multicolumn{1}{c}{$N$\commentc} \\
\hline
0 & 57076.9685 & 0.0014 & $-$0&0049 & 24 \\
53 & 57080.5915 & 0.0012 & 0&0068 & 16 \\
96 & 57083.5175 & 0.0092 & 0&0029 & 10 \\
97 & 57083.5841 & 0.0012 & 0&0014 & 17 \\
111 & 57084.5356 & 0.0023 & $-$0&0009 & 16 \\
112 & 57084.6042 & 0.0020 & $-$0&0006 & 12 \\
126 & 57085.5549 & 0.0019 & $-$0&0038 & 17 \\
141 & 57086.5797 & 0.0076 & $-$0&0010 & 17 \\
\hline
  \multicolumn{6}{l}{\commenta BJD$-$2400000.} \\
  \multicolumn{6}{l}{\commentb Against max $= 2457076.9734 + 0.068137 E$.} \\
  \multicolumn{6}{l}{\commentc Number of points used to determine the maximum.} \\
\end{tabular}
\end{center}
\end{table}

\subsection{V660 Herculis}\label{obj:v660her}

   For the explanation of the history of this object, see
\citet{Pdot5}.  The 2014 superoutburst was detected by
M. Moriyama at an unfiltered CCD magnitude of 14.4
on September 13 (vsnet-alert 17724).  Only one superhump
maximum was obtained: BJD 2456915.0538(8) ($N$=97).

\subsection{CT Hydrae}\label{obj:cthya}

   For history of CT Hya, refer to \citet{Pdot6}.
The 2015 superoutburst was visually detected by R. Stubbings
(vsnet-alert 18285).  The outburst was apparently
detected during the rising phase by P. Starr one
night before Stubbings' detection.
The times of superhump maxima
are listed in table \ref{tab:cthyaoc2015}.
A comparison of $O-C$ diagrams indicates that we observed
stage B and C superhumps, although observations were not
sufficient to determine individual periods.

\begin{figure}
  \begin{center}
    \FigureFile(88mm,70mm){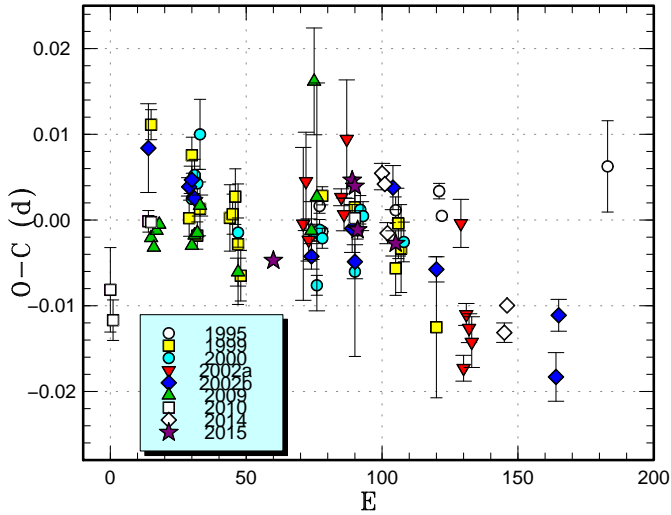}
  \end{center}
  \caption{Comparison of $O-C$ diagrams of CT Hya between different
  superoutbursts.  A period of 0.06650~d was used to draw this figure.
  Approximate cycle counts ($E$) after the maximum of the superoutburst
  were used.  Since the start of the 2014 superoutburst
  was not well constrained, we shifted the $O-C$ diagram
  to best fit the others.
  }
  \label{fig:cthyacomp4}
\end{figure}

\begin{table}
\caption{Superhump maxima of CT Hya (2015)}\label{tab:cthyaoc2015}
\begin{center}
\begin{tabular}{rp{55pt}p{40pt}r@{.}lr}
\hline
\multicolumn{1}{c}{$E$} & \multicolumn{1}{c}{max\commenta} & \multicolumn{1}{c}{error} & \multicolumn{2}{c}{$O-C$\commentb} & \multicolumn{1}{c}{$N$\commentc} \\
\hline
0 & 57067.0221 & 0.0010 & $-$0&0024 & 51 \\
29 & 57068.9599 & 0.0005 & 0&0045 & 65 \\
30 & 57069.0257 & 0.0011 & 0&0037 & 208 \\
31 & 57069.0872 & 0.0008 & $-$0&0015 & 177 \\
45 & 57070.0166 & 0.0015 & $-$0&0043 & 50 \\
\hline
  \multicolumn{6}{l}{\commenta BJD$-$2400000.} \\
  \multicolumn{6}{l}{\commentb Against max $= 2457067.0244 + 0.066587 E$.} \\
  \multicolumn{6}{l}{\commentc Number of points used to determine the maximum.} \\
\end{tabular}
\end{center}
\end{table}

\subsection{LY Hydrae}\label{obj:lyhya}

   LY Hya (=1329$-$294) was originally discovered as a CV
selected for blue color \citep{ech83lyhya}.  Double-peak
emission lines suggested a high orbital inclination
\citep{ech83lyhya}.  Although \citet{kun89lyhya}
reported a photometric period of 3.8~hr and
\citet{kub92lyhya} reported photometric and spectroscopic
observations indicating an orbital period of 0.13688~d,
these observations were not confirmed.
\citet{sti94lyhya} obtained an orbital period of
0.0748(5)~d and the Doppler tomogram suggested that
the object likely belong SU UMa-type dwarf novae.

   There was a report by T. Vanmunster that S. Howell
observed an outburst in 1996 around $V$=14.4
(cf. vsnet-obs 2047).
The details of this observation are not known.
It was only in 1998 April when a fresh outburst
was detected by R. Stubbings (vsnet-alert 1707).
The object rapidly faded during this outburst
(vsnet-alert 1713, 1714).
In 2000 April, another brightening at 16.0 mag
was reported by P. Schmeer (vsnet-alert 4559).
This outburst also quickly faded (vsnet-alert 4563).
Another outburst at 14.0 mag was reported visually
vy R. Stubbings in 2000 September (vsnet-outburst 462)
but no further observation was available for this
outburst.  In 2009 May, R. Stubbings detected
another outburst (vsnet-alert 11233) but it again
faded rapidly (vsnet-alert 11236).
After another faint outburst in 2011 February
(vsnet-outburst 12296), there was a bright outburst
reaching 13.1 mag in 2012 February (vsnet-alert 14206).
No time-series observations were, however,
obtained during this outburst.

\begin{figure}
  \begin{center}
    \FigureFile(88mm,110mm){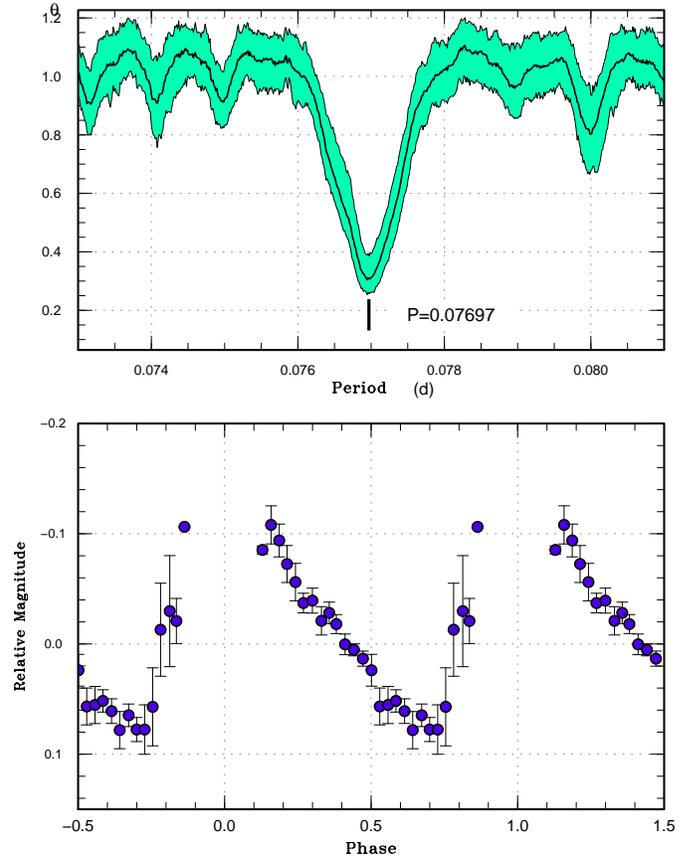}
  \end{center}
  \caption{Superhumps in LY Hya (2014).
     (Upper): PDM analysis.
     (Lower): Phase-averaged profile.}
  \label{fig:lyhyashpdm}
\end{figure}

   The 2014 superoutburst was the first one
during which time-series observations were obtained.
This outburst was detected at $V$=13.25 by
the ASAS-SN team on August 9 (vsnet-alert 17632).
Superhumps were finally detected
(vsnet-alert 17640, 17644, 17653).  Due to the unfavorable
seasonal condition, observations were limited to
short evening windows.  Despite that individual
superhump maxima were never recorded, a combined
light curve yielded an unmistakable superhump
signal (figure \ref{fig:lyhyashpdm}).
Since we observed only the later part of
the superoutburst until the rapid fading,
we likely recorded stage C superhumps.
In table \ref{tab:perlist}, we list 
this identification.

\subsection{MM Hydrae}\label{obj:mmhya}

   MM Hya was originally selected as a CV by the Palomer-Green 
survey \citep{gre82PGsurveyCV}.  Although \citet{mis95PGCV}
suggested it to be a WZ Sge-type dwarf nova based on the very
short orbital period, the object has been recognized as a rather
ordinary SU UMa-type dwarf nova \citet{pat03suumas}.
Past superhump observations were given in \citet{Pdot},
\citet{Pdot3}, \citet{Pdot4} and \citet{Pdot5}.

   The 2014 superoutburst was detected by R. Stubbings on
April 27 during its rising stage (vsnet-alert 17267).
Six days later, time-resolved observations detected
superhumps (vsnet-alert 17279).  The times of superhump
maxima are listed in table \ref{tab:mmhyaoc2014}.
A comparison of the $O-C$ diagrams between different
superoutbursts suggests that we observed stage B--C
superhumps during the 2014 superoutburst
(figure \ref{fig:mmhyacomp3}).

\begin{figure}
  \begin{center}
    \FigureFile(88mm,110mm){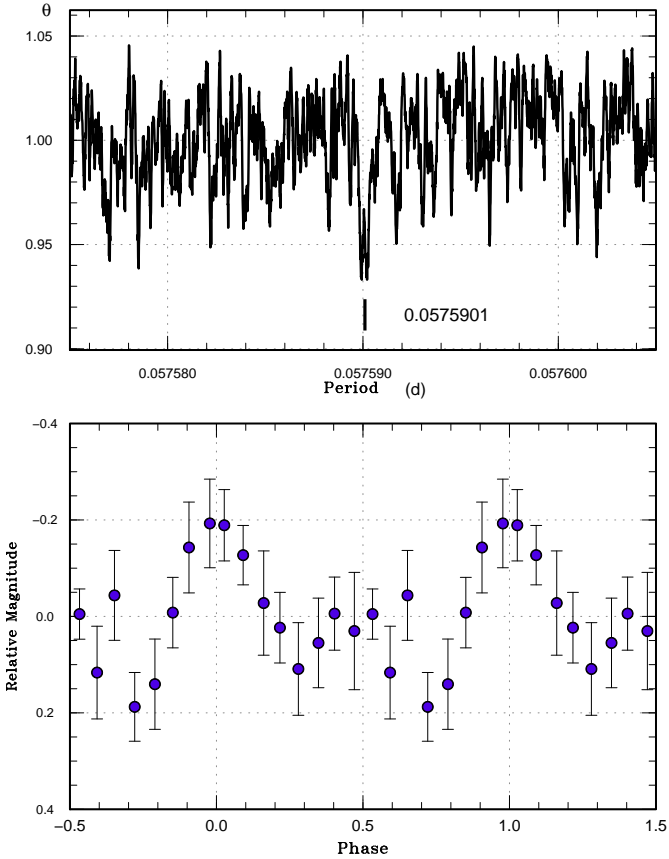}
  \end{center}
  \caption{Orbital variation of MM Hya in quiescence using the CRTS data.
     (Upper): PDM analysis.
     (Lower): Phase-averaged profile.}
  \label{fig:mmhyaporbpdm}
\end{figure}

\begin{figure}
  \begin{center}
    \FigureFile(88mm,70mm){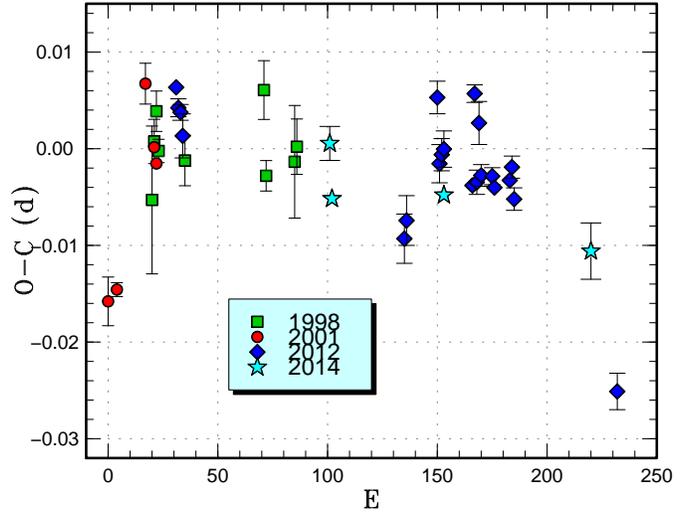}
  \end{center}
  \caption{Comparison of $O-C$ diagrams of MM Hya between different
  superoutbursts.  A period of 0.05892~d was used to draw this figure.
  Approximate cycle counts ($E$) after the appearance of superhumps
  were used.}
  \label{fig:mmhyacomp3}
\end{figure}

   We have also refined the orbital period to be
0.0575901(1) using the quiescent CRTS data
(figure \ref{fig:mmhyaporbpdm}; BJD 2453705--2456441).

   We have also examined the recent outbursts of MM Hya
(table \ref{tab:mmhyaoc2014}).  The $O-C$ diagram indicates
that the supercycle was long [386(3)~d] between 1998
and 2003, but it decreased to 330(2)~d.
Superoutbursts were not detected between 2007 and 2011
probably because superoutbursts occurred around solar
conjunctions since the supercycle is close to one year.
Although the shorter supercycle [347(8)~d] was also recorded
between 2011 and 2013, it again became longer in 2014.
The number of normal outbursts has been rather small,
although many of them may have escaped detection due to
the faintness, particularly for visual observations.

\begin{figure}
  \begin{center}
    \FigureFile(88mm,70mm){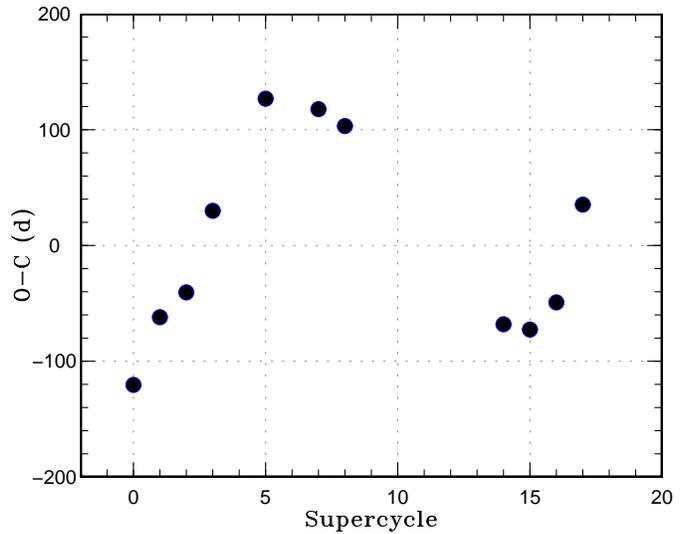}
  \end{center}
  \caption{$O-C$ diagram of superoutbursts in MM Hya.
  The ephemeris used was Max(JD)$=2451001+337.5E$.}
  \label{fig:mmhyasooc}
\end{figure}

\begin{table}
\caption{Superhump maxima of MM Hya (2014)}\label{tab:mmhyaoc2014}
\begin{center}
\begin{tabular}{rp{55pt}p{40pt}r@{.}lr}
\hline
\multicolumn{1}{c}{$E$} & \multicolumn{1}{c}{max\commenta} & \multicolumn{1}{c}{error} & \multicolumn{2}{c}{$O-C$\commentb} & \multicolumn{1}{c}{$N$\commentc} \\
\hline
0 & 56780.9808 & 0.0018 & 0&0026 & 122 \\
1 & 56781.0340 & 0.0008 & $-$0&0031 & 122 \\
52 & 56784.0393 & 0.0006 & 0&0008 & 108 \\
119 & 56787.9812 & 0.0029 & $-$0&0003 & 58 \\
\hline
  \multicolumn{6}{l}{\commenta BJD$-$2400000.} \\
  \multicolumn{6}{l}{\commentb Against max $= 2456780.9782 + 0.058851 E$.} \\
  \multicolumn{6}{l}{\commentc Number of points used to determine the maximum.} \\
\end{tabular}
\end{center}
\end{table}

\begin{table*}
\caption{List of recent outbursts of MM Hya.}\label{tab:mmhyaout}
\begin{center}
\begin{tabular}{cccccl}
\hline
Year & Month & max\commenta & magnitude & type & source \\
\hline
1998 & 3 & 50881 & 13.7 & super & J. Kemp, AAVSO; \citet{pat03suumas}; \citet{Pdot} \\
1999 & 4 & 51277 & 13.0 & super & vsnet-alert 2842, AAVSO \\
2000 & 4 & 51636 & 13.0 & super & vsnet-obs 27068, vsnet-alert 4565, AAVSO \\
2001 & 5 & 52044 & 13.0 & super & AAVSO; \citet{Pdot} \\
2002 & 11 & 52594 & 15.1\commentb & normal & AAVSO \\
2003 & 5 & 52816 & 13.2 & super & AAVSO, ASAS-3 \\
2004 & 1 & 53036 & 13.8 & normal & AAVSO, ASAS-3 \\
2004 & 7 & 53191 & 14.0\commentc & ? & vsnet-outburst 6354, AAVSO \\
2004 & 11 & 53320 & 14.1\commentb & normal & AAVSO \\
2005 & 5 & 53482 & 13.9 & super & ASAS-3, VSOLJ, AAVSO \\
2006 & 3 & 53805 & 13.0 & super & AAVSO \\
2006 & 10 & 54040 & 14.2 & normal & AAVSO, VSOLJ \\
2011 & 4 & 55659 & 13.1 & super & vsnet-alert 13113, AAVSO; \citet{Pdot3} \\
2012 & 3 & 55992 & 13.4 & super & AAVSO; \citet{Pdot4} \\
2012 & 5 & 56055 & 15.1\commentb & normal & AAVSO \\
2013 & 3 & 56353 & 13.3 & super & vsnet-alert 15487, AAVSO; \citet{Pdot5} \\
2014 & 4 & 56775 & 13.1 & super & vsnet-alert 17267; this paper \\
\hline
  \multicolumn{6}{l}{\commenta JD$-$2400000.} \\
  \multicolumn{6}{l}{\commentb CCD single detection.} \\
  \multicolumn{6}{l}{\commentc Single visual detection and 
     single positive detection by ASAS-3 5~d before.} \\
\end{tabular}
\end{center}
\end{table*}

\subsection{RZ Leonis Minoris}\label{obj:rzlmi}

   This object is one of the ER UMa-type dwarf novae,
and is renowned for its short supercycle length (19~d;
cf. \cite{rob95eruma}; \cite{nog95rzlmi}).
Two superhump maxima were obtained: BJD 2456725.7181(2)
($N=58$) and 2456725.7800(7) ($N=31$).

\subsection{BR Lupi}\label{obj:brlup}

   This object (=HV 4889) was discovered as a dwarf nova
(although it was referred to as an SS Cyg-type,
there was no distinction of subtypes at the time
of this discovery) by \citet{swo30abnorbrlup} with
a photographic range of 13.5 to fainter than 16.0
in the table.  \citet{swo30abnorbrlup} reported
ten outbursts and the photographic range was corrected
to 13.1 to fainter than 16.4 in \citet{pra41VScatalog}.
\citet{wal58CVchart} provided a finding chart.
\citet{vog82atlas} also provided a detailed finding chart.

   \citet{odo87brlup} reported on the detection of
superhumps during the 1986 outburst, establishing
the SU UMa-type nature.  \citet{mun98CVspec5} confirmed
the CV nature by spectroscopy.  \citet{men98brlup}
reported a radial-velocity study and photometry
in quiescence and during a superoutburst.

   We conducted a time-series photometry in quiescence
and superoutburst starting on 2014 July 31.
The times of superhump maxima are listed in table
\ref{tab:brlupoc2014}.  After BJD 2456878, the profile
became double-humped.  The two distinct hump maxima
at $E=194$ and $E=218$ may be traditional late superhumps
after an $\sim$0.5 phase jump.
The superhumps up to $E=24$ correspond to stage A
superhumps with growing amplitudes.

\begin{figure}
  \begin{center}
    \FigureFile(88mm,70mm){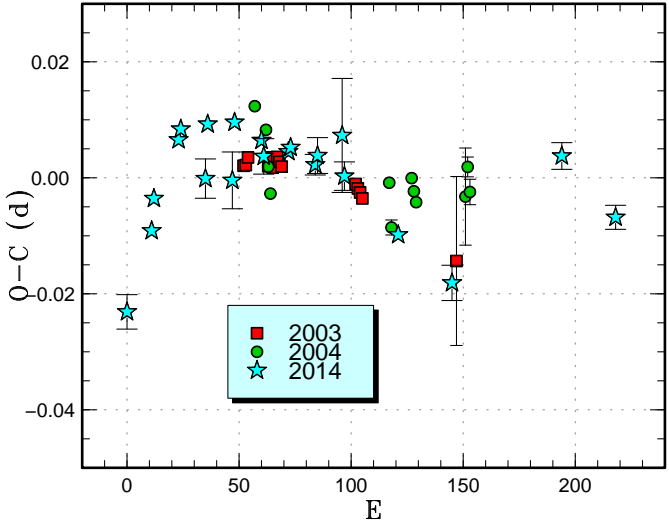}
  \end{center}
  \caption{Comparison of $O-C$ diagrams of BR Lup between different
  superoutbursts.  A period of 0.08228~d was used to draw this figure.
  Approximate cycle counts ($E$) after the appearance of superhumps
  were used.}
  \label{fig:brlupcomp2}
\end{figure}

   A comparison of $O-C$ diagrams between different
superoutbursts is shown in figure \ref{fig:brlupcomp2}.
The trends were the same except the late part of
the superoutburst, when superhump profiles became
less regular or doubly humped.  Since the mass-transfer rate
is expected to be large to account for the short
supercycle (140~d), the appearance of traditional
late superhumps reflecting the bright hot spot is not
a surprise.

\begin{figure}
  \begin{center}
    \FigureFile(88mm,110mm){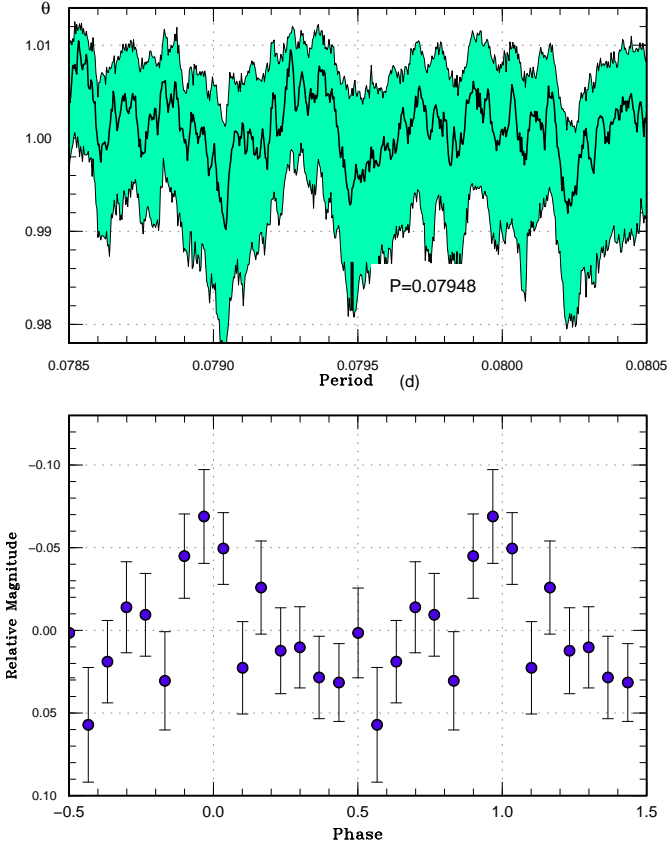}
  \end{center}
  \caption{Period analysis of BR Lup in quiescence
     (BJD 2456835--2456864).
     (Upper): PDM analysis.
     (Lower): Phase-averaged profile.}
  \label{fig:brlupquipdm}
\end{figure}

   The orbital period of this system has not been
well-established.  The period in the literature was
0.0795(5)~d in \citet{men98brlup}.
We observed this object in quiescence to determine
the orbital period (figure \ref{fig:brlupquipdm}).
The orbital signal was rather weak in contrast
to the reported 0.4-mag variation in \citet{men98brlup}.
We adopted here a period of 0.07948(2)~d 
because it is closer to the value by \citet{men98brlup}
and the estimated orbital period (0.0795~d) based on
the updated relation between the orbital and superhump periods
(equation 6 in \cite{Pdot3}).
There remains a possibility of 0.07904(2)~d.
The exact orbital period needs to be confirmed by 
further observations.
This orbital period gives $\epsilon^*$=0.0488(12)
for stage A superhumps and $q$=0.142(4).
This value needs to be established by confirming
the orbital period.

\begin{table}
\caption{Superhump maxima of BR Lup (2014)}\label{tab:brlupoc2014}
\begin{center}
\begin{tabular}{rp{55pt}p{40pt}r@{.}lr}
\hline
\multicolumn{1}{c}{$E$} & \multicolumn{1}{c}{max\commenta} & \multicolumn{1}{c}{error} & \multicolumn{2}{c}{$O-C$\commentb} & \multicolumn{1}{c}{$N$\commentc} \\
\hline
0 & 56869.5644 & 0.0030 & $-$0&0242 & 15 \\
11 & 56870.4835 & 0.0008 & $-$0&0101 & 11 \\
12 & 56870.5714 & 0.0006 & $-$0&0045 & 14 \\
23 & 56871.4865 & 0.0007 & 0&0058 & 15 \\
24 & 56871.5707 & 0.0007 & 0&0076 & 16 \\
35 & 56872.4672 & 0.0034 & $-$0&0007 & 6 \\
36 & 56872.5589 & 0.0007 & 0&0087 & 19 \\
47 & 56873.4543 & 0.0049 & $-$0&0008 & 6 \\
48 & 56873.5466 & 0.0006 & 0&0092 & 19 \\
60 & 56874.5308 & 0.0017 & 0&0062 & 15 \\
61 & 56874.6103 & 0.0031 & 0&0035 & 6 \\
72 & 56875.5162 & 0.0012 & 0&0045 & 14 \\
73 & 56875.5992 & 0.0016 & 0&0052 & 8 \\
84 & 56876.5013 & 0.0018 & 0&0024 & 14 \\
85 & 56876.5852 & 0.0031 & 0&0040 & 11 \\
96 & 56877.4937 & 0.0098 & 0&0077 & 11 \\
97 & 56877.5690 & 0.0025 & 0&0007 & 10 \\
121 & 56879.5336 & 0.0017 & $-$0&0091 & 14 \\
145 & 56881.5000 & 0.0030 & $-$0&0170 & 15 \\
194 & 56885.5536 & 0.0023 & 0&0056 & 20 \\
218 & 56887.5178 & 0.0021 & $-$0&0046 & 20 \\
\hline
  \multicolumn{6}{l}{\commenta BJD$-$2400000.} \\
  \multicolumn{6}{l}{\commentb Against max $= 2456869.5887 + 0.082265 E$.} \\
  \multicolumn{6}{l}{\commentc Number of points used to determine the maximum.} \\
\end{tabular}
\end{center}
\end{table}

\subsection{AY Lyrae}\label{obj:aylyr}

   The 2014 October superoutburst of this well-known
SU UMa-type dwarf nova was observed on two nights.
Three superhump maxima were recorded:
BJD 2456947.9590(7) ($N$=51), 2456948.0347(7) ($N$=56)
and 2456954.9360(10) ($N$=96).

\subsection{V453 Normae}\label{obj:v453nor}

   This object (=ASAS J160048$-$4846.2) was discovered as
a dwarf nova in 2005 June by \citet{poj05j1600iauc8539}.
Early observations
of superhumps was reported by \citet{mon05j1600iauc8540}.
\citet{ima06asas1600} reported the possible detection
of early superhumps and classified this object as a
WZ Sge-type dwarf nova.  \citet{soe09asas1600} provided
the detailed analysis of the superhumps, and detected
an increase in the superhump period which was later
associated with the increase of the amplitude.
\citet{Pdot} made an extended analysis after combination
with the AAVSO data, which generally confirmed the finding
by \citet{ima06asas1600}.

   After the 2005 superoutburst, there have been two
detections of apparent normal outbursts:
2006 May 20 (14.2 mag) and 2009 January 5 (13.8), both
detected by R. Stubbings.

   The next superoutburst was detected on 2014 April 23
at 12.4 mag by R. Stubbings (vsnet-alert 17254).
Growing superhumps were detected on April 25 and 26
(vsnet-alert 17262).

\begin{figure}
  \begin{center}
    \FigureFile(88mm,70mm){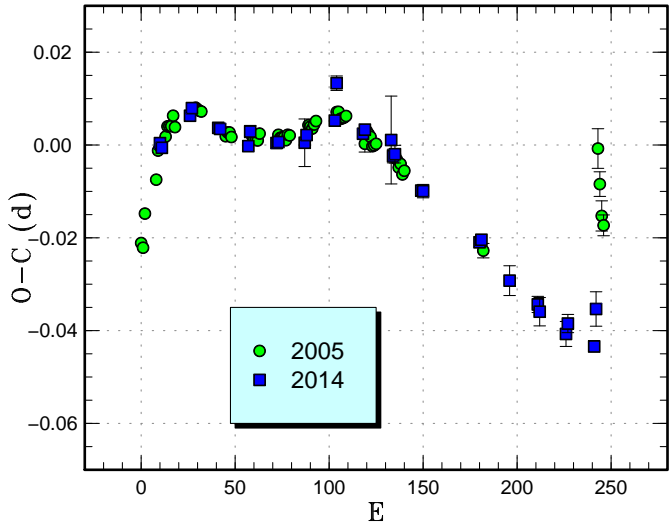}
  \end{center}
  \caption{Comparison of $O-C$ diagrams of V453 Nor between different
  superoutbursts.  A period of 0.06497~d was used to draw this figure.
  Approximate cycle counts ($E$) after the appearance of superhumps
  were used.  The 2014 superoutburst was shifted by 10 cycles
  to best match the 2005 one.}
  \label{fig:v453norcomp}
\end{figure}

   The times of superhump maxima are listed in table
\ref{tab:v453noroc2014}.  This table includes superhumps
observed after the rapid fading from the superoutburst.
Although the end phase of stage A was probably observed
on April 25 ($E=1$ and $E=2$), the period of stage A
superhumps could not be determined.  The $P_{\rm dot}$
of stage B superhumps was strikingly positive
[$+16.4(3.5) \times 10^{-5}$], similar to the 2005 value
[$+11.1(0.8) \times 10^{-5}$].  A very clear stage B-C
transition was recorded (cf. figure \ref{fig:v453norcomp}).
There was no phase jump during
the rapid fading phase, and a single period (stage C
superhumps) well explained the superhumps during the late
course of the plateau phase and the post-superoutburst
phase.  An analysis of observations after the rebrightening
(BJD 2456789.7--2456796) yielded weak modulations (0.05 mag)
with period of 0.06416(13)~d, which may be decaying
superhumps.

   The overall $O-C$ diagrams were very similar between
2005 and 2014 (figure \ref{fig:v453norcomp}) except the
appearance of humps with different phases during the
rebrightening in 2005 [see also discussion in \cite{Pdot}].

\begin{table}
\caption{Superhump maxima of V453 Nor (2014)}\label{tab:v453noroc2014}
\begin{center}
\begin{tabular}{rp{55pt}p{40pt}r@{.}lr}
\hline
\multicolumn{1}{c}{$E$} & \multicolumn{1}{c}{max\commenta} & \multicolumn{1}{c}{error} & \multicolumn{2}{c}{$O-C$\commentb} & \multicolumn{1}{c}{$N$\commentc} \\
\hline
0 & 56772.8084 & 0.0006 & $-$0&0138 & 30 \\
1 & 56772.8724 & 0.0005 & $-$0&0146 & 37 \\
16 & 56773.8538 & 0.0003 & $-$0&0046 & 35 \\
17 & 56773.9205 & 0.0010 & $-$0&0027 & 15 \\
31 & 56774.8258 & 0.0005 & $-$0&0041 & 22 \\
32 & 56774.8905 & 0.0004 & $-$0&0041 & 35 \\
47 & 56775.8614 & 0.0005 & $-$0&0047 & 35 \\
48 & 56775.9295 & 0.0013 & $-$0&0013 & 10 \\
62 & 56776.8366 & 0.0005 & $-$0&0009 & 33 \\
63 & 56776.9018 & 0.0005 & $-$0&0005 & 29 \\
77 & 56777.8112 & 0.0051 & 0&0023 & 10 \\
78 & 56777.8778 & 0.0008 & 0&0041 & 35 \\
93 & 56778.8555 & 0.0006 & 0&0104 & 35 \\
94 & 56778.9285 & 0.0016 & 0&0186 & 12 \\
108 & 56779.8272 & 0.0007 & 0&0106 & 33 \\
109 & 56779.8931 & 0.0006 & 0&0117 & 35 \\
123 & 56780.8004 & 0.0095 & 0&0124 & 16 \\
124 & 56780.8619 & 0.0015 & 0&0092 & 35 \\
125 & 56780.9273 & 0.0019 & 0&0098 & 15 \\
139 & 56781.8290 & 0.0010 & 0&0048 & 35 \\
140 & 56781.8939 & 0.0014 & 0&0049 & 35 \\
170 & 56783.8319 & 0.0009 & 0&0001 & 31 \\
171 & 56783.8975 & 0.0010 & 0&0009 & 29 \\
186 & 56784.8632 & 0.0032 & $-$0&0048 & 30 \\
201 & 56785.8326 & 0.0017 & $-$0&0068 & 27 \\
202 & 56785.8960 & 0.0031 & $-$0&0082 & 27 \\
216 & 56786.8008 & 0.0027 & $-$0&0101 & 27 \\
217 & 56786.8680 & 0.0020 & $-$0&0076 & 27 \\
231 & 56787.7727 & 0.0009 & $-$0&0097 & 16 \\
232 & 56787.8457 & 0.0037 & $-$0&0014 & 19 \\
\hline
  \multicolumn{6}{l}{\commenta BJD$-$2400000.} \\
  \multicolumn{6}{l}{\commentb Against max $= 2456772.8222 + 0.064762 E$.} \\
  \multicolumn{6}{l}{\commentc Number of points used to determine the maximum.} \\
\end{tabular}
\end{center}
\end{table}

\subsection{DT Octantis}\label{obj:dtoct}

\begin{figure}
  \begin{center}
    \FigureFile(88mm,70mm){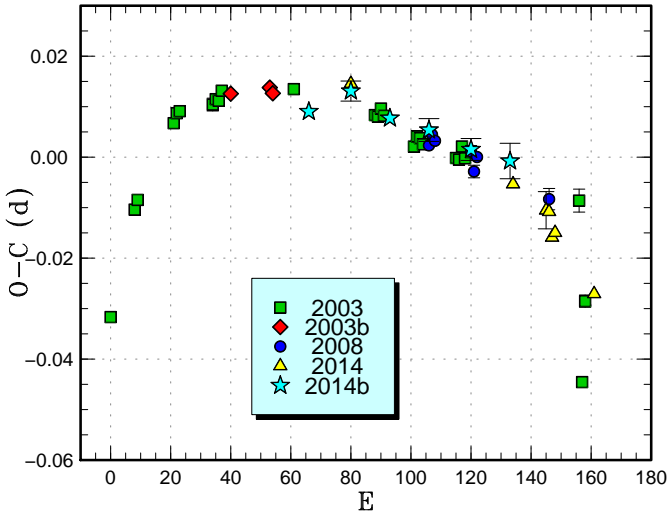}
  \end{center}
  \caption{Comparison of $O-C$ diagrams of DT Oct between different
  superoutbursts.  A period of 0.07485~d was used to draw this figure.
  Approximate cycle counts ($E$) after the start of the superoutburst
  were used.  Since the start of the 2014 and 2014b superoutbursts
  was not well constrained, we shifted the $O-C$ diagrams
  to best fit the others.
  }
  \label{fig:dtoctcomp3}
\end{figure}

   For the history of this object, see \citet{Pdot6}.
The 2014 December superoutburst was detected by
R. Stubbings and the ASAS-SN team (vsnet-alert 18079).
The times of superhump maxima are listed in table
\ref{tab:dtoctoc2014b}.  The period listed in table
\ref{tab:perlist} is probably a mixture of stage B and C
superhumps (figure \ref{fig:dtoctcomp3}).

\begin{table}
\caption{Superhump maxima of DT Oct (2014b)}\label{tab:dtoctoc2014b}
\begin{center}
\begin{tabular}{rp{55pt}p{40pt}r@{.}lr}
\hline
\multicolumn{1}{c}{$E$} & \multicolumn{1}{c}{max\commenta} & \multicolumn{1}{c}{error} & \multicolumn{2}{c}{$O-C$\commentb} & \multicolumn{1}{c}{$N$\commentc} \\
\hline
0 & 57012.5406 & 0.0010 & $-$0&0031 & 24 \\
14 & 57013.5926 & 0.0020 & 0&0035 & 11 \\
27 & 57014.5603 & 0.0013 & 0&0005 & 26 \\
40 & 57015.5309 & 0.0023 & 0&0005 & 18 \\
54 & 57016.5750 & 0.0022 & $-$0&0007 & 16 \\
67 & 57017.5458 & 0.0035 & $-$0&0007 & 25 \\
\hline
  \multicolumn{6}{l}{\commenta BJD$-$2400000.} \\
  \multicolumn{6}{l}{\commentb Against max $= 2457012.5437 + 0.074667 E$.} \\
  \multicolumn{6}{l}{\commentc Number of points used to determine the maximum.} \\
\end{tabular}
\end{center}
\end{table}

\subsection{UV Persei}\label{obj:uvper}

   UV Per is a famous dwarf nova known more than for
a century.  The object was discovered as a nova or 
a new variable star (87.1911) in 1911 November
by \citet{des12uvper}.
Since the object was initially suspected to be a nova,
\citet{wol12uvper} studied the counterpart and found
that the object is a southern component of the close
pair.  This finding was confirmed by \citet{des13uvper}.
\citet{nij14uvper} detected an outburst in 1914 June
lasting 14~d and concluded that the object must be
a U Gem-type variable.
\citet{har15uvper} recorded an outburst in 1915
September, which faded quickly.  This outburst was
also recorded by \citet{nij15uvper}, who confirmed
the shortness of the outburst.  \citet{har17uvper}
recorded a long outburst in 1916 December.
\citet{har17uvper} suggested that the intervals of
three outbursts were 442 and 473~d.  \citet{nij18uvper}
and \citet{har20uvper}
detected further outbursts in 1918 May and 1920 October,
respectively.  In the meantime, Harvard College
Observatory reported the intervals of outbursts
to be 214 and 299~d, suggesting the irregularity.
\citet{nij24uvper} summarized nine known outbursts
(five of them were long outbursts) and suggested
that the intervals could be expressed by the basic
period of 131.48~d.

   Since then, the object has long been monitored
by amateur observers (e.g. \cite{pet56uvper};
\cite{hor75uvper}).
\citet{pet56uvper} analyzed the outbursts in detail
and obtained a mean interval of 359.8~d.
\citet{pet56uvper} also classified outbursts into
two types: type I lasting for 12~d and type II lasting
for 4~d.  \citet{may66uvper} was another summary
article and reported a list of 32 outbursts between
1928 and 1966.  The existence of long and short
outbursts was clear from these data.
\citet{how78uvper} also presented a list of 63 outbursts
(dating back to Pickering's observation in 1896)
and discussed outburst statistics.  \citet{how78uvper}
estimated the mean cycle length of 320~d.

   The long cycle length of UV Per received attention,
and the object was considered to be a good candidate
for an SU UMa-type dwarf nova.  It was only in
1989 when \citet{uda89uvperiauc} succeeded in
detecting superhumps.  \citet{kat90uvper} reported
photometry in quiescence.  \citet{uda92uvper}
reported both during the 1989 superoutburst
and in quiescence.  The orbital period was
first reliably measured to be 0.06489(11)~d
by radial-velocity study by \citet{tho97uvpervyaqrv1504cyg}.
The 1989 superoutburst was also notable that this
was one of the first superoutbursts of SU UMa-type
dwarf novae followed by a rebrightening.\footnote{
   The first documented record of a rebrightening was
   VY Aqr by R. McNaught \citep{per86vyaqriauc4222}
   in 1986, which was later published \citep{pat93vyaqr}.
}
This rebrightening was best recorded by the VSOLJ
members [although there were several detections
in the present AAVSO data, they were not shown
in \citet{uda92uvper}].

   Although a number of superoutbursts were detected,
there have not been many papers.  \citet{pri03uvper}
was the only publication before \citet{Pdot} and dealt
with the 2003 superoutburst.  \citet{Pdot} reported
on the 2000, 2003 and 2007 superoutbursts and
\citet{Pdot2} reported on the 2010 superoutburst.
None of superoutbursts were better observed than
the 2003 one.

   The 2014 superoutburst was visually detected by
C. Chiselbrook (cf. vsnet-alert 18000).
Since the detection of the outburst was early enough
to record the appearance of superhumps
(vsnet-alert 18008, 18022).
There was a single post-superoutburst rebrightening
(vsnet-alert 18061).

\begin{figure}
  \begin{center}
    \FigureFile(88mm,70mm){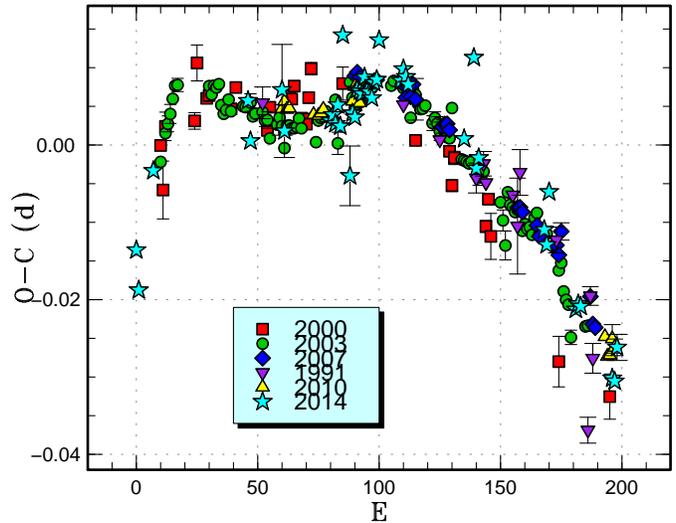}
  \end{center}
  \caption{Comparison of $O-C$ diagrams of UV Per between different
  superoutbursts.  A period of 0.06665~d was used to draw this figure.
  Approximate cycle counts ($E$) after the appearance of superhumps
  were used.  The superoutbursts before 2014 was shifted by
  10 cycles against the figure in \citet{Pdot2}
  to best match the start of the 2014 one.}
  \label{fig:uvpercomp3}
\end{figure}

The times of superhump maxima are listed in table
\ref{tab:uvperoc2014}.  Stages B and C can be recognized.
Although the maxima for $E \le 7$ correspond to stage A
superhumps, the period of stage A superhumps could not
be determined.  Since the early phase of stage B
superhumps were not observed, the value of $P_{\rm dot}$
in table \ref{tab:perlist} is not as reliable as
the 2003 one in \citet{Pdot}.
A comparison of $O-C$ diagrams between different
superoutbursts is given in figure \ref{fig:uvpercomp3}.

\begin{table}
\caption{Superhump maxima of UV Per (2014)}\label{tab:uvperoc2014}
\begin{center}
\begin{tabular}{rp{55pt}p{40pt}r@{.}lr}
\hline
\multicolumn{1}{c}{$E$} & \multicolumn{1}{c}{max\commenta} & \multicolumn{1}{c}{error} & \multicolumn{2}{c}{$O-C$\commentb} & \multicolumn{1}{c}{$N$\commentc} \\
\hline
0 & 56984.9961 & 0.0008 & $-$0&0253 & 79 \\
1 & 56985.0576 & 0.0013 & $-$0&0303 & 73 \\
7 & 56985.4729 & 0.0012 & $-$0&0142 & 229 \\
46 & 56988.0813 & 0.0003 & $-$0&0005 & 133 \\
47 & 56988.1428 & 0.0004 & $-$0&0056 & 134 \\
60 & 56989.0158 & 0.0059 & 0&0026 & 21 \\
61 & 56989.0772 & 0.0034 & $-$0&0026 & 23 \\
80 & 56990.3448 & 0.0004 & 0&0010 & 69 \\
81 & 56990.4123 & 0.0003 & 0&0019 & 66 \\
82 & 56990.4777 & 0.0004 & 0&0008 & 68 \\
83 & 56990.5468 & 0.0005 & 0&0034 & 64 \\
84 & 56990.6107 & 0.0005 & 0&0008 & 69 \\
85 & 56990.6892 & 0.0007 & 0&0126 & 14 \\
88 & 56990.8709 & 0.0039 & $-$0&0052 & 25 \\
89 & 56990.9457 & 0.0002 & 0&0030 & 248 \\
90 & 56991.0118 & 0.0002 & 0&0026 & 248 \\
91 & 56991.0820 & 0.0002 & 0&0063 & 249 \\
92 & 56991.1485 & 0.0006 & 0&0062 & 55 \\
93 & 56991.2148 & 0.0006 & 0&0060 & 68 \\
94 & 56991.2836 & 0.0005 & 0&0083 & 69 \\
95 & 56991.3484 & 0.0003 & 0&0066 & 191 \\
96 & 56991.4150 & 0.0004 & 0&0066 & 69 \\
97 & 56991.4808 & 0.0005 & 0&0059 & 66 \\
98 & 56991.5494 & 0.0004 & 0&0080 & 67 \\
99 & 56991.6165 & 0.0005 & 0&0086 & 69 \\
100 & 56991.6883 & 0.0005 & 0&0138 & 24 \\
109 & 56992.2829 & 0.0003 & 0&0096 & 107 \\
110 & 56992.3510 & 0.0003 & 0&0112 & 200 \\
111 & 56992.4167 & 0.0003 & 0&0104 & 154 \\
112 & 56992.4823 & 0.0004 & 0&0095 & 66 \\
135 & 56994.0082 & 0.0005 & 0&0052 & 84 \\
139 & 56994.2853 & 0.0007 & 0&0162 & 86 \\
\hline
  \multicolumn{6}{l}{\commenta BJD$-$2400000.} \\
  \multicolumn{6}{l}{\commentb Against max $= 2456985.0214 + 0.066531 E$.} \\
  \multicolumn{6}{l}{\commentc Number of points used to determine the maximum.} \\
\end{tabular}
\end{center}
\end{table}

\addtocounter{table}{-1}
\begin{table}
\caption{Superhump maxima of UV Per (2014) (continued)}
\begin{center}
\begin{tabular}{rp{55pt}p{40pt}r@{.}lr}
\hline
\multicolumn{1}{c}{$E$} & \multicolumn{1}{c}{max\commenta} & \multicolumn{1}{c}{error} & \multicolumn{2}{c}{$O-C$\commentb} & \multicolumn{1}{c}{$N$\commentc} \\
\hline
140 & 56994.3377 & 0.0003 & 0&0020 & 132 \\
141 & 56994.4057 & 0.0003 & 0&0034 & 111 \\
168 & 56996.1959 & 0.0008 & $-$0&0026 & 122 \\
169 & 56996.2606 & 0.0008 & $-$0&0045 & 144 \\
170 & 56996.3341 & 0.0010 & 0&0025 & 137 \\
181 & 56997.0521 & 0.0008 & $-$0&0114 & 86 \\
182 & 56997.1194 & 0.0005 & $-$0&0106 & 215 \\
183 & 56997.1858 & 0.0007 & $-$0&0108 & 202 \\
196 & 56998.0429 & 0.0010 & $-$0&0185 & 65 \\
197 & 56998.1092 & 0.0008 & $-$0&0188 & 62 \\
198 & 56998.1802 & 0.0017 & $-$0&0143 & 49 \\
\hline
  \multicolumn{6}{l}{\commenta BJD$-$2400000.} \\
  \multicolumn{6}{l}{\commentb Against max $= 2456985.0214 + 0.066531 E$.} \\
  \multicolumn{6}{l}{\commentc Number of points used to determine the maximum.} \\
\end{tabular}
\end{center}
\end{table}

\subsection{HY Piscium}\label{obj:hypsc}

   This object (=SDSSp J230351.64$+$010651.0) was originally
selected as a CV by the SDSS \citep{szk02SDSSCVs}.
\citet{szk02SDSSCVs} identified this object as a dwarf nova
by obtaining spectra both in outburst and quiescence.
A radial-velocity study by \citet{szk02SDSSCVs} yielded
an orbital period of 100$\pm$14 min.  The first secure
outburst since the discovery was detected on 2006 August 29
of 14.5 mag (magnitudes are unfiltered CCD ones
for this object) by I. Miller (cvnet-outburst 1327).
Although this outburst quickly faded, modulations with
a period of 0.07~d were reported during the declining
branch (vsnet-alert 9004).
In \citet{gan09SDSSCVs}, there was an orbital period
of 110.51(24)~min [0.07674(17)~d] by Dillon et al., but
this observation has not yet been published.
On 2010 January 3, J. Shears detected a bright outburst
at 13.6 mag (cvnet-outburst 3532).
Although this outburst was a long one, superhumps
were not confidently detected.

   The 2014 outburst was detected by the ASAS-SN
team (vsnet-alert 17560) at $V$=13.27 on July 28.
Although superhumps were securely detected,
the limited observation time in the morning sky
did not enable an unique selection of the superhump
period (vsnet-alert 17577, 17602).  
Based on the best observed part of the data
(BJD before 2456873), two periods of 0.07994(2)~d
and 0.07788(2)~d remained viable.  Since the period
0.080~d better expressed nightly observations,
we list the times of maxima based on this selection
of the period (table \ref{tab:hypscoc2014}).
Both periods, however, give anomalous $\epsilon$
if assume an orbital period of 0.07674(17)~d.
Both the orbital and superhump periods need to be
confirmed by further observations.
The later part of the data (BJD after 2456873)
gave a period of 0.07981(2)~d with a more irregular
profile.  We are not confident whether this period
corresponds to stage C superhump due to the lack
of observations.

\begin{figure}
  \begin{center}
    \FigureFile(88mm,110mm){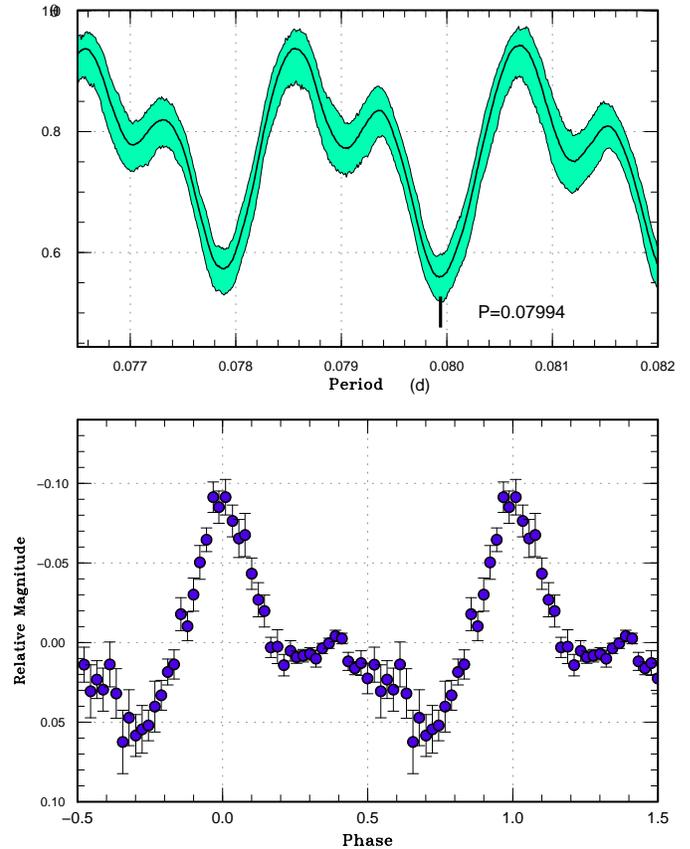}
  \end{center}
  \caption{Superhumps in HY Psc (2014) before BJD 2456873.
     (Upper): PDM analysis.
     (Lower): Phase-averaged profile.}
  \label{fig:hypscshpdm}
\end{figure}

\begin{table}
\caption{Superhump maxima of HY Psc (2014)}\label{tab:hypscoc2014}
\begin{center}
\begin{tabular}{rp{55pt}p{40pt}r@{.}lr}
\hline
\multicolumn{1}{c}{$E$} & \multicolumn{1}{c}{max\commenta} & \multicolumn{1}{c}{error} & \multicolumn{2}{c}{$O-C$\commentb} & \multicolumn{1}{c}{$N$\commentc} \\
\hline
0 & 56869.1790 & 0.0003 & $-$0&0022 & 109 \\
1 & 56869.2615 & 0.0002 & 0&0004 & 171 \\
38 & 56872.2221 & 0.0004 & 0&0036 & 145 \\
74 & 56875.0941 & 0.0011 & $-$0&0019 & 40 \\
\hline
  \multicolumn{6}{l}{\commenta BJD$-$2400000.} \\
  \multicolumn{6}{l}{\commentb Against max $= 2456869.1811 + 0.079930 E$.} \\
  \multicolumn{6}{l}{\commentc Number of points used to determine the maximum.} \\
\end{tabular}
\end{center}
\end{table}

\subsection{QW Serpentis}\label{obj:qwser}

   This object was identified as an SU UMa-type dwarf nova
in 2003 (\cite{pat03suumas}; \cite{ole03qwser}; \cite{nog04qwser}).
See \citet{Pdot5} for more history.

   The 2014 superoutburst was detected by MASTER-Tunka
on February 25 (vsnet-alert 16951).  The observations
on March 2 detected superhumps.  Only two times of
superhumps were obtained: BJD 2456718.5235(9) ($N=79$),
2456718.5991(15) ($N=77$).

\subsection{V418 Serpentis}\label{obj:v418ser}

\begin{figure}
  \begin{center}
    \FigureFile(88mm,110mm){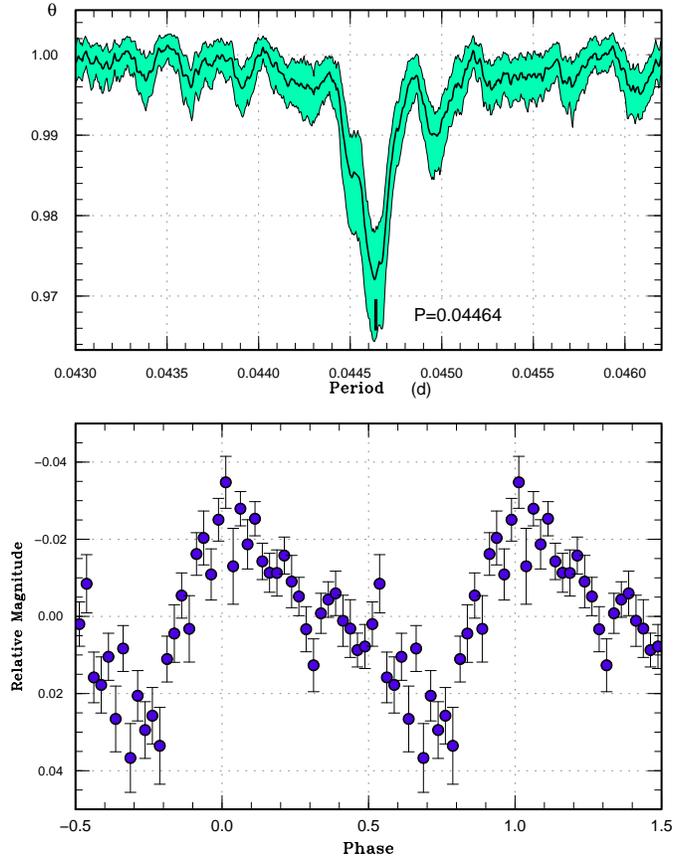}
  \end{center}
  \caption{Superhumps in V418 Ser (2014).  (Upper): PDM analysis.
     (Lower): Phase-averaged profile.}
  \label{fig:v418sershpdm}
\end{figure}

   This object (=ROTSE3 J151453.6$+$020934.2) was discovered
as a dwarf nova \citep{ryk04j1514j2215}.
Although the variable star designation was given
in \citet{NameList80b}, only little had been known
until the outburst detection by CRTS on 2014 May 21
(cf. vsnet-alert 17320).
Subsequent time-resolved photometry indicated that this
object is an SU UMa-type dwarf nova below the period minimum
by the detection of superhumps
(vsnet-alert 17321, 17322; figure \ref{fig:v418sershpdm}).
The times of superhump maxima are listed in table
\ref{tab:v418seroc2014}.  A very clear stage B-C
transition was recorded (vsnet-alert 17349, 17358).
Spectroscopic observation clarified that this object
contains hydrogen, ruling out the possibility of
an AM CVn-type object \citep{gar14v418seratel6287}.
The spectroscopic appearance was very similar to
SBS 1108+574 (\cite{lit13sbs1108}; \cite{car13sbs1108})
making this object a new member of EI Psc-type objects.

\begin{figure}
  \begin{center}
    \FigureFile(88mm,100mm){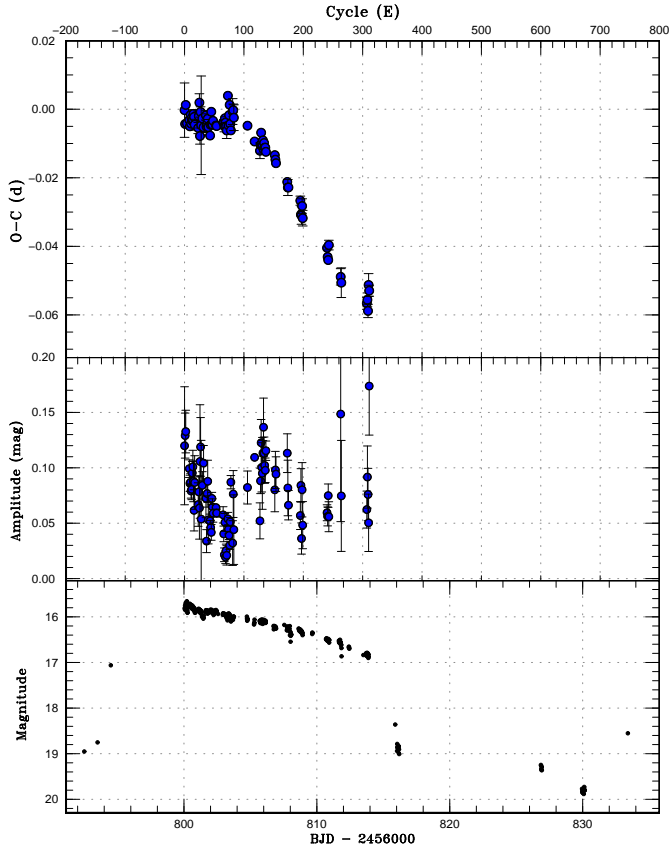}
  \end{center}
  \caption{$O-C$ diagram of superhumps in V418 Ser (2014).
     (Upper:) $O-C$ diagram.
     We used a period of 0.04467~d for calculating the $O-C$ residuals.
     (Middle:) Amplitudes of superhumps.  There was a slight tendency
     of regrowth of superhumps after the stage B--C transition.
     (Lower:) Light curve.  The data were binned to 0.015~d.
     The detection of an outburst on
     BJD 2456794 appears to be secure.  The seemingly rebrightening
     on BJD 2456833 may not be real since this magnitude was
     similar to the pre-outburst observations of the same
     observer.
  }
  \label{fig:v418serhumpall}
\end{figure}

   Among EI Psc-type objects, SBS 1108$+$574 was the first
in which distinct stages B and C were recorded
\citep{Pdot4}.  Another well-established object
is CSS J174033.5$+$414756 (\Ohtprep;
subsection \ref{obj:j1740}).
V418 Ser is the third object in this class of objects
in which distinct stages B and C are recorded
(figure \ref{fig:v418serhumpall}).  It was likely
the early phase of stage B was not observed.
There was an apparent increase in the superhump period
after the stage B-C transition, a phenomenon similar
to what is observed in hydrogen-rich CVs \citep{Pdot}
and in SBS 1108$+$574 \citep{Pdot4}.
The fractional decrease of the superhump period
($\sim$0.5\%) between stage B and C was also similar
to those in hydrogen-rich CVs \citep{Pdot}.
A comparison of these three objects has shown that
the stage transition in EI Psc-type objects are
similar to ordinary hydrogen-rich CVs.

   In the AAVSO database, one observer detected a rise to
$V$=16.02 on May 16.  The entire duration of the
superoutburst was 20--21~d, somewhat shorter than
the one in SBS 1108+574 \citep{Pdot4} and
CSS J174033.5$+$414756 (\Ohtprep).

\begin{table}
\caption{Superhump maxima of V418 Ser (2014)}\label{tab:v418seroc2014}
\begin{center}
\begin{tabular}{rp{55pt}p{40pt}r@{.}lr}
\hline
\multicolumn{1}{c}{$E$} & \multicolumn{1}{c}{max\commenta} & \multicolumn{1}{c}{error} & \multicolumn{2}{c}{$O-C$\commentb} & \multicolumn{1}{c}{$N$\commentc} \\
\hline
0 & 56800.0447 & 0.0079 & $-$0&0051 & 41 \\
1 & 56800.0854 & 0.0009 & $-$0&0089 & 75 \\
2 & 56800.1356 & 0.0008 & $-$0&0032 & 78 \\
4 & 56800.2196 & 0.0006 & $-$0&0082 & 56 \\
8 & 56800.3992 & 0.0004 & $-$0&0065 & 52 \\
9 & 56800.4420 & 0.0005 & $-$0&0082 & 55 \\
10 & 56800.4900 & 0.0005 & $-$0&0048 & 51 \\
11 & 56800.5322 & 0.0005 & $-$0&0071 & 48 \\
12 & 56800.5784 & 0.0006 & $-$0&0053 & 48 \\
13 & 56800.6223 & 0.0006 & $-$0&0059 & 100 \\
14 & 56800.6673 & 0.0008 & $-$0&0054 & 78 \\
15 & 56800.7137 & 0.0012 & $-$0&0036 & 78 \\
16 & 56800.7575 & 0.0021 & $-$0&0043 & 64 \\
17 & 56800.7999 & 0.0016 & $-$0&0064 & 74 \\
23 & 56801.0669 & 0.0016 & $-$0&0063 & 48 \\
24 & 56801.1151 & 0.0011 & $-$0&0027 & 90 \\
25 & 56801.1637 & 0.0026 & 0&0014 & 52 \\
26 & 56801.1986 & 0.0024 & $-$0&0081 & 57 \\
27 & 56801.2503 & 0.0015 & $-$0&0009 & 91 \\
28 & 56801.2911 & 0.0144 & $-$0&0047 & 39 \\
30 & 56801.3824 & 0.0008 & $-$0&0023 & 35 \\
32 & 56801.4692 & 0.0008 & $-$0&0046 & 35 \\
36 & 56801.6513 & 0.0014 & $-$0&0004 & 78 \\
37 & 56801.6927 & 0.0017 & $-$0&0035 & 120 \\
38 & 56801.7381 & 0.0009 & $-$0&0026 & 120 \\
39 & 56801.7844 & 0.0014 & $-$0&0009 & 83 \\
40 & 56801.8266 & 0.0022 & $-$0&0032 & 37 \\
43 & 56801.9581 & 0.0007 & $-$0&0051 & 44 \\
44 & 56802.0059 & 0.0009 & $-$0&0018 & 37 \\
45 & 56802.0544 & 0.0010 & 0&0022 & 44 \\
46 & 56802.0955 & 0.0004 & $-$0&0012 & 42 \\
47 & 56802.1400 & 0.0006 & $-$0&0012 & 44 \\
48 & 56802.1858 & 0.0007 & 0&0001 & 43 \\
\hline
  \multicolumn{6}{l}{\commenta BJD$-$2400000.} \\
  \multicolumn{6}{l}{\commentb Against max $= 2456800.0498 + 0.044498 E$.} \\
  \multicolumn{6}{l}{\commentc Number of points used to determine the maximum.} \\
\end{tabular}
\end{center}
\end{table}

\addtocounter{table}{-1}
\begin{table}
\caption{Superhump maxima of V418 Ser (2014) (continued)}
\begin{center}
\begin{tabular}{rp{55pt}p{40pt}r@{.}lr}
\hline
\multicolumn{1}{c}{$E$} & \multicolumn{1}{c}{max\commenta} & \multicolumn{1}{c}{error} & \multicolumn{2}{c}{$O-C$\commentb} & \multicolumn{1}{c}{$N$\commentc} \\
\hline
53 & 56802.4078 & 0.0003 & $-$0&0004 & 131 \\
54 & 56802.4523 & 0.0003 & $-$0&0004 & 102 \\
65 & 56802.9447 & 0.0007 & 0&0025 & 40 \\
66 & 56802.9894 & 0.0010 & 0&0027 & 43 \\
67 & 56803.0329 & 0.0016 & 0&0017 & 43 \\
68 & 56803.0799 & 0.0009 & 0&0043 & 42 \\
69 & 56803.1233 & 0.0017 & 0&0032 & 41 \\
70 & 56803.1669 & 0.0013 & 0&0022 & 41 \\
71 & 56803.2104 & 0.0023 & 0&0012 & 39 \\
73 & 56803.3098 & 0.0007 & 0&0117 & 129 \\
74 & 56803.3458 & 0.0005 & 0&0032 & 132 \\
75 & 56803.3935 & 0.0005 & 0&0064 & 132 \\
76 & 56803.4412 & 0.0007 & 0&0095 & 133 \\
77 & 56803.4803 & 0.0004 & 0&0042 & 130 \\
78 & 56803.5231 & 0.0004 & 0&0024 & 132 \\
81 & 56803.6626 & 0.0038 & 0&0085 & 76 \\
82 & 56803.7077 & 0.0018 & 0&0090 & 77 \\
83 & 56803.7501 & 0.0038 & 0&0070 & 64 \\
106 & 56804.7752 & 0.0010 & 0&0086 & 41 \\
118 & 56805.3066 & 0.0002 & 0&0061 & 91 \\
127 & 56805.7060 & 0.0023 & 0&0049 & 30 \\
128 & 56805.7524 & 0.0007 & 0&0068 & 47 \\
129 & 56805.8006 & 0.0006 & 0&0105 & 29 \\
130 & 56805.8421 & 0.0012 & 0&0076 & 43 \\
131 & 56805.8860 & 0.0020 & 0&0070 & 35 \\
132 & 56805.9324 & 0.0006 & 0&0088 & 56 \\
133 & 56805.9744 & 0.0012 & 0&0064 & 91 \\
134 & 56806.0209 & 0.0007 & 0&0084 & 137 \\
135 & 56806.0642 & 0.0010 & 0&0072 & 93 \\
136 & 56806.1089 & 0.0006 & 0&0074 & 42 \\
137 & 56806.1524 & 0.0004 & 0&0063 & 34 \\
152 & 56806.8214 & 0.0011 & 0&0079 & 35 \\
\hline
  \multicolumn{6}{l}{\commenta BJD$-$2400000.} \\
  \multicolumn{6}{l}{\commentb Against max $= 2456800.0498 + 0.044498 E$.} \\
  \multicolumn{6}{l}{\commentc Number of points used to determine the maximum.} \\
\end{tabular}
\end{center}
\end{table}

\addtocounter{table}{-1}
\begin{table}
\caption{Superhump maxima of V418 Ser (2014) (continued)}
\begin{center}
\begin{tabular}{rp{55pt}p{40pt}r@{.}lr}
\hline
\multicolumn{1}{c}{$E$} & \multicolumn{1}{c}{max\commenta} & \multicolumn{1}{c}{error} & \multicolumn{2}{c}{$O-C$\commentb} & \multicolumn{1}{c}{$N$\commentc} \\
\hline
153 & 56806.8648 & 0.0009 & 0&0068 & 40 \\
154 & 56806.9084 & 0.0009 & 0&0059 & 41 \\
173 & 56807.7517 & 0.0009 & 0&0037 & 35 \\
174 & 56807.7947 & 0.0023 & 0&0023 & 32 \\
175 & 56807.8394 & 0.0013 & 0&0025 & 41 \\
195 & 56808.7290 & 0.0013 & 0&0021 & 66 \\
196 & 56808.7696 & 0.0011 & $-$0&0019 & 83 \\
197 & 56808.8140 & 0.0026 & $-$0&0019 & 68 \\
198 & 56808.8614 & 0.0021 & 0&0010 & 55 \\
199 & 56808.9026 & 0.0023 & $-$0&0024 & 32 \\
240 & 56810.7254 & 0.0008 & $-$0&0040 & 74 \\
241 & 56810.7674 & 0.0010 & $-$0&0065 & 107 \\
242 & 56810.8111 & 0.0008 & $-$0&0072 & 88 \\
243 & 56810.8602 & 0.0014 & $-$0&0027 & 79 \\
263 & 56811.7443 & 0.0024 & $-$0&0085 & 16 \\
264 & 56811.7873 & 0.0043 & $-$0&0100 & 28 \\
307 & 56813.7021 & 0.0018 & $-$0&0086 & 31 \\
308 & 56813.7478 & 0.0020 & $-$0&0075 & 32 \\
309 & 56813.7892 & 0.0019 & $-$0&0105 & 32 \\
310 & 56813.8414 & 0.0033 & $-$0&0028 & 32 \\
311 & 56813.8844 & 0.0016 & $-$0&0043 & 32 \\
\hline
  \multicolumn{6}{l}{\commenta BJD$-$2400000.} \\
  \multicolumn{6}{l}{\commentb Against max $= 2456800.0498 + 0.044498 E$.} \\
  \multicolumn{6}{l}{\commentc Number of points used to determine the maximum.} \\
\end{tabular}
\end{center}
\end{table}

\subsection{V701 Tauri}\label{obj:v701tau}

   V701 Tau was discovered by \citet{era73v701tau} as
an eruptive object.  Since this ``eruption'' lasted more
than ten days, the object was considered to be
a dwarf nova and has been monitored by amateur observers.
The initial superoutburst was recorded in 1995 December
(vsnet-alert 303).  \citet{she07v701tau} further reported
the 2005 superoutburst and obtained a superhump period of
0.0690(2) d.  The 1995 superoutburst was analyzed in
\citep{Pdot} as well as the 2005 superoutburst.
Although there was also a well-recorded superoutburst in
2007 March (e.g. vsnet-outburst 7535), superhumps were
not sufficiently observed.

\begin{figure}
  \begin{center}
    \FigureFile(88mm,70mm){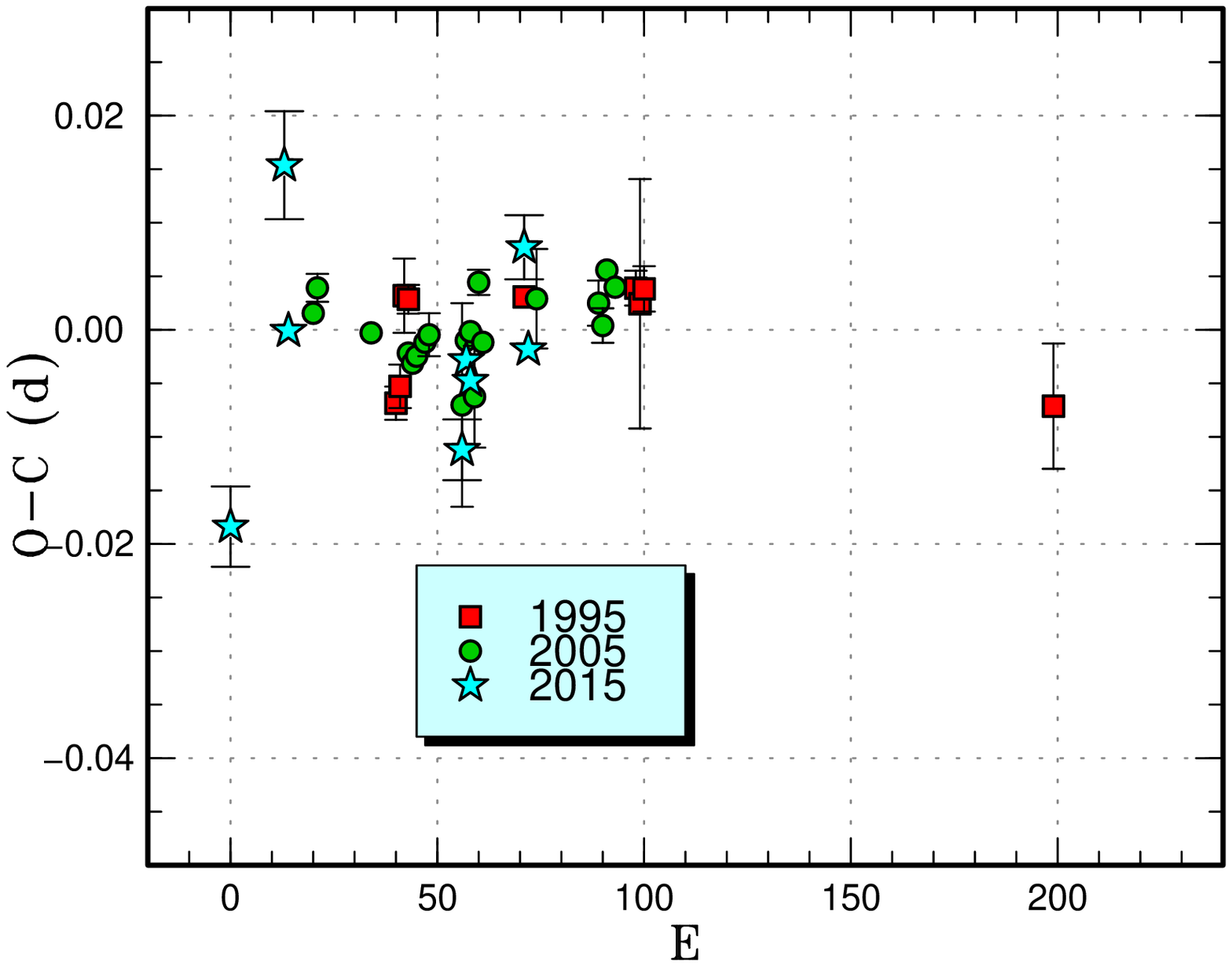}
  \end{center}
  \caption{Comparison of $O-C$ diagrams of V701 Tau between different
  superoutbursts.  A period of 0.06899~d was used to draw this figure.
  Approximate cycle counts ($E$) after the start of the superoutburst
  were used.  We assumed that the 2015 superoutburst was detected
  soon after the maximum and shifted other superoutbursts to
  get the best fit.}
  \label{fig:v701taucomp2}
\end{figure}

   The 2015 superoutburst was detected by K. Paxson
on February 16 (vsnet-alert 18301).  This is the first
secure superoutburst since 2007.  Subsequent observations
detected superhumps (vsnet-alert 18308, 18312).
The times of superhump maxima are listed in table
\ref{tab:v701tauoc2015}.  Although $E=0$ appears to be
a stage A superhump (figure \ref{fig:v701taucomp2}),
the period of stage A superhumps could not be determined.

\begin{table}
\caption{Superhump maxima of V701 Tau (2015)}\label{tab:v701tauoc2015}
\begin{center}
\begin{tabular}{rp{55pt}p{40pt}r@{.}lr}
\hline
\multicolumn{1}{c}{$E$} & \multicolumn{1}{c}{max\commenta} & \multicolumn{1}{c}{error} & \multicolumn{2}{c}{$O-C$\commentb} & \multicolumn{1}{c}{$N$\commentc} \\
\hline
0 & 57070.3621 & 0.0038 & $-$0&0147 & 68 \\
13 & 57071.2927 & 0.0050 & 0&0186 & 41 \\
14 & 57071.3462 & 0.0008 & 0&0031 & 75 \\
56 & 57074.2327 & 0.0028 & $-$0&0097 & 27 \\
57 & 57074.3101 & 0.0005 & $-$0&0014 & 77 \\
58 & 57074.3771 & 0.0010 & $-$0&0034 & 77 \\
71 & 57075.2865 & 0.0030 & 0&0086 & 37 \\
72 & 57075.3459 & 0.0006 & $-$0&0010 & 67 \\
\hline
  \multicolumn{6}{l}{\commenta BJD$-$2400000.} \\
  \multicolumn{6}{l}{\commentb Against max $= 2457070.3768 + 0.069030 E$.} \\
  \multicolumn{6}{l}{\commentc Number of points used to determine the maximum.} \\
\end{tabular}
\end{center}
\end{table}

\subsection{SU Ursae Majoris}\label{obj:suuma}

   This famous dwarf nova (=5.1908) was discovered
by \citet{cer08suuma}.  Hoffmeister identified this
object to be a U Gem-type object with cycle lengths
of 10--20~d.  \citet{elv43DNspec} observed the object
in outburst and obtained a spectrum B5 with H$\alpha$
emission.  \citet{pet60suuma} reported a summary of
outbursts and identified long and short outbursts.
According to this literature, Brun and Petit considered
a class of SU UMa-type variables and introduced
a concept of supermaxima in 1952.  \citet{isl74suuma}
also reported historical variations in 1926--1954
and proposed a mean cycle length between long outbursts
to be 170~d.  \citet{tho86suuma} determined the orbital
period by a radial-velocity study.
Although the object had long been
proposed to be the prototype of SU UMa-type dwarf novae,
it was only in 1989 when superhumps were detected
during a superoutburst \citet{uda90suuma}, after
unsuccessful reports before a superoutburst
\citep{bar81suuma} and during a normal outburst
\citep{uda88suuma}.  This delay of detection of superhumps
was partly caused by the decrease of frequency of
superoutbursts between 1980 and 1989, which prevented
scheduled observational campaigns (cf. \cite{uda90suuma}).

   We reported observations of the 1999, 2010 and 2013
superoutbursts in \citet{Pdot}, \citet{Pdot2} and
\citet{Pdot6}, respectively.

\begin{figure}
  \begin{center}
    \FigureFile(88mm,70mm){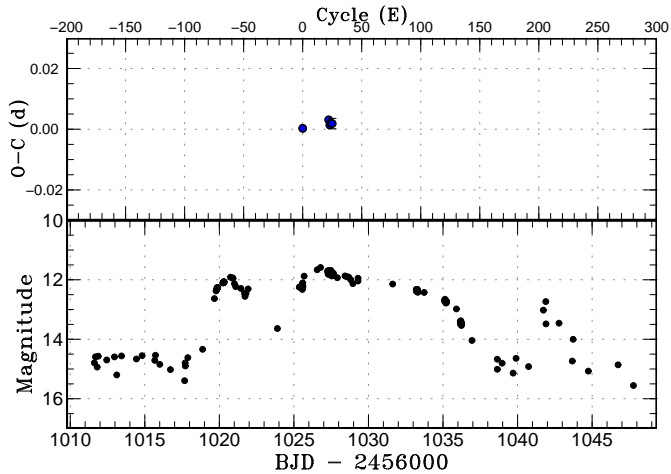}
  \end{center}
  \caption{$O-C$ diagram of superhumps in SU UMa (2014).
     (Upper): $O-C$ diagram.  A period of 0.07918~d
     was used to draw this figure.
     (Lower): Light curve.  The observations were binned to 0.01~d.
     A precursor outburst and a post-superoutburst rebrightening
     were clearly recorded.}
  \label{fig:suumahumpall}
\end{figure}

\begin{figure}
  \begin{center}
    \FigureFile(88mm,70mm){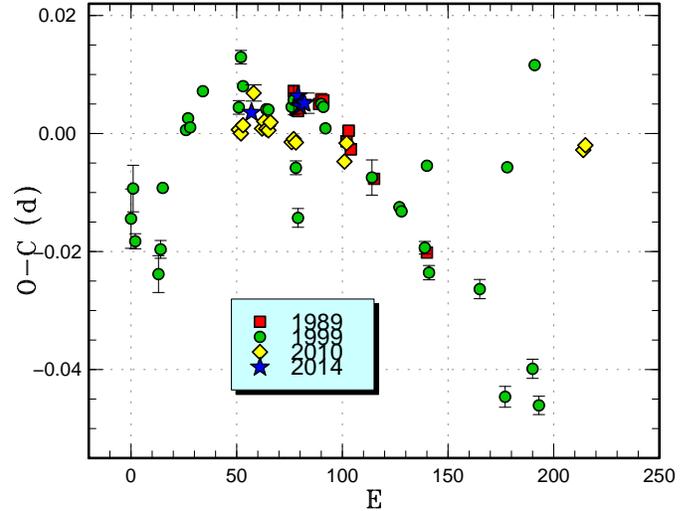}
  \end{center}
  \caption{Comparison of $O-C$ diagrams of SU UMa between different
  superoutbursts.  A period of 0.07918~d was used to draw this figure.
  Approximate cycle counts ($E$) after the start of the superoutburst
  were used.  For the 2014 superoutburst, the cycle count from
  the maximum of the precursor outburst was used.
  The two epochs in the 2010 data may be traditional
  late superhumps.}
  \label{fig:suumacomp2}
\end{figure}

   The rise to the 2014--2015 superoutburst took place on 
2014 December 27 according to the AAVSO data.
The outburst was a precursor outburst and the final rise
to the superoutburst was on 2015 January 1--2
(vsnet-alert 18142; figure \ref{fig:suumahumpall}).
There was a post-superoutburst rebrightening
on January 18--19 (figure \ref{fig:suumahumpall}).
Time-resolved photometry was obtained rather fragmentarily.
Although there were also observations in the late stage
of the plateau phase, superhumps became unclear.
Only the times of superhump maxima in the earlier stage
of the superoutburst are listed in table
\ref{tab:suumaoc2014}.
A comparison of $O-C$ diagrams suggests that the superhumps
started to grow around the maximum of the precursor outburst
in the 2014 superoutburst (figure \ref{fig:suumacomp2}).

\begin{table}
\caption{Superhump maxima of SU UMa (2014)}\label{tab:suumaoc2014}
\begin{center}
\begin{tabular}{rp{55pt}p{40pt}r@{.}lr}
\hline
\multicolumn{1}{c}{$E$} & \multicolumn{1}{c}{max\commenta} & \multicolumn{1}{c}{error} & \multicolumn{2}{c}{$O-C$\commentb} & \multicolumn{1}{c}{$N$\commentc} \\
\hline
0 & 57025.5893 & 0.0006 & $-$0&0001 & 36 \\
22 & 57027.3341 & 0.0010 & 0&0011 & 166 \\
23 & 57027.4115 & 0.0005 & $-$0&0006 & 146 \\
24 & 57027.4913 & 0.0006 & $-$0&0001 & 160 \\
25 & 57027.5704 & 0.0017 & $-$0&0003 & 49 \\
\hline
  \multicolumn{6}{l}{\commenta BJD$-$2400000.} \\
  \multicolumn{6}{l}{\commentb Against max $= 2457025.5894 + 0.079252 E$.} \\
  \multicolumn{6}{l}{\commentc Number of points used to determine the maximum.} \\
\end{tabular}
\end{center}
\end{table}

\subsection{CY Ursae Majoris}\label{obj:cyuma}

   This object was identified as an SU UMa-type dwarf nova
in 1988 (\cite{kat88cyuma}; \cite{kat97cyuma}).  The correct
superhump period was established by \citet{har95cyuma}.
See \citet{Pdot5} for more history.

   The 2014 superoutburst was detected by H. Maehara
(vsnet-alert 16967), which stared as a separate precursor
outburst on March 3.  The object once faded, but was detected
in outburst again after six days (vsnet-alert 16998).
This outburst turned out to be a genuine superoutburst.
Such a separate precursor was first observed in CY UMa.

   The times of superhump maxima are listed in table
\ref{tab:cyumaoc2014}.  Although $E=0$ corresponded to
the growing stage of superhumps (during the final rise
to the superoutburst), the period of stage A superhumps
was not meaningfully determined.  The transition from
stage B to C was rather smooth in this object and
the distinction between these stages is rather ambiguous.
The epochs for $E \ge 125$ correspond to the rapid
fading phase and post-superoutburst phase.
Since the observations were rather fragmentary in these
phases, we did not attempt to determine the period
after the superoutburst.

\begin{figure}
  \begin{center}
    \FigureFile(88mm,70mm){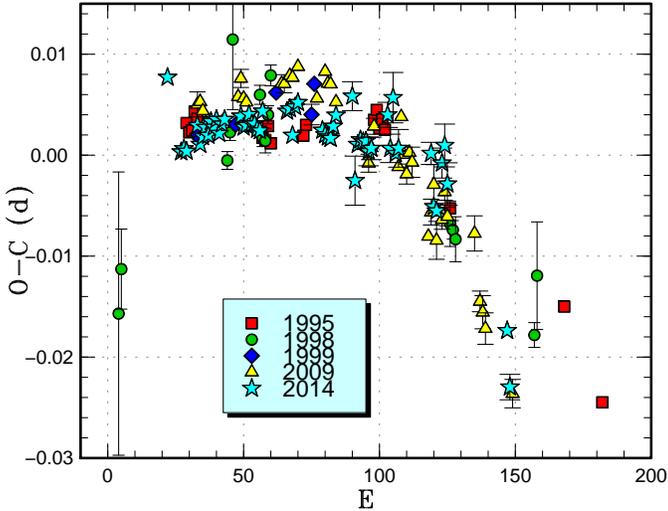}
  \end{center}
  \caption{Comparison of $O-C$ diagrams of CY UMa between different
  superoutbursts.  A period of 0.07212~d was used to draw this figure.
  Approximate cycle counts ($E$) after the start of the superoutburst
  were used.  The 2014 superoutburst was shifted by 22 cycles
  to best match the others.}
  \label{fig:cyumacomp2}
\end{figure}

   A comparison of the $O-C$ diagrams (figure \ref{fig:cyumacomp2})
indicates that the $O-C$ diagram of the 2014 superoutburst
well matched the others if we assume that superhumps
started to grow 22 cycles before $E=0$.  It implies that
the duration of stage A was 2.0~d (up to $E=5$), and it is
evident that superhumps started to grow after the
precursor outburst.

\begin{figure}
  \begin{center}
    \FigureFile(88mm,70mm){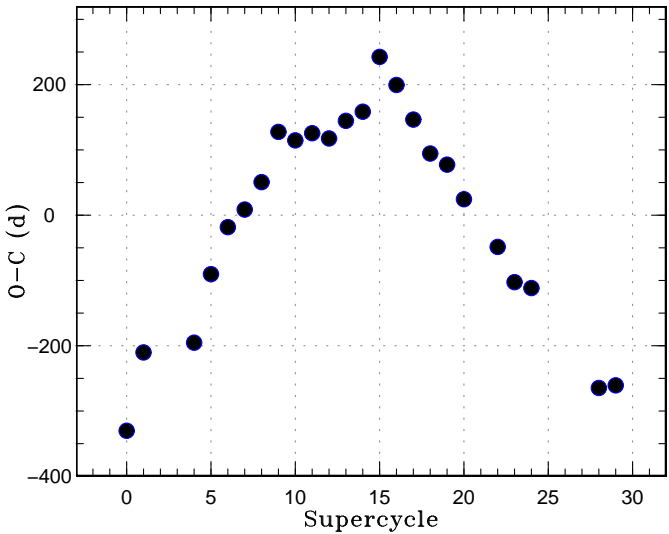}
  \end{center}
  \caption{$O-C$ diagram of superoutbursts in CY UMa.
  The ephemeris used was Max(JD)$=2447824+327.0E$.
  In 2003, the supercycle suddenly decreased from 362~d to 290~d.}
  \label{fig:cyumasooc}
\end{figure}

   Since there is no published statistics of outbursts of
CY UMa, we summarized the list of outbursts from the
readily available data (table \ref{tab:cyumaout}).
It has become evident that normal outbursts were more
frequently recorded in the late 1980s to early 1990s.
Most of superoutbursts were detected except two between
1989 and 1991, one between 2006 and 2008 and likely three
between 2010 and 2013.  These outburst detections heavily
relied on a few expert observers, and the variation of
the activity of these observers strongly affected
the detections.  The lack of superoutburst between 2010 and
2013 was probably a result of the decrease of visual observers
regularly watching this object.  This decrease has been
partly compensated by CCD monitoring, but it is not
still sufficient particularly around solar conjunctions.
It has also become evident that supercycle length
has experienced a dramatic change around 2003
(figure \ref{fig:cyumasooc}).
Before 2003 (JD 2452545), the mean supercycle was 362(3)~d.
After this, the value decreased to 290(1)~d.
Since this change was not apparently associated with
the increase of normal outbursts, it is less likely
to be a result of an increased mass-transfer rate.
It would be interesting to see the evolution of the
supercycle length in future.

\begin{table}
\caption{Superhump maxima of CY UMa (2014)}\label{tab:cyumaoc2014}
\begin{center}
\begin{tabular}{rp{55pt}p{40pt}r@{.}lr}
\hline
\multicolumn{1}{c}{$E$} & \multicolumn{1}{c}{max\commenta} & \multicolumn{1}{c}{error} & \multicolumn{2}{c}{$O-C$\commentb} & \multicolumn{1}{c}{$N$\commentc} \\
\hline
0 & 56726.1519 & 0.0006 & 0&0080 & 131 \\
5 & 56726.5052 & 0.0002 & 0&0003 & 101 \\
6 & 56726.5776 & 0.0002 & 0&0006 & 158 \\
7 & 56726.6495 & 0.0002 & 0&0003 & 137 \\
11 & 56726.9405 & 0.0005 & 0&0026 & 71 \\
12 & 56727.0107 & 0.0005 & 0&0007 & 79 \\
13 & 56727.0836 & 0.0003 & 0&0014 & 155 \\
14 & 56727.1567 & 0.0002 & 0&0022 & 177 \\
15 & 56727.2280 & 0.0002 & 0&0014 & 220 \\
16 & 56727.3015 & 0.0002 & 0&0027 & 238 \\
17 & 56727.3730 & 0.0002 & 0&0021 & 155 \\
18 & 56727.4459 & 0.0002 & 0&0028 & 149 \\
19 & 56727.5167 & 0.0002 & 0&0014 & 153 \\
20 & 56727.5897 & 0.0003 & 0&0022 & 154 \\
21 & 56727.6623 & 0.0005 & 0&0026 & 67 \\
27 & 56728.0953 & 0.0004 & 0&0026 & 127 \\
28 & 56728.1664 & 0.0005 & 0&0015 & 233 \\
29 & 56728.2387 & 0.0006 & 0&0016 & 205 \\
30 & 56728.3119 & 0.0004 & 0&0026 & 231 \\
31 & 56728.3830 & 0.0003 & 0&0017 & 127 \\
32 & 56728.4552 & 0.0002 & 0&0016 & 189 \\
33 & 56728.5268 & 0.0003 & 0&0010 & 199 \\
34 & 56728.5987 & 0.0004 & 0&0008 & 155 \\
35 & 56728.6728 & 0.0006 & 0&0027 & 46 \\
44 & 56729.3219 & 0.0007 & 0&0022 & 99 \\
45 & 56729.3942 & 0.0004 & 0&0024 & 116 \\
46 & 56729.4637 & 0.0003 & $-$0&0003 & 174 \\
47 & 56729.5387 & 0.0003 & 0&0025 & 161 \\
48 & 56729.6112 & 0.0003 & 0&0029 & 144 \\
57 & 56730.2575 & 0.0007 & $-$0&0004 & 137 \\
58 & 56730.3292 & 0.0005 & $-$0&0009 & 148 \\
59 & 56730.4011 & 0.0003 & $-$0&0012 & 120 \\
\hline
  \multicolumn{6}{l}{\commenta BJD$-$2400000.} \\
  \multicolumn{6}{l}{\commentb Against max $= 2456726.1440 + 0.072174 E$.} \\
  \multicolumn{6}{l}{\commentc Number of points used to determine the maximum.} \\
\end{tabular}
\end{center}
\end{table}

\addtocounter{table}{-1}
\begin{table}
\caption{Superhump maxima of CY UMa (2014) (continued)}
\begin{center}
\begin{tabular}{rp{55pt}p{40pt}r@{.}lr}
\hline
\multicolumn{1}{c}{$E$} & \multicolumn{1}{c}{max\commenta} & \multicolumn{1}{c}{error} & \multicolumn{2}{c}{$O-C$\commentb} & \multicolumn{1}{c}{$N$\commentc} \\
\hline
60 & 56730.4731 & 0.0004 & $-$0&0014 & 151 \\
61 & 56730.5464 & 0.0003 & $-$0&0002 & 155 \\
62 & 56730.6196 & 0.0004 & 0&0008 & 104 \\
68 & 56731.0542 & 0.0014 & 0&0024 & 41 \\
69 & 56731.1180 & 0.0024 & $-$0&0060 & 27 \\
70 & 56731.1937 & 0.0004 & $-$0&0025 & 235 \\
71 & 56731.2663 & 0.0004 & $-$0&0021 & 244 \\
72 & 56731.3382 & 0.0003 & $-$0&0023 & 230 \\
73 & 56731.4104 & 0.0007 & $-$0&0023 & 79 \\
74 & 56731.4815 & 0.0014 & $-$0&0034 & 57 \\
75 & 56731.5539 & 0.0011 & $-$0&0032 & 68 \\
81 & 56731.9899 & 0.0012 & $-$0&0002 & 54 \\
82 & 56732.0587 & 0.0014 & $-$0&0035 & 54 \\
83 & 56732.1359 & 0.0025 & 0&0014 & 127 \\
84 & 56732.2025 & 0.0009 & $-$0&0042 & 215 \\
85 & 56732.2751 & 0.0019 & $-$0&0037 & 71 \\
97 & 56733.1400 & 0.0011 & $-$0&0048 & 138 \\
98 & 56733.2069 & 0.0009 & $-$0&0102 & 167 \\
99 & 56733.2786 & 0.0004 & $-$0&0107 & 133 \\
101 & 56733.4276 & 0.0015 & $-$0&0060 & 18 \\
102 & 56733.5014 & 0.0021 & $-$0&0043 & 34 \\
103 & 56733.5697 & 0.0021 & $-$0&0082 & 35 \\
125 & 56735.1419 & 0.0007 & $-$0&0239 & 68 \\
126 & 56735.2084 & 0.0013 & $-$0&0296 & 110 \\
156 & 56737.4295 & 0.0014 & 0&0263 & 36 \\
170 & 56738.4478 & 0.0008 & 0&0342 & 38 \\
181 & 56739.2186 & 0.0006 & 0&0111 & 73 \\
182 & 56739.2815 & 0.0011 & 0&0018 & 76 \\
\hline
  \multicolumn{6}{l}{\commenta BJD$-$2400000.} \\
  \multicolumn{6}{l}{\commentb Against max $= 2456726.1440 + 0.072174 E$.} \\
  \multicolumn{6}{l}{\commentc Number of points used to determine the maximum.} \\
\end{tabular}
\end{center}
\end{table}

\begin{table*}
\caption{List of recent outbursts of CY UMa.}\label{tab:cyumaout}
\begin{center}
\begin{tabular}{cccccl}
\hline
Year & Month & max\commenta & magnitude & type & source \\
\hline
1988 & 1 & 47167 & 12.3 & super & VSOLJ; \citet{kat88cyuma}; \citet{kat97cyuma} \\
1988 & 5 & 47299 & 12.5 & normal & VSOLJ; \citet{wat89cyuma} \\
1988 & 10 & 47464 & 12.2 & normal & VSOLJ; \citet{wat89cyuma} \\
1989 & 1 & 47528 & 12.9 & normal & VSOLJ; \citet{wat89cyuma} \\
1989 & 3 & 47614 & 12.3 & super & P. Schmeer; \citet{wat89cyuma} \\
1989 & 6 & 47679 & 12.8 & normal & VSOLJ \\
1990 & 1 & 47915 & 12.5 & normal & VSOLJ \\
1990 & 4 & 47986 & 12.9 & normal & VSOLJ, AAVSO \\
1990 & 9 & 48156 & 12.3 & normal & VSOLJ, AAVSO \\
1990 & 12 & 48252 & 13.3 & normal & Schmeer \\
1991 & 3 & 48335 & 13.3 & normal & VSOLJ, AAVSO \\
1991 & 12 & 48610 & 12.8 & super & VSOLJ, AAVSO; \citet{kat95cyuma} \\
1993 & 2 & 49042 & 12.2 & super & VSOLJ, AAVSO \\
1993 & 6 & 49155 & 12.9 & normal & AAVSO \\
1994 & 3 & 49441 & 11.7 & super & VSOLJ, AAVSO \\
1995 & 3 & 49795 & 12.1 & super & VSOLJ, AAVSO; \citet{har95cyuma}; \citet{Pdot} \\
1995 & 5 & 49859 & 13.0 & normal & AAVSO \\
1995 & 7 & 49915 & 13.3 & normal & AAVSO \\
1996 & 3 & 50164 & 12.2 & super & AAVSO, VSOLJ \\
1996 & 6 & 50245 & 13.4 & normal & AAVSO \\
1996 & 12 & 50442 & 12.9 & normal & AAVSO \\
1997 & 3 & 50516 & 14.1 & ?\commentb & AAVSO \\
1997 & 3 & 50520 & 13.7 & ?\commentc & AAVSO \\
1997 & 4 & 50559 & 12.6 & precursor?\commentb & AAVSO \\
1997 & 4 & 50568 & 12.3 & super & AAVSO, VSOLJ \\
1997 & 10 & 50726 & 12.6 & normal & AAVSO \\
1998 & 3 & 50882 & 12.4 & super & AAVSO, VSOLJ; \citet{Pdot} \\
1998 & 5 & 50948 & l3.8 & normal & AAVSO \\
1998 & 12 & 51167 & 13.2 & normal\commentd & VSOLJ \\
1999 & 2 & 51220 & l2.5 & super & AAVSO, VSOLJ; \citet{kat99cyuma}; \citet{Pdot} \\
1999 & 6 & 51335 & 12.8 & normal & AAVSO \\
1999 & 12 & 51539 & 12.5 & super & AAVSO, VSOLJ \\
2000 & 2 & 51600 & 13.0 & normal & AAVSO \\
2000 & 4 & 51654 & 13.2 & normal & AAVSO \\
\hline
  \multicolumn{6}{l}{\commenta JD$-$2400000.} \\
  \multicolumn{6}{l}{\commentb CCD single detection.} \\
  \multicolumn{6}{l}{\commentc Detection by two visual observers, but contradict a CCD observation two hours later.} \\
  \multicolumn{6}{l}{\commentd Single visual detection.} \\
\end{tabular}
\end{center}
\end{table*}

\addtocounter{table}{-1}
\begin{table*}
\caption{List of recent outbursts of CY UMa (continued).}
\begin{center}
\begin{tabular}{cccccl}
\hline
Year & Month & max\commenta & magnitude & type & source \\
\hline
2000 & 6 & 51716 & 13.1 & normal & AAVSO \\
2000 & 12 & 51893 & 12.5 & super & AAVSO, VSOLJ \\
2001 & 6 & 52064 & 13.1 & normal & AAVSO \\
2001 & 11 & 52234 & 12.3 & super & AAVSO \\
2002 & 1 & 52300 & 13.9 & normal & AAVSO \\
2002 & 5 & 52422 & 12.9 & normal & AAVSO \\
2002 & 11 & 52601 & 12.9 & normal\commentb & AAVSO \\
2003 & 1 & 52645 & 12.4 & super & AAVSO, VSOLJ \\
2003 & 3 & 52717 & 12.7 & normal & AAVSO \\
2003 & 6 & 52818 & 12.9 & normal\commentb & AAVSO \\
2003 & 10 & 52929 & 12.3 & super & AAVSO \\
2003 & 12 & 52998 & 14.4\commentd & normal & AAVSO \\
2004 & 2 & 53055 & 13.0 & normal & AAVSO, VSOLJ \\
2004 & 4 & 53122 & 12.9 & normal & AAVSO \\
2004 & 7 & 53203 & 12.3 & super & AAVSO \\
2004 & 11 & 53315 & 12.3 & normal & AAVSO \\
2005 & 1 & 53398 & 12.9 & normal & AAVSO \\
2005 & 4 & 53478 & 12.5 & super & AAVSO, VSOLJ \\
2005 & 11 & 53676 & 12.5 & normal & AAVSO \\
2006 & 2 & 53788 & 12.8 & super & AAVSO \\
2006 & 6 & 53802 & 13.8 & normal & AAVSO \\
2006 & 11 & 54062 & 12.4 & super & vsnet-alert 9148; AAVSO \\
2007 & 4 & 54195 & 13.0 & normal & AAVSO \\
2008 & 6 & 54643 & l2.5 & super & AAVSO \\
2009 & 3 & 54916 & 12.6 & super & vsnet-alert 11127; AAVSO; \citet{Pdot} \\
2009 & 5 & 54972 & 12.6 & normal & AAVSO \\
2010 & 2 & 55234 & 12.3 & super & vsnet-alert 11811; AAVSO \\
2010 & 4 & 55302 & 13.2 & normal\commentb & AAVSO \\
2010 & 5 & 55330 & 15.1 & normal\commentb & AAVSO \\
2010 & 12 & 55558 & 13.5 & normal & AAVSO \\
2011 & 2 & 55615 & 13.3 & normal & AAVSO \\
2012 & 1 & 55944 & 14.6 & normal\commentb & AAVSO \\
2012 & 2 & 55986 & 13.1 & normal\commentb & AAVSO \\
2012 & 5 & 56078 & 13.7 & normal\commentd & AAVSO \\
2013 & 4 & 56389 & 12.5 & super & vsnet-alert 15596; AAVSO \\
2014 & 3 & 56720 & 12.6 & precursor + super & vsnet-alert 16967; this paper \\
\hline
  \multicolumn{6}{l}{\commenta JD$-$2400000.} \\
  \multicolumn{6}{l}{\commentb CCD single detection.} \\
  \multicolumn{6}{l}{\commentc Detection by two visual observers, but contradict a CCD observation two hours later.} \\
  \multicolumn{6}{l}{\commentd Single visual detection.} \\
\end{tabular}
\end{center}
\end{table*}

\subsection{QZ Virginis}\label{obj:qzvir}

   QZ Vir (former T Leo) is the second dwarf nova discovered
in the history.  See \citet{kat97tleo} for a summary of
the history.  In recent years, this object has shown
superoutbursts approximately once a year.

\begin{figure}
  \begin{center}
    \FigureFile(88mm,70mm){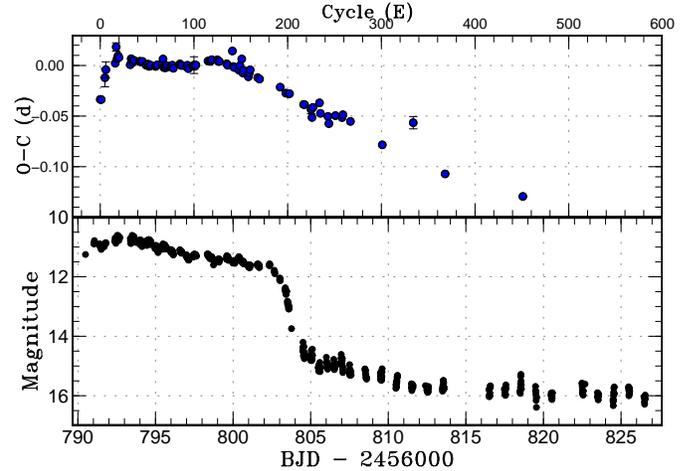}
  \end{center}
  \caption{$O-C$ diagram of superhumps in QZ Vir (2014).
     (Upper): $O-C$ diagram.  A period of 0.06038~d
     was used to draw this figure.
     (Lower): Light curve.  The observations were binned to 0.01~d.}
  \label{fig:qzvirhumpall}
\end{figure}

   The 2014 superoutburst was detected by R. Modic in 
relatively early phase (vsnet-alert 17291).  After further brightening
(vsnet-alert 17296), evolving superhumps were observed
(vsnet-alert 17301).  This part was later found to be
a precursor part of a superoutburst and fully developed
superhumps were recorded near the peak brightness.
The times of superhump maxima during the superoutburst
plateau are listed in table \ref{tab:qzviroc2014}.
All stages A--C were recorded (see also figure
\ref{fig:qzvirhumpall}).  After the plateau phase,
superhumps were also continuously observed
(table \ref{tab:qzviroc2014b}).  The superhumps in the
post-superoutburst phase were on a smooth continuation
of stage C superhumps, and there was no phase $\sim$0.5
jump as expected for ``traditional'' late superhumps.
This finding confirmed the earlier finding \citep{Pdot}.

   Although stage A superhumps were detected, the coverage
of the 2014 observation was not good enough to determine
the period sufficiently.  An analysis of the $O-C$ data
suggested a period of 0.0625(3)~d, which corresponds to
$q$=0.18(2).  However, since the baseline is only 1~d and the
coverage was not so good, we should not heavily rely
on this value.  The absence of stage A superhumps
during the 1993 superoutburst still remains a mystery
(\cite{kat97tleo}; \cite{Pdot}).  If we can shift the $O-C$
diagram of the 1993 superoutburst by 50 cycles,
the resultant $O-C$ diagram appears to fit the others
(figure \ref{fig:qzvirhumpall}; this figure used
the data in \cite{ohs11qzvir}).  This implies that
superhumps started to grow 3~d earlier than the precursor
outburst.  If superhumps had already grown at the time of
the precursor outburst, the lack of stage A superhumps
may be reconciled, although such early development
of superhumps may be a challenge to the standard thermal-tidal
disk instability model.

\begin{figure}
  \begin{center}
    \FigureFile(88mm,70mm){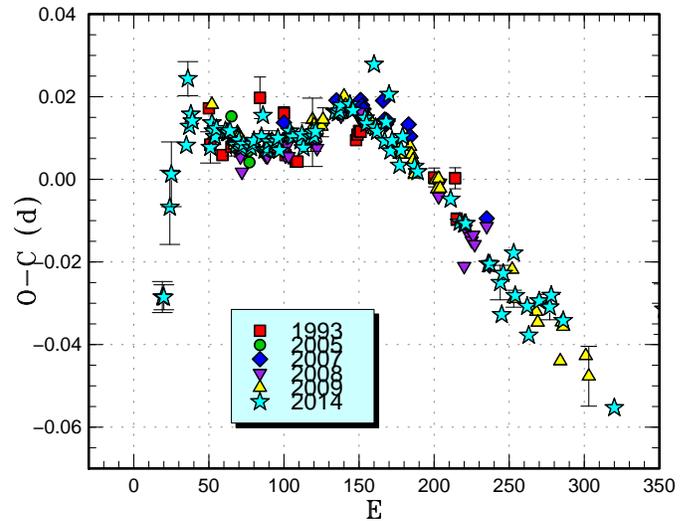}
  \end{center}
  \caption{Comparison of $O-C$ diagrams of QZ Vir between different
  superoutbursts.  A period of 0.06032~d was used to draw this figure.
  Approximate cycle counts ($E$) after the starts of outbursts
  were used.  Since the starts of the 2005, 2007 and 2008
  superoutbursts were not well constrained, we shifted
  the $O-C$ diagrams to best fit the others.
  The 1993 superoutburst was shifted by 50 cycles,
  which implies that superhumps started to grow 3~d before
  the precursor.}
  \label{fig:qzvircomp}
\end{figure}

\begin{table}
\caption{Superhump maxima of QZ Vir (2014)}\label{tab:qzviroc2014}
\begin{center}
\begin{tabular}{rp{55pt}p{40pt}r@{.}lr}
\hline
\multicolumn{1}{c}{$E$} & \multicolumn{1}{c}{max\commenta} & \multicolumn{1}{c}{error} & \multicolumn{2}{c}{$O-C$\commentb} & \multicolumn{1}{c}{$N$\commentc} \\
\hline
0 & 56791.4075 & 0.0038 & $-$0&0334 & 27 \\
1 & 56791.4677 & 0.0030 & $-$0&0335 & 23 \\
5 & 56791.7308 & 0.0090 & $-$0&0120 & 83 \\
6 & 56791.7991 & 0.0078 & $-$0&0041 & 78 \\
16 & 56792.4093 & 0.0017 & 0&0024 & 56 \\
17 & 56792.4858 & 0.0041 & 0&0185 & 26 \\
18 & 56792.5345 & 0.0005 & 0&0069 & 32 \\
19 & 56792.5978 & 0.0005 & 0&0097 & 19 \\
20 & 56792.6565 & 0.0018 & 0&0080 & 6 \\
32 & 56793.3739 & 0.0038 & 0&0010 & 22 \\
33 & 56793.4402 & 0.0008 & 0&0068 & 28 \\
34 & 56793.4985 & 0.0004 & 0&0048 & 22 \\
35 & 56793.5575 & 0.0006 & 0&0034 & 31 \\
36 & 56793.6197 & 0.0006 & 0&0052 & 25 \\
42 & 56793.9811 & 0.0005 & 0&0044 & 437 \\
43 & 56794.0410 & 0.0001 & 0&0039 & 1235 \\
44 & 56794.1013 & 0.0002 & 0&0038 & 1248 \\
45 & 56794.1621 & 0.0008 & 0&0043 & 137 \\
49 & 56794.3999 & 0.0011 & 0&0005 & 29 \\
51 & 56794.5217 & 0.0007 & 0&0016 & 32 \\
52 & 56794.5799 & 0.0003 & $-$0&0006 & 82 \\
53 & 56794.6406 & 0.0004 & $-$0&0003 & 100 \\
54 & 56794.7017 & 0.0002 & 0&0005 & 159 \\
59 & 56795.0025 & 0.0003 & $-$0&0007 & 339 \\
60 & 56795.0637 & 0.0002 & 0&0002 & 326 \\
61 & 56795.1248 & 0.0003 & 0&0009 & 474 \\
66 & 56795.4274 & 0.0003 & 0&0016 & 242 \\
67 & 56795.4929 & 0.0009 & 0&0067 & 25 \\
68 & 56795.5448 & 0.0009 & $-$0&0018 & 49 \\
69 & 56795.6053 & 0.0004 & $-$0&0016 & 110 \\
70 & 56795.6654 & 0.0005 & $-$0&0019 & 86 \\
71 & 56795.7278 & 0.0027 & 0&0001 & 26 \\
75 & 56795.9682 & 0.0018 & $-$0&0009 & 101 \\
76 & 56796.0294 & 0.0002 & $-$0&0002 & 450 \\
\hline
  \multicolumn{6}{l}{\commenta BJD$-$2400000.} \\
  \multicolumn{6}{l}{\commentb Against max $= 2456791.4409 + 0.060377 E$.} \\
  \multicolumn{6}{l}{\commentc Number of points used to determine the maximum.} \\
\end{tabular}
\end{center}
\end{table}

\addtocounter{table}{-1}
\begin{table}
\caption{Superhump maxima of QZ Vir (2014) (continued)}
\begin{center}
\begin{tabular}{rp{55pt}p{40pt}r@{.}lr}
\hline
\multicolumn{1}{c}{$E$} & \multicolumn{1}{c}{max\commenta} & \multicolumn{1}{c}{error} & \multicolumn{2}{c}{$O-C$\commentb} & \multicolumn{1}{c}{$N$\commentc} \\
\hline
77 & 56796.0906 & 0.0002 & 0&0006 & 513 \\
78 & 56796.1479 & 0.0009 & $-$0&0024 & 259 \\
85 & 56796.5748 & 0.0017 & 0&0019 & 26 \\
86 & 56796.6339 & 0.0006 & 0&0006 & 83 \\
87 & 56796.6941 & 0.0011 & 0&0004 & 81 \\
92 & 56796.9947 & 0.0009 & $-$0&0009 & 142 \\
93 & 56797.0566 & 0.0003 & 0&0006 & 485 \\
94 & 56797.1136 & 0.0007 & $-$0&0027 & 394 \\
99 & 56797.4179 & 0.0034 & $-$0&0004 & 18 \\
100 & 56797.4793 & 0.0083 & 0&0007 & 10 \\
101 & 56797.5386 & 0.0016 & $-$0&0004 & 41 \\
102 & 56797.6002 & 0.0006 & 0&0008 & 100 \\
115 & 56798.3891 & 0.0006 & 0&0048 & 132 \\
117 & 56798.5095 & 0.0016 & 0&0044 & 29 \\
118 & 56798.5702 & 0.0008 & 0&0048 & 84 \\
119 & 56798.6319 & 0.0005 & 0&0061 & 93 \\
125 & 56798.9941 & 0.0007 & 0&0060 & 65 \\
126 & 56799.0529 & 0.0006 & 0&0044 & 64 \\
127 & 56799.1133 & 0.0009 & 0&0045 & 39 \\
135 & 56799.5940 & 0.0003 & 0&0022 & 85 \\
136 & 56799.6529 & 0.0005 & 0&0007 & 88 \\
141 & 56799.9689 & 0.0016 & 0&0148 & 48 \\
142 & 56800.0139 & 0.0006 & $-$0&0005 & 76 \\
143 & 56800.0733 & 0.0006 & $-$0&0015 & 79 \\
148 & 56800.3726 & 0.0006 & $-$0&0041 & 114 \\
149 & 56800.4375 & 0.0012 & 0&0004 & 63 \\
151 & 56800.5648 & 0.0014 & 0&0070 & 55 \\
152 & 56800.6114 & 0.0007 & $-$0&0068 & 82 \\
153 & 56800.6743 & 0.0007 & $-$0&0043 & 82 \\
158 & 56800.9699 & 0.0012 & $-$0&0106 & 64 \\
159 & 56801.0339 & 0.0010 & $-$0&0069 & 104 \\
160 & 56801.0975 & 0.0012 & $-$0&0038 & 26 \\
168 & 56801.5729 & 0.0008 & $-$0&0114 & 67 \\
170 & 56801.6922 & 0.0010 & $-$0&0128 & 54 \\
\hline
  \multicolumn{6}{l}{\commenta BJD$-$2400000.} \\
  \multicolumn{6}{l}{\commentb Against max $= 2456791.4409 + 0.060377 E$.} \\
  \multicolumn{6}{l}{\commentc Number of points used to determine the maximum.} \\
\end{tabular}
\end{center}
\end{table}

\begin{table}
\caption{Superhump maxima of QZ Vir (2014) (post-superoutburst)}\label{tab:qzviroc2014b}
\begin{center}
\begin{tabular}{rp{55pt}p{40pt}r@{.}lr}
\hline
\multicolumn{1}{c}{$E$} & \multicolumn{1}{c}{max\commenta} & \multicolumn{1}{c}{error} & \multicolumn{2}{c}{$O-C$\commentb} & \multicolumn{1}{c}{$N$\commentc} \\
\hline
0 & 56803.0126 & 0.0002 & 0&0059 & 240 \\
6 & 56803.3689 & 0.0005 & 0&0022 & 131 \\
9 & 56803.5494 & 0.0007 & 0&0028 & 28 \\
10 & 56803.6099 & 0.0007 & 0&0033 & 19 \\
25 & 56804.5049 & 0.0010 & $-$0&0015 & 27 \\
26 & 56804.5652 & 0.0011 & $-$0&0013 & 27 \\
33 & 56804.9830 & 0.0042 & $-$0&0034 & 22 \\
34 & 56805.0355 & 0.0004 & $-$0&0109 & 331 \\
35 & 56805.1058 & 0.0005 & $-$0&0006 & 266 \\
42 & 56805.5330 & 0.0019 & 0&0067 & 28 \\
43 & 56805.5830 & 0.0015 & $-$0&0033 & 26 \\
51 & 56806.0630 & 0.0012 & $-$0&0033 & 60 \\
52 & 56806.1163 & 0.0021 & $-$0&0099 & 35 \\
59 & 56806.5468 & 0.0015 & 0&0007 & 27 \\
66 & 56806.9676 & 0.0031 & 0&0015 & 44 \\
67 & 56807.0307 & 0.0011 & 0&0047 & 64 \\
75 & 56807.5072 & 0.0015 & 0&0012 & 29 \\
109 & 56809.5370 & 0.0014 & $-$0&0086 & 27 \\
142 & 56811.5514 & 0.0061 & 0&0261 & 24 \\
176 & 56813.5537 & 0.0017 & $-$0&0112 & 19 \\
259 & 56818.5430 & 0.0012 & $-$0&0011 & 23 \\
\hline
  \multicolumn{6}{l}{\commenta BJD$-$2400000.} \\
  \multicolumn{6}{l}{\commentb Against max $= 2456803.0067 + 0.059990 E$.} \\
  \multicolumn{6}{l}{\commentc Number of points used to determine the maximum.} \\
\end{tabular}
\end{center}
\end{table}

\subsection{NSV 1436}\label{obj:nsv1436}

   NSV 1436 was one of variable stars (Ross 4) reported
in \citet{ros25nsv1436}.  Although \citet{ros25nsv1436}
gave a photographic range of 12--16 mag, the variability
type was not known.  T. Kato pointed out in 2000
that the object can be identified with an ROSAT
X-ray source (vsnet-chat 3326).\footnote{
  \citet{bro10nsv1436} incorrectly gave the first reference
of the ROSAT identification of this object.
}  Starting from 2000, a search for outburst started
mostly by visual observers.  This attempt had been
unsuccessful up to 2010.
In \citet{bro10nsv1436}, studied the historical
photographic material and characterized two outbursts
in 1904 and 1948.  The object was mostly in faint state
fainter than 15.8 mag and occasionally observed in outburst.
\citet{bro10nsv1436} suggested that the object might be
a recurrent nova.

   In 2011, a fresh outburst was reported by E. Muyllaert
at an unfiltered CCD magnitude of 13.49 on March 28,
and the outburst was confirmed by M. Linnolt
a visual magnitude of 12.8.  There was a suspected
fainter outburst on March 9 \citep{tem11nsv1436aan}.
Despite an intensive observing campaign, the object
rapidly faded.  \citet{osb11nsv1436} observed this outburst
with Swift satellite and found that the X-ray properties were
not those of a recurrent nova.  The optical light variation
suggested the dwarf nova classification \citep{osb11nsv1436}.

   A total of 14 outbursts were recorded between 2011 
and 2014 March, and the object was found to undergo outbursts 
more frequently than supposed.  \citet{pag14RNcand} also listed
this object as an ordinary dwarf nova.

\begin{figure}
  \begin{center}
    \FigureFile(88mm,110mm){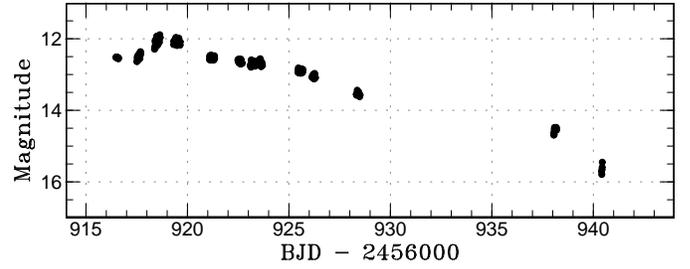}
  \end{center}
  \caption{Light curve of the superoutburst of NSV 1436
     (2014).  The data were binned to 0.01~d.
     The initial precursor part of the outburst and
     subsequent rise to the superoutburst maximum are
     clearly depicted.  The widths of the light curve
     mainly reflect the amplitudes of superhumps.}
  \label{fig:nsv1436lc}
\end{figure}

\begin{figure}
  \begin{center}
    \FigureFile(88mm,110mm){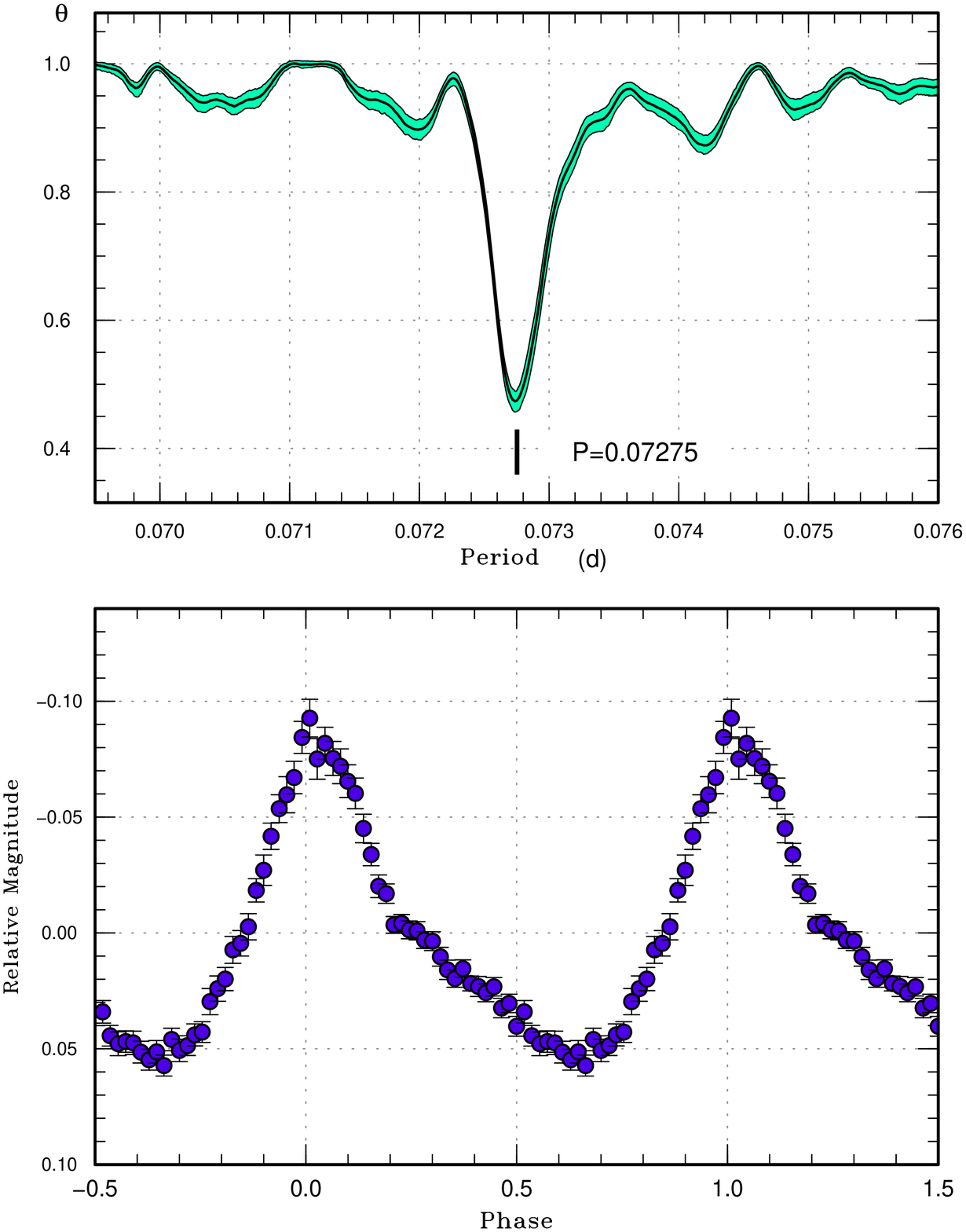}
  \end{center}
  \caption{Superhumps in NSV 1436 during the superoutburst
     plateau (2014).  (Upper): PDM analysis.
     (Lower): Phase-averaged profile.}
  \label{fig:nsv1436shpdm}
\end{figure}

   In 2014 September, a bright outburst of thie object
was visually detected by P. Schmeer (11.8 mag on September 15,
vsnet-alert 17729).  Subsequent observations finally
revealed the emergence of superhumps after a rise from
a precursor outburst (figure \ref{fig:nsv1436lc};
vsnet-alert 17736).  Further observations
confirmed superhump (vsnet-alert 17754, 17770;
figure \ref{fig:nsv1436shpdm}).

   The times of superhump maxima are listed in table
\ref{tab:nsv1436oc2014}.  The maxima for $E \le 2$ were likely
the final part of stage A.  Although there were observations
on the preceding night, the amplitudes were too small to
determine the period of stage A superhumps.
As is usual for an object with this moderate superhump
period, the transition between stages was not very sharp.
We could, however, identify stages B and C.  The resultant
behavior was typical for an SU UMa-type dwarf nova.

\begin{table}
\caption{Superhump maxima of NSV 1436 (2014)}\label{tab:nsv1436oc2014}
\begin{center}
\begin{tabular}{rp{55pt}p{40pt}r@{.}lr}
\hline
\multicolumn{1}{c}{$E$} & \multicolumn{1}{c}{max\commenta} & \multicolumn{1}{c}{error} & \multicolumn{2}{c}{$O-C$\commentb} & \multicolumn{1}{c}{$N$\commentc} \\
\hline
0 & 56917.5356 & 0.0004 & $-$0&0071 & 71 \\
1 & 56917.6102 & 0.0003 & $-$0&0053 & 77 \\
2 & 56917.6816 & 0.0004 & $-$0&0065 & 63 \\
12 & 56918.4124 & 0.0003 & $-$0&0030 & 80 \\
13 & 56918.4862 & 0.0003 & $-$0&0019 & 109 \\
14 & 56918.5589 & 0.0003 & $-$0&0019 & 109 \\
15 & 56918.6323 & 0.0003 & $-$0&0013 & 154 \\
26 & 56919.4332 & 0.0004 & $-$0&0002 & 113 \\
27 & 56919.5075 & 0.0003 & 0&0014 & 106 \\
28 & 56919.5782 & 0.0003 & $-$0&0007 & 95 \\
29 & 56919.6517 & 0.0057 & 0&0001 & 29 \\
49 & 56921.1041 & 0.0005 & $-$0&0019 & 127 \\
50 & 56921.1796 & 0.0005 & 0&0010 & 142 \\
51 & 56921.2519 & 0.0017 & 0&0005 & 50 \\
52 & 56921.3246 & 0.0009 & 0&0005 & 56 \\
68 & 56922.4948 & 0.0015 & 0&0071 & 53 \\
69 & 56922.5628 & 0.0007 & 0&0025 & 76 \\
70 & 56922.6393 & 0.0007 & 0&0062 & 76 \\
77 & 56923.1497 & 0.0007 & 0&0076 & 53 \\
78 & 56923.2208 & 0.0007 & 0&0060 & 55 \\
79 & 56923.2940 & 0.0005 & 0&0065 & 54 \\
82 & 56923.5114 & 0.0005 & 0&0058 & 77 \\
83 & 56923.5856 & 0.0003 & 0&0072 & 78 \\
84 & 56923.6569 & 0.0008 & 0&0058 & 77 \\
109 & 56925.4690 & 0.0005 & $-$0&0001 & 66 \\
110 & 56925.5404 & 0.0006 & $-$0&0014 & 78 \\
111 & 56925.6123 & 0.0004 & $-$0&0022 & 75 \\
112 & 56925.6895 & 0.0016 & 0&0023 & 36 \\
119 & 56926.1949 & 0.0008 & $-$0&0014 & 53 \\
120 & 56926.2684 & 0.0007 & $-$0&0005 & 53 \\
149 & 56928.3673 & 0.0008 & $-$0&0105 & 80 \\
150 & 56928.4361 & 0.0009 & $-$0&0144 & 76 \\
\hline
  \multicolumn{6}{l}{\commenta BJD$-$2400000.} \\
  \multicolumn{6}{l}{\commentb Against max $= 2456917.5428 + 0.072719 E$.} \\
  \multicolumn{6}{l}{\commentc Number of points used to determine the maximum.} \\
\end{tabular}
\end{center}
\end{table}

\subsection{NSV 4618}\label{obj:nsv4618}

\begin{figure}
  \begin{center}
    \FigureFile(88mm,110mm){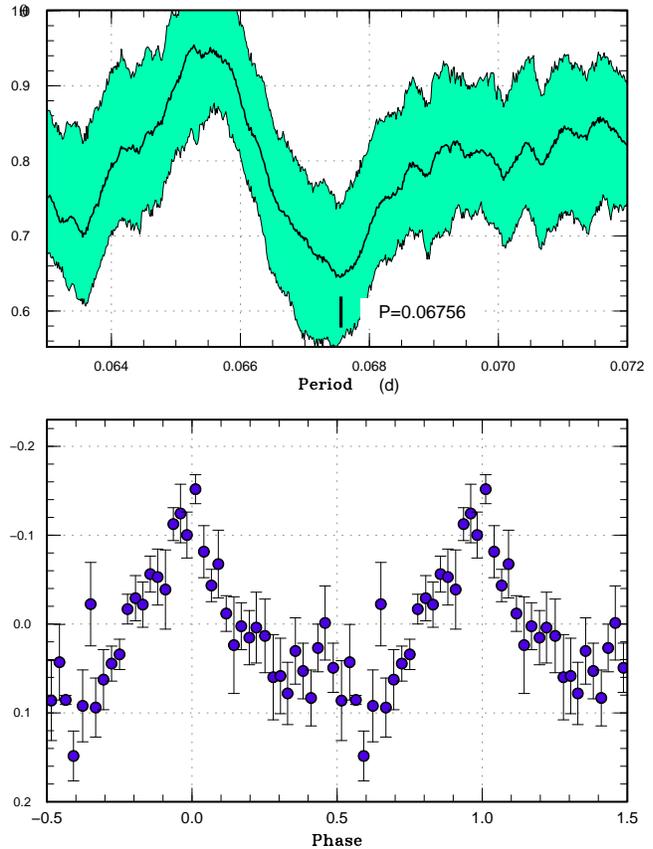}
  \end{center}
  \caption{Superhumps in NSV 4618 outside the eclipses (2015).
     Due to the limited data segment, the error is relatively
     large.  Post-superoutburst observations were not included
     in the analysis. 
     (Upper): PDM analysis.
     (Lower): Phase-averaged profile.}
  \label{fig:nsv4618shpdm}
\end{figure}

\begin{figure}
  \begin{center}
    \FigureFile(88mm,120mm){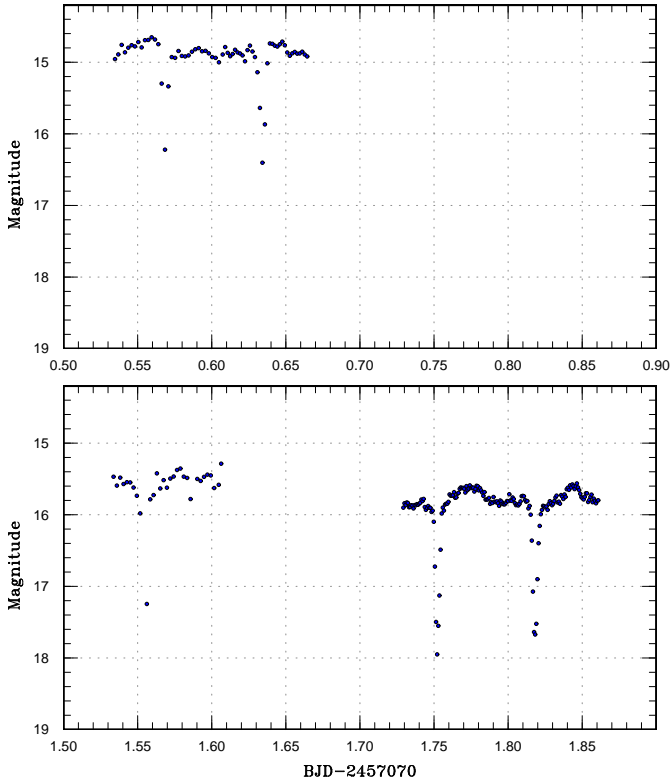}
  \end{center}
  \caption{Sample light curve of NSV 4618 in superoutburst.
  Superposition of superhumps and eclipse are well visible.}
  \label{fig:nsv4618lc}
\end{figure}

   NSV 4618 was discovered by \citet{luy38propermotion2}
as a variable object.  A photographic range of 14.0 to
fainter than 16 was reported.  Although the identity of
the object had long been unknown, CRTS Siding Spring Survey (SSS)
detected an outburst on 2013 April 13
(cf. vsnet-alert 15611).  The object was recorded in outburst
several times in ASAS-3 data.  The brightest outburst reached
$V$=13.3.  The eclipsing nature of this object was inferred
(vsnet-alert 15612, 15613, 15614).  Using the CRTS data,
an orbital period was determined (vsnet-alert 15615) and
the object was recognized as a candidate eclipsing SU UMa-type
dwarf nova.  The 2015 outburst was detected on February 14
by the ASAS-SN team (vsnet-alert 18297).  Subsequent observations
detected deep eclipses and superhumps
(vsnet-alert 18305, 18306; figures \ref{fig:nsv4618lc},
figure \ref{fig:nsv4618shpdm}).

   An MCMC analysis of both the CRTS data
and the present data in outburst yielded the following
orbital ephemeris:
\begin{equation}
{\rm Min(BJD)} = 2454884.13441(2) + 0.0657692860(6) E .
\label{equ:nsv4618ecl}
\end{equation}

   The times of superhump maxima outside the eclipses
are listed in table \ref{tab:nsv4618oc2015}.
Since the object faded three days after the initial
observation, it was likely we only observed the terminal
stage of the superoutburst.

\begin{table}
\caption{Superhump maxima of NSV 4618 (2015)}\label{tab:nsv4618oc2015}
\begin{center}
\begin{tabular}{rp{50pt}p{30pt}r@{.}lcr}
\hline
$E$ & max\commenta & error & \multicolumn{2}{c}{$O-C$\commentb} & phase\commentc & $N$\commentd \\
\hline
0 & 57070.5565 & 0.0014 & $-$0&0080 & 0.82 & 17 \\
1 & 57070.6404 & 0.0020 & 0&0084 & 0.09 & 24 \\
15 & 57071.5842 & 0.0128 & 0&0071 & 0.44 & 17 \\
18 & 57071.7727 & 0.0005 & $-$0&0069 & 0.31 & 56 \\
19 & 57071.8445 & 0.0006 & $-$0&0027 & 0.40 & 49 \\
30 & 57072.5944 & 0.0212 & 0&0047 & 0.80 & 19 \\
31 & 57072.6545 & 0.0036 & $-$0&0027 & 0.72 & 18 \\
\hline
  \multicolumn{7}{l}{\commenta BJD$-$2400000.} \\
  \multicolumn{7}{l}{\commentb Against max $= 2457070.5645 + 0.067506 E$.} \\
  \multicolumn{7}{l}{\commentc Orbital phase.} \\
  \multicolumn{7}{l}{\commentd Number of points used to determine the maximum.} \\
\end{tabular}
\end{center}
\end{table}

\subsection{1RXS J185310.0$+$594509}\label{obj:j1853}

   This object (hereafter 1RXS J185310) was originally
selected as a variable optical counterpart of an ROSAT
X-ray source \citep{den11ROSATCVs}.  The designation
DDE 14 was also given \citep{den11dde20dde21}.\footnote{
  For a list of DDE variables, see
  $<$http://hea.iki.rssi.ru/$\sim$denis/VarDDE.html$>$.
}
The object was found in outburst
on Palomar Observatory Sky Survey II infrared plate on
1992 June 28.  On 2011 May 8, J. Shears detected an outburst 
at an unfiltered CCD magnitude of 15.9 (all magnitudes 
are unfiltered for this object unless otherwise noted;
cvnet-outburst 4145).
On 2011 November 22, J. Shears also
detected an outburst at a magnitude of
15.1 (cvnet-outburst 4405).  During this outburst,
J. Shears detected superhump-like modulations, but it
was not confirmed because the observations were obtained
only for a night.  Two further outbursts were recorded
on 2012 December 5-11 (peak brightness at 15.16 on
December 9) and 2013 August 16 (16.6 mag) by J. Shears.
The 2013 August outburst was recorded by the ASAS-SN team
at $V$=15.18 on August 15 (vsnet-alert 16247).
Two faint outbursts in 2014 were detected by MASTER network
(vsnet-alert 17500).

\begin{figure}
  \begin{center}
    \FigureFile(88mm,110mm){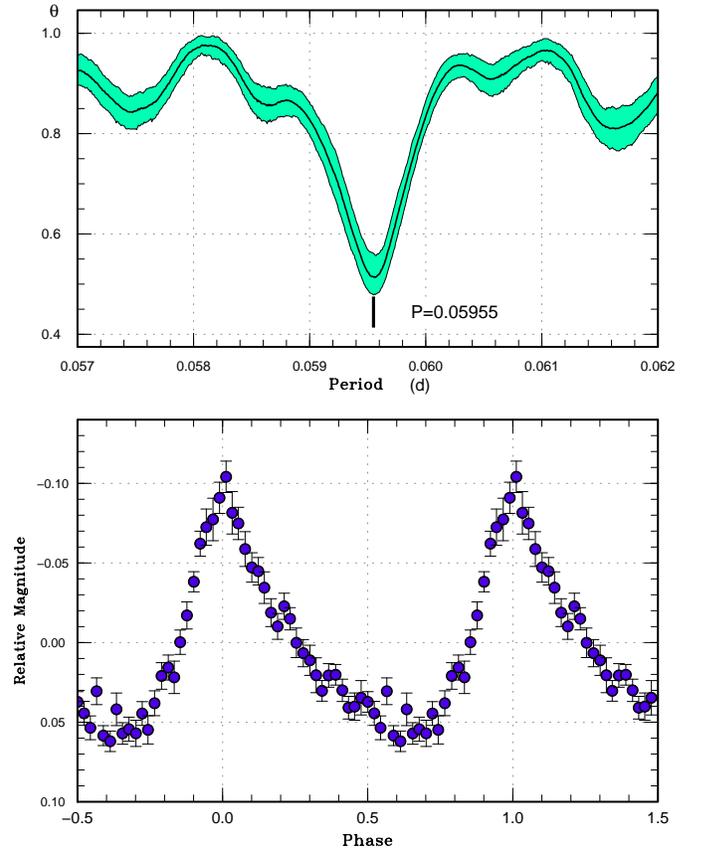}
  \end{center}
  \caption{Superhumps in 1RXS J185310 (2014).  (Upper): PDM analysis.
     (Lower): Phase-averaged profile.}
  \label{fig:j1853shpdm}
\end{figure}

   The 2014 July outburst was detected by J. Shears
at a magnitude of 14.8 on July 16 (vsnet-outburst 17115).
Subsequent observations detected superhumps
(vsnet-alert 17514, 17532; figure \ref{fig:j1853shpdm}).

   The times of superhump maxima are listed in table
\ref{tab:j1853oc2014}.
Although the superhump stage was not identified,
these superhumps were likely stage B superhumps
as judged from the large amplitudes.

   The two long outbursts in 2011 November and 2012
December must have been superoutbursts.
The interval between these two superoutburst was
379~d.  The interval between the 2012 and current
superoutbursts was 590~d.  The supercycle may be
around 190~d, which needs to be confirmed by further
observations.

\begin{table}
\caption{Superhump maxima of 1RXS J185310 (2014)}\label{tab:j1853oc2014}
\begin{center}
\begin{tabular}{rp{55pt}p{40pt}r@{.}lr}
\hline
\multicolumn{1}{c}{$E$} & \multicolumn{1}{c}{max\commenta} & \multicolumn{1}{c}{error} & \multicolumn{2}{c}{$O-C$\commentb} & \multicolumn{1}{c}{$N$\commentc} \\
\hline
0 & 56857.6878 & 0.0020 & 0&0016 & 54 \\
1 & 56857.7458 & 0.0003 & 0&0001 & 104 \\
2 & 56857.8060 & 0.0004 & 0&0008 & 104 \\
3 & 56857.8639 & 0.0005 & $-$0&0009 & 104 \\
14 & 56858.5174 & 0.0004 & $-$0&0021 & 64 \\
15 & 56858.5863 & 0.0009 & 0&0073 & 20 \\
23 & 56859.0538 & 0.0008 & $-$0&0014 & 22 \\
24 & 56859.1098 & 0.0033 & $-$0&0050 & 49 \\
25 & 56859.1747 & 0.0012 & 0&0005 & 95 \\
30 & 56859.4746 & 0.0028 & 0&0027 & 29 \\
31 & 56859.5276 & 0.0004 & $-$0&0037 & 64 \\
32 & 56859.5879 & 0.0006 & $-$0&0029 & 62 \\
57 & 56861.0800 & 0.0009 & 0&0011 & 57 \\
58 & 56861.1392 & 0.0012 & 0&0008 & 58 \\
59 & 56861.1978 & 0.0011 & $-$0&0001 & 19 \\
64 & 56861.4936 & 0.0024 & $-$0&0019 & 44 \\
65 & 56861.5583 & 0.0010 & 0&0033 & 63 \\
\hline
  \multicolumn{6}{l}{\commenta BJD$-$2400000.} \\
  \multicolumn{6}{l}{\commentb Against max $= 2456857.6862 + 0.059521 E$.} \\
  \multicolumn{6}{l}{\commentc Number of points used to determine the maximum.} \\
\end{tabular}
\end{center}
\end{table}

\subsection{1RXS J231935.0$+$364705}\label{obj:j2319}

   This object (=DDE 8, hereafter 1RXS J231935) was selected as
a likely dwarf nova during the course of identification
of the ROSAT sources \citep{den11ROSATCVs}.
The 2011 outburst was a superoutburst 
during which superhumps were detected \citep{Pdot4}.
The 2013 outburst also a superoutburst \citep{Pdot6}.
The 2005 outburst was also most likely a superoutburst
which was recorded by CRTS.  The 2010 May outburst was
also bright (CRTS data) but only single night observation
was available.

\begin{figure}
  \begin{center}
    \FigureFile(88mm,70mm){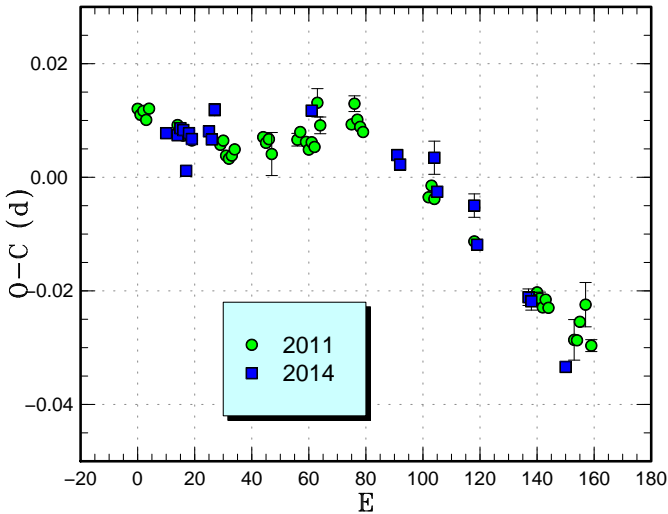}
  \end{center}
  \caption{Comparison of $O-C$ diagrams of 1RXS J231935 between different
  superoutbursts.  A period of 0.06600~d was used to draw this figure.
  Approximate cycle counts ($E$) after the start of observations
  were used.  The 2014 superoutburst was shifted by 10 cycles
  to best match the 2011 one.}
  \label{fig:j2319comp}
\end{figure}

   The 2014 outburst was detected by D. Denisenko
at an unfiltered CCD magnitude of 13.9 (vsnet-alert 17701).
Subsequent observations detected superhumps
(vsnet-alert 17703, 17717, 17739).
The times of superhump maxima are listed in table
\ref{tab:j2319oc2014}.  Although observations were not
dense enough in the middle part of the outburst,
stages B and C were recorded.  The $O-C$ diagrams
were very similar between the 2011 and 2014 superoutbursts
(figure \ref{fig:j2319comp}).

   The known outburst of this object is listed in
table \ref{tab:j2319out}.  The shortest interval between
outbursts was 74~d.  Although supercycle was not well
determined, the shortest known interval between 
superoutbursts was $\sim$420~d.

\begin{table}
\caption{Superhump maxima of 1RXS J231935 (2014)}\label{tab:j2319oc2014}
\begin{center}
\begin{tabular}{rp{55pt}p{40pt}r@{.}lr}
\hline
\multicolumn{1}{c}{$E$} & \multicolumn{1}{c}{max\commenta} & \multicolumn{1}{c}{error} & \multicolumn{2}{c}{$O-C$\commentb} & \multicolumn{1}{c}{$N$\commentc} \\
\hline
0 & 56908.0625 & 0.0002 & $-$0&0035 & 73 \\
4 & 56908.3262 & 0.0003 & $-$0&0030 & 100 \\
5 & 56908.3934 & 0.0003 & $-$0&0015 & 133 \\
6 & 56908.4591 & 0.0003 & $-$0&0017 & 131 \\
7 & 56908.5179 & 0.0008 & $-$0&0086 & 111 \\
8 & 56908.5905 & 0.0003 & $-$0&0018 & 207 \\
9 & 56908.6555 & 0.0004 & $-$0&0026 & 192 \\
15 & 56909.0529 & 0.0005 & $-$0&0001 & 130 \\
16 & 56909.1174 & 0.0004 & $-$0&0013 & 171 \\
17 & 56909.1886 & 0.0011 & 0&0041 & 52 \\
51 & 56911.4325 & 0.0009 & 0&0109 & 46 \\
81 & 56913.4047 & 0.0004 & 0&0092 & 71 \\
82 & 56913.4690 & 0.0008 & 0&0077 & 50 \\
94 & 56914.2622 & 0.0029 & 0&0114 & 36 \\
95 & 56914.3222 & 0.0004 & 0&0056 & 99 \\
108 & 56915.1778 & 0.0021 & 0&0058 & 39 \\
109 & 56915.2369 & 0.0007 & $-$0&0009 & 48 \\
127 & 56916.4156 & 0.0015 & $-$0&0065 & 24 \\
128 & 56916.4809 & 0.0016 & $-$0&0070 & 25 \\
140 & 56917.2614 & 0.0010 & $-$0&0161 & 49 \\
\hline
  \multicolumn{6}{l}{\commenta BJD$-$2400000.} \\
  \multicolumn{6}{l}{\commentb Against max $= 2456908.0660 + 0.065796 E$.} \\
  \multicolumn{6}{l}{\commentc Number of points used to determine the maximum.} \\
\end{tabular}
\end{center}
\end{table}

\begin{table*}
\caption{List of outbursts of 1RXS J231935}\label{tab:j2319out}
\begin{center}
\begin{tabular}{cccccl}
\hline
Year & Month & max\commenta & magnitude & type & source \\
\hline
1999 & 11 & 51496 & 13.7 & super & NSVS\commentb \\
2005 & 10 & 53675 & 13.9 & super & CRTS \\
2009 & 11 & 55140 & 14.5 & normal & \citet{den09j2319atel2282} \\ 
2010 &  5 & 55347 & 13.9 & super? & CRTS \\
2010 & 12 & 55557 & 14.2 & normal & vsnet-alert 12532 \\
2011 &  9 & 55834 & 13.7 & super & \citet{Pdot4} \\
2012 & 10 & 56230 & 14.4 & normal & vsnet-alert 15072 \\
2013 &  7 & 56489 & 14.4 & super? & vsnet-alert 15998 \\
2013 &  9 & 56563 & 15.2 & normal & vsnet-alert 16460 \\
2014 &  9 & 56908 & 14.0 & super & this paper \\
\hline
  \multicolumn{5}{l}{\commenta JD$-$2400000.} \\
  \multicolumn{5}{l}{\commentb NSVS 9022680, see \citet{den09j2319atel2282}.} \\
\end{tabular}
\end{center}
\end{table*}

\subsection{2QZ J130441.7$+$010330}\label{obj:j1304}

   This object (hereafter 2QZ J130441) was identified as a CV
during the course of the 2dF QSO Redshift Survey \citep{cro04qz7}.
Although little had been known about this object,
\citet{kat12DNSDSS} estimated the orbital period to be
0.064--0.069~d from the SDSS colors, suggesting 
an SU UMa-type dwarf nova.
ASAS-SN team detected this object in outburst on 2014
March 20 at 15.65 mag.

\begin{figure}
  \begin{center}
    \FigureFile(88mm,110mm){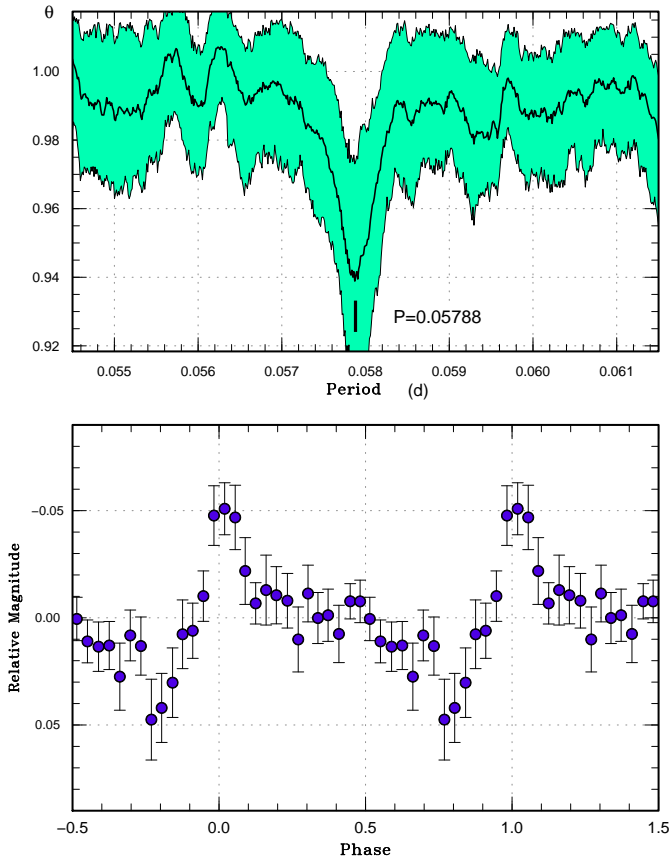}
  \end{center}
  \caption{Superhumps in 2QZ J130441 (2014).  (Upper): PDM analysis.
     (Lower): Phase-averaged profile.}
  \label{fig:j1304shpdm}
\end{figure}

   Subsequent observation detected superhumps
(vsnet-alert 17079, 17087; figure \ref{fig:j1304shpdm}).
The times of superhump maxima are listed in table
\ref{tab:j1304oc2014}.  Since the object started fading 
rapidly on the fifth night of the observation, we
probably observed only stage C superhumps.

\begin{table}
\caption{Superhump maxima of 2QZ J130441 (2014)}\label{tab:j1304oc2014}
\begin{center}
\begin{tabular}{rp{55pt}p{40pt}r@{.}lr}
\hline
\multicolumn{1}{c}{$E$} & \multicolumn{1}{c}{max\commenta} & \multicolumn{1}{c}{error} & \multicolumn{2}{c}{$O-C$\commentb} & \multicolumn{1}{c}{$N$\commentc} \\
\hline
0 & 56740.6716 & 0.0019 & 0&0061 & 20 \\
1 & 56740.7227 & 0.0058 & $-$0&0009 & 20 \\
2 & 56740.7843 & 0.0016 & 0&0027 & 21 \\
3 & 56740.8399 & 0.0040 & 0&0002 & 25 \\
17 & 56741.6542 & 0.0020 & 0&0016 & 21 \\
18 & 56741.7064 & 0.0017 & $-$0&0043 & 20 \\
19 & 56741.7653 & 0.0011 & $-$0&0035 & 20 \\
20 & 56741.8290 & 0.0015 & 0&0022 & 22 \\
34 & 56742.6373 & 0.0064 & $-$0&0025 & 22 \\
35 & 56742.6940 & 0.0029 & $-$0&0039 & 19 \\
36 & 56742.7534 & 0.0023 & $-$0&0025 & 21 \\
37 & 56742.8175 & 0.0017 & 0&0035 & 21 \\
38 & 56742.8690 & 0.0028 & $-$0&0031 & 21 \\
51 & 56743.6258 & 0.0021 & $-$0&0012 & 24 \\
52 & 56743.6816 & 0.0035 & $-$0&0034 & 21 \\
53 & 56743.7330 & 0.0031 & $-$0&0101 & 20 \\
54 & 56743.7959 & 0.0023 & $-$0&0053 & 19 \\
59 & 56744.0991 & 0.0018 & 0&0076 & 45 \\
60 & 56744.1743 & 0.0020 & 0&0247 & 63 \\
63 & 56744.3159 & 0.0014 & $-$0&0079 & 43 \\
69 & 56744.6750 & 0.0117 & 0&0028 & 22 \\
70 & 56744.7259 & 0.0047 & $-$0&0044 & 21 \\
71 & 56744.7946 & 0.0265 & 0&0062 & 19 \\
72 & 56744.8419 & 0.0030 & $-$0&0046 & 18 \\
\hline
  \multicolumn{6}{l}{\commenta BJD$-$2400000.} \\
  \multicolumn{6}{l}{\commentb Against max $= 2456740.6655 + 0.058069 E$.} \\
  \multicolumn{6}{l}{\commentc Number of points used to determine the maximum.} \\
\end{tabular}
\end{center}
\end{table}

\subsection{ASASSN-13cx}\label{obj:asassn13cx}

   This object was detected as a transient at $V$=15.47.
on 2013 September 14 by ASAS-SN team.  The coordinates are
\timeform{00h 02m 22.34s}, \timeform{+42D 42' 13.4''}
(the Initial Gaia Source List).
CRTS data detected six past outbursts.
This object was detected in outburst again by E. Muyllaert
on 2014 August 31 (cvnet-outburst 6073).
Noting that there were two faint measurements in the CRTS
data, an observational campaign started to verify
possible eclipses (vsnet-alert 17682).
The object immediately turned out to be an SU UMa-type
dwarf nova with deep eclipses (vsnet-alert 17684;
figure \ref{fig:asassn13cxlc}).

\begin{figure}
  \begin{center}
    \FigureFile(88mm,70mm){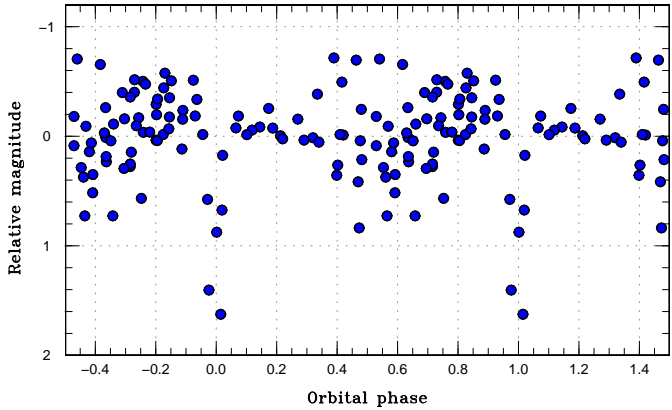}
  \end{center}
  \caption{Orbital light curve of ASASSN-13cx in the CRTS data.}
  \label{fig:asassn13cxqui}
\end{figure}

\begin{figure}
  \begin{center}
    \FigureFile(88mm,120mm){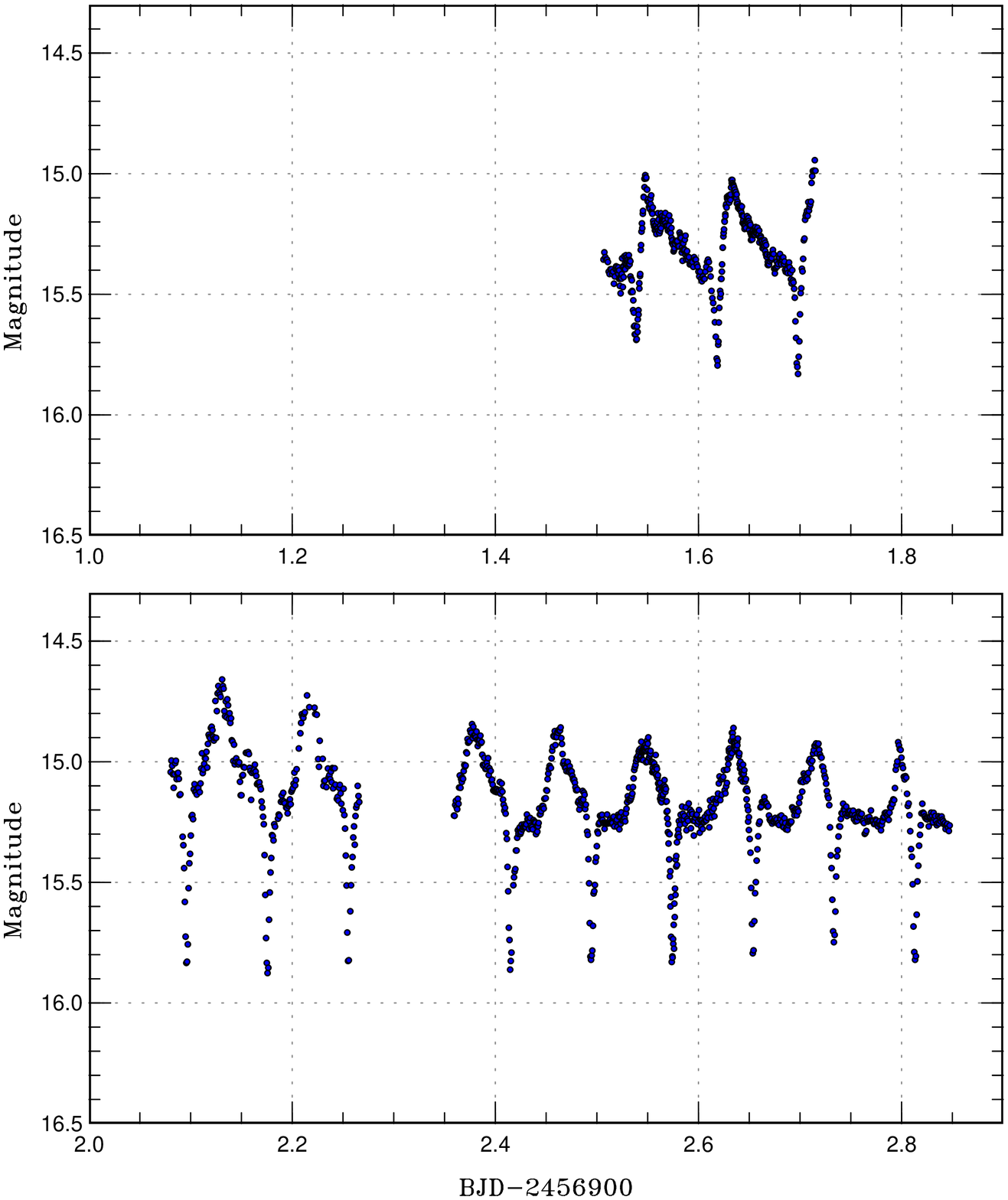}
  \end{center}
  \caption{Sample light curve of ASASSN-13cx in superoutburst.
  Superposition of superhumps and eclipse are well visible.}
  \label{fig:asassn13cxlc}
\end{figure}

   An MCMC analysis of both the CRTS data
and the present data in outburst yielded the following
orbital ephemeris (vsnet-alert 17691):
\begin{equation}
{\rm Min(BJD)} = 2456901.69823(5) + 0.079650075(5) E .
\label{equ:asassn13cxecl}
\end{equation}
The orbital light curve in quiescence based on this
ephemeris is in figure \ref{fig:asassn13cxqui}.
Orbital humps before eclipses were present.

\begin{figure}
  \begin{center}
    \FigureFile(88mm,110mm){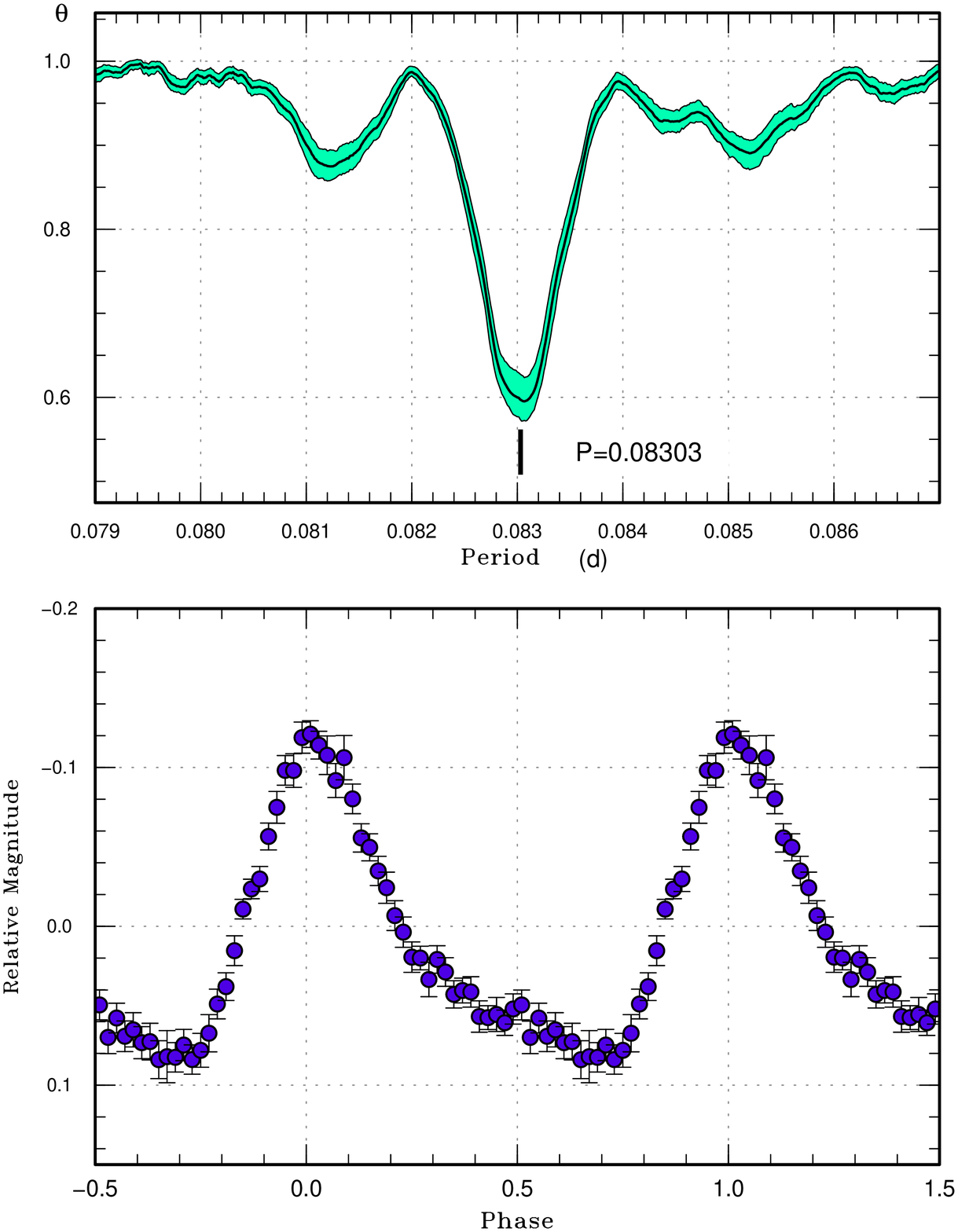}
  \end{center}
  \caption{Superhumps in ASASSN-13cx outside the eclipses (2014).
     (Upper): PDM analysis.
     (Lower): Phase-averaged profile.}
  \label{fig:asassn13cxshpdm}
\end{figure}

   The times of superhump maxima are listed in table
\ref{tab:asassn13cxoc2014}.  Stages B and C can be
identified.  Although $E \le 2$ appears to corresponds to
stage A superhumps, the period of stage A superhumps
could not be determined.  The mean superhump profile
outside the eclipses is shown in figure \ref{fig:asassn13cxshpdm}.

   Since the 2013 outburst was likely a superoutburst
as judged from the magnitude, the supercycle can be
estimated to be $\sim$350~d or possibly its half.

\begin{table}
\caption{Superhump maxima of ASASSN-13cx (2014)}\label{tab:asassn13cxoc2014}
\begin{center}
\begin{tabular}{rp{50pt}p{30pt}r@{.}lcr}
\hline
$E$ & max\commenta & error & \multicolumn{2}{c}{$O-C$\commentb} & phase\commentc & $N$\commentd \\
\hline
0 & 56901.5524 & 0.0006 & $-$0&0069 & 0.17 & 133 \\
1 & 56901.6357 & 0.0004 & $-$0&0064 & 0.22 & 136 \\
2 & 56901.7257 & 0.0012 & 0&0007 & 0.35 & 19 \\
7 & 56902.1313 & 0.0005 & $-$0&0079 & 0.44 & 80 \\
8 & 56902.2185 & 0.0005 & $-$0&0035 & 0.53 & 64 \\
10 & 56902.3819 & 0.0003 & $-$0&0059 & 0.58 & 80 \\
11 & 56902.4646 & 0.0004 & $-$0&0060 & 0.62 & 91 \\
12 & 56902.5500 & 0.0005 & $-$0&0034 & 0.69 & 123 \\
13 & 56902.6340 & 0.0008 & $-$0&0023 & 0.75 & 102 \\
14 & 56902.7172 & 0.0004 & $-$0&0019 & 0.79 & 73 \\
15 & 56902.8031 & 0.0009 & 0&0012 & 0.87 & 71 \\
34 & 56904.3712 & 0.0041 & $-$0&0047 & 0.56 & 26 \\
35 & 56904.4609 & 0.0004 & 0&0021 & 0.68 & 85 \\
36 & 56904.5450 & 0.0005 & 0&0034 & 0.74 & 87 \\
38 & 56904.7142 & 0.0075 & 0&0069 & 0.87 & 18 \\
47 & 56905.4597 & 0.0009 & 0&0068 & 0.23 & 215 \\
48 & 56905.5433 & 0.0008 & 0&0076 & 0.27 & 211 \\
49 & 56905.6253 & 0.0004 & 0&0068 & 0.30 & 86 \\
50 & 56905.7075 & 0.0006 & 0&0061 & 0.34 & 57 \\
52 & 56905.8692 & 0.0007 & 0&0021 & 0.37 & 71 \\
58 & 56906.3684 & 0.0012 & 0&0043 & 0.63 & 43 \\
59 & 56906.4487 & 0.0008 & 0&0018 & 0.64 & 54 \\
60 & 56906.5305 & 0.0013 & 0&0007 & 0.67 & 64 \\
61 & 56906.6176 & 0.0013 & 0&0049 & 0.76 & 54 \\
63 & 56906.7817 & 0.0006 & 0&0034 & 0.82 & 93 \\
64 & 56906.8728 & 0.0015 & 0&0116 & 0.97 & 80 \\
72 & 56907.5217 & 0.0018 & $-$0&0022 & 0.11 & 130 \\
73 & 56907.6051 & 0.0014 & $-$0&0016 & 0.16 & 116 \\
83 & 56908.4406 & 0.0024 & 0&0054 & 0.65 & 63 \\
84 & 56908.5243 & 0.0024 & 0&0062 & 0.70 & 53 \\
118 & 56911.3291 & 0.0027 & $-$0&0055 & 0.92 & 46 \\
130 & 56912.3164 & 0.0042 & $-$0&0124 & 0.31 & 39 \\
133 & 56912.5658 & 0.0077 & $-$0&0115 & 0.44 & 33 \\
\hline
  \multicolumn{7}{l}{\commenta BJD$-$2400000.} \\
  \multicolumn{7}{l}{\commentb Against max $= 2456901.5593 + 0.082842 E$.} \\
  \multicolumn{7}{l}{\commentc Orbital phase.} \\
  \multicolumn{7}{l}{\commentd Number of points used to determine the maximum.} \\
\end{tabular}
\end{center}
\end{table}

\subsection{ASASSN-14ag}\label{obj:asassn14ag}

\begin{figure}
  \begin{center}
    \FigureFile(88mm,70mm){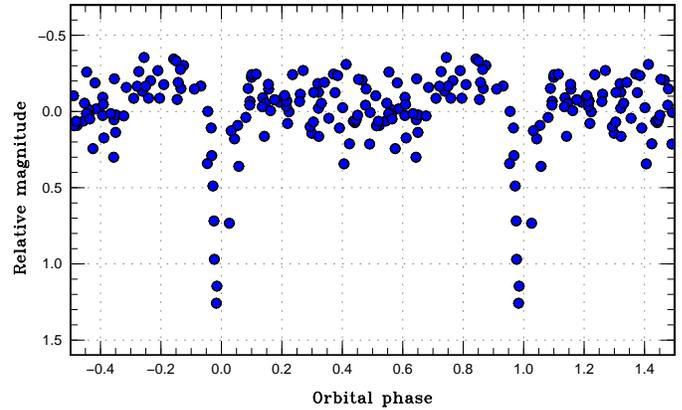}
  \end{center}
  \caption{Orbital light curve of ASASSN-14ag in the CRTS data.}
  \label{fig:asassn14agqui}
\end{figure}

\begin{figure}
  \begin{center}
    \FigureFile(88mm,120mm){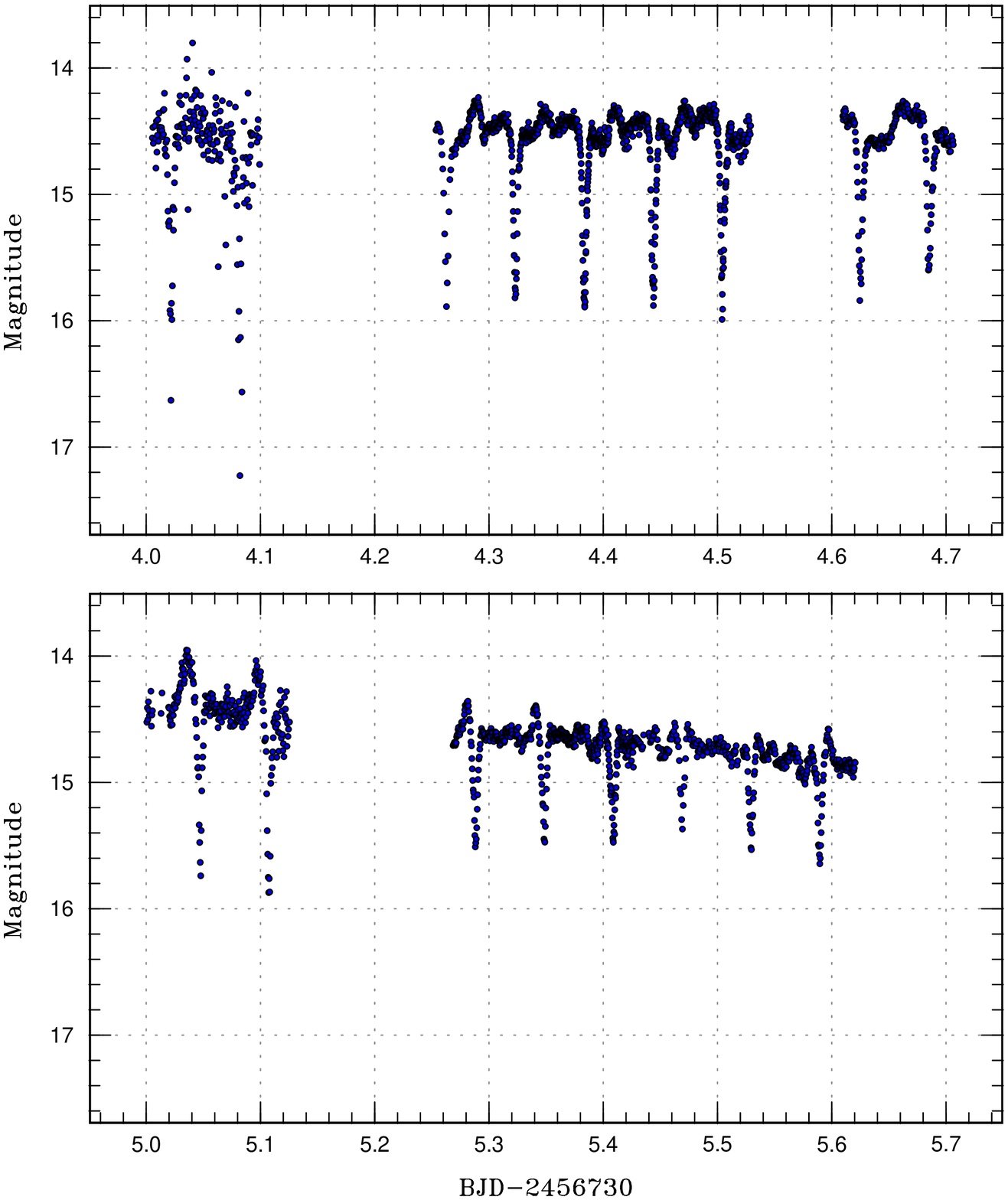}
  \end{center}
  \caption{Sample light curve of ASASSN-14ag in superoutburst.
  Superposition of superhumps and eclipse are well visible.}
  \label{fig:asassn14aglc}
\end{figure}

   This object was detected as a transient at $V$=13.5.
on 2014 March 14 by ASAS-SN team.  The coordinates are
\timeform{08h 13m 18.51s}, \timeform{-01D 03' 28.5''}
(2MASS position).
CRTS recorded five more outbursts in the past data.
One of the authors (TK)
noticed that the CRTS light curve showed evidence of eclipses,
and obtained a period below the period gap (vsnet-alert 17036;
figure \ref{fig:asassn14agqui}).
This suggestion was confirmed by photometry in outburst
(vsnet-alert 17041, 17042; figure \ref{fig:asassn14aglc}).
An MCMC analysis of both the CRTS data
and the present data in outburst yielded the following
orbital ephemeris:
\begin{equation}
{\rm Min(BJD)} = 2454413.99168(7) + 0.060310651(2) E .
\label{equ:asassn14agecl}
\end{equation}

\begin{figure}
  \begin{center}
    \FigureFile(88mm,110mm){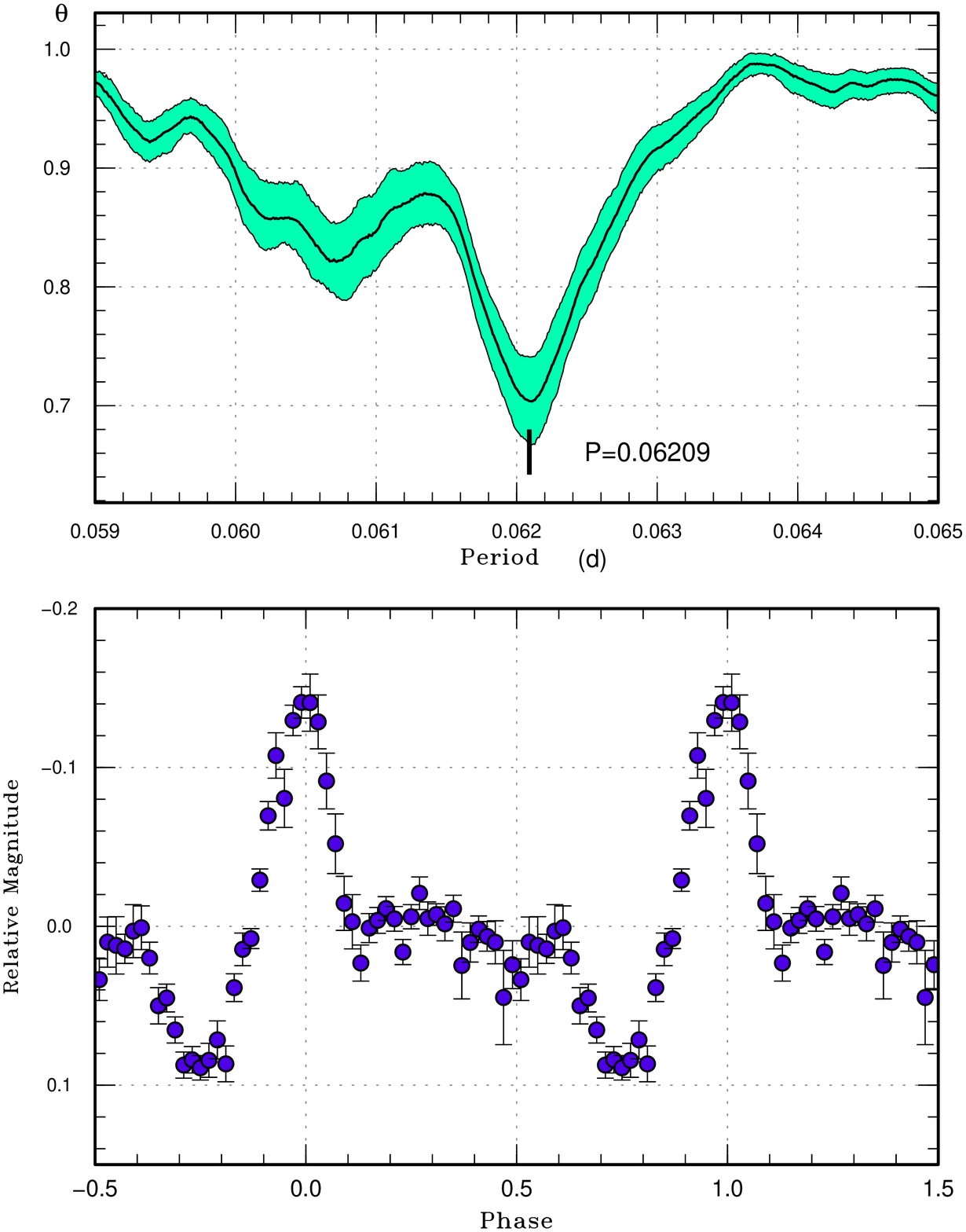}
  \end{center}
  \caption{Superhumps in ASASSN-14ag outside the eclipses (2014).
     (Upper): PDM analysis.
     (Lower): Phase-averaged profile.}
  \label{fig:asassn14agshpdm}
\end{figure}

   There was a strong beat phenomenon between the superhump
period and the orbital period (figure \ref{fig:asassn14aglc}).
The mean profile of superhumps outside the eclipses
is shown in figure \ref{fig:asassn14agshpdm}.
The times of superhump maxima determined outside the eclipses
are listed in table \ref{tab:asassn14agoc2014}.

   The relatively low outburst amplitude (2--3 mag in the 
CRTS data; the outburst data in ASAS-SN data appear to be
brighter than the real magnitude by $\sim$1 mag)
is probably due to the high orbital inclination.
Since such a bright, deeply eclipsing SU UMa-type dwarf nova
with a short orbital period is rare, further detailed
study will be promising in resolving the superhump light
source in different stages.

\begin{table}
\caption{Superhump maxima of ASASSN-14ag (2014)}\label{tab:asassn14agoc2014}
\begin{center}
\begin{tabular}{rp{50pt}p{30pt}r@{.}lcr}
\hline
$E$ & max\commenta & error & \multicolumn{2}{c}{$O-C$\commentb} & phase\commentc & $N$\commentd \\
\hline
0 & 56734.0403 & 0.0025 & $-$0&0027 & 0.31 & 85 \\
1 & 56734.0969 & 0.0025 & $-$0&0081 & 0.25 & 19 \\
4 & 56734.2909 & 0.0008 & $-$0&0003 & 0.46 & 166 \\
5 & 56734.3530 & 0.0011 & $-$0&0003 & 0.49 & 201 \\
6 & 56734.4172 & 0.0012 & 0&0018 & 0.56 & 222 \\
7 & 56734.4788 & 0.0011 & 0&0014 & 0.58 & 207 \\
8 & 56734.5431 & 0.0042 & 0&0036 & 0.64 & 30 \\
10 & 56734.6670 & 0.0004 & 0&0034 & 0.70 & 96 \\
16 & 56735.0392 & 0.0004 & 0&0033 & 0.87 & 77 \\
17 & 56735.0983 & 0.0011 & 0&0003 & 0.85 & 88 \\
20 & 56735.2854 & 0.0004 & 0&0012 & 0.95 & 79 \\
21 & 56735.3453 & 0.0007 & $-$0&0009 & 0.95 & 117 \\
22 & 56735.4083 & 0.0014 & $-$0&0000 & 0.99 & 132 \\
23 & 56735.4706 & 0.0009 & 0&0003 & 0.02 & 56 \\
24 & 56735.5349 & 0.0014 & 0&0025 & 0.09 & 80 \\
25 & 56735.5946 & 0.0006 & 0&0001 & 0.08 & 96 \\
36 & 56736.2739 & 0.0012 & $-$0&0032 & 0.34 & 100 \\
37 & 56736.3379 & 0.0008 & $-$0&0013 & 0.40 & 100 \\
38 & 56736.4001 & 0.0054 & $-$0&0011 & 0.43 & 111 \\
\hline
  \multicolumn{7}{l}{\commenta BJD$-$2400000.} \\
  \multicolumn{7}{l}{\commentb Against max $= 2456734.0430 + 0.062059 E$.} \\
  \multicolumn{7}{l}{\commentc Orbital phase.} \\
  \multicolumn{7}{l}{\commentd Number of points used to determine the maximum.} \\
\end{tabular}
\end{center}
\end{table}

\subsection{ASASSN-14aj}\label{obj:asassn14aj}

   This object was detected as a transient at $V$=15.1
on 2014 March 27 by ASAS-SN team (vsnet-alert 17101). 
The coordinates are
\timeform{17h 36m 15.09s}, \timeform{+07D 16' 56.0''}
(SDSS $g=21.7$ counterpart).
This object was also independently discovered on the same day
as TCP J17361506$+$0716560.\footnote{
$<$http://www.cbat.eps.harvard.edu/unconf/\\
followups/J17361506+0716560.html$>$.
}
There were also two past outbursts (vsnet-alert 17130).

   Subsequent observation detected superhumps (vsnet-alert 17119,
17128, 17131).
The times of superhump maxima are listed in table
\ref{tab:asassn14ajoc2014}.  Although we only observed
the late phase of the superoutburst, we left the stage
classification as an open question since the stages
tend to become unclear for long-$P_{\rm orb}$ systems.
If there were distinct stage transitions in this system,
we may have only observed stage C superhumps.

\begin{figure}
  \begin{center}
    \FigureFile(88mm,110mm){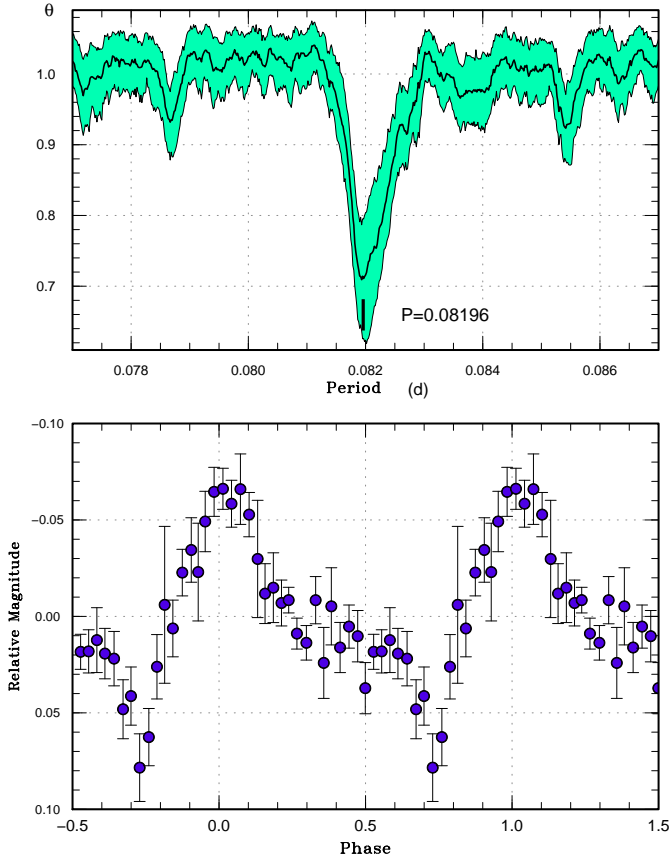}
  \end{center}
  \caption{Superhumps in ASASSN-14aj (2014).  (Upper): PDM analysis.
     (Lower): Phase-averaged profile.}
  \label{fig:asassn14ajshpdm}
\end{figure}

\begin{table}
\caption{Superhump maxima of ASASSN-14aj (2014)}\label{tab:asassn14ajoc2014}
\begin{center}
\begin{tabular}{rp{55pt}p{40pt}r@{.}lr}
\hline
\multicolumn{1}{c}{$E$} & \multicolumn{1}{c}{max\commenta} & \multicolumn{1}{c}{error} & \multicolumn{2}{c}{$O-C$\commentb} & \multicolumn{1}{c}{$N$\commentc} \\
\hline
0 & 56746.7697 & 0.0017 & $-$0&0057 & 15 \\
1 & 56746.8528 & 0.0009 & $-$0&0046 & 21 \\
12 & 56747.7625 & 0.0023 & 0&0036 & 16 \\
13 & 56747.8434 & 0.0011 & 0&0025 & 27 \\
25 & 56748.8266 & 0.0012 & 0&0022 & 27 \\
37 & 56749.8076 & 0.0010 & $-$0&0003 & 26 \\
38 & 56749.8930 & 0.0022 & 0&0031 & 11 \\
49 & 56750.7932 & 0.0016 & 0&0018 & 25 \\
50 & 56750.8692 & 0.0020 & $-$0&0042 & 19 \\
61 & 56751.7750 & 0.0013 & 0&0000 & 25 \\
62 & 56751.8612 & 0.0013 & 0&0043 & 24 \\
73 & 56752.7631 & 0.0040 & 0&0047 & 22 \\
98 & 56754.8000 & 0.0058 & $-$0&0074 & 26 \\
\hline
  \multicolumn{6}{l}{\commenta BJD$-$2400000.} \\
  \multicolumn{6}{l}{\commentb Against max $= 2456746.7754 + 0.081959 E$.} \\
  \multicolumn{6}{l}{\commentc Number of points used to determine the maximum.} \\
\end{tabular}
\end{center}
\end{table}

\subsection{ASASSN-14au}\label{obj:asassn14au}

\begin{figure}
  \begin{center}
    \FigureFile(88mm,70mm){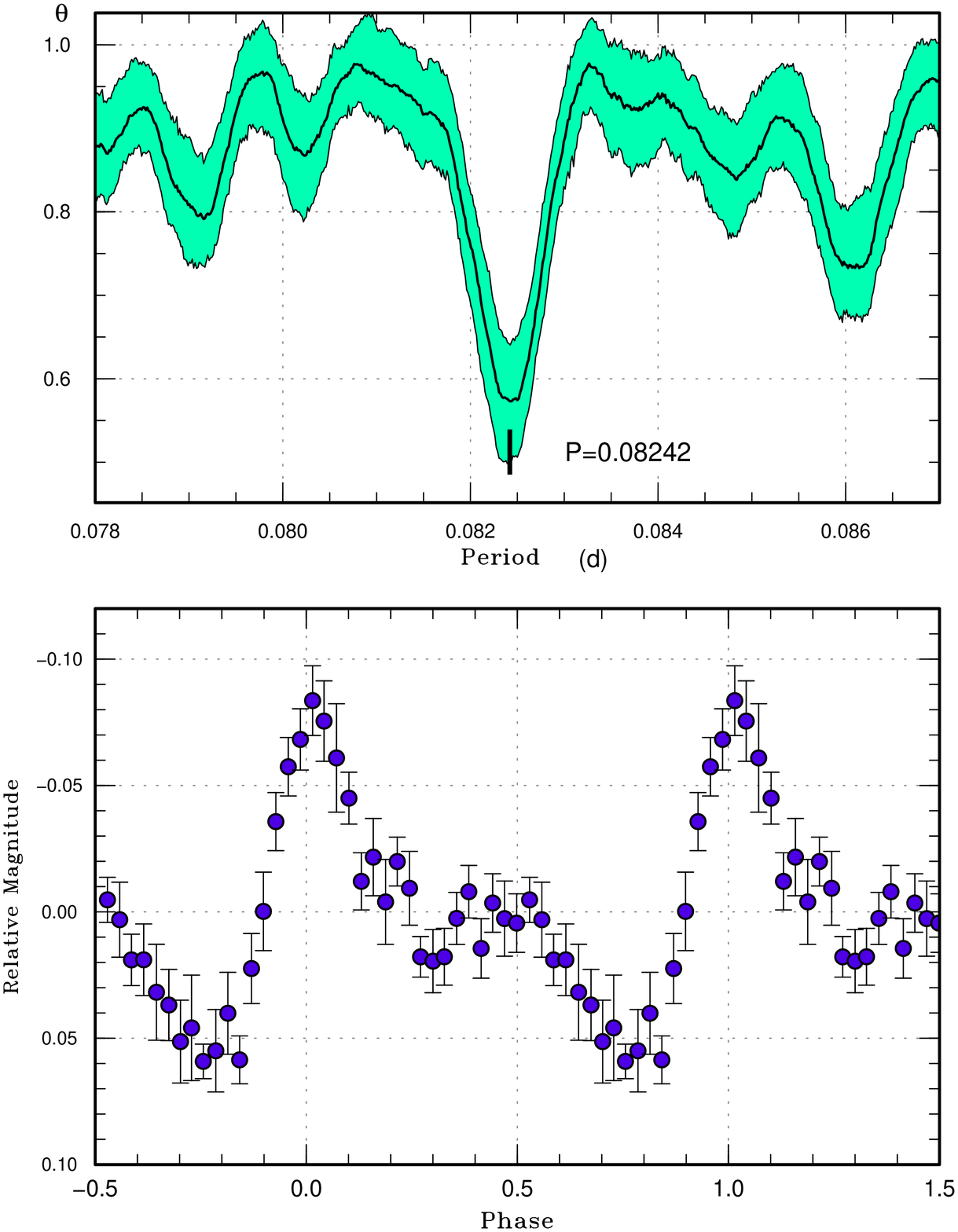}
  \end{center}
  \caption{Superhumps in ASASSN-14au (2014).  (Upper): PDM analysis.
     (Lower): Phase-averaged profile.}
  \label{fig:asassn14aushpdm}
\end{figure}

   This object was detected as a transient at $V$=16.4.
on 2014 May 8 by ASAS-SN team (vsnet-alert 17290). 
The coordinates are
\timeform{18h 17m 52.54s}, \timeform{+67D 08' 08.0''}.
There is a GALEX counterpart with an NUV magnitude
of 21.6.  Subsequent observations detected superhumps
(vsnet-alert 17295; figure \ref{fig:asassn14aushpdm}).
The times of superhump maxima are listed in table 
\ref{tab:asassn14auoc2014}.

\begin{table}
\caption{Superhump maxima of ASASSN-14au (2014)}\label{tab:asassn14auoc2014}
\begin{center}
\begin{tabular}{rp{55pt}p{40pt}r@{.}lr}
\hline
\multicolumn{1}{c}{$E$} & \multicolumn{1}{c}{max\commenta} & \multicolumn{1}{c}{error} & \multicolumn{2}{c}{$O-C$\commentb} & \multicolumn{1}{c}{$N$\commentc} \\
\hline
0 & 56790.6007 & 0.0020 & $-$0&0013 & 58 \\
1 & 56790.6816 & 0.0050 & $-$0&0029 & 30 \\
12 & 56791.5916 & 0.0012 & $-$0&0002 & 43 \\
13 & 56791.6810 & 0.0038 & 0&0067 & 23 \\
35 & 56793.4876 & 0.0016 & $-$0&0014 & 37 \\
36 & 56793.5703 & 0.0012 & $-$0&0012 & 43 \\
37 & 56793.6542 & 0.0025 & 0&0003 & 40 \\
\hline
  \multicolumn{6}{l}{\commenta BJD$-$2400000.} \\
  \multicolumn{6}{l}{\commentb Against max $= 2456790.6020 + 0.082486 E$.} \\
  \multicolumn{6}{l}{\commentc Number of points used to determine the maximum.} \\
\end{tabular}
\end{center}
\end{table}

\subsection{ASASSN-14aw}\label{obj:asassn14aw}

\begin{figure}
  \begin{center}
    \FigureFile(88mm,70mm){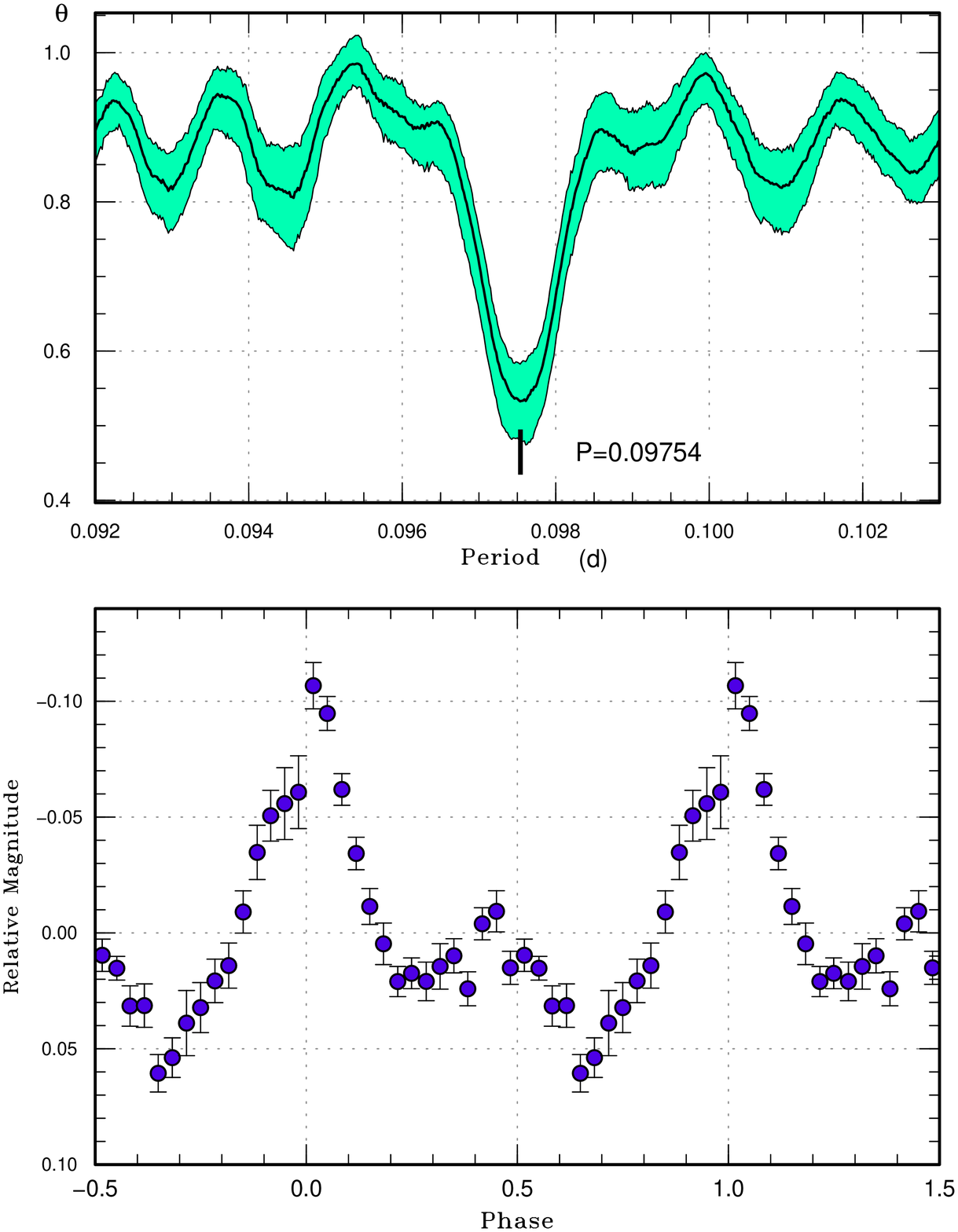}
  \end{center}
  \caption{Superhumps in ASASSN-14aw (2014).  (Upper): PDM analysis.
     (Lower): Phase-averaged profile.}
  \label{fig:asassn14awshpdm}
\end{figure}

   This object was detected as a transient at $V$=15.8.
on 2014 May 10 by ASAS-SN team (vsnet-alert 17290). 
The coordinates are
\timeform{18h 47m 02.57s}, \timeform{+53D 00' 30.6''}.
Subsequent observations detected superhumps
(vsnet-alert 17295; figure \ref{fig:asassn14awshpdm}).
The times of superhump maxima are listed in table 
\ref{tab:asassn14awoc2014}.
The superhump period indicates that this object is located
in the period gap.

\begin{table}
\caption{Superhump maxima of ASASSN-14aw (2014)}\label{tab:asassn14awoc2014}
\begin{center}
\begin{tabular}{rp{55pt}p{40pt}r@{.}lr}
\hline
\multicolumn{1}{c}{$E$} & \multicolumn{1}{c}{max\commenta} & \multicolumn{1}{c}{error} & \multicolumn{2}{c}{$O-C$\commentb} & \multicolumn{1}{c}{$N$\commentc} \\
\hline
0 & 56790.6159 & 0.0007 & $-$0&0013 & 105 \\
10 & 56791.5945 & 0.0011 & 0&0014 & 89 \\
29 & 56793.4530 & 0.0042 & 0&0057 & 58 \\
30 & 56793.5430 & 0.0010 & $-$0&0018 & 77 \\
31 & 56793.6383 & 0.0014 & $-$0&0041 & 63 \\
\hline
  \multicolumn{6}{l}{\commenta BJD$-$2400000.} \\
  \multicolumn{6}{l}{\commentb Against max $= 2456790.6172 + 0.097586 E$.} \\
  \multicolumn{6}{l}{\commentc Number of points used to determine the maximum.} \\
\end{tabular}
\end{center}
\end{table}

\subsection{ASASSN-14bh}\label{obj:asassn14bh}

\begin{figure}
  \begin{center}
    \FigureFile(88mm,70mm){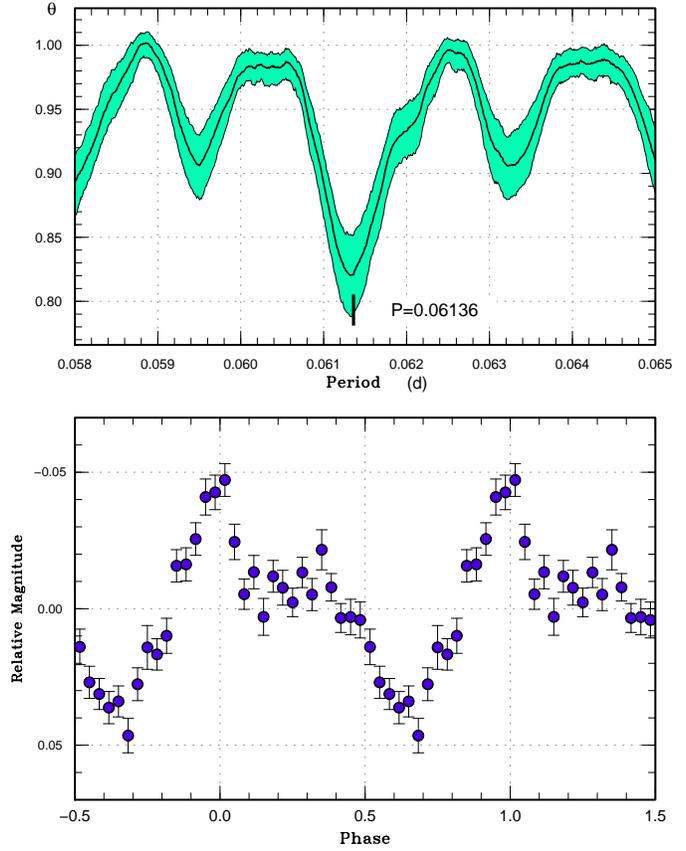}
  \end{center}
  \caption{Superhumps in ASASSN-14bh (2014).  (Upper): PDM analysis.
     (Lower): Phase-averaged profile.}
  \label{fig:asassn14bhshpdm}
\end{figure}

   This object was detected as a transient at $V$=14.66.
on 2014 May 21 by ASAS-SN team (vsnet-alert 17334). 
The coordinates given by the ASAS-SN team are
\timeform{13h 24m 40.0s}, \timeform{-19D 51' 31.4''}, which
is \timeform{3''} distant from a 20.4 mag (blue plate)
USNO B1.0 object.
Subsequent observations detected superhumps
(vsnet-alert 17359; figure \ref{fig:asassn14bhshpdm}).
The times of superhump maxima are listed in table 
\ref{tab:asassn14bhoc2014}.
The object started fading rapidly five nights after
the start of this observation.  We only observed the
final part of the superoutburst, and we most likely
observed stage C superhumps.

\begin{table}
\caption{Superhump maxima of ASASSN-14bh (2014)}\label{tab:asassn14bhoc2014}
\begin{center}
\begin{tabular}{rp{55pt}p{40pt}r@{.}lr}
\hline
\multicolumn{1}{c}{$E$} & \multicolumn{1}{c}{max\commenta} & \multicolumn{1}{c}{error} & \multicolumn{2}{c}{$O-C$\commentb} & \multicolumn{1}{c}{$N$\commentc} \\
\hline
0 & 56808.2910 & 0.0007 & $-$0&0035 & 139 \\
1 & 56808.3567 & 0.0009 & 0&0009 & 141 \\
2 & 56808.4156 & 0.0011 & $-$0&0016 & 135 \\
16 & 56809.2791 & 0.0014 & 0&0027 & 136 \\
17 & 56809.3420 & 0.0015 & 0&0043 & 46 \\
18 & 56809.4007 & 0.0014 & 0&0016 & 128 \\
32 & 56810.2569 & 0.0018 & $-$0&0013 & 140 \\
33 & 56810.3189 & 0.0031 & $-$0&0008 & 136 \\
34 & 56810.3773 & 0.0020 & $-$0&0037 & 140 \\
35 & 56810.4438 & 0.0023 & 0&0015 & 140 \\
\hline
  \multicolumn{6}{l}{\commenta BJD$-$2400000.} \\
  \multicolumn{6}{l}{\commentb Against max $= 2456808.2945 + 0.061368 E$.} \\
  \multicolumn{6}{l}{\commentc Number of points used to determine the maximum.} \\
\end{tabular}
\end{center}
\end{table}

\subsection{ASASSN-14cl}\label{obj:asassn14cl}

   This object was detected as a bright transient at $V$=10.66.
on 2014 June 14 by ASAS-SN team (\cite{sta14asassn14clatel6233};
vsnet-alert 17376).   The coordinates are \timeform{21h 54m 57.70s},
\timeform{+26D 41' 12.9''} (vsnet-alert 17429).
The quiescent counterpart in SDSS was
a $g=18.8$ star.  No previous outbursts are known.
The SDSS colors expected a short orbital
period of 0.066~d by using the neural network analysis
\citep{kat12DNSDSS}.  The large outburst amplitude suggested
a WZ Sge-type dwarf nova caught in the early stage
(vsnet-alert 17377).  \citet{tey14asassn14clatel6235}
spectroscopically confirmed that this object is an outbursting
dwarf nova.  The spectrum was characterized by broad Balmer
absorption lines with an emission core in H$\alpha$.
The He\textsc{ii} line was recorded in emission, which is
known to be a frequent signature of a WZ Sge-type
dwarf nova (cf. \cite{bab02wzsgeletter}).

\begin{figure}
  \begin{center}
    \FigureFile(88mm,110mm){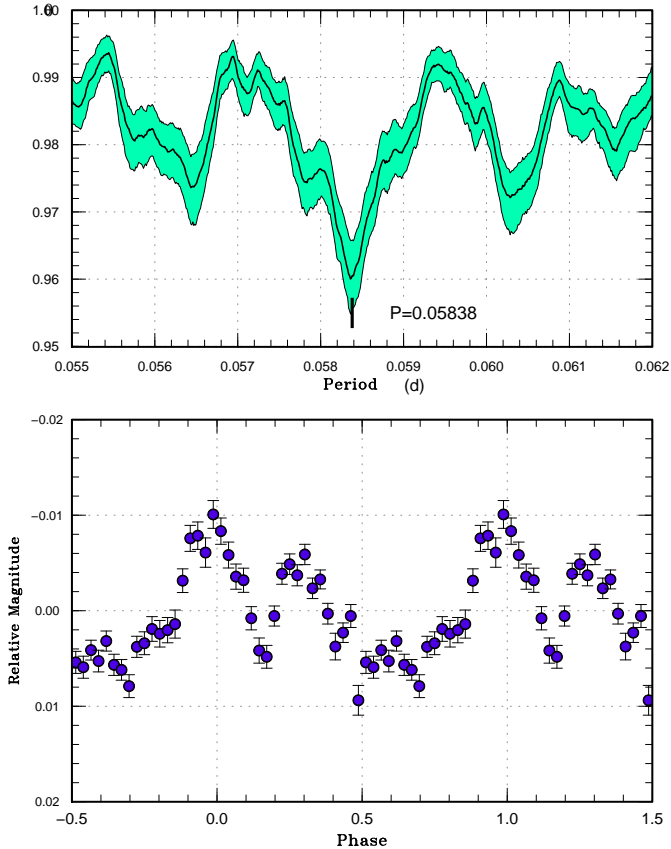}
  \end{center}
  \caption{Early superhumps in ASASSN-14cl (2014).
     (Upper): PDM analysis.
     (Lower): Phase-averaged profile}.
  \label{fig:asassn14cleshpdm}
\end{figure}

\begin{figure}
  \begin{center}
    \FigureFile(88mm,110mm){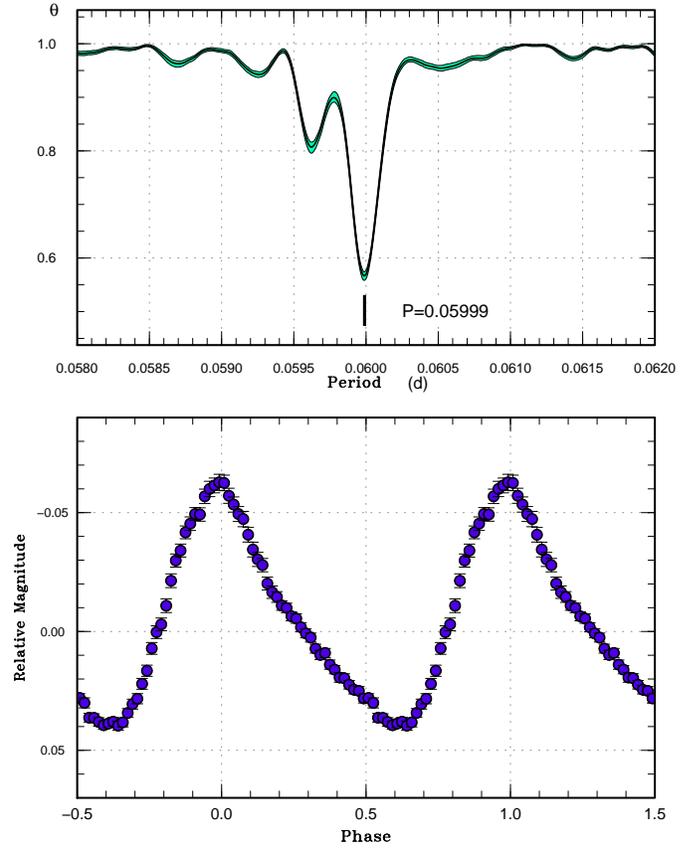}
  \end{center}
  \caption{Ordinary superhumps in ASASSN-14cl (2014) during
     the plateau phase (BJD 2456829--2456843).
     (Upper): PDM analysis.
     (Lower): Phase-averaged profile}.
  \label{fig:asassn14clshpdm}
\end{figure}

   Soon after the discovery, possible early superhumps
were detected (vsnet-alert 17383, 17384, 17387;
figure \ref{fig:asassn14cleshpdm}).
The object then started to show ordinary superhumps
(vsnet-alert 17392, 17398, 17413; figure
\ref{fig:asassn14clshpdm}).

\begin{figure}
  \begin{center}
    \FigureFile(88mm,70mm){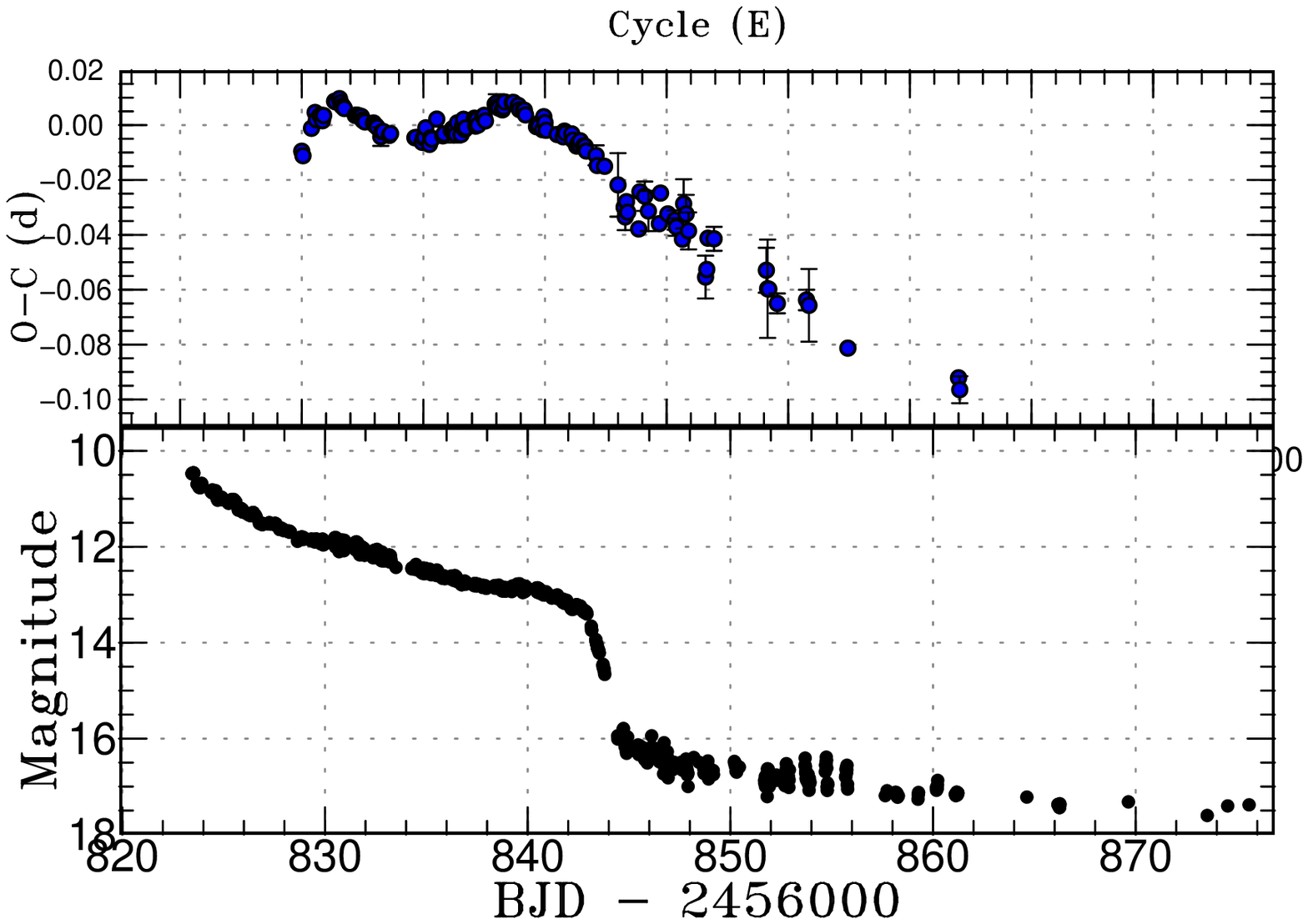}
  \end{center}
  \caption{$O-C$ diagram of superhumps in ASASSN-14cl (2014).
     (Upper): $O-C$ diagram.  A period of 0.06001~d
     was used to draw this figure.
     (Lower): Light curve.  The observations were binned to 0.012~d.}
  \label{fig:asassn14clhumpall}
\end{figure}

   The times of superhump maxima during the superoutburst
plateau are listed in table \ref{tab:asassn14cloc2014}.
The $O-C$ values showed very prominent stages A--C with
a definitely positive $P_{\rm dot}$ for stage B.
The times of superhumps in the post-superoutburst stage
are listed in table \ref{tab:asassn14cloc2014b}.
The post-superoutburst superhumps were on very good
extension of stage C superhumps
(figure \ref{fig:asassn14clhumpall}).

    The presence of distinct stage B with a definitely
positive $P_{\rm dot}$, the presence of stage C and
the persistence of stage C superhumps after the rapid
fading from the plateau phase indicate that ASASSN-14cl
resembles a rather ordinary large-amplitude SU UMat-type
dwarf nova rather than an extreme WZ Sge-type object.

\begin{table}
\caption{Superhump maxima of ASASSN-14cl (2014)}\label{tab:asassn14cloc2014}
\begin{center}
\begin{tabular}{rp{55pt}p{40pt}r@{.}lr}
\hline
\multicolumn{1}{c}{$E$} & \multicolumn{1}{c}{max\commenta} & \multicolumn{1}{c}{error} & \multicolumn{2}{c}{$O-C$\commentb} & \multicolumn{1}{c}{$N$\commentc} \\
\hline
0 & 56828.8437 & 0.0013 & $-$0&0122 & 207 \\
1 & 56828.9019 & 0.0006 & $-$0&0139 & 187 \\
8 & 56829.3319 & 0.0007 & $-$0&0039 & 99 \\
11 & 56829.5177 & 0.0007 & 0&0020 & 68 \\
12 & 56829.5750 & 0.0004 & $-$0&0007 & 63 \\
15 & 56829.7567 & 0.0017 & 0&0010 & 32 \\
16 & 56829.8167 & 0.0001 & 0&0011 & 358 \\
17 & 56829.8748 & 0.0001 & $-$0&0009 & 448 \\
18 & 56829.9367 & 0.0001 & 0&0010 & 250 \\
27 & 56830.4819 & 0.0004 & 0&0064 & 37 \\
28 & 56830.5417 & 0.0002 & 0&0062 & 52 \\
31 & 56830.7230 & 0.0002 & 0&0075 & 142 \\
32 & 56830.7815 & 0.0001 & 0&0061 & 277 \\
33 & 56830.8402 & 0.0001 & 0&0048 & 224 \\
34 & 56830.9001 & 0.0001 & 0&0047 & 214 \\
35 & 56830.9594 & 0.0002 & 0&0040 & 46 \\
44 & 56831.4970 & 0.0002 & 0&0017 & 110 \\
45 & 56831.5571 & 0.0003 & 0&0018 & 89 \\
46 & 56831.6170 & 0.0003 & 0&0018 & 85 \\
47 & 56831.6766 & 0.0012 & 0&0014 & 19 \\
48 & 56831.7368 & 0.0002 & 0&0016 & 92 \\
49 & 56831.7966 & 0.0001 & 0&0014 & 284 \\
50 & 56831.8553 & 0.0001 & 0&0002 & 291 \\
51 & 56831.9150 & 0.0001 & $-$0&0002 & 244 \\
52 & 56831.9747 & 0.0006 & $-$0&0005 & 49 \\
59 & 56832.3944 & 0.0002 & $-$0&0007 & 58 \\
60 & 56832.4537 & 0.0002 & $-$0&0014 & 84 \\
61 & 56832.5134 & 0.0002 & $-$0&0017 & 153 \\
62 & 56832.5728 & 0.0002 & $-$0&0022 & 111 \\
65 & 56832.7494 & 0.0034 & $-$0&0055 & 11 \\
66 & 56832.8112 & 0.0006 & $-$0&0037 & 18 \\
67 & 56832.8713 & 0.0006 & $-$0&0036 & 26 \\
72 & 56833.1700 & 0.0003 & $-$0&0049 & 125 \\
\hline
  \multicolumn{6}{l}{\commenta BJD$-$2400000.} \\
  \multicolumn{6}{l}{\commentb Against max $= 2456828.8559 + 0.059986 E$.} \\
  \multicolumn{6}{l}{\commentc Number of points used to determine the maximum.} \\
\end{tabular}
\end{center}
\end{table}

\addtocounter{table}{-1}
\begin{table}
\caption{Superhump maxima of ASASSN-14cl (2014) (continued)}
\begin{center}
\begin{tabular}{rp{55pt}p{40pt}r@{.}lr}
\hline
\multicolumn{1}{c}{$E$} & \multicolumn{1}{c}{max\commenta} & \multicolumn{1}{c}{error} & \multicolumn{2}{c}{$O-C$\commentb} & \multicolumn{1}{c}{$N$\commentc} \\
\hline
73 & 56833.2306 & 0.0002 & $-$0&0042 & 125 \\
93 & 56834.4293 & 0.0024 & $-$0&0053 & 69 \\
94 & 56834.4895 & 0.0002 & $-$0&0051 & 229 \\
99 & 56834.7877 & 0.0006 & $-$0&0068 & 224 \\
100 & 56834.8494 & 0.0003 & $-$0&0051 & 245 \\
101 & 56834.9094 & 0.0002 & $-$0&0050 & 228 \\
102 & 56834.9730 & 0.0005 & $-$0&0014 & 51 \\
105 & 56835.1470 & 0.0007 & $-$0&0074 & 88 \\
106 & 56835.2094 & 0.0002 & $-$0&0049 & 147 \\
107 & 56835.2689 & 0.0002 & $-$0&0054 & 138 \\
111 & 56835.5163 & 0.0004 & 0&0020 & 116 \\
116 & 56835.8102 & 0.0004 & $-$0&0040 & 193 \\
117 & 56835.8710 & 0.0005 & $-$0&0032 & 137 \\
123 & 56836.2322 & 0.0004 & $-$0&0019 & 121 \\
124 & 56836.2929 & 0.0006 & $-$0&0012 & 115 \\
125 & 56836.3510 & 0.0032 & $-$0&0031 & 53 \\
126 & 56836.4134 & 0.0005 & $-$0&0007 & 62 \\
127 & 56836.4709 & 0.0005 & $-$0&0032 & 60 \\
128 & 56836.5351 & 0.0019 & 0&0011 & 19 \\
131 & 56836.7108 & 0.0006 & $-$0&0032 & 55 \\
132 & 56836.7737 & 0.0004 & $-$0&0003 & 258 \\
133 & 56836.8364 & 0.0004 & 0&0025 & 216 \\
134 & 56836.8928 & 0.0003 & $-$0&0012 & 263 \\
135 & 56836.9534 & 0.0003 & $-$0&0006 & 94 \\
142 & 56837.3770 & 0.0010 & 0&0031 & 97 \\
143 & 56837.4341 & 0.0005 & 0&0002 & 119 \\
144 & 56837.4965 & 0.0006 & 0&0027 & 107 \\
145 & 56837.5545 & 0.0010 & 0&0006 & 32 \\
149 & 56837.7976 & 0.0007 & 0&0039 & 137 \\
150 & 56837.8581 & 0.0006 & 0&0043 & 129 \\
151 & 56837.9160 & 0.0006 & 0&0023 & 132 \\
159 & 56838.4023 & 0.0008 & 0&0087 & 91 \\
160 & 56838.4626 & 0.0032 & 0&0090 & 82 \\
\hline
  \multicolumn{6}{l}{\commenta BJD$-$2400000.} \\
  \multicolumn{6}{l}{\commentb Against max $= 2456828.8559 + 0.059986 E$.} \\
  \multicolumn{6}{l}{\commentc Number of points used to determine the maximum.} \\
\end{tabular}
\end{center}
\end{table}

\addtocounter{table}{-1}
\begin{table}
\caption{Superhump maxima of ASASSN-14cl (2014) (continued)}
\begin{center}
\begin{tabular}{rp{55pt}p{40pt}r@{.}lr}
\hline
\multicolumn{1}{c}{$E$} & \multicolumn{1}{c}{max\commenta} & \multicolumn{1}{c}{error} & \multicolumn{2}{c}{$O-C$\commentb} & \multicolumn{1}{c}{$N$\commentc} \\
\hline
161 & 56838.5214 & 0.0007 & 0&0078 & 160 \\
162 & 56838.5808 & 0.0014 & 0&0072 & 44 \\
165 & 56838.7602 & 0.0011 & 0&0067 & 23 \\
166 & 56838.8212 & 0.0013 & 0&0077 & 19 \\
167 & 56838.8831 & 0.0016 & 0&0096 & 16 \\
173 & 56839.2429 & 0.0004 & 0&0095 & 129 \\
174 & 56839.3030 & 0.0006 & 0&0096 & 35 \\
178 & 56839.5420 & 0.0004 & 0&0086 & 120 \\
179 & 56839.6004 & 0.0003 & 0&0071 & 86 \\
182 & 56839.7801 & 0.0003 & 0&0069 & 119 \\
183 & 56839.8402 & 0.0003 & 0&0070 & 117 \\
184 & 56839.8985 & 0.0003 & 0&0053 & 116 \\
193 & 56840.4343 & 0.0006 & 0&0012 & 123 \\
194 & 56840.4954 & 0.0003 & 0&0023 & 252 \\
195 & 56840.5544 & 0.0003 & 0&0014 & 228 \\
198 & 56840.7333 & 0.0012 & 0&0002 & 28 \\
199 & 56840.7980 & 0.0007 & 0&0050 & 62 \\
200 & 56840.8559 & 0.0007 & 0&0029 & 102 \\
201 & 56840.9131 & 0.0005 & 0&0001 & 85 \\
210 & 56841.4517 & 0.0017 & $-$0&0012 & 26 \\
215 & 56841.7509 & 0.0013 & $-$0&0019 & 34 \\
216 & 56841.8130 & 0.0020 & 0&0002 & 29 \\
217 & 56841.8722 & 0.0023 & $-$0&0006 & 20 \\
222 & 56842.1720 & 0.0017 & $-$0&0007 & 45 \\
223 & 56842.2298 & 0.0015 & $-$0&0029 & 27 \\
226 & 56842.4076 & 0.0008 & $-$0&0051 & 99 \\
227 & 56842.4685 & 0.0005 & $-$0&0041 & 153 \\
228 & 56842.5281 & 0.0008 & $-$0&0045 & 128 \\
229 & 56842.5895 & 0.0007 & $-$0&0031 & 48 \\
232 & 56842.7678 & 0.0012 & $-$0&0047 & 75 \\
233 & 56842.8276 & 0.0006 & $-$0&0049 & 128 \\
234 & 56842.8859 & 0.0008 & $-$0&0066 & 131 \\
242 & 56843.3644 & 0.0036 & $-$0&0080 & 45 \\
243 & 56843.4207 & 0.0012 & $-$0&0117 & 61 \\
249 & 56843.7805 & 0.0021 & $-$0&0119 & 86 \\
\hline
  \multicolumn{6}{l}{\commenta BJD$-$2400000.} \\
  \multicolumn{6}{l}{\commentb Against max $= 2456828.8559 + 0.059986 E$.} \\
  \multicolumn{6}{l}{\commentc Number of points used to determine the maximum.} \\
\end{tabular}
\end{center}
\end{table}

\begin{table}
\caption{Superhump maxima of ASASSN-14cl (2014) (post-superoutburst)}\label{tab:asassn14cloc2014b}
\begin{center}
\begin{tabular}{rp{55pt}p{40pt}r@{.}lr}
\hline
\multicolumn{1}{c}{$E$} & \multicolumn{1}{c}{max\commenta} & \multicolumn{1}{c}{error} & \multicolumn{2}{c}{$O-C$\commentb} & \multicolumn{1}{c}{$N$\commentc} \\
\hline
0 & 56844.4338 & 0.0116 & 0&0034 & 14 \\
5 & 56844.7257 & 0.0021 & $-$0&0034 & 81 \\
6 & 56844.7822 & 0.0048 & $-$0&0067 & 108 \\
7 & 56844.8478 & 0.0018 & $-$0&0009 & 76 \\
8 & 56844.9039 & 0.0026 & $-$0&0045 & 64 \\
17 & 56845.4379 & 0.0015 & $-$0&0082 & 42 \\
18 & 56845.5115 & 0.0007 & 0&0056 & 44 \\
22 & 56845.7499 & 0.0053 & 0&0050 & 14 \\
25 & 56845.9245 & 0.0073 & 0&0003 & 33 \\
34 & 56846.4600 & 0.0019 & $-$0&0020 & 27 \\
35 & 56846.5312 & 0.0010 & 0&0094 & 27 \\
41 & 56846.8836 & 0.0023 & 0&0034 & 45 \\
47 & 56847.2415 & 0.0035 & 0&0027 & 58 \\
48 & 56847.2989 & 0.0032 & 0&0004 & 56 \\
53 & 56847.5945 & 0.0012 & $-$0&0027 & 39 \\
54 & 56847.6675 & 0.0089 & 0&0104 & 9 \\
56 & 56847.7838 & 0.0070 & 0&0072 & 42 \\
58 & 56847.8976 & 0.0067 & 0&0016 & 51 \\
72 & 56848.7209 & 0.0078 & $-$0&0117 & 18 \\
73 & 56848.7837 & 0.0018 & $-$0&0086 & 49 \\
74 & 56848.8551 & 0.0027 & 0&0030 & 71 \\
79 & 56849.1549 & 0.0044 & 0&0041 & 16 \\
122 & 56851.7239 & 0.0082 & 0&0037 & 17 \\
123 & 56851.7772 & 0.0179 & $-$0&0028 & 15 \\
124 & 56851.8371 & 0.0026 & $-$0&0027 & 14 \\
131 & 56852.2519 & 0.0036 & $-$0&0061 & 55 \\
155 & 56853.6934 & 0.0037 & 0&0013 & 14 \\
157 & 56853.8115 & 0.0132 & $-$0&0001 & 15 \\
189 & 56855.7162 & 0.0024 & $-$0&0075 & 19 \\
280 & 56861.1663 & 0.0018 & 0&0051 & 58 \\
281 & 56861.2220 & 0.0049 & 0&0010 & 40 \\
\hline
  \multicolumn{6}{l}{\commenta BJD$-$2400000.} \\
  \multicolumn{6}{l}{\commentb Against max $= 2456844.4304 + 0.059753 E$.} \\
  \multicolumn{6}{l}{\commentc Number of points used to determine the maximum.} \\
\end{tabular}
\end{center}
\end{table}

\subsection{ASASSN-14cq}\label{obj:asassn14cq}

   This object was detected as a transient at $V$=13.72
on 2014 June 18 by ASAS-SN team (vsnet-alert 17386).
The coordinates are \timeform{15h 32m 00.29s},
\timeform{-28D 33' 57.2''} (vsnet-alert 17429).

\begin{figure}
  \begin{center}
    \FigureFile(88mm,110mm){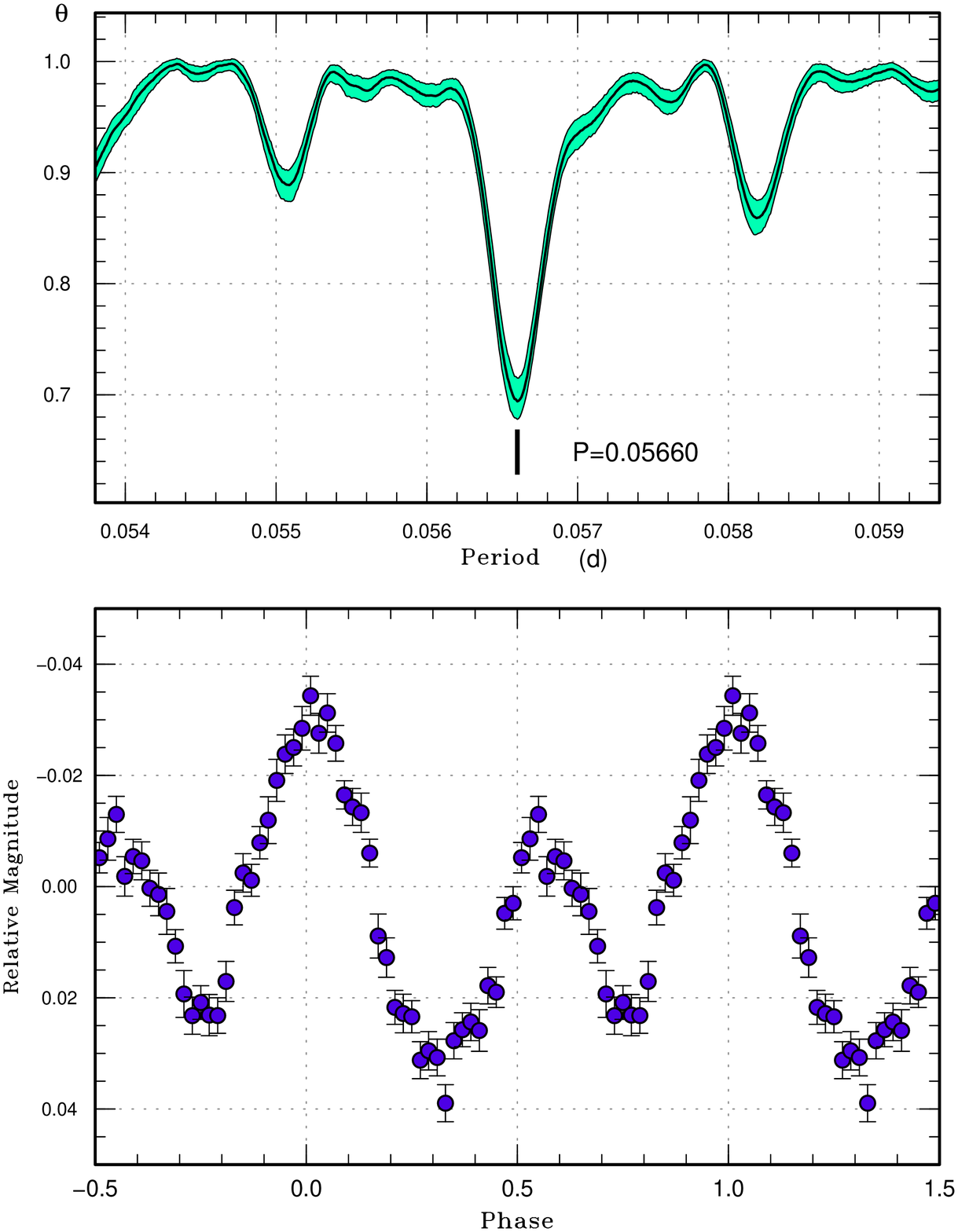}
  \end{center}
  \caption{Early superhumps in ASASSN-14cq (2014).
     (Upper): PDM analysis.
     (Lower): Phase-averaged profile}.
  \label{fig:asassn14cqeshpdm}
\end{figure}

\begin{figure}
  \begin{center}
    \FigureFile(88mm,110mm){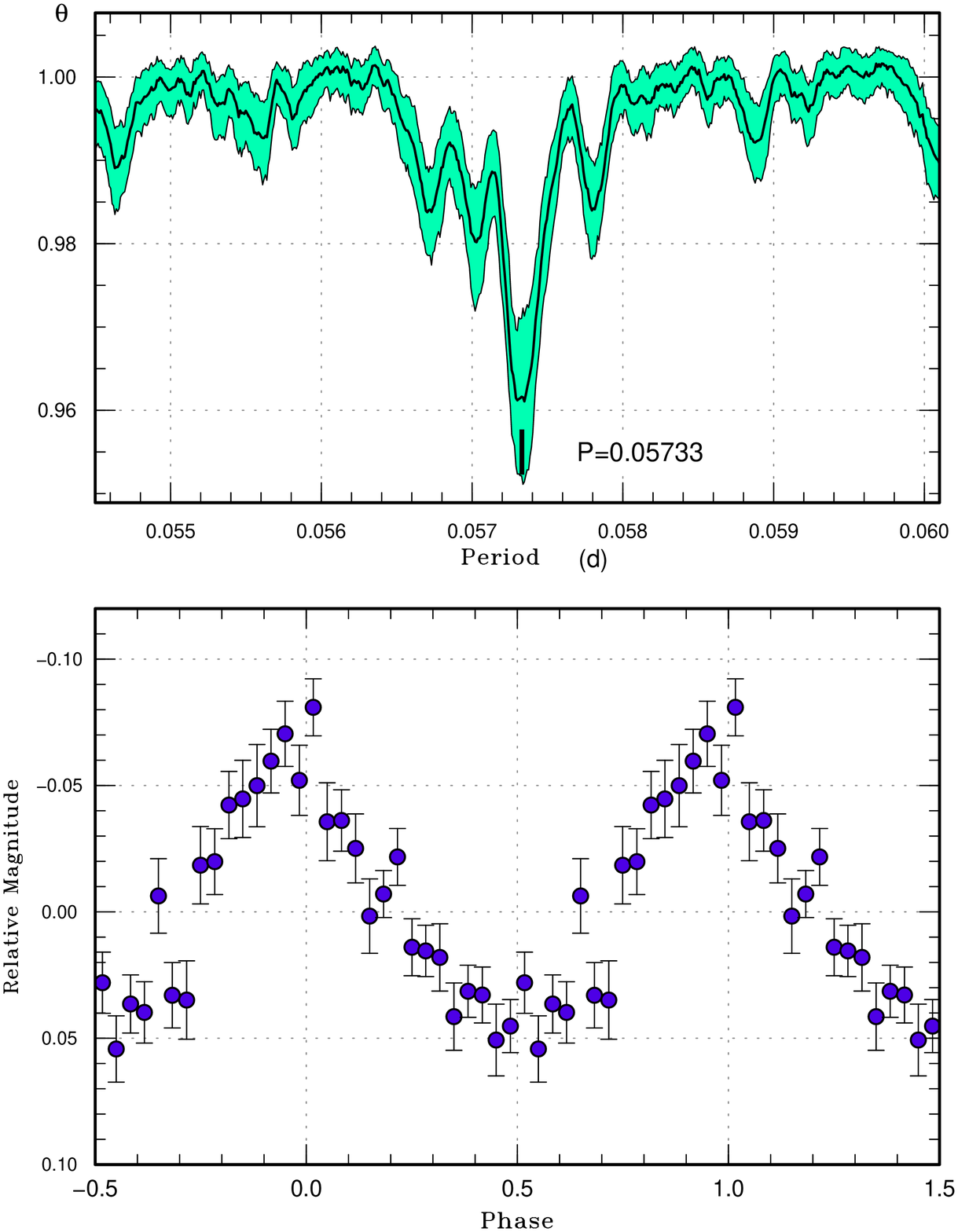}
  \end{center}
  \caption{Ordinary superhumps in ASASSN-14cq (2014).
     (Upper): PDM analysis.
     (Lower): Phase-averaged profile}.
  \label{fig:asassn14cqshpdm}
\end{figure}

   Subsequent observations recorded prominent early superhumps
(vsnet-alert 17403, 17405; figure \ref{fig:asassn14cqeshpdm}).
Although ordinary superhumps (figure \ref{fig:asassn14cqshpdm})
started to be observed on June 26 (vsnet-alert 17426),
the early phase of the development of superhumps
fell in the 1.8-d gap in the observation.
The times of superhump maxima are listed in table
\ref{tab:asassn14cqoc2014}.  Although there were short
gaps in the observation in the late stage of the
superoutburst, the times of maxima for $34 \le E \le 194$
appear to be expressed by a single, positive $P_{\rm dot}$.
We identified this part as stage B.  There was some hint
of stage C superhump after $E=194$, but the stage was not
well observed because the object became too faint.
Although $E=0$ corresponds to one of stage A superhumps,
the period of stage A superhumps (and $q$) could not be
estimated.

\begin{table}
\caption{Superhump maxima of ASASSN-14cq (2014)}\label{tab:asassn14cqoc2014}
\begin{center}
\begin{tabular}{rp{55pt}p{40pt}r@{.}lr}
\hline
\multicolumn{1}{c}{$E$} & \multicolumn{1}{c}{max\commenta} & \multicolumn{1}{c}{error} & \multicolumn{2}{c}{$O-C$\commentb} & \multicolumn{1}{c}{$N$\commentc} \\
\hline
0 & 56833.3147 & 0.0040 & $-$0&0019 & 59 \\
34 & 56835.2705 & 0.0003 & 0&0035 & 132 \\
35 & 56835.3281 & 0.0004 & 0&0037 & 111 \\
36 & 56835.3838 & 0.0002 & 0&0021 & 130 \\
37 & 56835.4422 & 0.0003 & 0&0030 & 132 \\
38 & 56835.4988 & 0.0008 & 0&0023 & 19 \\
39 & 56835.5558 & 0.0005 & 0&0020 & 15 \\
40 & 56835.6149 & 0.0011 & 0&0037 & 11 \\
41 & 56835.6724 & 0.0005 & 0&0039 & 19 \\
42 & 56835.7277 & 0.0007 & 0&0017 & 11 \\
52 & 56836.3009 & 0.0004 & 0&0013 & 132 \\
53 & 56836.3598 & 0.0004 & 0&0028 & 132 \\
54 & 56836.4162 & 0.0004 & 0&0019 & 131 \\
55 & 56836.4730 & 0.0004 & 0&0013 & 140 \\
56 & 56836.5306 & 0.0013 & 0&0016 & 37 \\
57 & 56836.5876 & 0.0007 & 0&0011 & 15 \\
68 & 56837.2147 & 0.0010 & $-$0&0027 & 77 \\
69 & 56837.2747 & 0.0004 & $-$0&0001 & 132 \\
70 & 56837.3316 & 0.0004 & $-$0&0006 & 132 \\
71 & 56837.3873 & 0.0005 & $-$0&0022 & 129 \\
72 & 56837.4473 & 0.0008 & 0&0004 & 128 \\
73 & 56837.5025 & 0.0007 & $-$0&0018 & 84 \\
74 & 56837.5610 & 0.0007 & $-$0&0006 & 15 \\
75 & 56837.6181 & 0.0005 & $-$0&0009 & 11 \\
76 & 56837.6739 & 0.0007 & $-$0&0024 & 17 \\
90 & 56838.4779 & 0.0014 & $-$0&0015 & 12 \\
91 & 56838.5345 & 0.0010 & $-$0&0024 & 15 \\
92 & 56838.5923 & 0.0008 & $-$0&0019 & 13 \\
93 & 56838.6487 & 0.0009 & $-$0&0029 & 16 \\
94 & 56838.7067 & 0.0007 & $-$0&0023 & 15 \\
108 & 56839.5077 & 0.0014 & $-$0&0043 & 17 \\
109 & 56839.5671 & 0.0011 & $-$0&0023 & 15 \\
\hline
  \multicolumn{6}{l}{\commenta BJD$-$2400000.} \\
  \multicolumn{6}{l}{\commentb Against max $= 2456833.3166 + 0.057365 E$.} \\
  \multicolumn{6}{l}{\commentc Number of points used to determine the maximum.} \\
\end{tabular}
\end{center}
\end{table}

\addtocounter{table}{-1}
\begin{table}
\caption{Superhump maxima of ASASSN-14cq (2014) (continued)}
\begin{center}
\begin{tabular}{rp{55pt}p{40pt}r@{.}lr}
\hline
\multicolumn{1}{c}{$E$} & \multicolumn{1}{c}{max\commenta} & \multicolumn{1}{c}{error} & \multicolumn{2}{c}{$O-C$\commentb} & \multicolumn{1}{c}{$N$\commentc} \\
\hline
111 & 56839.6814 & 0.0009 & $-$0&0027 & 16 \\
125 & 56840.4861 & 0.0006 & $-$0&0011 & 14 \\
126 & 56840.5423 & 0.0008 & $-$0&0022 & 14 \\
127 & 56840.6021 & 0.0007 & 0&0001 & 10 \\
128 & 56840.6572 & 0.0009 & $-$0&0021 & 15 \\
139 & 56841.2865 & 0.0010 & $-$0&0038 & 92 \\
141 & 56841.4044 & 0.0010 & $-$0&0007 & 101 \\
142 & 56841.4585 & 0.0009 & $-$0&0039 & 124 \\
143 & 56841.5176 & 0.0009 & $-$0&0022 & 15 \\
144 & 56841.5717 & 0.0016 & $-$0&0054 & 14 \\
145 & 56841.6307 & 0.0008 & $-$0&0039 & 12 \\
146 & 56841.6865 & 0.0015 & $-$0&0054 & 15 \\
192 & 56844.3392 & 0.0016 & 0&0085 & 119 \\
193 & 56844.3934 & 0.0010 & 0&0053 & 132 \\
194 & 56844.4534 & 0.0020 & 0&0080 & 96 \\
225 & 56846.2213 & 0.0030 & $-$0&0025 & 132 \\
226 & 56846.2855 & 0.0019 & 0&0044 & 131 \\
227 & 56846.3427 & 0.0018 & 0&0043 & 132 \\
\hline
  \multicolumn{6}{l}{\commenta BJD$-$2400000.} \\
  \multicolumn{6}{l}{\commentb Against max $= 2456833.3166 + 0.057365 E$.} \\
  \multicolumn{6}{l}{\commentc Number of points used to determine the maximum.} \\
\end{tabular}
\end{center}
\end{table}

\subsection{ASASSN-14dm}\label{obj:asassn14dm}

\begin{figure}
  \begin{center}
    \FigureFile(88mm,110mm){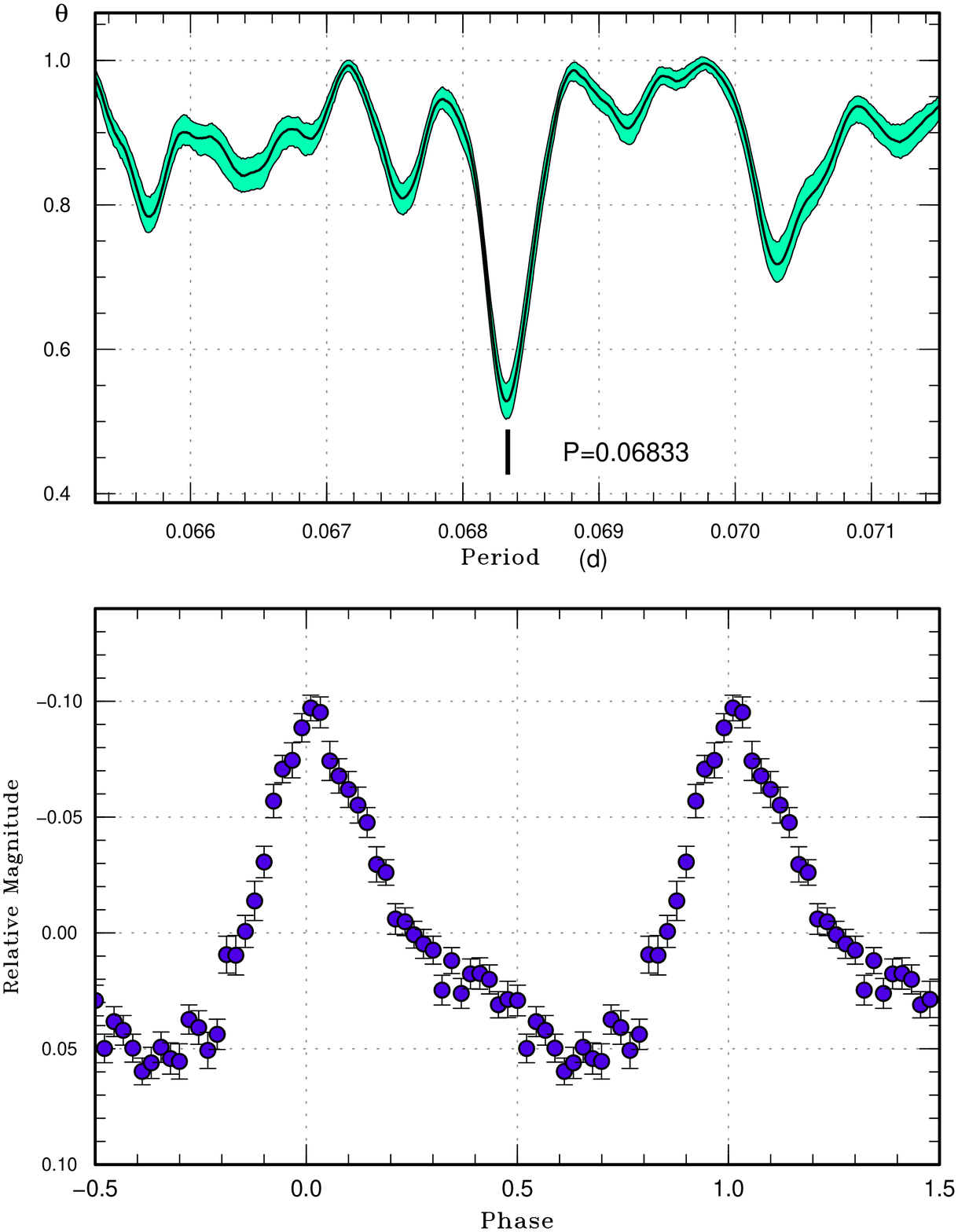}
  \end{center}
  \caption{Superhumps in ASASSN-14dm (2014).
     (Upper): PDM analysis.
     (Lower): Phase-averaged profile}.
  \label{fig:asassn14dmshpdm}
\end{figure}

   This object was detected as a transient at $V$=15.02
on 2014 July 1 by ASAS-SN team (vsnet-alert 17434).
The coordinates are \timeform{14h 09m 32.07s},
\timeform{-29D 17' 04.5''}.
(GSC 2.3.2 position).
The object has a GALEX UV counterpart with an NUV
magnitude of 20.6.
Subsequent observations detected superhumps
(vsnet-alert 17452; figure \ref{fig:asassn14dmshpdm}).
The times of superhump maxima are listed in table
\ref{tab:asassn14dmoc2014}.  The $O-C$ values suggest
that we recorded stage B-C transition around $E=75$.

\begin{table}
\caption{Superhump maxima of ASASSN-14dm (2014)}\label{tab:asassn14dmoc2014}
\begin{center}
\begin{tabular}{rp{55pt}p{40pt}r@{.}lr}
\hline
\multicolumn{1}{c}{$E$} & \multicolumn{1}{c}{max\commenta} & \multicolumn{1}{c}{error} & \multicolumn{2}{c}{$O-C$\commentb} & \multicolumn{1}{c}{$N$\commentc} \\
\hline
0 & 56842.2076 & 0.0003 & 0&0003 & 116 \\
1 & 56842.2752 & 0.0002 & $-$0&0003 & 157 \\
2 & 56842.3417 & 0.0003 & $-$0&0022 & 153 \\
31 & 56844.3220 & 0.0005 & $-$0&0026 & 154 \\
32 & 56844.3910 & 0.0005 & $-$0&0019 & 157 \\
33 & 56844.4662 & 0.0018 & 0&0050 & 43 \\
73 & 56847.1864 & 0.0043 & $-$0&0068 & 50 \\
74 & 56847.2667 & 0.0007 & 0&0051 & 157 \\
75 & 56847.3351 & 0.0010 & 0&0053 & 157 \\
76 & 56847.4008 & 0.0010 & 0&0026 & 134 \\
90 & 56848.3549 & 0.0007 & 0&0005 & 151 \\
91 & 56848.4250 & 0.0047 & 0&0023 & 54 \\
103 & 56849.2399 & 0.0008 & $-$0&0024 & 157 \\
104 & 56849.3058 & 0.0010 & $-$0&0048 & 146 \\
\hline
  \multicolumn{6}{l}{\commenta BJD$-$2400000.} \\
  \multicolumn{6}{l}{\commentb Against max $= 2456842.2073 + 0.068301 E$.} \\
  \multicolumn{6}{l}{\commentc Number of points used to determine the maximum.} \\
\end{tabular}
\end{center}
\end{table}

\subsection{ASASSN-14do}\label{obj:asassn14do}

   This object was detected as a transient at $V$=15.08
on 2014 July 1 by ASAS-SN team (vsnet-alert 17447).
The coordinates are \timeform{20h 37m 06.79s},
\timeform{-30D 12' 21.7''}.  The object has been
spectroscopically confirmed to be a CV
\citep{pri14asassn14doatel6293}.

\begin{figure}
  \begin{center}
    \FigureFile(88mm,110mm){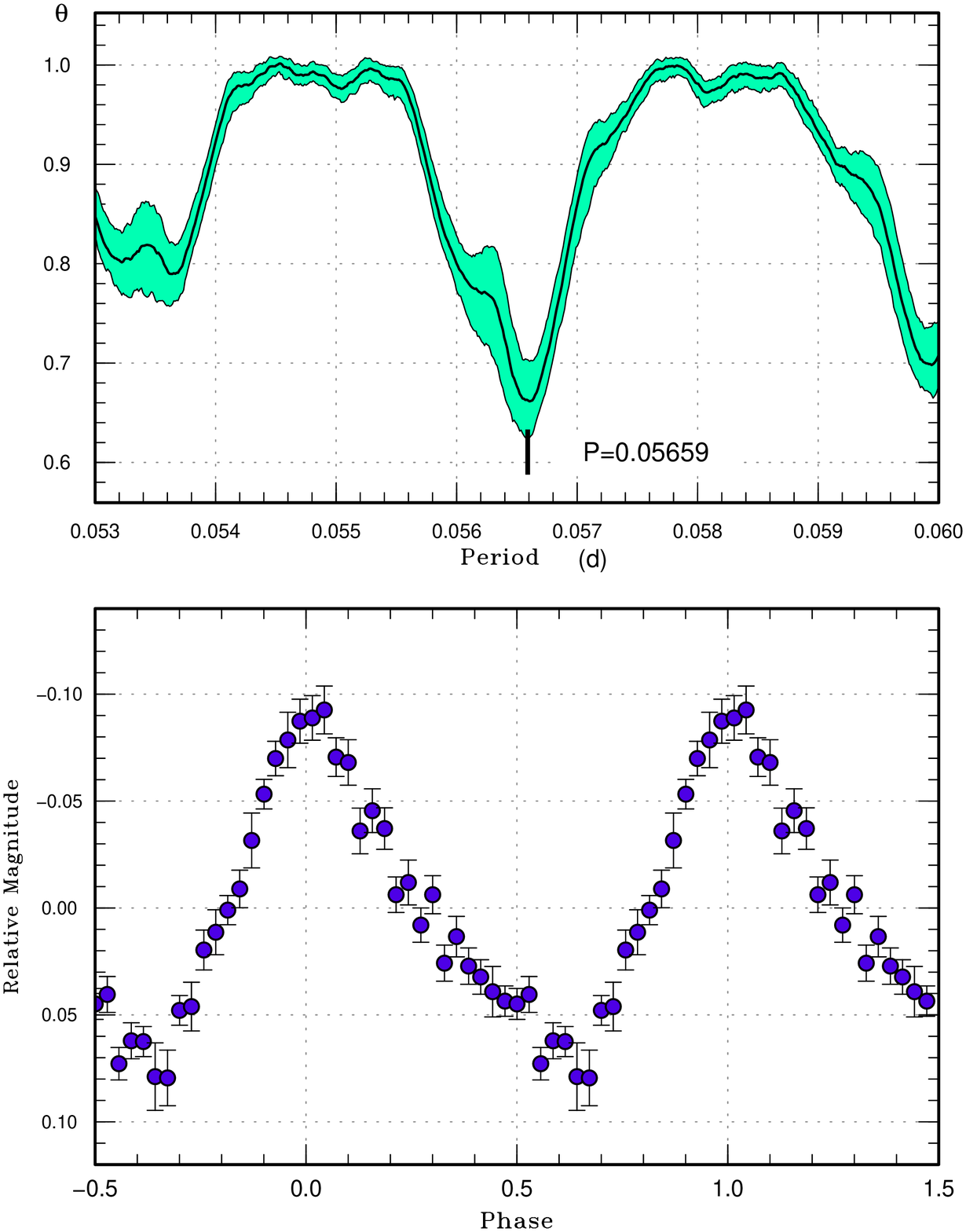}
  \end{center}
  \caption{Superhumps in ASASSN-14do (2014).
     (Upper): PDM analysis.
     (Lower): Phase-averaged profile}.
  \label{fig:asassn14doshpdm}
\end{figure}

   Although initial observations did not show strong
superhumps (vsnet-alert 17464), the object soon developed
superhumps (vsnet-alert 17467, 17476, 17506;
figure \ref{fig:asassn14doshpdm}).
The times of superhump maxima are listed in table
\ref{tab:asassn14dooc2014}.  Although $E=0$ corresponds
to stage A superhumps, the period of stage A superhumps
could not have been determined.

\begin{table}
\caption{Superhump maxima of ASASSN-14do (2014)}\label{tab:asassn14dooc2014}
\begin{center}
\begin{tabular}{rp{55pt}p{40pt}r@{.}lr}
\hline
\multicolumn{1}{c}{$E$} & \multicolumn{1}{c}{max\commenta} & \multicolumn{1}{c}{error} & \multicolumn{2}{c}{$O-C$\commentb} & \multicolumn{1}{c}{$N$\commentc} \\
\hline
0 & 56845.7565 & 0.0029 & $-$0&0083 & 19 \\
14 & 56846.5539 & 0.0005 & $-$0&0028 & 130 \\
15 & 56846.6156 & 0.0005 & 0&0023 & 130 \\
17 & 56846.7277 & 0.0016 & 0&0013 & 19 \\
18 & 56846.7866 & 0.0016 & 0&0036 & 17 \\
19 & 56846.8439 & 0.0013 & 0&0043 & 8 \\
30 & 56847.4636 & 0.0005 & 0&0018 & 131 \\
31 & 56847.5202 & 0.0005 & 0&0019 & 131 \\
32 & 56847.5770 & 0.0005 & 0&0021 & 130 \\
33 & 56847.6333 & 0.0005 & 0&0018 & 130 \\
34 & 56847.6920 & 0.0008 & 0&0039 & 90 \\
47 & 56848.4187 & 0.0058 & $-$0&0047 & 60 \\
48 & 56848.4764 & 0.0010 & $-$0&0036 & 130 \\
49 & 56848.5365 & 0.0006 & $-$0&0001 & 130 \\
50 & 56848.5938 & 0.0006 & 0&0007 & 131 \\
51 & 56848.6498 & 0.0062 & 0&0001 & 45 \\
52 & 56848.7094 & 0.0010 & 0&0032 & 12 \\
53 & 56848.7601 & 0.0017 & $-$0&0028 & 16 \\
54 & 56848.8136 & 0.0027 & $-$0&0058 & 15 \\
105 & 56851.7056 & 0.0029 & 0&0012 & 14 \\
\hline
  \multicolumn{6}{l}{\commenta BJD$-$2400000.} \\
  \multicolumn{6}{l}{\commentb Against max $= 2456845.7648 + 0.056568 E$.} \\
  \multicolumn{6}{l}{\commentc Number of points used to determine the maximum.} \\
\end{tabular}
\end{center}
\end{table}

\subsection{ASASSN-14dw}\label{obj:asassn14dw}

\begin{figure}
  \begin{center}
    \FigureFile(88mm,110mm){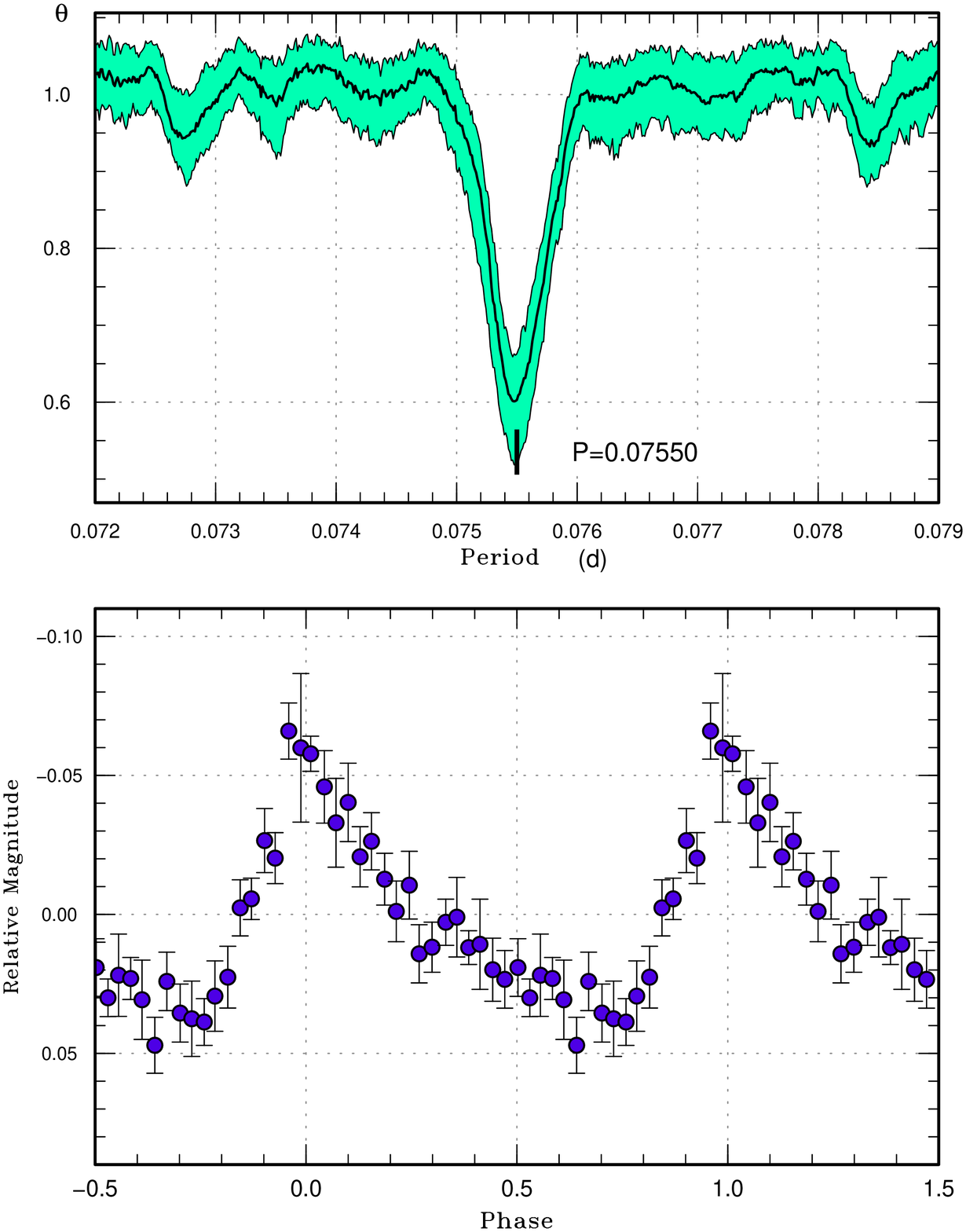}
  \end{center}
  \caption{Superhumps in ASASSN-14dw during the superoutburst
     plateau (2014).
     (Upper): PDM analysis.
     (Lower): Phase-averaged profile}.
  \label{fig:asassn14dwshpdm}
\end{figure}

   This object was detected as a transient at $V$=14.23
on 2014 July 10 by ASAS-SN team (vsnet-alert 17487).
The coordinates are \timeform{13h 43m 37.18s},
\timeform{-44D 26' 42.0''}
(2MASS position).  The object further brightened
to $V$=13.4 on July 12 in the ASAS-SN data.
Time-resolved photometry detected superhumps
(vsnet-alert 17495, 17498; figure \ref{fig:asassn14dwshpdm}).
The times of superhump maxima are listed in table
\ref{tab:asassn14dwoc2014}.
Both stages B and C can be identified.
The object started fading rapidly on July 22.
The duration of the superoutburst was $\sim$12~d,
which was not particularly long as in objects with
similar $P_{\rm SH}$, such as TT Boo \citep{ole04ttboo}.
During the post-superoutburst period, the object stayed
around 17 mag for at least eight days.  Although individual
superhump maxima were not measured, an analysis of
the entire post-superoutburst data yielded a strong
period of 0.07498(6)~d.
A PDM analysis of the combined data of stage C during
the superoutburst plateau and post-superoutburst yielded
a period of 0.07491(3)~d.  This period can express well
both observations of stage C during
the superoutburst plateau and post-superoutburst
without a phase shift.  It is most likely that there
was no phase jump around the termination of the superoutburst.

\begin{table}
\caption{Superhump maxima of ASASSN-14dw (2014)}\label{tab:asassn14dwoc2014}
\begin{center}
\begin{tabular}{rp{55pt}p{40pt}r@{.}lr}
\hline
\multicolumn{1}{c}{$E$} & \multicolumn{1}{c}{max\commenta} & \multicolumn{1}{c}{error} & \multicolumn{2}{c}{$O-C$\commentb} & \multicolumn{1}{c}{$N$\commentc} \\
\hline
0 & 56852.4711 & 0.0011 & $-$0&0080 & 10 \\
1 & 56852.5542 & 0.0022 & $-$0&0003 & 14 \\
14 & 56853.5352 & 0.0013 & $-$0&0003 & 17 \\
15 & 56853.6093 & 0.0011 & $-$0&0016 & 17 \\
27 & 56854.5197 & 0.0013 & 0&0034 & 16 \\
28 & 56854.5943 & 0.0006 & 0&0025 & 16 \\
53 & 56856.4832 & 0.0009 & 0&0051 & 14 \\
54 & 56856.5588 & 0.0012 & 0&0053 & 16 \\
67 & 56857.5379 & 0.0010 & 0&0035 & 16 \\
68 & 56857.6115 & 0.0012 & 0&0017 & 14 \\
80 & 56858.5150 & 0.0019 & $-$0&0002 & 17 \\
81 & 56858.5899 & 0.0021 & $-$0&0008 & 17 \\
93 & 56859.4895 & 0.0025 & $-$0&0066 & 15 \\
94 & 56859.5679 & 0.0053 & $-$0&0036 & 17 \\
\hline
  \multicolumn{6}{l}{\commenta BJD$-$2400000.} \\
  \multicolumn{6}{l}{\commentb Against max $= 2456852.4791 + 0.075451 E$.} \\
  \multicolumn{6}{l}{\commentc Number of points used to determine the maximum.} \\
\end{tabular}
\end{center}
\end{table}

\subsection{ASASSN-14eh}\label{obj:asassn14eh}

\begin{figure}
  \begin{center}
    \FigureFile(88mm,110mm){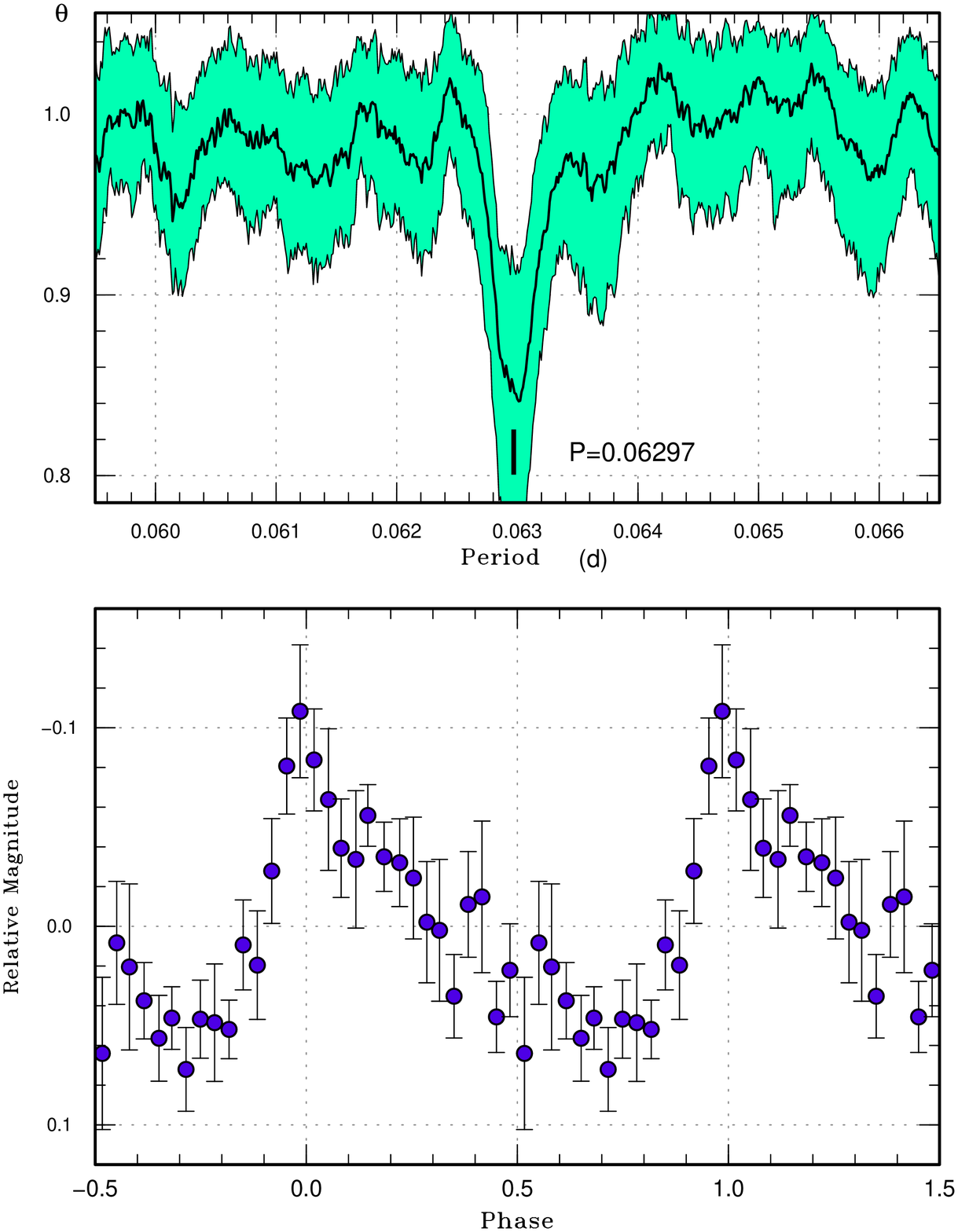}
  \end{center}
  \caption{Superhumps in ASASSN-14eh (2014).
     (Upper): PDM analysis.
     (Lower): Phase-averaged profile}.
  \label{fig:asassn14ehshpdm}
\end{figure}

   This object was detected as a transient at $V$=16.7
on 2014 July 21 by ASAS-SN team (vsnet-alert 17533).
The coordinates are \timeform{20h 32m 37.49s},
\timeform{+03D 05' 25.3''}
(GSC 2.3.2 position).
The object has a GALEX UV counterpart with an NUV
magnitude of 22.4.  The object further brightened
to $V$=14.94 on July 23.  On July 28, the object started
to show prominent superhumps (vsnet-alert 17568;
figure \ref{fig:asassn14ehshpdm})
accompanied by a slight rise of the brightness.
The times of superhump maxima are listed in
table \ref{tab:asassn14ehoc2014}.  Although there was
possibility of early superhumps or stage A superhumps
before July 28, we could not detect variations securely
due to the limited observational coverage.
The superoutburst lasted at least 13~d.

\begin{table}
\caption{Superhump maxima of ASASSN-14eh (2014)}\label{tab:asassn14ehoc2014}
\begin{center}
\begin{tabular}{rp{55pt}p{40pt}r@{.}lr}
\hline
\multicolumn{1}{c}{$E$} & \multicolumn{1}{c}{max\commenta} & \multicolumn{1}{c}{error} & \multicolumn{2}{c}{$O-C$\commentb} & \multicolumn{1}{c}{$N$\commentc} \\
\hline
0 & 56866.6506 & 0.0011 & 0&0001 & 7 \\
1 & 56866.7139 & 0.0007 & 0&0005 & 12 \\
16 & 56867.6592 & 0.0005 & 0&0022 & 10 \\
17 & 56867.7214 & 0.0013 & 0&0015 & 10 \\
33 & 56868.7278 & 0.0017 & 0&0014 & 7 \\
48 & 56869.6684 & 0.0012 & $-$0&0016 & 12 \\
49 & 56869.7314 & 0.0034 & $-$0&0015 & 9 \\
64 & 56870.6711 & 0.0022 & $-$0&0055 & 13 \\
65 & 56870.7339 & 0.0017 & $-$0&0056 & 7 \\
80 & 56871.6850 & 0.0024 & 0&0019 & 14 \\
86 & 56872.0644 & 0.0020 & 0&0039 & 71 \\
87 & 56872.1225 & 0.0030 & $-$0&0009 & 66 \\
96 & 56872.6933 & 0.0026 & 0&0037 & 12 \\
\hline
  \multicolumn{6}{l}{\commenta BJD$-$2400000.} \\
  \multicolumn{6}{l}{\commentb Against max $= 2456866.6505 + 0.062907 E$.} \\
  \multicolumn{6}{l}{\commentc Number of points used to determine the maximum.} \\
\end{tabular}
\end{center}
\end{table}

\subsection{ASASSN-14eq}\label{obj:asassn14eq}

\begin{figure}
  \begin{center}
    \FigureFile(88mm,110mm){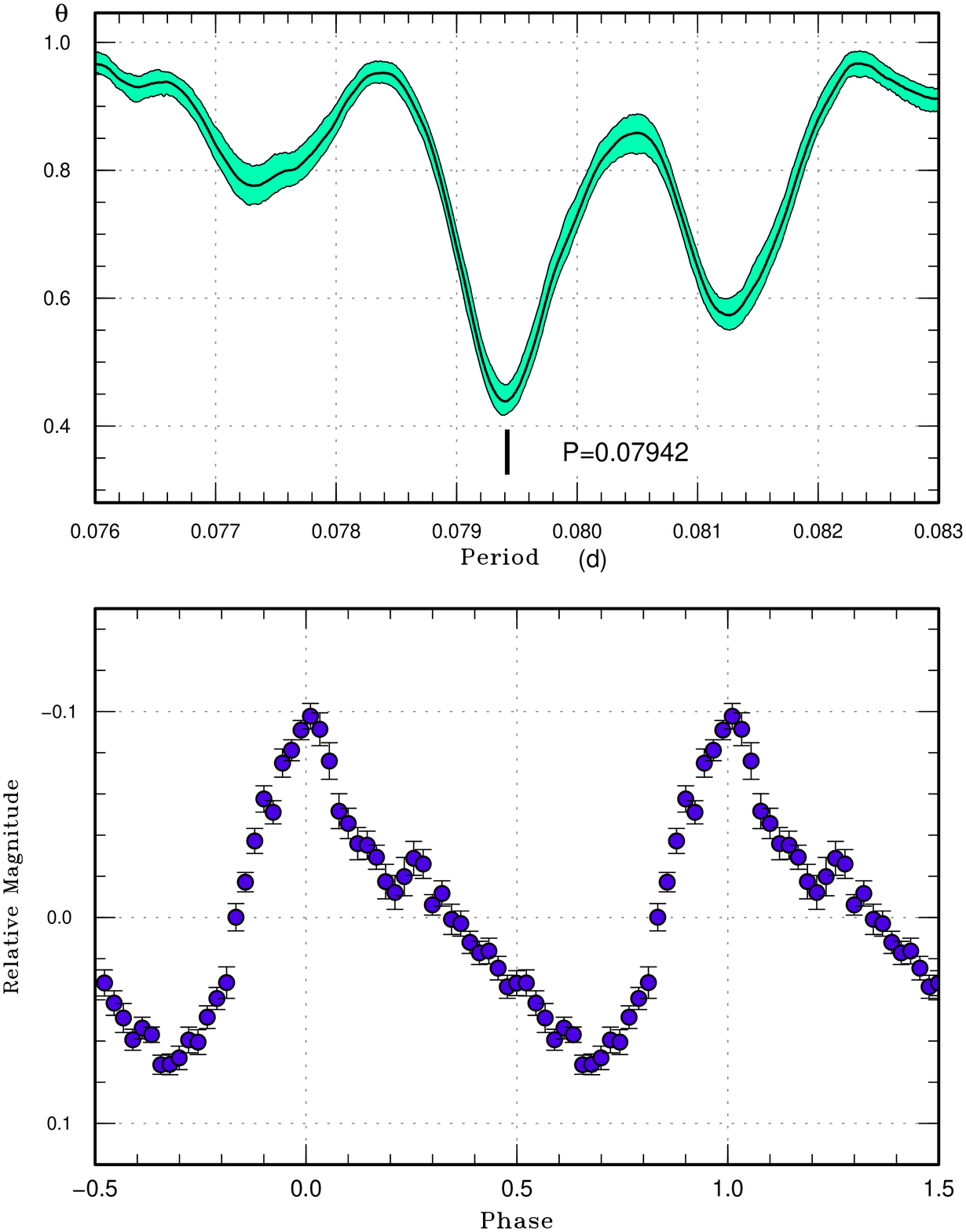}
  \end{center}
  \caption{Superhumps in ASASSN-14eq (2014).
     (Upper): PDM analysis.
     (Lower): Phase-averaged profile}.
  \label{fig:asassn14eqshpdm}
\end{figure}

   This object was detected as a transient at $V$=13.53
on 2014 July 28 by ASAS-SN team (vsnet-alert 17561).
The coordinates are \timeform{00h 21m 30.92s},
\timeform{-57D 19' 22.0''} (2MASS position).
The object has a GALEX UV counterpart with an NUV
magnitude of 18.1.  ASAS-3 data indicated
at least five past outbursts and at least one of them
(2005 November-December one) looked like a superoutburst.
Subsequent observations detected superhumps
(vsnet-alert 17576; figure \ref{fig:asassn14eqshpdm}).
The times of superhump maxima are listed in table
\ref{tab:asassn14eqoc2014}.
The large negative $P_{\rm dot}$ for the entire observation
appears to be a result of a stage transition.
We could not distinguish whether it was stage A-B or
stage B-C transition due to the lack of observations
before the ASAS-SN detection.

   A list of known outbursts is given in table
\ref{tab:asassn14eqout}.  It looks likely only superoutbursts
were recorded by ASAS-3.  The shortest interval between
the outbursts was $\sim$410~d, which may be
the supercycle.  Since the object is relatively bright,
the object would also be a good target for visual monitoring.

\begin{table}
\caption{Superhump maxima of ASASSN-14eq (2014)}\label{tab:asassn14eqoc2014}
\begin{center}
\begin{tabular}{rp{55pt}p{40pt}r@{.}lr}
\hline
\multicolumn{1}{c}{$E$} & \multicolumn{1}{c}{max\commenta} & \multicolumn{1}{c}{error} & \multicolumn{2}{c}{$O-C$\commentb} & \multicolumn{1}{c}{$N$\commentc} \\
\hline
0 & 56868.5150 & 0.0040 & $-$0&0050 & 64 \\
1 & 56868.5970 & 0.0005 & $-$0&0025 & 185 \\
2 & 56868.6762 & 0.0005 & $-$0&0027 & 148 \\
12 & 56869.4779 & 0.0003 & 0&0043 & 183 \\
13 & 56869.5612 & 0.0004 & 0&0082 & 184 \\
14 & 56869.6333 & 0.0004 & 0&0008 & 184 \\
50 & 56872.4948 & 0.0007 & 0&0015 & 183 \\
51 & 56872.5706 & 0.0007 & $-$0&0022 & 184 \\
52 & 56872.6499 & 0.0008 & $-$0&0024 & 164 \\
\hline
  \multicolumn{6}{l}{\commenta BJD$-$2400000.} \\
  \multicolumn{6}{l}{\commentb Against max $= 2456868.5200 + 0.079467 E$.} \\
  \multicolumn{6}{l}{\commentc Number of points used to determine the maximum.} \\
\end{tabular}
\end{center}
\end{table}

\begin{table}
\caption{List of known outbursts of ASASSN-14eq.}\label{tab:asassn14eqout}
\begin{center}
\begin{tabular}{cccccl}
\hline
Year & Month & max\commenta & magnitude & type & source \\
\hline
2001 & 9 & 52168 & 14.0 & ? & ASAS-3 \\
2002 & 10 & 52578 & 13.3 & ? & ASAS-3 \\
2004 & 1 & 53005 & 13.6 & ? & ASAS-3 \\
2005 & 11 & 53702 & 13.5 & super & ASAS-3 \\
2007 & 5 & 54231 & 13.5: & super? & ASAS-3 \\
2014 & 7 & 56867 & 13.5 & super & ASAS-SN \\
\hline
  \multicolumn{6}{l}{\commenta JD$-$2400000.} \\
\end{tabular}
\end{center}
\end{table}

\subsection{ASASSN-14fr}\label{obj:asassn14fr}

\begin{figure}
  \begin{center}
    \FigureFile(88mm,70mm){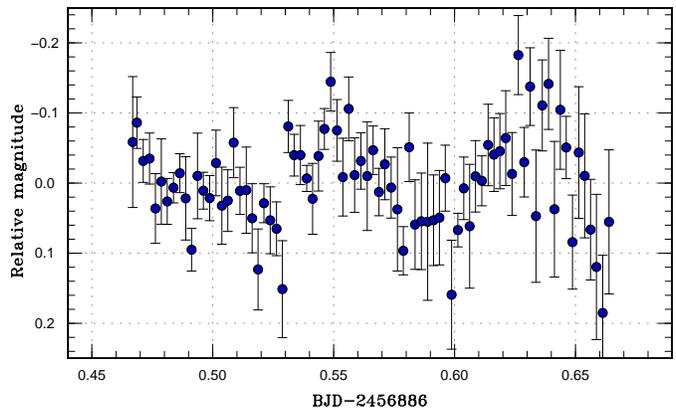}
  \end{center}
  \caption{Superhumps in ASASSN-14fr (2014).
  The data were binned to 0.0025~d.
  }
  \label{fig:asassn14frshlc}
\end{figure}

   This object was detected as a transient at $V$=15.9
on 2014 August 16 by ASAS-SN team (vsnet-alert 17628).
The coordinates are \timeform{21h 59m 06.90s},
\timeform{-35D 29' 56.8''}
(The Initial Gaia Source List).
The object was spectroscopically confirmed as
a CV in outburst \citep{hod14asassn14fratel6407}.
Based on a single-night
observation, a period of 0.073~d was reported
(vsnet-alert 17631; figure \ref{fig:asassn14frshlc}).
The times of superhump maxima
from these data are BJD 2456886.5052(18) ($N$=169)
and 2456886.6289(26) ($N$=169).  The observation
was not long enough to constrain the period
better than the initial report.

\subsection{ASASSN-14gd}\label{obj:asassn14gd}

\begin{figure}
  \begin{center}
    \FigureFile(88mm,110mm){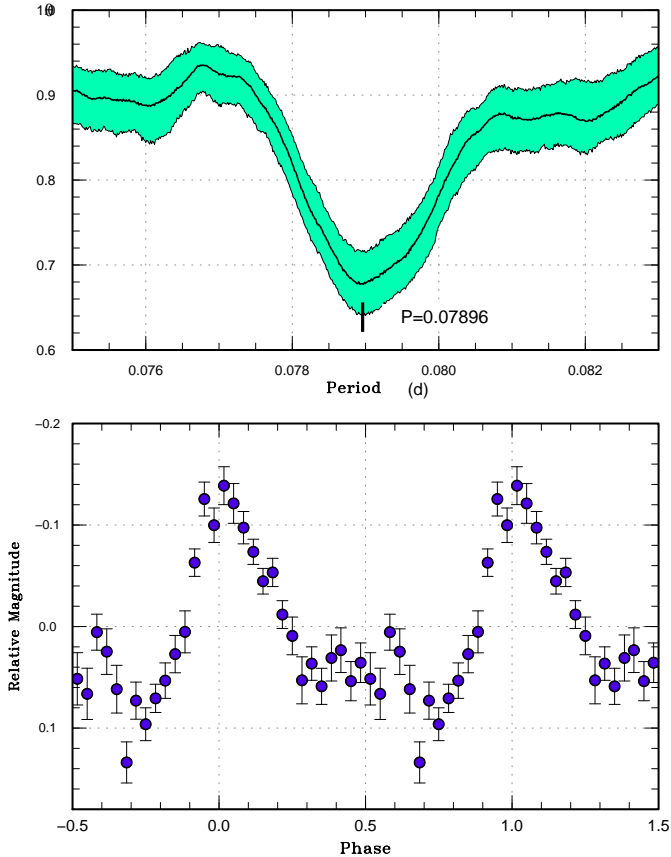}
  \end{center}
  \caption{Superhumps in ASASSN-14gd (2014).
     (Upper): PDM analysis.
     (Lower): Phase-averaged profile}.
  \label{fig:asassn14gdshpdm}
\end{figure}

   This object was detected as a transient at $V$=16.23
on 2014 August 23 by ASAS-SN team (vsnet-alert 17670).
The coordinates are \timeform{19h 26m 08.92s},
\timeform{-65D 26' 21.0''} (GSC 2.3.2 position).
There is a GALEX counterpart with an NUV magnitude
of 21.9.  Superhumps were detected in observations
on two nights (vsnet-alert 17669;
figure \ref{fig:asassn14gdshpdm}).
The times of superhump maxima are listed in table
\ref{tab:asassn14gdoc2014}.  The superhump period
in table \ref{tab:perlist} was obtained by the PDM method.

\begin{table}
\caption{Superhump maxima of ASASSN-14gd (2014)}\label{tab:asassn14gdoc2014}
\begin{center}
\begin{tabular}{rp{55pt}p{40pt}r@{.}lr}
\hline
\multicolumn{1}{c}{$E$} & \multicolumn{1}{c}{max\commenta} & \multicolumn{1}{c}{error} & \multicolumn{2}{c}{$O-C$\commentb} & \multicolumn{1}{c}{$N$\commentc} \\
\hline
0 & 56895.3042 & 0.0011 & 0&0010 & 182 \\
1 & 56895.3813 & 0.0013 & $-$0&0010 & 179 \\
12 & 56896.2528 & 0.0008 & 0&0003 & 177 \\
13 & 56896.3298 & 0.0009 & $-$0&0018 & 180 \\
14 & 56896.4121 & 0.0010 & 0&0014 & 148 \\
\hline
  \multicolumn{6}{l}{\commenta BJD$-$2400000.} \\
  \multicolumn{6}{l}{\commentb Against max $= 2456895.3032 + 0.079109 E$.} \\
  \multicolumn{6}{l}{\commentc Number of points used to determine the maximum.} \\
\end{tabular}
\end{center}
\end{table}

\subsection{ASASSN-14gq}\label{obj:asassn14gq}

\begin{figure}
  \begin{center}
    \FigureFile(88mm,70mm){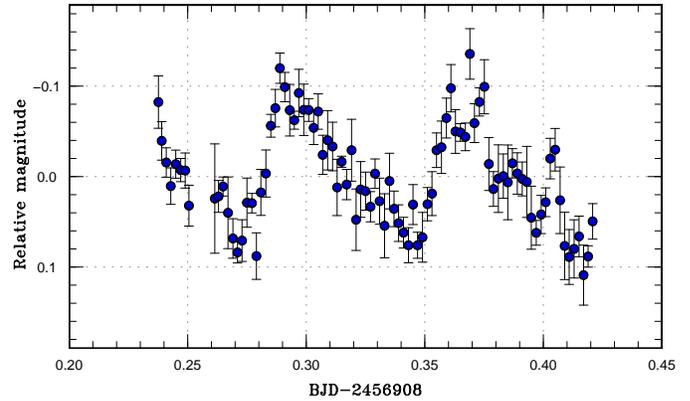}
  \end{center}
  \caption{Superhumps in ASASSN-14gq (2014).
  The data were binned to 0.002~d.
  }
  \label{fig:asassn14gqshlc}
\end{figure}

   This object was detected as a transient at $V$=13.92
on 2014 September 4 by ASAS-SN team (vsnet-alert 17695).
The coordinates are \timeform{19h 27m 10.15s},
\timeform{-48D 47' 52.3''} (2MASS position).
At least six past outbursts were recorded by CRTS.
Superhumps with a period of $\sim$0.072~d were
recorded (vsnet-alert 17731; figure \ref{fig:asassn14gqshlc}x).
Two superhump maxima were recorded: BJD 2456908.2970(7) ($N$=164)
and 2456908.3687(8) ($N$=166).  Although there were
observation on BJD 2456916, the amplitudes of superhumps
were too small to determine the times of maxima.

\subsection{ASASSN-14gx}\label{obj:asassn14gx}

\begin{figure}
  \begin{center}
    \FigureFile(88mm,110mm){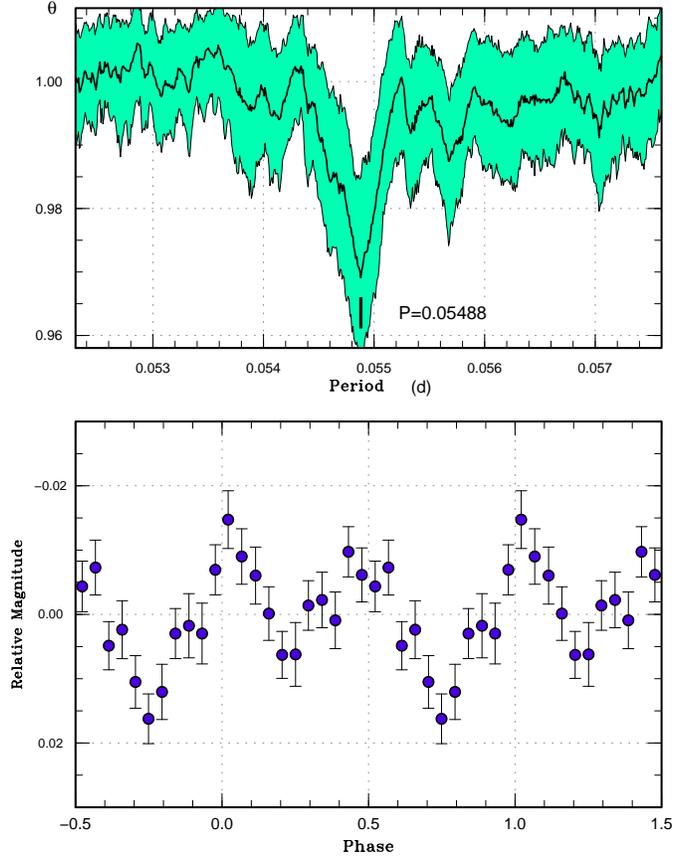}
  \end{center}
  \caption{Early superhumps in ASASSN-14gx (2014).
     (Upper): PDM analysis.
     (Lower): Phase-averaged profile}.
  \label{fig:asassn14gxeshpdm}
\end{figure}

\begin{figure}
  \begin{center}
    \FigureFile(88mm,110mm){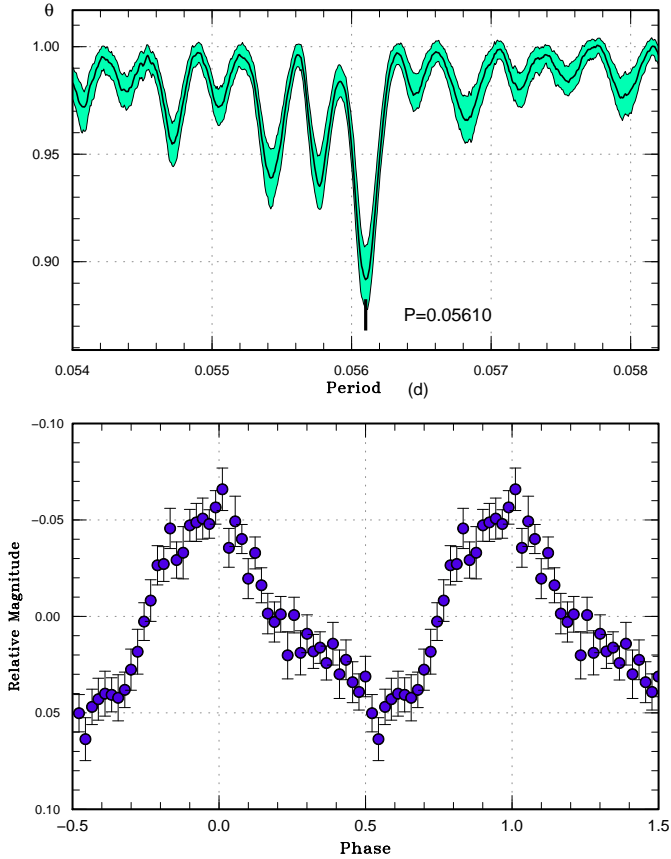}
  \end{center}
  \caption{Ordinary superhumps in ASASSN-14gx during
     the plateau phase (2014).
     (Upper): PDM analysis.
     (Lower): Phase-averaged profile}.
  \label{fig:asassn14gxshpdm}
\end{figure}

\begin{figure}
  \begin{center}
    \FigureFile(88mm,70mm){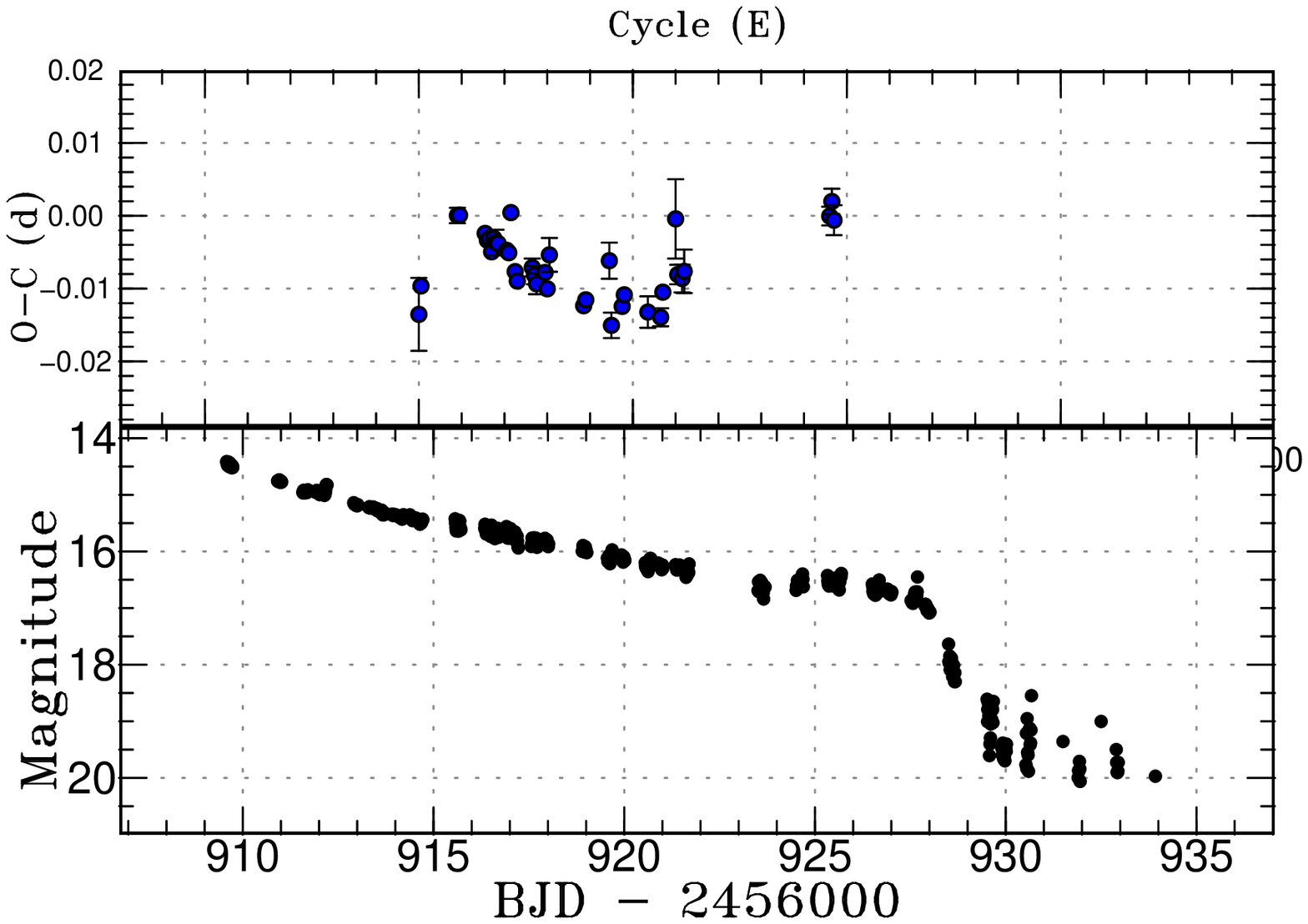}
  \end{center}
  \caption{$O-C$ diagram of superhumps in ASASSN-14gx (2014).
     (Upper): $O-C$ diagram.  A period of 0.05609~d
     was used to draw this figure.
     (Lower): Light curve.  The observations were binned to 0.011~d.}
  \label{fig:asassn14gxhumpall}
\end{figure}

   This object was detected as a transient at $V$=14.9
on 2014 September 6 by ASAS-SN team (vsnet-alert 17702).
The object further brightened to $V$=13.8 on September 8.
The coordinates are \timeform{23h 51m 23.92s},
\timeform{-30D 27' 25.1''} (GSC 2.3.2 position).
A signature of early superhumps was immediately
reported (vsnet-alert 17715).  The period of early superhumps
was updated to 0.05488(3)~d
(figure \ref{fig:asassn14gxeshpdm}).
On September 15, the object showed ordinary superhumps
(vsnet-alert 17738, 17748; figure \ref{fig:asassn14gxshpdm}).
During the terminal stage of the plateau phase,
the object slightly brightened (vsnet-alert 17768;
figure \ref{fig:asassn14gxhumpall}).
The times of superhumps maxima are listed in table
\ref{tab:asassn14gxoc2014}.
The maxima $E \le 1$ correspond to stage A superhumps.
Although stage A superhumps were recorded, we could
not determine the period due to the shortness
of the observation.
Despite the brightening in the later part, there was
no apparent indication of stage B-C transition
and maxima up to $E$=194 were well expressed by
an ephemeris up to $E$=124.  We cannot, however,
exclude a discontinuous period change due to the gap
of 70 cycles.  A PDM analysis after
the brightening (BJD 2456924--2456928) yielded
a possible period of 0.05592(9)~d.  Although
this period may be that of stage C superhumps,
observations were not sufficient to confirm this
identification. 
Despite the large outburst amplitude, the behavior of
this object did not look like that of an extreme
WZ Sge-type dwarf nova.

\begin{table}
\caption{Superhump maxima of ASASSN-14gx (2014)}\label{tab:asassn14gxoc2014}
\begin{center}
\begin{tabular}{rp{55pt}p{40pt}r@{.}lr}
\hline
\multicolumn{1}{c}{$E$} & \multicolumn{1}{c}{max\commenta} & \multicolumn{1}{c}{error} & \multicolumn{2}{c}{$O-C$\commentb} & \multicolumn{1}{c}{$N$\commentc} \\
\hline
0 & 56914.6027 & 0.0050 & $-$0&0065 & 144 \\
1 & 56914.6627 & 0.0008 & $-$0&0026 & 14 \\
18 & 56915.6259 & 0.0011 & 0&0070 & 14 \\
19 & 56915.6821 & 0.0005 & 0&0070 & 15 \\
31 & 56916.3527 & 0.0006 & 0&0044 & 101 \\
32 & 56916.4078 & 0.0006 & 0&0035 & 90 \\
33 & 56916.4642 & 0.0008 & 0&0038 & 90 \\
34 & 56916.5184 & 0.0007 & 0&0019 & 91 \\
35 & 56916.5763 & 0.0007 & 0&0037 & 131 \\
36 & 56916.6319 & 0.0018 & 0&0032 & 64 \\
37 & 56916.6878 & 0.0010 & 0&0030 & 15 \\
41 & 56916.9112 & 0.0002 & 0&0021 & 47 \\
42 & 56916.9670 & 0.0003 & 0&0017 & 55 \\
43 & 56917.0286 & 0.0010 & 0&0072 & 21 \\
45 & 56917.1327 & 0.0009 & $-$0&0009 & 42 \\
46 & 56917.1875 & 0.0009 & $-$0&0022 & 41 \\
53 & 56917.5819 & 0.0012 & $-$0&0004 & 15 \\
54 & 56917.6370 & 0.0013 & $-$0&0014 & 14 \\
55 & 56917.6919 & 0.0014 & $-$0&0026 & 15 \\
59 & 56917.9178 & 0.0005 & $-$0&0011 & 55 \\
60 & 56917.9717 & 0.0004 & $-$0&0034 & 56 \\
61 & 56918.0324 & 0.0023 & 0&0013 & 13 \\
77 & 56918.9229 & 0.0005 & $-$0&0058 & 56 \\
78 & 56918.9798 & 0.0005 & $-$0&0050 & 56 \\
89 & 56919.6021 & 0.0025 & 0&0003 & 15 \\
90 & 56919.6493 & 0.0017 & $-$0&0086 & 14 \\
95 & 56919.9324 & 0.0007 & $-$0&0060 & 56 \\
96 & 56919.9901 & 0.0006 & $-$0&0044 & 56 \\
107 & 56920.6047 & 0.0022 & $-$0&0068 & 14 \\
113 & 56920.9405 & 0.0012 & $-$0&0076 & 51 \\
114 & 56921.0000 & 0.0006 & $-$0&0041 & 46 \\
\hline
  \multicolumn{6}{l}{\commenta BJD$-$2400000.} \\
  \multicolumn{6}{l}{\commentb Against max $= 2456914.6092 + 0.056096 E$.} \\
  \multicolumn{6}{l}{\commentc Number of points used to determine the maximum.} \\
\end{tabular}
\end{center}
\end{table}

\addtocounter{table}{-1}
\begin{table}
\caption{Superhump maxima of ASASSN-14gx (2014) (continued)}
\begin{center}
\begin{tabular}{rp{55pt}p{40pt}r@{.}lr}
\hline
\multicolumn{1}{c}{$E$} & \multicolumn{1}{c}{max\commenta} & \multicolumn{1}{c}{error} & \multicolumn{2}{c}{$O-C$\commentb} & \multicolumn{1}{c}{$N$\commentc} \\
\hline
120 & 56921.3467 & 0.0055 & 0&0059 & 76 \\
121 & 56921.3951 & 0.0013 & $-$0&0018 & 127 \\
123 & 56921.5067 & 0.0019 & $-$0&0023 & 129 \\
124 & 56921.5638 & 0.0030 & $-$0&0013 & 141 \\
192 & 56925.3855 & 0.0013 & 0&0058 & 129 \\
193 & 56925.4436 & 0.0017 & 0&0078 & 129 \\
194 & 56925.4971 & 0.0021 & 0&0052 & 136 \\
\hline
  \multicolumn{6}{l}{\commenta BJD$-$2400000.} \\
  \multicolumn{6}{l}{\commentb Against max $= 2456914.6092 + 0.056096 E$.} \\
  \multicolumn{6}{l}{\commentc Number of points used to determine the maximum.} \\
\end{tabular}
\end{center}
\end{table}

\subsection{ASASSN-14hk}\label{obj:asassn14hk}

\begin{figure}
  \begin{center}
    \FigureFile(88mm,110mm){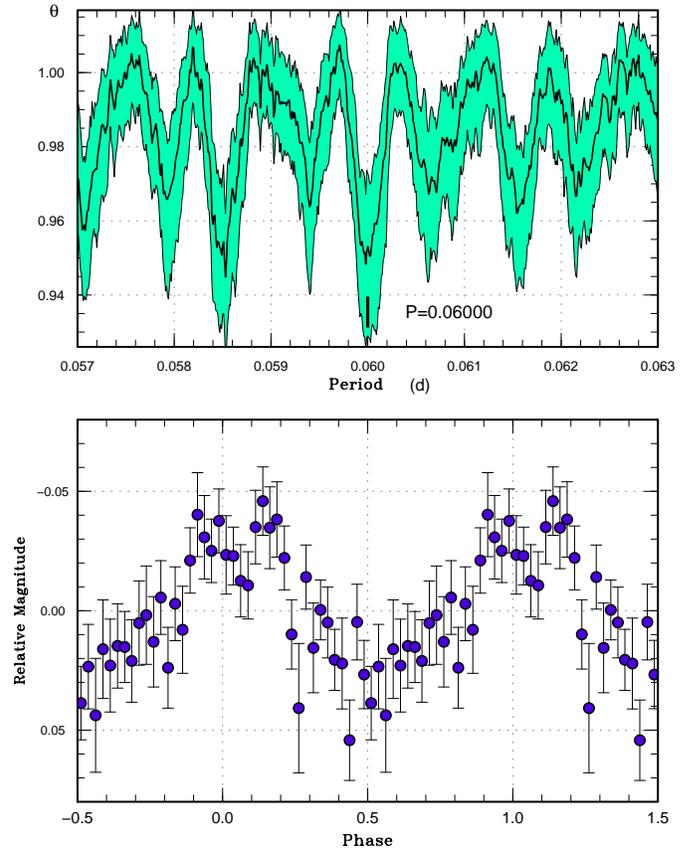}
  \end{center}
  \caption{Superhumps in ASASSN-14hk after BJD 2456923 (2014).
     (Upper): PDM analysis.
     (Lower): Phase-averaged profile}.
  \label{fig:asassn14hkshpdm}
\end{figure}

   This object was detected as a transient at $V$=14.61
on 2014 September 16 by ASAS-SN team \citep{sta14asassn14hkatel6479}.
The coordinates are \timeform{01h 39m 35.62s},
\timeform{+35D 38' 40.8''} (GSC 2.3.2 position).
There is a GALEX counterpart with an NUV magnitude
of 21.9.  The large outburst amplitude ($\sim$6 mag) made
the object a good candidate for an SU UMa-type dwarf nova.
Subsequent observations initially detect no apparent
modulations on September 20.  Superhumps were present
on September 22 (BJD 2456921, vsnet-alert 17773).
The superhumps took 4--6~d to develop.
Although there remained some ambiguity in period
selection, we selected the superhump period
to best match the result from the long single-night observation
on BJD 2456921 (figure \ref{fig:asassn14hkshpdm}).
The times of superhump maxima are listed in table
\ref{tab:asassn14hkoc2014}.  The epochs for $E \le 118$
were not very certain due to the low signal-to-noise ratio.

\begin{table}
\caption{Superhump maxima of ASASSN-14hk (2014)}\label{tab:asassn14hkoc2014}
\begin{center}
\begin{tabular}{rp{55pt}p{40pt}r@{.}lr}
\hline
\multicolumn{1}{c}{$E$} & \multicolumn{1}{c}{max\commenta} & \multicolumn{1}{c}{error} & \multicolumn{2}{c}{$O-C$\commentb} & \multicolumn{1}{c}{$N$\commentc} \\
\hline
0 & 56921.0890 & 0.0035 & 0&0026 & 51 \\
1 & 56921.1463 & 0.0026 & $-$0&0002 & 71 \\
2 & 56921.2101 & 0.0026 & 0&0037 & 120 \\
3 & 56921.2638 & 0.0016 & $-$0&0026 & 128 \\
36 & 56923.2428 & 0.0010 & $-$0&0036 & 45 \\
37 & 56923.3049 & 0.0008 & $-$0&0016 & 45 \\
118 & 56928.1686 & 0.0050 & 0&0021 & 61 \\
135 & 56929.1861 & 0.0018 & $-$0&0005 & 27 \\
\hline
  \multicolumn{6}{l}{\commenta BJD$-$2400000.} \\
  \multicolumn{6}{l}{\commentb Against max $= 2456921.0864 + 0.060001 E$.} \\
  \multicolumn{6}{l}{\commentc Number of points used to determine the maximum.} \\
\end{tabular}
\end{center}
\end{table}

\subsection{ASASSN-14hl}\label{obj:asassn14hl}

\begin{figure}
  \begin{center}
    \FigureFile(88mm,70mm){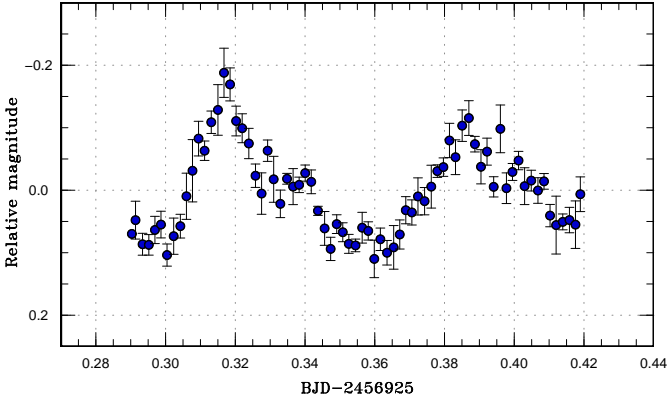}
  \end{center}
  \caption{Superhumps in ASASSN-14hl (2014).
  The data were binned to 0.0018~d.
  }
  \label{fig:asassn14hlshlc}
\end{figure}

   This object was detected as a transient at $V$=15.42
on 2014 September 18 by ASAS-SN team (vsnet-alert 17746).
The coordinates are \timeform{23h 49m 55.12s},
\timeform{-60D 54' 17.1''} (the Initial Gaia Source List).
Two superhumps were detected from a single-night
observation (vsnet-alert 17772,
figure \ref{fig:asassn14hlshlc}): BJD 2456925.3196(6)
($N$=157) and 2456925.3885(7) ($N$=158).
The superhump period is 0.0685(5)~d.

\subsection{ASASSN-14hs}\label{obj:asassn14hs}

\begin{figure}
  \begin{center}
    \FigureFile(88mm,110mm){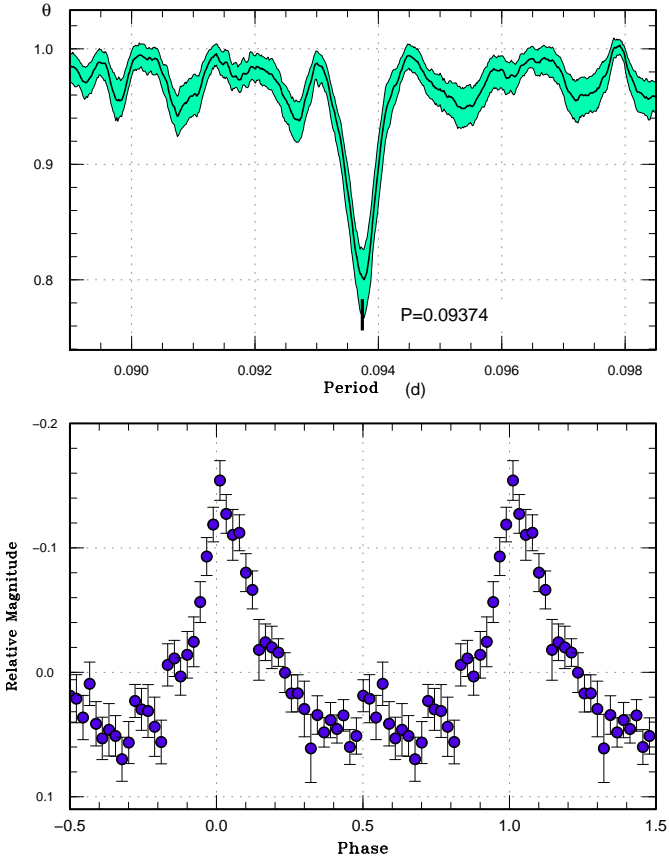}
  \end{center}
  \caption{Superhumps in ASASSN-14hs (2014).
     (Upper): PDM analysis.
     (Lower): Phase-averaged profile}.
  \label{fig:asassn14hsshpdm}
\end{figure}

   This object was detected as a transient at $V$=13.99
on 2014 September 26 by ASAS-SN team (vsnet-alert 17777).
The coordinates are \timeform{22h 42m 58.04s},
\timeform{-19D 45' 51.5''} (the Initial Gaia Source List).
There is a GALEX counterpart with an NUV magnitude
of 20.0.  Subsequent observations
detected superhumps (vsnet-alert 17785, 17802, 17809;
figure \ref{fig:asassn14hsshpdm}).
The period suggests that the object is located
in the lower edge of the period gap.
The times of superhumps maxima are listed in table
\ref{tab:asassn14hsoc2014}.  Except $E \le 1$,
the maxima could be expressed by a relatively
constant period.  The last two epochs correspond to
the rapidly fading part.  Since $O-C$ variations
in long-$P_{\rm orb}$ systems strongly depends on
object, we could not identify the superhump stage
confidently in this system.

\begin{table}
\caption{Superhump maxima of ASASSN-14hs (2014)}\label{tab:asassn14hsoc2014}
\begin{center}
\begin{tabular}{rp{55pt}p{40pt}r@{.}lr}
\hline
\multicolumn{1}{c}{$E$} & \multicolumn{1}{c}{max\commenta} & \multicolumn{1}{c}{error} & \multicolumn{2}{c}{$O-C$\commentb} & \multicolumn{1}{c}{$N$\commentc} \\
\hline
0 & 56929.0606 & 0.0005 & $-$0&0092 & 97 \\
1 & 56929.1540 & 0.0005 & $-$0&0096 & 118 \\
10 & 56930.0118 & 0.0020 & 0&0047 & 51 \\
11 & 56930.1008 & 0.0069 & $-$0&0001 & 9 \\
26 & 56931.5084 & 0.0118 & 0&0016 & 8 \\
27 & 56931.6067 & 0.0009 & 0&0062 & 18 \\
27 & 56931.6071 & 0.0010 & 0&0065 & 18 \\
37 & 56932.5417 & 0.0006 & 0&0039 & 22 \\
38 & 56932.6277 & 0.0065 & $-$0&0038 & 10 \\
55 & 56934.2313 & 0.0069 & 0&0064 & 97 \\
56 & 56934.3167 & 0.0005 & $-$0&0020 & 193 \\
57 & 56934.4112 & 0.0021 & $-$0&0012 & 65 \\
66 & 56935.2541 & 0.0008 & $-$0&0019 & 162 \\
67 & 56935.3500 & 0.0011 & 0&0003 & 195 \\
69 & 56935.5386 & 0.0014 & 0&0014 & 39 \\
70 & 56935.6408 & 0.0018 & 0&0099 & 41 \\
80 & 56936.5526 & 0.0027 & $-$0&0156 & 47 \\
81 & 56936.6670 & 0.0070 & 0&0050 & 22 \\
88 & 56937.3185 & 0.0025 & 0&0005 & 216 \\
90 & 56937.5213 & 0.0080 & 0&0158 & 30 \\
101 & 56938.5340 & 0.0027 & $-$0&0026 & 29 \\
102 & 56938.6143 & 0.0032 & $-$0&0160 & 50 \\
\hline
  \multicolumn{6}{l}{\commenta BJD$-$2400000.} \\
  \multicolumn{6}{l}{\commentb Against max $= 2456929.0698 + 0.093730 E$.} \\
  \multicolumn{6}{l}{\commentc Number of points used to determine the maximum.} \\
\end{tabular}
\end{center}
\end{table}

\subsection{ASASSN-14ia}\label{obj:asassn14ia}

\begin{figure}
  \begin{center}
    \FigureFile(88mm,110mm){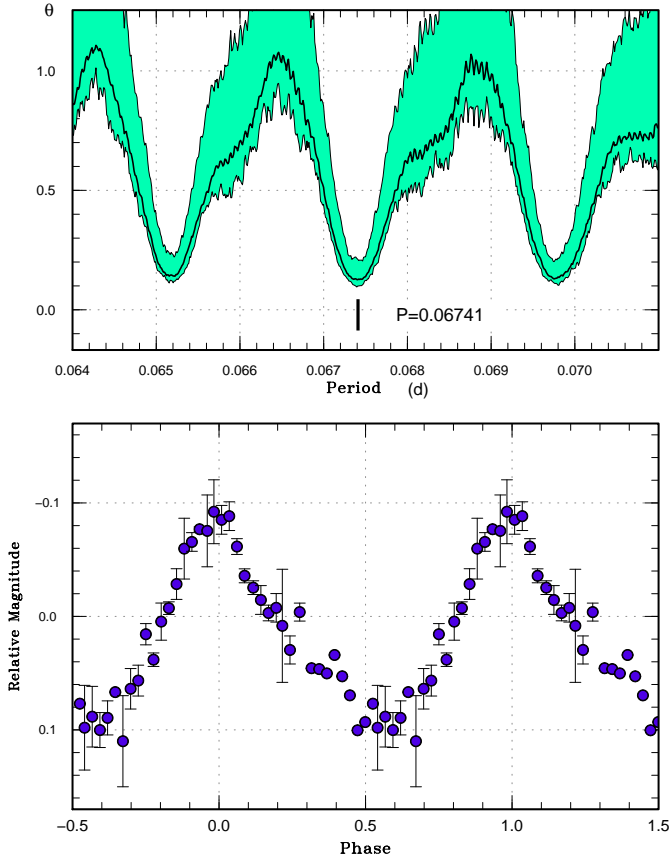}
  \end{center}
  \caption{Superhumps in ASASSN-14ia (2014).
     (Upper): PDM analysis.
     (Lower): Phase-averaged profile}.
  \label{fig:asassn14iashpdm}
\end{figure}

   This object was first detected as a transient at $V$=14.3
on 2014 September 19 by ASAS-SN team but was recognized
at $V$=15.6 on September 24 (vsnet-alert 17786).
The coordinates are \timeform{06h 40m 20.89s},
\timeform{+44D 21' 26.4''} (the Initial Gaia Source List).
Only rather fragmentary
observations on two nights were obtained.
The resultant superhump maxima were BJD 2456930.2213(29)
($N$=15), 2456930.2910(6) ($N$=24) and
2456932.2452(9) ($N$=19).
The superhump period was not very well constrained.
In figure \ref{fig:asassn14iashpdm}, we selected
one of the possible aliases.

\subsection{ASASSN-14id}\label{obj:asassn14id}

\begin{figure}
  \begin{center}
    \FigureFile(88mm,110mm){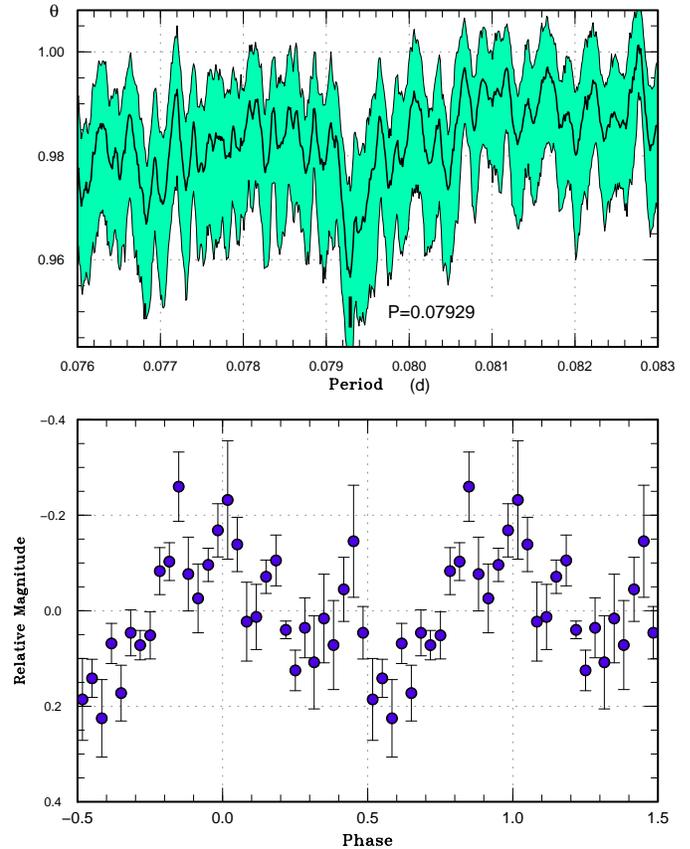}
  \end{center}
  \caption{Superhumps in ASASSN-14id (2014).
     (Upper): PDM analysis.  A weak signal near
     0.0769~d is likely the orbital one.
     (Lower): Phase-averaged profile}.
  \label{fig:asassn14idshpdm}
\end{figure}

\begin{figure}
  \begin{center}
    \FigureFile(88mm,70mm){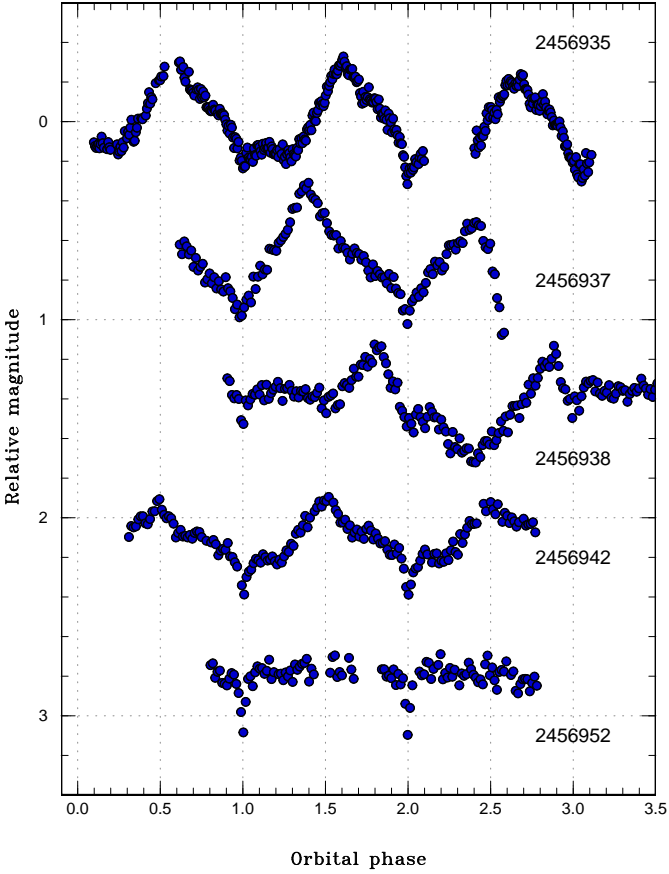}
  \end{center}
  \caption{Superhumps and eclipses in ASASSN-14id (2014).
  The orbital phase refers to equation (\ref{equ:asassn14idecl}).
  Shallow signatures of eclipses were also detected in
  all observations.
  The data were binned to 0.0025~d.
  }
  \label{fig:asassn14idshlc}
\end{figure}

   This object was detected as a transient at $V$=14.3
on 2014 September 29 by ASAS-SN team (vsnet-alert 17791).
The coordinates are \timeform{07h 26m 37.23s},
\timeform{+83D 32' 18.6''} (SDSS $g$=20.8 counterpart).
Subsequent observations detected large-amplitude
superhumps (vsnet-alert 17814, 17826;
figures \ref{fig:asassn14idshpdm}, \ref{fig:asassn14idshlc}).
The times superhumps maxima are listed in table
\ref{tab:asassn14idoc2014}.
Although the PDM figure is not so clear, it is probably
a result of beat modulation because this object is eclipsing
(as shown later).  The period selection is secure from
the $O-C$ analysis of the best observed segment of $E \le 24$.
The superhump stage was not clear from these
observations.

   On BJD 2456942 and 2456952, shallow eclipses
($\sim$0.3 mag) were observed.  Since superhumps
were strongly seen on BJD 2456942, we subtracted
the superhump signal.  After combining with
the data on BJD 2456952, we applied an MCMC analysis
(cf. \cite{Pdot4}) to determine the eclipse ephemeris.
The result was
\begin{equation}
{\rm Min(BJD)} = 2456942.3720(5) + 0.076857(4) E .
\label{equ:asassn14idecl}
\end{equation}
Although there were several possibilities of the orbital
period due to the long gap between two nights,
this period was selected because it gives
a reasonably acceptable $\epsilon$ of 0.033.
This period selection is confirmed by the detection
of shallow eclipse-like features at zero orbital phase
on earlier nights (figure \ref{fig:asassn14idshlc}).
Since the eclipse profile was likely affected by
superhumps, this ephemeris is not intended for
long-term prediction of eclipse times.
The observed large superhump amplitudes on the
initial two nights were most likely a result of
a high orbital inclination.

\begin{table}
\caption{Superhump maxima of ASASSN-14id (2014)}\label{tab:asassn14idoc2014}
\begin{center}
\begin{tabular}{rp{50pt}p{30pt}r@{.}lcr}
\hline
$E$ & max\commenta & error & \multicolumn{2}{c}{$O-C$\commentb} & phase\commentc & $N$\commentd \\
\hline
0 & 56935.3475 & 0.0003 & $-$0&0012 & 0.60 & 75 \\
1 & 56935.4267 & 0.0002 & $-$0&0014 & 0.63 & 103 \\
2 & 56935.5066 & 0.0002 & $-$0&0009 & 0.67 & 77 \\
24 & 56937.2530 & 0.0003 & $-$0&0006 & 0.40 & 52 \\
25 & 56937.3256 & 0.0008 & $-$0&0074 & 0.34 & 36 \\
36 & 56938.2071 & 0.0016 & 0&0012 & 0.81 & 46 \\
37 & 56938.2945 & 0.0005 & 0&0092 & 0.95 & 46 \\
48 & 56939.1624 & 0.0038 & 0&0041 & 0.24 & 31 \\
88 & 56942.3328 & 0.0007 & $-$0&0002 & 0.49 & 40 \\
89 & 56942.4127 & 0.0005 & 0&0004 & 0.53 & 54 \\
90 & 56942.4908 & 0.0006 & $-$0&0009 & 0.55 & 40 \\
137 & 56946.2196 & 0.0024 & $-$0&0023 & 0.06 & 26 \\
\hline
  \multicolumn{7}{l}{\commenta BJD$-$2400000.} \\
  \multicolumn{7}{l}{\commentb Against max $= 2456935.3488 + 0.079366 E$.} \\
  \multicolumn{7}{l}{\commentc Orbital phase.} \\
  \multicolumn{7}{l}{\commentd Number of points used to determine the maximum.} \\
\end{tabular}
\end{center}
\end{table}

\subsection{ASASSN-14it}\label{obj:asassn14it}

\begin{figure}
  \begin{center}
    \FigureFile(88mm,70mm){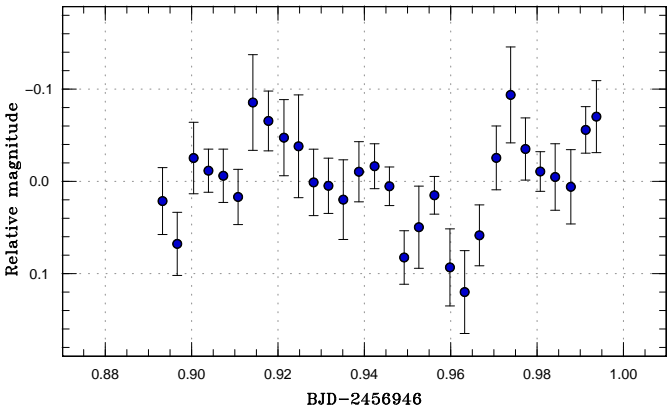}
  \end{center}
  \caption{Possible superhumps in ASASSN-14it (2014).
  The data were binned to 0.0035~d.
  }
  \label{fig:asassn14itshlc}
\end{figure}

   This object was detected as a transient at $V$=15.6
on 2014 October 10 by ASAS-SN team.
The coordinates are \timeform{20h 11m 09.13s},
\timeform{+53D 39' 01.7''} (R. Nesci, vsnet-alert 17840).
The absence of a quiescent counterpart in past
catalogs suggested a large ($\ge$6 mag) outburst amplitude.
Only a short run on a single night was obtained.
Although possible superhumps were detected
(figure \ref{fig:asassn14itshlc}), the period was 
difficult to estimate due to the shortness of
the observation.

\subsection{ASASSN-14iv}\label{obj:asassn14iv}

\begin{figure}
  \begin{center}
    \FigureFile(88mm,110mm){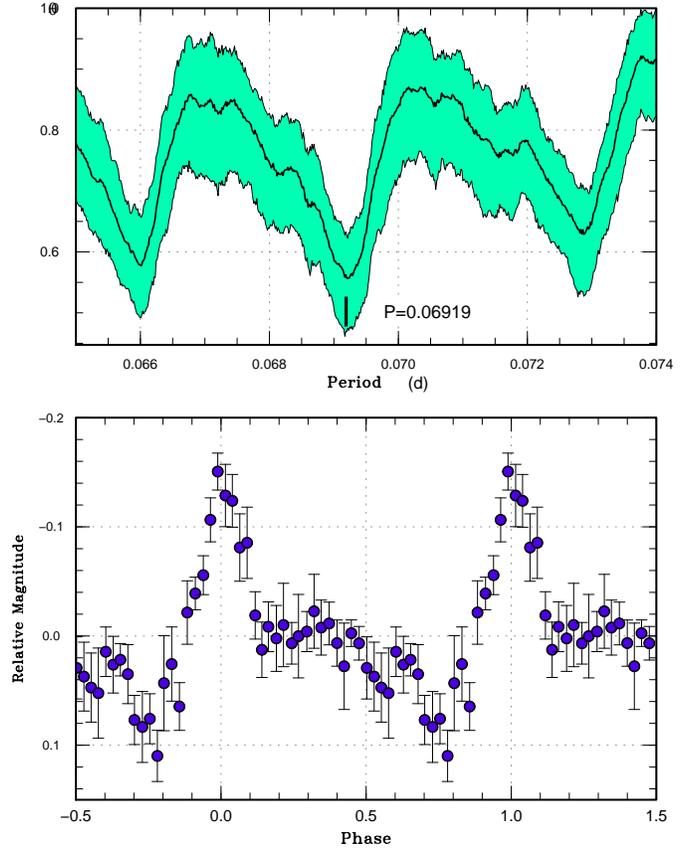}
  \end{center}
  \caption{Superhumps in ASASSN-14iv (2014).
     (Upper): PDM analysis.  The selection of the alias
     was based on $O-C$ analysis.
     (Lower): Phase-averaged profile}.
  \label{fig:asassn14ivshpdm}
\end{figure}

   This object was detected as a transient at $V$=15.6
on 2014 October 10 by ASAS-SN team.
The coordinates are \timeform{20h 35m 09.75s},
\timeform{+09D 20' 46.4''} (SDSS $g$=20.7 counterpart).
Subsequent observations detected superhumps
(vsnet-alert 17844, 17850; figure \ref{fig:asassn14ivshpdm}).
The times of superhump maxima are listed in table
\ref{tab:asassn14ivoc2014}.

\begin{table}
\caption{Superhump maxima of ASASSN-14iv (2014)}\label{tab:asassn14ivoc2014}
\begin{center}
\begin{tabular}{rp{55pt}p{40pt}r@{.}lr}
\hline
\multicolumn{1}{c}{$E$} & \multicolumn{1}{c}{max\commenta} & \multicolumn{1}{c}{error} & \multicolumn{2}{c}{$O-C$\commentb} & \multicolumn{1}{c}{$N$\commentc} \\
\hline
0 & 56944.5019 & 0.0255 & $-$0&0073 & 9 \\
1 & 56944.5829 & 0.0019 & 0&0042 & 33 \\
2 & 56944.6516 & 0.0030 & 0&0035 & 15 \\
21 & 56945.9698 & 0.0010 & 0&0028 & 72 \\
22 & 56946.0333 & 0.0018 & $-$0&0031 & 71 \\
\hline
  \multicolumn{6}{l}{\commenta BJD$-$2400000.} \\
  \multicolumn{6}{l}{\commentb Against max $= 2456944.5092 + 0.069418 E$.} \\
  \multicolumn{6}{l}{\commentc Number of points used to determine the maximum.} \\
\end{tabular}
\end{center}
\end{table}

\subsection{ASASSN-14je}\label{obj:asassn14je}

\begin{figure}
  \begin{center}
    \FigureFile(88mm,110mm){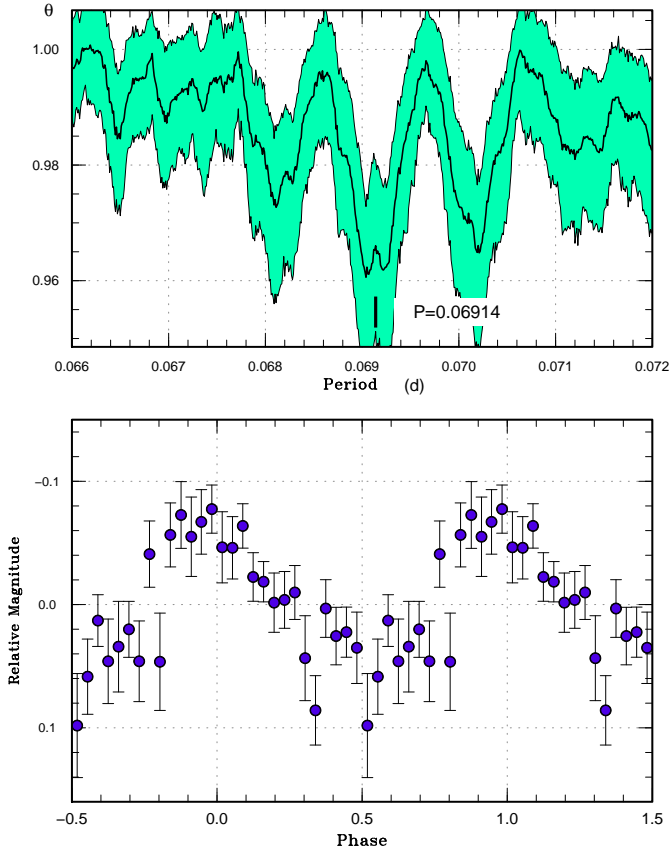}
  \end{center}
  \caption{Superhumps in ASASSN-14je (2014).
     (Upper): PDM analysis.
     (Lower): Phase-averaged profile}.
  \label{fig:asassn14jeshpdm}
\end{figure}

   This object was detected as a transient at $V$=14.26
on 2014 October 20 by ASAS-SN team (vsnet-alert 17880).
The coordinates are \timeform{06h 32m 27.68s},
\timeform{-56D 38' 52.6''} (the Initial Gaia Source List).
Subsequent observations detected superhumps
(vsnet-alert 17883, 17889; figure \ref{fig:asassn14jeshpdm}).
The times of superhumps maxima are listed in table
\ref{tab:asassn14jeoc2014}.  The maxima for $E \ge 73$
correspond to the rapid fading phase.
Since the object started fading rapidly within five days
of the start of our outburst, the superhumps we observed
were most likely stage C superhumps.

\begin{table}
\caption{Superhump maxima of ASASSN-14je (2014)}\label{tab:asassn14jeoc2014}
\begin{center}
\begin{tabular}{rp{55pt}p{40pt}r@{.}lr}
\hline
\multicolumn{1}{c}{$E$} & \multicolumn{1}{c}{max\commenta} & \multicolumn{1}{c}{error} & \multicolumn{2}{c}{$O-C$\commentb} & \multicolumn{1}{c}{$N$\commentc} \\
\hline
0 & 56952.4237 & 0.0037 & $-$0&0029 & 85 \\
1 & 56952.4958 & 0.0003 & 0&0002 & 159 \\
2 & 56952.5637 & 0.0004 & $-$0&0010 & 160 \\
3 & 56952.6372 & 0.0013 & 0&0035 & 33 \\
14 & 56953.3908 & 0.0037 & $-$0&0027 & 84 \\
15 & 56953.4654 & 0.0004 & 0&0028 & 160 \\
16 & 56953.5328 & 0.0005 & 0&0011 & 160 \\
17 & 56953.6002 & 0.0012 & $-$0&0005 & 76 \\
73 & 56957.4603 & 0.0025 & $-$0&0083 & 159 \\
74 & 56957.5456 & 0.0057 & 0&0079 & 160 \\
\hline
  \multicolumn{6}{l}{\commenta BJD$-$2400000.} \\
  \multicolumn{6}{l}{\commentb Against max $= 2456952.4266 + 0.069070 E$.} \\
  \multicolumn{6}{l}{\commentc Number of points used to determine the maximum.} \\
\end{tabular}
\end{center}
\end{table}

\subsection{ASASSN-14jf}\label{obj:asassn14jf}

\begin{figure}
  \begin{center}
    \FigureFile(88mm,110mm){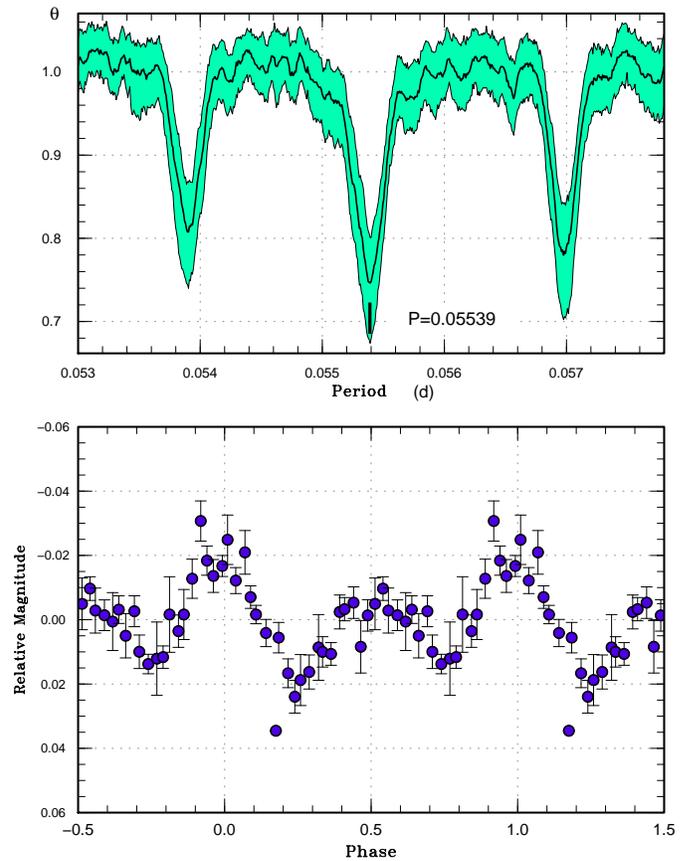}
  \end{center}
  \caption{Early superhumps in ASASSN-14jf (2014).
     (Upper): PDM analysis.
     (Lower): Phase-averaged profile}.
  \label{fig:asassn14jfeshpdm}
\end{figure}

\begin{figure}
  \begin{center}
    \FigureFile(88mm,110mm){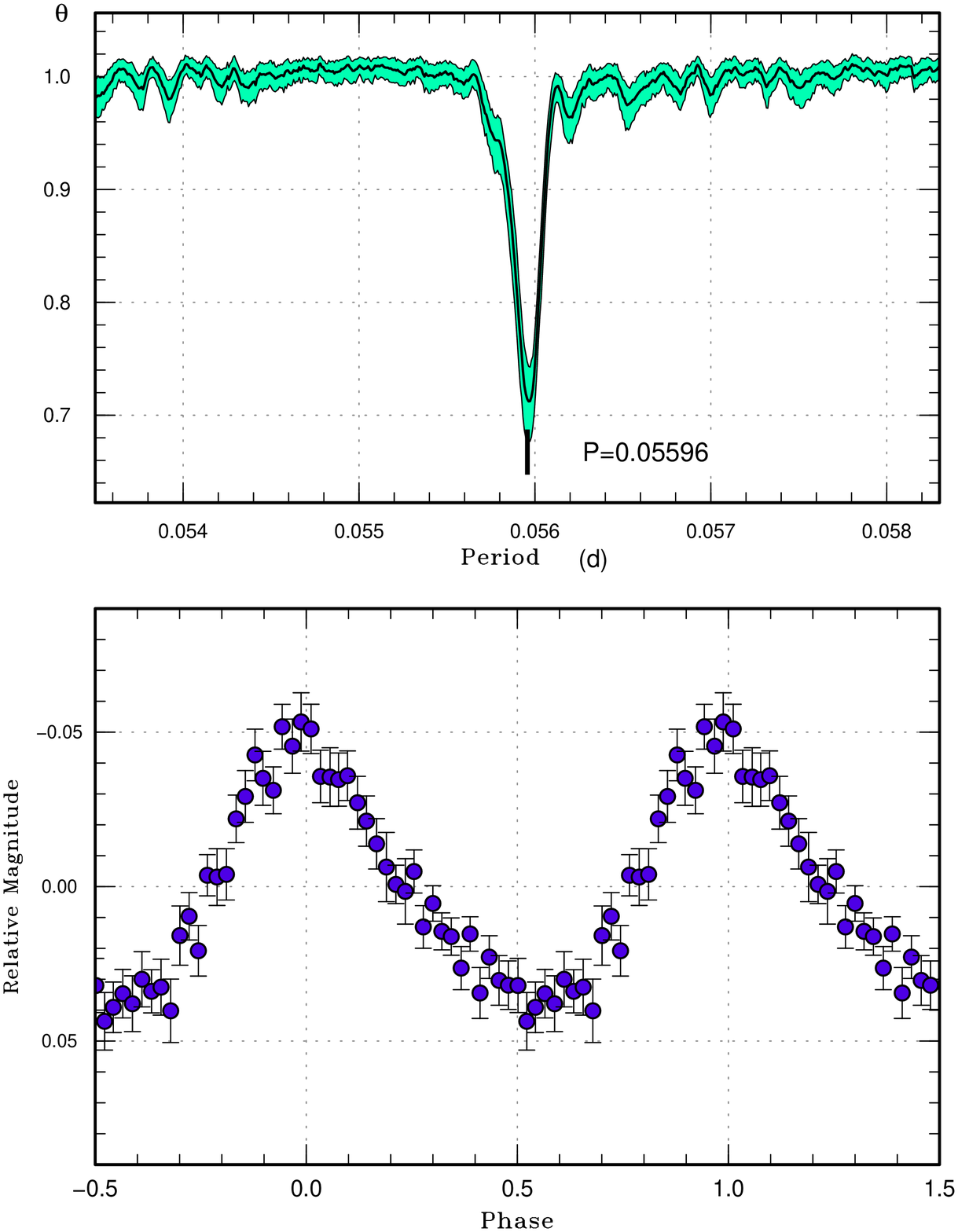}
  \end{center}
  \caption{Ordinary superhumps in ASASSN-14jf (2014).
     (Upper): PDM analysis.
     (Lower): Phase-averaged profile}.
  \label{fig:asassn14jfshpdm}
\end{figure}

\begin{figure}
  \begin{center}
    \FigureFile(88mm,70mm){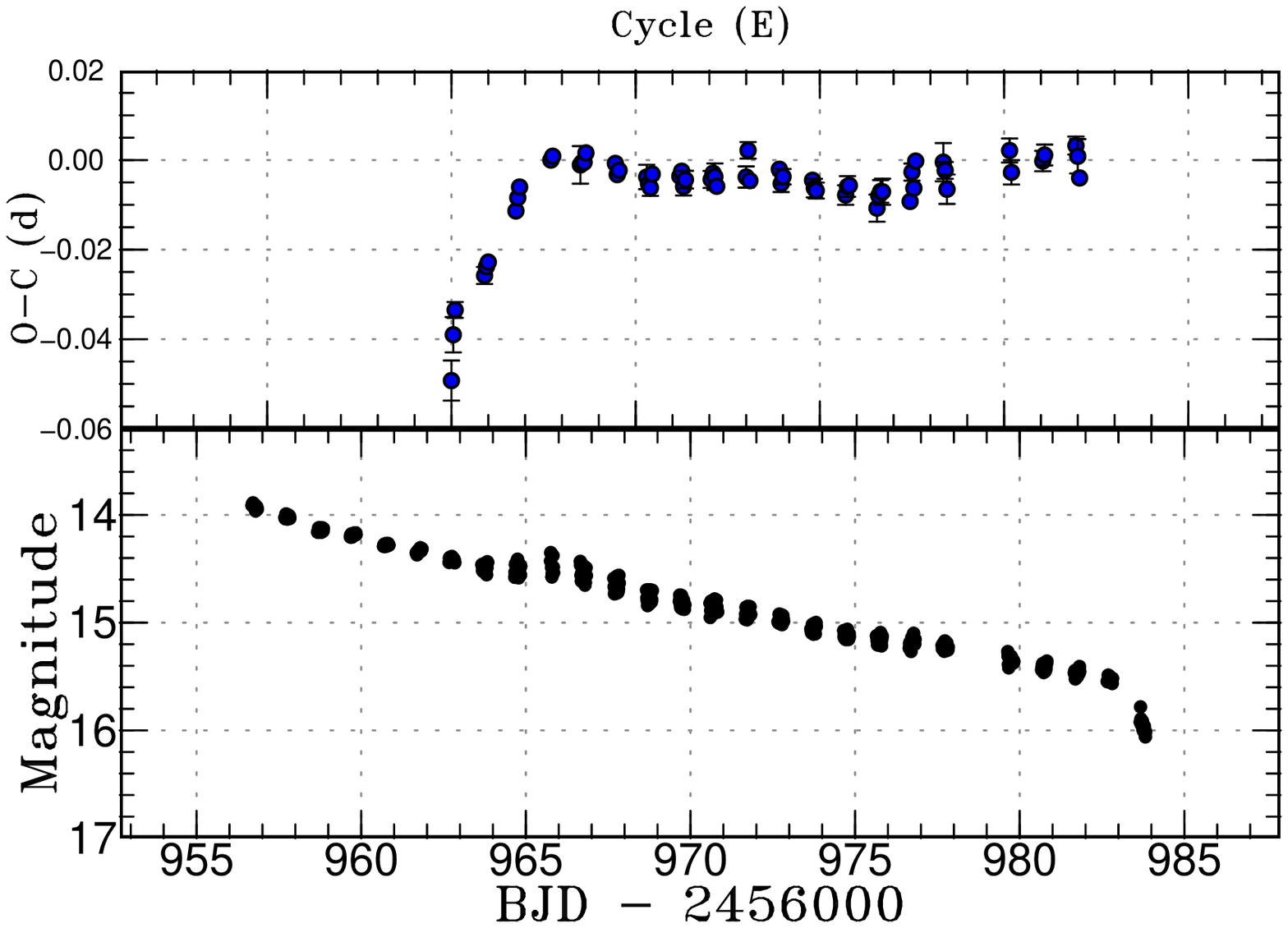}
  \end{center}
  \caption{$O-C$ diagram of superhumps in ASASSN-14jf (2014).
     (Upper): $O-C$ diagram.  A period of 0.05595~d
     was used to draw this figure.
     (Lower): Light curve.  The observations were binned to 0.011~d.}
  \label{fig:asassn14jfhumpall}
\end{figure}

   This object was detected as a transient at $V$=13.3
on 2014 October 22 by ASAS-SN team
\citet{sim14asassn14jfatel6608}.
The coordinates are \timeform{06h 06m 02.61s},
\timeform{-60D 39' 41.5''} (the Initial Gaia Source List).
There is a GALEX counterpart with an NUV magnitude
of 20.9.  The large outburst amplitude ($\sim$7.9 mag) suggested
a WZ Sge-type dwarf nova.
Subsequent observations indeed detected early superhumps
(vsnet-alert 17897, 17898, 17901, 17923;
figure \ref{fig:asassn14jfeshpdm}).
Nine days after the outburst detection, this object
started to show ordinary superhumps (vsnet-alert 17928,
17930, 17938, 17960; figure \ref{fig:asassn14jfshpdm}).
The times of superhump maxima are listed in table
\ref{tab:asassn14jfoc2014}.  Stages A and B can be
clearly identified (figure \ref{fig:asassn14jfhumpall}).
There was no evidence of transition
to stage C superhumps during the superoutburst plateau,
as is usual the case in WZ Sge-type dwarf novae.
The period of stage A superhumps corresponds to
$\epsilon^*$=0.026(2) and $q$=0.070(5).

\begin{table}
\caption{Superhump maxima of ASASSN-14jf (2014)}\label{tab:asassn14jfoc2014}
\begin{center}
\begin{tabular}{rp{55pt}p{40pt}r@{.}lr}
\hline
\multicolumn{1}{c}{$E$} & \multicolumn{1}{c}{max\commenta} & \multicolumn{1}{c}{error} & \multicolumn{2}{c}{$O-C$\commentb} & \multicolumn{1}{c}{$N$\commentc} \\
\hline
0 & 56962.6916 & 0.0045 & $-$0&0347 & 16 \\
1 & 56962.7577 & 0.0039 & $-$0&0245 & 16 \\
2 & 56962.8192 & 0.0018 & $-$0&0190 & 20 \\
18 & 56963.7221 & 0.0019 & $-$0&0121 & 16 \\
19 & 56963.7800 & 0.0011 & $-$0&0102 & 17 \\
20 & 56963.8370 & 0.0015 & $-$0&0093 & 14 \\
35 & 56964.6877 & 0.0010 & 0&0014 & 17 \\
36 & 56964.7466 & 0.0010 & 0&0043 & 16 \\
37 & 56964.8049 & 0.0007 & 0&0067 & 19 \\
54 & 56965.7621 & 0.0008 & 0&0118 & 16 \\
55 & 56965.8189 & 0.0005 & 0&0126 & 21 \\
70 & 56966.6562 & 0.0042 & 0&0100 & 10 \\
71 & 56966.7127 & 0.0009 & 0&0105 & 17 \\
72 & 56966.7687 & 0.0010 & 0&0104 & 16 \\
73 & 56966.8267 & 0.0005 & 0&0125 & 19 \\
89 & 56967.7196 & 0.0014 & 0&0093 & 19 \\
90 & 56967.7731 & 0.0009 & 0&0068 & 19 \\
91 & 56967.8299 & 0.0010 & 0&0077 & 17 \\
106 & 56968.6677 & 0.0027 & 0&0055 & 11 \\
107 & 56968.7221 & 0.0010 & 0&0039 & 19 \\
108 & 56968.7772 & 0.0018 & 0&0030 & 19 \\
109 & 56968.8362 & 0.0016 & 0&0060 & 14 \\
124 & 56969.6750 & 0.0011 & 0&0047 & 17 \\
125 & 56969.7321 & 0.0012 & 0&0058 & 20 \\
126 & 56969.7846 & 0.0020 & 0&0023 & 20 \\
127 & 56969.8421 & 0.0020 & 0&0038 & 6 \\
141 & 56970.6254 & 0.0019 & 0&0032 & 18 \\
142 & 56970.6828 & 0.0015 & 0&0045 & 16 \\
143 & 56970.7379 & 0.0030 & 0&0037 & 17 \\
144 & 56970.7917 & 0.0009 & 0&0015 & 20 \\
160 & 56971.6890 & 0.0024 & 0&0027 & 17 \\
161 & 56971.7509 & 0.0018 & 0&0087 & 17 \\
162 & 56971.8001 & 0.0017 & 0&0018 & 17 \\
\hline
  \multicolumn{6}{l}{\commenta BJD$-$2400000.} \\
  \multicolumn{6}{l}{\commentb Against max $= 2456962.7263 + 0.056000 E$.} \\
  \multicolumn{6}{l}{\commentc Number of points used to determine the maximum.} \\
\end{tabular}
\end{center}
\end{table}

\begin{table}
\caption{Superhump maxima of ASASSN-14jf (2014) (continued)}
\begin{center}
\begin{tabular}{rp{55pt}p{40pt}r@{.}lr}
\hline
\multicolumn{1}{c}{$E$} & \multicolumn{1}{c}{max\commenta} & \multicolumn{1}{c}{error} & \multicolumn{2}{c}{$O-C$\commentb} & \multicolumn{1}{c}{$N$\commentc} \\
\hline
178 & 56972.6979 & 0.0016 & 0&0036 & 16 \\
179 & 56972.7506 & 0.0019 & 0&0003 & 16 \\
180 & 56972.8081 & 0.0017 & 0&0018 & 17 \\
196 & 56973.7025 & 0.0011 & 0&0002 & 17 \\
197 & 56973.7567 & 0.0021 & $-$0&0016 & 16 \\
198 & 56973.8121 & 0.0018 & $-$0&0022 & 14 \\
214 & 56974.7063 & 0.0022 & $-$0&0040 & 16 \\
215 & 56974.7641 & 0.0023 & $-$0&0021 & 15 \\
216 & 56974.8203 & 0.0014 & $-$0&0019 & 10 \\
231 & 56975.6545 & 0.0030 & $-$0&0078 & 11 \\
232 & 56975.7133 & 0.0015 & $-$0&0050 & 17 \\
233 & 56975.7701 & 0.0025 & $-$0&0041 & 15 \\
234 & 56975.8260 & 0.0029 & $-$0&0043 & 9 \\
249 & 56976.6631 & 0.0015 & $-$0&0072 & 15 \\
250 & 56976.7256 & 0.0019 & $-$0&0007 & 17 \\
251 & 56976.7780 & 0.0014 & $-$0&0043 & 16 \\
252 & 56976.8400 & 0.0016 & 0&0017 & 5 \\
267 & 56977.6790 & 0.0043 & 0&0007 & 16 \\
268 & 56977.7331 & 0.0019 & $-$0&0012 & 17 \\
269 & 56977.7848 & 0.0033 & $-$0&0054 & 16 \\
303 & 56979.6958 & 0.0027 & 0&0015 & 16 \\
304 & 56979.7469 & 0.0027 & $-$0&0034 & 15 \\
321 & 56980.7005 & 0.0023 & $-$0&0018 & 17 \\
322 & 56980.7578 & 0.0023 & $-$0&0005 & 16 \\
339 & 56981.7111 & 0.0020 & 0&0008 & 16 \\
340 & 56981.7646 & 0.0038 & $-$0&0016 & 16 \\
341 & 56981.8158 & 0.0014 & $-$0&0065 & 10 \\
\hline
  \multicolumn{6}{l}{\commenta BJD$-$2400000.} \\
  \multicolumn{6}{l}{\commentb Against max $= 2456962.7263 + 0.056000 E$.} \\
  \multicolumn{6}{l}{\commentc Number of points used to determine the maximum.} \\
\end{tabular}
\end{center}
\end{table}

\subsection{ASASSN-14jq}\label{obj:asassn14jq}

\begin{figure}
  \begin{center}
    \FigureFile(88mm,110mm){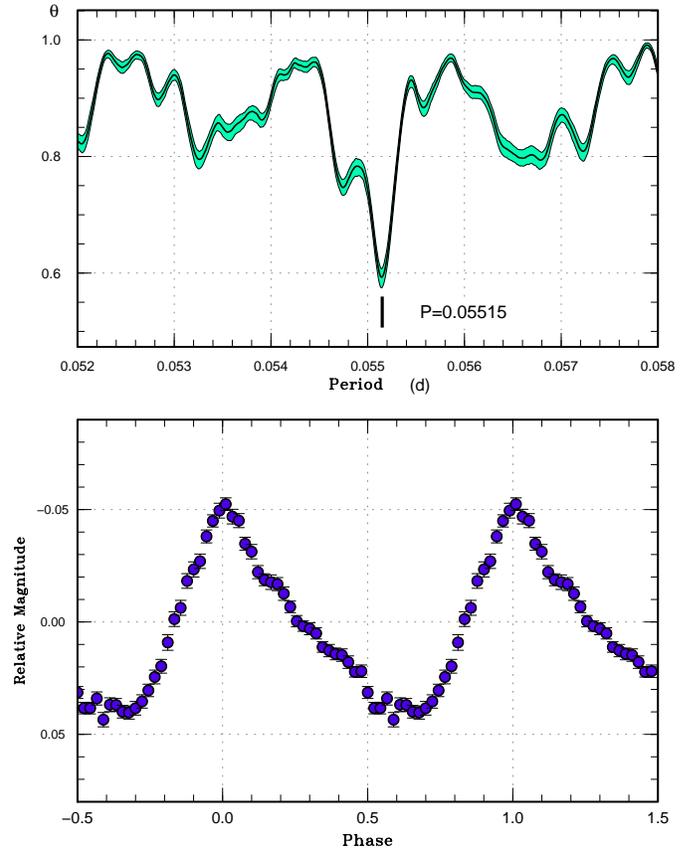}
  \end{center}
  \caption{Ordinary superhumps in ASASSN-14jq during
     the superoutburst plateau (2014).
     (Upper): PDM analysis.
     (Lower): Phase-averaged profile}.
  \label{fig:asassn14jqshpdm}
\end{figure}

\begin{figure}
  \begin{center}
    \FigureFile(88mm,100mm){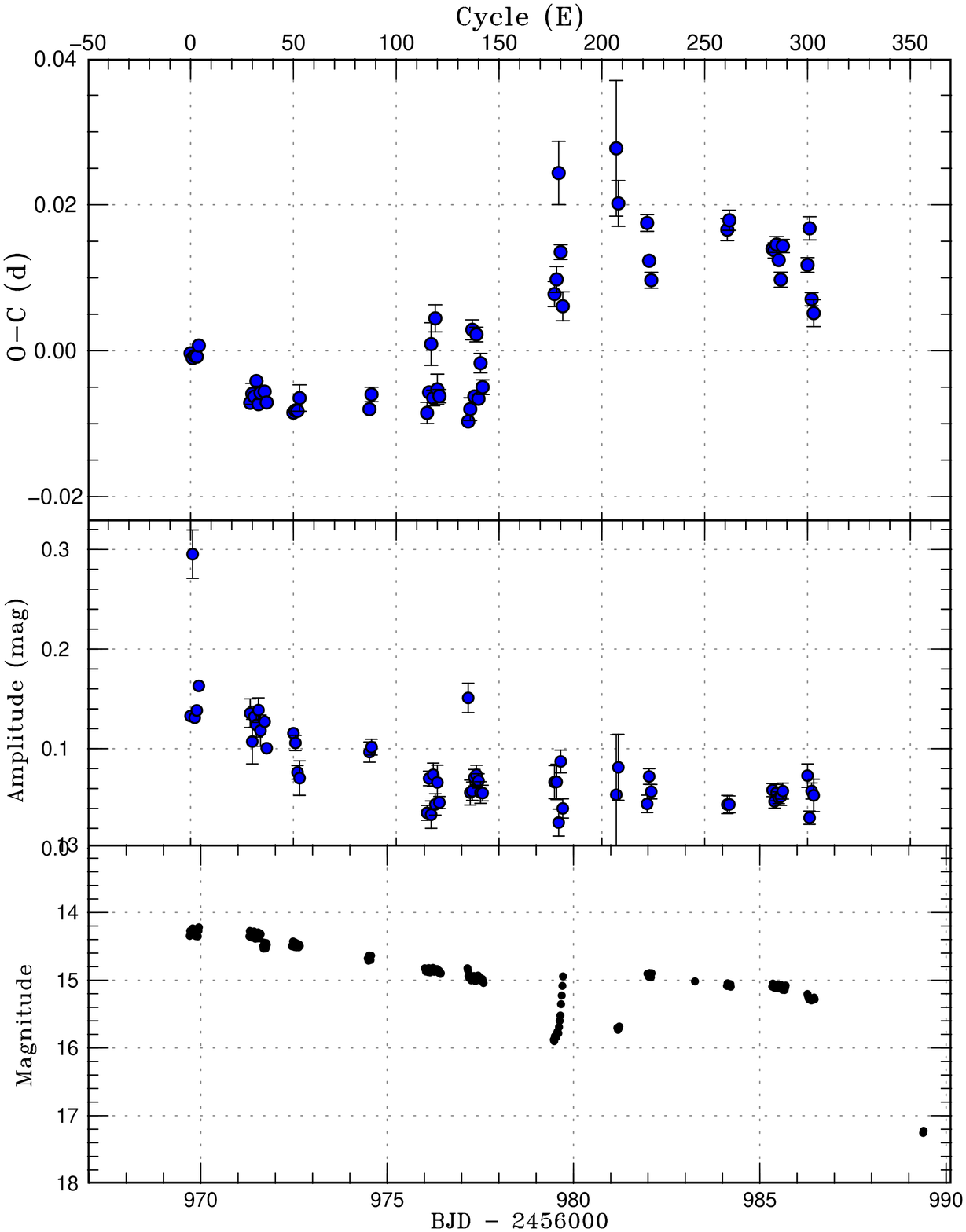}
  \end{center}
  \caption{$O-C$ diagram of superhumps in ASASSN-14jq (2014).
     (Upper:) $O-C$ diagram.
     We used a period of 0.05518~d for calculating the $O-C$ residuals.
     (Middle:) Amplitudes of superhumps.
     (Lower:) Light curve.  The data were binned to 0.011~d.
  }
  \label{fig:asassn14jqhumpall}
\end{figure}

\begin{figure}
  \begin{center}
    \FigureFile(88mm,110mm){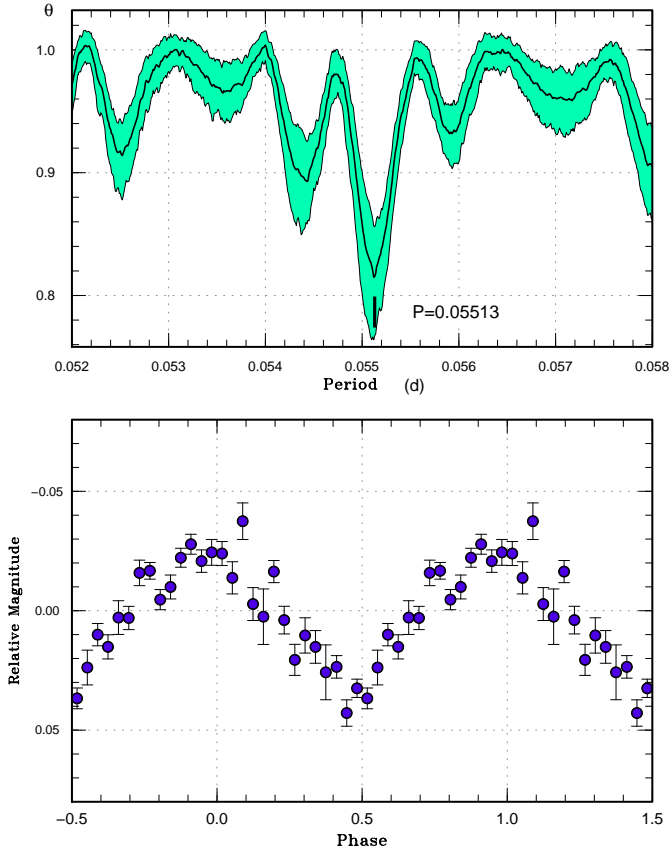}
  \end{center}
  \caption{Ordinary superhumps in ASASSN-14jq during
     the rebrightening (2014).
     (Upper): PDM analysis.
     (Lower): Phase-averaged profile}.
  \label{fig:asassn14jqshpdmreb}
\end{figure}

   This object was detected as a transient at $V$=13.74
on 2014 November 6 by ASAS-SN team (vsnet-alert 17943).
The coordinates are \timeform{02h 21m 02.78s},
\timeform{+73D 22' 45.2''} (SDSS $g$=20.5 counterpart).
Subsequent observations detected superhumps
(vsnet-alert 17957, 17974; figure \ref{fig:asassn14jqshpdm}).
The object then faded after the plateau phase
(a dip around BJD 2456979, vsnet-alert 17978).
The object faded again (BJD 2456982) and
a long rebrightening followed, which lasted
more than 5~d (vsnet-alert 17997, 18010;
figure \ref{fig:asassn14jqhumpall}).
During this rebrightening, the superhumps grew
again (figure \ref{fig:asassn14jqshpdmreb}).

   The times of superhump maxima during the superoutburst
plateau are listed in table \ref{tab:asassn14jqoc2014}.
During the dip, the superhump signal became very weak
(See the middle panel of figure \ref{fig:asassn14jqhumpall}.
The amplitudes are shown in magnitudes and the pulsed
flux during the dip remarkably decreased).
The times of superhump maxima during the dip and
the rebrightening are listed in table
\ref{tab:asassn14jqoc2014reb}.  The maxima for
$E \le 4$ correspond to maxima observed during the dip
and the rising phase from the dip.
There was no marked period variation compared to
the superhumps observed during the superoutburst
plateau (figure \ref{fig:asassn14jqshpdmreb}).
The $O-C$ values between the plateau and the rebrightening
showed a jump by a phase $\sim$0.3 (figure
\ref{fig:asassn14jqhumpall}).  Since the periods
were almost the same between the plateau and
the rebrightening, it was unlikely a continuous
period change gave rise to this shift.
Since the amplitudes of superhumps grew again
during the rebrightening, it is most likely
these superhumps were newly excited during
the rebrightening.

   The last ASAS-SN observation was 25~d before
the outburst detection, and it was very likely
the true maximum was missed.  Although early superhump
were not recorded, we consider that the object
is likely a WZ Sge-type dwarf nova because
it showed a dip and a long rebrightening,
characteristic to WZ Sge-type dwarf novae
(type-A rebrightening, cf. \cite{ima06tss0222};
\cite{Pdot}).

   There has been an attempt to explain WZ Sge-type
rebrightenings by an enhanced mass-transfer
(originally proposed by \cite{pat02wzsge}
and modeled by \cite{bua02suumamodel}).
This model predicts the shrinkage of the disk radius
(see figure 5 in \cite{bua02suumamodel}) by
an addition of matter with low-angular momentum.
The present observation did not detect noticeable
period variation, indicating that the disk radius
was almost constant (we consider that we can
reasonably neglect the pressure effect since
the amplitudes of superhumps were very small
around the dip and during the rebrightening).
This observation does not favor the enhanced
mass-transfer model.

\begin{table}
\caption{Superhump maxima of ASASSN-14jq during the plateau
   phase (2014)}\label{tab:asassn14jqoc2014}
\begin{center}
\begin{tabular}{rp{55pt}p{40pt}r@{.}lr}
\hline
\multicolumn{1}{c}{$E$} & \multicolumn{1}{c}{max\commenta} & \multicolumn{1}{c}{error} & \multicolumn{2}{c}{$O-C$\commentb} & \multicolumn{1}{c}{$N$\commentc} \\
\hline
0 & 56969.7257 & 0.0001 & 0&0041 & 569 \\
1 & 56969.7802 & 0.0002 & 0&0034 & 165 \\
2 & 56969.8356 & 0.0002 & 0&0037 & 400 \\
3 & 56969.8907 & 0.0001 & 0&0036 & 569 \\
4 & 56969.9474 & 0.0003 & 0&0051 & 320 \\
29 & 56971.3191 & 0.0007 & $-$0&0027 & 50 \\
30 & 56971.3755 & 0.0014 & $-$0&0014 & 61 \\
31 & 56971.4303 & 0.0006 & $-$0&0018 & 65 \\
32 & 56971.4876 & 0.0007 & 0&0003 & 63 \\
33 & 56971.5396 & 0.0006 & $-$0&0028 & 64 \\
34 & 56971.5964 & 0.0009 & $-$0&0013 & 50 \\
36 & 56971.7069 & 0.0001 & $-$0&0011 & 568 \\
37 & 56971.7606 & 0.0002 & $-$0&0026 & 430 \\
50 & 56972.4765 & 0.0003 & $-$0&0040 & 150 \\
51 & 56972.5320 & 0.0005 & $-$0&0037 & 121 \\
52 & 56972.5871 & 0.0007 & $-$0&0037 & 123 \\
53 & 56972.6441 & 0.0018 & $-$0&0019 & 51 \\
87 & 56974.5186 & 0.0007 & $-$0&0034 & 60 \\
88 & 56974.5758 & 0.0010 & $-$0&0014 & 31 \\
115 & 56976.0632 & 0.0015 & $-$0&0039 & 115 \\
116 & 56976.1212 & 0.0007 & $-$0&0010 & 116 \\
117 & 56976.1830 & 0.0029 & 0&0056 & 137 \\
118 & 56976.2308 & 0.0011 & $-$0&0018 & 145 \\
119 & 56976.2969 & 0.0019 & 0&0091 & 87 \\
120 & 56976.3423 & 0.0021 & $-$0&0006 & 65 \\
121 & 56976.3966 & 0.0009 & $-$0&0015 & 57 \\
135 & 56977.1656 & 0.0007 & $-$0&0050 & 40 \\
136 & 56977.2225 & 0.0016 & $-$0&0033 & 148 \\
137 & 56977.2885 & 0.0014 & 0&0076 & 106 \\
138 & 56977.3346 & 0.0008 & $-$0&0016 & 109 \\
139 & 56977.3982 & 0.0010 & 0&0069 & 53 \\
140 & 56977.4446 & 0.0007 & $-$0&0019 & 57 \\
141 & 56977.5047 & 0.0013 & 0&0030 & 57 \\
142 & 56977.5566 & 0.0010 & $-$0&0003 & 54 \\
\hline
  \multicolumn{6}{l}{\commenta BJD$-$2400000.} \\
  \multicolumn{6}{l}{\commentb Against max $= 2456969.7216 + 0.055178 E$.} \\
  \multicolumn{6}{l}{\commentc Number of points used to determine the maximum.} \\
\end{tabular}
\end{center}
\end{table}

\begin{table}
\caption{Superhump maxima of ASASSN-14jq during the dip
   and rebrightening (2014)}\label{tab:asassn14jqoc2014reb}
\begin{center}
\begin{tabular}{rp{55pt}p{40pt}r@{.}lr}
\hline
\multicolumn{1}{c}{$E$} & \multicolumn{1}{c}{max\commenta} & \multicolumn{1}{c}{error} & \multicolumn{2}{c}{$O-C$\commentb} & \multicolumn{1}{c}{$N$\commentc} \\
\hline
0 & 56979.5006 & 0.0017 & $-$0&0076 & 60 \\
1 & 56979.5578 & 0.0018 & $-$0&0056 & 58 \\
2 & 56979.6276 & 0.0043 & 0&0091 & 59 \\
3 & 56979.6719 & 0.0010 & $-$0&0018 & 59 \\
4 & 56979.7197 & 0.0020 & $-$0&0092 & 36 \\
30 & 56981.1760 & 0.0093 & 0&0131 & 61 \\
31 & 56981.2236 & 0.0031 & 0&0055 & 75 \\
45 & 56981.9935 & 0.0011 & 0&0032 & 106 \\
46 & 56982.0435 & 0.0008 & $-$0&0020 & 115 \\
47 & 56982.0960 & 0.0011 & $-$0&0046 & 88 \\
84 & 56984.1446 & 0.0015 & 0&0032 & 54 \\
85 & 56984.2010 & 0.0014 & 0&0045 & 52 \\
106 & 56985.3559 & 0.0008 & 0&0010 & 57 \\
107 & 56985.4109 & 0.0010 & 0&0008 & 54 \\
108 & 56985.4669 & 0.0011 & 0&0017 & 57 \\
109 & 56985.5199 & 0.0008 & $-$0&0004 & 55 \\
110 & 56985.5724 & 0.0010 & $-$0&0031 & 53 \\
111 & 56985.6322 & 0.0009 & 0&0015 & 57 \\
123 & 56986.2918 & 0.0010 & $-$0&0008 & 49 \\
124 & 56986.3520 & 0.0016 & 0&0043 & 54 \\
125 & 56986.3974 & 0.0009 & $-$0&0054 & 57 \\
126 & 56986.4507 & 0.0018 & $-$0&0073 & 46 \\
\hline
  \multicolumn{6}{l}{\commenta BJD$-$2400000.} \\
  \multicolumn{6}{l}{\commentb Against max $= 2456979.5082 + 0.055157 E$.} \\
  \multicolumn{6}{l}{\commentc Number of points used to determine the maximum.} \\
\end{tabular}
\end{center}
\end{table}

\subsection{ASASSN-14jv}\label{obj:asassn14jv}

\begin{figure}
  \begin{center}
    \FigureFile(88mm,110mm){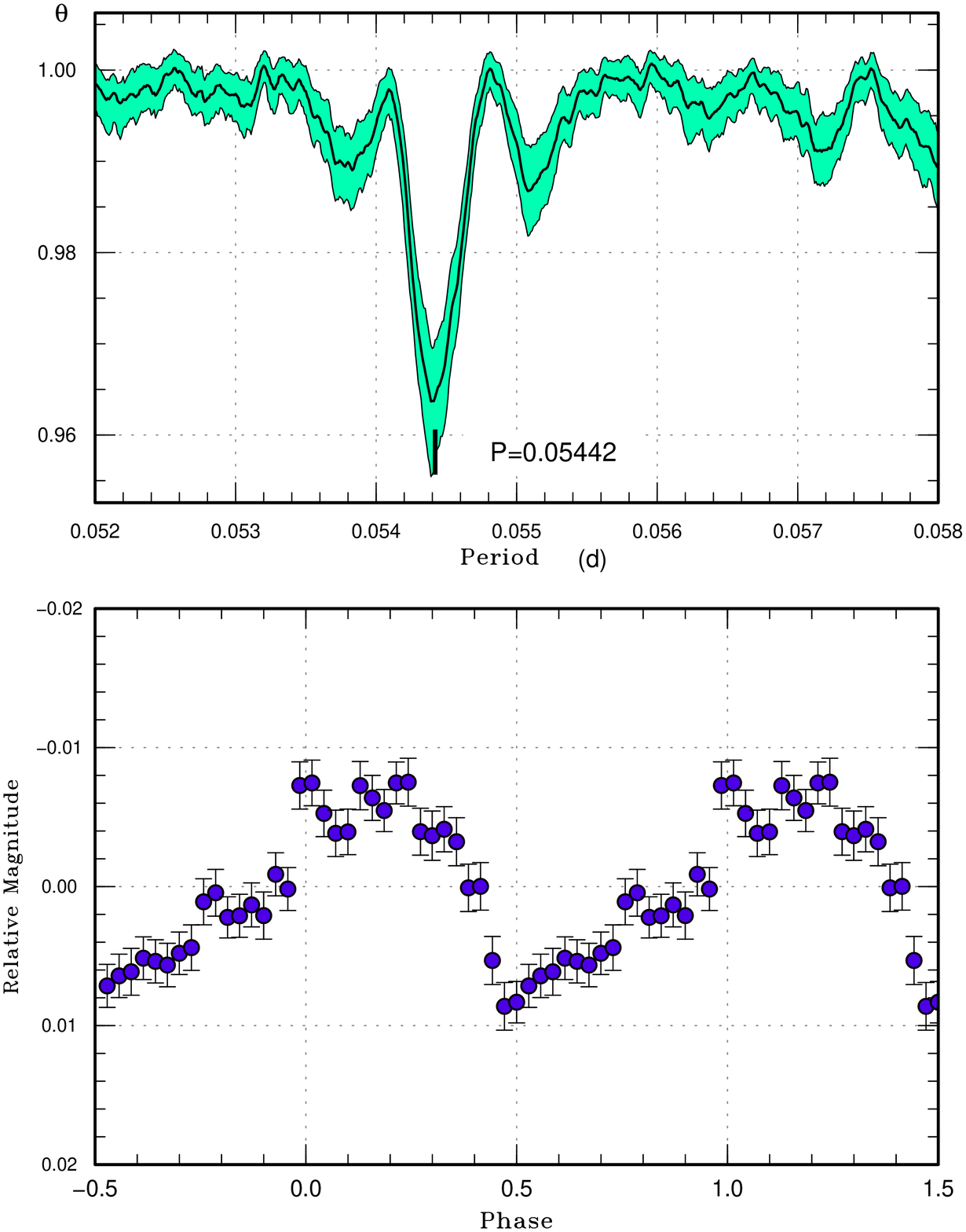}
  \end{center}
  \caption{Early superhumps in ASASSN-14jv (2014).
     (Upper): PDM analysis.
     (Lower): Phase-averaged profile}.
  \label{fig:asassn14jveshpdm}
\end{figure}

\begin{figure}
  \begin{center}
    \FigureFile(88mm,110mm){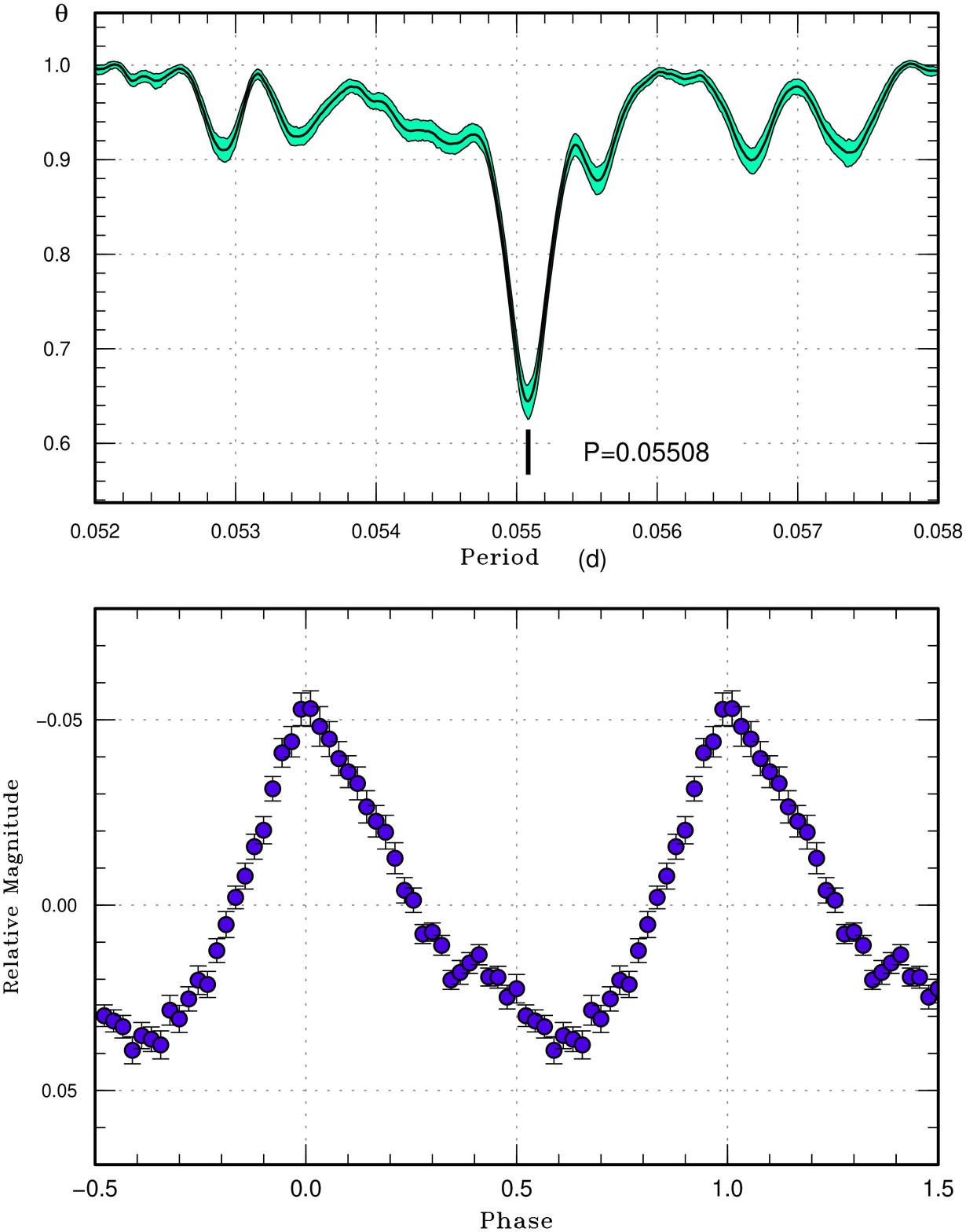}
  \end{center}
  \caption{Ordinary superhumps in ASASSN-14jv during
     the superoutburst plateau (2014).
     (Upper): PDM analysis.
     (Lower): Phase-averaged profile}.
  \label{fig:asassn14jvshpdm}
\end{figure}

   This object was detected as a transient at $V$=11.3
on 2014 November 9 by ASAS-SN team
\citet{sha14asassn14jvatel6676}.
The coordinates are \timeform{18h 53m 28.81s},
\timeform{+42D 03' 43.3''} (the Initial Gaia Source List).
\citet{ber14asassn14jvatel6684} spectroscopically
confirmed the object to be an outbursting dwarf nova.
Early superhumps were immediately reported
(vsnet-alert 17950, 17953, 17956, 17966;
figure \ref{fig:asassn14jveshpdm}).
Following this phase, ordinary superhumps
appeared (vsnet-alert 17973, 17985, 18015;
figure \ref{fig:asassn14jvshpdm}).

\begin{figure}
  \begin{center}
    \FigureFile(88mm,70mm){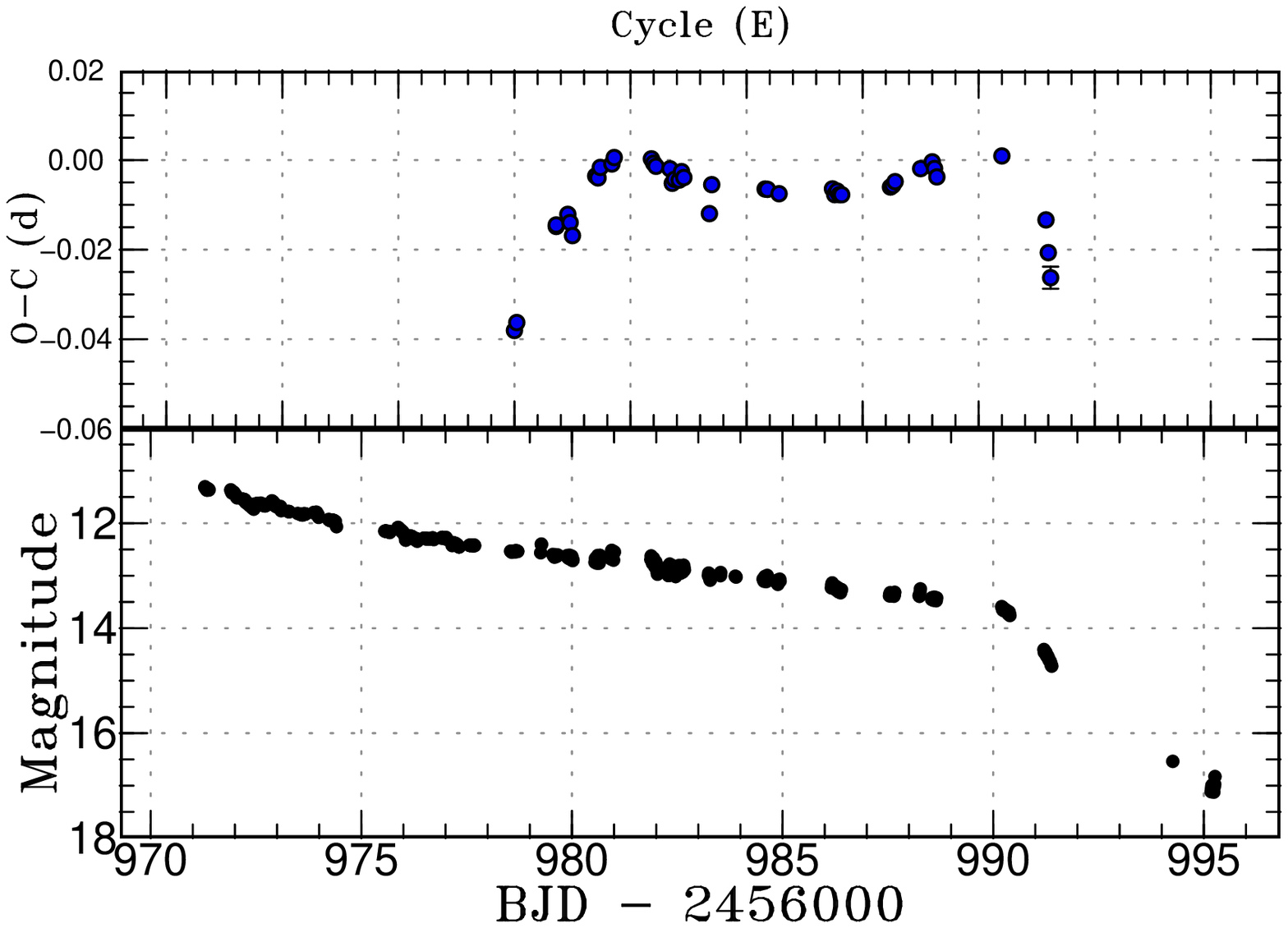}
  \end{center}
  \caption{$O-C$ diagram of superhumps in ASASSN-14jv (2014).
     (Upper): $O-C$ diagram.  A period of 0.05510~d
     was used to draw this figure.
     (Lower): Light curve.  The observations were binned to 0.011~d.}
  \label{fig:asassn14jvhumpall}
\end{figure}

   The times of superhump maxima are listed in table
\ref{tab:asassn14jvoc2014}.  Both stages A and B
were clearly recorded.  Instead of stage C superhumps,
either a phase shift or a large decrease in the period
was recorded during the rapid fading phase
(figure \ref{fig:asassn14jvhumpall}).
This phenomenon appears similar to what was observed
in FL Psc and GW Lib \citep{Pdot}.

   The period of early superhumps was determined
to be 0.054417(15)~d.  The period of stage A superhumps
corresponds to $\epsilon^*$=0.0278(9) and $q$=0.074(3).

\begin{table}
\caption{Superhump maxima of ASASSN-14jv (2014)}\label{tab:asassn14jvoc2014}
\begin{center}
\begin{tabular}{rp{55pt}p{40pt}r@{.}lr}
\hline
\multicolumn{1}{c}{$E$} & \multicolumn{1}{c}{max\commenta} & \multicolumn{1}{c}{error} & \multicolumn{2}{c}{$O-C$\commentb} & \multicolumn{1}{c}{$N$\commentc} \\
\hline
0 & 56978.5970 & 0.0017 & $-$0&0281 & 182 \\
1 & 56978.6539 & 0.0015 & $-$0&0263 & 185 \\
18 & 56979.6121 & 0.0005 & $-$0&0052 & 103 \\
18 & 56979.6124 & 0.0005 & $-$0&0048 & 103 \\
23 & 56979.8903 & 0.0004 & $-$0&0026 & 80 \\
24 & 56979.9435 & 0.0004 & $-$0&0044 & 103 \\
25 & 56979.9957 & 0.0004 & $-$0&0074 & 96 \\
35 & 56980.5600 & 0.0003 & 0&0058 & 96 \\
36 & 56980.6147 & 0.0001 & 0&0053 & 108 \\
37 & 56980.6722 & 0.0004 & 0&0076 & 52 \\
42 & 56980.9484 & 0.0002 & 0&0083 & 116 \\
43 & 56981.0050 & 0.0003 & 0&0097 & 69 \\
59 & 56981.8863 & 0.0003 & 0&0091 & 138 \\
60 & 56981.9404 & 0.0003 & 0&0081 & 165 \\
61 & 56981.9948 & 0.0003 & 0&0074 & 155 \\
67 & 56982.3249 & 0.0002 & 0&0068 & 78 \\
68 & 56982.3767 & 0.0003 & 0&0035 & 68 \\
69 & 56982.4327 & 0.0004 & 0&0044 & 87 \\
71 & 56982.5427 & 0.0005 & 0&0042 & 60 \\
72 & 56982.5997 & 0.0002 & 0&0060 & 107 \\
73 & 56982.6535 & 0.0003 & 0&0047 & 90 \\
84 & 56983.2515 & 0.0001 & $-$0&0036 & 83 \\
85 & 56983.3131 & 0.0005 & 0&0029 & 32 \\
108 & 56984.5794 & 0.0003 & 0&0014 & 103 \\
109 & 56984.6345 & 0.0003 & 0&0014 & 108 \\
114 & 56984.9090 & 0.0013 & 0&0003 & 92 \\
137 & 56986.1774 & 0.0003 & 0&0009 & 75 \\
138 & 56986.2312 & 0.0003 & $-$0&0004 & 181 \\
139 & 56986.2871 & 0.0008 & 0&0004 & 140 \\
140 & 56986.3414 & 0.0003 & $-$0&0004 & 90 \\
141 & 56986.3965 & 0.0005 & $-$0&0005 & 57 \\
\hline
  \multicolumn{6}{l}{\commenta BJD$-$2400000.} \\
  \multicolumn{6}{l}{\commentb Against max $= 2456978.6251 + 0.055119 E$.} \\
  \multicolumn{6}{l}{\commentc Number of points used to determine the maximum.} \\
\end{tabular}
\end{center}
\end{table}

\addtocounter{table}{-1}
\begin{table}
\caption{Superhump maxima of ASASSN-14jv (2014) (continued)}
\begin{center}
\begin{tabular}{rp{55pt}p{40pt}r@{.}lr}
\hline
\multicolumn{1}{c}{$E$} & \multicolumn{1}{c}{max\commenta} & \multicolumn{1}{c}{error} & \multicolumn{2}{c}{$O-C$\commentb} & \multicolumn{1}{c}{$N$\commentc} \\
\hline
162 & 56987.5553 & 0.0004 & 0&0008 & 97 \\
163 & 56987.6108 & 0.0004 & 0&0012 & 107 \\
164 & 56987.6667 & 0.0014 & 0&0020 & 45 \\
175 & 56988.2757 & 0.0004 & 0&0047 & 25 \\
180 & 56988.5527 & 0.0008 & 0&0061 & 60 \\
181 & 56988.6063 & 0.0006 & 0&0046 & 68 \\
182 & 56988.6595 & 0.0012 & 0&0027 & 37 \\
210 & 56990.2070 & 0.0009 & 0&0069 & 35 \\
229 & 56991.2396 & 0.0016 & $-$0&0078 & 103 \\
230 & 56991.2874 & 0.0013 & $-$0&0151 & 124 \\
231 & 56991.3369 & 0.0025 & $-$0&0207 & 49 \\
\hline
  \multicolumn{6}{l}{\commenta BJD$-$2400000.} \\
  \multicolumn{6}{l}{\commentb Against max $= 2456978.6251 + 0.055119 E$.} \\
  \multicolumn{6}{l}{\commentc Number of points used to determine the maximum.} \\
\end{tabular}
\end{center}
\end{table}

\subsection{ASASSN-14kf}\label{obj:asassn14kf}

\begin{figure}
  \begin{center}
    \FigureFile(88mm,110mm){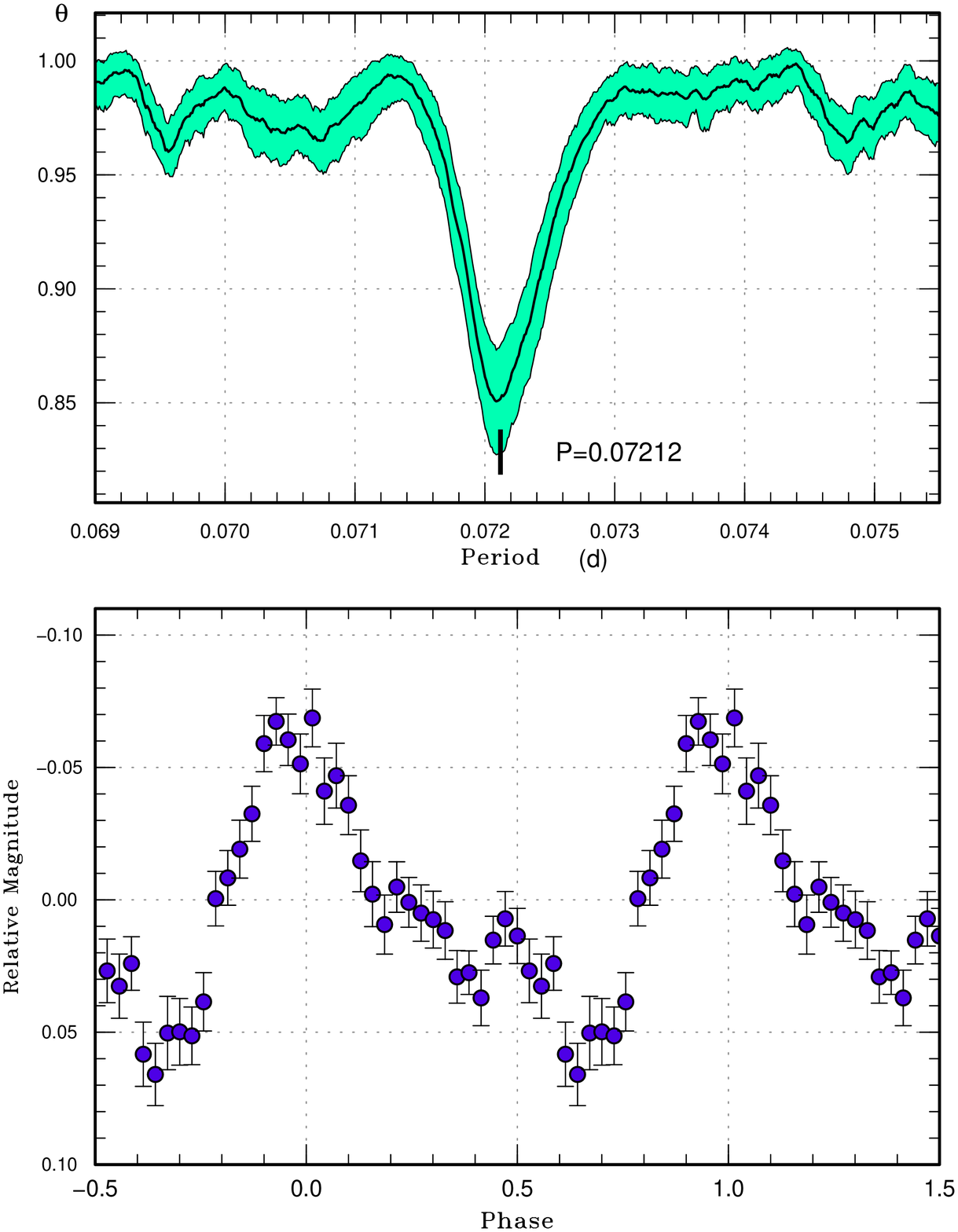}
  \end{center}
  \caption{Superhumps in ASASSN-14kf (2014).
     (Upper): PDM analysis.
     (Lower): Phase-averaged profile}.
  \label{fig:asassn14kfshpdm}
\end{figure}

   This object was detected as a transient at $V$=15.36
on 2014 November 11 by ASAS-SN team (vsnet-alert 17880).
The coordinates are \timeform{05h 13m 06.59s},
\timeform{-26D 19' 52.0''} (the Initial Gaia Source List).
Superhumps were immediately detected (vsnet-alert 17992;
figure \ref{fig:asassn14kfshpdm}).
The times of superhump maxima are listed in table
\ref{tab:asassn14kfoc2014}.
Although $E$=0 corresponds to stage A superhump,
the period of stage A superhumps could not be determined.

\begin{table}
\caption{Superhump maxima of ASASSN-14kf (2014)}\label{tab:asassn14kfoc2014}
\begin{center}
\begin{tabular}{rp{55pt}p{40pt}r@{.}lr}
\hline
\multicolumn{1}{c}{$E$} & \multicolumn{1}{c}{max\commenta} & \multicolumn{1}{c}{error} & \multicolumn{2}{c}{$O-C$\commentb} & \multicolumn{1}{c}{$N$\commentc} \\
\hline
0 & 56978.5438 & 0.0009 & $-$0&0030 & 158 \\
13 & 56979.4857 & 0.0009 & 0&0013 & 167 \\
14 & 56979.5566 & 0.0008 & 0&0000 & 167 \\
27 & 56980.4963 & 0.0010 & 0&0021 & 166 \\
28 & 56980.5662 & 0.0011 & $-$0&0001 & 166 \\
54 & 56982.4440 & 0.0039 & 0&0024 & 145 \\
55 & 56982.5161 & 0.0017 & 0&0024 & 165 \\
56 & 56982.5836 & 0.0018 & $-$0&0022 & 139 \\
69 & 56983.5252 & 0.0025 & 0&0017 & 166 \\
70 & 56983.5910 & 0.0042 & $-$0&0046 & 126 \\
\hline
  \multicolumn{6}{l}{\commenta BJD$-$2400000.} \\
  \multicolumn{6}{l}{\commentb Against max $= 2456978.5468 + 0.072127 E$.} \\
  \multicolumn{6}{l}{\commentc Number of points used to determine the maximum.} \\
\end{tabular}
\end{center}
\end{table}

\subsection{ASASSN-14kk}\label{obj:asassn14kk}

\begin{figure}
  \begin{center}
    \FigureFile(88mm,110mm){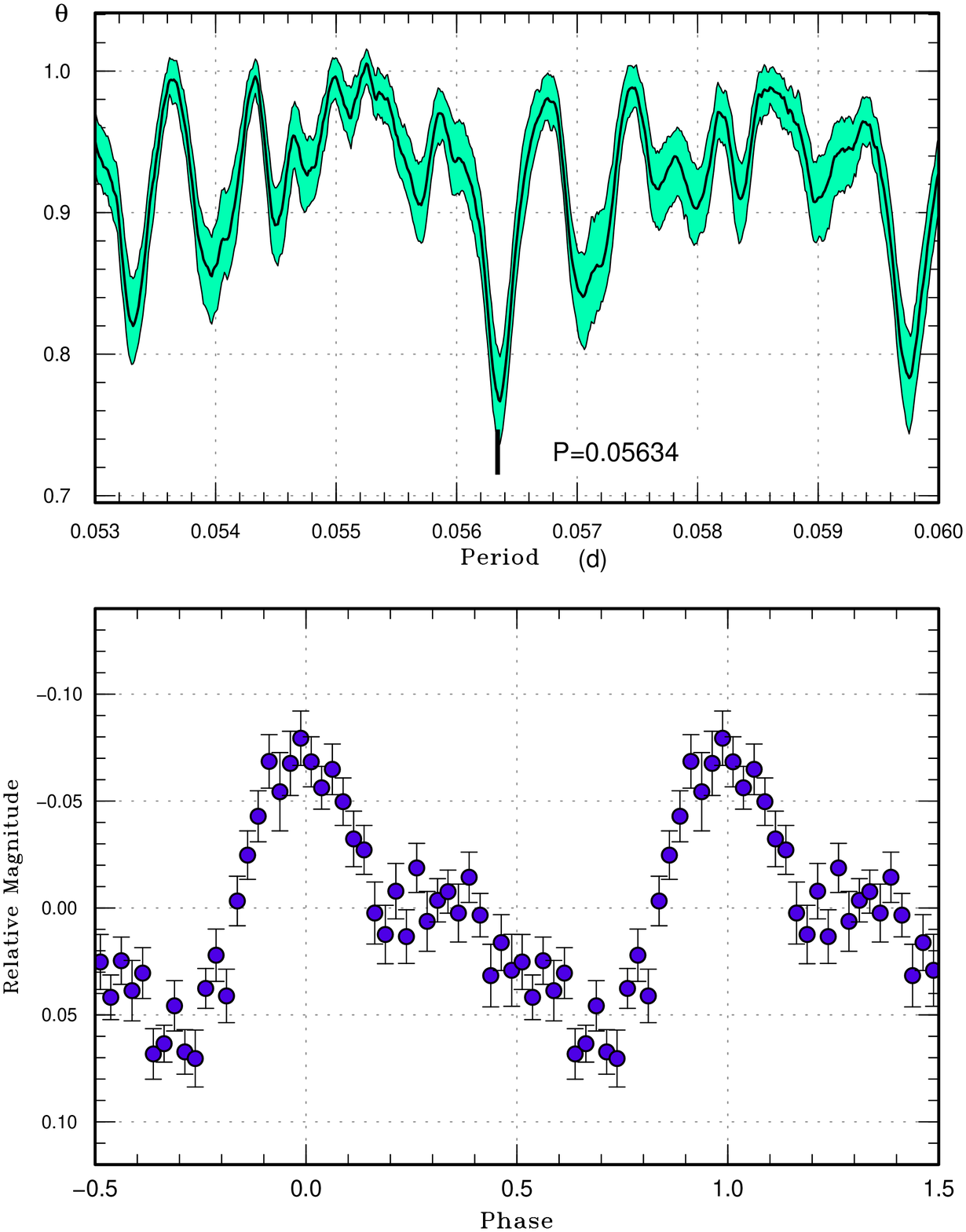}
  \end{center}
  \caption{Superhumps in ASASSN-14kk (2014).
     (Upper): PDM analysis.
     (Lower): Phase-averaged profile}.
  \label{fig:asassn14kkshpdm}
\end{figure}

   This object was detected as a transient at $V$=16.17
on 2014 November 12 by ASAS-SN team (vsnet-alert 17988).
The coordinates are \timeform{01h 32m 02.77s},
\timeform{-10D 43' 57.8''} (SDSS $g$=20.8 counterpart).
The object brightened to $V$=15.87 on November 17
(vsnet-alert 17989).
Observations starting on November 16 detected superhumps
(vsnet-alert 18011; figure \ref{fig:asassn14kkshpdm}).
The times of superhump maxima are listed in table
\ref{tab:asassn14kkoc2014}.

\begin{table}
\caption{Superhump maxima of ASASSN-14kk (2014)}\label{tab:asassn14kkoc2014}
\begin{center}
\begin{tabular}{rp{55pt}p{40pt}r@{.}lr}
\hline
\multicolumn{1}{c}{$E$} & \multicolumn{1}{c}{max\commenta} & \multicolumn{1}{c}{error} & \multicolumn{2}{c}{$O-C$\commentb} & \multicolumn{1}{c}{$N$\commentc} \\
\hline
0 & 56978.4042 & 0.0009 & $-$0&0071 & 80 \\
1 & 56978.4650 & 0.0006 & $-$0&0026 & 129 \\
16 & 56979.3286 & 0.0041 & 0&0156 & 130 \\
17 & 56979.3704 & 0.0030 & 0&0010 & 129 \\
18 & 56979.4197 & 0.0021 & $-$0&0060 & 77 \\
69 & 56982.3067 & 0.0014 & 0&0065 & 84 \\
70 & 56982.3501 & 0.0009 & $-$0&0065 & 128 \\
71 & 56982.4164 & 0.0032 & 0&0035 & 74 \\
87 & 56983.3139 & 0.0063 & $-$0&0007 & 47 \\
88 & 56983.3692 & 0.0012 & $-$0&0019 & 129 \\
89 & 56983.4254 & 0.0015 & $-$0&0020 & 124 \\
\hline
  \multicolumn{6}{l}{\commenta BJD$-$2400000.} \\
  \multicolumn{6}{l}{\commentb Against max $= 2456978.4112 + 0.056361 E$.} \\
  \multicolumn{6}{l}{\commentc Number of points used to determine the maximum.} \\
\end{tabular}
\end{center}
\end{table}

\subsection{ASASSN-14ku}\label{obj:asassn14ku}

\begin{figure}
  \begin{center}
    \FigureFile(88mm,110mm){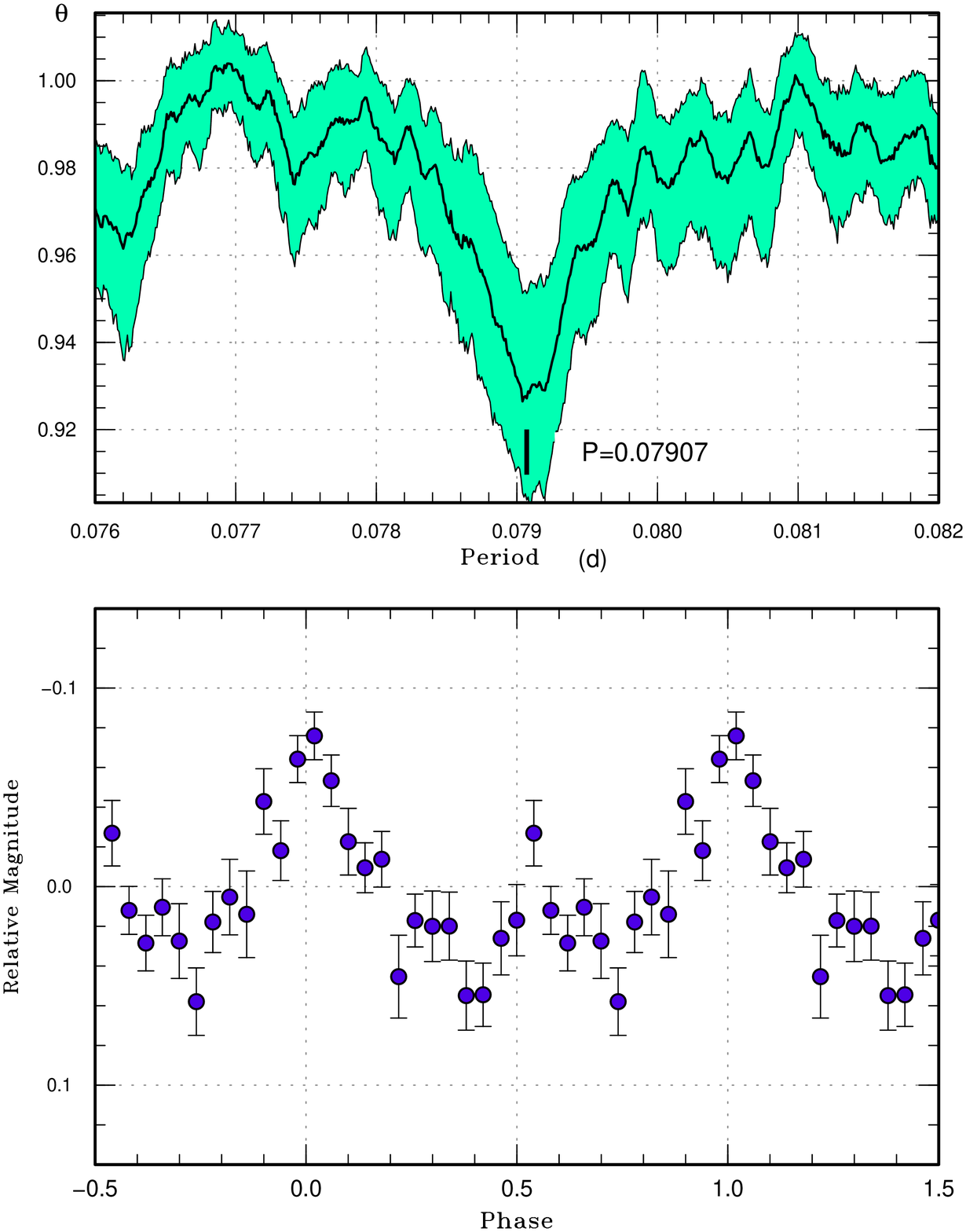}
  \end{center}
  \caption{Superhumps in ASASSN-14ku (2014).
     (Upper): PDM analysis.
     (Lower): Phase-averaged profile}.
  \label{fig:asassn14kushpdm}
\end{figure}

   This object was detected as a transient at $V$=15.45
on 2014 November 23 by ASAS-SN team.
The coordinates are \timeform{23h 43m 38.18s},
\timeform{-58D 52' 47.3''} (the Initial Gaia Source List).
Subsequent observations detected superhumps
(vsnet-alert 18019, 18023, 18028; figure \ref{fig:asassn14kushpdm}).
The times of superhump maxima are listed in table
\ref{tab:asassn14kuoc2014}.
The superhump stage is unknown.

\begin{table}
\caption{Superhump maxima of ASASSN-14ku (2014)}\label{tab:asassn14kuoc2014}
\begin{center}
\begin{tabular}{rp{55pt}p{40pt}r@{.}lr}
\hline
\multicolumn{1}{c}{$E$} & \multicolumn{1}{c}{max\commenta} & \multicolumn{1}{c}{error} & \multicolumn{2}{c}{$O-C$\commentb} & \multicolumn{1}{c}{$N$\commentc} \\
\hline
0 & 56988.3239 & 0.0022 & 0&0021 & 133 \\
1 & 56988.3986 & 0.0017 & $-$0&0022 & 183 \\
3 & 56988.5632 & 0.0024 & 0&0042 & 29 \\
13 & 56989.3492 & 0.0016 & $-$0&0005 & 182 \\
14 & 56989.4234 & 0.0021 & $-$0&0053 & 140 \\
16 & 56989.5877 & 0.0038 & 0&0009 & 28 \\
25 & 56990.3036 & 0.0073 & 0&0052 & 82 \\
26 & 56990.3772 & 0.0040 & $-$0&0003 & 135 \\
28 & 56990.5300 & 0.0145 & $-$0&0056 & 15 \\
29 & 56990.6175 & 0.0048 & 0&0028 & 22 \\
38 & 56991.3279 & 0.0083 & 0&0017 & 142 \\
41 & 56991.5551 & 0.0059 & $-$0&0083 & 27 \\
54 & 56992.5967 & 0.0059 & 0&0053 & 29 \\
\hline
  \multicolumn{6}{l}{\commenta BJD$-$2400000.} \\
  \multicolumn{6}{l}{\commentb Against max $= 2456988.3218 + 0.079066 E$.} \\
  \multicolumn{6}{l}{\commentc Number of points used to determine the maximum.} \\
\end{tabular}
\end{center}
\end{table}

\subsection{ASASSN-14lk}\label{obj:asassn14lk}

   This object was detected as a transient at $V$=13.48
on 2014 December 1 by ASAS-SN team (vsnet-alert 18032).
The coordinates are \timeform{20h 09m 24.09s},
\timeform{-63D 26' 22.8''} (the Initial Gaia Source List).
No previous outbursts were recorded in ASAS-3 data.
Although the object may be identical with NSV 12802
(=HV 9672) and NLTT 161-48 (vsnet-alert 18032),
the identification requires further investigation because
no finding chart is available for these objects.

\begin{figure}
  \begin{center}
    \FigureFile(88mm,110mm){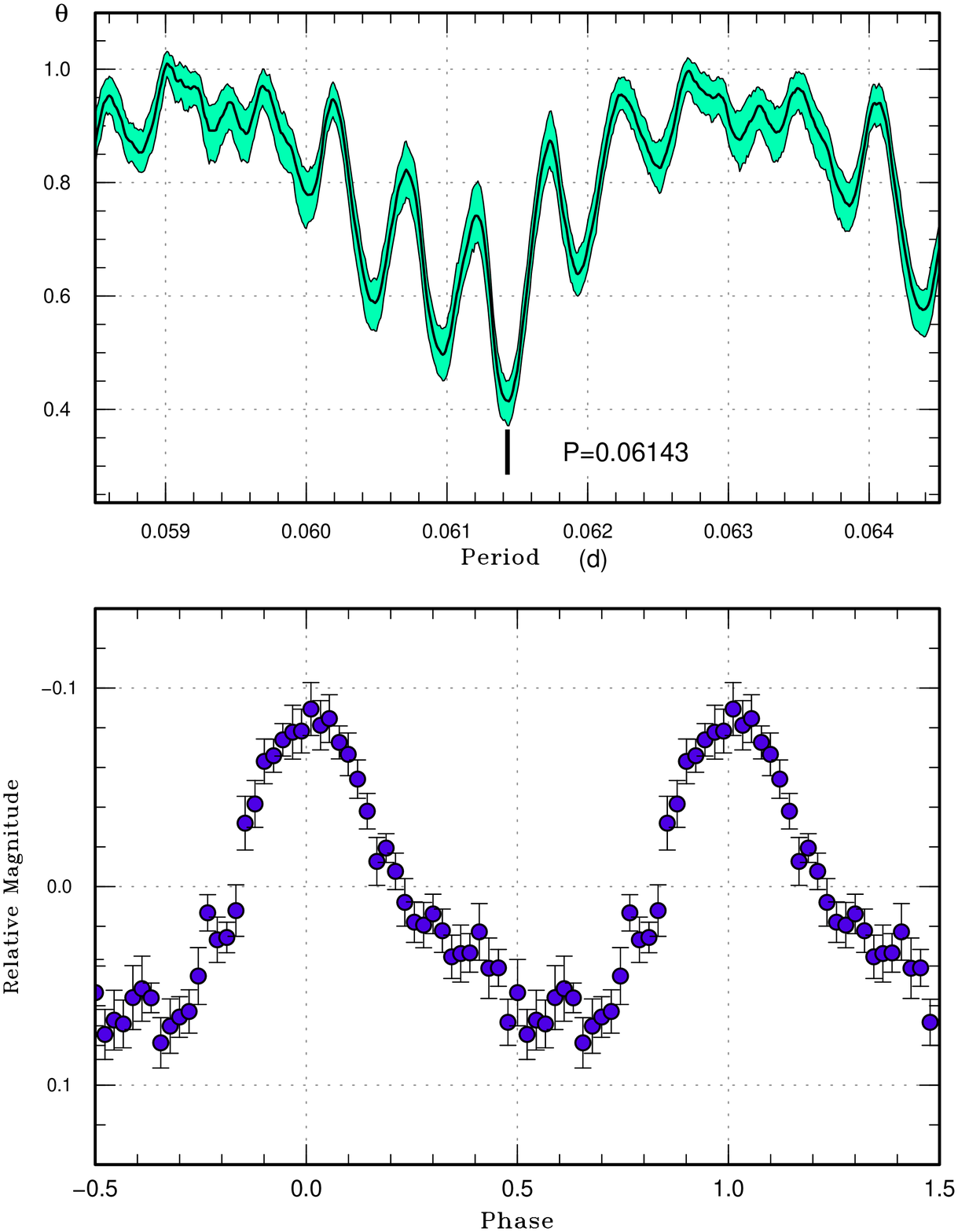}
  \end{center}
  \caption{Superhumps in ASASSN-14lk (2014).
     (Upper): PDM analysis.
     (Lower): Phase-averaged profile}.
  \label{fig:asassn14lkshpdm}
\end{figure}

   Superhumps were immediately recorded
(vsnet-alert 18038; figure \ref{fig:asassn14lkshpdm}).
The times of superhump maxima are listed in table
\ref{tab:asassn14lkoc2014}.
Although the superhump stage is not very clear,
the large superhump amplitudes suggest that at least
early observations recorded stage B superhumps.
The resultant $O-C$ values and mean period are
likely a mixture of stage B and C superhumps.
The selection of the period is based on the initial
two nights.  If the object showed anomalous period
variation, the cycle counts of the later part may
contain errors due to the observational gap.

\begin{table}
\caption{Superhump maxima of ASASSN-14lk (2014)}\label{tab:asassn14lkoc2014}
\begin{center}
\begin{tabular}{rp{55pt}p{40pt}r@{.}lr}
\hline
\multicolumn{1}{c}{$E$} & \multicolumn{1}{c}{max\commenta} & \multicolumn{1}{c}{error} & \multicolumn{2}{c}{$O-C$\commentb} & \multicolumn{1}{c}{$N$\commentc} \\
\hline
0 & 56997.3080 & 0.0003 & 0&0018 & 137 \\
16 & 56998.2864 & 0.0004 & $-$0&0027 & 122 \\
17 & 56998.3494 & 0.0005 & $-$0&0012 & 134 \\
81 & 57002.2863 & 0.0013 & 0&0041 & 58 \\
130 & 57005.2903 & 0.0026 & $-$0&0021 & 94 \\
\hline
  \multicolumn{6}{l}{\commenta BJD$-$2400000.} \\
  \multicolumn{6}{l}{\commentb Against max $= 2456997.3062 + 0.061432 E$.} \\
  \multicolumn{6}{l}{\commentc Number of points used to determine the maximum.} \\
\end{tabular}
\end{center}
\end{table}

\subsection{ASASSN-14mc}\label{obj:asassn14mc}

\begin{figure}
  \begin{center}
    \FigureFile(88mm,110mm){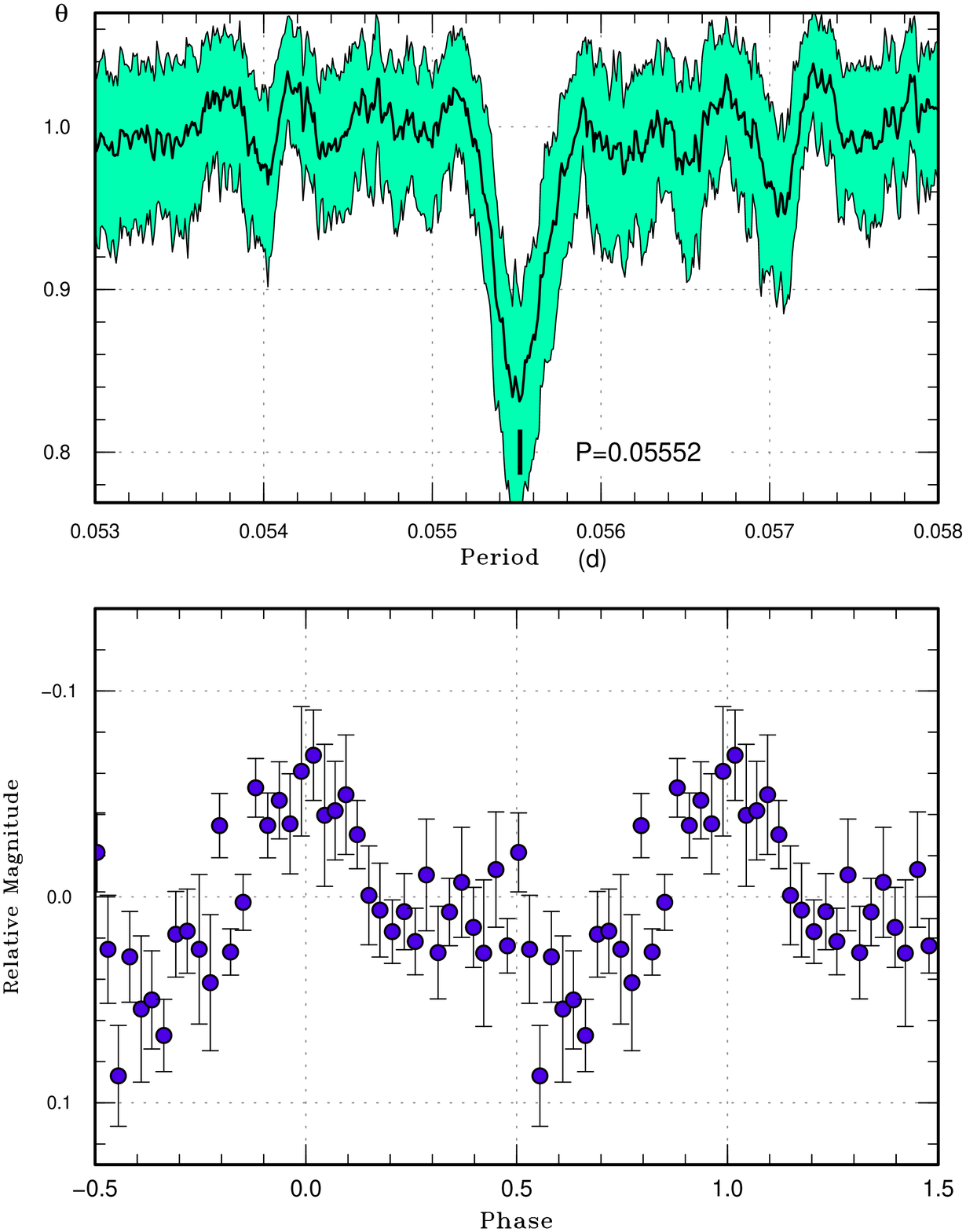}
  \end{center}
  \caption{Superhumps in ASASSN-14mc (2014).
     (Upper): PDM analysis.
     (Lower): Phase-averaged profile}.
  \label{fig:asassn14mcshpdm}
\end{figure}

   This object was detected as a transient at $V$=16.5
on 2014 December 14 by ASAS-SN team.  The object
further brightened to $V$=14.3 on December 15
(vsnet-alert 18063).
The coordinates are \timeform{04h 47m 23.27s},
\timeform{-36D 56' 01.8''} (ASAS-SN position).
Superhumps were detected 10~d after the outburst
detection (vsnet-alert 18104, 18129;
figure \ref{fig:asassn14mcshpdm}).

   The times of superhump maxima are listed in table
\ref{tab:asassn14mcoc2014}.  The maxima for $E \le 1$
correspond to stage A superhumps.  Although stage A
superhumps were detected, neither $O-C$ analysis
nor PDM method could not yield a reliable period
of stage A superhumps.   Although this object is
likely a WZ Sge-type dwarf nova, no significant signal of
early superhumps was detected in the earlier part
of the observation.  This was probably due
to a low orbital inclination.

\begin{table}
\caption{Superhump maxima of ASASSN-14mc (2014)}\label{tab:asassn14mcoc2014}
\begin{center}
\begin{tabular}{rp{55pt}p{40pt}r@{.}lr}
\hline
\multicolumn{1}{c}{$E$} & \multicolumn{1}{c}{max\commenta} & \multicolumn{1}{c}{error} & \multicolumn{2}{c}{$O-C$\commentb} & \multicolumn{1}{c}{$N$\commentc} \\
\hline
0 & 57015.6156 & 0.0035 & $-$0&0074 & 28 \\
1 & 57015.6675 & 0.0035 & $-$0&0110 & 30 \\
18 & 57016.6291 & 0.0009 & 0&0062 & 30 \\
19 & 57016.6841 & 0.0009 & 0&0057 & 26 \\
36 & 57017.6277 & 0.0018 & 0&0049 & 30 \\
37 & 57017.6828 & 0.0021 & 0&0044 & 25 \\
55 & 57018.6826 & 0.0012 & 0&0042 & 21 \\
73 & 57019.6801 & 0.0027 & 0&0017 & 21 \\
91 & 57020.6747 & 0.0014 & $-$0&0036 & 22 \\
109 & 57021.6741 & 0.0022 & $-$0&0042 & 19 \\
127 & 57022.6774 & 0.0035 & $-$0&0009 & 25 \\
\hline
  \multicolumn{6}{l}{\commenta BJD$-$2400000.} \\
  \multicolumn{6}{l}{\commentb Against max $= 2457015.6229 + 0.055554 E$.} \\
  \multicolumn{6}{l}{\commentc Number of points used to determine the maximum.} \\
\end{tabular}
\end{center}
\end{table}

\subsection{ASASSN-14md}\label{obj:asassn14md}

\begin{figure}
  \begin{center}
    \FigureFile(88mm,110mm){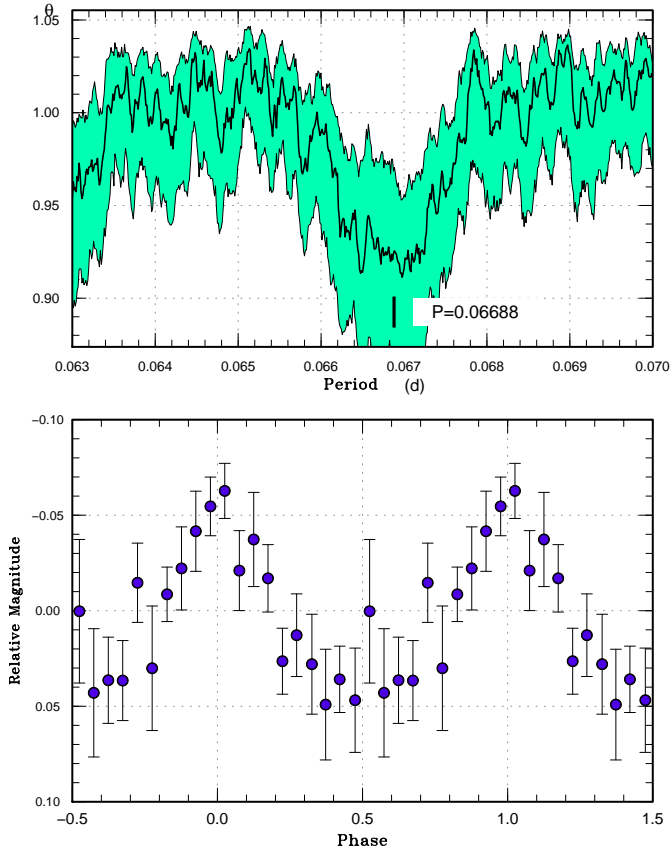}
  \end{center}
  \caption{Superhumps in ASASSN-14md (2014).
     The data after BJD 2457014 were used.
     (Upper): PDM analysis.
     (Lower): Phase-averaged profile}.
  \label{fig:asassn14mdshpdm}
\end{figure}

   This object was detected as a transient at $V$=15.7
on 2014 December 15 by ASAS-SN team (vsnet-alert 18074).
The coordinates are \timeform{04h 43m 45.41s},
\timeform{-21D 08' 34.1''} (refined astrometry,
vsnet-alert 18078).
Superhumps appeared accompanied by brightening
of the brightness of the object (vsnet-alert 18095,
18099, 18105, 18130; figure \ref{fig:asassn14mdshpdm}).
The times of superhump maxima are listed in table
\ref{tab:asassn14mdoc2014}.  Stages A and B can be
clearly recognized.  Since the object is very faint
($V \sim 17$), no secure superhumps were detected
in the later part of the superoutburst.
The alias selection was made to express the $O-C$
values for the best observed data on BJD 2457014
and 2457015.

\begin{table}
\caption{Superhump maxima of ASASSN-14md (2014)}\label{tab:asassn14mdoc2014}
\begin{center}
\begin{tabular}{rp{55pt}p{40pt}r@{.}lr}
\hline
\multicolumn{1}{c}{$E$} & \multicolumn{1}{c}{max\commenta} & \multicolumn{1}{c}{error} & \multicolumn{2}{c}{$O-C$\commentb} & \multicolumn{1}{c}{$N$\commentc} \\
\hline
0 & 57012.6905 & 0.0045 & $-$0&0119 & 23 \\
1 & 57012.7571 & 0.0024 & $-$0&0128 & 21 \\
2 & 57012.8216 & 0.0033 & $-$0&0159 & 17 \\
14 & 57013.6598 & 0.0029 & 0&0111 & 32 \\
28 & 57014.6129 & 0.0023 & 0&0178 & 16 \\
29 & 57014.6787 & 0.0027 & 0&0160 & 34 \\
43 & 57015.6226 & 0.0012 & 0&0136 & 31 \\
44 & 57015.6875 & 0.0015 & 0&0109 & 29 \\
58 & 57016.6258 & 0.0015 & 0&0028 & 35 \\
59 & 57016.6891 & 0.0018 & $-$0&0014 & 26 \\
73 & 57017.6235 & 0.0021 & $-$0&0134 & 35 \\
74 & 57017.6878 & 0.0075 & $-$0&0167 & 25 \\
\hline
  \multicolumn{6}{l}{\commenta BJD$-$2400000.} \\
  \multicolumn{6}{l}{\commentb Against max $= 2457012.7023 + 0.067597 E$.} \\
  \multicolumn{6}{l}{\commentc Number of points used to determine the maximum.} \\
\end{tabular}
\end{center}
\end{table}

\subsection{ASASSN-14mh}\label{obj:asassn14mh}

\begin{figure}
  \begin{center}
    \FigureFile(88mm,110mm){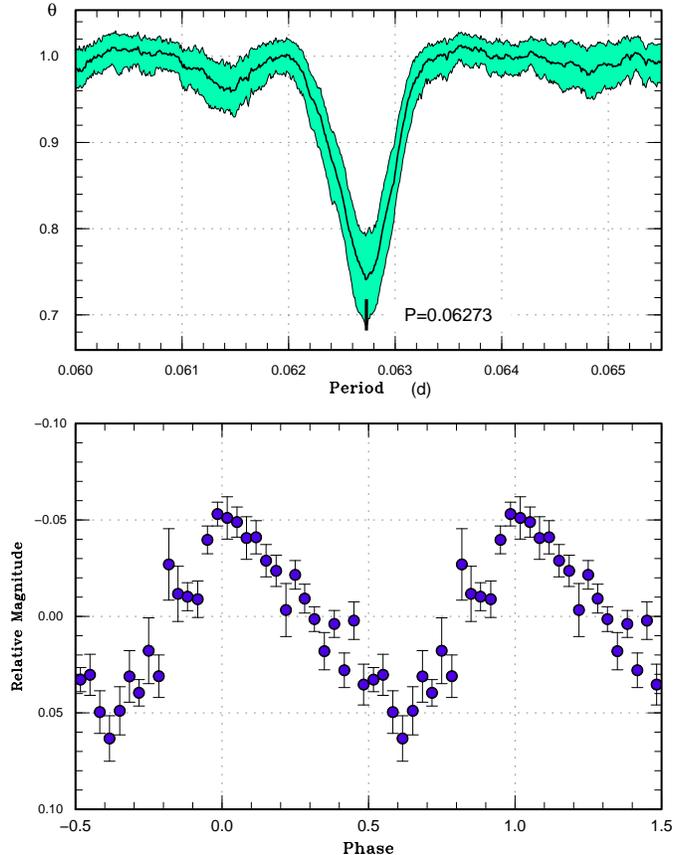}
  \end{center}
  \caption{Superhumps in ASASSN-14mh (2014).
     The data between BJD 2457015 and 2457024 were used.
     (Upper): PDM analysis.
     (Lower): Phase-averaged profile}.
  \label{fig:asassn14mhshpdm}
\end{figure}

   This object was detected as a transient at $V$=15.1
on 2014 December 20 by ASAS-SN team (vsnet-alert 18074).
The coordinates are \timeform{11h 27m 25.99s},
\timeform{-28D 22' 11.8''} (the Initial Gaia Source List).
The object initially showed rapid fading and subsequent
brightening (vsnet-alert 18094), which turned out to be
a precursor outburst and rise to a superoutburst
accompanied by superhumps (vsnet-alert 18098;
figure \ref{fig:asassn14mhshpdm}).
The times of superhump maxima are listed in table
\ref{tab:asassn14mhoc2014}.  The maxima for $E \le 1$
correspond to stage A superhumps.

\begin{table}
\caption{Superhump maxima of ASASSN-14mh (2014)}\label{tab:asassn14mhoc2014}
\begin{center}
\begin{tabular}{rp{55pt}p{40pt}r@{.}lr}
\hline
\multicolumn{1}{c}{$E$} & \multicolumn{1}{c}{max\commenta} & \multicolumn{1}{c}{error} & \multicolumn{2}{c}{$O-C$\commentb} & \multicolumn{1}{c}{$N$\commentc} \\
\hline
0 & 57014.7329 & 0.0020 & $-$0&0029 & 28 \\
1 & 57014.7916 & 0.0017 & $-$0&0070 & 28 \\
16 & 57015.7409 & 0.0005 & 0&0007 & 28 \\
17 & 57015.8051 & 0.0006 & 0&0021 & 27 \\
32 & 57016.7468 & 0.0008 & 0&0020 & 28 \\
33 & 57016.8087 & 0.0006 & 0&0012 & 28 \\
47 & 57017.6909 & 0.0051 & 0&0045 & 11 \\
48 & 57017.7507 & 0.0008 & 0&0015 & 28 \\
49 & 57017.8125 & 0.0008 & 0&0006 & 28 \\
56 & 57018.2584 & 0.0068 & 0&0070 & 43 \\
57 & 57018.3187 & 0.0018 & 0&0045 & 64 \\
65 & 57018.8100 & 0.0021 & $-$0&0064 & 23 \\
73 & 57019.3181 & 0.0021 & $-$0&0005 & 54 \\
79 & 57019.6931 & 0.0054 & $-$0&0022 & 16 \\
80 & 57019.7568 & 0.0018 & $-$0&0013 & 28 \\
96 & 57020.7613 & 0.0013 & $-$0&0013 & 27 \\
111 & 57021.7051 & 0.0017 & 0&0009 & 25 \\
112 & 57021.7623 & 0.0010 & $-$0&0047 & 27 \\
128 & 57022.7724 & 0.0023 & 0&0009 & 15 \\
143 & 57023.7155 & 0.0076 & 0&0023 & 15 \\
144 & 57023.7741 & 0.0050 & $-$0&0019 & 13 \\
\hline
  \multicolumn{6}{l}{\commenta BJD$-$2400000.} \\
  \multicolumn{6}{l}{\commentb Against max $= 2457014.7358 + 0.062779 E$.} \\
  \multicolumn{6}{l}{\commentc Number of points used to determine the maximum.} \\
\end{tabular}
\end{center}
\end{table}

\subsection{ASASSN-14mj}\label{obj:asassn14mj}

\begin{figure}
  \begin{center}
    \FigureFile(88mm,110mm){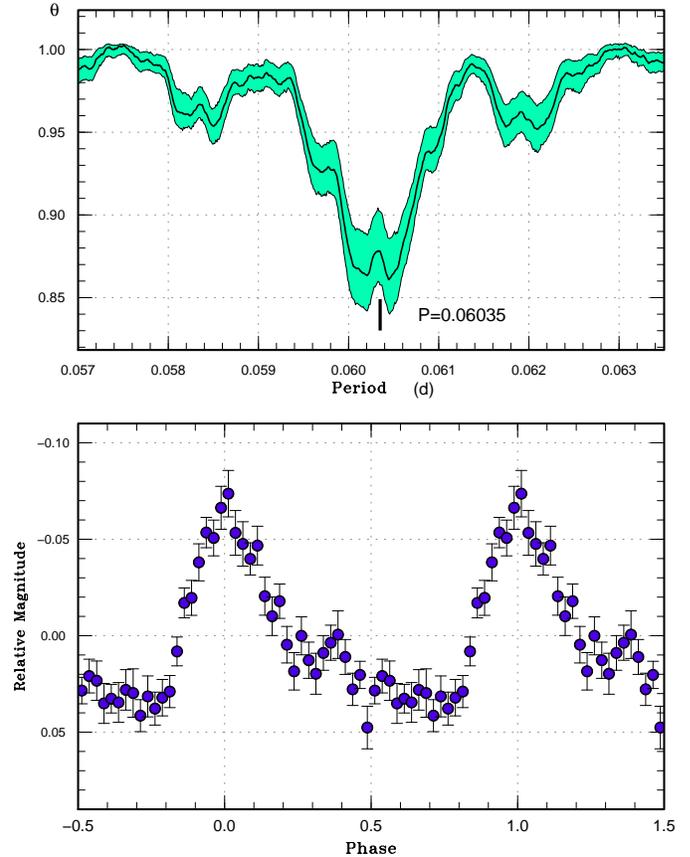}
  \end{center}
  \caption{Superhumps in ASASSN-14mj (2014).
     The data after BJD 2457019 were used.
     (Upper): PDM analysis.
     (Lower): Phase-averaged profile}.
  \label{fig:asassn14mjshpdm}
\end{figure}

   This object was detected as a transient at $V$=13.7
on 2014 December 21 by ASAS-SN team (vsnet-alert 18092).
The coordinates are \timeform{06h 43m 35.15s},
\timeform{+74D 10' 15.5''} (GSC 2.3.2 position).
Although the object was originally suspected to be
a WZ Sge-type dwarf nova based on the apparent large
outburst amplitude, it showed development of superhumps
as in ordinary SU UMa-type dwarf novae
(vsnet-alert 18103, 18114, 18118, 18125;
figure \ref{fig:asassn14mjshpdm}).
The object significantly brightened as superhumps grew.
The times of superhump maxima are listed in table
\ref{tab:asassn14mjoc2014}.  Although the
epochs for $E \le 5$ correspond to stage A
superhumps, we could not determine the period
of stage A superhumps.

\begin{table}
\caption{Superhump maxima of ASASSN-14mj (2014)}\label{tab:asassn14mjoc2014}
\begin{center}
\begin{tabular}{rp{55pt}p{40pt}r@{.}lr}
\hline
\multicolumn{1}{c}{$E$} & \multicolumn{1}{c}{max\commenta} & \multicolumn{1}{c}{error} & \multicolumn{2}{c}{$O-C$\commentb} & \multicolumn{1}{c}{$N$\commentc} \\
\hline
0 & 57019.1881 & 0.0029 & $-$0&0094 & 199 \\
4 & 57019.4351 & 0.0013 & $-$0&0041 & 56 \\
5 & 57019.4915 & 0.0010 & $-$0&0082 & 61 \\
18 & 57020.2919 & 0.0003 & 0&0063 & 54 \\
19 & 57020.3512 & 0.0002 & 0&0053 & 63 \\
20 & 57020.4123 & 0.0002 & 0&0059 & 65 \\
21 & 57020.4718 & 0.0003 & 0&0050 & 64 \\
22 & 57020.5326 & 0.0003 & 0&0053 & 45 \\
33 & 57021.1948 & 0.0006 & 0&0026 & 59 \\
34 & 57021.2538 & 0.0005 & 0&0011 & 59 \\
35 & 57021.3153 & 0.0006 & 0&0022 & 58 \\
45 & 57021.9209 & 0.0012 & 0&0034 & 122 \\
51 & 57022.2796 & 0.0005 & $-$0&0006 & 102 \\
52 & 57022.3397 & 0.0006 & $-$0&0009 & 117 \\
53 & 57022.3997 & 0.0003 & $-$0&0014 & 61 \\
54 & 57022.4588 & 0.0004 & $-$0&0027 & 59 \\
55 & 57022.5208 & 0.0003 & $-$0&0012 & 62 \\
56 & 57022.5808 & 0.0002 & $-$0&0016 & 61 \\
57 & 57022.6404 & 0.0003 & $-$0&0025 & 62 \\
58 & 57022.6994 & 0.0024 & $-$0&0039 & 22 \\
62 & 57022.9399 & 0.0040 & $-$0&0052 & 47 \\
63 & 57023.0103 & 0.0018 & 0&0047 & 44 \\
\hline
  \multicolumn{6}{l}{\commenta BJD$-$2400000.} \\
  \multicolumn{6}{l}{\commentb Against max $= 2457019.1975 + 0.060445 E$.} \\
  \multicolumn{6}{l}{\commentc Number of points used to determine the maximum.} \\
\end{tabular}
\end{center}
\end{table}

\subsection{ASASSN-15ah}\label{obj:asassn15ah}

\begin{figure}
  \begin{center}
    \FigureFile(88mm,110mm){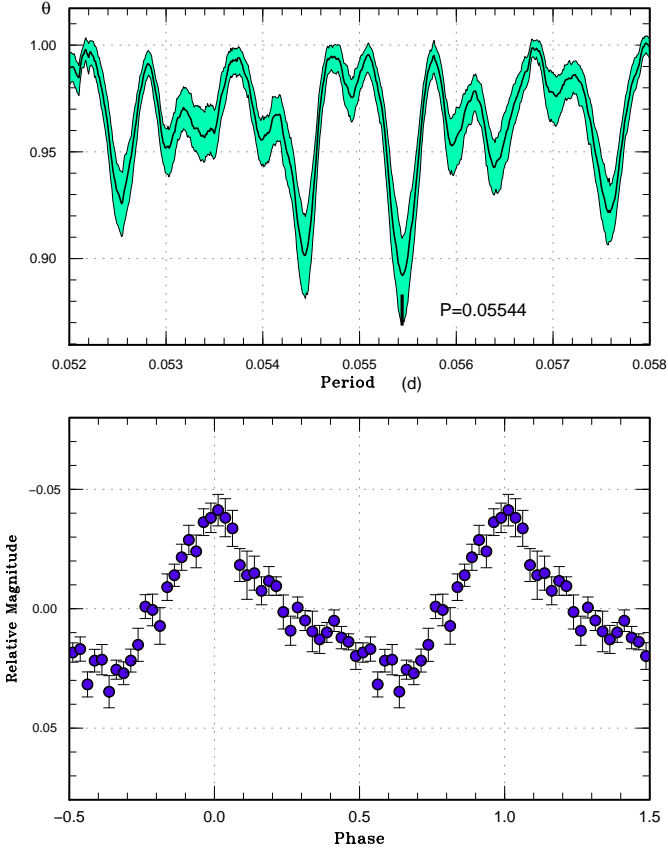}
  \end{center}
  \caption{Superhumps in ASASSN-15ah (2015).
     The data after BJD 2457038 were used.
     (Upper): PDM analysis.
     (Lower): Phase-averaged profile}.
  \label{fig:asassn15ahshpdm}
\end{figure}

   This object was detected as a transient at $V$=13.69
on 2015 January 7 by ASAS-SN team (vsnet-alert 18153).
The coordinates are \timeform{06h 00m 30.00s},
\timeform{-32D 07' 35.6''} (GSC 2.3.2 position).
Although the outburst amplitude suggested a WZ Sge-type
dwarf nova, no evident early superhumps were detected.
Eight days after the detection, this object started to
show ordinary superhumps (vsnet-alert 18189, 18200;
figure \ref{fig:asassn15ahshpdm}).
The times of superhump maxima are listed in table
\ref{tab:asassn15ahoc2015}.  The epochs for $E \le 2$
correspond to stage A superhumps.
Despite the large outburst amplitude, the shortness
of the segment before the appearance of ordinary
superhumps suggests that ASASSN-15ah is not
an extreme WZ Sge-type dwarf nova.

\begin{table}
\caption{Superhump maxima of ASASSN-15ah (2015)}\label{tab:asassn15ahoc2015}
\begin{center}
\begin{tabular}{rp{55pt}p{40pt}r@{.}lr}
\hline
\multicolumn{1}{c}{$E$} & \multicolumn{1}{c}{max\commenta} & \multicolumn{1}{c}{error} & \multicolumn{2}{c}{$O-C$\commentb} & \multicolumn{1}{c}{$N$\commentc} \\
\hline
0 & 57038.3679 & 0.0006 & $-$0&0028 & 128 \\
1 & 57038.4223 & 0.0006 & $-$0&0040 & 128 \\
2 & 57038.4741 & 0.0033 & $-$0&0077 & 33 \\
35 & 57040.3212 & 0.0003 & 0&0069 & 127 \\
36 & 57040.3760 & 0.0002 & 0&0061 & 127 \\
37 & 57040.4306 & 0.0003 & 0&0052 & 127 \\
38 & 57040.4873 & 0.0003 & 0&0064 & 128 \\
85 & 57043.0859 & 0.0035 & $-$0&0049 & 70 \\
89 & 57043.3102 & 0.0004 & $-$0&0027 & 127 \\
90 & 57043.3662 & 0.0003 & $-$0&0023 & 127 \\
107 & 57044.3073 & 0.0007 & $-$0&0052 & 95 \\
108 & 57044.3739 & 0.0021 & 0&0059 & 78 \\
109 & 57044.4210 & 0.0013 & $-$0&0026 & 56 \\
139 & 57046.0969 & 0.0033 & 0&0075 & 54 \\
144 & 57046.3651 & 0.0008 & $-$0&0020 & 128 \\
145 & 57046.4185 & 0.0007 & $-$0&0041 & 127 \\
\hline
  \multicolumn{6}{l}{\commenta BJD$-$2400000.} \\
  \multicolumn{6}{l}{\commentb Against max $= 2457038.3708 + 0.055530 E$.} \\
  \multicolumn{6}{l}{\commentc Number of points used to determine the maximum.} \\
\end{tabular}
\end{center}
\end{table}

\subsection{ASASSN-15ap}\label{obj:asassn15ap}

\begin{figure}
  \begin{center}
    \FigureFile(88mm,110mm){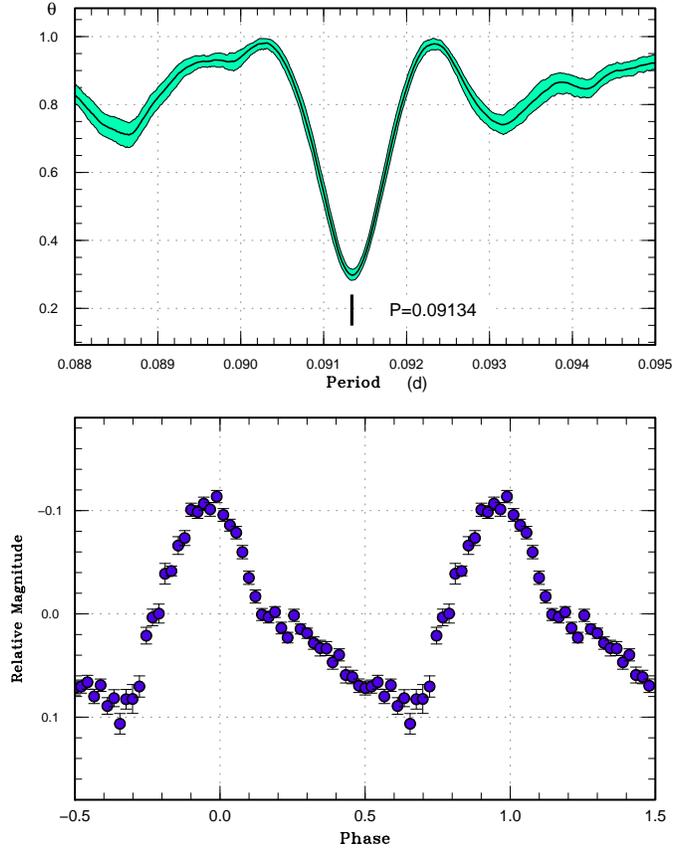}
  \end{center}
  \caption{Superhumps in ASASSN-15ap (2015).
     (Upper): PDM analysis.
     (Lower): Phase-averaged profile}.
  \label{fig:asassn15apshpdm}
\end{figure}

   This object was detected as a transient at $V$=17.0
on 2015 January 8 by ASAS-SN team.  The brightened
to $V$=15.55 on January 11.
The coordinates are \timeform{06h 37m 47.51s},
\timeform{-25D 41' 20.1''} (the Initial Gaia Source List).
Superhumps were observed since our initial observation
on January 17 (vsnet-alert 18198, 18222;
figure \ref{fig:asassn15apshpdm}).
The times of superhump maxima are listed in table
\ref{tab:asassn15apoc2015}.
The superhump stage is not known.  The superhump
period was relatively constant for this long-$P_{\rm SH}$
system.

\begin{table}
\caption{Superhump maxima of ASASSN-15ap (2015)}\label{tab:asassn15apoc2015}
\begin{center}
\begin{tabular}{rp{55pt}p{40pt}r@{.}lr}
\hline
\multicolumn{1}{c}{$E$} & \multicolumn{1}{c}{max\commenta} & \multicolumn{1}{c}{error} & \multicolumn{2}{c}{$O-C$\commentb} & \multicolumn{1}{c}{$N$\commentc} \\
\hline
0 & 57038.4067 & 0.0010 & $-$0&0058 & 122 \\
1 & 57038.5068 & 0.0042 & 0&0031 & 28 \\
22 & 57040.4233 & 0.0008 & 0&0014 & 124 \\
23 & 57040.5166 & 0.0004 & 0&0034 & 211 \\
55 & 57043.4353 & 0.0004 & $-$0&0009 & 166 \\
56 & 57043.5259 & 0.0005 & $-$0&0016 & 211 \\
66 & 57044.4410 & 0.0006 & 0&0001 & 166 \\
67 & 57044.5326 & 0.0006 & 0&0004 & 210 \\
\hline
  \multicolumn{6}{l}{\commenta BJD$-$2400000.} \\
  \multicolumn{6}{l}{\commentb Against max $= 2457038.4125 + 0.091340 E$.} \\
  \multicolumn{6}{l}{\commentc Number of points used to determine the maximum.} \\
\end{tabular}
\end{center}
\end{table}

\subsection{ASASSN-15aq}\label{obj:asassn15aq}

\begin{figure}
  \begin{center}
    \FigureFile(88mm,110mm){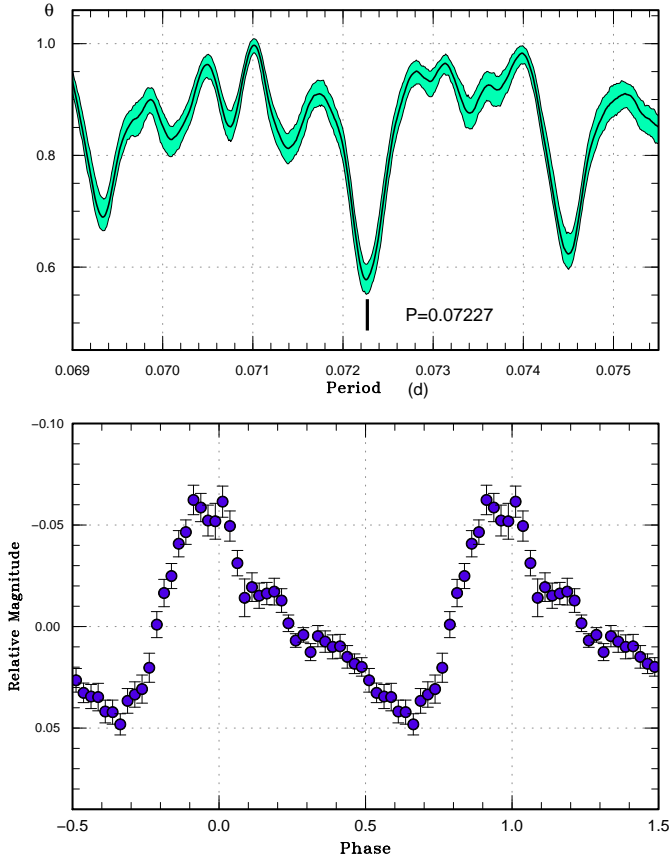}
  \end{center}
  \caption{Superhumps in ASASSN-15aq during the plateau phase (2015).
     (Upper): PDM analysis.
     (Lower): Phase-averaged profile}.
  \label{fig:asassn15aqshpdm}
\end{figure}

   This object was detected as a transient at $V$=14.2
on 2015 January 13 by ASAS-SN team (vsnet-alert 18175).
The coordinates are \timeform{00h 42m 14.24s},
\timeform{-56D 09' 19.9''} (2MASS position).
There were several outbursts in the ASAS-3.  The 2007
outburst looked like superoutbursts (vsnet-alert 18175).
Superhumps were immediately detected
(vsnet-alert 18186, 18199; figure \ref{fig:asassn15aqshpdm}).
The times of superhump maxima are listed in table
\ref{tab:asassn15aqoc2015}.  Both stages B and C
can be recognized.  It was likely that the outburst
detection was not early enough to detect stage A
superhumps.

\begin{table}
\caption{Superhump maxima of ASASSN-15aq (2015)}\label{tab:asassn15aqoc2015}
\begin{center}
\begin{tabular}{rp{55pt}p{40pt}r@{.}lr}
\hline
\multicolumn{1}{c}{$E$} & \multicolumn{1}{c}{max\commenta} & \multicolumn{1}{c}{error} & \multicolumn{2}{c}{$O-C$\commentb} & \multicolumn{1}{c}{$N$\commentc} \\
\hline
0 & 57037.5159 & 0.0076 & $-$0&0110 & 9 \\
1 & 57037.6044 & 0.0028 & 0&0054 & 9 \\
11 & 57038.3208 & 0.0003 & 0&0001 & 167 \\
12 & 57038.3946 & 0.0005 & 0&0017 & 80 \\
14 & 57038.5341 & 0.0015 & $-$0&0032 & 15 \\
39 & 57040.3398 & 0.0004 & $-$0&0016 & 166 \\
40 & 57040.4175 & 0.0010 & 0&0039 & 59 \\
42 & 57040.5562 & 0.0007 & $-$0&0017 & 37 \\
56 & 57041.5701 & 0.0021 & 0&0018 & 26 \\
80 & 57043.3113 & 0.0005 & 0&0110 & 167 \\
108 & 57045.3254 & 0.0004 & 0&0044 & 167 \\
125 & 57046.5455 & 0.0016 & $-$0&0024 & 23 \\
139 & 57047.5499 & 0.0030 & $-$0&0084 & 24 \\
\hline
  \multicolumn{6}{l}{\commenta BJD$-$2400000.} \\
  \multicolumn{6}{l}{\commentb Against max $= 2457037.5269 + 0.072168 E$.} \\
  \multicolumn{6}{l}{\commentc Number of points used to determine the maximum.} \\
\end{tabular}
\end{center}
\end{table}

\subsection{ASASSN-15aw}\label{obj:asassn15aw}

\begin{figure}
  \begin{center}
    \FigureFile(88mm,70mm){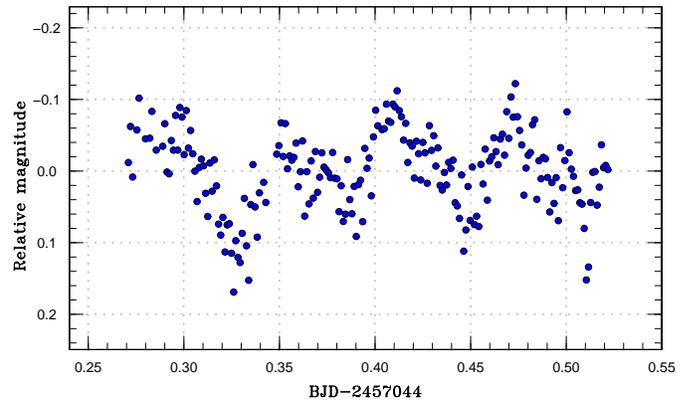}
  \end{center}
  \caption{Superhumps in ASASSN-15aw (2015).}
  \label{fig:asassn15awshlc}
\end{figure}

   This object was detected as a transient at $V$=15.8
on 2015 January 15 by ASAS-SN team.
The coordinates are \timeform{01h 57m 46.13s},
\timeform{+51D 10' 24.1''} (The Initial Gaia Source List).
There is a GALEX counterpart with an NUV magnitude
of 21.1.  MASTER team also detected this object
(MASTER OT J015746.16$+$511023.2) three days later
\citep{bal15asassn15awatel6967}.
Superhumps were recorded in single-night observations
(vsnet-alert 18203; figure \ref{fig:asassn15awshlc}).
Three superhump maxima were
recorded: BJD 2457044.3558(18) ($N$=39), 2457044.4104(9)
($N$=42) and 2457044.4720(10) ($N$=43).
The superhump period determined by the PDM method
is listed in table \ref{tab:perlist}.

\subsection{ASASSN-15bg}\label{obj:asassn15bg}

\begin{figure}
  \begin{center}
    \FigureFile(88mm,110mm){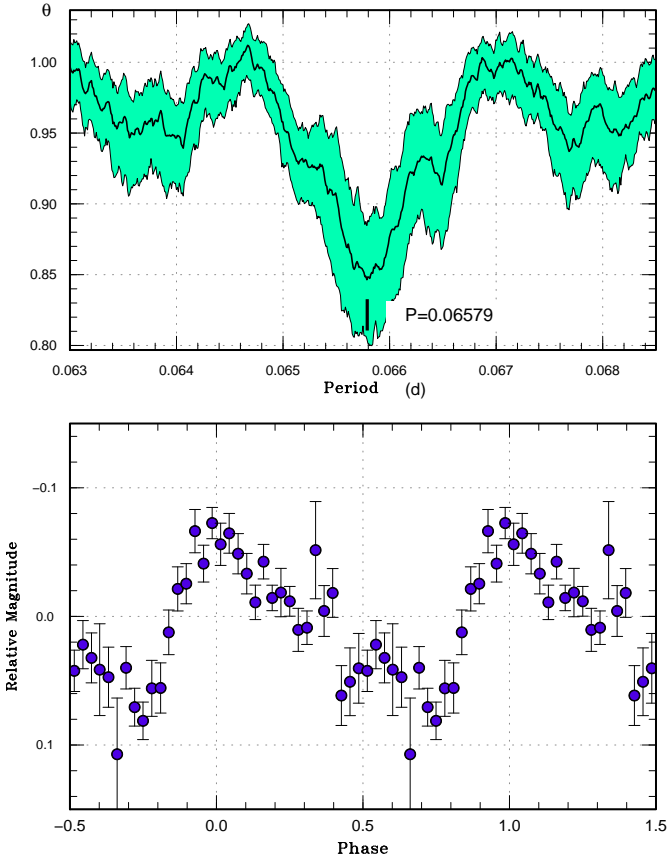}
  \end{center}
  \caption{Superhumps in ASASSN-15bg (2015).
     (Upper): PDM analysis.
     (Lower): Phase-averaged profile}.
  \label{fig:asassn15bgshpdm}
\end{figure}

   This object was detected as a transient at $V$=15.6
on 2015 January 20 by ASAS-SN team.
The coordinates are \timeform{01h 27m 54.61s},
\timeform{-58D 17' 16.8''} (the Initial Gaia Source List).
Subsequent observations detected superhumps
(vsnet-alert 18223; figure \ref{fig:asassn15bgshpdm}).
The times of superhump maxima are listed in table
\ref{tab:asassn15bgoc2015}.

\begin{table}
\caption{Superhump maxima of ASASSN-15bg (2015)}\label{tab:asassn15bgoc2015}
\begin{center}
\begin{tabular}{rp{55pt}p{40pt}r@{.}lr}
\hline
\multicolumn{1}{c}{$E$} & \multicolumn{1}{c}{max\commenta} & \multicolumn{1}{c}{error} & \multicolumn{2}{c}{$O-C$\commentb} & \multicolumn{1}{c}{$N$\commentc} \\
\hline
0 & 57044.3073 & 0.0022 & $-$0&0019 & 73 \\
1 & 57044.3782 & 0.0013 & 0&0034 & 104 \\
15 & 57045.2932 & 0.0041 & $-$0&0011 & 108 \\
16 & 57045.3583 & 0.0014 & $-$0&0016 & 148 \\
31 & 57046.3456 & 0.0010 & 0&0006 & 151 \\
32 & 57046.4112 & 0.0019 & 0&0006 & 103 \\
\hline
  \multicolumn{6}{l}{\commenta BJD$-$2400000.} \\
  \multicolumn{6}{l}{\commentb Against max $= 2457044.3092 + 0.065669 E$.} \\
  \multicolumn{6}{l}{\commentc Number of points used to determine the maximum.} \\
\end{tabular}
\end{center}
\end{table}

\subsection{ASASSN-15bp}\label{obj:asassn15bp}

\begin{figure}
  \begin{center}
    \FigureFile(88mm,110mm){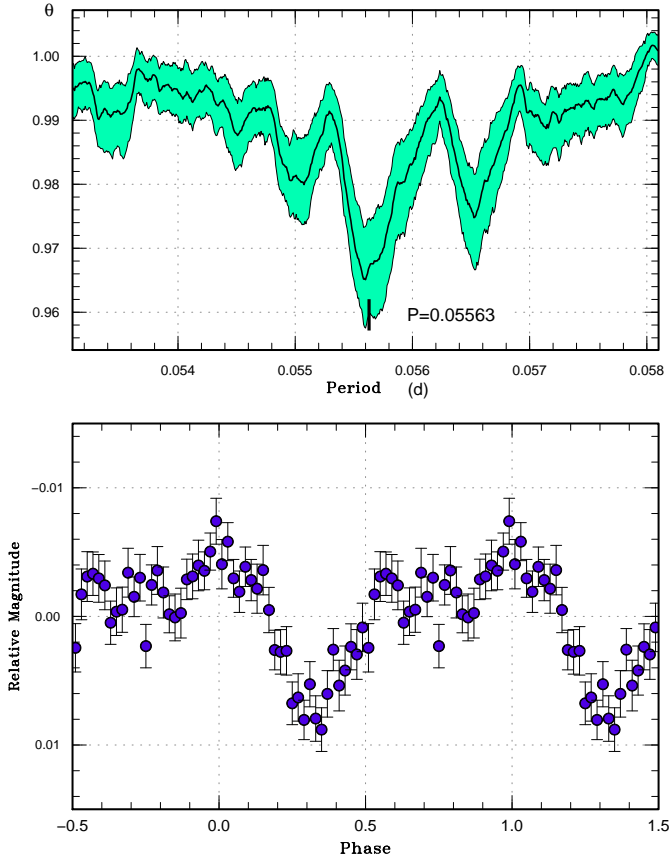}
  \end{center}
  \caption{Early superhumps in ASASSN-15bp (2015).
     (Upper): PDM analysis.
     (Lower): Phase-averaged profile}.
  \label{fig:asassn15bpeshpdm}
\end{figure}

\begin{figure}
  \begin{center}
    \FigureFile(88mm,110mm){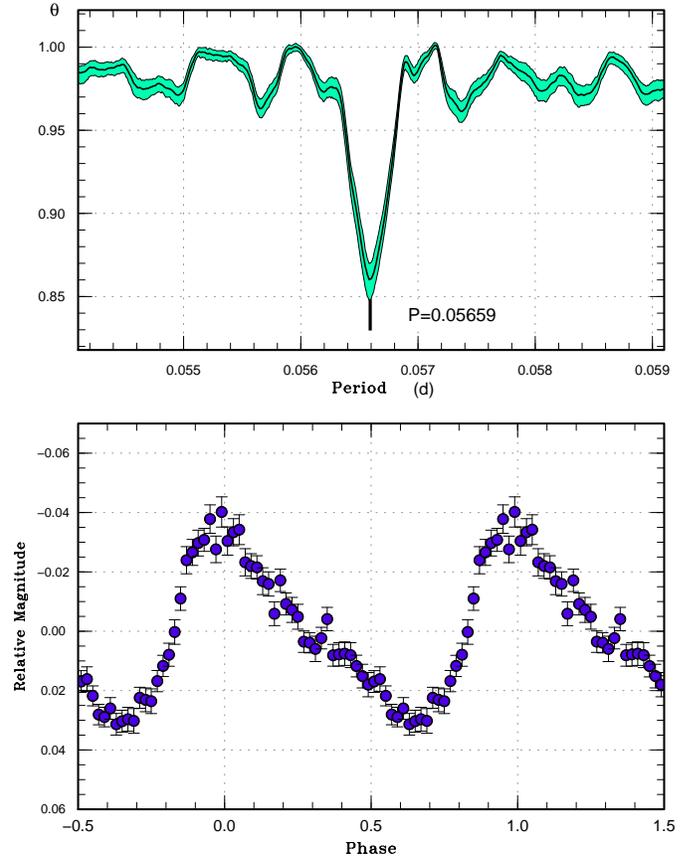}
  \end{center}
  \caption{Ordinary superhumps in ASASSN-15bp (2015).
     The data for BJD 2457056--2457069 were used.
     (Upper): PDM analysis.
     (Lower): Phase-averaged profile}.
  \label{fig:asassn15bpshpdm}
\end{figure}

\begin{figure}
  \begin{center}
    \FigureFile(88mm,100mm){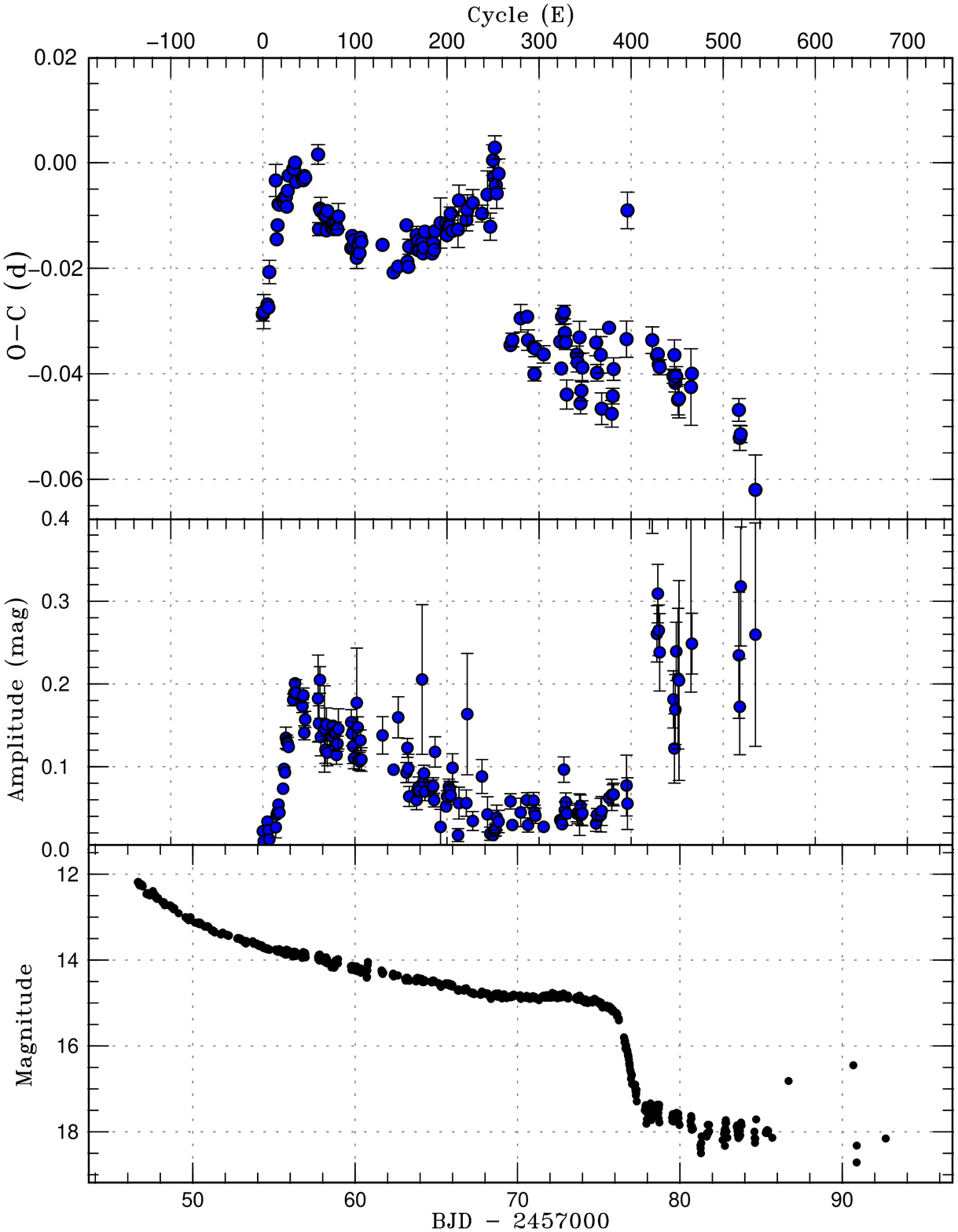}
  \end{center}
  \caption{$O-C$ diagram of superhumps in ASASSN-15bp (2015).
     (Upper:) $O-C$ diagram.
     We used a period of 0.05670~d for calculating the $O-C$ residuals.
     (Middle:) Amplitudes of superhumps.
     (Lower:) Light curve.  The data were binned to 0.018~d.
  }
  \label{fig:asassn15bphumpall}
\end{figure}

   This object was detected as a transient at $V$=11.9
on 2015 January 23 by ASAS-SN team
\citep{sim15asassn15bpatel6981}.
The coordinates are \timeform{12h 12m 40.43s},
\timeform{+04D 16' 55.8''} (SDSS $g$=20.5 counterpart).
There is a GALEX counterpart with a NUV magnitude
of 20.9.  The object was spectroscopically confirmed
as an outbursting CV \citep{wil15asassn15bpatel6992}.
The object was immediately identified as a WZ Sge-type
dwarf nova by the detection of early superhumps
(vsnet-alert 18219, 18325, 18241; figure \ref{fig:asassn15bpeshpdm}).
The object then started to show ordinary
superhumps (vsnet-alert 18250, 18255, 18262, 18269,
18277, 18300, 18311; figure \ref{fig:asassn15bpshpdm}).
The object then started fading rapidly
on February 23 (31~d after the outburst detection,
vsnet-alert 18331).
Although there was some hint of post-superoutburst
rebrightenings (figure \ref{fig:asassn15bphumpall}),
they were not confirmed by independent observers.

   The times of superhump maxima during the plateau phase
are listed in table \ref{tab:asassn15bpoc2015}.
The times $E \le 35$ are clearly growing superhumps
(cf. figure \ref{fig:asassn15bphumpall}).
Using the data for $E \le 33$ (BJD 2457054--2457056),
we could determine the period of stage A superhumps as
0.05731(3)~d with the PDM method.  The maxima for
$35 \le E \le 256$ represent characteristic stage B
superhumps with a small positive $P_{\rm dot}$.
As is usual for a WZ Sge-type dwarf nova, this object
did not show a smooth transition to stage C superhumps.
There was an apparent phase jump after $E$=256.
This jump corresponds to the later part of the plateau
phase, and it looks like the superhumps observed after
this phase jump appear to be on a smooth extension
of early stage B superhumps
(figure \ref{fig:asassn15bphumpall}), just as seen in GW Lib
during the 2007 superoutburst (figure 33 in \cite{Pdot}).
There is difference, however, in that this jump occurred
during the rapid fading in GW Lib while it occurred
during the plateau phase in ASASSN-15bp.
The superhump period after $E=256$ (BJD 2457069--2457076.8)
was determined to be 0.05666(1)~d by the PDM method.
We listed this period as stage C superhumps
in table \ref{tab:perlist}.
The times of superhump maxima in the post-superoutburst
stage are listed in table \ref{tab:asassn15bppost}.
The period was determined as 0.05651(2)~d with
the PDM method.

\begin{figure}
  \begin{center}
    \FigureFile(88mm,100mm){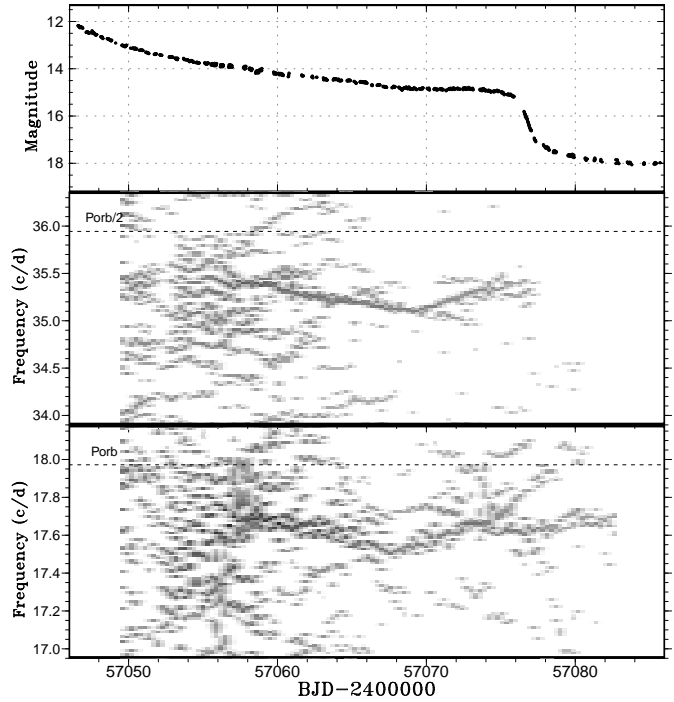}
  \end{center}
  \caption{Lasso period analysis of ASASSN-15bp (2015).
  (Upper:) Light curve.  The data were were binned to 0.018~d.
  (Middle:) First harmonics of the superhump and orbital signals.
  (Lower:) Fundamental of the superhump and orbital signal.
  The orbital signal was only weakly detected during the earliest
  stage (early superhumps).
  The signal of positive superhumps with
  variable frequency (around 17.6 c/d) was recorded during the superoutburst
  plateau post-superoutburst stage.  It is obvious that
  the frequency decreased between BJD 2457055 and 2457059,
  the increased, in agreement with the $O-C$ analysis.
  Superhump signals before BJD 2457055 were noisy due to
  rapidly variable periods and contamination of early superhumps
  in analyzed windows.
  $\log \lambda=-7.8$ was used.  The width of 
  the sliding window and the time step used are 6~d and 0.3~d,
  respectively.
  }
  \label{fig:asassn15bplasso}
\end{figure}

   The variation of the superhump period is also
well visualized in two-dimensional period analysis
(figure \ref{fig:asassn15bplasso}) using Lasso.

   The period of stage A superhumps corresponds to
$\epsilon^*$=0.0293(6) and $q$=0.079(2).
Following the formulation in \citet{kat13qfromstageA},
the period of superhump in the post-superoutburst
phase is found to correspond to a disk radius of
0.355(5)$A$, where $A$ is the binary separation.
This value is close to those recorded in WZ Sge-type
dwarf novae without multiple rebrightenings
\citep{kat13qfromstageA}.

\begin{table}
\caption{Superhump maxima of ASASSN-15bp (2015)}\label{tab:asassn15bpoc2015}
\begin{center}
\begin{tabular}{rp{55pt}p{40pt}r@{.}lr}
\hline
\multicolumn{1}{c}{$E$} & \multicolumn{1}{c}{max\commenta} & \multicolumn{1}{c}{error} & \multicolumn{2}{c}{$O-C$\commentb} & \multicolumn{1}{c}{$N$\commentc} \\
\hline
0 & 57054.2881 & 0.0013 & $-$0&0233 & 196 \\
1 & 57054.3453 & 0.0032 & $-$0&0228 & 324 \\
5 & 57054.5734 & 0.0010 & $-$0&0211 & 60 \\
6 & 57054.6295 & 0.0012 & $-$0&0216 & 66 \\
7 & 57054.6930 & 0.0022 & $-$0&0148 & 62 \\
14 & 57055.1072 & 0.0030 & 0&0030 & 65 \\
15 & 57055.1528 & 0.0010 & $-$0&0081 & 214 \\
16 & 57055.2121 & 0.0010 & $-$0&0053 & 278 \\
17 & 57055.2729 & 0.0008 & $-$0&0013 & 192 \\
18 & 57055.3294 & 0.0008 & $-$0&0013 & 121 \\
22 & 57055.5572 & 0.0006 & $-$0&0000 & 70 \\
23 & 57055.6133 & 0.0004 & $-$0&0006 & 73 \\
24 & 57055.6711 & 0.0004 & 0&0006 & 60 \\
25 & 57055.7278 & 0.0006 & 0&0007 & 69 \\
26 & 57055.7827 & 0.0006 & $-$0&0011 & 102 \\
27 & 57055.8424 & 0.0004 & 0&0020 & 172 \\
28 & 57055.9020 & 0.0003 & 0&0049 & 105 \\
33 & 57056.1867 & 0.0003 & 0&0065 & 118 \\
34 & 57056.2433 & 0.0002 & 0&0065 & 131 \\
35 & 57056.3013 & 0.0002 & 0&0079 & 123 \\
36 & 57056.3544 & 0.0006 & 0&0043 & 92 \\
43 & 57056.7520 & 0.0003 & 0&0055 & 45 \\
44 & 57056.8083 & 0.0003 & 0&0052 & 45 \\
45 & 57056.8658 & 0.0004 & 0&0060 & 45 \\
46 & 57056.9222 & 0.0005 & 0&0058 & 40 \\
60 & 57057.7204 & 0.0019 & 0&0112 & 18 \\
61 & 57057.7629 & 0.0012 & $-$0&0029 & 19 \\
62 & 57057.8235 & 0.0006 & 0&0010 & 22 \\
63 & 57057.8798 & 0.0026 & 0&0007 & 11 \\
67 & 57058.1057 & 0.0015 & 0&0001 & 18 \\
68 & 57058.1624 & 0.0010 & 0&0002 & 35 \\
69 & 57058.2163 & 0.0011 & $-$0&0026 & 30 \\
\hline
  \multicolumn{6}{l}{\commenta BJD$-$2400000.} \\
  \multicolumn{6}{l}{\commentb Against max $= 2457054.3114 + 0.056629 E$.} \\
  \multicolumn{6}{l}{\commentc Number of points used to determine the maximum.} \\
\end{tabular}
\end{center}
\end{table}

\addtocounter{table}{-1}
\begin{table}
\caption{Superhump maxima of ASASSN-15bp (2015) (continued)}
\begin{center}
\begin{tabular}{rp{55pt}p{40pt}r@{.}lr}
\hline
\multicolumn{1}{c}{$E$} & \multicolumn{1}{c}{max\commenta} & \multicolumn{1}{c}{error} & \multicolumn{2}{c}{$O-C$\commentb} & \multicolumn{1}{c}{$N$\commentc} \\
\hline
70 & 57058.2766 & 0.0009 & 0&0011 & 29 \\
75 & 57058.5569 & 0.0003 & $-$0&0018 & 81 \\
76 & 57058.6145 & 0.0002 & $-$0&0008 & 92 \\
77 & 57058.6701 & 0.0002 & $-$0&0018 & 90 \\
78 & 57058.7275 & 0.0002 & $-$0&0010 & 83 \\
79 & 57058.7846 & 0.0006 & $-$0&0006 & 60 \\
80 & 57058.8404 & 0.0007 & $-$0&0014 & 64 \\
81 & 57058.8969 & 0.0005 & $-$0&0015 & 62 \\
82 & 57058.9560 & 0.0025 & 0&0010 & 24 \\
96 & 57059.7438 & 0.0006 & $-$0&0041 & 42 \\
97 & 57059.8028 & 0.0007 & $-$0&0017 & 94 \\
98 & 57059.8574 & 0.0012 & $-$0&0037 & 105 \\
99 & 57059.9154 & 0.0007 & $-$0&0024 & 44 \\
102 & 57060.0822 & 0.0020 & $-$0&0055 & 28 \\
103 & 57060.1404 & 0.0006 & $-$0&0039 & 179 \\
104 & 57060.1982 & 0.0009 & $-$0&0027 & 191 \\
105 & 57060.2532 & 0.0007 & $-$0&0044 & 257 \\
106 & 57060.3128 & 0.0007 & $-$0&0014 & 152 \\
107 & 57060.3687 & 0.0011 & $-$0&0021 & 80 \\
130 & 57061.6722 & 0.0012 & $-$0&0011 & 87 \\
142 & 57062.3474 & 0.0004 & $-$0&0054 & 62 \\
147 & 57062.6320 & 0.0005 & $-$0&0040 & 15 \\
156 & 57063.1501 & 0.0009 & 0&0045 & 119 \\
157 & 57063.1999 & 0.0007 & $-$0&0024 & 145 \\
158 & 57063.2557 & 0.0012 & $-$0&0033 & 148 \\
159 & 57063.3162 & 0.0014 & 0&0006 & 109 \\
167 & 57063.7720 & 0.0016 & 0&0034 & 59 \\
168 & 57063.8259 & 0.0004 & 0&0007 & 167 \\
169 & 57063.8843 & 0.0012 & 0&0025 & 18 \\
170 & 57063.9392 & 0.0003 & 0&0007 & 110 \\
173 & 57064.1107 & 0.0016 & 0&0023 & 31 \\
174 & 57064.1654 & 0.0010 & 0&0004 & 119 \\
175 & 57064.2232 & 0.0008 & 0&0015 & 125 \\
\hline
  \multicolumn{6}{l}{\commenta BJD$-$2400000.} \\
  \multicolumn{6}{l}{\commentb Against max $= 2457054.3114 + 0.056629 E$.} \\
  \multicolumn{6}{l}{\commentc Number of points used to determine the maximum.} \\
\end{tabular}
\end{center}
\end{table}

\addtocounter{table}{-1}
\begin{table}
\caption{Superhump maxima of ASASSN-15bp (2015) (continued)}
\begin{center}
\begin{tabular}{rp{55pt}p{40pt}r@{.}lr}
\hline
\multicolumn{1}{c}{$E$} & \multicolumn{1}{c}{max\commenta} & \multicolumn{1}{c}{error} & \multicolumn{2}{c}{$O-C$\commentb} & \multicolumn{1}{c}{$N$\commentc} \\
\hline
176 & 57064.2829 & 0.0011 & 0&0047 & 120 \\
184 & 57064.7324 & 0.0009 & 0&0011 & 58 \\
185 & 57064.7911 & 0.0010 & 0&0031 & 65 \\
186 & 57064.8465 & 0.0010 & 0&0019 & 65 \\
187 & 57064.9067 & 0.0012 & 0&0055 & 28 \\
193 & 57065.2485 & 0.0048 & 0&0075 & 27 \\
199 & 57065.5886 & 0.0007 & 0&0079 & 58 \\
200 & 57065.6431 & 0.0008 & 0&0057 & 58 \\
202 & 57065.7584 & 0.0009 & 0&0078 & 48 \\
203 & 57065.8148 & 0.0010 & 0&0075 & 58 \\
204 & 57065.8739 & 0.0013 & 0&0100 & 60 \\
206 & 57065.9841 & 0.0010 & 0&0069 & 45 \\
212 & 57066.3245 & 0.0034 & 0&0076 & 43 \\
213 & 57066.3867 & 0.0029 & 0&0132 & 21 \\
221 & 57066.8366 & 0.0021 & 0&0100 & 70 \\
222 & 57066.8953 & 0.0028 & 0&0120 & 19 \\
228 & 57067.2368 & 0.0025 & 0&0138 & 78 \\
238 & 57067.8018 & 0.0016 & 0&0124 & 30 \\
244 & 57068.1456 & 0.0045 & 0&0165 & 77 \\
247 & 57068.3096 & 0.0026 & 0&0106 & 177 \\
250 & 57068.4923 & 0.0029 & 0&0234 & 59 \\
251 & 57068.5458 & 0.0017 & 0&0203 & 56 \\
252 & 57068.6080 & 0.0023 & 0&0259 & 56 \\
253 & 57068.6577 & 0.0023 & 0&0189 & 59 \\
254 & 57068.7128 & 0.0028 & 0&0174 & 24 \\
256 & 57068.8299 & 0.0028 & 0&0213 & 71 \\
269 & 57069.5345 & 0.0009 & $-$0&0103 & 46 \\
271 & 57069.6489 & 0.0013 & $-$0&0092 & 59 \\
280 & 57070.1633 & 0.0026 & $-$0&0045 & 35 \\
287 & 57070.5605 & 0.0011 & $-$0&0037 & 118 \\
\hline
  \multicolumn{6}{l}{\commenta BJD$-$2400000.} \\
  \multicolumn{6}{l}{\commentb Against max $= 2457054.3114 + 0.056629 E$.} \\
  \multicolumn{6}{l}{\commentc Number of points used to determine the maximum.} \\
\end{tabular}
\end{center}
\end{table}

\addtocounter{table}{-1}
\begin{table}
\caption{Superhump maxima of ASASSN-15bp (2015) (continued)}
\begin{center}
\begin{tabular}{rp{55pt}p{40pt}r@{.}lr}
\hline
\multicolumn{1}{c}{$E$} & \multicolumn{1}{c}{max\commenta} & \multicolumn{1}{c}{error} & \multicolumn{2}{c}{$O-C$\commentb} & \multicolumn{1}{c}{$N$\commentc} \\
\hline
288 & 57070.6128 & 0.0019 & $-$0&0080 & 112 \\
294 & 57070.9516 & 0.0012 & $-$0&0090 & 60 \\
295 & 57071.0033 & 0.0013 & $-$0&0140 & 63 \\
296 & 57071.0648 & 0.0015 & $-$0&0091 & 81 \\
305 & 57071.5740 & 0.0016 & $-$0&0096 & 121 \\
323 & 57072.5970 & 0.0012 & $-$0&0059 & 116 \\
324 & 57072.6486 & 0.0011 & $-$0&0109 & 150 \\
325 & 57072.7151 & 0.0015 & $-$0&0010 & 145 \\
327 & 57072.8294 & 0.0013 & 0&0000 & 77 \\
328 & 57072.8821 & 0.0017 & $-$0&0039 & 101 \\
329 & 57072.9370 & 0.0013 & $-$0&0056 & 103 \\
330 & 57072.9839 & 0.0028 & $-$0&0154 & 64 \\
341 & 57073.6151 & 0.0016 & $-$0&0071 & 37 \\
342 & 57073.6704 & 0.0018 & $-$0&0085 & 37 \\
344 & 57073.7885 & 0.0031 & $-$0&0036 & 32 \\
345 & 57073.8327 & 0.0020 & $-$0&0160 & 86 \\
346 & 57073.8918 & 0.0019 & $-$0&0136 & 98 \\
347 & 57073.9529 & 0.0027 & $-$0&0091 & 74 \\
362 & 57074.8081 & 0.0025 & $-$0&0033 & 61 \\
363 & 57074.8591 & 0.0016 & $-$0&0090 & 58 \\
367 & 57075.0893 & 0.0035 & $-$0&0053 & 45 \\
368 & 57075.1358 & 0.0030 & $-$0&0154 & 44 \\
376 & 57075.6047 & 0.0005 & 0&0005 & 154 \\
379 & 57075.7585 & 0.0025 & $-$0&0156 & 35 \\
380 & 57075.8186 & 0.0015 & $-$0&0122 & 24 \\
381 & 57075.8804 & 0.0021 & $-$0&0070 & 17 \\
395 & 57076.6799 & 0.0034 & $-$0&0004 & 27 \\
396 & 57076.7610 & 0.0035 & 0&0241 & 21 \\
\hline
  \multicolumn{6}{l}{\commenta BJD$-$2400000.} \\
  \multicolumn{6}{l}{\commentb Against max $= 2457054.3114 + 0.056629 E$.} \\
  \multicolumn{6}{l}{\commentc Number of points used to determine the maximum.} \\
\end{tabular}
\end{center}
\end{table}

\begin{table}
\caption{Superhump maxima of ASASSN-15bp (2015) (post-superoutburst)}\label{tab:asassn15bppost}
\begin{center}
\begin{tabular}{rp{55pt}p{40pt}r@{.}lr}
\hline
\multicolumn{1}{c}{$E$} & \multicolumn{1}{c}{max\commenta} & \multicolumn{1}{c}{error} & \multicolumn{2}{c}{$O-C$\commentb} & \multicolumn{1}{c}{$N$\commentc} \\
\hline
0 & 57078.2673 & 0.0025 & 0&0022 & 15 \\
5 & 57078.5479 & 0.0009 & 0&0002 & 51 \\
6 & 57078.6048 & 0.0009 & 0&0007 & 43 \\
7 & 57078.6596 & 0.0008 & $-$0&0011 & 45 \\
8 & 57078.7157 & 0.0014 & $-$0&0015 & 44 \\
23 & 57079.5645 & 0.0013 & $-$0&0005 & 42 \\
24 & 57079.6252 & 0.0029 & 0&0037 & 41 \\
25 & 57079.6767 & 0.0017 & $-$0&0014 & 50 \\
26 & 57079.7346 & 0.0019 & $-$0&0001 & 28 \\
28 & 57079.8436 & 0.0029 & $-$0&0041 & 14 \\
29 & 57079.9005 & 0.0037 & $-$0&0037 & 12 \\
42 & 57080.6398 & 0.0073 & 0&0008 & 10 \\
43 & 57080.6990 & 0.0011 & 0&0035 & 48 \\
94 & 57083.5838 & 0.0021 & 0&0057 & 37 \\
95 & 57083.6352 & 0.0024 & 0&0005 & 40 \\
96 & 57083.6926 & 0.0016 & 0&0014 & 40 \\
112 & 57084.5893 & 0.0066 & $-$0&0063 & 20 \\
\hline
  \multicolumn{6}{l}{\commenta BJD$-$2400000.} \\
  \multicolumn{6}{l}{\commentb Against max $= 2457078.2650 + 0.056523 E$.} \\
  \multicolumn{6}{l}{\commentc Number of points used to determine the maximum.} \\
\end{tabular}
\end{center}
\end{table}

\subsection{ASASSN-15bu}\label{obj:asassn15bu}

\begin{figure}
  \begin{center}
    \FigureFile(88mm,110mm){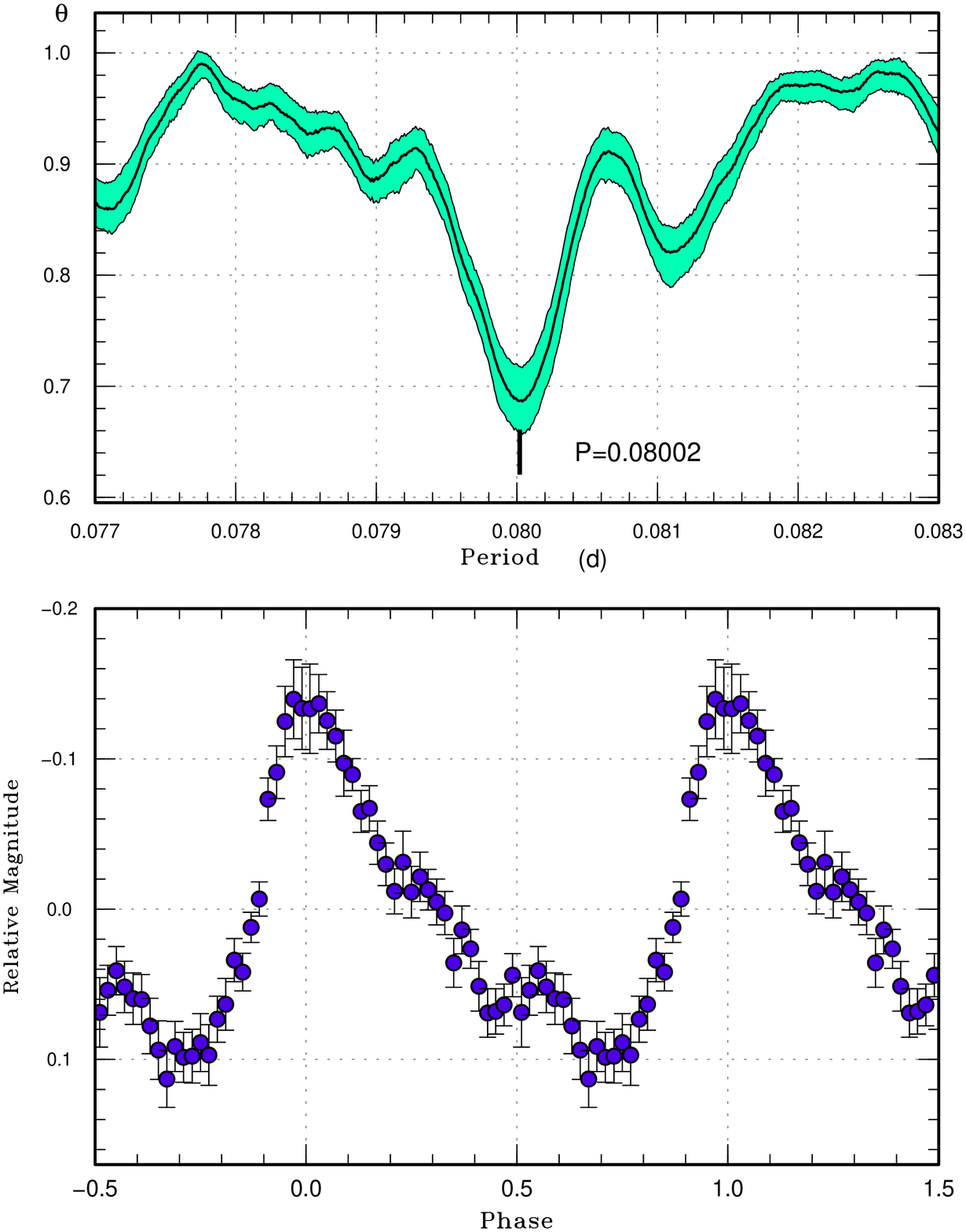}
  \end{center}
  \caption{Superhumps in ASASSN-15bu outside the eclipses (2015).
     (Upper): PDM analysis.
     (Lower): Phase-averaged profile.}
  \label{fig:asassn15bushpdm}
\end{figure}

   This object was detected as a transient at $V$=15.8
on 2015 January 22 by ASAS-SN team.
The coordinates are \timeform{02h 54m 43.77s},
\timeform{+22D 44' 01.9''} (2MASS position).
There is a GALEX counterpart with a NUV magnitude
of 19.7.  The object was suspected to be an eclipsing 
SU UMa-type dwarf nova based on apparent eclipses
in the CRTS data (vsnet-alert 18228).
Although the outburst amplitude
was small, the outburst turned out to be a superoutburst
with superhumps (vsnet-alert 18234, 18242;
figure \ref{fig:asassn15bushpdm}).

   An MCMC analysis of both the CRTS data
and the present data in outburst yielded the following
orbital ephemeris:
\begin{equation}
{\rm Min(BJD)} = 2455129.74232(3) + 0.076819040(1) E .
\label{equ:asassn15buecl}
\end{equation}

\begin{figure}
  \begin{center}
    \FigureFile(88mm,120mm){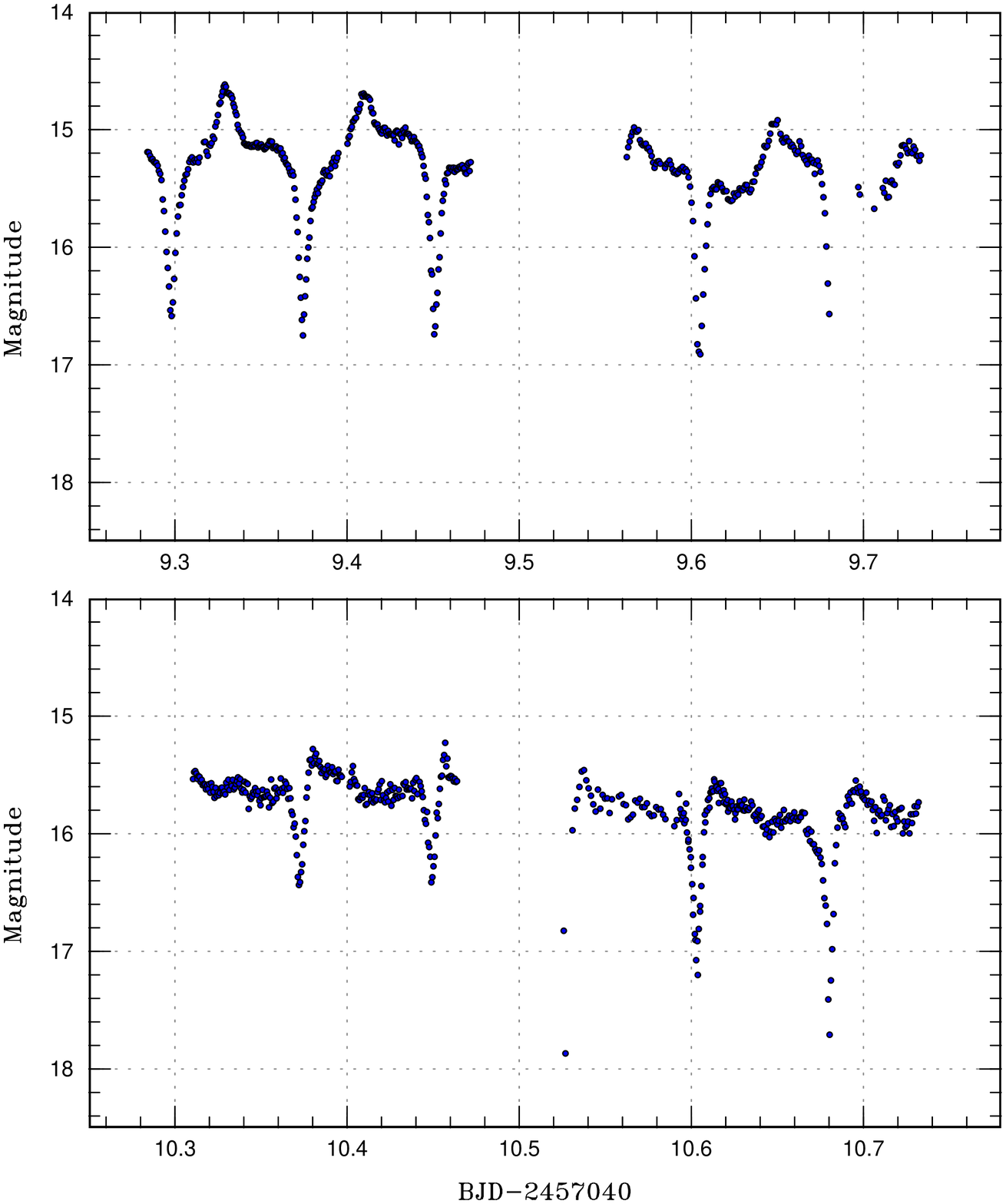}
  \end{center}
  \caption{Sample light curve of ASASSN-15bu in superoutburst.
  Superposition of superhumps and eclipse are well visible.}
  \label{fig:asassn15bulc}
\end{figure}

   Like ASASSN-14ag, the object showed a strong 
beat phenomenon between the superhump
period and the orbital period (figure \ref{fig:asassn15bulc}).

   The times of superhump maxima outside
the eclipses are listed in table
\ref{tab:asassn15buoc2015}.  Since the object started
fading rapidly $\sim$9~d after the ASAS-SN detection,
it was likely that we observed the only the later
part of the superoutburst.  The superhumps were possibly
mostly stage C superhumps although the superhump amplitudes
(sometimes reaching 0.6 mag) were much larger than
those of typical stage C superhumps in non-eclipsing
systems.

\begin{table}
\caption{Superhump maxima of ASASSN-15bu (2015)}\label{tab:asassn15buoc2015}
\begin{center}
\begin{tabular}{rp{50pt}p{30pt}r@{.}lcr}
\hline
$E$ & max\commenta & error & \multicolumn{2}{c}{$O-C$\commentb} & phase\commentc & $N$\commentd \\
\hline
0 & 57049.3319 & 0.0006 & $-$0&0021 & 0.46 & 88 \\
1 & 57049.4138 & 0.0005 & $-$0&0003 & 0.53 & 92 \\
3 & 57049.5672 & 0.0015 & $-$0&0070 & 0.52 & 39 \\
4 & 57049.6540 & 0.0004 & $-$0&0002 & 0.65 & 60 \\
5 & 57049.7334 & 0.0016 & $-$0&0008 & 0.69 & 30 \\
12 & 57050.2921 & 0.0020 & $-$0&0025 & 0.96 & 23 \\
13 & 57050.3811 & 0.0008 & 0&0065 & 0.12 & 74 \\
14 & 57050.4557 & 0.0017 & 0&0010 & 0.09 & 41 \\
15 & 57050.5329 & 0.0096 & $-$0&0019 & 0.09 & 22 \\
16 & 57050.6134 & 0.0011 & $-$0&0013 & 0.14 & 74 \\
17 & 57050.7007 & 0.0014 & 0&0059 & 0.28 & 66 \\
26 & 57051.4174 & 0.0005 & 0&0021 & 0.61 & 84 \\
28 & 57051.5777 & 0.0006 & 0&0024 & 0.70 & 93 \\
29 & 57051.6536 & 0.0008 & $-$0&0019 & 0.68 & 76 \\
53 & 57053.5811 & 0.0046 & 0&0045 & 0.78 & 71 \\
55 & 57053.7416 & 0.0022 & 0&0049 & 0.87 & 60 \\
64 & 57054.4507 & 0.0033 & $-$0&0064 & 0.10 & 38 \\
76 & 57055.4257 & 0.0015 & 0&0080 & 0.79 & 66 \\
79 & 57055.6572 & 0.0005 & $-$0&0007 & 0.80 & 66 \\
80 & 57055.7278 & 0.0007 & $-$0&0101 & 0.72 & 52 \\
\hline
  \multicolumn{7}{l}{\commenta BJD$-$2400000.} \\
  \multicolumn{7}{l}{\commentb Against max $= 2457049.3340 + 0.080049 E$.} \\
  \multicolumn{7}{l}{\commentc Orbital phase.} \\
  \multicolumn{7}{l}{\commentd Number of points used to determine the maximum.} \\
\end{tabular}
\end{center}
\end{table}

\subsection{ASASSN-15bv}\label{obj:asassn15bv}

\begin{figure}
  \begin{center}
    \FigureFile(88mm,70mm){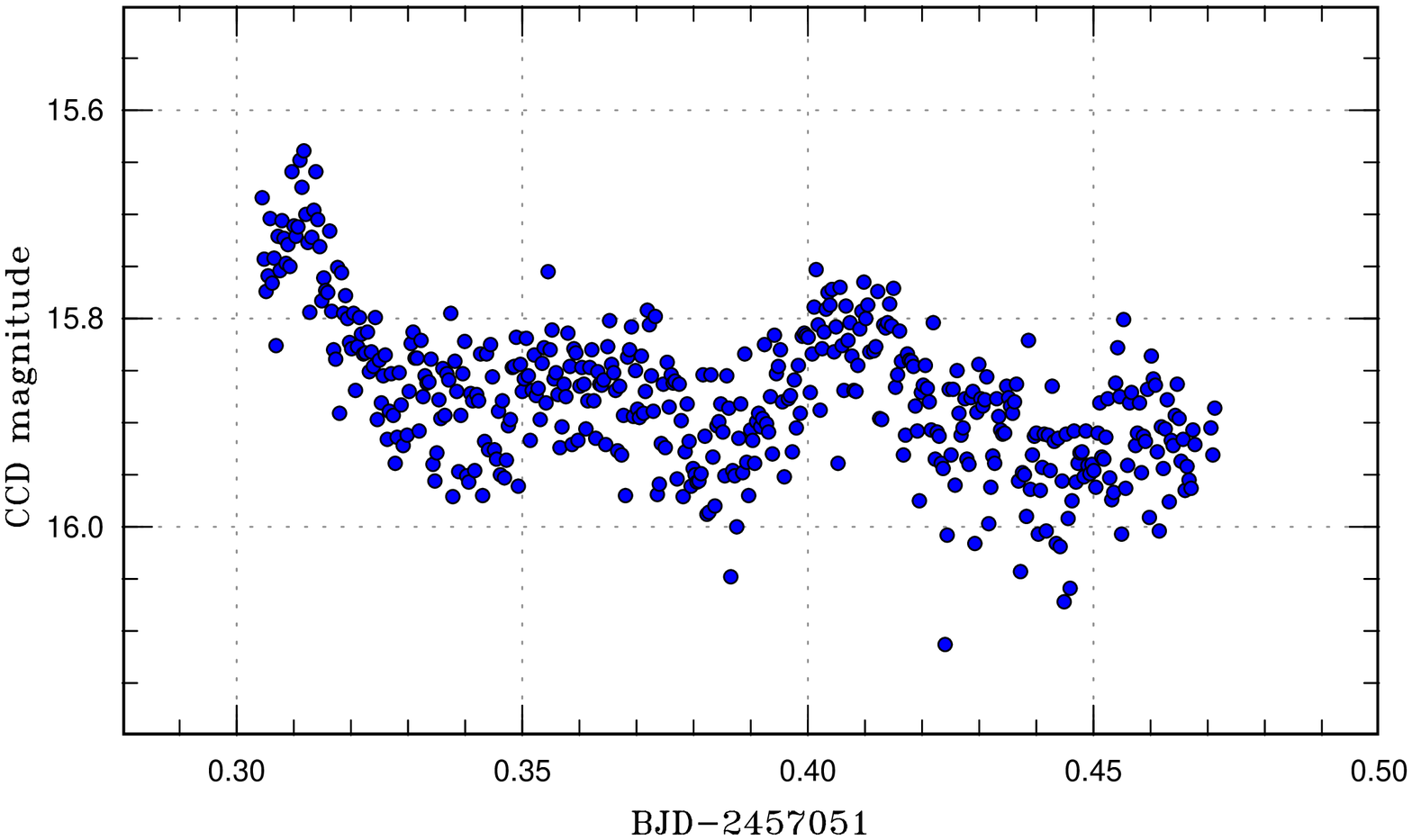}
  \end{center}
  \caption{Likely superhumps in ASASSN-15bv (2015).}
  \label{fig:asassn15bvshlc}
\end{figure}

   This object was detected as a transient at $V$=15.5
on 2015 January 22 by ASAS-SN team.
The coordinates are \timeform{06h 25m 01.75s},
\timeform{-02D 47' 58.6''} (the Initial Gaia Source List).
A single-night observations detected likely superhumps
(vsnet-alert 18244, figure \ref{fig:asassn15bvshlc}).
Although a period of 0.101(1)~d can be determined,
the reality of superhumps needs to be confirmed by future
observations.

\subsection{ASASSN-15dq}\label{obj:asassn15dq}

\begin{figure}
  \begin{center}
    \FigureFile(88mm,110mm){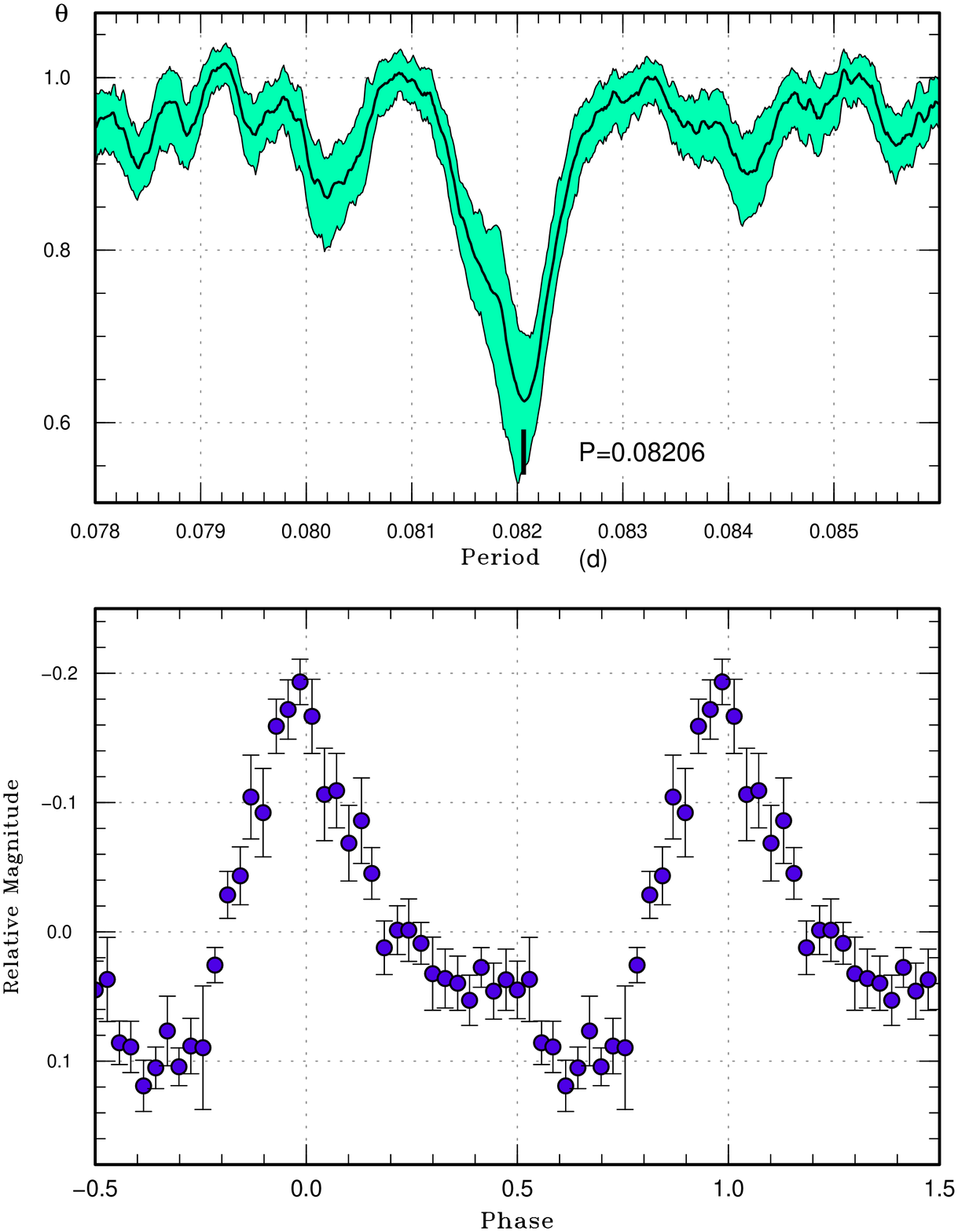}
  \end{center}
  \caption{Superhumps in ASASSN-15dq during the superoutburst
     plateau (2015).
     (Upper): PDM analysis.
     (Lower): Phase-averaged profile}.
  \label{fig:asassn15dqshpdm}
\end{figure}

   This object was detected as a transient at $V$=14.5
on 2015 February 22 by ASAS-SN team.
The coordinates are \timeform{09h 46m 06.91s},
\timeform{-00D 56' 01.6''} (SDSS $g$=20.4 counterpart).
CRTS also detected the object four days later
(cf. vsnet-alert 18348).
The SDSS colors suggested an SU UMa-type dwarf nova
(vsnet-alert 18329).
Superhumps were immediately detected
(vsnet-alert 18333, 18352, 18362; figure \ref{fig:asassn15dqshpdm}).
The times of superhump maxima are listed in table
\ref{tab:asassn15dqoc2015}.  Although superhump stage
is now known, no strong period variation was detected
except $E \ge 61$.
The object rapidly faded on March 7, 13~d after the
outburst detection.

\begin{table}
\caption{Superhump maxima of ASASSN-15dq (2015)}\label{tab:asassn15dqoc2015}
\begin{center}
\begin{tabular}{rp{55pt}p{40pt}r@{.}lr}
\hline
\multicolumn{1}{c}{$E$} & \multicolumn{1}{c}{max\commenta} & \multicolumn{1}{c}{error} & \multicolumn{2}{c}{$O-C$\commentb} & \multicolumn{1}{c}{$N$\commentc} \\
\hline
0 & 57078.5776 & 0.0003 & $-$0&0031 & 76 \\
13 & 57079.6433 & 0.0009 & $-$0&0020 & 22 \\
25 & 57080.6286 & 0.0007 & 0&0006 & 18 \\
36 & 57081.5292 & 0.0005 & 0&0004 & 73 \\
37 & 57081.6157 & 0.0012 & 0&0050 & 31 \\
46 & 57082.3530 & 0.0008 & 0&0054 & 91 \\
47 & 57082.4331 & 0.0038 & 0&0036 & 35 \\
61 & 57083.5670 & 0.0040 & $-$0&0090 & 9 \\
62 & 57083.6569 & 0.0022 & $-$0&0009 & 19 \\
\hline
  \multicolumn{6}{l}{\commenta BJD$-$2400000.} \\
  \multicolumn{6}{l}{\commentb Against max $= 2457078.5807 + 0.081889 E$.} \\
  \multicolumn{6}{l}{\commentc Number of points used to determine the maximum.} \\
\end{tabular}
\end{center}
\end{table}

\subsection{CRTS J081936.1$+$191540}\label{obj:j0819}

\begin{figure}
  \begin{center}
    \FigureFile(88mm,110mm){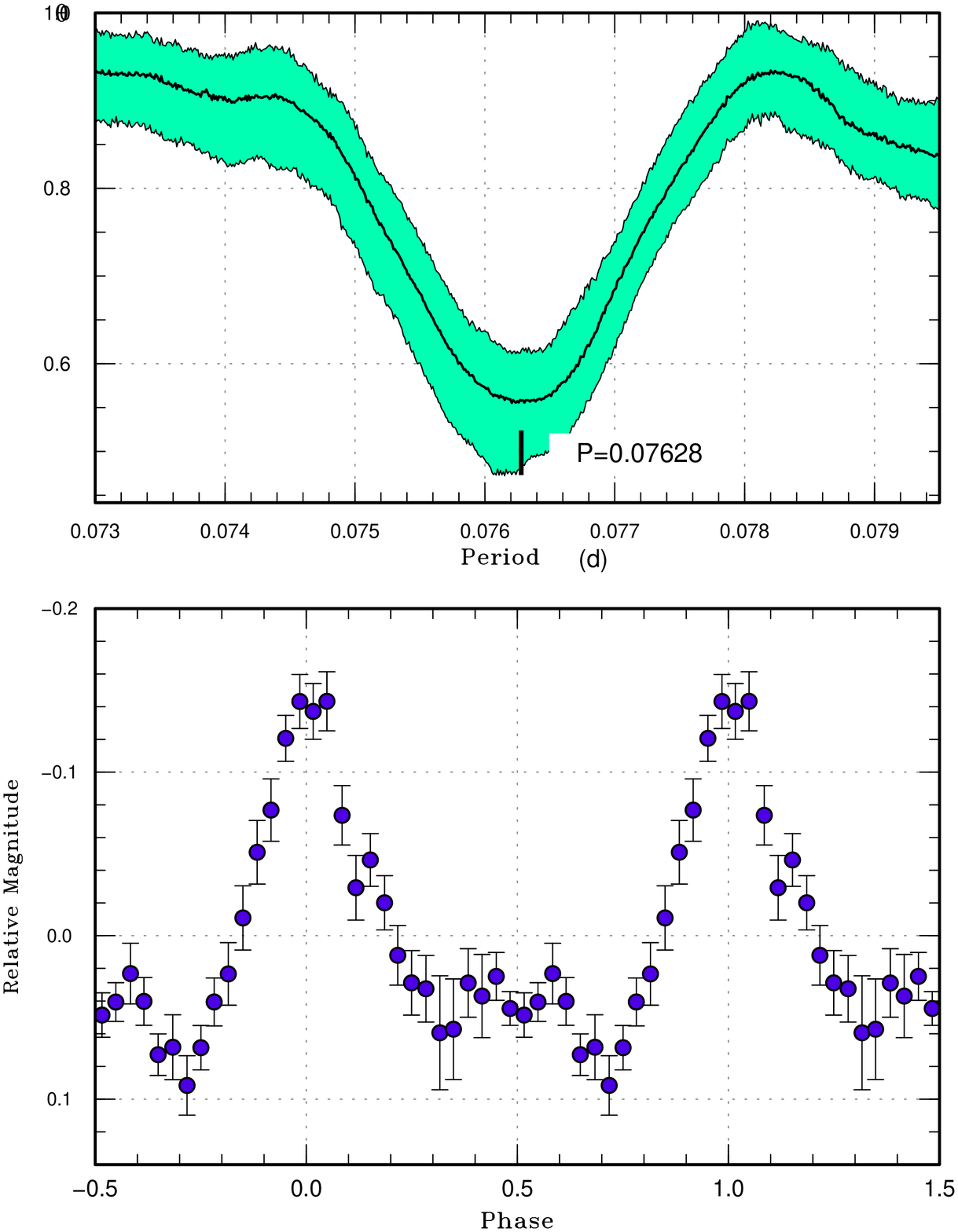}
  \end{center}
  \caption{Superhumps in CRTS J081936 during the superoutburst
     plateau (2014).  (Upper): PDM analysis.
     (Lower): Phase-averaged profile}.
  \label{fig:j0819shpdm}
\end{figure}

   This object (hereafter CRTS J081936) was originally
detected as a CV by CRTS on 2010 February 2
(=CSS100202:081936$+$191540).  Although it was a faint
(17.35 mag), there was another brighter outburst (16.33 mag)
on 2011 February 4.
The object was given a CRTS designation in \citet{dra14CRTSCVs}.
The 2013 bright (15.02 mag) outburst was
by MASTER network (vsnet-alert 15562).  Subsequent observations
detected superhumps (vsnet-alert 15565, 15574; figure
\ref{fig:j0819shpdm}).  The times of superhump maxima are listed
in table \ref{tab:j0819oc2013}.  The superhump stage is
unknown.

\begin{table}
\caption{Superhump maxima of CRTS J081936 (2013)}\label{tab:j0819oc2013}
\begin{center}
\begin{tabular}{rp{55pt}p{40pt}r@{.}lr}
\hline
\multicolumn{1}{c}{$E$} & \multicolumn{1}{c}{max\commenta} & \multicolumn{1}{c}{error} & \multicolumn{2}{c}{$O-C$\commentb} & \multicolumn{1}{c}{$N$\commentc} \\
\hline
0 & 56384.4396 & 0.0005 & 0&0008 & 89 \\
1 & 56384.5141 & 0.0014 & $-$0&0009 & 81 \\
13 & 56385.4304 & 0.0007 & $-$0&0001 & 90 \\
14 & 56385.5069 & 0.0012 & 0&0001 & 87 \\
\hline
  \multicolumn{6}{l}{\commenta BJD$-$2400000.} \\
  \multicolumn{6}{l}{\commentb Against max $= 2456384.4387 + 0.076285 E$.} \\
  \multicolumn{6}{l}{\commentc Number of points used to determine the maximum.} \\
\end{tabular}
\end{center}
\end{table}

\subsection{CRTS J172038.7$+$183802}\label{obj:j1720}

\begin{figure}
  \begin{center}
    \FigureFile(88mm,70mm){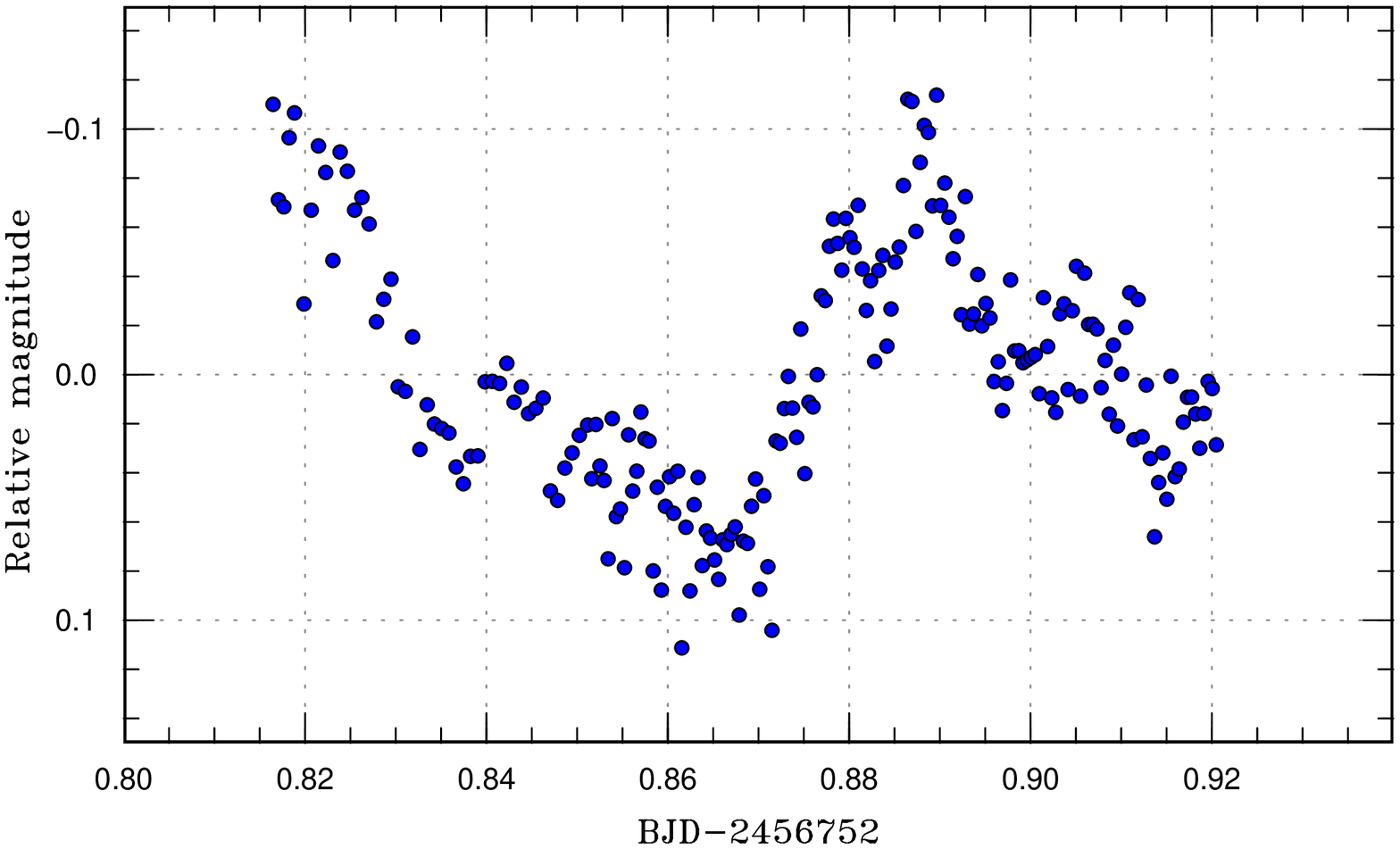}
  \end{center}
  \caption{Superhumps in CRTS J172038 (2014).}
  \label{fig:j1720shlc}
\end{figure}

   This object (hereafter CRTS J172038) was originally
detected as a CV by CRTS on 2009 September 29
(=CSS090929:172039$+$183802).
The object was given a CRTS designation in \citet{dra14CRTSCVs}.
MASTER network detected a new outburst on 2014 April 3
(vsnet-alert 17148).  Subsequent observations detected
superhumps (vsnet-alert 17154).  Only single-night observation
was obtained (figure \ref{fig:j1720shlc}) and only
an approximate superhump period of 0.06~d was obtained.

\subsection{CSS J174033.5$+$414756}\label{obj:j1740}

\subsubsection{Introduction}

   This object (=CSS130418:174033$+$414756, hereafter CSS J174033)
was discovered by the CRTS team in 2013 April
(cf. vsnet-alert 15629).  The object was also detected
by the ASAS-SN team (ASASSN-13ad: \cite{pri13j1740asassn13adatel4999};
\cite{nes13j1740ibvs6059}).  The 2013 outburst
showed a sequence of early superhumps and ordinary
superhumps.  The object was identified as a WZ Sge-type
object having a period below the period minimum.
The details of the 2013 outburst is reported in
\Ohtprep.

\subsubsection{2014 Outburst}

   The 2014 outburst was detected by C. Chiselbrook
on August 6 (vsnet-alert 17610).  Since there was
a 6~d gap in observation before this detection
and since the brightness was fainter by 0.7 mag than
the peak brightness of the 2013 outburst,
it was likely the peak of the 2013 outburst was missed
for several days.  It was a surprise that an object
showing a WZ Sge-type outburst repeated a superoutburst
less than 500~d after the previous outburst
(vsnet-alert 17610).

   As in the 2013 outburst, there was a deep dip
following the plateau (BJD 2456886--2456889;
figure \ref{fig:j1740humpall}.  After this dip,
the object showed a plateau-type rebrightening
as in AL Com in 1995
(\cite{pat96alcom}; \cite{how96alcom}; \cite{nog97alcom})
and 2001 \citep{ish02wzsgeletter}.
Such a dip was recorded in a WZ Sge-type dwarf nova
IK Leo (=OT J102146.4+234926; \cite{uem08j1021},
\cite{gol07j1021}, \cite{Pdot}), which was regarded
as a temporary dip in a plateau rather than
a dip preceding a rebrightening.
After this dip, the plateau continued for 15~d.
There was a short dip (BJD 2456902) before the termination
of the entire plateau.  Such a dip was recorded
in KK Cnc (=OT J080714.2+113812; \cite{Pdot}).

\subsubsection{Superhumps}

   The times of superhumps before the initial dip
are listed in table \ref{tab:j1740oc2014}.
As in the 2013 superoutburst, there were stage B superhumps
($E \le 166$) with a positive $P_{\rm dot}$
of $+2.0(0.3) \times 10^{-5}$.  The value was similar
to the 2013 one [$+1.6(0.1) \times 10^{-5}$].
The phases of early superhumps and stage A superhumps
were not covered by the present observation since
the detection was apparently several days after
the peak brightness.

   Before entering the dip, the superhump signal decreased
and another period appeared ($E \le 176$).  The new period
is 0.04505(6)~d, which is identical to the orbital period
within errors.  We consider that the orbital signal
was somehow enhanced during the decline branch to
the dip minimum.

\begin{figure}
  \begin{center}
    \FigureFile(88mm,100mm){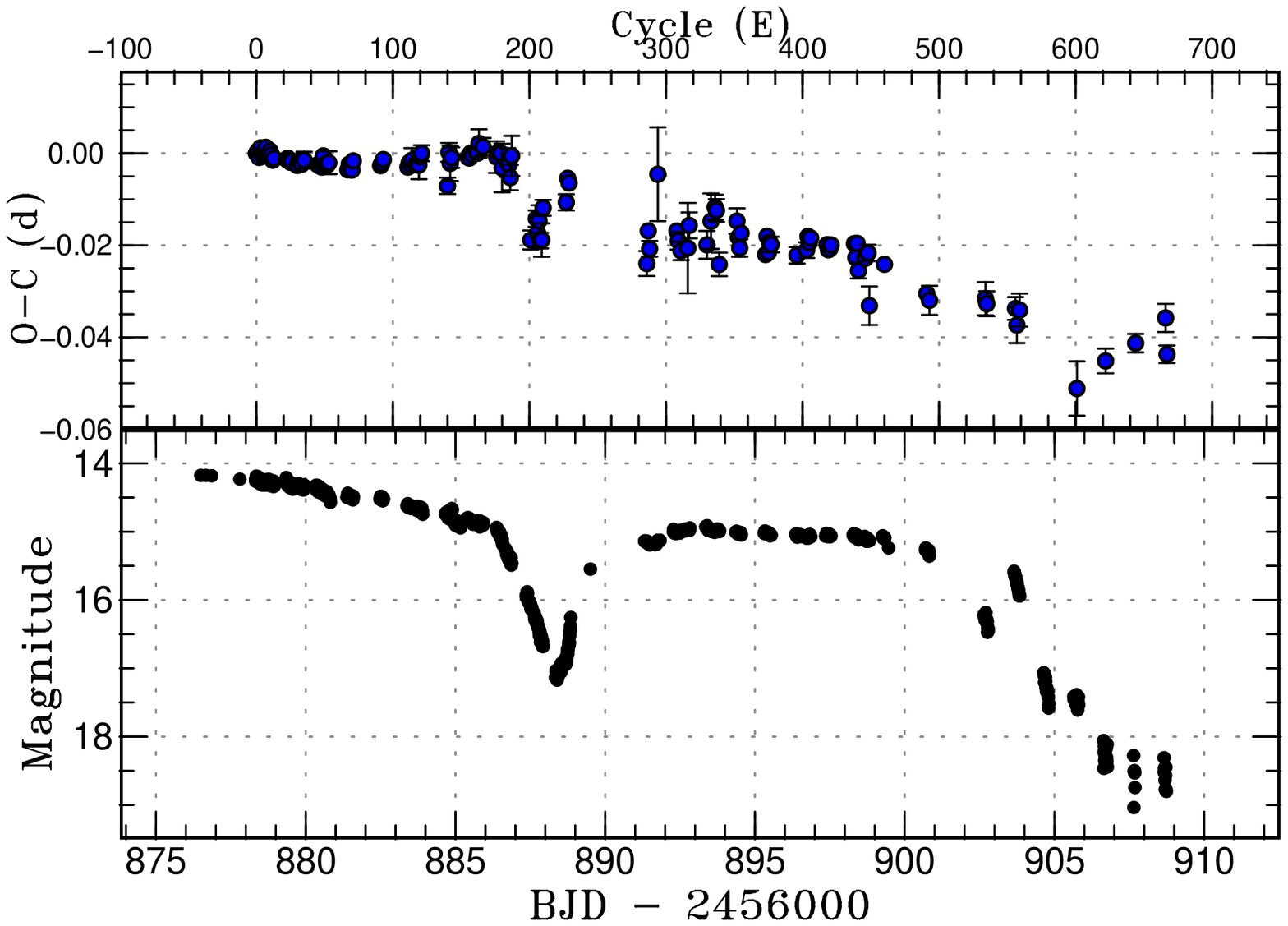}
  \end{center}
  \caption{$O-C$ diagram of superhumps in CSS J174033 (2014).
     (Upper:) $O-C$ diagram.
     We used a period of 0.045592~d for calculating the $O-C$ residuals.
     (Middle:) Amplitudes of superhumps.
     (Lower:) Light curve.  The data were binned to 0.009~d.
  }
  \label{fig:j1740humpall}
\end{figure}

   The times of superhump after tis dip are listed
in table \ref{tab:j1740oc2014b}.  After the dip,
the period was almost constant and slightly shorter
than the period of stage B superhumps.
The situation is clear in figure \ref{fig:j1740humpall}
that superhumps after the dip are on a smooth continuation
of superhumps before the dip.  Since the period was
almost constant and shorter, we identified this period
as that of stage C superhumps.  The cycle counts
listed in table \ref{tab:perlist} refer to table
\ref{tab:j1740oc2014b}.  Although these superhumps
were phenomenologically identified as stage C superhumps,
the variation of the period from stage B to C was
smaller (0.14\%) compared to $\sim$0.5\% in ordinary
hydrogen-rich SU UMa-type dwarf novae \citep{Pdot}.
This phenomenon may be different from stage C superhumps
in ordinary hydrogen-rich SU UMa-type dwarf novae.
Although the entire light curve is similar to
long-lasting rebrightenings in WZ Sge-type dwarf novae
(such as AL Com, figure 13 in \cite{Pdot6}),
the $O-C$ behavior is different and resembles those
of SU UMa-type dwarf novae with stage B-C transition.
This difference needs to be further examined.

\subsubsection{Comparison of superoutbursts}

\begin{figure}
  \begin{center}
    \FigureFile(88mm,70mm){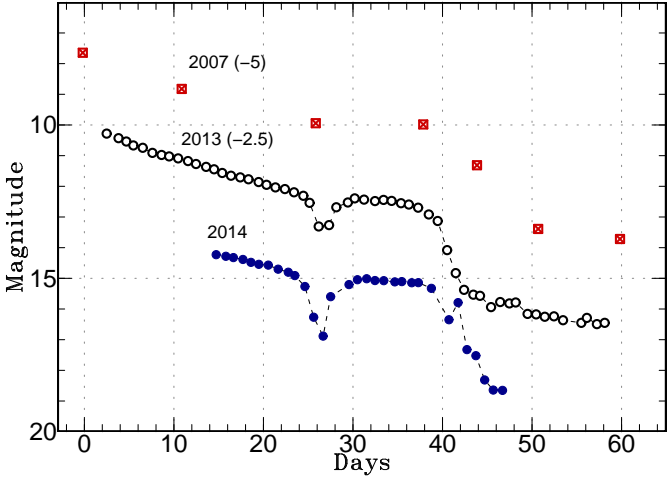}
  \end{center}
  \caption{Comparison of superoutbursts of CSS J174033.
  The data were binned to 1~d and shifted in magnitude.
  The dashed lines are added to aid recognizing the variation.
  The data for the 2013 superoutburst were from \Ohtprep
  The data for the 2007 superoutburst were
  from the CRTS public data.}
  \label{fig:j1740lccomp}
\end{figure}

   In figure \ref{fig:j1740lccomp}, three known (super)outbursts
are compared.  The zero point in abscissa refers to the
CRTS detection for the 2013 outburst.
The object was at 14.0 mag in this detection, and the outburst
was most likely detected during its rising phase.
We therefore consider that zero point for the 2013 outburst
refers to the start of the outburst within an error of 1~d.
The 2007 outburst was detected at a much brighter magnitude
(12.7).  We consider that the 2007 outburst was detected
near its peak brightness since the light curves well agree
with the 2013 one following this assumption.
The situation for the 2014 outburst is less clear.
Although there was a negative observation 7~d before
the initial detection by the same observer
(there were four negative observations starting from
11~d before the initial detection).  Since all these
observations were visual observations and the detection
limit was close to the magnitude of the initial detection,
we consider that there was a chance that the early
part of the outburst was not detected by visual 
observations.  If we consider that the actual outburst
started 14~d before the initial detection, the magnitude,
the epoch of the dip and the final fading agree with
the 2013 ones.  We adopted this assumption in figure
\ref{fig:j1740lccomp}.  If this assumption is correct,
the outburst light curves in CSS J174033 are highly
reproducible.

\begin{table}
\caption{Superhump maxima of CSS J174033 (2014)}\label{tab:j1740oc2014}
\begin{center}
\begin{tabular}{rp{55pt}p{40pt}r@{.}lr}
\hline
\multicolumn{1}{c}{$E$} & \multicolumn{1}{c}{max\commenta} & \multicolumn{1}{c}{error} & \multicolumn{2}{c}{$O-C$\commentb} & \multicolumn{1}{c}{$N$\commentc} \\
\hline
0 & 56878.3516 & 0.0002 & $-$0&0003 & 93 \\
1 & 56878.3977 & 0.0004 & 0&0003 & 175 \\
2 & 56878.4419 & 0.0004 & $-$0&0011 & 180 \\
3 & 56878.4896 & 0.0006 & 0&0010 & 165 \\
4 & 56878.5339 & 0.0009 & $-$0&0002 & 52 \\
5 & 56878.5790 & 0.0007 & $-$0&0007 & 71 \\
6 & 56878.6249 & 0.0005 & $-$0&0003 & 93 \\
7 & 56878.6720 & 0.0006 & 0&0013 & 99 \\
8 & 56878.7159 & 0.0005 & $-$0&0004 & 216 \\
9 & 56878.7612 & 0.0004 & $-$0&0007 & 185 \\
10 & 56878.8081 & 0.0006 & 0&0006 & 150 \\
11 & 56878.8528 & 0.0005 & $-$0&0002 & 152 \\
12 & 56878.8972 & 0.0009 & $-$0&0014 & 125 \\
13 & 56878.9432 & 0.0010 & $-$0&0010 & 64 \\
22 & 56879.3533 & 0.0004 & $-$0&0009 & 24 \\
23 & 56879.3991 & 0.0014 & $-$0&0006 & 73 \\
24 & 56879.4444 & 0.0008 & $-$0&0009 & 48 \\
25 & 56879.4895 & 0.0007 & $-$0&0014 & 49 \\
26 & 56879.5352 & 0.0015 & $-$0&0012 & 48 \\
30 & 56879.7167 & 0.0006 & $-$0&0019 & 145 \\
31 & 56879.7633 & 0.0005 & $-$0&0009 & 145 \\
32 & 56879.8087 & 0.0006 & $-$0&0010 & 136 \\
33 & 56879.8537 & 0.0011 & $-$0&0016 & 119 \\
35 & 56879.9458 & 0.0018 & $-$0&0006 & 32 \\
45 & 56880.4008 & 0.0014 & $-$0&0013 & 96 \\
46 & 56880.4463 & 0.0009 & $-$0&0014 & 95 \\
47 & 56880.4916 & 0.0009 & $-$0&0016 & 58 \\
48 & 56880.5370 & 0.0006 & $-$0&0017 & 59 \\
49 & 56880.5851 & 0.0008 & 0&0007 & 104 \\
50 & 56880.6287 & 0.0007 & $-$0&0012 & 80 \\
51 & 56880.6745 & 0.0009 & $-$0&0010 & 172 \\
52 & 56880.7199 & 0.0007 & $-$0&0011 & 185 \\
\hline
  \multicolumn{6}{l}{\commenta BJD$-$2400000.} \\
  \multicolumn{6}{l}{\commentb Against max $= 2456878.3519 + 0.045560 E$.} \\
  \multicolumn{6}{l}{\commentc Number of points used to determine the maximum.} \\
\end{tabular}
\end{center}
\end{table}

\addtocounter{table}{-1}
\begin{table}
\caption{Superhump maxima of CSS J174033 (2014) (continued)}
\begin{center}
\begin{tabular}{rp{55pt}p{40pt}r@{.}lr}
\hline
\multicolumn{1}{c}{$E$} & \multicolumn{1}{c}{max\commenta} & \multicolumn{1}{c}{error} & \multicolumn{2}{c}{$O-C$\commentb} & \multicolumn{1}{c}{$N$\commentc} \\
\hline
53 & 56880.7659 & 0.0025 & $-$0&0006 & 69 \\
67 & 56881.4027 & 0.0005 & $-$0&0017 & 48 \\
68 & 56881.4495 & 0.0004 & $-$0&0005 & 48 \\
69 & 56881.4943 & 0.0006 & $-$0&0012 & 49 \\
70 & 56881.5394 & 0.0005 & $-$0&0017 & 48 \\
71 & 56881.5870 & 0.0013 & 0&0003 & 27 \\
91 & 56882.4978 & 0.0007 & $-$0&0000 & 88 \\
92 & 56882.5440 & 0.0006 & 0&0006 & 87 \\
93 & 56882.5903 & 0.0007 & 0&0014 & 67 \\
111 & 56883.4093 & 0.0005 & 0&0003 & 47 \\
112 & 56883.4559 & 0.0007 & 0&0013 & 47 \\
113 & 56883.5012 & 0.0006 & 0&0010 & 46 \\
114 & 56883.5476 & 0.0026 & 0&0019 & 17 \\
117 & 56883.6836 & 0.0011 & 0&0011 & 31 \\
118 & 56883.7292 & 0.0013 & 0&0013 & 40 \\
119 & 56883.7745 & 0.0031 & 0&0010 & 41 \\
120 & 56883.8219 & 0.0016 & 0&0028 & 40 \\
121 & 56883.8683 & 0.0017 & 0&0036 & 40 \\
140 & 56884.7274 & 0.0018 & $-$0&0029 & 32 \\
141 & 56884.7803 & 0.0020 & 0&0044 & 31 \\
142 & 56884.8235 & 0.0038 & 0&0021 & 10 \\
143 & 56884.8703 & 0.0022 & 0&0033 & 10 \\
155 & 56885.4175 & 0.0008 & 0&0038 & 49 \\
156 & 56885.4629 & 0.0006 & 0&0037 & 49 \\
157 & 56885.5096 & 0.0007 & 0&0048 & 46 \\
158 & 56885.5548 & 0.0008 & 0&0044 & 48 \\
161 & 56885.6922 & 0.0012 & 0&0051 & 78 \\
162 & 56885.7375 & 0.0013 & 0&0049 & 86 \\
163 & 56885.7853 & 0.0031 & 0&0071 & 57 \\
166 & 56885.9213 & 0.0019 & 0&0064 & 53 \\
176 & 56886.3750 & 0.0034 & 0&0045 & 23 \\
177 & 56886.4216 & 0.0008 & 0&0055 & 44 \\
\hline
  \multicolumn{6}{l}{\commenta BJD$-$2400000.} \\
  \multicolumn{6}{l}{\commentb Against max $= 2456878.3519 + 0.045560 E$.} \\
  \multicolumn{6}{l}{\commentc Number of points used to determine the maximum.} \\
\end{tabular}
\end{center}
\end{table}

\addtocounter{table}{-1}
\begin{table}
\caption{Superhump maxima of CSS J174033 (2014) (continued)}
\begin{center}
\begin{tabular}{rp{55pt}p{40pt}r@{.}lr}
\hline
\multicolumn{1}{c}{$E$} & \multicolumn{1}{c}{max\commenta} & \multicolumn{1}{c}{error} & \multicolumn{2}{c}{$O-C$\commentb} & \multicolumn{1}{c}{$N$\commentc} \\
\hline
178 & 56886.4671 & 0.0011 & 0&0055 & 49 \\
179 & 56886.5126 & 0.0011 & 0&0055 & 48 \\
180 & 56886.5550 & 0.0053 & 0&0023 & 48 \\
184 & 56886.7385 & 0.0031 & 0&0036 & 67 \\
185 & 56886.7836 & 0.0023 & 0&0031 & 66 \\
186 & 56886.8264 & 0.0027 & 0&0004 & 46 \\
187 & 56886.8767 & 0.0044 & 0&0051 & 28 \\
201 & 56887.4968 & 0.0021 & $-$0&0127 & 114 \\
205 & 56887.6838 & 0.0018 & $-$0&0080 & 63 \\
206 & 56887.7262 & 0.0011 & $-$0&0111 & 79 \\
207 & 56887.7745 & 0.0034 & $-$0&0084 & 75 \\
208 & 56887.8158 & 0.0017 & $-$0&0126 & 72 \\
209 & 56887.8615 & 0.0036 & $-$0&0125 & 53 \\
210 & 56887.9141 & 0.0017 & $-$0&0055 & 35 \\
\hline
  \multicolumn{6}{l}{\commenta BJD$-$2400000.} \\
  \multicolumn{6}{l}{\commentb Against max $= 2456878.3519 + 0.045560 E$.} \\
  \multicolumn{6}{l}{\commentc Number of points used to determine the maximum.} \\
\end{tabular}
\end{center}
\end{table}

\begin{table}
\caption{Superhump maxima of CSS J174033 (2014) (after dip)}\label{tab:j1740oc2014b}
\begin{center}
\begin{tabular}{rp{55pt}p{40pt}r@{.}lr}
\hline
\multicolumn{1}{c}{$E$} & \multicolumn{1}{c}{max\commenta} & \multicolumn{1}{c}{error} & \multicolumn{2}{c}{$O-C$\commentb} & \multicolumn{1}{c}{$N$\commentc} \\
\hline
0 & 56888.6903 & 0.0017 & $-$0&0020 & 73 \\
1 & 56888.7412 & 0.0016 & 0&0033 & 94 \\
2 & 56888.7857 & 0.0012 & 0&0023 & 81 \\
59 & 56891.3669 & 0.0027 & $-$0&0108 & 25 \\
60 & 56891.4196 & 0.0010 & $-$0&0036 & 24 \\
61 & 56891.4613 & 0.0017 & $-$0&0074 & 24 \\
67 & 56891.7511 & 0.0102 & 0&0093 & 21 \\
81 & 56892.3770 & 0.0013 & $-$0&0020 & 68 \\
82 & 56892.4205 & 0.0019 & $-$0&0041 & 76 \\
84 & 56892.5095 & 0.0020 & $-$0&0061 & 71 \\
89 & 56892.7381 & 0.0098 & $-$0&0051 & 21 \\
90 & 56892.7886 & 0.0028 & $-$0&0001 & 21 \\
103 & 56893.3771 & 0.0030 & $-$0&0033 & 59 \\
106 & 56893.5190 & 0.0060 & 0&0021 & 46 \\
109 & 56893.6588 & 0.0027 & 0&0054 & 12 \\
110 & 56893.7037 & 0.0025 & 0&0047 & 22 \\
112 & 56893.7831 & 0.0026 & $-$0&0068 & 21 \\
125 & 56894.3853 & 0.0028 & 0&0036 & 48 \\
126 & 56894.4273 & 0.0011 & 0&0002 & 72 \\
127 & 56894.4706 & 0.0019 & $-$0&0021 & 70 \\
128 & 56894.5194 & 0.0013 & 0&0012 & 57 \\
146 & 56895.3354 & 0.0007 & $-$0&0020 & 23 \\
147 & 56895.3850 & 0.0016 & 0&0020 & 62 \\
148 & 56895.4271 & 0.0013 & $-$0&0014 & 70 \\
149 & 56895.4748 & 0.0016 & 0&0008 & 62 \\
150 & 56895.5199 & 0.0017 & 0&0004 & 43 \\
169 & 56896.3838 & 0.0018 & $-$0&0004 & 45 \\
176 & 56896.7041 & 0.0017 & 0&0013 & 32 \\
177 & 56896.7527 & 0.0013 & 0&0043 & 45 \\
178 & 56896.7969 & 0.0010 & 0&0030 & 42 \\
179 & 56896.8435 & 0.0014 & 0&0041 & 46 \\
\hline
  \multicolumn{6}{l}{\commenta BJD$-$2400000.} \\
  \multicolumn{6}{l}{\commentb Against max $= 2456888.6924 + 0.045514 E$.} \\
  \multicolumn{6}{l}{\commentc Number of points used to determine the maximum.} \\
\end{tabular}
\end{center}
\end{table}

\addtocounter{table}{-1}
\begin{table}
\caption{Superhump maxima of CSS J174033 (2014) (after dip, continued)}
\begin{center}
\begin{tabular}{rp{55pt}p{40pt}r@{.}lr}
\hline
\multicolumn{1}{c}{$E$} & \multicolumn{1}{c}{max\commenta} & \multicolumn{1}{c}{error} & \multicolumn{2}{c}{$O-C$\commentb} & \multicolumn{1}{c}{$N$\commentc} \\
\hline
191 & 56897.3892 & 0.0013 & 0&0036 & 44 \\
192 & 56897.4337 & 0.0010 & 0&0026 & 29 \\
193 & 56897.4795 & 0.0008 & 0&0029 & 48 \\
194 & 56897.5259 & 0.0014 & 0&0038 & 43 \\
211 & 56898.3013 & 0.0006 & 0&0054 & 25 \\
212 & 56898.3438 & 0.0016 & 0&0024 & 32 \\
213 & 56898.3925 & 0.0010 & 0&0056 & 63 \\
214 & 56898.4322 & 0.0017 & $-$0&0002 & 62 \\
219 & 56898.6628 & 0.0012 & 0&0028 & 19 \\
221 & 56898.7552 & 0.0018 & 0&0041 & 66 \\
222 & 56898.7893 & 0.0042 & $-$0&0073 & 53 \\
233 & 56899.2998 & 0.0016 & 0&0026 & 25 \\
264 & 56900.7068 & 0.0010 & $-$0&0014 & 18 \\
266 & 56900.7965 & 0.0032 & $-$0&0027 & 22 \\
307 & 56902.6661 & 0.0036 & 0&0009 & 20 \\
308 & 56902.7106 & 0.0027 & $-$0&0001 & 22 \\
329 & 56903.6670 & 0.0025 & 0&0005 & 62 \\
330 & 56903.7090 & 0.0039 & $-$0&0031 & 68 \\
332 & 56903.8034 & 0.0036 & 0&0003 & 60 \\
374 & 56905.7012 & 0.0059 & $-$0&0135 & 45 \\
395 & 56906.6647 & 0.0027 & $-$0&0058 & 20 \\
417 & 56907.6716 & 0.0020 & $-$0&0002 & 20 \\
439 & 56908.6801 & 0.0031 & 0&0070 & 43 \\
440 & 56908.7178 & 0.0019 & $-$0&0009 & 40 \\
\hline
  \multicolumn{6}{l}{\commenta BJD$-$2400000.} \\
  \multicolumn{6}{l}{\commentb Against max $= 2456888.6924 + 0.045514 E$.} \\
  \multicolumn{6}{l}{\commentc Number of points used to determine the maximum.} \\
\end{tabular}
\end{center}
\end{table}

\subsection{CRTS J202731.2$-$224002}\label{obj:j2027}

\begin{figure}
  \begin{center}
    \FigureFile(88mm,110mm){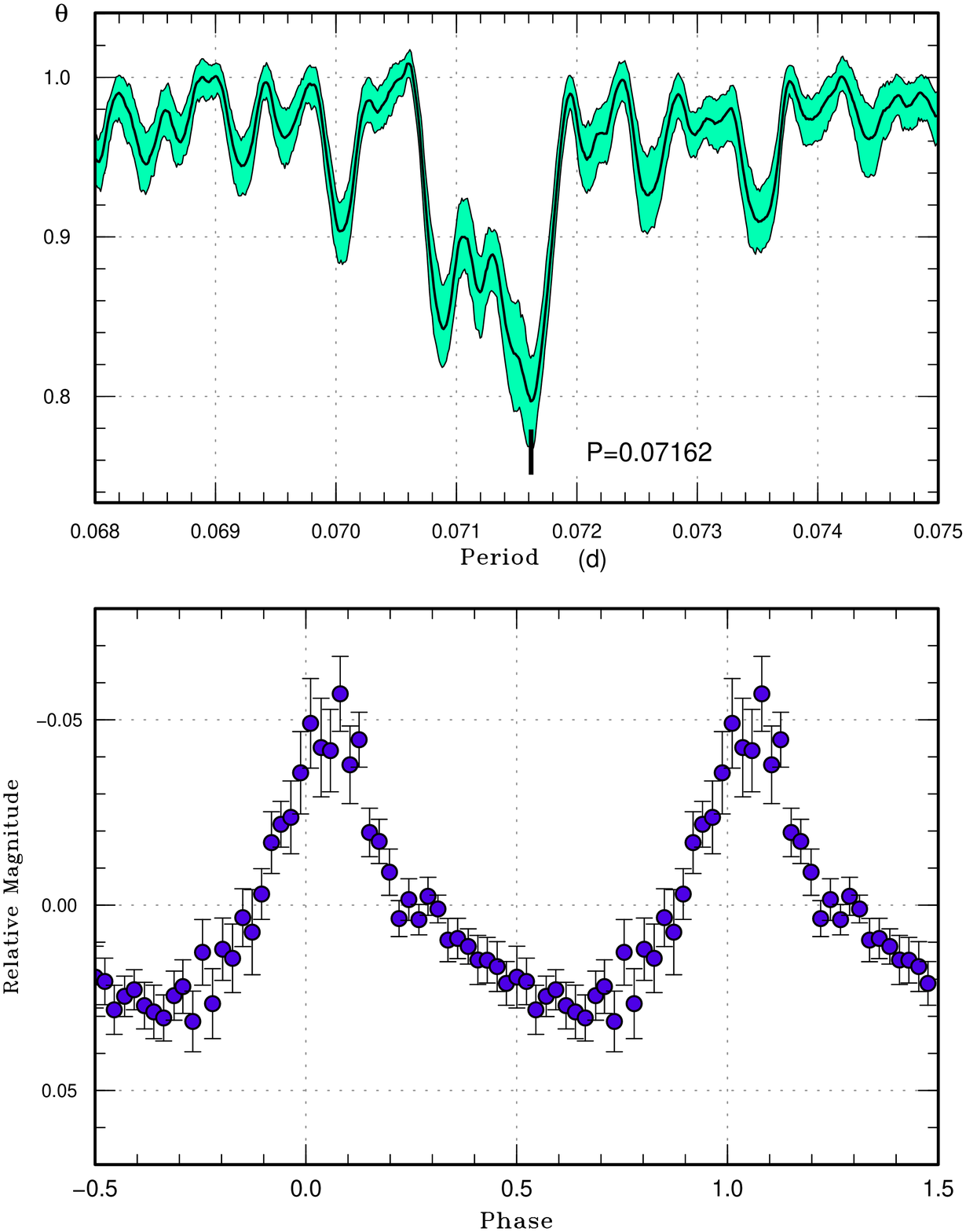}
  \end{center}
  \caption{Superhumps in CRTS J202731 during the superoutburst
     plateau (2014).  (Upper): PDM analysis.
     (Lower): Phase-averaged profile}.
  \label{fig:j2027shpdm}
\end{figure}

   This object (=SSS110515:202731$-$224002, hereafter
CRTS J202731) is a dwarf nova discovered by CRTS
in 2011 May \citep{bre14CRTSCVs}.
The 2014 outburst was detected by the ASAS-SN team
(vsnet-alert 17857).  Nine past outbursts were 
detected by ASAS-3 and the brightest outburst reached
$V$=12.2 (vsnet-alert 17858, 17859).
Subsequent observations detected superhumps
(vsnet-alert 17861, 17868, 17877; figure \ref{fig:j2027shpdm}).
The times of superhump maxima are listed in table
\ref{tab:j2027oc2014}.  The values for $E \le 191$
represent post-superoutburst observations.
Although stage B was not very clearly identified,
a sharp variation in the $O-C$ values signified
stage B-C transition around $E=78$.
The post-superoutburst data did not yield a significant
signal of superhumps except for the initial two nights.

   A list of recent outbursts is given in table \ref{tab:j2027out}.
The intervals of known superoutburst were 473~d, 415~d
and 619~d.  Given the relatively frequent normal outbursts,
the true supercycle appears to be 206--237~d.

\begin{table}
\caption{Superhump maxima of CRTS J202731 (2014)}\label{tab:j2027oc2014}
\begin{center}
\begin{tabular}{rp{55pt}p{40pt}r@{.}lr}
\hline
\multicolumn{1}{c}{$E$} & \multicolumn{1}{c}{max\commenta} & \multicolumn{1}{c}{error} & \multicolumn{2}{c}{$O-C$\commentb} & \multicolumn{1}{c}{$N$\commentc} \\
\hline
0 & 56948.9418 & 0.0002 & $-$0&0129 & 78 \\
1 & 56949.0140 & 0.0005 & $-$0&0121 & 53 \\
8 & 56949.5127 & 0.0044 & $-$0&0126 & 10 \\
9 & 56949.5858 & 0.0008 & $-$0&0108 & 19 \\
23 & 56950.5865 & 0.0006 & $-$0&0085 & 21 \\
36 & 56951.5151 & 0.0055 & $-$0&0071 & 13 \\
37 & 56951.5860 & 0.0034 & $-$0&0075 & 19 \\
42 & 56951.9434 & 0.0071 & $-$0&0067 & 11 \\
70 & 56953.9357 & 0.0012 & $-$0&0112 & 18 \\
78 & 56954.5287 & 0.0016 & 0&0113 & 19 \\
83 & 56954.8765 & 0.0019 & 0&0025 & 23 \\
84 & 56954.9597 & 0.0005 & 0&0144 & 78 \\
92 & 56955.5259 & 0.0013 & 0&0101 & 29 \\
93 & 56955.6000 & 0.0005 & 0&0128 & 47 \\
94 & 56955.6715 & 0.0011 & 0&0130 & 22 \\
106 & 56956.5225 & 0.0015 & 0&0082 & 29 \\
107 & 56956.5951 & 0.0006 & 0&0095 & 47 \\
108 & 56956.6672 & 0.0021 & 0&0102 & 25 \\
121 & 56957.5938 & 0.0006 & 0&0098 & 47 \\
122 & 56957.6638 & 0.0028 & 0&0085 & 26 \\
126 & 56957.9572 & 0.0010 & 0&0165 & 53 \\
127 & 56958.0132 & 0.0010 & 0&0013 & 41 \\
135 & 56958.5920 & 0.0009 & 0&0095 & 45 \\
136 & 56958.6630 & 0.0028 & 0&0092 & 26 \\
139 & 56958.8925 & 0.0021 & 0&0247 & 40 \\
140 & 56958.9507 & 0.0008 & 0&0116 & 52 \\
149 & 56959.5872 & 0.0009 & 0&0063 & 48 \\
150 & 56959.6540 & 0.0011 & 0&0018 & 24 \\
163 & 56960.5779 & 0.0010 & $-$0&0014 & 49 \\
164 & 56960.6500 & 0.0013 & $-$0&0006 & 29 \\
191 & 56962.5409 & 0.0029 & $-$0&0353 & 45 \\
192 & 56962.6166 & 0.0038 & $-$0&0309 & 38 \\
205 & 56963.5409 & 0.0020 & $-$0&0337 & 36 \\
\hline
  \multicolumn{6}{l}{\commenta BJD$-$2400000.} \\
  \multicolumn{6}{l}{\commentb Against max $= 2456948.9548 + 0.071316 E$.} \\
  \multicolumn{6}{l}{\commentc Number of points used to determine the maximum.} \\
\end{tabular}
\end{center}
\end{table}

\begin{table*}
\caption{List of recent outbursts of CRTS J202731}\label{tab:j2027out}
\begin{center}
\begin{tabular}{cccccl}
\hline
Year & Month & max\commenta & magnitude & type & source \\
\hline
2001 & 5 & 52053 & 13.0 & super & ASAS-3 \\
2001 & 9 & 52157 & 13.4 & normal?\commentb & ASAS-3 \\
2002 & 9 & 52526 & 12.2 & super & ASAS-3 \\
2003 & 5 & 52770 & 12.8 & normal?\commentb & ASAS-3 \\
2003 & 10 & 52941 & 12.7 & super & ASAS-3 \\
2005 & 7 & 53560 & 12.2 & super & ASAS-3 \\
2007 & 11 & 54409 & 13.0 & normal?\commentb & ASAS-3 \\
2008 & 8 & 54686 & 13.0 & normal & ASAS-3 \\
2009 & 5 & 54976 & 13.0 & normal & ASAS-3 \\
2011 & 5 & 55697 & 13.0 & normal? & CRTS \\
2014 & 10 & 56949 & 12.5 & super & this paper \\
\hline
  \multicolumn{5}{l}{\commenta JD$-$2400000.} \\
  \multicolumn{5}{l}{\commentb Single detection.} \\
\end{tabular}
\end{center}
\end{table*}

\subsection{CRTS J214738.4$+$244554}\label{obj:j2147}

   This object (=CSS111004:214738$+$244554, hereafter
CRTS J214738) is a dwarf nova discovered by CRTS
\citep{bre14CRTSCVs}.  The 2011 superoutburst was
relatively well observed and a positive $P_{\rm dot}$
for stage B superhumps was recorded in spite of
the long superhump period \citep{Pdot4}.

   The 2014 superoutburst was detected by the ASAS-SN team
at $V$=12.81 on September 11 (vsnet-alert 17728).
On September 14, low-amplitude superhumps were detected
despite that already $\sim$3~d had passed since
the outburst detection (vsnet-alert 17752).
The superhumps further evolved $\sim$4~d later
(vsnet-alert 17751, 17766).
The times of superhump maxima are listed in table
\ref{tab:j2147oc2014}.  We assigned negative $O-C$ for
$E \le 2$ assuming that they are stage A superhumps.
These counts may be in error by one cycle.

\begin{figure}
  \begin{center}
    \FigureFile(88mm,70mm){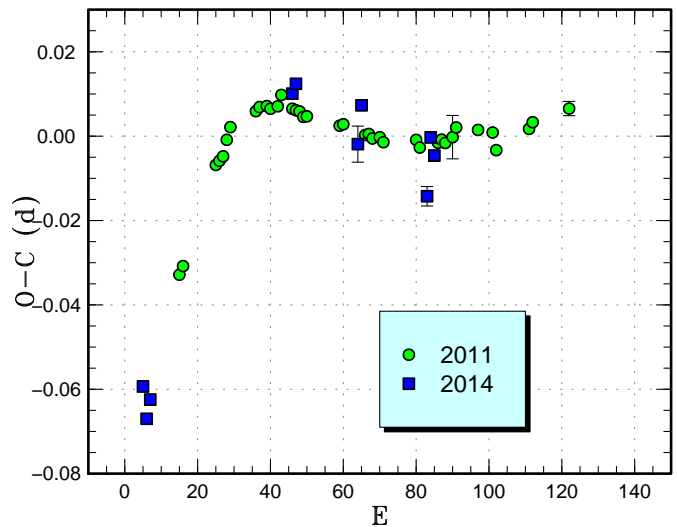}
  \end{center}
  \caption{Comparison of $O-C$ diagrams of CRTS J214738
  between different superoutbursts.
  A period of 0.09723~d was used to draw this figure.
  Approximate cycle counts ($E$) after the start of the superoutburst
  were used.  We have shifted the 2011 $O-C$ diagram by
  10 cycles to best match the 2014 one.  This shift
  suggests that the 2011 superoutburst started 1~d
  earlier than the CRTS detection.}
  \label{fig:j2147comp}
\end{figure}

   Although the observations were not sufficient,
there was a suggestion of slow evolution of superhumps
both in 2011 superoutburst (stage A lasting for more than
21 cycles) and the 2014 one.
This indication appears to have been confirmed
by a comparison of $O-C$ diagrams
(figure \ref{fig:j2147comp}).
Such a slow evolution with long stage A was hypothesized
for objects in the period gap as a consequence of
nearly critical condition of tidal instability \citep{Pdot6}.
CRTS J214738 may be another example supporting
this interpretation.

\begin{table}
\caption{Superhump maxima of CRTS J214738 (2014)}\label{tab:j2147oc2014}
\begin{center}
\begin{tabular}{rp{55pt}p{40pt}r@{.}lr}
\hline
\multicolumn{1}{c}{$E$} & \multicolumn{1}{c}{max\commenta} & \multicolumn{1}{c}{error} & \multicolumn{2}{c}{$O-C$\commentb} & \multicolumn{1}{c}{$N$\commentc} \\
\hline
0 & 56915.3175 & 0.0009 & $-$0&0071 & 187 \\
1 & 56915.4071 & 0.0013 & $-$0&0156 & 186 \\
2 & 56915.5089 & 0.0009 & $-$0&0118 & 125 \\
41 & 56919.3734 & 0.0003 & 0&0306 & 92 \\
42 & 56919.4730 & 0.0007 & 0&0321 & 37 \\
59 & 56921.1116 & 0.0043 & 0&0047 & 32 \\
60 & 56921.2180 & 0.0013 & 0&0131 & 79 \\
78 & 56922.9466 & 0.0023 & $-$0&0224 & 71 \\
79 & 56923.0577 & 0.0005 & $-$0&0092 & 200 \\
80 & 56923.1507 & 0.0006 & $-$0&0143 & 181 \\
\hline
  \multicolumn{6}{l}{\commenta BJD$-$2400000.} \\
  \multicolumn{6}{l}{\commentb Against max $= 2456915.3247 + 0.098004 E$.} \\
  \multicolumn{6}{l}{\commentc Number of points used to determine the maximum.} \\
\end{tabular}
\end{center}
\end{table}

\subsection{MASTER OT J031600.08$+$175824.4}\label{obj:j0316}

\begin{figure}
  \begin{center}
    \FigureFile(88mm,110mm){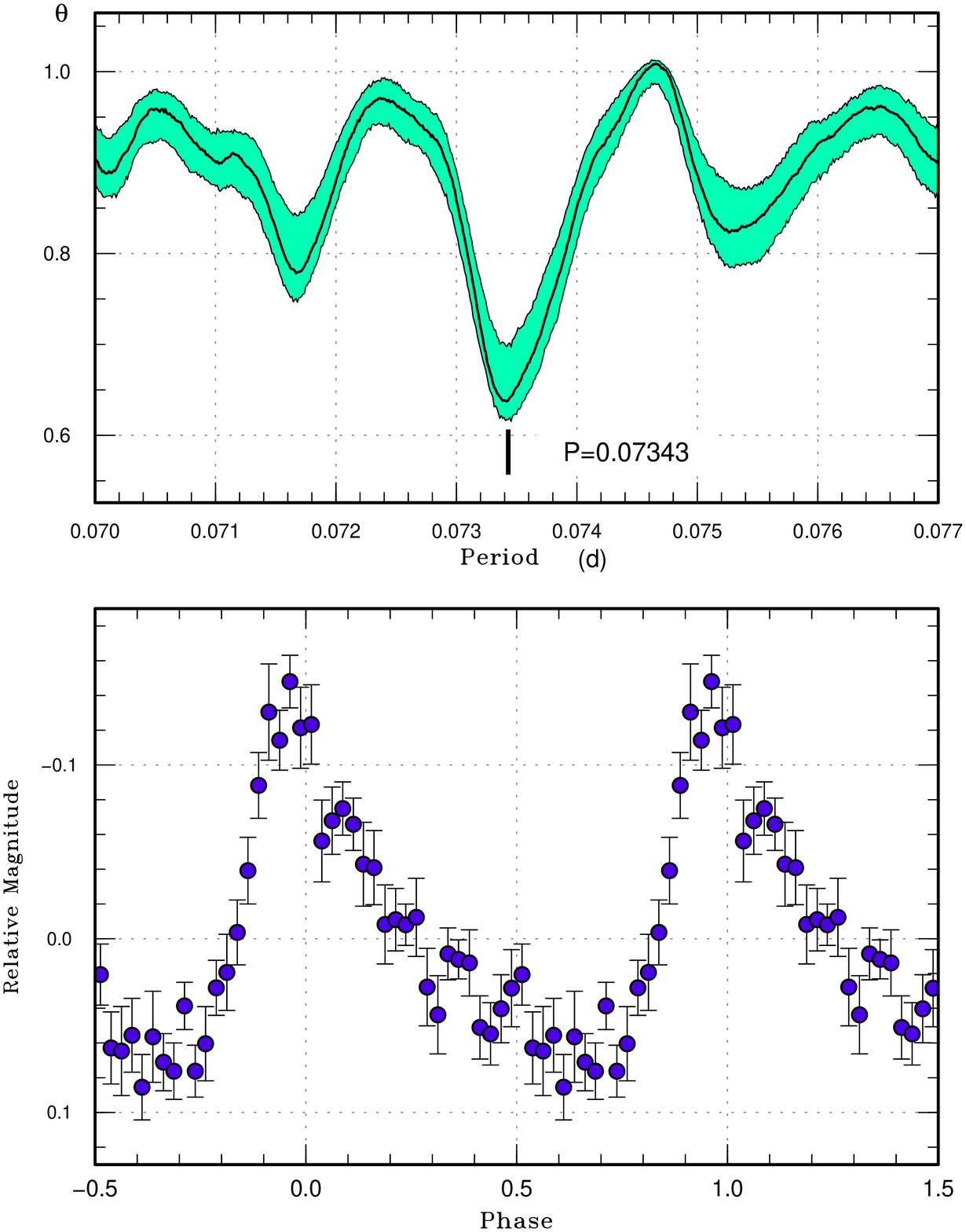}
  \end{center}
  \caption{Superhumps in MASTER J031600 (2014).  (Upper): PDM analysis.
     (Lower): Phase-averaged profile}.
  \label{fig:j0316shpdm}
\end{figure}

   This optical transient (hereafter MASTER J031600)
was detected on 2014 December 26 at a magnitude of 15.1
\citep{shu14j0316atel6851}.  The outburst was detected
in the early stage and subsequent development of
superhumps was recorded (vsnet-alert 18112, 18138).
On December 28, there were only small-amplitude variations.
On 2015 January 2, fully developed superhumps were
observed (figure \ref{fig:j0316shpdm}).
The times of superhump maxima are listed in table
\ref{tab:j0316oc2014}.  The $O-C$ data suggest that we
observed the early phase of stage B superhumps.

\begin{table}
\caption{Superhump maxima of MASTER J031600 (2014)}\label{tab:j0316oc2014}
\begin{center}
\begin{tabular}{rp{55pt}p{40pt}r@{.}lr}
\hline
\multicolumn{1}{c}{$E$} & \multicolumn{1}{c}{max\commenta} & \multicolumn{1}{c}{error} & \multicolumn{2}{c}{$O-C$\commentb} & \multicolumn{1}{c}{$N$\commentc} \\
\hline
0 & 57025.2070 & 0.0023 & $-$0&0034 & 36 \\
1 & 57025.2844 & 0.0008 & 0&0006 & 73 \\
2 & 57025.3578 & 0.0009 & 0&0005 & 49 \\
30 & 57027.4221 & 0.0032 & 0&0086 & 48 \\
31 & 57027.4852 & 0.0025 & $-$0&0018 & 75 \\
42 & 57028.2964 & 0.0014 & 0&0017 & 76 \\
43 & 57028.3730 & 0.0012 & 0&0048 & 67 \\
44 & 57028.4308 & 0.0024 & $-$0&0109 & 76 \\
\hline
  \multicolumn{6}{l}{\commenta BJD$-$2400000.} \\
  \multicolumn{6}{l}{\commentb Against max $= 2457025.2104 + 0.073438 E$.} \\
  \multicolumn{6}{l}{\commentc Number of points used to determine the maximum.} \\
\end{tabular}
\end{center}
\end{table}

\subsection{MASTER OT J043915.60$+$424232.3}\label{obj:j0439}

\begin{figure}
  \begin{center}
    \FigureFile(88mm,110mm){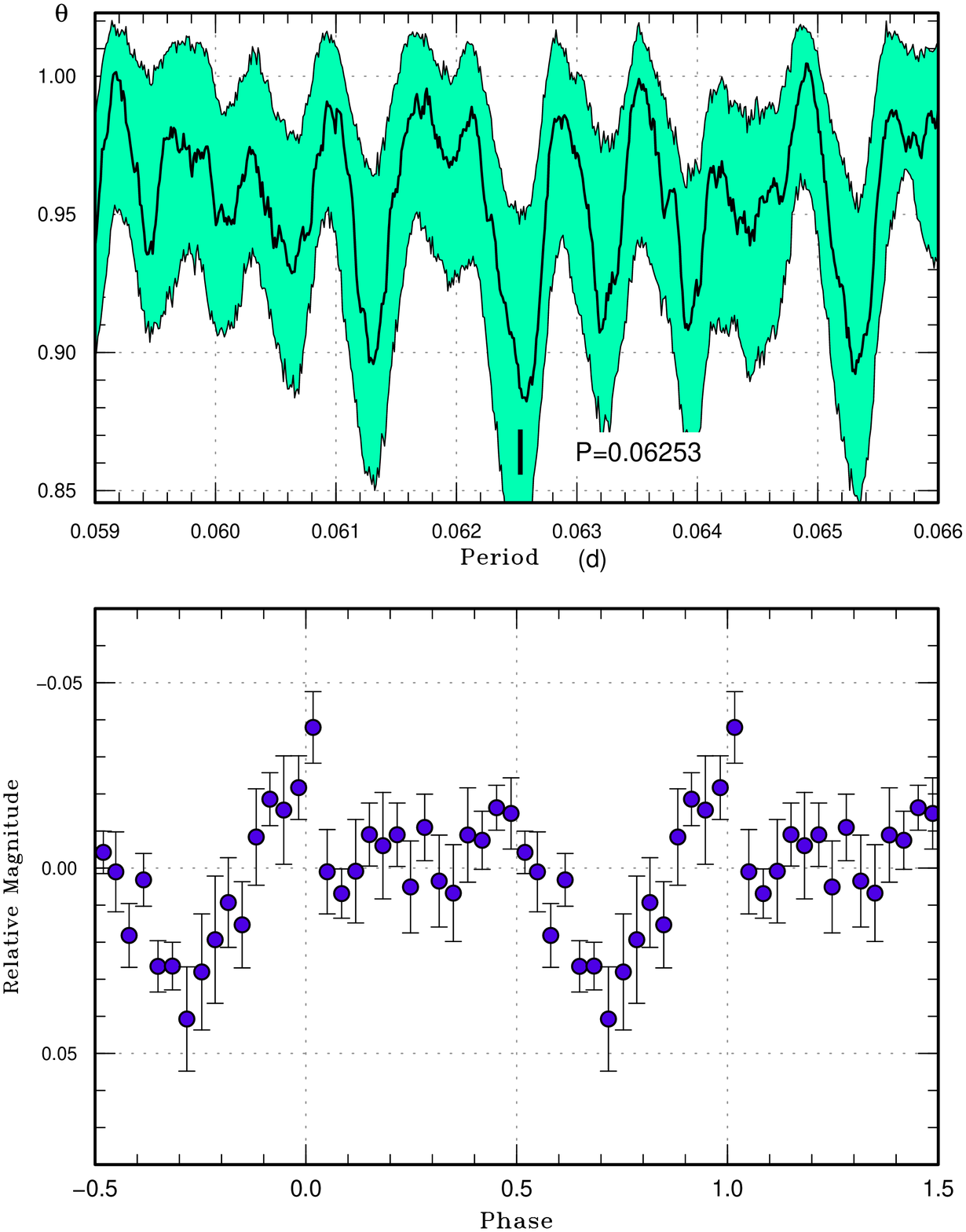}
  \end{center}
  \caption{Superhumps in MASTER J043915 (2014).  (Upper): PDM analysis.
     (Lower): Phase-averaged profile}.
  \label{fig:j0439shpdm}
\end{figure}

   This optical transient (hereafter MASTER J043915)
was detected on 2014 January 21 at a magnitude of 15.7
\citep{bal14j0439atel5787}.  The second outburst was
detected also by the D. Denisenko on 2014 December 26
at a magnitude of 15.7 (vsnet-alert 18106).
Subsequent observations detected superhumps
(vsnet-alert 18127, 18140; figure \ref{fig:j0439shpdm}).
The alias selection was based on the period [0.0625(4)~d]
derived from the continuous observation on the first night.
The times of superhump maxima are listed in table
\ref{tab:j0439oc2014}.  The superhump stage is unknown.
The magnitude on 2014 January 21 would suggest
a superoutburst.  If this is the case, the supercycle
is around 340~d.

\begin{table}
\caption{Superhump maxima of MASTER J043915 (2014)}\label{tab:j0439oc2014}
\begin{center}
\begin{tabular}{rp{55pt}p{40pt}r@{.}lr}
\hline
\multicolumn{1}{c}{$E$} & \multicolumn{1}{c}{max\commenta} & \multicolumn{1}{c}{error} & \multicolumn{2}{c}{$O-C$\commentb} & \multicolumn{1}{c}{$N$\commentc} \\
\hline
0 & 57019.0953 & 0.0018 & 0&0017 & 47 \\
1 & 57019.1564 & 0.0010 & 0&0003 & 46 \\
2 & 57019.2164 & 0.0017 & $-$0&0021 & 27 \\
48 & 57022.0914 & 0.0045 & 0&0001 & 45 \\
\hline
  \multicolumn{6}{l}{\commenta BJD$-$2400000.} \\
  \multicolumn{6}{l}{\commentb Against max $= 2457019.0936 + 0.062452 E$.} \\
  \multicolumn{6}{l}{\commentc Number of points used to determine the maximum.} \\
\end{tabular}
\end{center}
\end{table}

\subsection{MASTER OT J055845.55$+$391533.4}\label{obj:j0558}

\begin{figure}
  \begin{center}
    \FigureFile(88mm,70mm){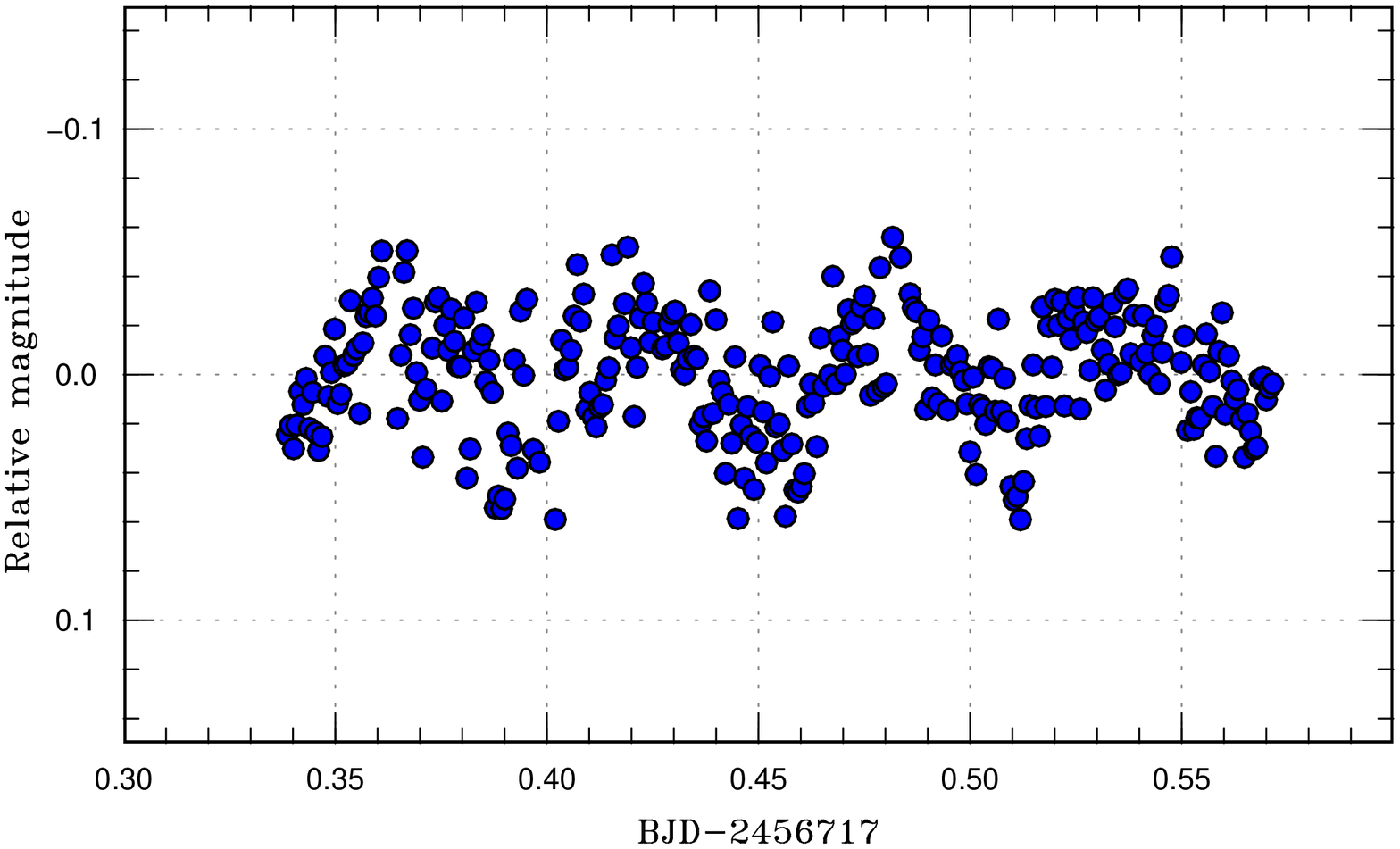}
  \end{center}
  \caption{Superhumps in MASTER J055845 (2014).}
  \label{fig:j0558shlc}
\end{figure}

   This optical transient (hereafter MASTER J055845)
was detected on 2014 February 19 at a magnitude of 14.4
\citep{yec14j0558atel5905}.  A retrospective examination
of the MASTER data indicated that the object was already
bright at 13.9 mag on February 13.  There were also two
outbursts (2011 November 28 and 2012 November 19) 
recorded in the MASTER data.  Single-night observations
on February 28 detected superhumps (vsnet-alert 16955;
figure \ref{fig:j0558shlc}).
The best superhump period using the PDM method was
0.0563(4) d.  The times of superhump maxima are
listed in table \ref{tab:j0558oc2014}.
Since the object had been in outburst
already at least for 15~d, this observation most likely
recorded stage C superhumps.  If the three outbursts
were all superoutbursts, the supercycle would be
an order of 360--450~d.

\begin{table}
\caption{Superhump maxima of MASTER J055845 (2014)}\label{tab:j0558oc2014}
\begin{center}
\begin{tabular}{rp{55pt}p{40pt}r@{.}lr}
\hline
\multicolumn{1}{c}{$E$} & \multicolumn{1}{c}{max\commenta} & \multicolumn{1}{c}{error} & \multicolumn{2}{c}{$O-C$\commentb} & \multicolumn{1}{c}{$N$\commentc} \\
\hline
0 & 56717.3638 & 0.0016 & $-$0&0005 & 55 \\
1 & 56717.4203 & 0.0019 & $-$0&0005 & 54 \\
2 & 56717.4796 & 0.0011 & 0&0024 & 54 \\
3 & 56717.5322 & 0.0017 & $-$0&0015 & 57 \\
\hline
  \multicolumn{6}{l}{\commenta BJD$-$2400000.} \\
  \multicolumn{6}{l}{\commentb Against max $= 2456717.3643 + 0.056443 E$.} \\
  \multicolumn{6}{l}{\commentc Number of points used to determine the maximum.} \\
\end{tabular}
\end{center}
\end{table}

\subsection{MASTER OT J085854.16$-$274030.7}\label{obj:j0858}

\begin{figure}
  \begin{center}
    \FigureFile(88mm,100mm){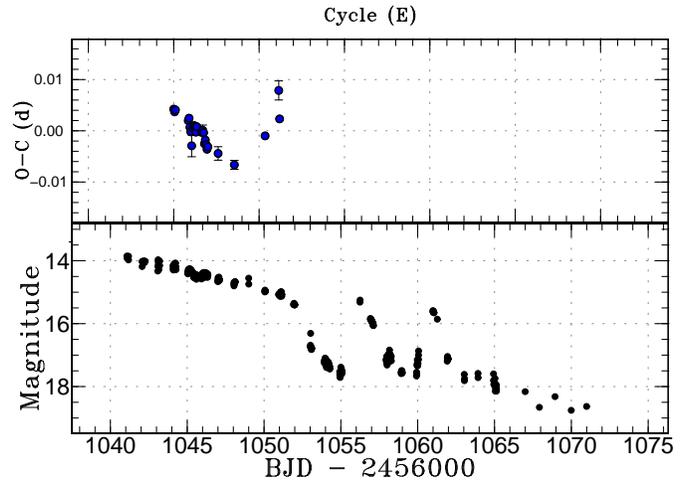}
  \end{center}
  \caption{$O-C$ diagram of superhumps in MASTER J085854 (2015).
     (Upper:) $O-C$ diagram.
     We used a period of 0.05556~d for calculating the $O-C$ residuals.
     (Lower:) Light curve.  The data were binned to 0.011~d.
  }
  \label{fig:j0858humpall}
\end{figure}

\begin{figure}
  \begin{center}
    \FigureFile(88mm,110mm){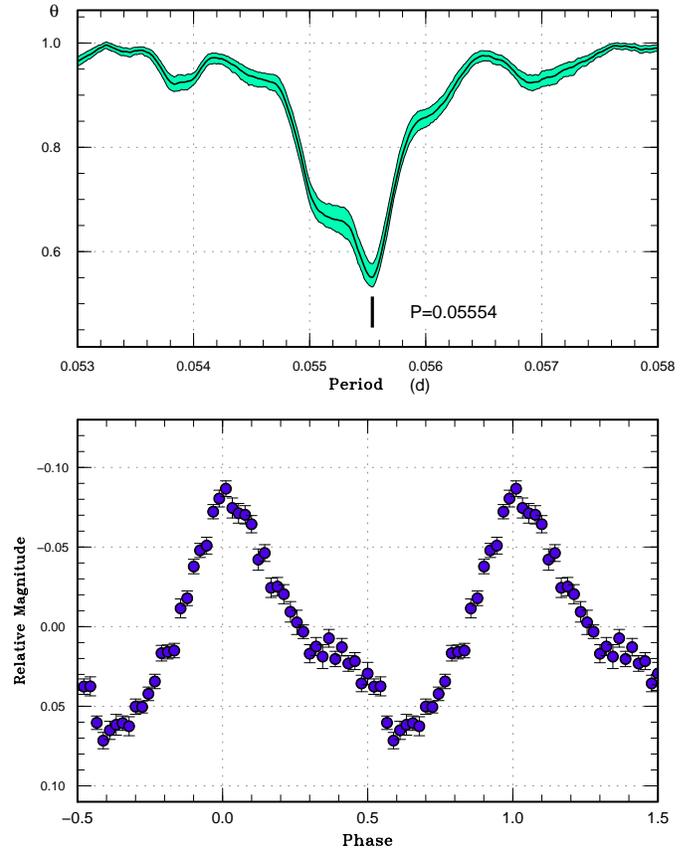}
  \end{center}
  \caption{Superhumps in MASTER J085854 during the plateau phase (2015).
     (Upper): PDM analysis.
     (Lower): Phase-averaged profile}.
  \label{fig:j0858shpdm}
\end{figure}

   This optical transient (hereafter MASTER J085854)
was detected on 2015 January 17 at a magnitude of 13.7
\citep{bal15j0858atel6946}.  Four days after the detection,
the object started to show short-period superhumps
(vsnet-alert 18210, 18220; figure \ref{fig:j0858shpdm}).
After the superoutburst plateau ended, the object showed
two rebrightenings (vsnet-alert 18254, 18266;
figure \ref{fig:j0858humpall}).

   The times of superhumps during the plateau phase
are listed in table \ref{tab:j0858oc2015}.
Although observations started relatively early,
we could not confidently detect stage A superhumps.
The resultant $P_{\rm dot}$ for stage B superhumps
was relatively large.

   WZ Sge-type dwarf novae with multiple rebrightenings
have been recently systematically studied by
\citet{nak13j2112j2037}, \Nakataprep.
\citet{mro13OGLEDN2} also reported two new
objects (OGLE-GD-DN-001, OGLE-GD-DN-014) with multiple
rebrightenings.  All these objects were known to have
orbital periods (or superhump periods) longer than
$\gtrsim$0.06~d.  MASTER J085854 is the only known
system showing multiple rebrightenings and with
a very short superhump period.

   \Nakataprep\ suggest that WZ Sge-type dwarf novae
with multiple rebrightenings have higher mass ratios
than period bouncers, which also appear to be
consistent with the shortness of stage A, i.e.
superhumps quickly grow.  This also likely applies
to MASTER J085854 as judged from the quick appearance
of stage B superhumps.  Since there was possibly a
past outburst in the CRTS data \citep{bal15j0858atel6946},
continued monitoring of this object will clarify
the supercycle length and where there is a phase
of early superhumps.

\begin{table}
\caption{Superhump maxima of MASTER J085854 (2015)}\label{tab:j0858oc2015}
\begin{center}
\begin{tabular}{rp{55pt}p{40pt}r@{.}lr}
\hline
\multicolumn{1}{c}{$E$} & \multicolumn{1}{c}{max\commenta} & \multicolumn{1}{c}{error} & \multicolumn{2}{c}{$O-C$\commentb} & \multicolumn{1}{c}{$N$\commentc} \\
\hline
0 & 57044.1262 & 0.0003 & 0&0042 & 89 \\
1 & 57044.1813 & 0.0002 & 0&0037 & 115 \\
2 & 57044.2372 & 0.0002 & 0&0040 & 98 \\
17 & 57045.0685 & 0.0007 & 0&0020 & 43 \\
18 & 57045.1246 & 0.0004 & 0&0024 & 100 \\
19 & 57045.1783 & 0.0002 & 0&0006 & 117 \\
20 & 57045.2330 & 0.0003 & $-$0&0002 & 117 \\
21 & 57045.2858 & 0.0021 & $-$0&0029 & 22 \\
23 & 57045.4003 & 0.0004 & 0&0004 & 79 \\
24 & 57045.4565 & 0.0002 & 0&0010 & 128 \\
25 & 57045.5117 & 0.0002 & 0&0007 & 128 \\
26 & 57045.5663 & 0.0002 & $-$0&0003 & 127 \\
27 & 57045.6229 & 0.0005 & 0&0008 & 70 \\
33 & 57045.9552 & 0.0004 & $-$0&0003 & 46 \\
34 & 57046.0111 & 0.0010 & 0&0001 & 38 \\
35 & 57046.0662 & 0.0004 & $-$0&0004 & 212 \\
36 & 57046.1196 & 0.0004 & $-$0&0026 & 248 \\
37 & 57046.1759 & 0.0006 & $-$0&0018 & 236 \\
38 & 57046.2306 & 0.0003 & $-$0&0027 & 153 \\
39 & 57046.2852 & 0.0004 & $-$0&0036 & 129 \\
40 & 57046.3413 & 0.0005 & $-$0&0031 & 84 \\
52 & 57047.0067 & 0.0013 & $-$0&0044 & 53 \\
71 & 57048.0601 & 0.0009 & $-$0&0066 & 35 \\
107 & 57050.0659 & 0.0007 & $-$0&0010 & 95 \\
123 & 57050.9638 & 0.0019 & 0&0079 & 57 \\
124 & 57051.0138 & 0.0007 & 0&0023 & 58 \\
\hline
  \multicolumn{6}{l}{\commenta BJD$-$2400000.} \\
  \multicolumn{6}{l}{\commentb Against max $= 2457044.1220 + 0.055560 E$.} \\
  \multicolumn{6}{l}{\commentc Number of points used to determine the maximum.} \\
\end{tabular}
\end{center}
\end{table}

\subsection{MASTER OT J105545.20$+$573109.7}\label{obj:j1055}

\begin{figure}
  \begin{center}
    \FigureFile(88mm,110mm){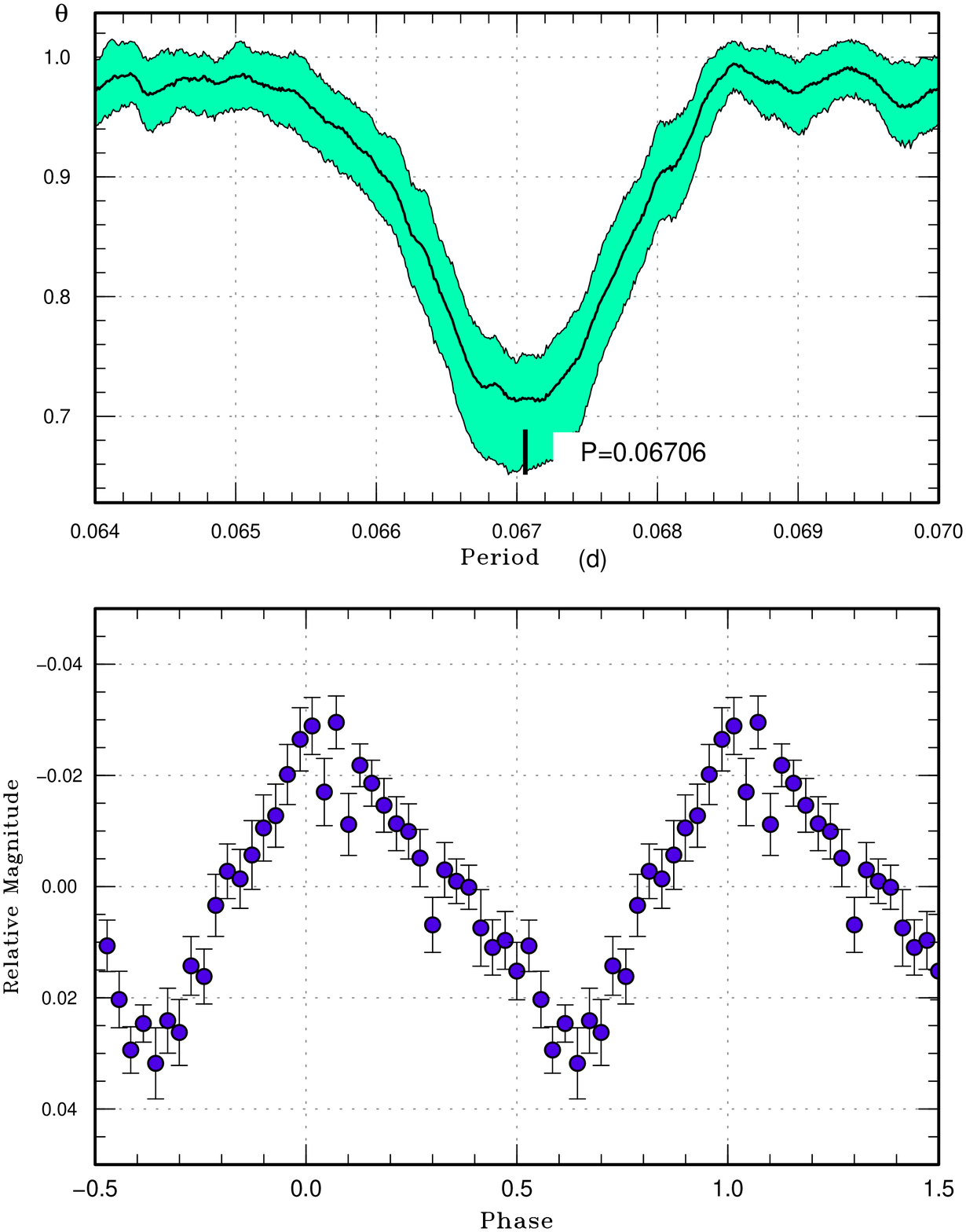}
  \end{center}
  \caption{Superhumps in MASTER J105545 (2014).  (Upper): PDM analysis.
     (Lower): Phase-averaged profile}.
  \label{fig:j1055shpdm}
\end{figure}

   This optical transient (hereafter MASTER J105545)
was detected on 2014 March 4 at a magnitude of 15.5
\citep{vla14j1055atel5950}.  Subsequent observations
detected superhumps (vsnet-alert 17007, 17018; 
figure \ref{fig:j1055shpdm}).
The times of superhump maxima are listed
in table \ref{tab:j1055oc2014}.
Although the stage of the superoutburst was unknown,
the small amplitudes of superhumps may suggest that
we observed the later stage of the superoutburst.

\begin{table}
\caption{Superhump maxima of MASTER J105545 (2014)}\label{tab:j1055oc2014}
\begin{center}
\begin{tabular}{rp{55pt}p{40pt}r@{.}lr}
\hline
\multicolumn{1}{c}{$E$} & \multicolumn{1}{c}{max\commenta} & \multicolumn{1}{c}{error} & \multicolumn{2}{c}{$O-C$\commentb} & \multicolumn{1}{c}{$N$\commentc} \\
\hline
0 & 56726.4942 & 0.0010 & $-$0&0024 & 72 \\
14 & 56727.4377 & 0.0016 & 0&0040 & 19 \\
15 & 56727.5009 & 0.0006 & 0&0002 & 33 \\
16 & 56727.5645 & 0.0007 & $-$0&0031 & 60 \\
17 & 56727.6369 & 0.0011 & 0&0024 & 94 \\
18 & 56727.7015 & 0.0008 & 0&0001 & 65 \\
29 & 56728.4369 & 0.0016 & $-$0&0009 & 60 \\
30 & 56728.5070 & 0.0013 & 0&0022 & 73 \\
31 & 56728.5703 & 0.0016 & $-$0&0014 & 35 \\
45 & 56729.5094 & 0.0119 & 0&0006 & 26 \\
46 & 56729.5741 & 0.0026 & $-$0&0016 & 33 \\
\hline
  \multicolumn{6}{l}{\commenta BJD$-$2400000.} \\
  \multicolumn{6}{l}{\commentb Against max $= 2456726.4966 + 0.066937 E$.} \\
  \multicolumn{6}{l}{\commentc Number of points used to determine the maximum.} \\
\end{tabular}
\end{center}
\end{table}

\subsection{OT J030929.8$+$263804}\label{obj:j0309}

\begin{figure}
  \begin{center}
    \FigureFile(88mm,110mm){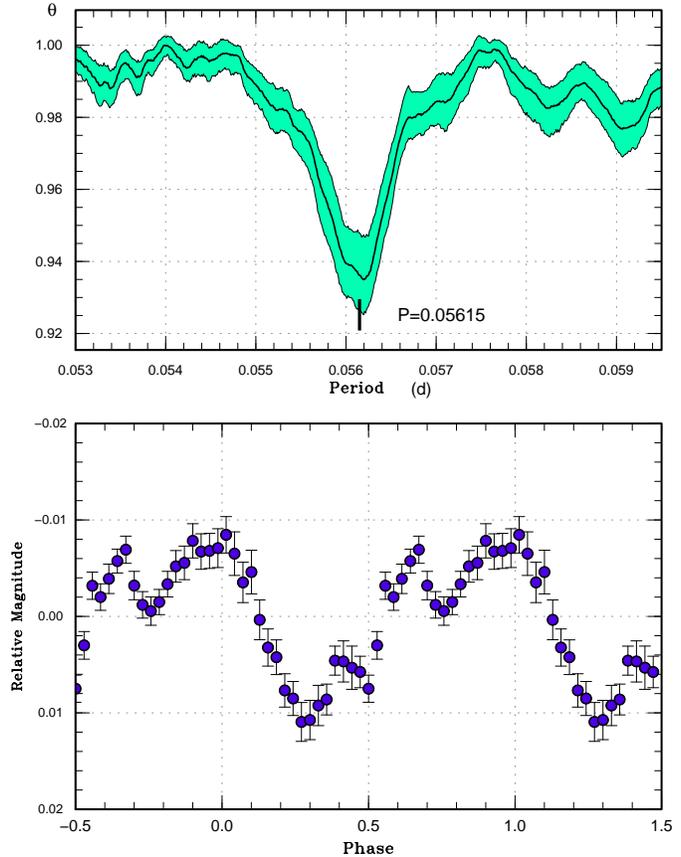}
  \end{center}
  \caption{Early superhumps in OT J030929 (2014).  (Upper): PDM analysis.
     (Lower): Phase-averaged profile}.
  \label{fig:j0309eshpdm}
\end{figure}

\begin{figure}
  \begin{center}
    \FigureFile(88mm,110mm){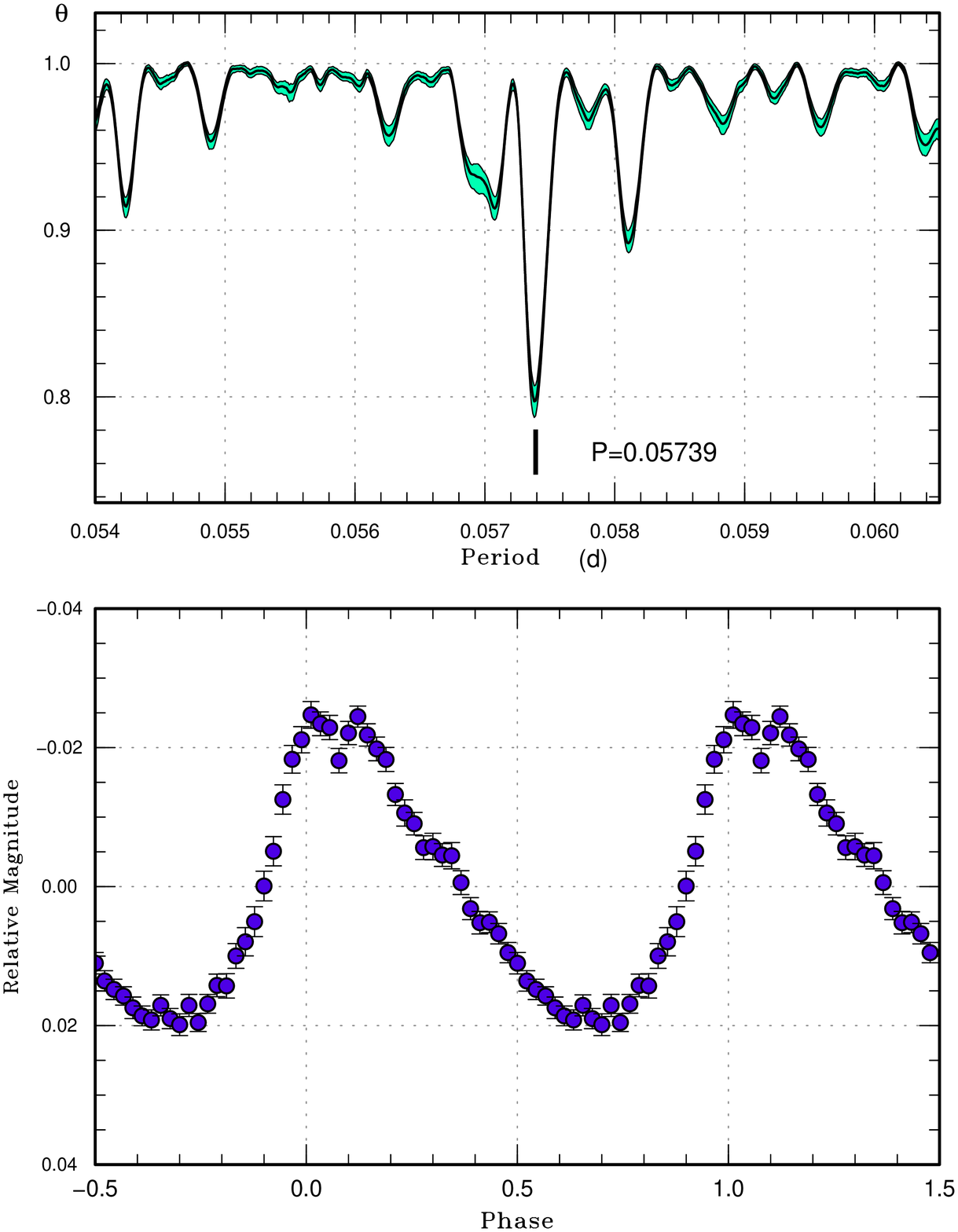}
  \end{center}
  \caption{Ordinary superhumps in OT J030929 (2014).  (Upper): PDM analysis.
     (Lower): Phase-averaged profile}.
  \label{fig:j0309shpdm}
\end{figure}

   This object (=PNV J03093063$+$2638031, hereafter
OT J030929) was announced as a possible nova
detected by S. Ueda.\footnote{
  $<$http://www.cbat.eps.harvard.edu/unconf/\\
followups/J03093063+2638031.html$>$.
  Also note that the reported coordinates were somewhat
  different from the final values.
}  Based on the presence of a blue counterpart and
the relatively small outburst amplitude, the object
was considered to be a WZ Sge-type dwarf nova
rather than a nova (vsnet-alert 17907, 17908).
Spectroscopic observations also indicated a dwarf nova
(vsnet-alert 17912, 17919).
Soon after the announcement, early superhumps were detected
(vsnet-alert 17918, 17920, 17926, 17931;
figure \ref{fig:j0309eshpdm}).  The period was determined
to be 0.05615(2)~d.
The object then showed ordinary superhumps
(vsnet-alert 17934, 17959, 17972; 
figure \ref{fig:j0309shpdm}).

   The times of superhump maxima during the outburst
plateau are listed in table \ref{tab:j2309oc2014}.
Although stage A superhumps were recorded ($E \le 36$),
there was an observational gap in the early stage B.
There was a clear stage B-C transition around $E$=197.
The period of stage A superhumps corresponds to
$\epsilon^*$=0.0291(3) and $q$=0.078(1).

\begin{figure}
  \begin{center}
    \FigureFile(88mm,100mm){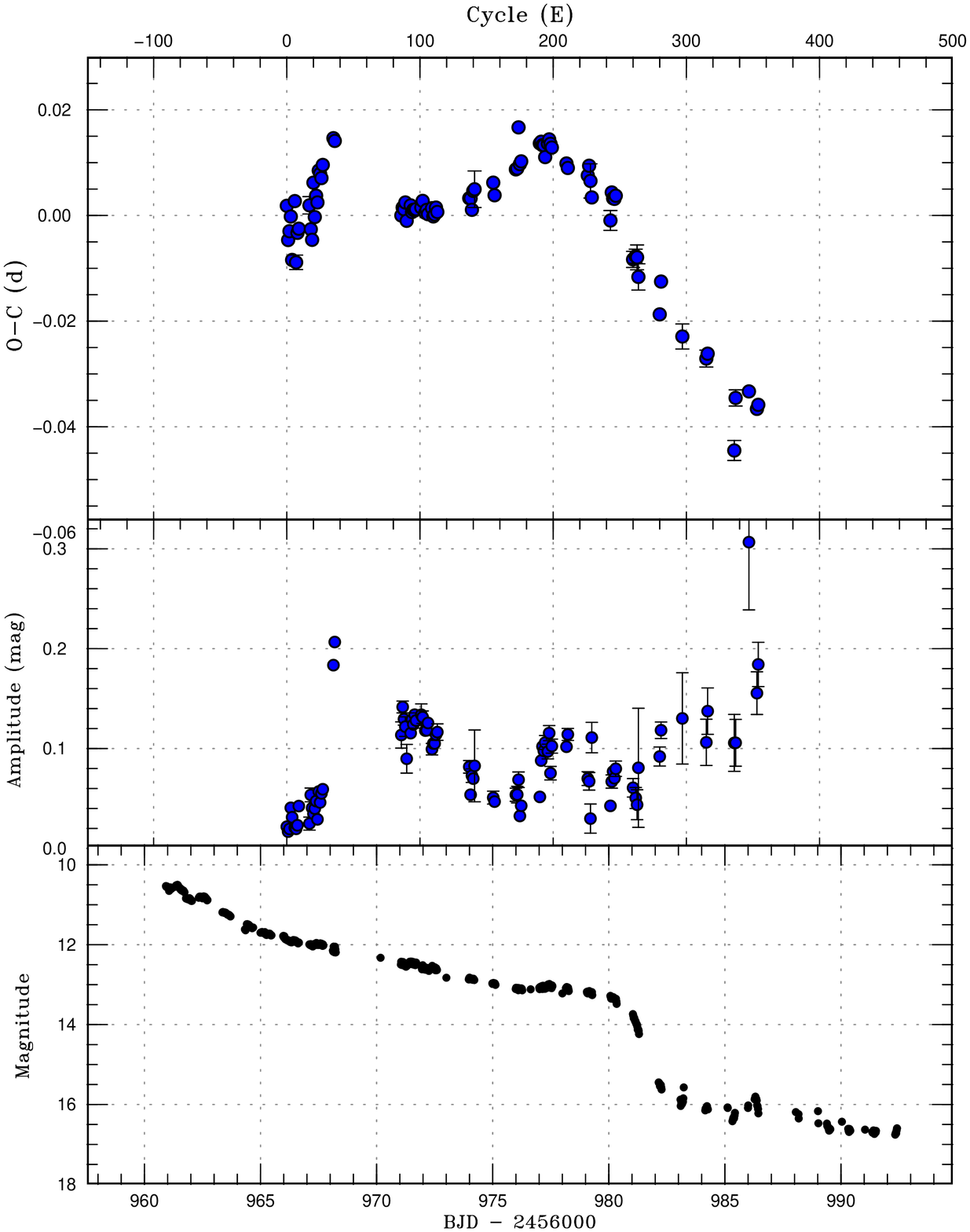}
  \end{center}
  \caption{$O-C$ diagram of superhumps in OT J030929 (2014).
     (Upper:) $O-C$ diagram.
     We used a period of 0.05736~d for calculating the $O-C$ residuals.
     (Middle:) Amplitudes of superhumps.
     (Lower:) Light curve.  The data were binned to 0.011~d.
  }
  \label{fig:j0309humpall}
\end{figure}

\begin{figure}
  \begin{center}
    \FigureFile(88mm,70mm){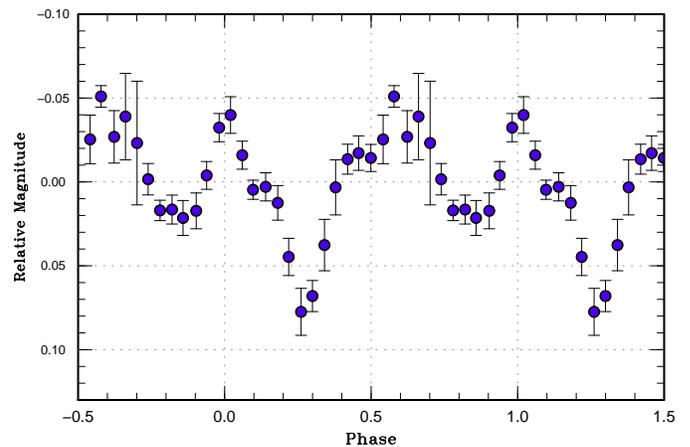}
  \end{center}
  \caption{Late-stage post-superoutburst superhumps in 
     OT J030929 (2014).
     The data after BJD 2456987 were used.
     The phase corresponds to the times of maxima
     in table \ref{tab:j0309ocpost}.}
  \label{fig:j0309shlate}
\end{figure}

   Superhumps after the rapid fading became less
clear.  The times before BJD 2456987 are listed in
table \ref{tab:j0309ocpost}.  These post-superoutburst
superhumps were on a smooth extension of
stage C superhumps (figure \ref{fig:j0309humpall}).
After this, the profile of superhumps became double-humped
(figure \ref{fig:j0309shlate}).  The secondary
humps even became stronger than the original superhumps.
Although this phenomenon bore some resemblance to
traditional late superhumps, this change
in the profile occurred more than $\sim$4~d
after the rapid decline, which is not the usual case
in traditional late superhumps.

   The relatively large $P_{\rm dot}$ of stage B 
superhumps and the transition to stage C superhumps
during the plateau phase were similar to SU UMa-type
dwarf novae rather than extreme WZ Sge-type
dwarf novae.  The object may have intermediate
properties between WZ Sge-type dwarf novae
and ordinary SU UMa-type dwarf novae.

\begin{table}
\caption{Superhump maxima of OT J030929 (2014)}\label{tab:j2309oc2014}
\begin{center}
\begin{tabular}{rp{55pt}p{40pt}r@{.}lr}
\hline
\multicolumn{1}{c}{$E$} & \multicolumn{1}{c}{max\commenta} & \multicolumn{1}{c}{error} & \multicolumn{2}{c}{$O-C$\commentb} & \multicolumn{1}{c}{$N$\commentc} \\
\hline
0 & 56966.1342 & 0.0009 & 0&0003 & 202 \\
1 & 56966.1851 & 0.0009 & $-$0&0062 & 291 \\
2 & 56966.2441 & 0.0008 & $-$0&0046 & 290 \\
3 & 56966.3042 & 0.0005 & $-$0&0018 & 209 \\
4 & 56966.3534 & 0.0008 & $-$0&0100 & 102 \\
6 & 56966.4792 & 0.0009 & 0&0011 & 102 \\
7 & 56966.5250 & 0.0014 & $-$0&0105 & 105 \\
8 & 56966.5879 & 0.0010 & $-$0&0050 & 108 \\
9 & 56966.6461 & 0.0008 & $-$0&0042 & 82 \\
17 & 56967.1094 & 0.0016 & 0&0001 & 89 \\
18 & 56967.1622 & 0.0009 & $-$0&0045 & 71 \\
19 & 56967.2176 & 0.0011 & $-$0&0065 & 85 \\
20 & 56967.2857 & 0.0009 & 0&0043 & 177 \\
21 & 56967.3366 & 0.0006 & $-$0&0022 & 193 \\
22 & 56967.3980 & 0.0007 & 0&0018 & 201 \\
23 & 56967.4541 & 0.0007 & 0&0005 & 208 \\
24 & 56967.5175 & 0.0004 & 0&0065 & 139 \\
25 & 56967.5742 & 0.0006 & 0&0059 & 103 \\
26 & 56967.6308 & 0.0006 & 0&0051 & 103 \\
27 & 56967.6906 & 0.0006 & 0&0076 & 96 \\
35 & 56968.1546 & 0.0001 & 0&0125 & 121 \\
36 & 56968.2114 & 0.0001 & 0&0119 & 115 \\
86 & 56971.0653 & 0.0009 & $-$0&0030 & 40 \\
87 & 56971.1242 & 0.0003 & $-$0&0016 & 119 \\
88 & 56971.1812 & 0.0003 & $-$0&0019 & 110 \\
89 & 56971.2398 & 0.0006 & $-$0&0006 & 103 \\
90 & 56971.2937 & 0.0011 & $-$0&0041 & 94 \\
93 & 56971.4687 & 0.0003 & $-$0&0013 & 66 \\
94 & 56971.5248 & 0.0002 & $-$0&0025 & 62 \\
95 & 56971.5826 & 0.0003 & $-$0&0021 & 67 \\
96 & 56971.6398 & 0.0003 & $-$0&0023 & 66 \\
97 & 56971.6974 & 0.0005 & $-$0&0020 & 49 \\
\hline
  \multicolumn{6}{l}{\commenta BJD$-$2400000.} \\
  \multicolumn{6}{l}{\commentb Against max $= 2456966.1339 + 0.057377 E$.} \\
  \multicolumn{6}{l}{\commentc Number of points used to determine the maximum.} \\
\end{tabular}
\end{center}
\end{table}

\addtocounter{table}{-1}
\begin{table}
\caption{Superhump maxima of OT J030929 (2014) (continued)}
\begin{center}
\begin{tabular}{rp{55pt}p{40pt}r@{.}lr}
\hline
\multicolumn{1}{c}{$E$} & \multicolumn{1}{c}{max\commenta} & \multicolumn{1}{c}{error} & \multicolumn{2}{c}{$O-C$\commentb} & \multicolumn{1}{c}{$N$\commentc} \\
\hline
101 & 56971.9272 & 0.0005 & $-$0&0018 & 43 \\
102 & 56971.9858 & 0.0005 & $-$0&0006 & 31 \\
104 & 56972.0983 & 0.0003 & $-$0&0028 & 95 \\
105 & 56972.1562 & 0.0002 & $-$0&0023 & 109 \\
106 & 56972.2127 & 0.0002 & $-$0&0031 & 112 \\
109 & 56972.3860 & 0.0012 & $-$0&0021 & 51 \\
110 & 56972.4418 & 0.0002 & $-$0&0036 & 101 \\
111 & 56972.4995 & 0.0002 & $-$0&0033 & 112 \\
112 & 56972.5582 & 0.0003 & $-$0&0020 & 119 \\
113 & 56972.6147 & 0.0006 & $-$0&0028 & 36 \\
137 & 56973.9939 & 0.0005 & $-$0&0007 & 51 \\
138 & 56974.0513 & 0.0008 & $-$0&0007 & 149 \\
139 & 56974.1065 & 0.0007 & $-$0&0029 & 150 \\
140 & 56974.1674 & 0.0006 & 0&0007 & 83 \\
141 & 56974.2251 & 0.0035 & 0&0010 & 10 \\
155 & 56975.0294 & 0.0009 & 0&0020 & 94 \\
156 & 56975.0843 & 0.0008 & $-$0&0004 & 96 \\
172 & 56976.0070 & 0.0009 & 0&0042 & 94 \\
173 & 56976.0645 & 0.0011 & 0&0043 & 95 \\
174 & 56976.1297 & 0.0007 & 0&0121 & 75 \\
175 & 56976.1801 & 0.0009 & 0&0051 & 106 \\
176 & 56976.2379 & 0.0006 & 0&0056 & 120 \\
190 & 56977.0444 & 0.0008 & 0&0088 & 167 \\
191 & 56977.1021 & 0.0003 & 0&0091 & 298 \\
192 & 56977.1588 & 0.0003 & 0&0084 & 348 \\
193 & 56977.2161 & 0.0003 & 0&0084 & 271 \\
194 & 56977.2712 & 0.0006 & 0&0061 & 113 \\
196 & 56977.3885 & 0.0006 & 0&0087 & 57 \\
197 & 56977.4467 & 0.0005 & 0&0094 & 60 \\
198 & 56977.5032 & 0.0006 & 0&0086 & 60 \\
199 & 56977.5598 & 0.0005 & 0&0079 & 46 \\
210 & 56978.1878 & 0.0003 & 0&0047 & 122 \\
211 & 56978.2443 & 0.0004 & 0&0038 & 120 \\
\hline
  \multicolumn{6}{l}{\commenta BJD$-$2400000.} \\
  \multicolumn{6}{l}{\commentb Against max $= 2456966.1339 + 0.057377 E$.} \\
  \multicolumn{6}{l}{\commentc Number of points used to determine the maximum.} \\
\end{tabular}
\end{center}
\end{table}

\addtocounter{table}{-1}
\begin{table}
\caption{Superhump maxima of OT J030929 (2014) (continued)}
\begin{center}
\begin{tabular}{rp{55pt}p{40pt}r@{.}lr}
\hline
\multicolumn{1}{c}{$E$} & \multicolumn{1}{c}{max\commenta} & \multicolumn{1}{c}{error} & \multicolumn{2}{c}{$O-C$\commentb} & \multicolumn{1}{c}{$N$\commentc} \\
\hline
226 & 56979.1033 & 0.0007 & 0&0021 & 103 \\
227 & 56979.1625 & 0.0010 & 0&0039 & 142 \\
228 & 56979.2170 & 0.0032 & 0&0010 & 68 \\
229 & 56979.2712 & 0.0010 & $-$0&0021 & 123 \\
243 & 56980.0699 & 0.0019 & $-$0&0067 & 67 \\
244 & 56980.1325 & 0.0007 & $-$0&0014 & 150 \\
245 & 56980.1888 & 0.0003 & $-$0&0026 & 244 \\
246 & 56980.2460 & 0.0004 & $-$0&0027 & 212 \\
247 & 56980.3040 & 0.0007 & $-$0&0021 & 101 \\
260 & 56981.0376 & 0.0015 & $-$0&0144 & 82 \\
262 & 56981.1530 & 0.0013 & $-$0&0138 & 106 \\
263 & 56981.2101 & 0.0024 & $-$0&0140 & 69 \\
264 & 56981.2637 & 0.0025 & $-$0&0178 & 30 \\
\hline
  \multicolumn{6}{l}{\commenta BJD$-$2400000.} \\
  \multicolumn{6}{l}{\commentb Against max $= 2456966.1339 + 0.057377 E$.} \\
  \multicolumn{6}{l}{\commentc Number of points used to determine the maximum.} \\
\end{tabular}
\end{center}
\end{table}

\begin{table}
\caption{Superhump maxima of OT J030929 (2014) (post-superoutburst)}\label{tab:j0309ocpost}
\begin{center}
\begin{tabular}{rp{55pt}p{40pt}r@{.}lr}
\hline
\multicolumn{1}{c}{$E$} & \multicolumn{1}{c}{max\commenta} & \multicolumn{1}{c}{error} & \multicolumn{2}{c}{$O-C$\commentb} & \multicolumn{1}{c}{$N$\commentc} \\
\hline
0 & 56982.1744 & 0.0008 & $-$0&0020 & 63 \\
1 & 56982.2380 & 0.0005 & 0&0045 & 63 \\
17 & 56983.1453 & 0.0024 & $-$0&0011 & 55 \\
35 & 56984.1737 & 0.0016 & 0&0002 & 58 \\
36 & 56984.2319 & 0.0012 & 0&0014 & 63 \\
56 & 56985.3608 & 0.0019 & $-$0&0110 & 59 \\
57 & 56985.4281 & 0.0015 & $-$0&0007 & 45 \\
67 & 56986.0030 & 0.0006 & 0&0035 & 10 \\
73 & 56986.3438 & 0.0010 & 0&0020 & 59 \\
74 & 56986.4019 & 0.0009 & 0&0031 & 54 \\
\hline
  \multicolumn{6}{l}{\commenta BJD$-$2400000.} \\
  \multicolumn{6}{l}{\commentb Against max $= 2456982.1764 + 0.057060 E$.} \\
  \multicolumn{6}{l}{\commentc Number of points used to determine the maximum.} \\
\end{tabular}
\end{center}
\end{table}

\subsection{OT J064833.4$+$065624}\label{obj:j0648}

   This object (=PNV J06483343$+$0656236, hereafter
OT J064833) was announced as a possible nova
detected at an unfiltered CCD magnitude of 11.6 on 
2014 November 22 by S. Kaneko.\footnote{
  $<$http://www.cbat.eps.harvard.edu/unconf/\\
followups/J06483343+0656236.html$>$.
}  The presence of a relatively bright quiescent
counterpart made the object a good candidate
for a dwarf nova (vsnet-alert 18001).
Spectroscopic observation by M. Fujii
indeed confirmed the dwarf nova-type nature
(vsnet-alert 18004).

\begin{figure}
  \begin{center}
    \FigureFile(88mm,110mm){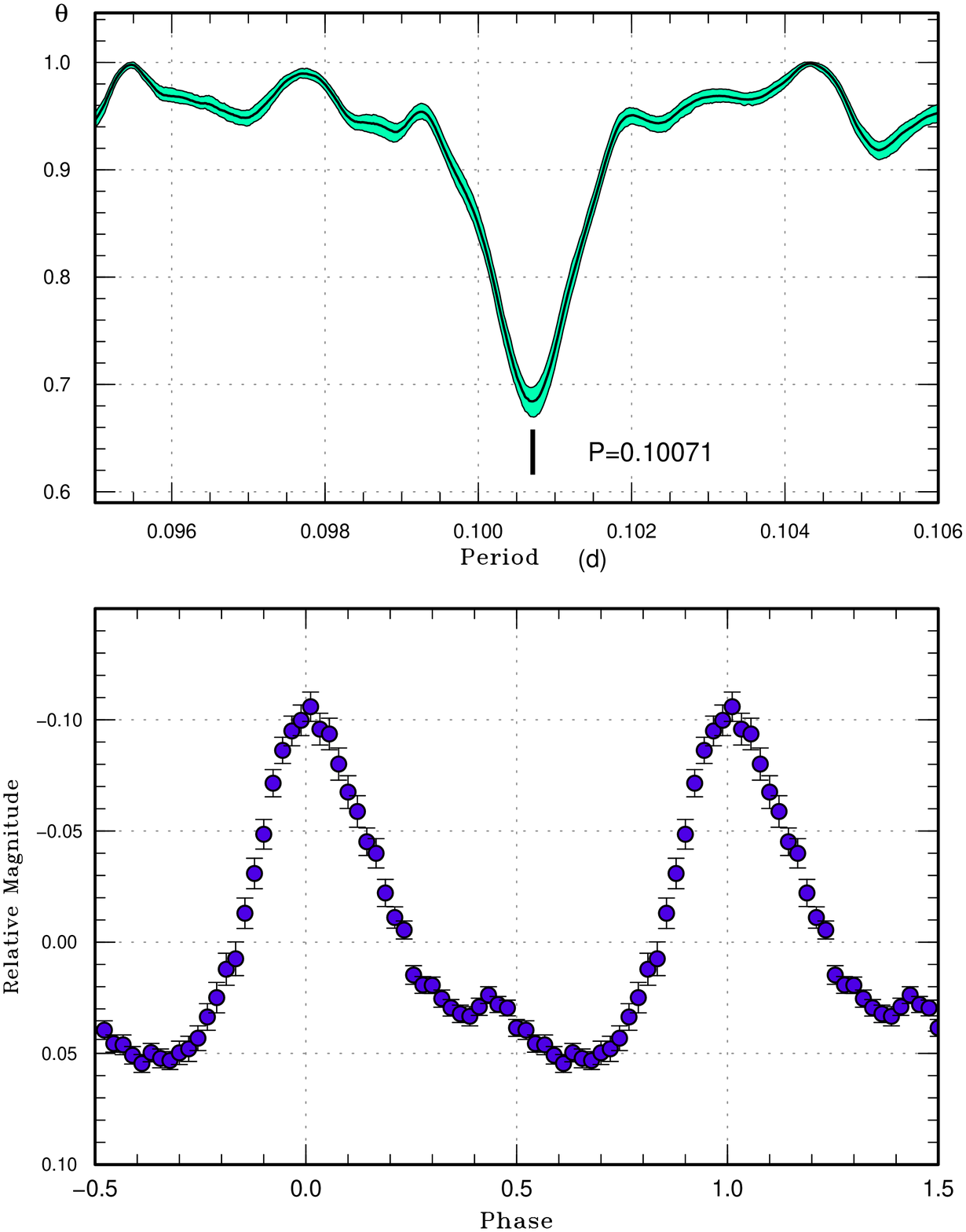}
  \end{center}
  \caption{Superhumps in OT J064833 during the superoutburst
     plateau (2014).  (Upper): PDM analysis.
     (Lower): Phase-averaged profile}.
  \label{fig:j0648shpdm}
\end{figure}

\begin{figure}
  \begin{center}
    \FigureFile(88mm,100mm){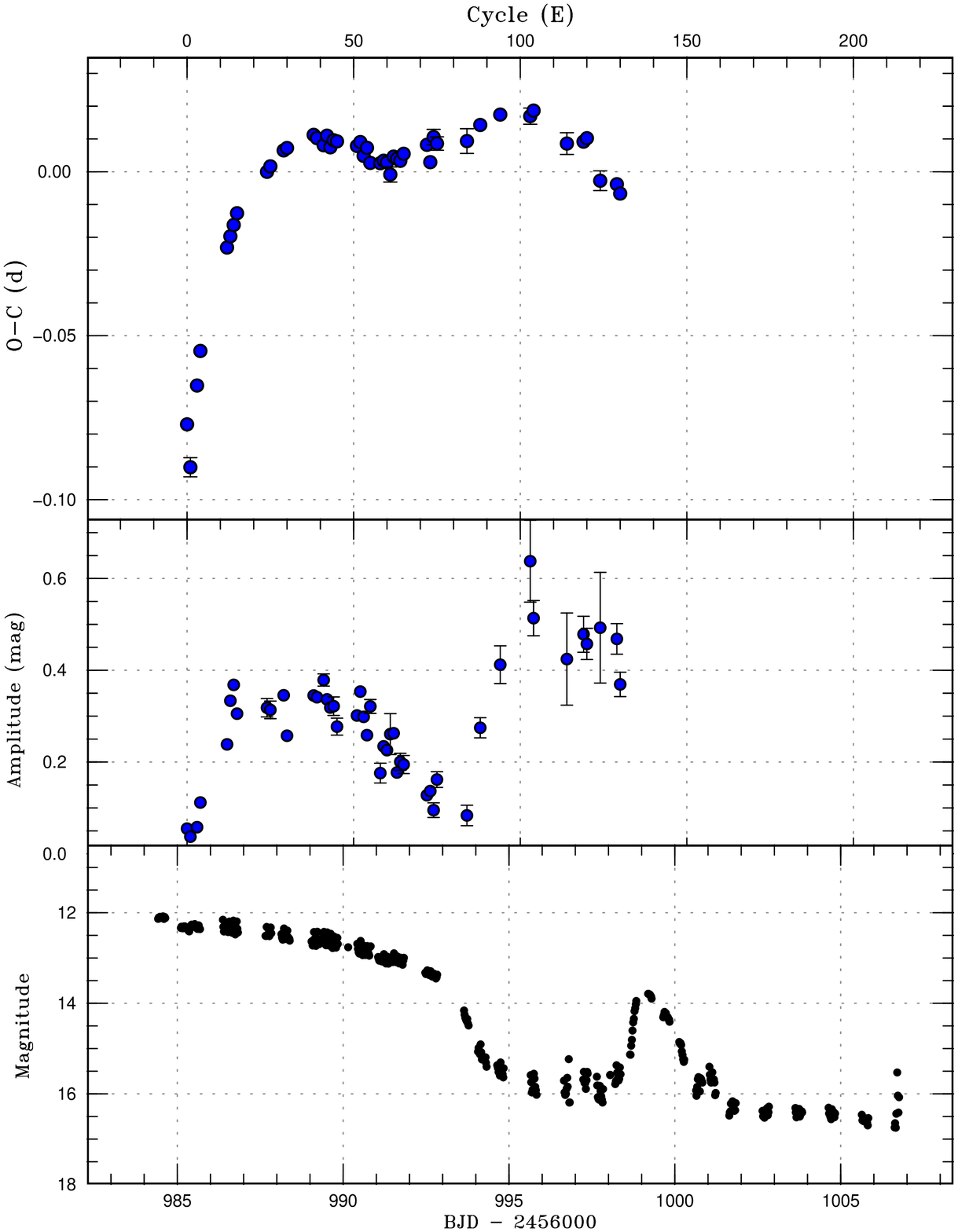}
  \end{center}
  \caption{$O-C$ diagram of superhumps in OT J064833 (2014).
     (Upper:) $O-C$ diagram.
     We used a period of 0.01005~d for calculating the $O-C$ residuals.
     (Middle:) Amplitudes of superhumps.
     (Lower:) Light curve.  The data were binned to 0.020~d.
  }
  \label{fig:j0648humpall}
\end{figure}

   Subsequent observations detected growing
superhumps with a long period (vsnet-alert 18012,
18014, 18018, 18027; figure \ref{fig:j0648shpdm}).
The times of superhump maxima are listed in table
\ref{tab:j0648oc2014}.  Stage A superhumps are
clearly detected ($E \le 15$; figure \ref{fig:j0648humpall}).
The epochs after $E=84$ are post-superoutburst
superhumps, and they are not used to determine
the periods in table \ref{tab:perlist}.
There was no indication of a phase jump
at the end of the superoutburst plateau.

   The most notable feature of this object
is the long-lasting stage A despite the long
superhump period, which makes this object
as a dwarf nova in the period gap.
In \citet{Pdot6}, we have argued that in some
long-$P_{\rm orb}$ dwarf novae, the condition
of tidal instability is critically reached
and it may take longer time to develop superhumps.
This object appears to fit this interpretation.

   The object also showed a rebrightening
(vsnet-alert 18040; figure \ref{fig:j0648humpall}),
which is relatively rare in long-$P_{\rm orb}$ systems.
Well-documented rebrightenings in long-$P_{\rm orb}$ 
systems include V725 Aql \citep{uem01v725aql} and QZ Ser
(\Ohtprep, vsnet-alert 15567, 15620).
The 1997 superoutburst of EF Peg may have also
been accompanied by a rebrightening (vsnet-alert 1344).
All these objects are known to have long
recurrence times and are hence considered
to have low mass-transfer rates.
Given the lack of past detections of outbursts
in OT J064833, this object also likely have
a low mass-transfer rate.  The features resembling
short-$P_{\rm orb}$ (rebrightenings, positive
$P_{\rm dot}$ and absence of phase jump at
the end of superoutburst) may be a result of
low mass-transfer rates and await future
theoretical investigation.

   Since the object has a bright quiescent counterpart
(15.9 mag in $J$), it is a promising candidate
to detect the secondary.  Accurate measurement
of the orbital period will lead to determination
of $q$ by using the stage A superhump method
and would clarify the evolutionary state
of this object.

\begin{table}
\caption{Superhump maxima of OT J064833 (2014)}\label{tab:j0648oc2014}
\begin{center}
\begin{tabular}{rp{55pt}p{40pt}r@{.}lr}
\hline
\multicolumn{1}{c}{$E$} & \multicolumn{1}{c}{max\commenta} & \multicolumn{1}{c}{error} & \multicolumn{2}{c}{$O-C$\commentb} & \multicolumn{1}{c}{$N$\commentc} \\
\hline
0 & 56985.2170 & 0.0011 & $-$0&0538 & 419 \\
1 & 56985.3043 & 0.0029 & $-$0&0672 & 253 \\
3 & 56985.5302 & 0.0007 & $-$0&0430 & 353 \\
4 & 56985.6411 & 0.0004 & $-$0&0329 & 308 \\
12 & 56986.4763 & 0.0006 & $-$0&0042 & 143 \\
13 & 56986.5802 & 0.0002 & $-$0&0012 & 220 \\
14 & 56986.6841 & 0.0002 & 0&0019 & 279 \\
15 & 56986.7881 & 0.0006 & 0&0051 & 46 \\
24 & 56987.7048 & 0.0008 & 0&0145 & 27 \\
25 & 56987.8069 & 0.0007 & 0&0158 & 31 \\
29 & 56988.2136 & 0.0002 & 0&0192 & 415 \\
30 & 56988.3147 & 0.0004 & 0&0195 & 285 \\
38 & 56989.1224 & 0.0002 & 0&0206 & 429 \\
39 & 56989.2218 & 0.0002 & 0&0193 & 234 \\
41 & 56989.4205 & 0.0006 & 0&0163 & 62 \\
42 & 56989.5239 & 0.0002 & 0&0189 & 87 \\
43 & 56989.6208 & 0.0004 & 0&0150 & 83 \\
44 & 56989.7234 & 0.0008 & 0&0168 & 27 \\
45 & 56989.8235 & 0.0009 & 0&0161 & 26 \\
51 & 56990.4248 & 0.0006 & 0&0125 & 62 \\
52 & 56990.5265 & 0.0003 & 0&0133 & 208 \\
53 & 56990.6227 & 0.0003 & 0&0087 & 202 \\
54 & 56990.7256 & 0.0005 & 0&0108 & 136 \\
55 & 56990.8215 & 0.0007 & 0&0058 & 27 \\
58 & 56991.1227 & 0.0009 & 0&0047 & 67 \\
59 & 56991.2239 & 0.0004 & 0&0050 & 324 \\
60 & 56991.3238 & 0.0003 & 0&0042 & 349 \\
61 & 56991.4206 & 0.0023 & 0&0001 & 29 \\
62 & 56991.5265 & 0.0005 & 0&0052 & 113 \\
63 & 56991.6263 & 0.0007 & 0&0042 & 112 \\
64 & 56991.7261 & 0.0011 & 0&0032 & 27 \\
65 & 56991.8287 & 0.0016 & 0&0050 & 23 \\
\hline
  \multicolumn{6}{l}{\commenta BJD$-$2400000.} \\
  \multicolumn{6}{l}{\commentb Against max $= 2456985.2707 + 0.100816 E$.} \\
  \multicolumn{6}{l}{\commentc Number of points used to determine the maximum.} \\
\end{tabular}
\end{center}
\end{table}

\addtocounter{table}{-1}
\begin{table}
\caption{Superhump maxima of OT J064833 (2014) (continued)}
\begin{center}
\begin{tabular}{rp{55pt}p{40pt}r@{.}lr}
\hline
\multicolumn{1}{c}{$E$} & \multicolumn{1}{c}{max\commenta} & \multicolumn{1}{c}{error} & \multicolumn{2}{c}{$O-C$\commentb} & \multicolumn{1}{c}{$N$\commentc} \\
\hline
72 & 56992.5346 & 0.0006 & 0&0051 & 281 \\
73 & 56992.6298 & 0.0008 & $-$0&0005 & 131 \\
74 & 56992.7379 & 0.0023 & 0&0068 & 28 \\
75 & 56992.8364 & 0.0021 & 0&0044 & 18 \\
84 & 56993.7412 & 0.0038 & 0&0019 & 27 \\
88 & 56994.1479 & 0.0013 & 0&0053 & 143 \\
94 & 56994.7537 & 0.0013 & 0&0063 & 40 \\
103 & 56995.6573 & 0.0025 & 0&0025 & 18 \\
104 & 56995.7595 & 0.0010 & 0&0039 & 41 \\
114 & 56996.7539 & 0.0033 & $-$0&0098 & 28 \\
119 & 56997.2567 & 0.0008 & $-$0&0111 & 107 \\
120 & 56997.3582 & 0.0012 & $-$0&0104 & 97 \\
124 & 56997.7471 & 0.0030 & $-$0&0248 & 37 \\
129 & 56998.2483 & 0.0009 & $-$0&0277 & 98 \\
130 & 56998.3458 & 0.0009 & $-$0&0310 & 79 \\
\hline
  \multicolumn{6}{l}{\commenta BJD$-$2400000.} \\
  \multicolumn{6}{l}{\commentb Against max $= 2456985.2707 + 0.100816 E$.} \\
  \multicolumn{6}{l}{\commentc Number of points used to determine the maximum.} \\
\end{tabular}
\end{center}
\end{table}

\subsection{OT J213806.6$+$261957}\label{obj:j2138}

\subsubsection{Introduction}

   This object (OT J213806) was discovered as a possible
nova independently by D.-A. Yi \citep{yam10j2138cbet2273} and 
S. Kaneko \citep{nak10j2138cbet2275} in 2010.  Spectroscopic
observations demonstrated that the dwarf nova-type
nature of the object (\cite{gra10j2138cbet2275}
\cite{tov10j2138cbet2283}, vsnet-alert 11987).
There was a known outburst in 1942 \citep{hud10j2138atel2619}.
The object has a bright X-ray counterpart
(1RXS J213807.1$+$261958) and has a close visual
companion \citep{yam10j2138cbet2273} which dominates
when the object is near quiescence.
Although the 2010 superoutburst was well observed,
the phase of early superhumps was not very well
covered by observations \citep{Pdot2}.  
The 2010 superoutburst was also analyzed by
\citet{cho12j2138} and \citet{zem13j2138}.
Although the orbital period is not well determined yet,
a possible period of (0.054523~d) was
reported \citep{Pdot2}.  \citet{cho12j2138} reported
a period of 0.05435~d after the object started fading.
\citet{mit14j2138} reported a detailed spectroscopic
study in quiescence and obtained Doppler tomograms.
\citet{mit14j2138} did not obtain the orbital period
but assumed a period reported by \citet{cho12j2138}.

\subsubsection{2014 Superoutburst}

   The 2014 superoutburst occurred only four years
after the 2010 one, and the interval was surprisingly
short for a WZ Sge-type dwarf nova.
The outburst was detected visually by C. Chiselbrook
at a magnitude of 9.7 on October 22.
The same observed reported that the object was
fainter than 13.8 on the night before
(AAVSO observations).  Approximately 1.5~d after
this detection, initial CCD observations started.
The object already showed ordinary superhumps
(vsnet-alert 17891, 17893).

   The times of superhump maxima during the superoutburst
plateau are listed in table \ref{tab:j2138oc2014}.
As in the 2010 superoutburst, stages A-B-C were
clearly present.  The present observation
recorded stage A superhumps much better than in
the 2010 one.  A PDM analysis before BJD 2456954.7
yielded a period of 0.0568(3)~d, which is adopted
in table \ref{tab:perlist}.  This period corresponds
to $\epsilon^*$=0.041(4) and $q$=0.12(2) assuming
the orbital period of 0.054523~d.  The values are
0.044(4) and 0.13(2), respectively, if we adopt
the period of 0.05435~d.

\subsubsection{Object classification}

   The 2014 observation indicated that the duration
of early superhumps was less than 1~d, indicating
that the 2014 superoutburst was of an ordinary
SU UMa-type dwarf nova rather than a WZ Sge-type
dwarf nova.  Since the peak brightness of the 2014
superoutburst was fainter than the 2010 one
at least by 1.0 mag, it may be that the 2014
superoutburst was much less powerful as in
the 2010 one, and that the 2:1 resonance was not
reached during this superoutburst.
The 2010 superoutburst, however, was not very well
observed before the appearance of ordinary
superhumps, and the existence of early superhumps
was not very certain.  It may be that the 2010
superoutburst did not show early superhumps
at all, as was likely in the 2014 one.
If it is the case, the object should be better
classified as an SU UMa-type dwarf nova
rather than a WZ Sge-type one.
The large $q$ value (although this value has
a large uncertainty), the short outburst
recurrence time, the large positive
$P_{\rm dot}$ of stage B superhumps, 
the short evolutionary time of stage A superhumps
and the presence of stage C superhumps
are better reconciled with this interpretation.
We probably need to wait another bright outburst
as in 2010 to test the presence of early superhumps.

\subsubsection{Comparison of 2010 and 2014 superoutbursts}

\begin{figure}
  \begin{center}
    \FigureFile(88mm,70mm){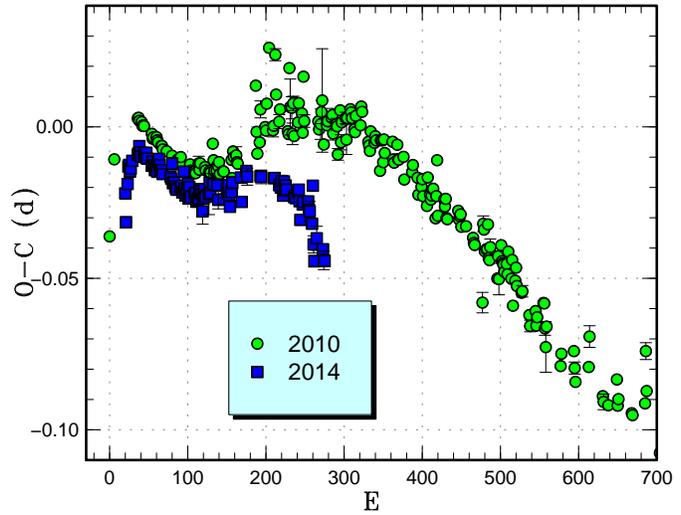}
  \end{center}
  \caption{Comparison of $O-C$ diagrams of OT J213806 between different
  superoutbursts.  A period of 0.05513~d was used to draw this figure.
  Approximate cycle counts ($E$) after the appearance of superhumps
  were used.  The 2014 superoutburst was shifted by 20 cycles
  to best match the 2010 one.  The 2014 superoutburst was
  artificially shifted by 0.01~d to avoid excessive overlaps
  with the 2010 diagram.}
  \label{fig:j2138comp}
\end{figure}

   A comparison of $O-C$ diagrams
(figure \ref{fig:j2138comp}) indicates that
the duration of stage B was different between
these superoutbursts: stage B was longer
in a brighter superoutburst in 2010.
The behavior after transition to stage C
was also different.
Such features was not observed in other SU UMa-type
dwarf novae and this result and would provide
a clue in understanding the origin of superhumps
in different stages.

\begin{figure}
  \begin{center}
    \FigureFile(88mm,70mm){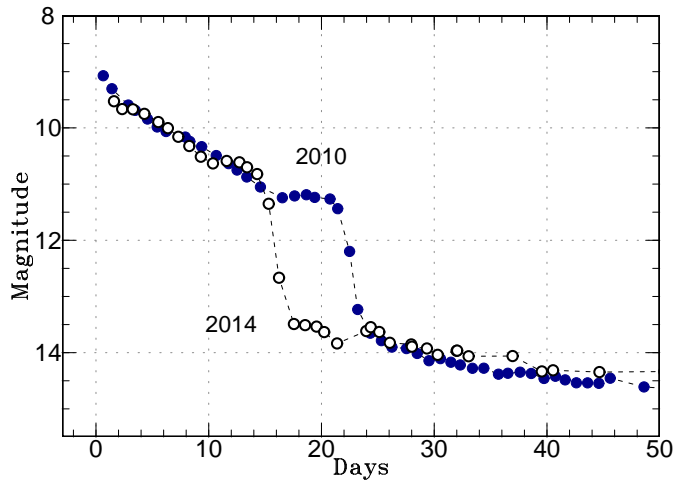}
  \end{center}
  \caption{Comparison of superoutbursts of OT J213806.
  The data were binned to 1~d and shifted in magnitude.
  The dashed lines are added to aid recognizing the variation.
  The duration of the 2014 superoutburst was shorter
  than the 2010 one although the magnitudes were
  almost similar during the linearly fading part,
  which corresponds to the phase of stage B superhumps.
  }
  \label{fig:j2138lccomp}
\end{figure}

   The duration of the superoutburst was different
(figure \ref{fig:j2138lccomp}).  The magnitudes
during the linearly fading part, which corresponds
to stage B superhumps, were almost similar between
these superoutbursts despite the difference
of the peak brightness.

\begin{table}
\caption{Superhump maxima of OT J213806 (2014)}\label{tab:j2138oc2014}
\begin{center}
\begin{tabular}{rp{55pt}p{40pt}r@{.}lr}
\hline
\multicolumn{1}{c}{$E$} & \multicolumn{1}{c}{max\commenta} & \multicolumn{1}{c}{error} & \multicolumn{2}{c}{$O-C$\commentb} & \multicolumn{1}{c}{$N$\commentc} \\
\hline
0 & 56954.0563 & 0.0005 & $-$0&0076 & 160 \\
1 & 56954.1018 & 0.0003 & $-$0&0171 & 305 \\
3 & 56954.2249 & 0.0016 & $-$0&0042 & 53 \\
4 & 56954.2854 & 0.0004 & 0&0012 & 150 \\
5 & 56954.3403 & 0.0008 & 0&0011 & 150 \\
6 & 56954.3942 & 0.0003 & $-$0&0002 & 59 \\
7 & 56954.4505 & 0.0005 & 0&0010 & 60 \\
9 & 56954.5633 & 0.0006 & 0&0037 & 22 \\
16 & 56954.9512 & 0.0002 & 0&0059 & 102 \\
17 & 56955.0067 & 0.0002 & 0&0064 & 160 \\
18 & 56955.0642 & 0.0001 & 0&0088 & 391 \\
19 & 56955.1173 & 0.0001 & 0&0069 & 537 \\
20 & 56955.1711 & 0.0001 & 0&0056 & 430 \\
24 & 56955.3927 & 0.0009 & 0&0068 & 86 \\
25 & 56955.4466 & 0.0002 & 0&0055 & 157 \\
26 & 56955.5015 & 0.0003 & 0&0054 & 142 \\
27 & 56955.5582 & 0.0003 & 0&0070 & 39 \\
28 & 56955.6114 & 0.0005 & 0&0051 & 28 \\
34 & 56955.9398 & 0.0005 & 0&0030 & 115 \\
35 & 56955.9964 & 0.0004 & 0&0045 & 171 \\
36 & 56956.0501 & 0.0002 & 0&0031 & 257 \\
37 & 56956.1054 & 0.0001 & 0&0033 & 345 \\
38 & 56956.1589 & 0.0004 & 0&0018 & 226 \\
39 & 56956.2150 & 0.0007 & 0&0027 & 211 \\
40 & 56956.2706 & 0.0005 & 0&0032 & 376 \\
41 & 56956.3237 & 0.0006 & 0&0013 & 357 \\
42 & 56956.3822 & 0.0008 & 0&0047 & 284 \\
43 & 56956.4382 & 0.0007 & 0&0057 & 101 \\
44 & 56956.4913 & 0.0008 & 0&0036 & 10 \\
45 & 56956.5447 & 0.0003 & 0&0019 & 127 \\
46 & 56956.5995 & 0.0003 & 0&0016 & 122 \\
47 & 56956.6551 & 0.0003 & 0&0022 & 124 \\
\hline
  \multicolumn{6}{l}{\commenta BJD$-$2400000.} \\
  \multicolumn{6}{l}{\commentb Against max $= 2456954.0638 + 0.055087 E$.} \\
  \multicolumn{6}{l}{\commentc Number of points used to determine the maximum.} \\
\end{tabular}
\end{center}
\end{table}

\addtocounter{table}{-1}
\begin{table}
\caption{Superhump maxima of OT J213806 (2014) (continued)}
\begin{center}
\begin{tabular}{rp{55pt}p{40pt}r@{.}lr}
\hline
\multicolumn{1}{c}{$E$} & \multicolumn{1}{c}{max\commenta} & \multicolumn{1}{c}{error} & \multicolumn{2}{c}{$O-C$\commentb} & \multicolumn{1}{c}{$N$\commentc} \\
\hline
48 & 56956.7089 & 0.0005 & 0&0009 & 83 \\
58 & 56957.2593 & 0.0004 & 0&0004 & 44 \\
59 & 56957.3137 & 0.0005 & $-$0&0002 & 52 \\
60 & 56957.3738 & 0.0021 & 0&0047 & 114 \\
61 & 56957.4226 & 0.0004 & $-$0&0016 & 57 \\
62 & 56957.4778 & 0.0003 & $-$0&0014 & 66 \\
63 & 56957.5331 & 0.0003 & $-$0&0012 & 110 \\
64 & 56957.5868 & 0.0003 & $-$0&0027 & 132 \\
65 & 56957.6411 & 0.0003 & $-$0&0034 & 74 \\
66 & 56957.6960 & 0.0003 & $-$0&0035 & 65 \\
71 & 56957.9724 & 0.0003 & $-$0&0026 & 59 \\
72 & 56958.0281 & 0.0003 & $-$0&0020 & 60 \\
73 & 56958.0811 & 0.0003 & $-$0&0041 & 56 \\
74 & 56958.1363 & 0.0007 & $-$0&0039 & 37 \\
75 & 56958.1982 & 0.0009 & 0&0028 & 65 \\
76 & 56958.2477 & 0.0003 & $-$0&0028 & 247 \\
77 & 56958.3006 & 0.0004 & $-$0&0050 & 379 \\
78 & 56958.3559 & 0.0006 & $-$0&0047 & 202 \\
79 & 56958.4126 & 0.0006 & $-$0&0031 & 180 \\
80 & 56958.4698 & 0.0004 & $-$0&0010 & 26 \\
81 & 56958.5201 & 0.0005 & $-$0&0058 & 98 \\
82 & 56958.5790 & 0.0004 & $-$0&0020 & 41 \\
83 & 56958.6311 & 0.0005 & $-$0&0050 & 65 \\
84 & 56958.6858 & 0.0004 & $-$0&0054 & 65 \\
89 & 56958.9627 & 0.0003 & $-$0&0039 & 57 \\
90 & 56959.0181 & 0.0003 & $-$0&0035 & 142 \\
91 & 56959.0703 & 0.0003 & $-$0&0064 & 337 \\
92 & 56959.1256 & 0.0002 & $-$0&0062 & 351 \\
93 & 56959.1812 & 0.0024 & $-$0&0057 & 209 \\
94 & 56959.2368 & 0.0004 & $-$0&0052 & 212 \\
95 & 56959.2950 & 0.0004 & $-$0&0021 & 111 \\
96 & 56959.3489 & 0.0005 & $-$0&0033 & 56 \\
\hline
  \multicolumn{6}{l}{\commenta BJD$-$2400000.} \\
  \multicolumn{6}{l}{\commentb Against max $= 2456954.0638 + 0.055087 E$.} \\
  \multicolumn{6}{l}{\commentc Number of points used to determine the maximum.} \\
\end{tabular}
\end{center}
\end{table}

\addtocounter{table}{-1}
\begin{table}
\caption{Superhump maxima of OT J213806 (2014) (continued)}
\begin{center}
\begin{tabular}{rp{55pt}p{40pt}r@{.}lr}
\hline
\multicolumn{1}{c}{$E$} & \multicolumn{1}{c}{max\commenta} & \multicolumn{1}{c}{error} & \multicolumn{2}{c}{$O-C$\commentb} & \multicolumn{1}{c}{$N$\commentc} \\
\hline
97 & 56959.4047 & 0.0008 & $-$0&0026 & 63 \\
98 & 56959.4588 & 0.0014 & $-$0&0036 & 36 \\
99 & 56959.5082 & 0.0042 & $-$0&0093 & 43 \\
100 & 56959.5675 & 0.0006 & $-$0&0050 & 100 \\
101 & 56959.6258 & 0.0004 & $-$0&0018 & 118 \\
102 & 56959.6788 & 0.0003 & $-$0&0039 & 148 \\
103 & 56959.7335 & 0.0004 & $-$0&0043 & 82 \\
107 & 56959.9549 & 0.0068 & $-$0&0032 & 13 \\
108 & 56960.0138 & 0.0006 & 0&0006 & 106 \\
109 & 56960.0656 & 0.0006 & $-$0&0027 & 121 \\
112 & 56960.2335 & 0.0007 & $-$0&0000 & 66 \\
113 & 56960.2890 & 0.0010 & 0&0003 & 161 \\
118 & 56960.5642 & 0.0013 & 0&0001 & 127 \\
119 & 56960.6147 & 0.0030 & $-$0&0045 & 40 \\
126 & 56961.0007 & 0.0012 & $-$0&0041 & 59 \\
130 & 56961.2246 & 0.0020 & $-$0&0006 & 203 \\
131 & 56961.2783 & 0.0011 & $-$0&0020 & 442 \\
132 & 56961.3362 & 0.0008 & 0&0009 & 249 \\
133 & 56961.3872 & 0.0017 & $-$0&0032 & 222 \\
134 & 56961.4392 & 0.0012 & $-$0&0063 & 57 \\
136 & 56961.5544 & 0.0027 & $-$0&0013 & 38 \\
137 & 56961.6126 & 0.0020 & 0&0019 & 11 \\
148 & 56962.2208 & 0.0006 & 0&0040 & 246 \\
149 & 56962.2678 & 0.0010 & $-$0&0040 & 175 \\
155 & 56962.6089 & 0.0013 & 0&0065 & 104 \\
156 & 56962.6621 & 0.0028 & 0&0046 & 25 \\
173 & 56963.5994 & 0.0008 & 0&0055 & 128 \\
174 & 56963.6542 & 0.0022 & 0&0052 & 87 \\
190 & 56964.5360 & 0.0018 & 0&0056 & 22 \\
191 & 56964.5912 & 0.0012 & 0&0057 & 24 \\
197 & 56964.9195 & 0.0004 & 0&0035 & 128 \\
199 & 56965.0290 & 0.0007 & 0&0029 & 173 \\
200 & 56965.0864 & 0.0006 & 0&0051 & 143 \\
\hline
  \multicolumn{6}{l}{\commenta BJD$-$2400000.} \\
  \multicolumn{6}{l}{\commentb Against max $= 2456954.0638 + 0.055087 E$.} \\
  \multicolumn{6}{l}{\commentc Number of points used to determine the maximum.} \\
\end{tabular}
\end{center}
\end{table}

\addtocounter{table}{-1}
\begin{table}
\caption{Superhump maxima of OT J213806 (2014) (continued)}
\begin{center}
\begin{tabular}{rp{55pt}p{40pt}r@{.}lr}
\hline
\multicolumn{1}{c}{$E$} & \multicolumn{1}{c}{max\commenta} & \multicolumn{1}{c}{error} & \multicolumn{2}{c}{$O-C$\commentb} & \multicolumn{1}{c}{$N$\commentc} \\
\hline
202 & 56965.1918 & 0.0016 & 0&0003 & 155 \\
203 & 56965.2516 & 0.0005 & 0&0051 & 247 \\
204 & 56965.3037 & 0.0006 & 0&0021 & 245 \\
205 & 56965.3605 & 0.0007 & 0&0038 & 247 \\
206 & 56965.4143 & 0.0011 & 0&0025 & 242 \\
216 & 56965.9631 & 0.0003 & 0&0005 & 317 \\
217 & 56966.0187 & 0.0003 & 0&0009 & 367 \\
218 & 56966.0730 & 0.0004 & 0&0002 & 316 \\
221 & 56966.2406 & 0.0017 & 0&0025 & 173 \\
222 & 56966.2965 & 0.0014 & 0&0033 & 277 \\
223 & 56966.3476 & 0.0008 & $-$0&0006 & 279 \\
224 & 56966.3966 & 0.0008 & $-$0&0067 & 140 \\
227 & 56966.5671 & 0.0008 & $-$0&0015 & 125 \\
228 & 56966.6231 & 0.0008 & $-$0&0006 & 99 \\
234 & 56966.9540 & 0.0009 & $-$0&0002 & 61 \\
235 & 56967.0064 & 0.0006 & $-$0&0029 & 61 \\
236 & 56967.0610 & 0.0032 & $-$0&0034 & 21 \\
239 & 56967.2223 & 0.0004 & $-$0&0073 & 102 \\
240 & 56967.4001 & 0.0019 & 0&1154 & 121 \\
241 & 56967.3257 & 0.0029 & $-$0&0141 & 176 \\
242 & 56967.3753 & 0.0019 & $-$0&0196 & 105 \\
245 & 56967.5483 & 0.0040 & $-$0&0119 & 24 \\
253 & 56967.9857 & 0.0019 & $-$0&0152 & 36 \\
254 & 56968.0372 & 0.0031 & $-$0&0188 & 41 \\
255 & 56968.0921 & 0.0018 & $-$0&0189 & 33 \\
\hline
  \multicolumn{6}{l}{\commenta BJD$-$2400000.} \\
  \multicolumn{6}{l}{\commentb Against max $= 2456954.0638 + 0.055087 E$.} \\
  \multicolumn{6}{l}{\commentc Number of points used to determine the maximum.} \\
\end{tabular}
\end{center}
\end{table}

\subsection{OT J230523.1$-$022546}\label{obj:j2305}

\begin{figure}
  \begin{center}
    \FigureFile(88mm,110mm){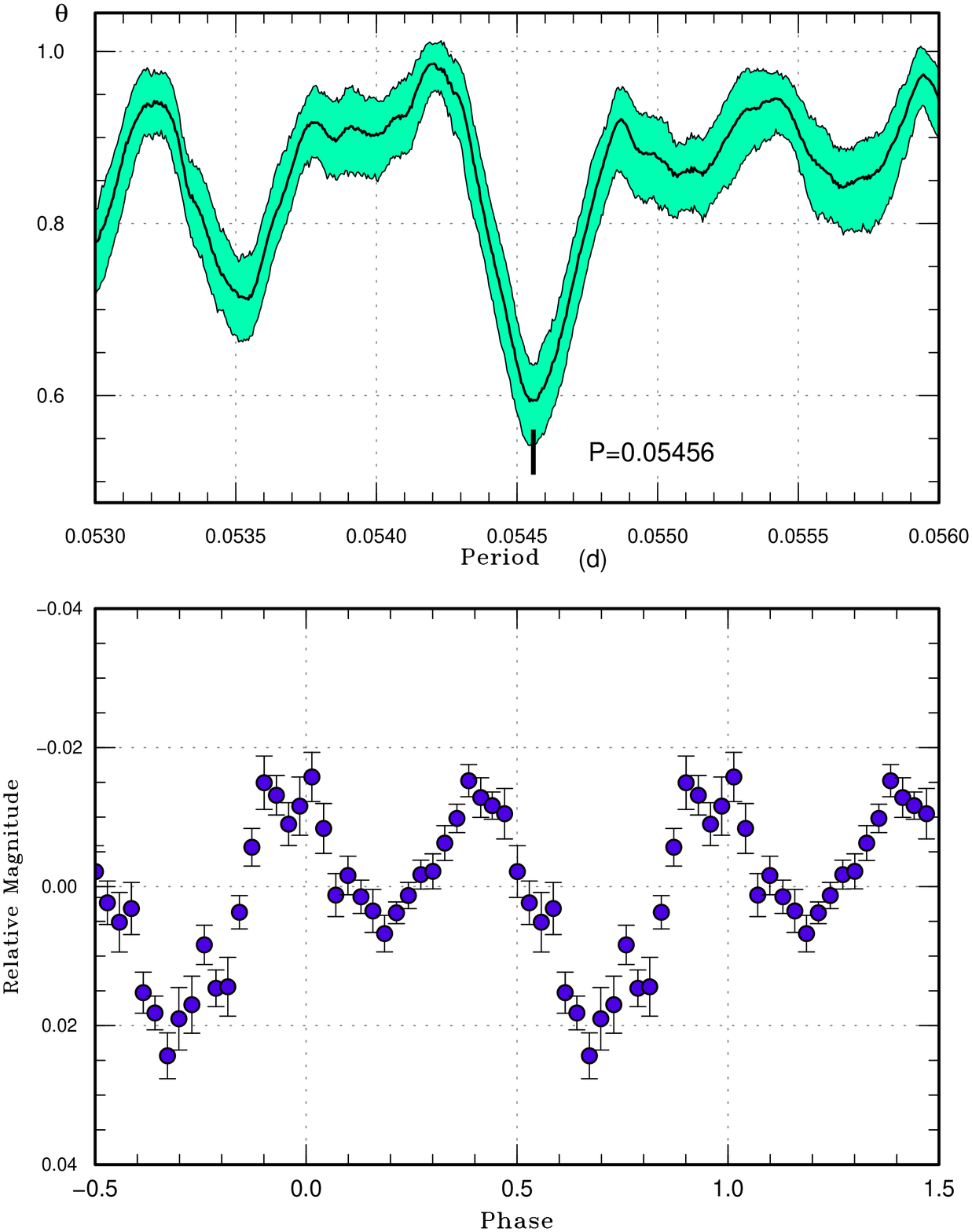}
  \end{center}
  \caption{Early superhumps in OT J230523 (2014).  (Upper): PDM analysis.
     (Lower): Phase-averaged profile.}
  \label{fig:j2305eshpdm}
\end{figure}

\begin{figure}
  \begin{center}
    \FigureFile(88mm,110mm){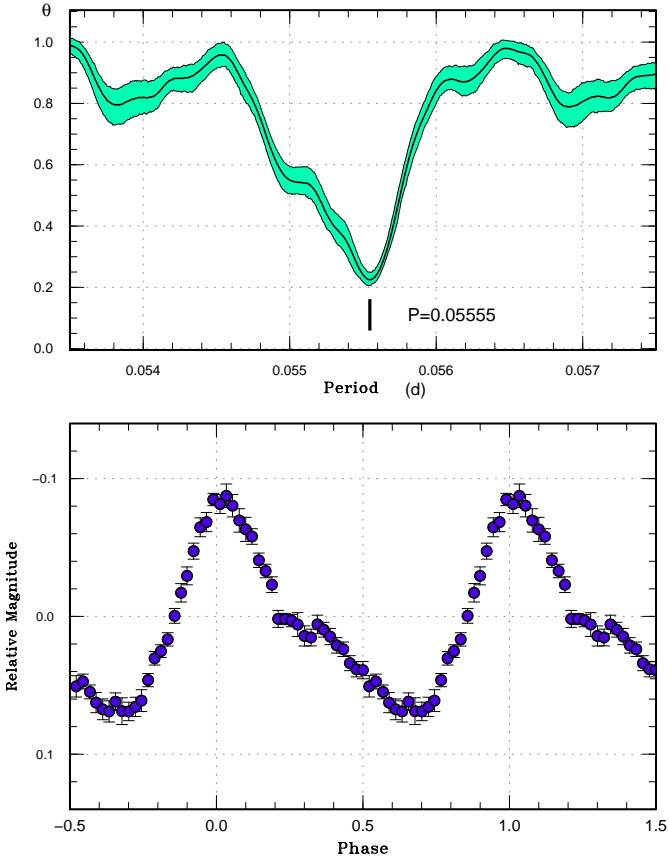}
  \end{center}
  \caption{Ordinary superhumps in OT J230523 (2014)
     for the interval BJD 2456865--2456871.  (Upper): PDM analysis.
     (Lower): Phase-averaged profile.}
  \label{fig:j2305shpdm}
\end{figure}

   This object (=PNV J23052314$-$0225455, hereafter
OT J230523) was discovered as a transient at an unfiltered
CCD magnitude of 12.3 on 2014 July 19 by M. Mukai.\footnote{
  $<$http://www.cbat.eps.harvard.edu/unconf/\\
followups/J23052314$-$0225455.html$>$.
}  Since there was a 19.7-mag ($g$) SDSS blue counterpart
(see also vsnet-alert 17512),
the object was immediately considered as a dwarf nova.
G. Masi also obtained low-resolution spectrum which did not
indicate strong an H$\alpha$ emission line.
Subsequent observations detected early superhumps
(vsnet-alert 17513, 17528, 17571; figure \ref{fig:j2305eshpdm}),
indicating that the object is a WZ Sge-type dwarf nova.
On July 25, the object started to show growing
ordinary superhumps (vsnet-alert 17550).
The period of fully grown superhumps turned out
to be very short (vsnet-alert 17566).
The mean profile of stage B superhumps is shown in
figure \ref{fig:j2305shpdm}.

\begin{figure}
  \begin{center}
    \FigureFile(88mm,100mm){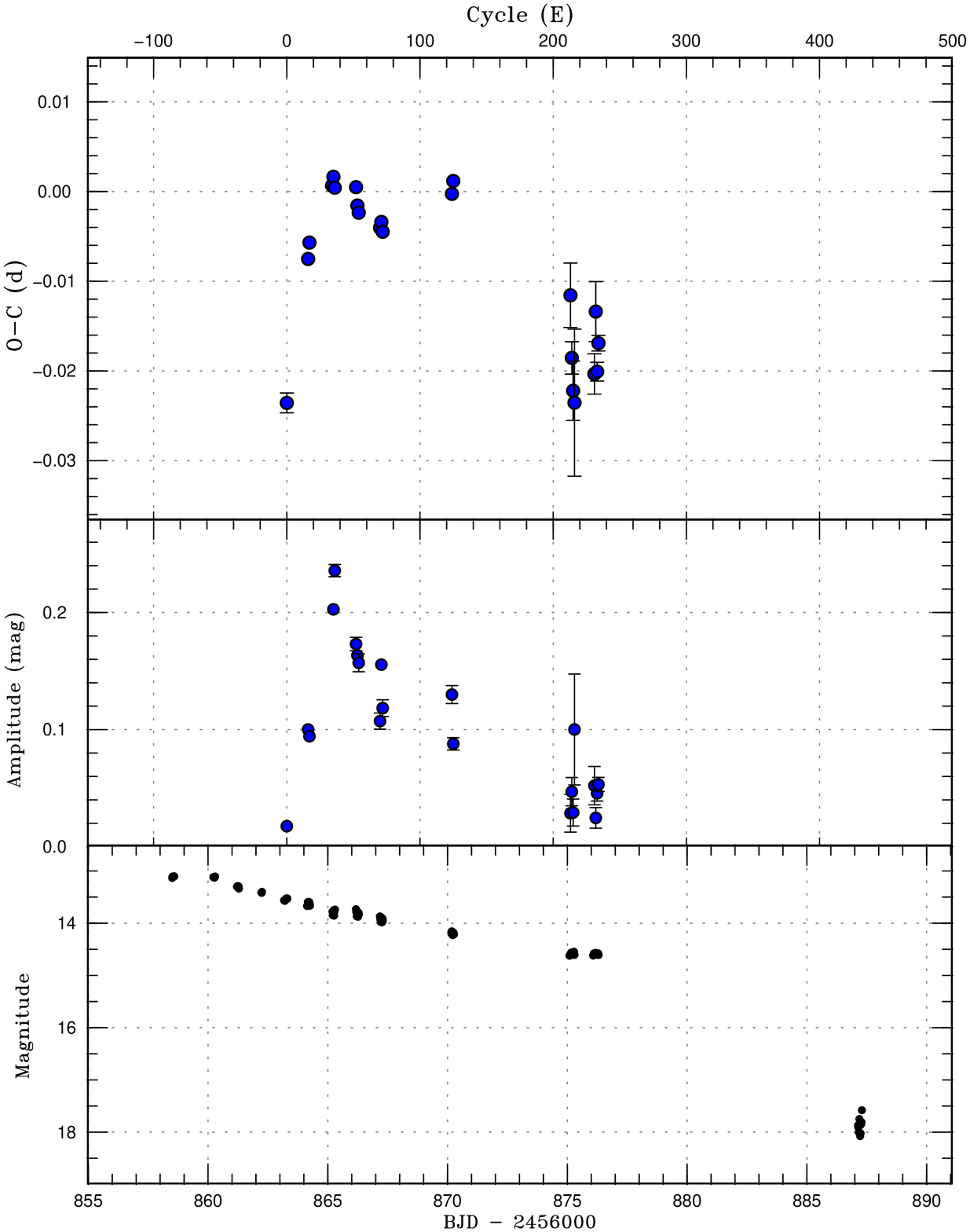}
  \end{center}
  \caption{$O-C$ diagram of superhumps in OT J230523 (2014).
     (Upper:) $O-C$ diagram.
     We used a period of 0.05560~d for calculating the $O-C$ residuals.
     (Middle:) Amplitudes of superhumps.
     (Lower:) Light curve.  The data were binned to 0.011~d.
  }
  \label{fig:j2305humpall}
\end{figure}

   The times of superhump maxima are listed in table
\ref{tab:j2305oc2014}.  There was clear stage A for
$E \le 17$.  Although superhumps after $E=213$ were
likely stage C superhumps, the epoch when stage B-C
transition occurred was not determined due to the gap
in the observation (see figure \ref{fig:j2305humpall}).

   The period of stage A superhumps [0.05663(4)~d]
corresponds to $\epsilon^*$ of 0.037.  This value gives
$q$=0.102(2), which is relatively large for
a WZ Sge-type dwarf nova.  The relatively large $P_{\rm dot}$
of stage B superhumps [$+8.2(1.3) \times 10^{-5}$],
the presence of stage C and relatively short duration
of stage A indicate that OT J230523 is not an extreme
WZ Sge-type and but an object close to ordinary
SU UMa-type dwarf novae.  Thus identification is consistent
with a relatively large $q$.

\begin{table}
\caption{Superhump maxima of OT J230523 (2014)}\label{tab:j2305oc2014}
\begin{center}
\begin{tabular}{rp{55pt}p{40pt}r@{.}lr}
\hline
\multicolumn{1}{c}{$E$} & \multicolumn{1}{c}{max\commenta} & \multicolumn{1}{c}{error} & \multicolumn{2}{c}{$O-C$\commentb} & \multicolumn{1}{c}{$N$\commentc} \\
\hline
0 & 56863.2644 & 0.0011 & $-$0&0226 & 117 \\
16 & 56864.1701 & 0.0002 & $-$0&0054 & 118 \\
17 & 56864.2275 & 0.0002 & $-$0&0036 & 119 \\
34 & 56865.1791 & 0.0007 & 0&0039 & 28 \\
35 & 56865.2356 & 0.0002 & 0&0050 & 119 \\
36 & 56865.2900 & 0.0001 & 0&0039 & 97 \\
52 & 56866.1797 & 0.0003 & 0&0050 & 39 \\
53 & 56866.2332 & 0.0002 & 0&0030 & 61 \\
54 & 56866.2880 & 0.0003 & 0&0023 & 57 \\
70 & 56867.1760 & 0.0006 & 0&0017 & 40 \\
71 & 56867.2322 & 0.0002 & 0&0024 & 61 \\
72 & 56867.2867 & 0.0004 & 0&0014 & 58 \\
124 & 56870.1821 & 0.0005 & 0&0092 & 29 \\
125 & 56870.2392 & 0.0004 & 0&0107 & 41 \\
213 & 56875.1192 & 0.0036 & 0&0039 & 94 \\
214 & 56875.1679 & 0.0018 & $-$0&0030 & 95 \\
215 & 56875.2198 & 0.0033 & $-$0&0066 & 79 \\
216 & 56875.2741 & 0.0082 & $-$0&0078 & 44 \\
231 & 56876.1113 & 0.0022 & $-$0&0036 & 88 \\
232 & 56876.1738 & 0.0033 & 0&0034 & 130 \\
233 & 56876.2227 & 0.0010 & $-$0&0032 & 185 \\
234 & 56876.2815 & 0.0009 & 0&0000 & 199 \\
\hline
  \multicolumn{6}{l}{\commenta BJD$-$2400000.} \\
  \multicolumn{6}{l}{\commentb Against max $= 2456863.2870 + 0.055532 E$.} \\
  \multicolumn{6}{l}{\commentc Number of points used to determine the maximum.} \\
\end{tabular}
\end{center}
\end{table}

\subsection{PNV J17292916$+$0054043}\label{obj:j1729}

   This object (hereafter PNV J172929) was discovered as
a possible nova on May 22 at an unfiltered CCD magnitude
of 12.1 by H. Nishimura.\footnote{
$<$http://www.cbat.eps.harvard.edu/unconf/\\
followups/J17292916+0054043.html$>$.
}
S. Kiyota reported that the blue color suggested
a dwarf nova-type outburst.  There was also a faint
($g$=21.5) SDSS counterpart (vsnet-alert 17325).
Spectroscopic observation by K. Ayani indicated
the dwarf nova-type spectrum showing narrow emission
lines of H$\alpha$, H$\beta$ and He\textsc{ii} 4686
on a blue continuum (see above URL).  The amplitude
and the presence of He\textsc{ii} emission line
suggested the WZ Sge-type classification
(vsnet-alert 17327).

\begin{figure}
  \begin{center}
    \FigureFile(88mm,110mm){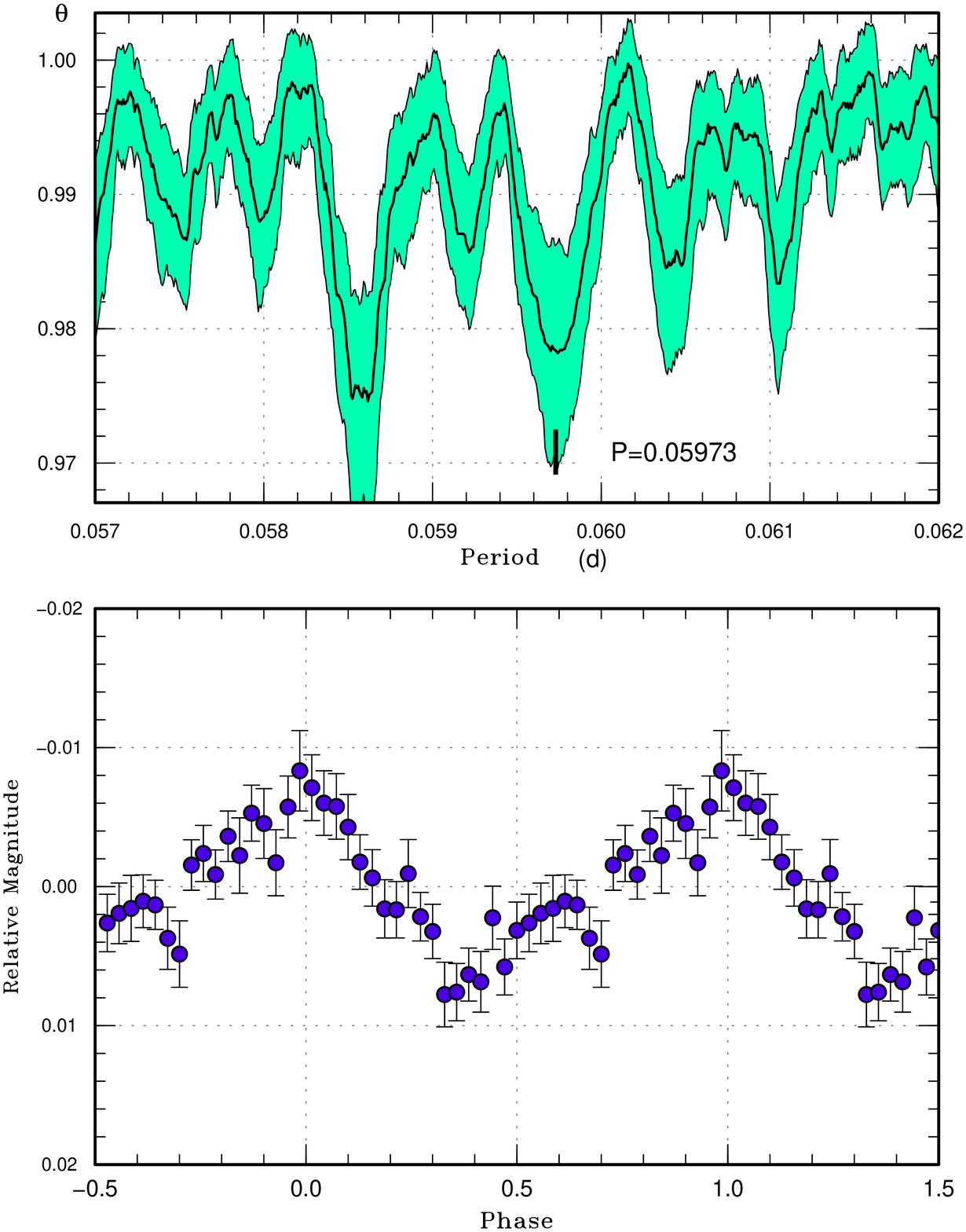}
  \end{center}
  \caption{Early superhumps in PNV J172929 (2014).  (Upper): PDM analysis.
     The selection of the period was based on
     comparison with the period of ordinary superhumps.
     (Lower): Phase-averaged profile.}
  \label{fig:j1729eshpdm}
\end{figure}

\begin{figure}
  \begin{center}
    \FigureFile(88mm,110mm){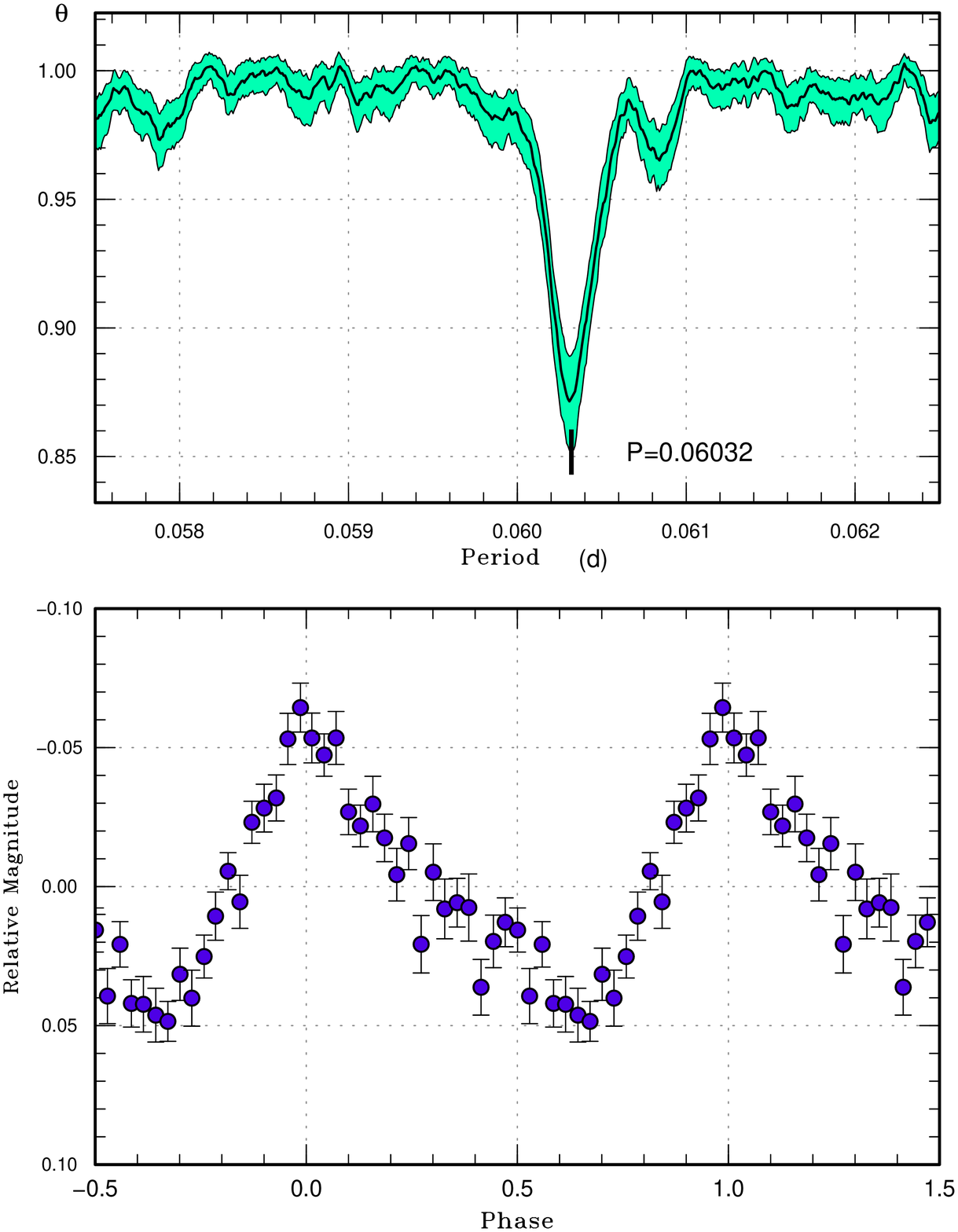}
  \end{center}
  \caption{Ordinary superhumps in PNV J172929 (2014).  (Upper): PDM analysis.
     (Lower): Phase-averaged profile.}
  \label{fig:j1729shpdm}
\end{figure}

\begin{figure}
  \begin{center}
    \FigureFile(88mm,70mm){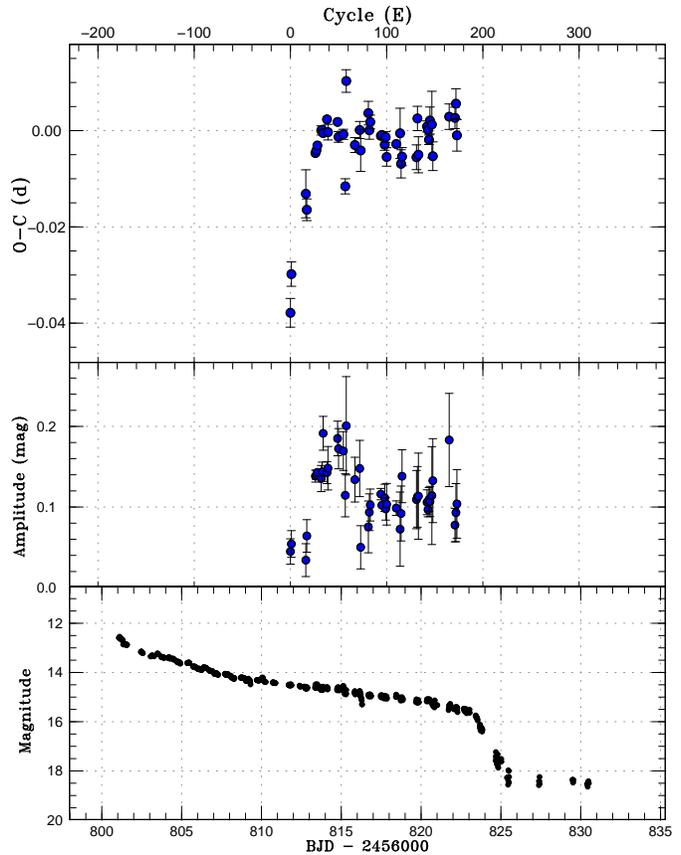}
  \end{center}
  \caption{$O-C$ diagram of superhumps in PNV J172929 (2014).
     (Upper): $O-C$ diagram.  A period of 0.06028~d
     was used to draw this figure.
     (Lower): Light curve.  The observations were binned to 0.012~d.}
  \label{fig:j1729humpall}
\end{figure}

   Subsequent observations detected possible
early superhumps (vsnet-alert 17337;
figure \ref{fig:j1729eshpdm}).
The object developed ordinary superhumps
12~d after the discovery (vsnet-alert 17365;
figure \ref{fig:j1729shpdm}).
The times of maxima of ordinary superhumps are listed in
table \ref{tab:j1729oc2014}.  After a relatively long
stage A ($E \le 28$), stage B appeared
(figure \ref{fig:j1729humpall}).
There was no strong hint of transition to stage C
superhumps during the superoutburst plateau.

   The signal of early superhump was weak and
the data were not sufficient to uniquely determine
the period.  Although a signal around 0.0586~d
was slightly stronger, we have selected a period
of 0.05973(3)~d because this value gave a more
reasonable value of $\epsilon^*$.  The period
of stage A superhumps corresponds to
$\epsilon^*$=0.0273(5) and $q$=0.073(2).

   The orbital period is apparently longer than
the period minimum and the system parameters
resemble those of period bouncers although
$q$ is not so extreme.
This object appears to be similar to recently
identified objects with superhump periods
of 0.058--0.061~d having properties intermediate
between ordinary hydrogen-rich dwarf novae
and period bouncers (\Nakataprep).
There was no indication of rebrightenings in
PNV J172929, which makes a difference from
the objects mentioned in \Nakataprep.

\begin{table}
\caption{Superhump maxima of PNV J172929 (2014)}\label{tab:j1729oc2014}
\begin{center}
\begin{tabular}{rp{55pt}p{40pt}r@{.}lr}
\hline
\multicolumn{1}{c}{$E$} & \multicolumn{1}{c}{max\commenta} & \multicolumn{1}{c}{error} & \multicolumn{2}{c}{$O-C$\commentb} & \multicolumn{1}{c}{$N$\commentc} \\
\hline
0 & 56811.7769 & 0.0030 & $-$0&0283 & 14 \\
1 & 56811.8452 & 0.0025 & $-$0&0203 & 14 \\
16 & 56812.7661 & 0.0050 & $-$0&0048 & 15 \\
17 & 56812.8231 & 0.0023 & $-$0&0082 & 16 \\
26 & 56813.3774 & 0.0004 & 0&0029 & 75 \\
27 & 56813.4382 & 0.0003 & 0&0034 & 83 \\
28 & 56813.4995 & 0.0003 & 0&0044 & 68 \\
32 & 56813.7437 & 0.0010 & 0&0072 & 16 \\
33 & 56813.8042 & 0.0006 & 0&0073 & 25 \\
34 & 56813.8638 & 0.0009 & 0&0065 & 22 \\
38 & 56814.1077 & 0.0006 & 0&0090 & 43 \\
39 & 56814.1653 & 0.0016 & 0&0063 & 16 \\
49 & 56814.7703 & 0.0009 & 0&0077 & 17 \\
50 & 56814.8274 & 0.0011 & 0&0045 & 30 \\
55 & 56815.1294 & 0.0010 & 0&0047 & 77 \\
57 & 56815.2391 & 0.0016 & $-$0&0063 & 85 \\
58 & 56815.3213 & 0.0024 & 0&0155 & 37 \\
67 & 56815.8505 & 0.0015 & 0&0016 & 28 \\
72 & 56816.1550 & 0.0018 & 0&0043 & 101 \\
73 & 56816.2111 & 0.0044 & $-$0&0000 & 115 \\
81 & 56816.7010 & 0.0025 & 0&0071 & 13 \\
82 & 56816.7578 & 0.0019 & 0&0035 & 16 \\
83 & 56816.8198 & 0.0014 & 0&0051 & 29 \\
94 & 56817.4800 & 0.0004 & 0&0015 & 47 \\
95 & 56817.5404 & 0.0003 & 0&0016 & 47 \\
98 & 56817.7192 & 0.0011 & $-$0&0007 & 18 \\
99 & 56817.7811 & 0.0011 & 0&0008 & 18 \\
100 & 56817.8373 & 0.0019 & $-$0&0034 & 17 \\
110 & 56818.4428 & 0.0007 & $-$0&0014 & 33 \\
114 & 56818.6861 & 0.0052 & 0&0005 & 11 \\
115 & 56818.7400 & 0.0029 & $-$0&0059 & 19 \\
116 & 56818.8018 & 0.0019 & $-$0&0045 & 17 \\
\hline
  \multicolumn{6}{l}{\commenta BJD$-$2400000.} \\
  \multicolumn{6}{l}{\commentb Against max $= 2456811.8052 + 0.060354 E$.} \\
  \multicolumn{6}{l}{\commentc Number of points used to determine the maximum.} \\
\end{tabular}
\end{center}
\end{table}

\addtocounter{table}{-1}
\begin{table}
\caption{Superhump maxima of PNV J172929 (2014) (continued)}
\begin{center}
\begin{tabular}{rp{55pt}p{40pt}r@{.}lr}
\hline
\multicolumn{1}{c}{$E$} & \multicolumn{1}{c}{max\commenta} & \multicolumn{1}{c}{error} & \multicolumn{2}{c}{$O-C$\commentb} & \multicolumn{1}{c}{$N$\commentc} \\
\hline
131 & 56819.7059 & 0.0026 & $-$0&0057 & 18 \\
132 & 56819.7742 & 0.0025 & 0&0022 & 18 \\
133 & 56819.8270 & 0.0037 & $-$0&0054 & 18 \\
142 & 56820.3754 & 0.0012 & $-$0&0002 & 30 \\
143 & 56820.4349 & 0.0007 & $-$0&0010 & 31 \\
144 & 56820.4932 & 0.0010 & $-$0&0031 & 32 \\
145 & 56820.5574 & 0.0028 & 0&0008 & 11 \\
147 & 56820.6772 & 0.0069 & $-$0&0001 & 10 \\
148 & 56820.7309 & 0.0030 & $-$0&0068 & 18 \\
165 & 56821.7638 & 0.0026 & 0&0001 & 18 \\
171 & 56822.1253 & 0.0022 & $-$0&0005 & 216 \\
172 & 56822.1885 & 0.0031 & 0&0023 & 97 \\
173 & 56822.2422 & 0.0033 & $-$0&0043 & 98 \\
\hline
  \multicolumn{6}{l}{\commenta BJD$-$2400000.} \\
  \multicolumn{6}{l}{\commentb Against max $= 2456811.8052 + 0.060354 E$.} \\
  \multicolumn{6}{l}{\commentc Number of points used to determine the maximum.} \\
\end{tabular}
\end{center}
\end{table}

\subsection{PTF1 J071912.13$+$485834.0}\label{obj:j0719}

   This object (hereafter PTF1 J071912) is an AM CVn-type
object detected in outburst by Palomar Transient Factory (PTF)
\citep{lev11j0719}.
\citet{lev11j0719} identified the orbital period
of 26.77(2) min [0.01859(1)~d] and reported superoutbursts
recurring with intervals longer than 65~d together
with normal outbursts.
\citet{lev11j0719} recorded superhumps with amplitudes
of $\sim$0.1 mag.  Due to the shortness of the observation,
the only approximated period of 26--27 min was obtained.

\begin{figure}
  \begin{center}
    \FigureFile(88mm,110mm){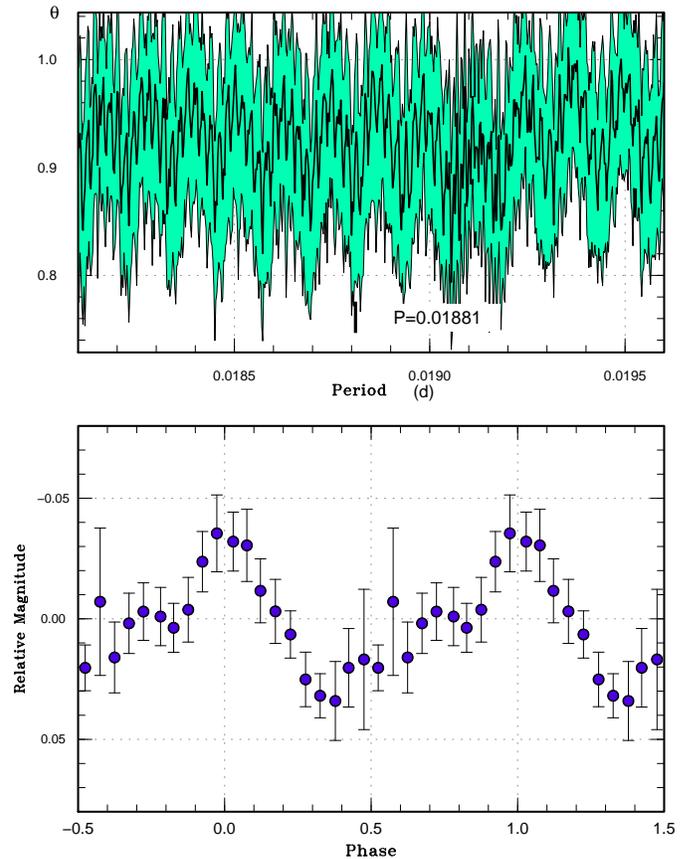}
  \end{center}
  \caption{Superhumps in PTF1 J071912 (2014).  (Upper): PDM analysis.
     (Lower): Phase-averaged profile.}
  \label{fig:j0719shpdm}
\end{figure}

   The 2014 outburst was detected by the ASAS-SN team
on August 31 at $V$=15.57 (vsnet-alert 17685).
Subsequent observations detected superhumps
(vsnet-alert 17688).  This single-night observation
determined the period of 0.0186(3)~d, better than
in \citet{lev11j0719}.  We obtained time-resolved
observations three nights later.  Although there was
a long gap, we have selected a period of 0.01881(1)~d
among alias periods based on the knowledge that
superhump periods of AM CVn-type objects are 0.67--2.18\%
longer than the orbital periods \citep{pea07amcvnSH}\footnote{
   The superhump period in \citet{pea07amcvnSH} was an old
   value.  A more updated period 0.018728~d \citep{kat04v803cen}
   corresponds to 1.3\%.
}, corresponding to 0.01871--0.01900~d in the present case.
figure \ref{fig:j0719shpdm}).
Table \ref{tab:j0719oc2014} is based on this identification.

\begin{table}
\caption{Superhump maxima of PTF1 J071912 (2014)}\label{tab:j0719oc2014}
\begin{center}
\begin{tabular}{rp{55pt}p{40pt}r@{.}lr}
\hline
\multicolumn{1}{c}{$E$} & \multicolumn{1}{c}{max\commenta} & \multicolumn{1}{c}{error} & \multicolumn{2}{c}{$O-C$\commentb} & \multicolumn{1}{c}{$N$\commentc} \\
\hline
0 & 56902.6058 & 0.0006 & 0&0010 & 32 \\
1 & 56902.6244 & 0.0008 & 0&0009 & 36 \\
2 & 56902.6423 & 0.0006 & $-$0&0000 & 36 \\
3 & 56902.6593 & 0.0012 & $-$0&0019 & 33 \\
153 & 56905.4829 & 0.0019 & $-$0&0009 & 9 \\
155 & 56905.5218 & 0.0014 & 0&0004 & 9 \\
158 & 56905.5772 & 0.0018 & $-$0&0007 & 10 \\
159 & 56905.5979 & 0.0006 & 0&0012 & 7 \\
\hline
  \multicolumn{6}{l}{\commenta BJD$-$2400000.} \\
  \multicolumn{6}{l}{\commentb Against max $= 2456902.6047 + 0.018818 E$.} \\
  \multicolumn{6}{l}{\commentc Number of points used to determine the maximum.} \\
\end{tabular}
\end{center}
\end{table}

\subsection{SDSS J033449.86$-$071047.8}\label{obj:j0334}

   This object (hereafter SDSS J033449) was selected as a CV
during the course of the SDSS \citep{szk07SDSSCV6}.
The 2009 superoutburst was reported in \citet{Pdot}.
The 2014 superoutburst was detected by the ASAS-SN team
(vsnet-alert 17711).  Single-night observation was
available and the following superhump maxima were
obtained: BJD 2456909.1948(33) ($N$=69),
2456909.2738(4) ($N$=143) and 2456909.3576(11) ($N$=57).
Two other outbursts (likely normal outbursts) were
recorded on 2013 February 16 (15.5 mag, unfiltered CCD,
vsnet-outburst 15132) and 2014 March 26 (16.6 mag,
unfiltered CCD, AAVSO data).

\subsection{SDSS J081408.42$+$090759.1}\label{obj:j0814}

\begin{figure}
  \begin{center}
    \FigureFile(88mm,110mm){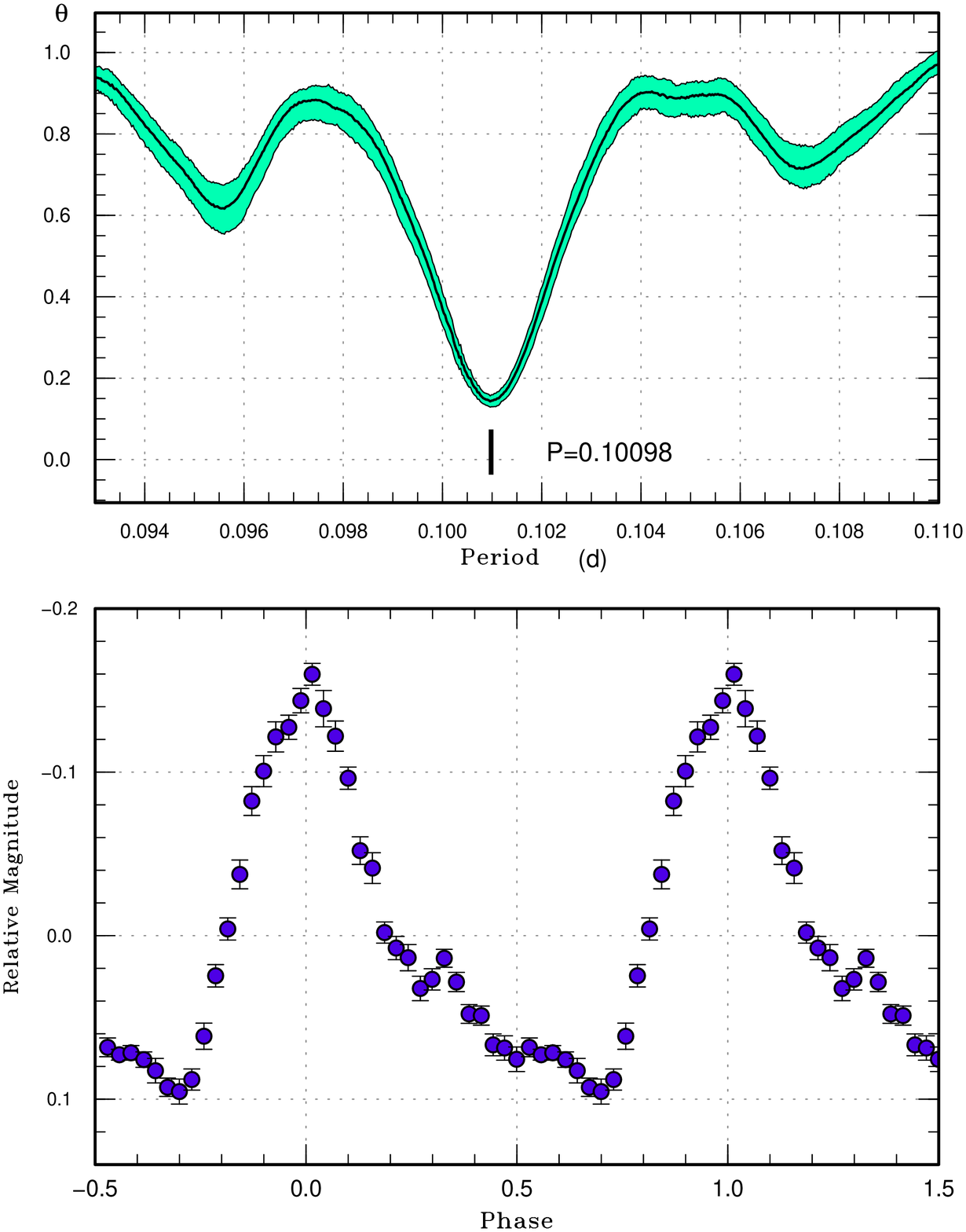}
  \end{center}
  \caption{Superhumps in SDSS J081408 (2014).  (Upper): PDM analysis.
     (Lower): Phase-averaged profile.}
  \label{fig:j081408shpdm}
\end{figure}

   This object (hereafter SDSS J081408) was selected as a CV
by its variability by \citet{wil10newCVs}.  \citet{kat12DNSDSS}
estimated the orbital period to be 0.11--0.14~d from the SDSS
colors of the quiescent counterpart.
CRTS detected an outburst on 2014 March 8.
Subsequent observations detected superhumps
(vsnet-alert 16994, 17000; figure \ref{fig:j081408shpdm}).
The resultant superhump period qualified this object
as an SU UMa-type dwarf nova in the period gap.
The times of superhump maxima are listed in
table \ref{tab:j0814oc2014}.  There was a large decrease
in the superhump period, which is often seen in long-$P_{\rm orb}$
systems [cf. section 4.10 in \citet{Pdot}].

\begin{table}
\caption{Superhump maxima of SDSS J081408 (2014)}\label{tab:j0814oc2014}
\begin{center}
\begin{tabular}{rp{55pt}p{40pt}r@{.}lr}
\hline
\multicolumn{1}{c}{$E$} & \multicolumn{1}{c}{max\commenta} & \multicolumn{1}{c}{error} & \multicolumn{2}{c}{$O-C$\commentb} & \multicolumn{1}{c}{$N$\commentc} \\
\hline
0 & 56724.7195 & 0.0003 & $-$0&0022 & 101 \\
11 & 56725.8337 & 0.0003 & 0&0018 & 97 \\
16 & 56726.3384 & 0.0007 & 0&0019 & 88 \\
17 & 56726.4390 & 0.0004 & 0&0015 & 113 \\
18 & 56726.5377 & 0.0008 & $-$0&0007 & 96 \\
27 & 56727.4444 & 0.0007 & $-$0&0023 & 58 \\
\hline
  \multicolumn{6}{l}{\commenta BJD$-$2400000.} \\
  \multicolumn{6}{l}{\commentb Against max $= 2456724.7217 + 0.100929 E$.} \\
  \multicolumn{6}{l}{\commentc Number of points used to determine the maximum.} \\
\end{tabular}
\end{center}
\end{table}

\subsection{SDSS J090221.35$+$381941.9}\label{obj:j0902}

   This object (hereafter SDSS J090221) is an AM CVn-type
object identified by \citet{rau10HeDN}.  The object underwent
a superoutburst preceded by a precursor in 2014 March.
This superoutburst made SDSS J090221 the longest
$P_{\rm orb}$ object which ever showed outbursts
among AM CVn-type objects.  The analysis of the superhumps,
evolution of the outburst, estimation of $q$ from stage A
superhumps and their implications were discussed in
\citet{kat14j0902}.  Here, we present materials used
in \citet{kat14j0902}.  Since the evolution of the outburst
was so complex, we show here the times of superhump maxima
separately in different stages of the outburst
(tables \ref{tab:j090221oc2014}, \ref{tab:j090221oc2014b},
\ref{tab:j090221oc2014c}, \ref{tab:j090221oc2014d}).
The $E$ values given in table \ref{tab:perlist} refer to
those in table \ref{tab:j090221oc2014c}.

   The enlarged $O-C$ diagram of figure 1 in \citet{kat14j0902}
illustrates the stage B--C transition as well as
a phase hump associated with the ``dip" phenomenon.

   Mean superhump profiles of stages B and C are shown
in figures \ref{fig:j0902shpdm1} and \ref{fig:j0902shpdm2},
respectively.  The mean profile of stage A superhumps
was already shown in \citet{kat14j0902}.  The small
amplitudes of superhumps are consistent with the small $q$
estimated from stage A superhumps \citet{kat14j0902}.
Superhumps persisted after the rapid fading, and the mean
profile is shown in figure \ref{fig:j0902shpdm3}.
After the rebrightening, the superhump signal became
weaker (full amplitude 0.05 mag) and the period could not
be uniquely determined due to the small number of observations.
The variations, however, could be well expressed by
a period close to 0.0336~d.

\begin{figure}
  \begin{center}
    \FigureFile(88mm,70mm){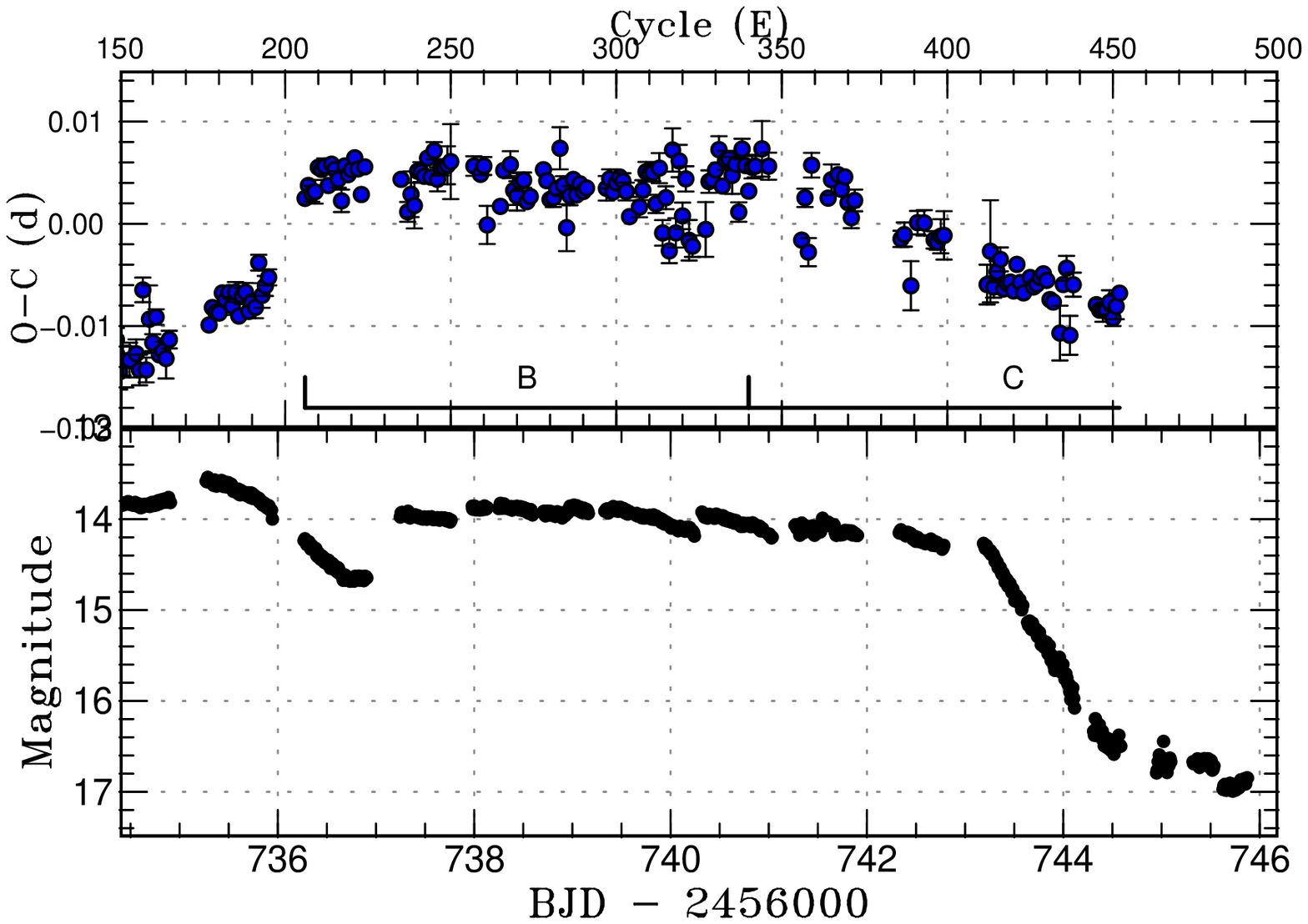}
  \end{center}
  \caption{$O-C$ diagram of superhumps in SDSS J090221 (2014)
     showing the stage B--C transition.
     (Upper): $O-C$ diagram.  A period of 0.03372~d
     was used to draw this figure.
     (Lower): Light curve.  The observations were binned to 0.01~d.}
  \label{fig:j0902humpbc}
\end{figure}

\begin{figure}
  \begin{center}
    \FigureFile(88mm,110mm){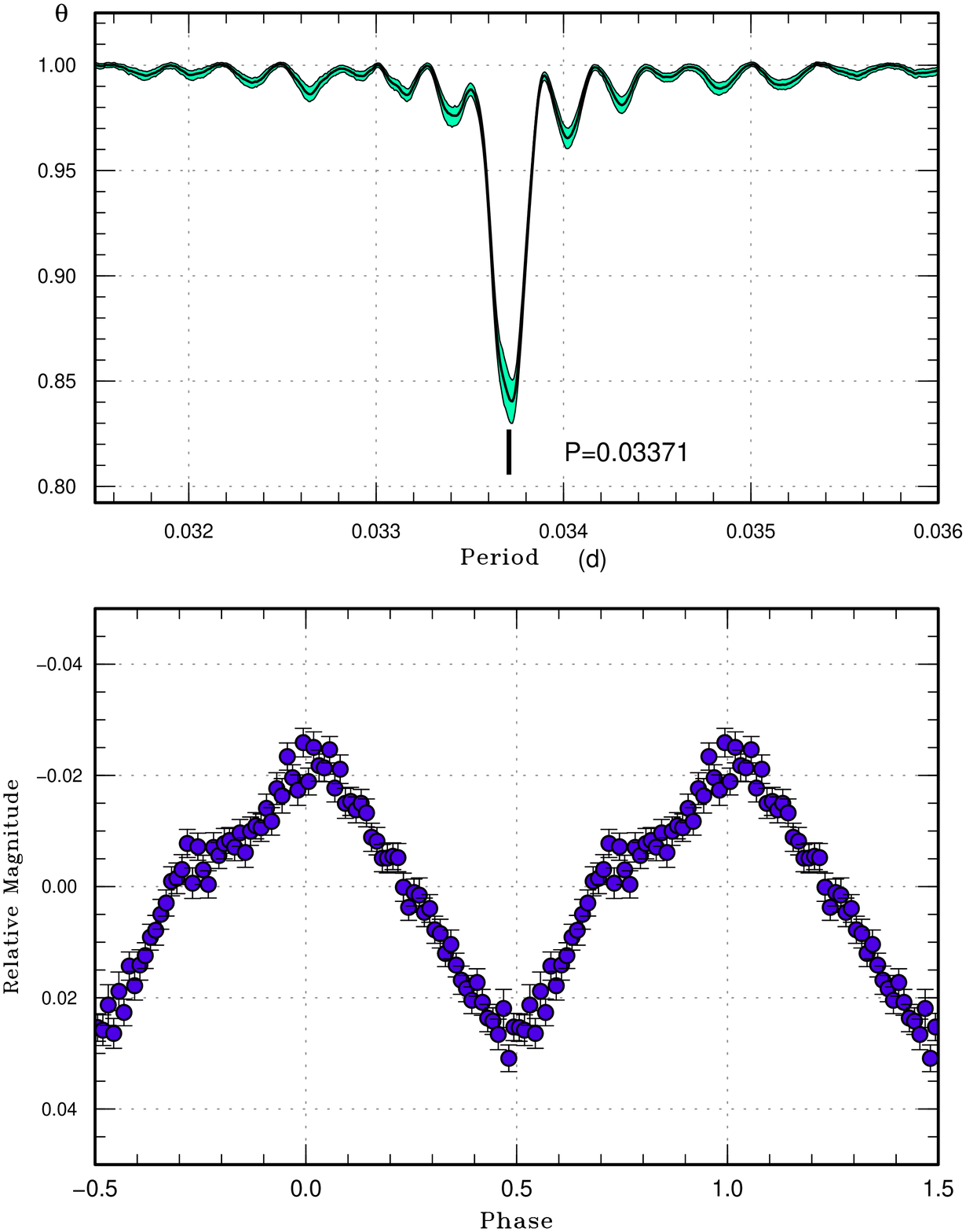}
  \end{center}
  \caption{Stage B superhumps in SDSS J090221 (2014).  (Upper): PDM analysis.
     (Lower): Phase-averaged profile.}
  \label{fig:j0902shpdm1}
\end{figure}

\begin{figure}
  \begin{center}
    \FigureFile(88mm,110mm){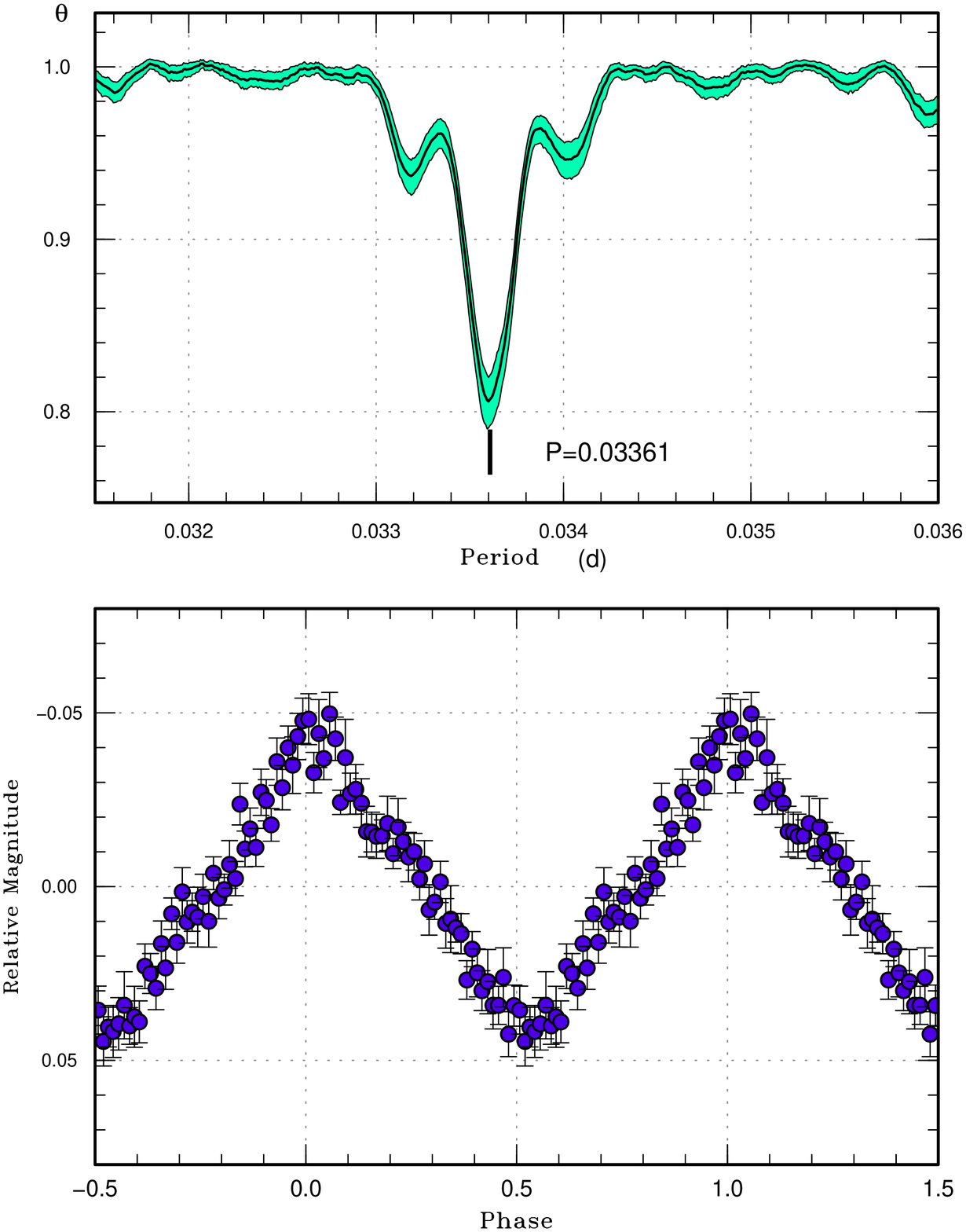}
  \end{center}
  \caption{Stage C superhumps in SDSS J090221 (2014).  (Upper): PDM analysis.
     (Lower): Phase-averaged profile.}
  \label{fig:j0902shpdm2}
\end{figure}

\begin{figure}
  \begin{center}
    \FigureFile(88mm,110mm){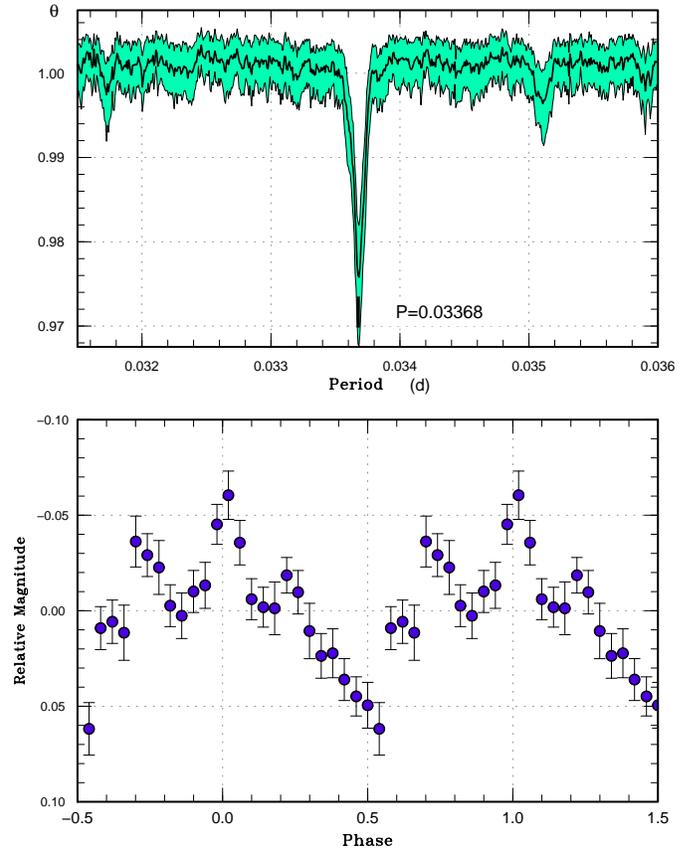}
  \end{center}
  \caption{Superhumps in the post-superoutburst stage before
     the rebrightening in SDSS J090221 (2014).
     (Upper): PDM analysis.
     (Lower): Phase-averaged profile.}
  \label{fig:j0902shpdm3}
\end{figure}

\begin{table}
\caption{Superhump maxima of SDSS J090221 (2014) (slow rising part)}\label{tab:j090221oc2014}
\begin{center}
\begin{tabular}{rp{55pt}p{40pt}r@{.}lr}
\hline
\multicolumn{1}{c}{$E$} & \multicolumn{1}{c}{max\commenta} & \multicolumn{1}{c}{error} & \multicolumn{2}{c}{$O-C$\commentb} & \multicolumn{1}{c}{$N$\commentc} \\
\hline
0 & 56729.2778 & 0.0059 & 0&0016 & 18 \\
1 & 56729.3088 & 0.0015 & $-$0&0016 & 18 \\
2 & 56729.3462 & 0.0014 & 0&0018 & 12 \\
3 & 56729.3805 & 0.0009 & 0&0019 & 18 \\
5 & 56729.4408 & 0.0027 & $-$0&0061 & 14 \\
29 & 56730.2652 & 0.0012 & $-$0&0005 & 17 \\
30 & 56730.2994 & 0.0011 & $-$0&0004 & 18 \\
31 & 56730.3356 & 0.0021 & 0&0016 & 16 \\
32 & 56730.3661 & 0.0010 & $-$0&0020 & 25 \\
33 & 56730.4020 & 0.0012 & $-$0&0002 & 27 \\
34 & 56730.4373 & 0.0018 & 0&0010 & 16 \\
35 & 56730.4716 & 0.0011 & 0&0011 & 21 \\
36 & 56730.5050 & 0.0016 & 0&0004 & 31 \\
37 & 56730.5415 & 0.0028 & 0&0028 & 25 \\
38 & 56730.5742 & 0.0016 & 0&0014 & 18 \\
58 & 56731.2540 & 0.0011 & $-$0&0012 & 19 \\
59 & 56731.2894 & 0.0010 & 0&0000 & 17 \\
60 & 56731.3215 & 0.0013 & $-$0&0019 & 11 \\
61 & 56731.3570 & 0.0010 & $-$0&0005 & 16 \\
64 & 56731.4581 & 0.0013 & $-$0&0018 & 13 \\
90 & 56732.3503 & 0.0011 & 0&0033 & 39 \\
91 & 56732.3852 & 0.0012 & 0&0041 & 40 \\
92 & 56732.4176 & 0.0008 & 0&0024 & 40 \\
93 & 56732.4518 & 0.0015 & 0&0024 & 27 \\
94 & 56732.4836 & 0.0009 & 0&0001 & 27 \\
95 & 56732.5165 & 0.0016 & $-$0&0012 & 23 \\
96 & 56732.5524 & 0.0013 & 0&0007 & 60 \\
97 & 56732.5858 & 0.0006 & $-$0&0001 & 115 \\
98 & 56732.6193 & 0.0009 & $-$0&0007 & 73 \\
99 & 56732.6537 & 0.0009 & $-$0&0004 & 80 \\
100 & 56732.6862 & 0.0014 & $-$0&0020 & 23 \\
101 & 56732.7183 & 0.0021 & $-$0&0040 & 15 \\
102 & 56732.7545 & 0.0011 & $-$0&0020 & 14 \\
\hline
  \multicolumn{6}{l}{\commenta BJD$-$2400000.} \\
  \multicolumn{6}{l}{\commentb Against max $= 2456729.2763 + 0.034120 E$.} \\
  \multicolumn{6}{l}{\commentc Number of points used to determine the maximum.} \\
\end{tabular}
\end{center}
\end{table}

\begin{table}
\caption{Superhump maxima of SDSS J090221 (2014) (rise to superoutburst)}\label{tab:j090221oc2014b}
\begin{center}
\begin{tabular}{rp{55pt}p{40pt}r@{.}lr}
\hline
\multicolumn{1}{c}{$E$} & \multicolumn{1}{c}{max\commenta} & \multicolumn{1}{c}{error} & \multicolumn{2}{c}{$O-C$\commentb} & \multicolumn{1}{c}{$N$\commentc} \\
\hline
0 & 56733.3333 & 0.0018 & 0&0026 & 32 \\
1 & 56733.3633 & 0.0006 & $-$0&0012 & 65 \\
3 & 56733.4342 & 0.0015 & 0&0022 & 65 \\
4 & 56733.4672 & 0.0010 & 0&0013 & 65 \\
5 & 56733.4996 & 0.0014 & $-$0&0001 & 61 \\
6 & 56733.5372 & 0.0036 & 0&0037 & 40 \\
7 & 56733.5660 & 0.0005 & $-$0&0013 & 30 \\
10 & 56733.6707 & 0.0041 & 0&0021 & 31 \\
11 & 56733.7031 & 0.0021 & 0&0007 & 37 \\
12 & 56733.7366 & 0.0011 & 0&0004 & 93 \\
14 & 56733.8041 & 0.0014 & 0&0003 & 109 \\
15 & 56733.8406 & 0.0037 & 0&0030 & 106 \\
30 & 56734.3442 & 0.0011 & $-$0&0004 & 37 \\
31 & 56734.3748 & 0.0017 & $-$0&0036 & 64 \\
32 & 56734.4096 & 0.0015 & $-$0&0026 & 65 \\
34 & 56734.4771 & 0.0017 & $-$0&0027 & 52 \\
36 & 56734.5451 & 0.0014 & $-$0&0023 & 74 \\
37 & 56734.5773 & 0.0015 & $-$0&0039 & 69 \\
38 & 56734.6188 & 0.0012 & 0&0038 & 64 \\
39 & 56734.6447 & 0.0012 & $-$0&0041 & 69 \\
40 & 56734.6834 & 0.0033 & 0&0008 & 91 \\
41 & 56734.7148 & 0.0008 & $-$0&0016 & 79 \\
42 & 56734.7510 & 0.0008 & 0&0009 & 59 \\
43 & 56734.7811 & 0.0007 & $-$0&0029 & 75 \\
44 & 56734.8151 & 0.0009 & $-$0&0026 & 61 \\
45 & 56734.8481 & 0.0020 & $-$0&0034 & 39 \\
46 & 56734.8837 & 0.0009 & $-$0&0016 & 38 \\
58 & 56735.2898 & 0.0002 & $-$0&0011 & 139 \\
59 & 56735.3252 & 0.0003 & 0&0005 & 198 \\
60 & 56735.3585 & 0.0003 & $-$0&0000 & 206 \\
61 & 56735.3921 & 0.0005 & $-$0&0002 & 236 \\
62 & 56735.4278 & 0.0004 & 0&0017 & 98 \\
63 & 56735.4608 & 0.0004 & 0&0009 & 190 \\
\hline
  \multicolumn{6}{l}{\commenta BJD$-$2400000.} \\
  \multicolumn{6}{l}{\commentb Against max $= 2456733.3307 + 0.033797 E$.} \\
  \multicolumn{6}{l}{\commentc Number of points used to determine the maximum.} \\
\end{tabular}
\end{center}
\end{table}

\addtocounter{table}{-1}
\begin{table}
\caption{Superhump maxima of SDSS J090221 (2014) (rise to superoutburst, continued)}
\begin{center}
\begin{tabular}{rp{55pt}p{40pt}r@{.}lr}
\hline
\multicolumn{1}{c}{$E$} & \multicolumn{1}{c}{max\commenta} & \multicolumn{1}{c}{error} & \multicolumn{2}{c}{$O-C$\commentb} & \multicolumn{1}{c}{$N$\commentc} \\
\hline
64 & 56735.4953 & 0.0006 & 0&0016 & 152 \\
65 & 56735.5277 & 0.0009 & 0&0002 & 136 \\
66 & 56735.5628 & 0.0010 & 0&0015 & 52 \\
67 & 56735.5941 & 0.0005 & $-$0&0010 & 50 \\
68 & 56735.6298 & 0.0008 & 0&0009 & 45 \\
69 & 56735.6639 & 0.0009 & 0&0012 & 56 \\
70 & 56735.6957 & 0.0006 & $-$0&0007 & 80 \\
71 & 56735.7304 & 0.0007 & 0&0002 & 41 \\
72 & 56735.7636 & 0.0011 & $-$0&0005 & 39 \\
73 & 56735.8017 & 0.0008 & 0&0038 & 39 \\
74 & 56735.8322 & 0.0012 & 0&0005 & 43 \\
75 & 56735.8668 & 0.0010 & 0&0014 & 43 \\
76 & 56735.9014 & 0.0008 & 0&0021 & 21 \\
\hline
  \multicolumn{6}{l}{\commenta BJD$-$2400000.} \\
  \multicolumn{6}{l}{\commentb Against max $= 2456733.3307 + 0.033797 E$.} \\
  \multicolumn{6}{l}{\commentc Number of points used to determine the maximum.} \\
\end{tabular}
\end{center}
\end{table}

\begin{table}
\caption{Superhump maxima of SDSS J090221 (2014) (superoutburst plateau)}\label{tab:j090221oc2014c}
\begin{center}
\begin{tabular}{rp{55pt}p{40pt}r@{.}lr}
\hline
\multicolumn{1}{c}{$E$} & \multicolumn{1}{c}{max\commenta} & \multicolumn{1}{c}{error} & \multicolumn{2}{c}{$O-C$\commentb} & \multicolumn{1}{c}{$N$\commentc} \\
\hline
0 & 56736.2800 & 0.0006 & $-$0&0048 & 109 \\
1 & 56736.3150 & 0.0007 & $-$0&0035 & 111 \\
2 & 56736.3479 & 0.0008 & $-$0&0042 & 145 \\
3 & 56736.3818 & 0.0012 & $-$0&0040 & 111 \\
4 & 56736.4180 & 0.0006 & $-$0&0016 & 34 \\
5 & 56736.4515 & 0.0007 & $-$0&0017 & 47 \\
6 & 56736.4855 & 0.0008 & $-$0&0013 & 36 \\
7 & 56736.5173 & 0.0006 & $-$0&0032 & 42 \\
8 & 56736.5531 & 0.0006 & $-$0&0010 & 101 \\
9 & 56736.5864 & 0.0007 & $-$0&0015 & 96 \\
10 & 56736.6191 & 0.0008 & $-$0&0024 & 79 \\
11 & 56736.6507 & 0.0011 & $-$0&0045 & 135 \\
12 & 56736.6879 & 0.0005 & $-$0&0010 & 166 \\
13 & 56736.7207 & 0.0007 & $-$0&0018 & 164 \\
14 & 56736.7549 & 0.0005 & $-$0&0013 & 167 \\
15 & 56736.7898 & 0.0005 & $-$0&0001 & 166 \\
16 & 56736.8224 & 0.0004 & $-$0&0011 & 135 \\
17 & 56736.8537 & 0.0006 & $-$0&0036 & 126 \\
18 & 56736.8901 & 0.0005 & $-$0&0008 & 102 \\
29 & 56737.2598 & 0.0004 & $-$0&0015 & 37 \\
31 & 56737.3240 & 0.0010 & $-$0&0045 & 37 \\
32 & 56737.3595 & 0.0022 & $-$0&0028 & 75 \\
33 & 56737.3921 & 0.0022 & $-$0&0038 & 75 \\
34 & 56737.4292 & 0.0007 & $-$0&0004 & 120 \\
35 & 56737.4628 & 0.0010 & $-$0&0005 & 142 \\
36 & 56737.4961 & 0.0007 & $-$0&0008 & 139 \\
37 & 56737.5316 & 0.0007 & 0&0010 & 135 \\
38 & 56737.5635 & 0.0007 & $-$0&0008 & 137 \\
39 & 56737.5998 & 0.0008 & 0&0018 & 164 \\
40 & 56737.6307 & 0.0012 & $-$0&0009 & 57 \\
41 & 56737.6656 & 0.0008 & 0&0003 & 196 \\
42 & 56737.6993 & 0.0006 & 0&0004 & 182 \\
43 & 56737.7332 & 0.0019 & 0&0006 & 203 \\
\hline
  \multicolumn{6}{l}{\commenta BJD$-$2400000.} \\
  \multicolumn{6}{l}{\commentb Against max $= 2456736.2848 + 0.033669 E$.} \\
  \multicolumn{6}{l}{\commentc Number of points used to determine the maximum.} \\
\end{tabular}
\end{center}
\end{table}

\addtocounter{table}{-1}
\begin{table}
\caption{Superhump maxima of SDSS J090221 (2014) (superoutburst plateau, continued)}
\begin{center}
\begin{tabular}{rp{55pt}p{40pt}r@{.}lr}
\hline
\multicolumn{1}{c}{$E$} & \multicolumn{1}{c}{max\commenta} & \multicolumn{1}{c}{error} & \multicolumn{2}{c}{$O-C$\commentb} & \multicolumn{1}{c}{$N$\commentc} \\
\hline
44 & 56737.7673 & 0.0037 & 0&0010 & 48 \\
51 & 56738.0029 & 0.0010 & 0&0010 & 19 \\
53 & 56738.0696 & 0.0006 & 0&0003 & 25 \\
54 & 56738.1041 & 0.0009 & 0&0011 & 21 \\
55 & 56738.1321 & 0.0019 & $-$0&0046 & 15 \\
59 & 56738.2687 & 0.0005 & $-$0&0026 & 36 \\
60 & 56738.3060 & 0.0004 & 0&0010 & 37 \\
62 & 56738.3740 & 0.0013 & 0&0017 & 111 \\
63 & 56738.4052 & 0.0014 & $-$0&0008 & 106 \\
64 & 56738.4384 & 0.0014 & $-$0&0013 & 72 \\
65 & 56738.4734 & 0.0009 & 0&0001 & 73 \\
66 & 56738.5073 & 0.0007 & 0&0003 & 67 \\
67 & 56738.5390 & 0.0005 & $-$0&0017 & 73 \\
68 & 56738.5732 & 0.0006 & $-$0&0012 & 36 \\
72 & 56738.7107 & 0.0007 & 0&0017 & 66 \\
73 & 56738.7433 & 0.0004 & 0&0006 & 75 \\
74 & 56738.7752 & 0.0005 & $-$0&0011 & 75 \\
75 & 56738.8091 & 0.0009 & $-$0&0009 & 75 \\
76 & 56738.8437 & 0.0007 & 0&0000 & 75 \\
77 & 56738.8814 & 0.0021 & 0&0040 & 70 \\
78 & 56738.9115 & 0.0009 & 0&0004 & 75 \\
79 & 56738.9411 & 0.0023 & $-$0&0037 & 74 \\
80 & 56738.9779 & 0.0009 & $-$0&0005 & 24 \\
81 & 56739.0132 & 0.0005 & 0&0012 & 74 \\
82 & 56739.0454 & 0.0006 & $-$0&0003 & 85 \\
83 & 56739.0802 & 0.0004 & 0&0008 & 74 \\
84 & 56739.1132 & 0.0005 & 0&0001 & 69 \\
85 & 56739.1473 & 0.0006 & 0&0005 & 68 \\
91 & 56739.3496 & 0.0012 & 0&0008 & 73 \\
92 & 56739.3842 & 0.0009 & 0&0018 & 75 \\
93 & 56739.4166 & 0.0005 & 0&0005 & 94 \\
94 & 56739.4512 & 0.0006 & 0&0015 & 101 \\
95 & 56739.4855 & 0.0007 & 0&0021 & 101 \\
\hline
  \multicolumn{6}{l}{\commenta BJD$-$2400000.} \\
  \multicolumn{6}{l}{\commentb Against max $= 2456736.2848 + 0.033669 E$.} \\
  \multicolumn{6}{l}{\commentc Number of points used to determine the maximum.} \\
\end{tabular}
\end{center}
\end{table}

\addtocounter{table}{-1}
\begin{table}
\caption{Superhump maxima of SDSS J090221 (2014) (superoutburst plateau, continued)}
\begin{center}
\begin{tabular}{rp{55pt}p{40pt}r@{.}lr}
\hline
\multicolumn{1}{c}{$E$} & \multicolumn{1}{c}{max\commenta} & \multicolumn{1}{c}{error} & \multicolumn{2}{c}{$O-C$\commentb} & \multicolumn{1}{c}{$N$\commentc} \\
\hline
96 & 56739.5188 & 0.0008 & 0&0018 & 102 \\
97 & 56739.5516 & 0.0009 & 0&0008 & 35 \\
98 & 56739.5828 & 0.0007 & $-$0&0016 & 22 \\
101 & 56739.6849 & 0.0010 & $-$0&0005 & 81 \\
102 & 56739.7203 & 0.0010 & 0&0012 & 78 \\
103 & 56739.7558 & 0.0010 & 0&0030 & 84 \\
104 & 56739.7893 & 0.0007 & 0&0029 & 85 \\
105 & 56739.8232 & 0.0008 & 0&0031 & 78 \\
106 & 56739.8538 & 0.0009 & 0&0000 & 84 \\
107 & 56739.8910 & 0.0015 & 0&0036 & 75 \\
108 & 56739.9184 & 0.0012 & $-$0&0027 & 136 \\
109 & 56739.9556 & 0.0011 & 0&0008 & 160 \\
110 & 56739.9841 & 0.0012 & $-$0&0044 & 172 \\
111 & 56740.0277 & 0.0021 & 0&0056 & 219 \\
112 & 56740.0533 & 0.0014 & $-$0&0025 & 183 \\
113 & 56740.0941 & 0.0016 & 0&0046 & 196 \\
114 & 56740.1224 & 0.0013 & $-$0&0007 & 215 \\
115 & 56740.1598 & 0.0012 & 0&0029 & 164 \\
116 & 56740.1875 & 0.0020 & $-$0&0030 & 122 \\
117 & 56740.2206 & 0.0011 & $-$0&0035 & 122 \\
121 & 56740.3571 & 0.0027 & $-$0&0017 & 98 \\
122 & 56740.3955 & 0.0011 & 0&0030 & 114 \\
123 & 56740.4294 & 0.0007 & 0&0033 & 180 \\
124 & 56740.4641 & 0.0007 & 0&0043 & 175 \\
125 & 56740.4998 & 0.0013 & 0&0063 & 195 \\
126 & 56740.5300 & 0.0007 & 0&0028 & 206 \\
127 & 56740.5662 & 0.0009 & 0&0054 & 207 \\
128 & 56740.6001 & 0.0008 & 0&0056 & 238 \\
129 & 56740.6322 & 0.0018 & 0&0040 & 108 \\
130 & 56740.6670 & 0.0006 & 0&0051 & 211 \\
131 & 56740.6960 & 0.0010 & 0&0005 & 309 \\
132 & 56740.7359 & 0.0010 & 0&0067 & 300 \\
133 & 56740.7680 & 0.0005 & 0&0052 & 249 \\
\hline
  \multicolumn{6}{l}{\commenta BJD$-$2400000.} \\
  \multicolumn{6}{l}{\commentb Against max $= 2456736.2848 + 0.033669 E$.} \\
  \multicolumn{6}{l}{\commentc Number of points used to determine the maximum.} \\
\end{tabular}
\end{center}
\end{table}

\addtocounter{table}{-1}
\begin{table}
\caption{Superhump maxima of SDSS J090221 (2014) (superoutburst plateau, continued)}
\begin{center}
\begin{tabular}{rp{55pt}p{40pt}r@{.}lr}
\hline
\multicolumn{1}{c}{$E$} & \multicolumn{1}{c}{max\commenta} & \multicolumn{1}{c}{error} & \multicolumn{2}{c}{$O-C$\commentb} & \multicolumn{1}{c}{$N$\commentc} \\
\hline
134 & 56740.7992 & 0.0006 & 0&0027 & 233 \\
135 & 56740.8353 & 0.0011 & 0&0051 & 216 \\
136 & 56740.8691 & 0.0011 & 0&0053 & 189 \\
138 & 56740.9382 & 0.0027 & 0&0070 & 72 \\
140 & 56741.0040 & 0.0013 & 0&0055 & 49 \\
150 & 56741.3340 & 0.0004 & $-$0&0013 & 76 \\
151 & 56741.3718 & 0.0009 & 0&0029 & 97 \\
152 & 56741.4002 & 0.0014 & $-$0&0023 & 117 \\
153 & 56741.4425 & 0.0012 & 0&0062 & 85 \\
158 & 56741.6078 & 0.0007 & 0&0032 & 117 \\
159 & 56741.6434 & 0.0015 & 0&0051 & 84 \\
161 & 56741.7113 & 0.0005 & 0&0057 & 104 \\
162 & 56741.7436 & 0.0007 & 0&0043 & 104 \\
163 & 56741.7785 & 0.0005 & 0&0056 & 102 \\
164 & 56741.8097 & 0.0007 & 0&0031 & 102 \\
165 & 56741.8420 & 0.0010 & 0&0017 & 102 \\
166 & 56741.8774 & 0.0011 & 0&0034 & 102 \\
180 & 56742.3457 & 0.0008 & 0&0004 & 41 \\
181 & 56742.3798 & 0.0012 & 0&0008 & 54 \\
183 & 56742.4423 & 0.0024 & $-$0&0041 & 39 \\
185 & 56742.5159 & 0.0011 & 0&0022 & 25 \\
187 & 56742.5833 & 0.0012 & 0&0023 & 34 \\
190 & 56742.6828 & 0.0009 & 0&0008 & 64 \\
191 & 56742.7163 & 0.0008 & 0&0006 & 62 \\
192 & 56742.7506 & 0.0017 & 0&0012 & 64 \\
193 & 56742.7844 & 0.0024 & 0&0014 & 32 \\
206 & 56743.2179 & 0.0019 & $-$0&0028 & 17 \\
207 & 56743.2549 & 0.0050 & 0&0005 & 13 \\
208 & 56743.2851 & 0.0010 & $-$0&0029 & 32 \\
209 & 56743.3204 & 0.0010 & $-$0&0014 & 30 \\
210 & 56743.3553 & 0.0012 & $-$0&0001 & 30 \\
211 & 56743.3861 & 0.0006 & $-$0&0030 & 71 \\
212 & 56743.4204 & 0.0004 & $-$0&0024 & 93 \\
\hline
  \multicolumn{6}{l}{\commenta BJD$-$2400000.} \\
  \multicolumn{6}{l}{\commentb Against max $= 2456736.2848 + 0.033669 E$.} \\
  \multicolumn{6}{l}{\commentc Number of points used to determine the maximum.} \\
\end{tabular}
\end{center}
\end{table}

\addtocounter{table}{-1}
\begin{table}
\caption{Superhump maxima of SDSS J090221 (2014) (superoutburst plateau, continued)}
\begin{center}
\begin{tabular}{rp{55pt}p{40pt}r@{.}lr}
\hline
\multicolumn{1}{c}{$E$} & \multicolumn{1}{c}{max\commenta} & \multicolumn{1}{c}{error} & \multicolumn{2}{c}{$O-C$\commentb} & \multicolumn{1}{c}{$N$\commentc} \\
\hline
213 & 56743.4542 & 0.0004 & $-$0&0022 & 97 \\
214 & 56743.4871 & 0.0005 & $-$0&0030 & 92 \\
215 & 56743.5234 & 0.0005 & $-$0&0003 & 89 \\
216 & 56743.5553 & 0.0004 & $-$0&0021 & 37 \\
217 & 56743.5880 & 0.0006 & $-$0&0030 & 22 \\
219 & 56743.6570 & 0.0004 & $-$0&0014 & 70 \\
220 & 56743.6898 & 0.0002 & $-$0&0023 & 64 \\
221 & 56743.7238 & 0.0005 & $-$0&0020 & 66 \\
222 & 56743.7582 & 0.0003 & $-$0&0013 & 72 \\
223 & 56743.7922 & 0.0003 & $-$0&0009 & 70 \\
224 & 56743.8253 & 0.0004 & $-$0&0015 & 72 \\
225 & 56743.8571 & 0.0003 & $-$0&0033 & 54 \\
226 & 56743.8906 & 0.0005 & $-$0&0035 & 72 \\
228 & 56743.9550 & 0.0027 & $-$0&0064 & 15 \\
229 & 56743.9935 & 0.0005 & $-$0&0016 & 39 \\
230 & 56744.0288 & 0.0012 & 0&0000 & 83 \\
231 & 56744.0560 & 0.0019 & $-$0&0065 & 95 \\
232 & 56744.0946 & 0.0012 & $-$0&0015 & 92 \\
239 & 56744.3287 & 0.0006 & $-$0&0031 & 45 \\
240 & 56744.3619 & 0.0007 & $-$0&0036 & 72 \\
241 & 56744.3956 & 0.0011 & $-$0&0035 & 97 \\
242 & 56744.4293 & 0.0006 & $-$0&0035 & 99 \\
243 & 56744.4638 & 0.0012 & $-$0&0026 & 63 \\
244 & 56744.4960 & 0.0008 & $-$0&0041 & 45 \\
245 & 56744.5309 & 0.0013 & $-$0&0029 & 35 \\
246 & 56744.5659 & 0.0006 & $-$0&0016 & 24 \\
\hline
  \multicolumn{6}{l}{\commenta BJD$-$2400000.} \\
  \multicolumn{6}{l}{\commentb Against max $= 2456736.2848 + 0.033669 E$.} \\
  \multicolumn{6}{l}{\commentc Number of points used to determine the maximum.} \\
\end{tabular}
\end{center}
\end{table}

\begin{table}
\caption{Superhump maxima of SDSS J090221 (2014) (rapid fading)}\label{tab:j090221oc2014d}
\begin{center}
\begin{tabular}{rp{55pt}p{40pt}r@{.}lr}
\hline
\multicolumn{1}{c}{$E$} & \multicolumn{1}{c}{max\commenta} & \multicolumn{1}{c}{error} & \multicolumn{2}{c}{$O-C$\commentb} & \multicolumn{1}{c}{$N$\commentc} \\
\hline
0 & 56754.9713 & 0.0008 & $-$0&0009 & 64 \\
1 & 56755.0049 & 0.0007 & $-$0&0011 & 64 \\
2 & 56755.0385 & 0.0009 & $-$0&0014 & 65 \\
3 & 56755.0758 & 0.0013 & 0&0021 & 65 \\
4 & 56755.1058 & 0.0028 & $-$0&0017 & 23 \\
9 & 56755.2813 & 0.0007 & 0&0046 & 32 \\
10 & 56755.3099 & 0.0010 & $-$0&0007 & 45 \\
11 & 56755.3451 & 0.0007 & 0&0007 & 70 \\
12 & 56755.3784 & 0.0006 & 0&0002 & 72 \\
13 & 56755.4114 & 0.0012 & $-$0&0007 & 87 \\
14 & 56755.4461 & 0.0007 & 0&0002 & 74 \\
15 & 56755.4803 & 0.0008 & 0&0005 & 61 \\
16 & 56755.5134 & 0.0010 & $-$0&0002 & 39 \\
22 & 56755.7168 & 0.0007 & 0&0002 & 72 \\
23 & 56755.7495 & 0.0004 & $-$0&0009 & 72 \\
24 & 56755.7831 & 0.0008 & $-$0&0011 & 70 \\
22 & 56755.7169 & 0.0007 & 0&0003 & 70 \\
\hline
  \multicolumn{6}{l}{\commenta BJD$-$2400000.} \\
  \multicolumn{6}{l}{\commentb Against max $= 2456754.9722 + 0.033836 E$.} \\
  \multicolumn{6}{l}{\commentc Number of points used to determine the maximum.} \\
\end{tabular}
\end{center}
\end{table}

\subsection{SDSS J120231.01$+$450349.1}\label{obj:j1202}

   This object (hereafter SDSS J120231) was selected as
a CV during the course of SDSS \citep{szk06SDSSCV5}
The SDSS spectrum was indicative of a low mass-transfer
rate with absorption lines of a white dwarf.  The spectrum
itself resembled that of a WZ Sge-type dwarf nova.

   \citet{qui06j1202atel787} detected an outburst on 2006
March 21.15 UT at 13.4 mag with ROTSE-IIIb telescope.
This outburst faded to 14.8 mag over the ten days.
The quiescent SDSS colors also suggested a short
(0.056~d) orbital period \citep{kat12DNSDSS}.

   The 2014 outburst was detected by ASAS-SN on May 27
at $V$=13.88.  The object was not detected in outburst
on May 25 (vsnet-alert 17341).

   Subsequent observations recorded well-developed
superhumps (vsnet-alert 17348).  Considering that
the outburst detection was made sufficiently early,
the lack of a stage with early superhumps indicates
that this object is an ordinary SU UMa-type dwarf nova
rather than a WZ Sge-type object as suggested from
spectroscopy.

\begin{figure}
  \begin{center}
    \FigureFile(88mm,110mm){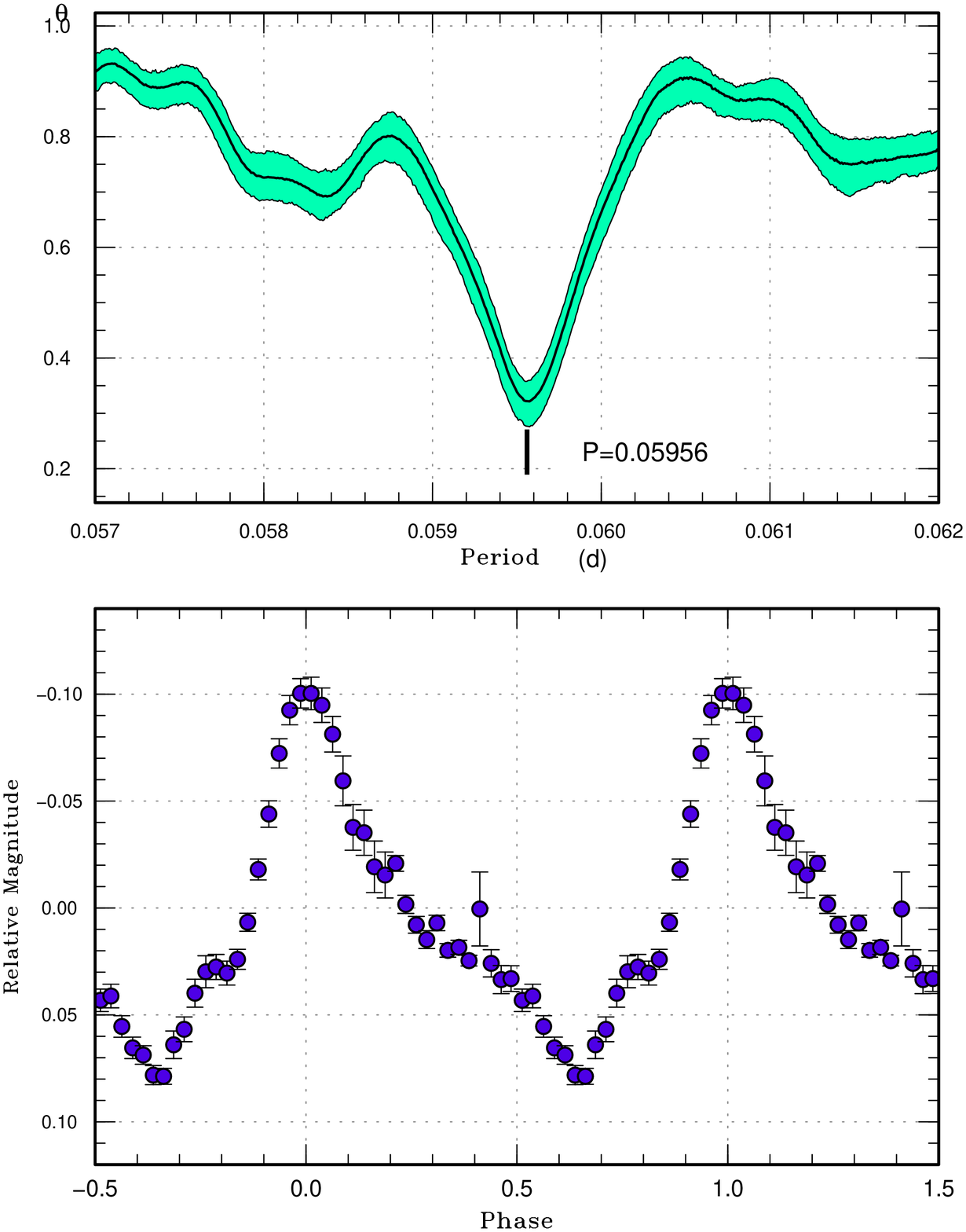}
  \end{center}
  \caption{Superhumps in SDSS J120231 (2014, stage B).
     (Upper): PDM analysis.
     (Lower): Phase-averaged profile.}
  \label{fig:j1202shpdm1}
\end{figure}

\begin{figure}
  \begin{center}
    \FigureFile(88mm,110mm){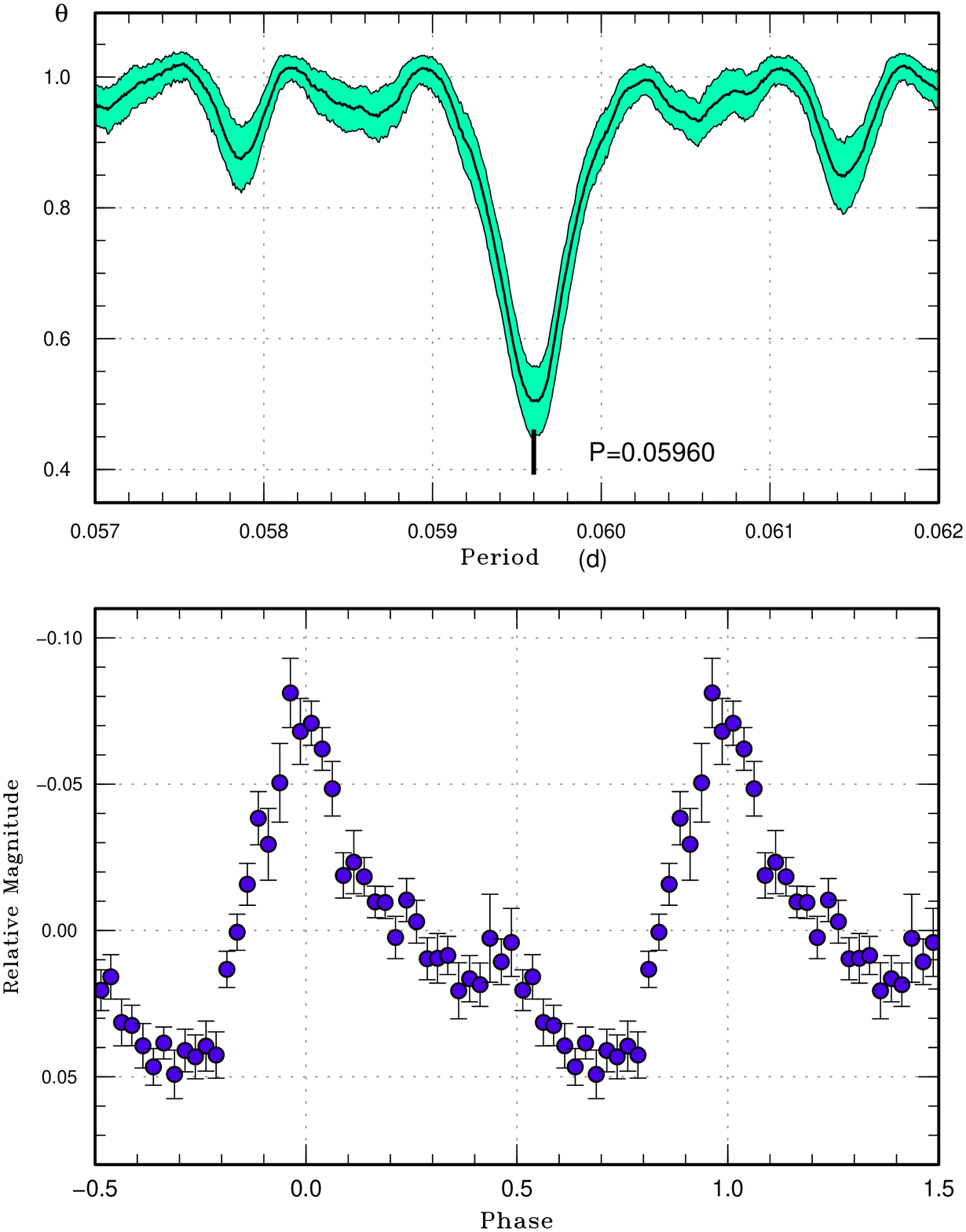}
  \end{center}
  \caption{Superhumps in SDSS J120231 (2014, stage C).
     (Upper): PDM analysis.
     (Lower): Phase-averaged profile.}
  \label{fig:j1202shpdm2}
\end{figure}

   The times of superhump maxima are listed in table
\ref{tab:j1202oc2014}.  Although there was a gap in
the observation, the entire observation can be well
interpreted by stage B with a large positive $P_{\rm dot}$
and a transition to stage C.  The $P_{\rm dot}$ of
$+9.0(1.3) \times 10^{-5}$ in stage B is typical 
for this $P_{\rm SH}$.  The profiles of superhumps
are given separately for stages A and B
(figures \ref{fig:j1202shpdm1}, \ref{fig:j1202shpdm2})
since the large change in the period, which is supposed to be 
present in the later part of stage B but was not covered
by observation, caused a split in the superhump signal
if the combined data are used.  It would be worth
mentioning that the superhump periods of early stage B
and stage C are almost the same (as described in
\cite{Pdot}).

\begin{table}
\caption{Superhump maxima of SDSS J120231 (2014)}\label{tab:j1202oc2014}
\begin{center}
\begin{tabular}{rp{55pt}p{40pt}r@{.}lr}
\hline
\multicolumn{1}{c}{$E$} & \multicolumn{1}{c}{max\commenta} & \multicolumn{1}{c}{error} & \multicolumn{2}{c}{$O-C$\commentb} & \multicolumn{1}{c}{$N$\commentc} \\
\hline
0 & 56806.4104 & 0.0002 & 0&0022 & 94 \\
1 & 56806.4700 & 0.0002 & 0&0020 & 92 \\
2 & 56806.5297 & 0.0003 & 0&0019 & 68 \\
16 & 56807.3663 & 0.0004 & 0&0013 & 62 \\
17 & 56807.4272 & 0.0002 & 0&0024 & 143 \\
18 & 56807.4826 & 0.0003 & $-$0&0021 & 124 \\
19 & 56807.5441 & 0.0005 & $-$0&0004 & 87 \\
51 & 56809.4493 & 0.0003 & $-$0&0088 & 47 \\
52 & 56809.5098 & 0.0002 & $-$0&0081 & 61 \\
53 & 56809.5711 & 0.0008 & $-$0&0065 & 33 \\
134 & 56814.4280 & 0.0004 & 0&0066 & 49 \\
135 & 56814.4894 & 0.0005 & 0&0081 & 53 \\
150 & 56815.3838 & 0.0006 & 0&0055 & 36 \\
151 & 56815.4452 & 0.0006 & 0&0071 & 52 \\
152 & 56815.5047 & 0.0004 & 0&0068 & 41 \\
167 & 56816.3978 & 0.0009 & 0&0029 & 20 \\
168 & 56816.4560 & 0.0009 & 0&0013 & 53 \\
169 & 56816.5139 & 0.0018 & $-$0&0006 & 31 \\
184 & 56817.4124 & 0.0006 & 0&0009 & 52 \\
185 & 56817.4698 & 0.0007 & $-$0&0015 & 54 \\
200 & 56818.3587 & 0.0010 & $-$0&0096 & 19 \\
201 & 56818.4223 & 0.0018 & $-$0&0058 & 28 \\
202 & 56818.4822 & 0.0013 & $-$0&0057 & 21 \\
\hline
  \multicolumn{6}{l}{\commenta BJD$-$2400000.} \\
  \multicolumn{6}{l}{\commentb Against max $= 2456806.4082 + 0.059800 E$.} \\
  \multicolumn{6}{l}{\commentc Number of points used to determine the maximum.} \\
\end{tabular}
\end{center}
\end{table}

\subsection{SDSS J140037.99$+$572341.3}\label{obj:j1400}

\begin{figure}
  \begin{center}
    \FigureFile(88mm,110mm){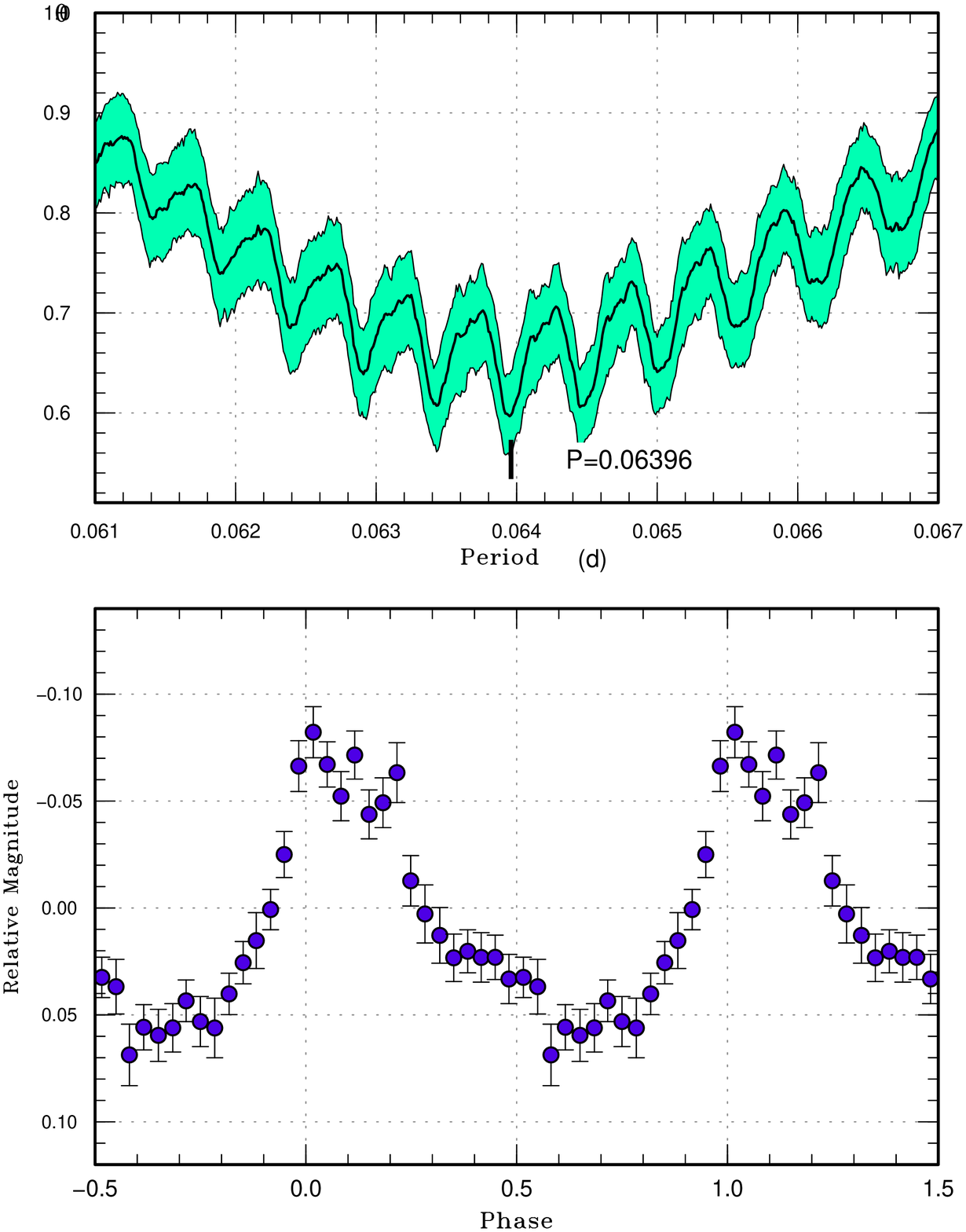}
  \end{center}
  \caption{Superhumps in SDSS J140037 during the plateau
     phase (2015).
     (Upper): PDM analysis.  The shown period is one of possible
     aliases, which has been selected based on the long, continuous
     data on the first night.
     (Lower): Phase-averaged profile.}
  \label{fig:j1400shpdm}
\end{figure}

   This object (hereafter SDSS J140037) was selected
from SDSS and spectroscopically confirmed as a CV
by \citet{car13SDSSamcvn}.  No previous outburst was
known.  The 2015 outburst was detected by the ASAS-SN team
on February 2 (vsnet-alert 18257).  Superhumps were immediately
detected (vsnet-alert 18263, 18265; figure \ref{fig:j1400shpdm}).
The times of superhump maxima are listed in table
\ref{tab:j1400oc2015}.  There remains ambiguity in cycle
count between $E=8$ and $E=123$.  Among the possible aliases
(see figure \ref{fig:j1400shpdm}), we have selected the most
likely period based on the period determined
by the PDM method from the observations on the first night
was 0.0639(2)~d.  This selection, however, may not be true
if there were significant variations in the period during
the 8~d gap.
The object already faded from the superoutburst
on February 17, 15~d after the initial detection.

\begin{table}
\caption{Superhump maxima of SDSS J140037 (2015)}\label{tab:j1400oc2015}
\begin{center}
\begin{tabular}{rp{55pt}p{40pt}r@{.}lr}
\hline
\multicolumn{1}{c}{$E$} & \multicolumn{1}{c}{max\commenta} & \multicolumn{1}{c}{error} & \multicolumn{2}{c}{$O-C$\commentb} & \multicolumn{1}{c}{$N$\commentc} \\
\hline
0 & 57057.4814 & 0.0013 & 0&0005 & 73 \\
1 & 57057.5429 & 0.0015 & $-$0&0018 & 99 \\
2 & 57057.6082 & 0.0010 & $-$0&0006 & 81 \\
5 & 57057.8007 & 0.0006 & 0&0001 & 48 \\
6 & 57057.8627 & 0.0005 & $-$0&0019 & 67 \\
7 & 57057.9250 & 0.0008 & $-$0&0035 & 61 \\
8 & 57057.9998 & 0.0009 & 0&0073 & 19 \\
123 & 57065.3426 & 0.0023 & $-$0&0046 & 37 \\
124 & 57065.4116 & 0.0018 & 0&0005 & 35 \\
125 & 57065.4789 & 0.0014 & 0&0039 & 35 \\
\hline
  \multicolumn{6}{l}{\commenta BJD$-$2400000.} \\
  \multicolumn{6}{l}{\commentb Against max $= 2457057.4808 + 0.063954 E$.} \\
  \multicolumn{6}{l}{\commentc Number of points used to determine the maximum.} \\
\end{tabular}
\end{center}
\end{table}

\subsection{SDSS J172325.99$+$330414.1}\label{obj:j1723}

\begin{figure}
  \begin{center}
    \FigureFile(88mm,110mm){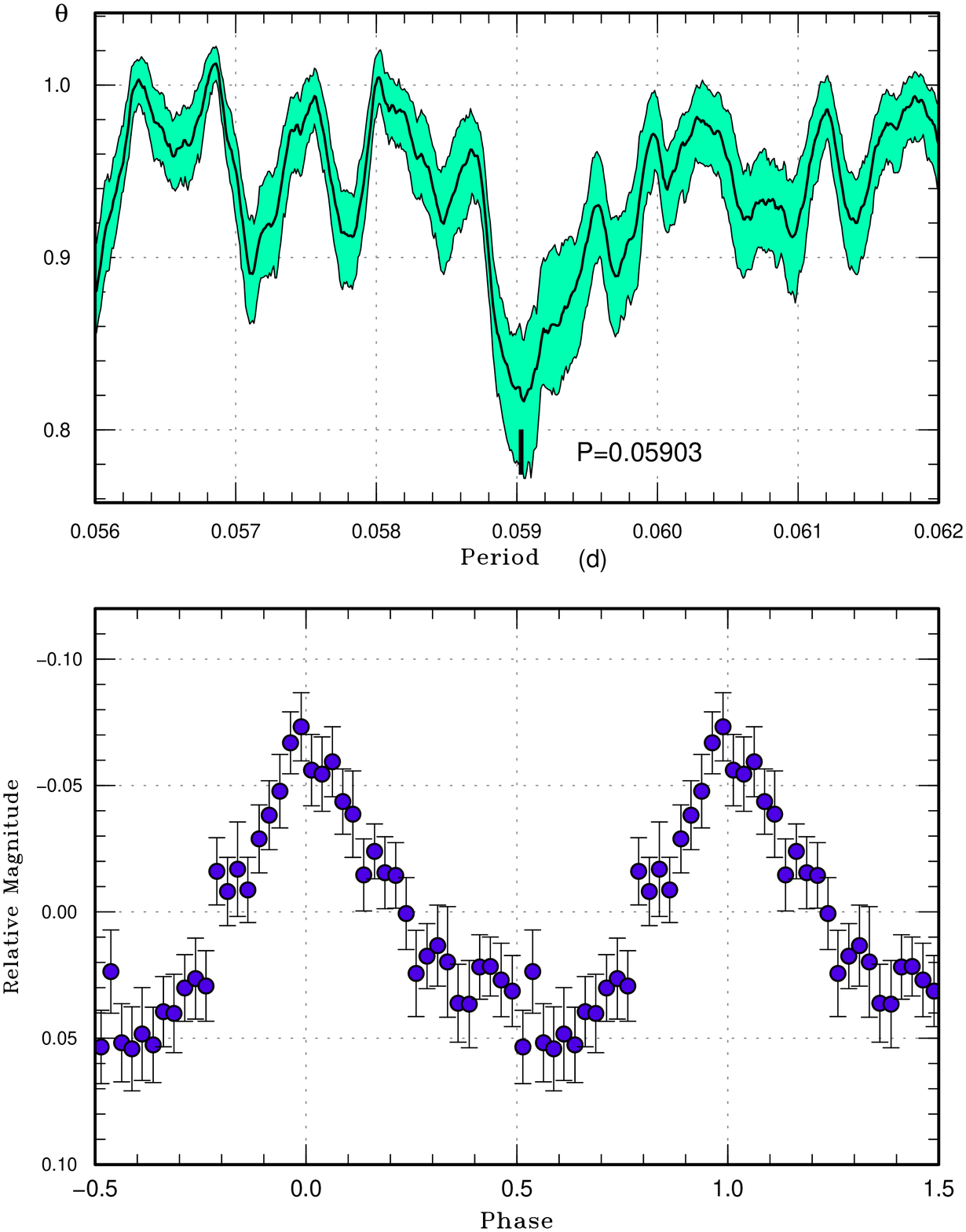}
  \end{center}
  \caption{Superhumps in SDSS J172325 (2014).  The data between
     BJD 2456916.5 and BJD 2456928 were used.
     (Upper): PDM analysis.
     (Lower): Phase-averaged profile.}
  \label{fig:j1723shpdm}
\end{figure}

   This object (hereafter SDSS J172325) was reported in
outburst by the ASAS-SN team (=ASASSN-14gz,
\cite{dav14j1723atel6455}) at $V$=14.2 on 2014 September 9.
Since the spectrum in quiescence was already obtained
in the SDSS archive and was clearly recognized as a CV,
we call this object by the SDSS name.
The large outburst amplitude (7.7 mag) was suggestive
of a WZ Sge-type dwarf nova \citep{dav14j1723atel6455}.
On September 16, this object started to show ordinary
superhumps (vsnet-alert 17737, 17755, 17800;
figure \ref{fig:j1723shpdm}).
The times of superhump maxima are listed in table
\ref{tab:j1723oc2014}.  Although $E \le 1$ most likely
corresponded to stage A superhumps as judged from
the growing amplitude, the period of stage A superhumps
was not determined due to the lack of the data.

\begin{figure}
  \begin{center}
    \FigureFile(88mm,70mm){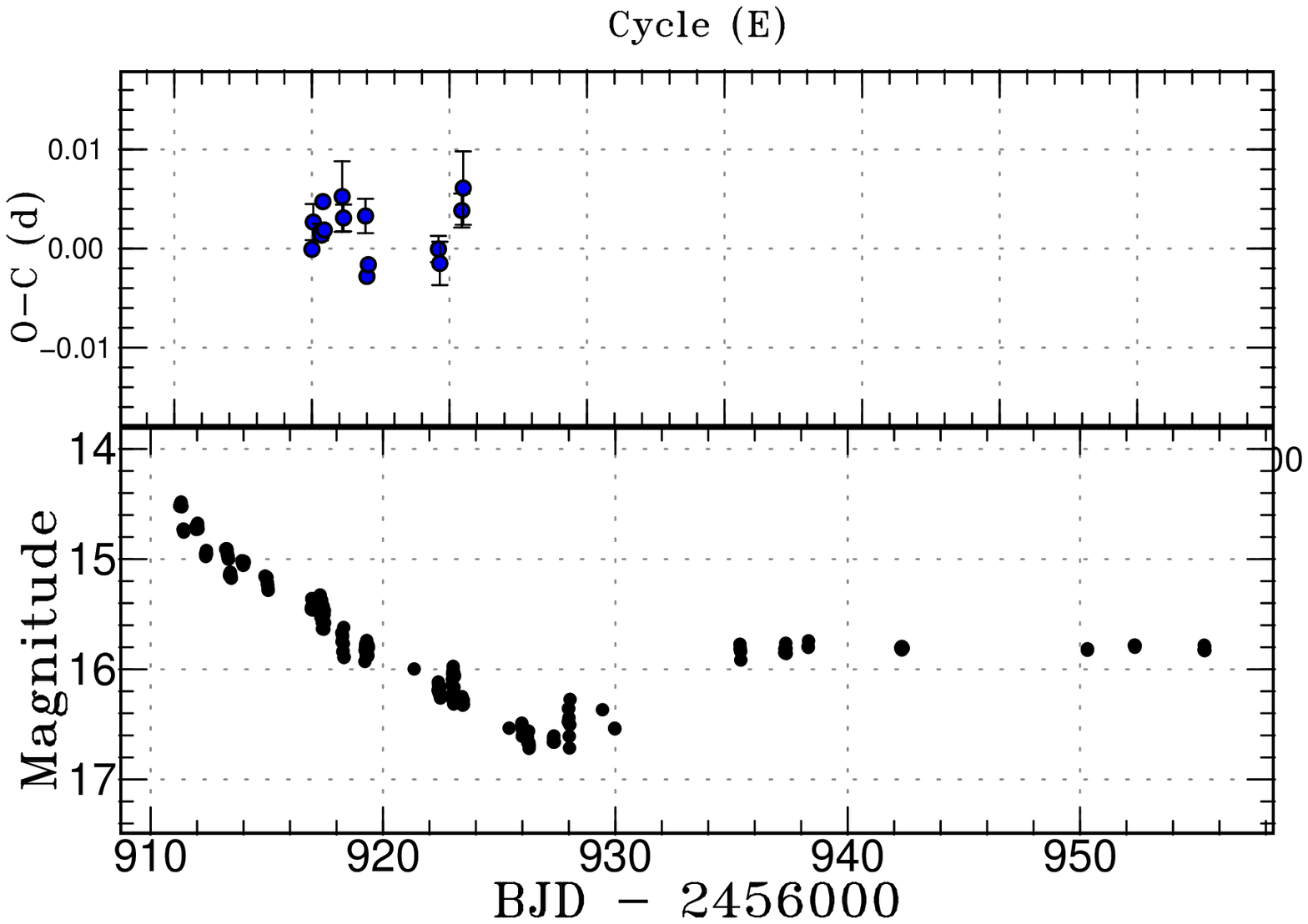}
  \end{center}
  \caption{$O-C$ diagram of superhumps in SDSS J172325 (2014).
     (Upper): $O-C$ diagram.  A period of 0.05920~d
     was used to draw this figure.
     (Lower): Light curve.  The observations were binned to 0.011~d.}
  \label{fig:j1723humpall}
\end{figure}

   After fading to $V$=16.6 on September 26, the object
underwent a rebrightening to $V$=15.8 on October 4
(vsnet-alert 17824) and remained at this level at least
until October 24 (figure \ref{fig:j1723humpall}).

   There was no convincing signal of early superhumps
before the appearance of ordinary superhumps.

\begin{table}
\caption{Superhump maxima of SDSS J172325 (2014)}\label{tab:j1723oc2014}
\begin{center}
\begin{tabular}{rp{55pt}p{40pt}r@{.}lr}
\hline
\multicolumn{1}{c}{$E$} & \multicolumn{1}{c}{max\commenta} & \multicolumn{1}{c}{error} & \multicolumn{2}{c}{$O-C$\commentb} & \multicolumn{1}{c}{$N$\commentc} \\
\hline
0 & 56916.9409 & 0.0005 & $-$0&0018 & 66 \\
1 & 56917.0029 & 0.0018 & 0&0009 & 19 \\
6 & 56917.2979 & 0.0008 & $-$0&0001 & 37 \\
7 & 56917.3568 & 0.0004 & $-$0&0004 & 61 \\
8 & 56917.4193 & 0.0005 & 0&0029 & 89 \\
9 & 56917.4757 & 0.0005 & 0&0001 & 51 \\
22 & 56918.2486 & 0.0036 & 0&0034 & 31 \\
23 & 56918.3057 & 0.0014 & 0&0013 & 34 \\
39 & 56919.2531 & 0.0017 & 0&0014 & 49 \\
40 & 56919.3062 & 0.0007 & $-$0&0047 & 58 \\
41 & 56919.3666 & 0.0005 & $-$0&0035 & 24 \\
92 & 56922.3873 & 0.0013 & $-$0&0020 & 62 \\
93 & 56922.4451 & 0.0022 & $-$0&0035 & 60 \\
109 & 56923.3976 & 0.0017 & 0&0018 & 40 \\
110 & 56923.4591 & 0.0037 & 0&0041 & 29 \\
\hline
  \multicolumn{6}{l}{\commenta BJD$-$2400000.} \\
  \multicolumn{6}{l}{\commentb Against max $= 2456916.9428 + 0.059202 E$.} \\
  \multicolumn{6}{l}{\commentc Number of points used to determine the maximum.} \\
\end{tabular}
\end{center}
\end{table}

\subsection{SDSS J173047.59$+$554518.5}\label{obj:j1730}

   This object (hereafter SDSS J173047) was selected as
an AM CVn-type object from the SDSS photometric catalog
by \citet{car14j1730}.  \citet{car14j1730} obtained time-resolved
spectroscopy and found an orbital period of 35.2(2) min,
possibly allowing one-day aliases.  Although the emission lines
showed double peaks, \citet{car14j1730} suggested a low
orbital inclination since the lines were narrow.
A single outburst was detected in the past CRTS data.

\begin{figure}
  \begin{center}
    \FigureFile(88mm,110mm){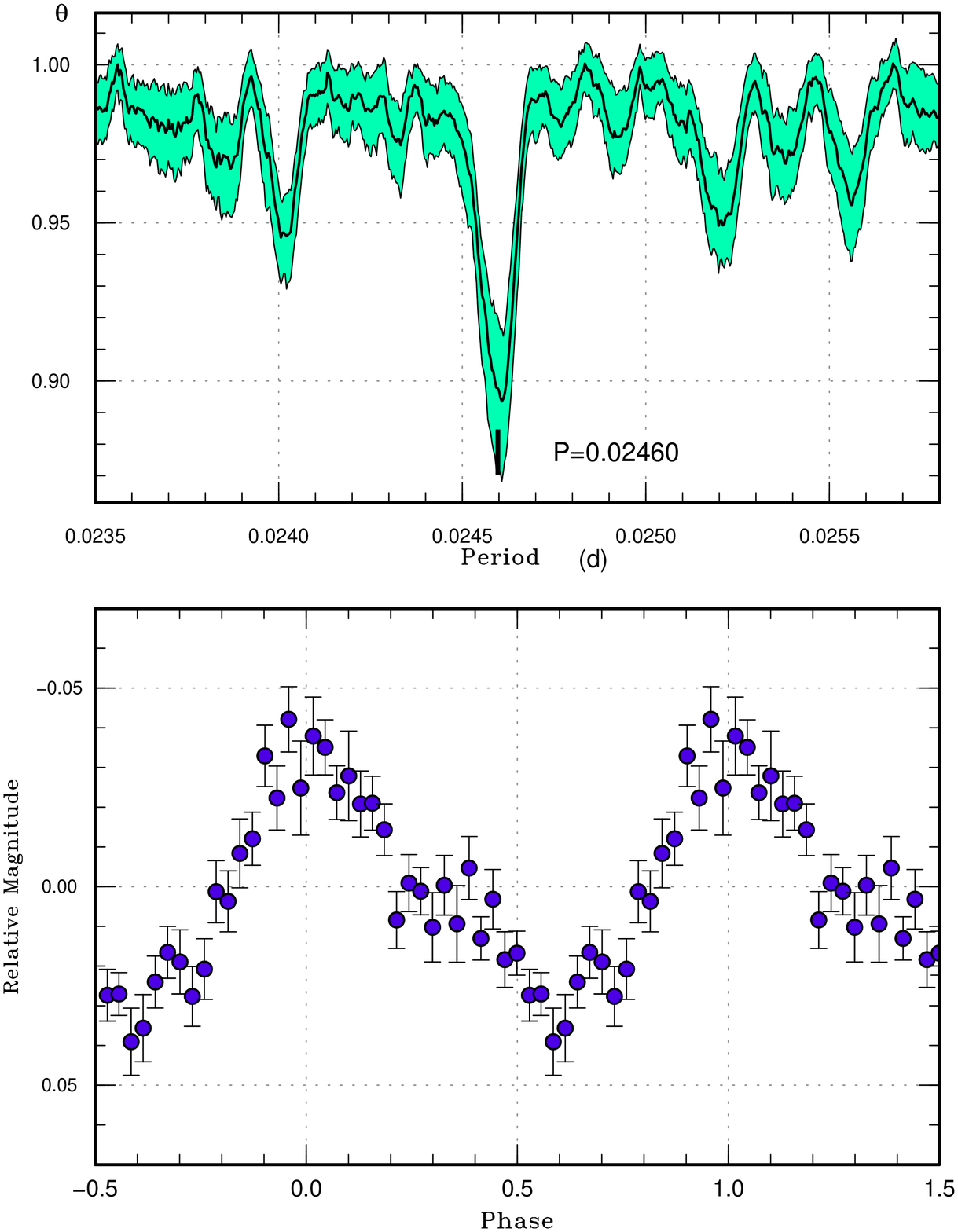}
  \end{center}
  \caption{Superhumps in SDSS J173047 (2014).  The data between
     BJD 2456753 and BJD 2456758 were used.
     (Upper): PDM analysis.
     (Lower): Phase-averaged profile.}
  \label{fig:j173047shpdm}
\end{figure}

\begin{figure}
  \begin{center}
    \FigureFile(88mm,70mm){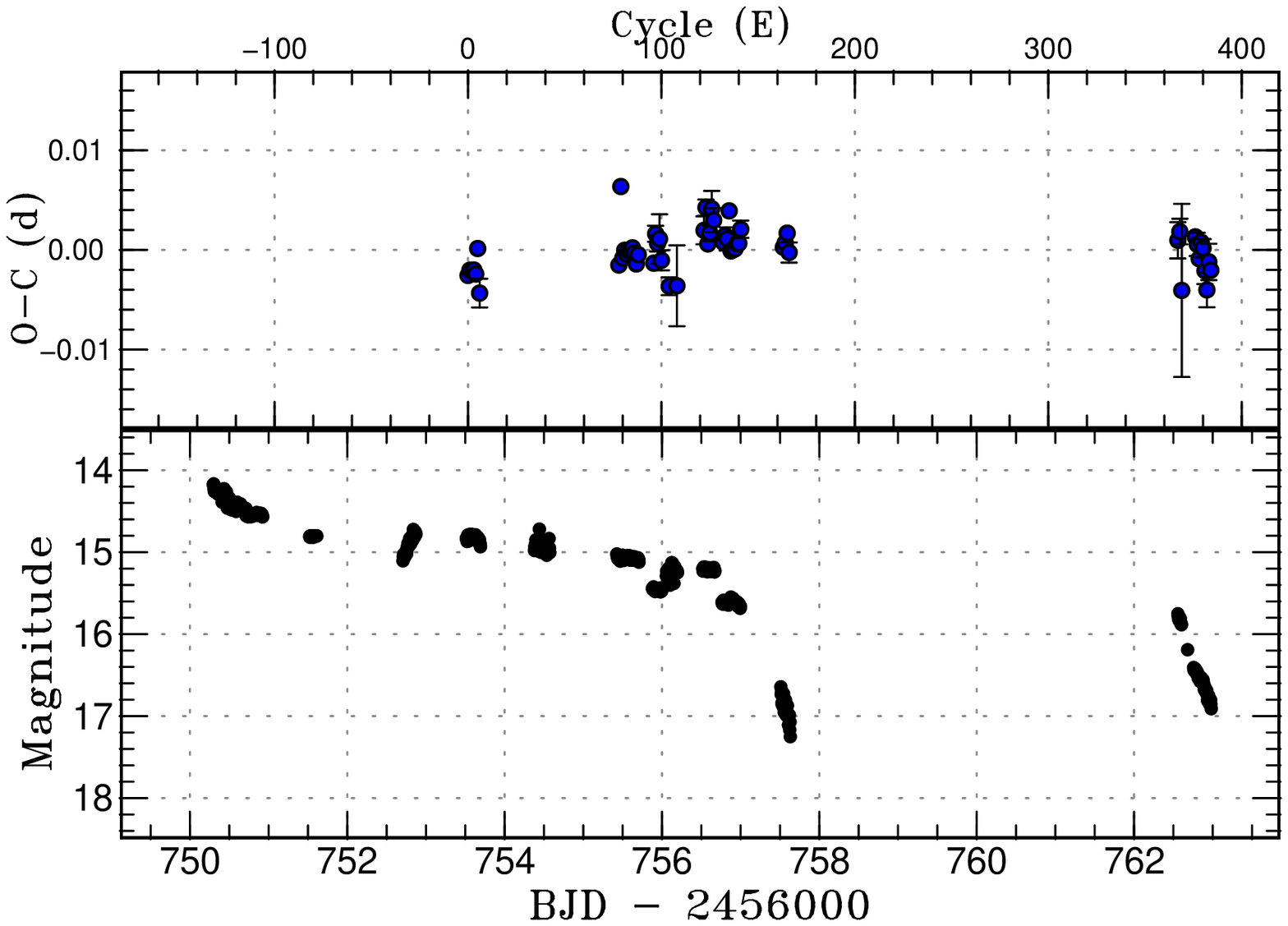}
  \end{center}
  \caption{$O-C$ diagram of superhumps in SDSS J173047 (2014).
     (Upper): $O-C$ diagram.  A period of 0.024586~d
     was used to draw this figure.
     (Lower): Light curve.  The observations were binned to 0.005~d.}
  \label{fig:j173047humpall}
\end{figure}

   The 2014 outburst was detected by CRTS on April 2 at 14.5 mag
(cvnet-outburst 5889).  On the first night of the observation, 
there were no significant modulations (vsnet-alert 17143).
On April 6, the object was found to show prominent superhumps
(vsnet-alert 17160; figure \ref{fig:j173047shpdm}).
These superhumps developed somewhere between April 4 and 6,
and we probably missed stage A superhumps.
The object started fading rapidly on April 10
(lower panel of figure \ref{fig:j173047humpall}).
The object was again observed bright on April 15,
accompanied by rapid fading.

   The times of superhump are listed in table
\ref{tab:j1730oc2014}.  During the rebrightening, superhumps
were still strongly seen.  The epochs $E \ge 367$ in table
\ref{tab:j1730oc2014} represent these superhumps during
the rebrightening, although the continuity of the phase
or the cycle count between $E=166$ and $E=367$ is unclear.
The mean superhump period of 0.024597(6)~d confirmed
the alias selection in \citet{car14j1730}.
The fractional superhump excess is 0.006(6), where the
source of the error is mostly due to the uncertainty
of the orbital period.  Future refinement of the orbital
period is desired to estimate $q$ from the fractional
superhump excess.

   The outburst behavior of SDSS J173047 is characterized
by the short duration of the plateau phase, consisting
only 5$\pm$1~d after the appearance of superhumps.
This behavior is similar to another AM CVn-type object,
SDSS J012940.05$+$384210.4 (\cite{Pdot2}; \cite{she12j0129}).

\begin{table}
\caption{Superhump maxima of SDSS J173047 (2014)}\label{tab:j1730oc2014}
\begin{center}
\begin{tabular}{rp{55pt}p{40pt}r@{.}lr}
\hline
\multicolumn{1}{c}{$E$} & \multicolumn{1}{c}{max\commenta} & \multicolumn{1}{c}{error} & \multicolumn{2}{c}{$O-C$\commentb} & \multicolumn{1}{c}{$N$\commentc} \\
\hline
0 & 56753.5318 & 0.0003 & $-$0&0025 & 39 \\
1 & 56753.5570 & 0.0004 & $-$0&0020 & 38 \\
2 & 56753.5814 & 0.0004 & $-$0&0021 & 39 \\
3 & 56753.6061 & 0.0006 & $-$0&0020 & 39 \\
4 & 56753.6303 & 0.0004 & $-$0&0024 & 33 \\
5 & 56753.6575 & 0.0007 & 0&0002 & 39 \\
6 & 56753.6776 & 0.0014 & $-$0&0043 & 39 \\
78 & 56755.4506 & 0.0007 & $-$0&0015 & 13 \\
79 & 56755.4830 & 0.0002 & 0&0064 & 12 \\
80 & 56755.5004 & 0.0005 & $-$0&0008 & 13 \\
81 & 56755.5258 & 0.0005 & $-$0&0000 & 51 \\
82 & 56755.5500 & 0.0004 & $-$0&0004 & 53 \\
83 & 56755.5749 & 0.0004 & $-$0&0001 & 46 \\
84 & 56755.5995 & 0.0005 & $-$0&0001 & 29 \\
85 & 56755.6244 & 0.0005 & 0&0002 & 30 \\
86 & 56755.6485 & 0.0006 & $-$0&0003 & 31 \\
87 & 56755.6720 & 0.0005 & $-$0&0014 & 31 \\
88 & 56755.6975 & 0.0006 & $-$0&0005 & 31 \\
96 & 56755.8933 & 0.0007 & $-$0&0013 & 26 \\
97 & 56755.9209 & 0.0008 & 0&0016 & 25 \\
98 & 56755.9445 & 0.0008 & 0&0007 & 25 \\
99 & 56755.9695 & 0.0025 & 0&0011 & 25 \\
100 & 56755.9920 & 0.0010 & $-$0&0010 & 24 \\
104 & 56756.0877 & 0.0009 & $-$0&0036 & 26 \\
108 & 56756.1860 & 0.0041 & $-$0&0036 & 23 \\
122 & 56756.5356 & 0.0014 & 0&0018 & 49 \\
123 & 56756.5625 & 0.0008 & 0&0041 & 46 \\
124 & 56756.5836 & 0.0006 & 0&0006 & 33 \\
125 & 56756.6092 & 0.0008 & 0&0016 & 27 \\
126 & 56756.6361 & 0.0018 & 0&0039 & 25 \\
127 & 56756.6599 & 0.0012 & 0&0031 & 23 \\
\hline
  \multicolumn{6}{l}{\commenta BJD$-$2400000.} \\
  \multicolumn{6}{l}{\commentb Against max $= 2456753.5344 + 0.024586 E$.} \\
  \multicolumn{6}{l}{\commentc Number of points used to determine the maximum.} \\
\end{tabular}
\end{center}
\end{table}

\addtocounter{table}{-1}
\begin{table}
\caption{Superhump maxima of SDSS J173047 (2014) (continued)}
\begin{center}
\begin{tabular}{rp{55pt}p{40pt}r@{.}lr}
\hline
\multicolumn{1}{c}{$E$} & \multicolumn{1}{c}{max\commenta} & \multicolumn{1}{c}{error} & \multicolumn{2}{c}{$O-C$\commentb} & \multicolumn{1}{c}{$N$\commentc} \\
\hline
132 & 56756.7805 & 0.0005 & 0&0008 & 26 \\
133 & 56756.8058 & 0.0006 & 0&0015 & 25 \\
134 & 56756.8300 & 0.0012 & 0&0011 & 23 \\
135 & 56756.8574 & 0.0007 & 0&0040 & 26 \\
136 & 56756.8780 & 0.0006 & $-$0&0001 & 26 \\
137 & 56756.9027 & 0.0006 & 0&0000 & 25 \\
138 & 56756.9274 & 0.0007 & 0&0001 & 26 \\
139 & 56756.9525 & 0.0006 & 0&0006 & 25 \\
140 & 56756.9771 & 0.0006 & 0&0007 & 25 \\
141 & 56757.0031 & 0.0009 & 0&0021 & 6 \\
163 & 56757.5421 & 0.0010 & 0&0002 & 24 \\
164 & 56757.5671 & 0.0010 & 0&0007 & 27 \\
165 & 56757.5927 & 0.0011 & 0&0017 & 23 \\
166 & 56757.6152 & 0.0015 & $-$0&0004 & 27 \\
367 & 56762.5584 & 0.0018 & 0&0010 & 18 \\
368 & 56762.5839 & 0.0013 & 0&0019 & 31 \\
369 & 56762.6026 & 0.0087 & $-$0&0040 & 22 \\
376 & 56762.7800 & 0.0006 & 0&0014 & 25 \\
377 & 56762.8039 & 0.0012 & 0&0006 & 22 \\
378 & 56762.8270 & 0.0007 & $-$0&0008 & 26 \\
379 & 56762.8533 & 0.0004 & 0&0008 & 26 \\
380 & 56762.8772 & 0.0009 & 0&0002 & 26 \\
381 & 56762.8996 & 0.0013 & $-$0&0020 & 26 \\
382 & 56762.9222 & 0.0017 & $-$0&0040 & 24 \\
383 & 56762.9496 & 0.0018 & $-$0&0012 & 26 \\
384 & 56762.9734 & 0.0006 & $-$0&0020 & 25 \\
\hline
  \multicolumn{6}{l}{\commenta BJD$-$2400000.} \\
  \multicolumn{6}{l}{\commentb Against max $= 2456753.5344 + 0.024586 E$.} \\
  \multicolumn{6}{l}{\commentc Number of points used to determine the maximum.} \\
\end{tabular}
\end{center}
\end{table}

\subsection{TCP J16054809$+$2405338}\label{obj:j1605}

\begin{figure}
  \begin{center}
    \FigureFile(88mm,110mm){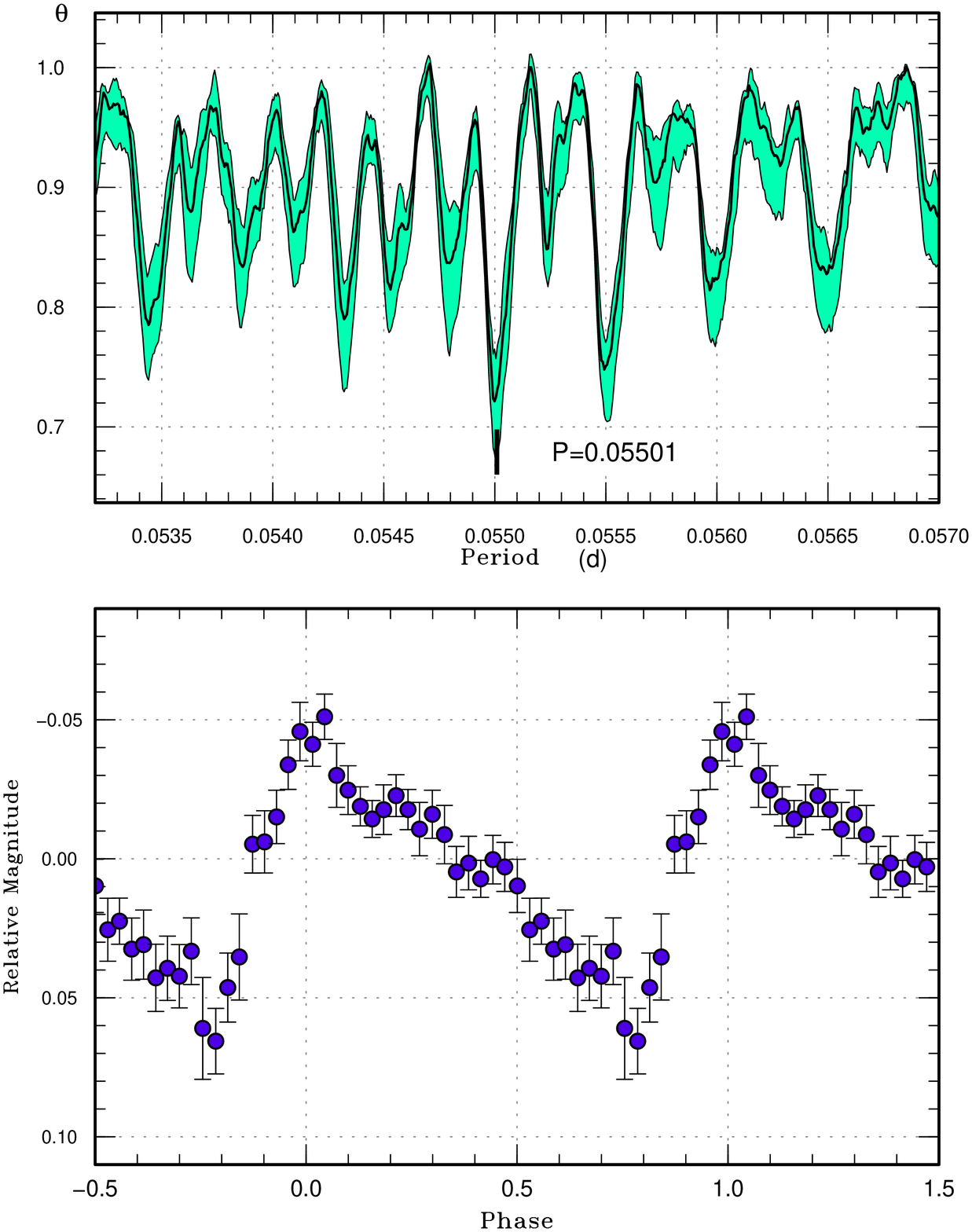}
  \end{center}
  \caption{Superhumps in SDSS J173047 (2014).  The data after
     BJD 2457019 were used.
     (Upper): PDM analysis.
     (Lower): Phase-averaged profile.}
  \label{fig:j1605shpdm}
\end{figure}

   This object (hereafter TCP J160548) is a transient
discovered by H. Nishimura at an unfiltered CCD magnitude
of 12.6 on 2014 December 20.  The object was not detected
on December 2 and 3.\footnote{
  $<$http://www.cbat.eps.harvard.edu/unconf/\\
followups/J16054809+2405338.html$>$.
}
The object was readily identified with a blue SDSS
object and a GALEX source (vsnet-alert 18085).
The object was also detected in outburst by
ASAS-SN survey (vsnet-alert 18107, 18109).
The object was also selected as a candidate for
an AM CVn-type object from SDSS colors \citep{car13SDSSamcvn}.
Although the object received attention, the difficult
visibility in the morning sky and unstable weather
hindered early observations.
On December 21, apparent large-amplitude variations
were recorded (vsnet-alert 18102).
Although short-term variations were recorded
on December 26 and 27 (vsnet-alert 18126),
the nature of the variation was unclear.
It was only on 2015 January 1, when clear superhumps
were detected (vsnet-alert 18131, 18134;
figure \ref{fig:j1605shpdm}).
Although the data were rather fragmentary and with
long gaps, a period of 0.05501~d was found to
express observations after BJD 2457019 (December 27).
The times of superhump maxima based on this period
are listed in table \ref{tab:j1605oc2014}.
Although there was a possibility of stage A superhumps
or early superhumps before BJD 2457019, this could not
be confirmed due to the lack of observations.

\begin{table}
\caption{Superhump maxima of TCP J160548 (2014)}\label{tab:j1605oc2014}
\begin{center}
\begin{tabular}{rp{55pt}p{40pt}r@{.}lr}
\hline
\multicolumn{1}{c}{$E$} & \multicolumn{1}{c}{max\commenta} & \multicolumn{1}{c}{error} & \multicolumn{2}{c}{$O-C$\commentb} & \multicolumn{1}{c}{$N$\commentc} \\
\hline
0 & 57019.3296 & 0.0009 & 0&0087 & 29 \\
79 & 57023.6644 & 0.0007 & $-$0&0006 & 70 \\
80 & 57023.7192 & 0.0003 & $-$0&0009 & 101 \\
81 & 57023.7720 & 0.0016 & $-$0&0030 & 32 \\
187 & 57029.6010 & 0.0010 & $-$0&0029 & 54 \\
188 & 57029.6547 & 0.0008 & $-$0&0042 & 56 \\
189 & 57029.7077 & 0.0029 & $-$0&0062 & 45 \\
315 & 57036.6494 & 0.0015 & 0&0068 & 42 \\
316 & 57036.6998 & 0.0012 & 0&0023 & 41 \\
\hline
  \multicolumn{6}{l}{\commenta BJD$-$2400000.} \\
  \multicolumn{6}{l}{\commentb Against max $= 2457019.3209 + 0.054989 E$.} \\
  \multicolumn{6}{l}{\commentc Number of points used to determine the maximum.} \\
\end{tabular}
\end{center}
\end{table}

\section{Discussion}\label{sec:discuss}

\subsection{Statistics of Objects}

\begin{figure}
  \begin{center}
    \FigureFile(88mm,70mm){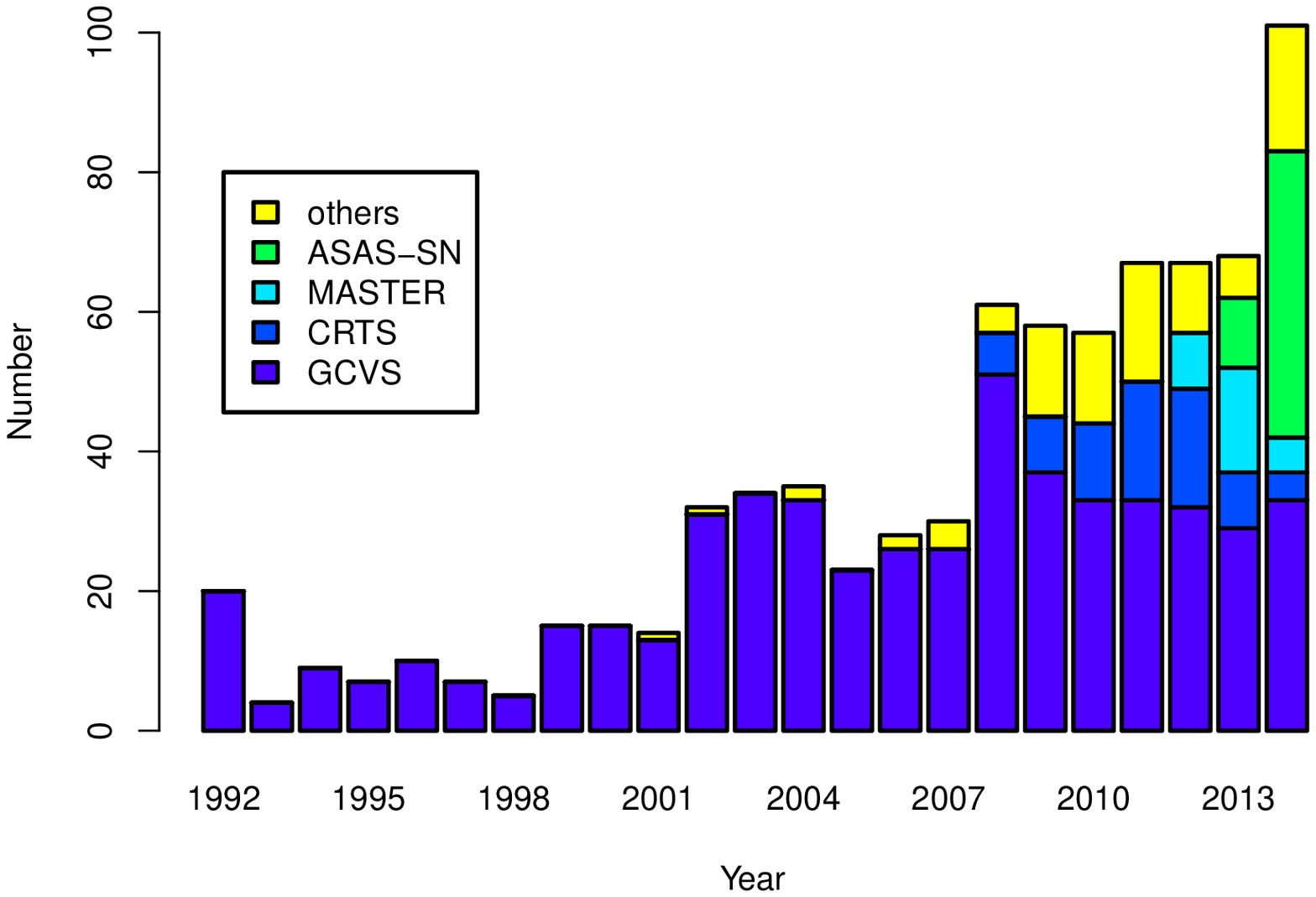}
  \end{center}
  \caption{Object categories in our survey.  Superoutbursts
  with measured superhump periods are included.
  The year represents the year of outburst.
  The year 1992 represents outbursts up to 1992 and the year
  2014 includes the outbursts in 2015, respectively.
  The category GCVS includes the objects named in the General
  Catalog of Variable Stars \citet{GCVS} in the latest version
  and objects named in New Catalog of Suspected Variable Stars
  (NSV).  The categories CRTS, MASTER, ASAS-SN represent
  objects which were discovered in respective surveys.
  A small fraction of objects discovered by the CRTS are already
  named in GCVS and are included in the category GCVS.
  }
  \label{fig:objsource}
\end{figure}

   Up to \citet{Pdot}, a large fraction of objects studied in
our survey were known variable stars in the General
Catalog of Variable Stars (GCVS: \cite{GCVS}).
Since 2007, the CRTS started to produce new transients,
most of which were initially supernovae.  The first CRTS CV
studied in our survey was CRTS J021110.2$+$171624 in 2008
\citep{djo08atel1416}.  The CRTS then became the dominant
source of new CVs for the subsequent five years
(figure \ref{fig:objsource}).  Since the main targets
of the CRTS are moving Solar system objects, the observation
strategy (approximately once in 10~d) was not necessarily
suitable for detecting CV outbursts in the early phase.
Starting from 2012, the MASTER network started to produce
new transients.  The first MASTER CV studied in our survey
was MASTER OT J072948.66$+$593824.4 \citep{bal12j0729atl3935}.
Since the observation strategy was more suitable for
early detection of CV outbursts, the contribution of
outburst detections by the MASTER network increased up to
2013.  A large number of newly discovered WZ Sge-type
dwarf novae was a remarkable product of this survey
(e.g. \cite{Pdot4}; \cite{Pdot5}; \cite{nak13j2112j2037}).
Starting from 2013, the ASAS-SN system \citep{ASASSN}
started to produce new transients.
The first ASAS-SN CV we studied was
ASASSN-13cf and the number of ASAS-SN CVs rapidly increased
thanks to the high frequency of observations (observations
up to every night) and large sky coverage up to 15000
square degrees.  As a result, the fraction of ASAS-SN CVs
and detections of CV outbursts by the ASAS-SN team
dramatically increased.  If we restrict the objects to those
studied in our series of papers, the number of objects designated by
the ASAS-SN survey has already surpassed the number registered
in the GCVS.  It is likely that the majority of CVs
within the reach of small telescopes have ASAS-SN names
(currently 452 at the time of writing in
2015 June) in the near future unless the GCVS names
(currently 581, including NSV) are given more quickly.
The ASAS-SN transients are mostly not very faint as
in the CRTS ones and are reachable by relatively
small telescopes.  This present statistics would provide
a clue for strategy of surveys in maximizing the scientific
gain for small telescopes.

\subsection{Period Distribution}

\begin{figure}
  \begin{center}
    \FigureFile(88mm,135mm){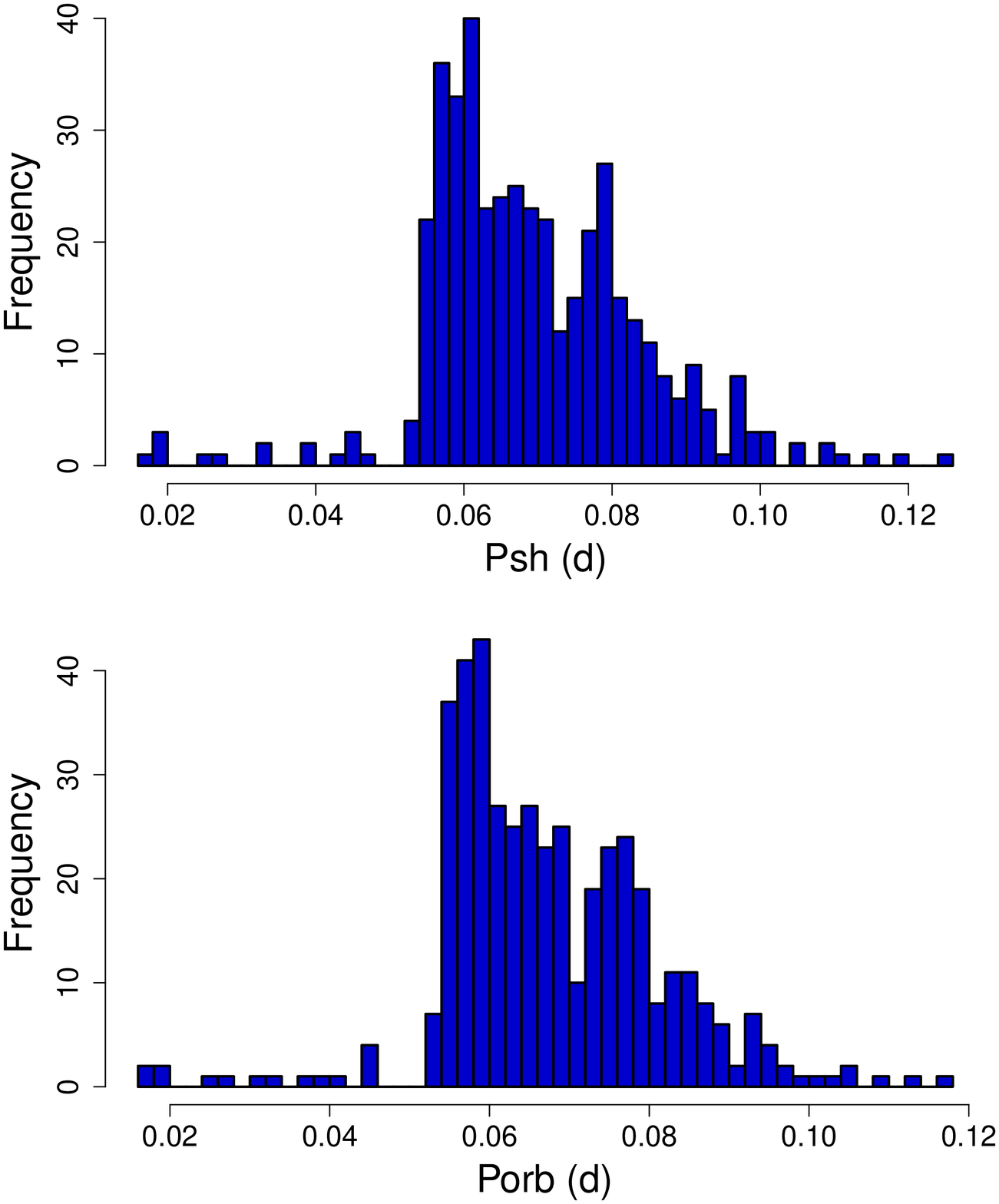}
  \end{center}
  \caption{Distribution of superhump periods in this survey.
  The data are from \citet{Pdot}, \citet{Pdot2}, \citet{Pdot3},
  \citet{Pdot4}, \citet{Pdot5}, \citet{Pdot6} and this paper.
  The mean values are used when multiple superoutbursts
  were observed.
  (Upper) distribution of superhump periods.
  (Lower) distribution of orbital periods.  For objects with
  superhump periods shorter than 0.053~d, the orbital periods
  were assumed to be 1\% shorter than superhump periods.
  For objects with superhump periods longer than 0.053~d,
  we used the new calibration in \citet{Pdot3} to estimate
  orbital periods.
  }
  \label{fig:phist}
\end{figure}

   In figure \ref{fig:phist}, we give distribution of
superhump and estimated orbital periods,
greatly updated since \citet{Pdot}. 
Since most of non-magnetic CVs below the period gap
are SU UMa-type dwarf novae, this figure is expected to
well represent the distribution of superhump periods
of non-magnetic CVs.  It is quite noticeable that
there is a sharp cut-off at a period of 0.053~d
(the objects below this period are either AM CVn-type
systems and EI Psc-type objects).  This finding
has strengthened the modern observational identification
of the period minimum (e.g. \cite{kni06CVsecondary};
\cite{gan09SDSSCVs}; \cite{kni11CVdonor}) and there
is no indication of systems below this period,
contrary to what was suggested by \citet{uem10shortPCV}.
Since there have been a sizeable number of genuine
candidates for period bouncers (\cite{kat13j1222};
\cite{kat13qfromstageA}; \cite{nak14j0754j2304}),
it is unlikely that the mass-transfer rate quickly
decreases as systems approach the period minimum
and this identification of the period minimum appears
to be secure.

   There is a peak of distribution just above
the period minimum and it is compatible with
the period spike \citep{gan09SDSSCVs}.
The distribution monotonically decreases towards
the longer period, and, rather surprisingly,
there is no clear indication of a sudden decrease
of the number at the supposed lower edge of the period gap.
This tendency is the same as reported in
\citet{pav14nyser}.  This result may even cast
a doubt against the presence of the period gap,
at least for SU UMa-type dwarf novae and the lower edge of
the gap (we should note, our surveys of superhumps
systems are not sensitive to the upper edge of
the period gap, above which almost all dwarf novae
are not SU UMa-type stars).

\subsection{Period Derivatives during Stage B}\label{sec:stagebpdot}

\begin{figure*}
  \begin{center}
    \FigureFile(160mm,110mm){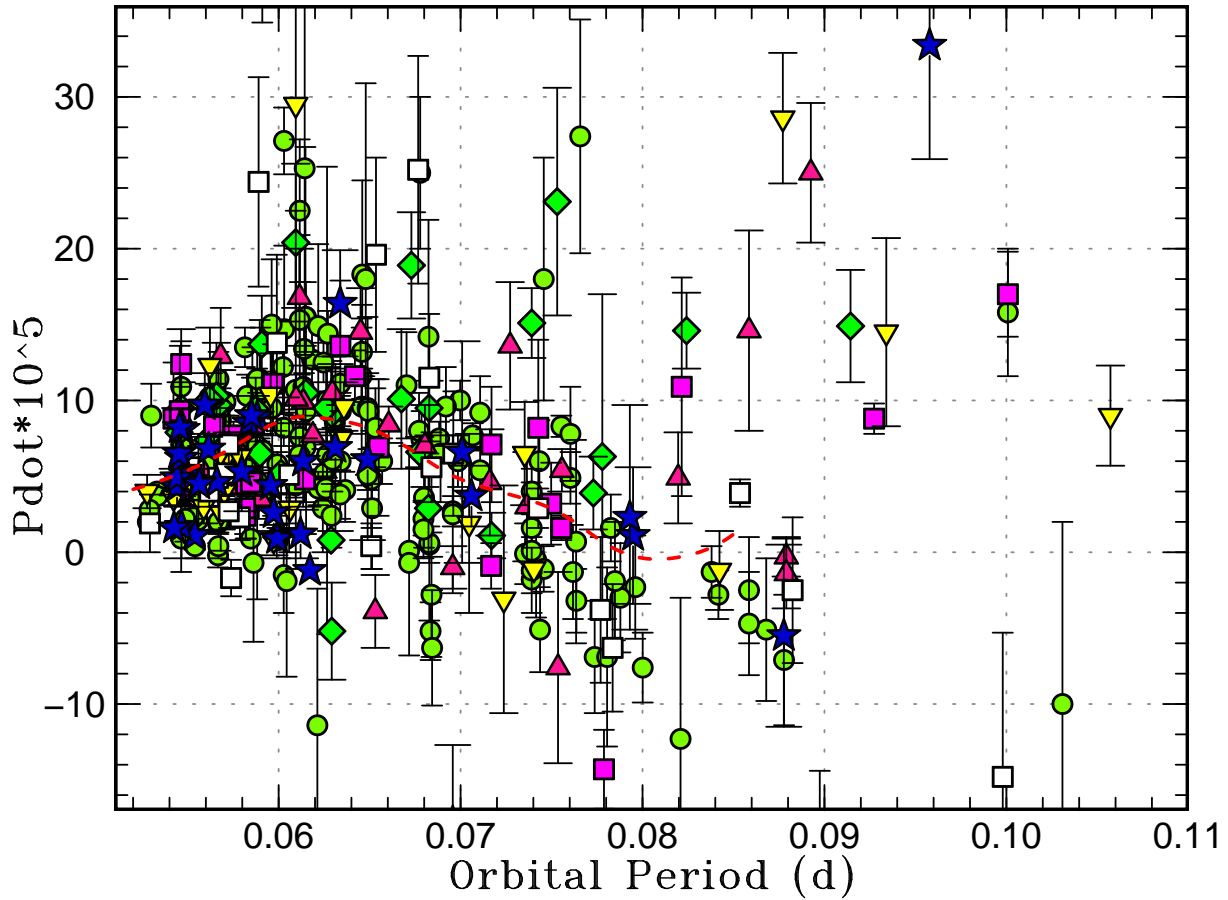}
  \end{center}
  \caption{$P_{\rm dot}$ for stage B versus $P_{\rm orb}$.
  Filled circles, filled diamonds, filled triangles, filled squares,
  filled lower-pointed triangles, open square
  and filled stars represent samples in 
  \citet{Pdot}, \citet{Pdot2}, \citet{Pdot3},
  \citet{Pdot4}, \citet{Pdot5}, \citet{Pdot6}
  and this paper, respectively.
  The curve represents the spline-smoothed global trend.
  }
  \label{fig:pdotporb7}
\end{figure*}

   Figure \ref{fig:pdotporb7} represents updated relation
between $P_{\rm dot}$ for stage B versus $P_{\rm orb}$.
The objects studied in this paper are heavily concentrated
in the region of $P_{\rm orb} < 0.065$ d, and most
of the object showed positive $P_{\rm dot}$ as in earlier
studies.  The longer-period systems studied in this
work generally followed the trend shown in previous works.
OT J064833 showed an exceptionally large $P_{\rm dot}$.
As discussed in subsection \ref{obj:j0648}, this object
has features common to short-$P_{\rm orb}$ systems
and warrants further study.

\subsection{Mass Ratios from Stage A Superhumps}\label{sec:stagea}

   We have proposed that the precession rate of
stage A superhumps represents the dynamical precession
rate of at the 3:1 resonance, thereby enabling
estimation of mass ratios without experimental
calibration \citep{kat13qfromstageA}.
We list new estimates for the binary mass ratio from
stage A superhumps in table \ref{tab:newqstageA}.
This table includes two objects in \Nakataprep.
In table \ref{tab:pera}, we list stage A superhumps
recorded in the present study.

\begin{figure*}
  \begin{center}
    \FigureFile(160mm,110mm){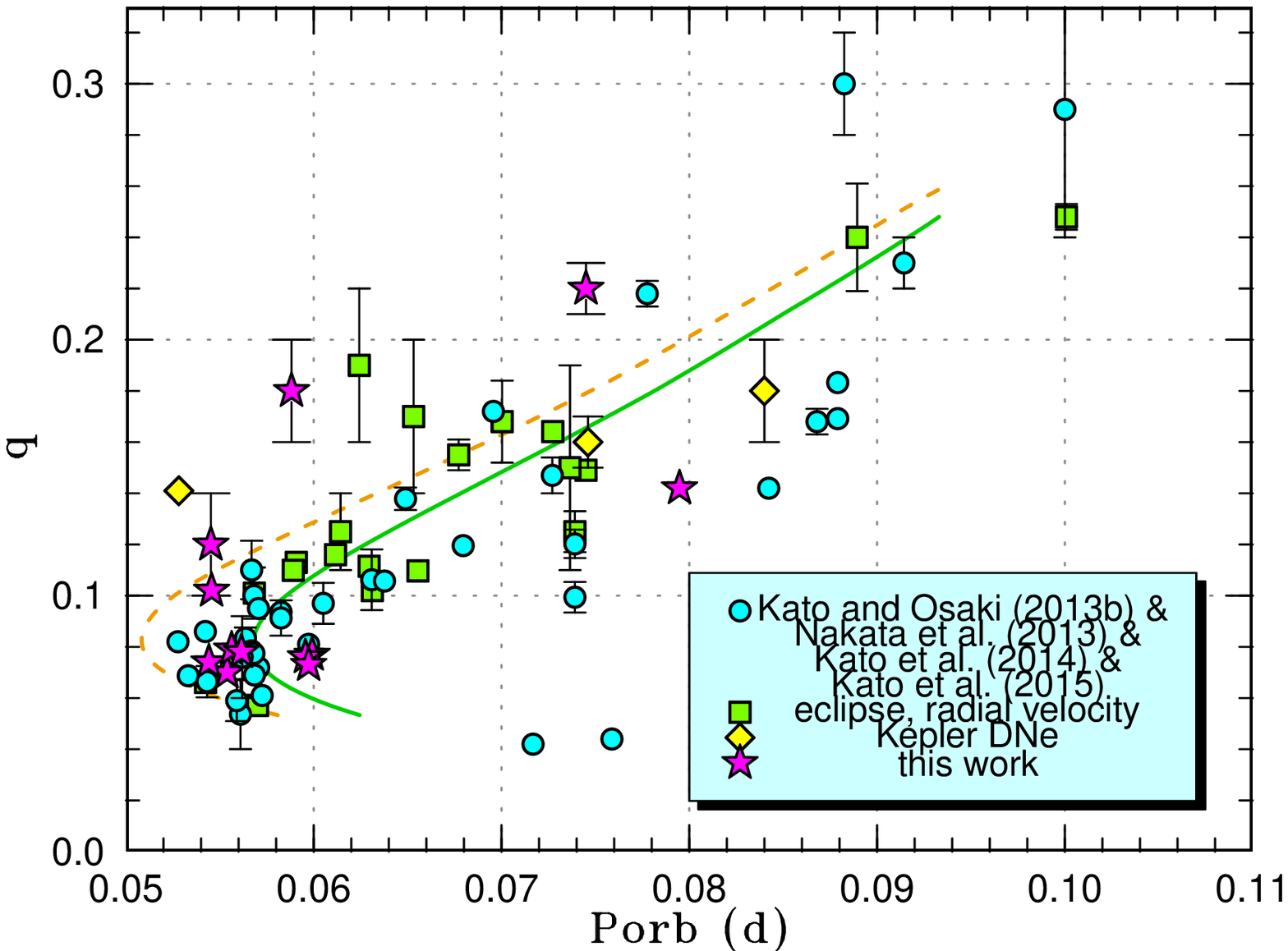}
  \end{center}
  \caption{Mass ratio versus orbital period.
  The dashed and solid curves represent the standard and optimal
  evolutionary tracks in \citet{kni11CVdonor}, respectively.
  The filled circles, filled squares, filled stars, filled diamonds
  represent $q$ values from a combination of the estimates
  from stage A superhumps published in four preceding
  sources (\cite{kat13qfromstageA}; \cite{nak13j2112j2037};
  \cite{Pdot5}; \cite{Pdot6}), known $q$ values from quiescent eclipses or 
  radial-velocity study (see \cite{kat13qfromstageA} for
  the data source), $q$ estimated in this work and dwarf novae
  in the Kepler data (see text for the complete reference),
  respectively.  The objects in ``this work'' includes
  two objects in \Nakataprep}
  \label{fig:qall4}
\end{figure*}

   A updated distribution of mass ratios is shown in
figure \ref{fig:qall4}.
The Kepler DNe shown in this figures are
V516 Lyr \citep{kat13j1939v585lyrv516lyr},
KIC 7524178 \citep{kat13j1922} and the unusual
short-$P_{\rm orb}$ object GALEX J194419.33$+$491257.0
in the field of KIC 11412044 \citep{kat14j1944}
(located at $P_{\rm orb}$=0.05282~d, $q$=0.14).
The present study has strengthened the concentration
of WZ Sge-type dwarf novae around $q=0.07$
just above the period minimum.

\begin{table}
\caption{New estimates for the binary mass ratio from stage A superhumps}\label{tab:newqstageA}
\begin{center}
\begin{tabular}{ccc}
\hline
Object         & $\varepsilon^*$ (stage A) & $q$ from stage A \\
\hline
Z Cha          & 0.0707(4)  & 0.22(1) \\
BR Lup         & 0.0488(12) & 0.142(4) \\
QZ Vir         & 0.059(5)   & 0.18(2) \\
ASASSN-14cv    & 0.0286(3)  & 0.077(1) \\
ASASSN-14jf    & 0.026(2)   & 0.070(5) \\
ASASSN-14jv    & 0.0278(9)  & 0.074(3) \\
ASASSN-15bp    & 0.0293(6)  & 0.079(2) \\
OT J030929     & 0.0291(3)  & 0.078(1) \\
OT J213806     & 0.041(4)   & 0.12(2)  \\
OT J230523     & 0.0366(7)  & 0.102(2) \\
PNV J171442    & 0.0284(3)  & 0.076(1) \\
PNV J172929    & 0.0273(5)  & 0.073(2) \\
\hline
\end{tabular}
\end{center}
\end{table}

\begin{table}
\caption{Superhump Periods during Stage A}\label{tab:pera}
\begin{center}
\begin{tabular}{cccc}
\hline
Object & Year & period (d) & err \\
\hline
Z Cha & 2014 & 0.08017 & 0.00031 \\
BR Lup & 2014 & 0.08357 & 0.00011 \\
QZ Vir & 2014 & 0.06248 & 0.00034 \\
ASASSN-14cl & 2014 & 0.06082 & 0.00014 \\
ASASSN-14cv & 2014 & 0.06168 & 0.00002 \\
ASASSN-14jf & 2014 & 0.05687 & 0.00009 \\
ASASSN-14jv & 2014 & 0.05597 & 0.00006 \\
ASASSN-14md & 2014 & 0.06943 & 0.00035 \\
ASASSN-15bp & 2015 & 0.05731 & 0.00003 \\
OT J030929 & 2014 & 0.05783 & 0.00008 \\
OT J060009 & 2014 & 0.06458 & 0.00005 \\
OT J064833 & 2014 & 0.10519 & 0.00040 \\
OT J213806 & 2014 & 0.05684 & 0.00026 \\
OT J230523 & 2014 & 0.05663 & 0.00004 \\
PNV J171442 & 2014 & 0.06130 & 0.00002 \\
PNV J172929 & 2014 & 0.06141 & 0.00009 \\
SDSS J090221 & 2014 & 0.03409 & 0.00001 \\
\hline
\end{tabular}
\end{center}
\end{table}

\subsection{Development of Superhumps Following Precursor Outburst}

   In \citet{Pdot3}, we showed from the Kepler data of
V1504 Cyg and V344 Lyr that superhumps start to develop
following precursor outbursts.  This finding was examined
in more detail using the Kepler data of the same objects
in \citet{osa13v1504cygKepler}, \citet{osa13v344lyrv1504cyg}.
The same phenomenon was also confirmed in the Kepler
data of the background dwarf nova of KIC 4378554 and V516 Lyr
\citep{kat13j1939v585lyrv516lyr}.  These findings have
strengthened the universal application of the TTI model
to various SU UMa-type dwarf novae.

   In this paper, we have observed similar evolution of
superhumps following the precursor outburst in Z Cha.
Although only one superhump was recorded, superhumps
likely started to grow just before the final rise
to the main superoutburst also in CY UMa.
An $O-C$ analysis of SU UMa also suggested that
superhumps likely started to appear around the maximum
of the precursor outburst.  All of these findings are
in good agreement with what the TTI model predicts
(cf. \cite{osa13v1504cygKepler}).

\subsection{AM CVn-Type Objects and Related Objects}

   In recent years, an increasing number of AM CVn-type objects
[for recent reviews of AM CVn-type objects, see 
e.g. \citet{nel05amcvnreview}; \citet{sol10amcvnreview}]
have been recorded in outbursts and superhumps were
detected during outbursts:
CR Boo (observation in atypical state in \cite{pat97crboo},
more regular superoutburst in \cite{kat01crboo});
V803 Cen (observation in atypical state in \cite{pat00v803cen});
KL Dra \citep{woo02kldra};
V406 Hya \citep{nog04v406hya};
SDSS J012940.05$+$384210.4 (\cite{Pdot2}; \cite{she12j0129});
YZ LMi (partial coverage in \cite{cop11j0926});
PTF1 J071912.13$+$485834.0 \citep{lev11j0719};
CP Eri (\cite{arm12cperi}; \cite{Pdot6});
SDSS J172102.48$+$273301.2 \citep{Pdot4};
CSS J045019.7$-$093113 \citep{wou13j0450atel4726}.

   In \citet{Pdot4}, we also studied superhumps and outburst
patterns in an AM CVn star CR Boo and found that
the basic $O-C$ variation of superhumps in
AM CVn-type superoutburst is the same as in hydrogen-rich
systems.  \Isogaiprep\ also detected stage A superhumps
in CR Boo, first time in AM CVn-type objects with
typical supercycles.

   In this paper, we added new examples of AM CVn-type
systems PTF1 J071912,
SDSS J090221 (see also \cite{kat14j0902})
and SDSS J173047.
During the present survey, two further objects
ASASSN-14ei and ASASSN-14mv have been identified
as AM CVn-type objects showing superoutbursts and
multiple rebrightenings (\Isogaiprep; for spectroscopic
identifications, see \cite{pri14asassn14eiatel6475},
see also vsnet-alert 18160).
ASASSN-14cn was also found to be an eclipsing AM CVn-type
object in outburst (vsnet-alert 17879; superhumps
are yet to be detected).
ASASSN-14fv was also found to be an outbursting
AM CVn-type object (\cite{wag14asassn14fvatel6669}
for spectroscopic identification; superhumps
are yet to be detected).  These data suggest that
about 8\% of objects showing dwarf nova-type outbursts
are AM CVn-type objects.  This fraction is much larger
than the hitherto known statistics (about 1\% in
RKcat Edition 7.21 \cite{RKCat}).
AM CVn-type objects may be more populous than
have been considered.

   There have also been an increasing number of
EI Psc-type objects (CVs containing but depleted hydrogen
with orbital period below the period minimum):
V485 Cen (\cite{aug95v485cen}; \cite{ole97v485cen});
EI Psc (\cite{tho02j2329}; \cite{uem02j2329letter};
\cite{ski02j2329});
CRTS J102842.9$-$081927 (\cite{Pdot}; \cite{Pdot4}; \cite{Pdot5});
CRTS J112253.3$-$111037 \citep{Pdot2};
SBS 1108$+$574 (\cite{Pdot4}; \cite{lit13sbs1108};
\cite{car13sbs1108});
CSS J174033.5$+$414756 (\Ohtprep).
Candidate systems include CRTS J233313.0$-$155744
\citep{wou11j2333atel}.

   We have added another object of this class, V418 Ser,
in this paper.  In all well-observed cases, the development
of superhumps in these systems follow the same pattern
as in hydrogen-rich systems.  Although the number of
these objects is smaller than AM CVn-type objects,
these object may be more abundant than have been considered.

\subsection{Long-Term Variation of Supercycles}

   In this paper, we studied long-term variation of
supercycles in MM Hya (subsection \ref{obj:mmhya}) and
CY UMa (subsection \ref{obj:cyuma}).
In MM Hya, variations of the supercycle in the range of
330~d and 386~d were recorded.
In CY UMa, a sudden decrease of the supercycle from
362(3)~d to 290(1)~d was observed in 2003.
In both systems, the supercycle tended to be constant
for several to ten years, and there was a tendency
of a sudden switch to a different period.

   Similar systematic variations of cycle lengths
were studied in SS Cyg-type dwarf novae
[cf. AR And, UU Aql, RU Peg: \citet{and90DNcycle}].
\citet{and90DNcycle} suggested the Solar-type
activity as a possible cause, but the phenomenon
still remained a puzzle.  In recent years,
\citet{zem13eruma} recorded variations of supercycles
from 43.6 to 59.2~d in ER UMa.  The variations of
the supercycle in ER UMa was an order similar to
those in the two systems studied in this paper.
Although there was a possibility that a disk tilt,
which is supposed to produce negative superhumps
(\cite{woo07negSH}; \cite{mon12negSHSPH}),
could affect the outburst properties, \citet{zem13eruma}
could not find a correlation between the appearance
of negative superhumps and the supercycle length.
Since a disk tilt is less likely to occur in
systems with lower mass-transfer rate like MM Hya
and CY UMa supposing a theoretical interpretation
by \citet{mon10disktilt}, we consider it less likely
that a disk tilt can explain the variations of
the supercycle commonly seen in a variety of
dwarf novae.  The mechanism needs to be sought further. 

\section{Summary}\label{sec:summary}

   In addition to results of observations superhumps of
the objects studied in this paper, the major findings we obtained
can be summarized as follows.

\begin{itemize}

\item The contribution of various surveys in detecting
SU UMa-type dwarf novae have dramatically changed
in the last ten years.  The advantage of a wide-field
survey on nightly basis (such as ASAS-SN survey)
has become obvious.

\item Thanks to the increase of samples of SU UMa-type
dwarf novae, we could clarify the distribution of
orbital periods in SU UMa-type dwarf novae,
which approximate the CV population with short orbital
periods.  There is a sharp cut-off at a period of 0.053~d,
which is considered to be the period minimum.
There is a high concentration of objects just above
the period minimum, which can be interpreted as
the ``period spike''.  The distribution monotonically
decreases towards the longer period, and there is
little indication of the period gap.

\item We observed a precursor outburst in Z Cha
and found that superhumps developed just following
the precursor outburst.  $O-C$ analyses in CY UMa
and YZ Cnc also support this finding.  The agreement of
the time of growing superhumps with the rising branch
following the precursor outburst has been confirmed
in all SU UMa-type systems so far studied.
This finding provides a strong support to the
thermal-tidal disk instability model which predicts
that a superoutburst develops as a result of
the development of superhumps.

\item We detected possible negative superhumps in
Z Cha.  During the phase of negative superhumps,
the outburst cycle apparently lengthened.
This finding seems to support a suggestion
that a disk tilt suppresses normal outbursts.

\item We studied secular variations of the quiescent
brightness and the amplitude of orbital humps
throughout one supercycle in Z Cha.
The quiescent brightness decreased as the system
approached the next superoutburst.
There was no enhanced orbital humps just before
the superoutburst.

\item We studied long-term trends in supercycles
in MM Hya and CY UMa and found systematic variations
of supercycles of $\sim$20\%.  This degree and
characteristics os variations are similar to 
those recorded in other SS Cyg-type dwarf novae
and the SU UMa-type system ER UMa.  A disk tilt
is unlikely a common source of this variation.

\item The WZ Sge-type object ASASSN-15bp showed
a phase jump of superhumps during the plateau phase.

\item A sizable number of AM CVn-type objects
(PTF1 J071912, SDSS J090221, SDSS J173047) were studied
in this paper and four more AM CVn-type objects
(ASASSN-14cn, ASASSN-14ei, ASASSN-14fv and ASASSN-14mv)
were found in this period.  This number suggests
that about 8\% of objects showing dwarf nova-type outbursts
are AM CVn-type objects.

\item We have added another EI Psc-type object
V418 Ser in this paper.  These object may be more 
abundant than have been considered.

\item CSS J174033, an EI Psc-type object,
showed a similar type of superoutbursts in 2012
and 2014, comprising of a dip and the second plateau
phase.  This finding suggests that the same type
of superoutbursts tend to be reproduced
in EI Psc-type objects as in hydrogen-rich systems.

\item OT J213806, a WZ Sge-type object, exhibited
a remarkably different peak brightness and
$O-C$ diagrams between the 2010 and 2014
superoutbursts.  The fainter superoutburst in 2014
was shorter and the difference was most striking
in the later part of the plateau phase.

\item MASTER J085854 showed two post-superoutburst
rebrightenings.  This object had a rather exceptionally
short superhump period among the objects showing
multiple rebrightenings.

\item Four deeply eclipsing SU UMa-type dwarf novae
were identified (ASASSN-13cx, ASASSN-14ag, ASASSN-15bu,
NSV 4618).
ASASSN-14id also showed shallow eclipses.

\end{itemize}

\section*{Acknowledgements}

This work was supported by the Grant-in-Aid
``Initiative for High-Dimensional Data-Driven Science through Deepening
of Sparse Modeling'' (25120007) 
from the Ministry of Education, Culture, Sports, 
Science and Technology (MEXT) of Japan.
The authors are grateful to observers of VSNET Collaboration and
VSOLJ observers who supplied vital data.
We acknowledge with thanks the variable star
observations from the AAVSO International Database contributed by
observers worldwide and used in this research.  We are also grateful
to the VSOLJ database.
This work is deeply indebted to outburst detections and announcement
by a number of variable star observers worldwide, including participants of
CVNET and BAA VSS alert.
The CCD operation of the Bronberg Observatory is partly sponsored by
the Center for Backyard Astrophysics.
We are grateful to the Catalina Real-time Transient Survey
team for making their real-time
detection of transient objects available to the public.
R. Modic acknowledge the Bradford Robotic Telescope
for the detection of the 2014 outburst of QZ Vir.
The work by A. Sklyanov is partially performed according to 
the Russian Government Program of Competitive Growth of 
Kazan Federal University.


\begin{thebibliography}{}

\bibitem[{Andronov}, {Shakun}(1990)]{and90DNcycle}
  {Andronov}, I.~L., \& {Shakun}, L.~I.\ 1990, Ap\&SS, 169, 237

\bibitem[{Armstrong} et~al.(2012)]{arm12cperi}
  {Armstrong}, E., {Patterson}, J., \& {Kemp}, J.\ 2012, MNRAS, 421, 2310

\bibitem[Augusteijn(1995)]{aug95v485cen}
  Augusteijn, T.\ 1995, in Cataclysmic Variables, ed. A. Bianchini, M. della
  Valle, \& M. Orio (Dordrecht: Kluwer Academic Publishers), p.~129

\bibitem[{Baade}(1928)]{baa28VS}
  {Baade}, W.\ 1928, Astron.\ Nachr., 232, 65

\bibitem[Baba et~al.(2002)]{bab02wzsgeletter}
  Baba, H., {et~al.}\ 2002, PASJ, 54, L7

\bibitem[Bailey(1979)]{bai79zcha}
  Bailey, J.\ 1979, MNRAS, 187, 645

\bibitem[Bailey, Ward(1981)]{bai81oycar}
  Bailey, J., \& Ward, M.\ 1981, MNRAS, 194, 17

\bibitem[{Balanutsa} et~al.(2014)]{bal14j0439atel5787}
  {Balanutsa}, P., {et~al.}\ 2014, Astron.\ Telegram, 5787

\bibitem[{Balanutsa} et~al.(2015a)]{bal15j0858atel6946}
  {Balanutsa}, P., {et~al.}\ 2015a, Astron.\ Telegram, 6946

\bibitem[{Balanutsa} et~al.(2015b)]{bal15asassn15awatel6967}
  {Balanutsa}, P., {et~al.}\ 2015b, Astron.\ Telegram, 6967

\bibitem[{Balanutsa} et~al.(2012)]{bal12j0729atl3935}
  {Balanutsa}, P., {et~al.}\ 2012, Astron.\ Telegram, 3935

\bibitem[Barwig, Schoembs(1981)]{bar81suuma}
  Barwig, H., \& Schoembs, R.\ 1981, IBVS, 1989

\bibitem[{Beljawsky}(1927)]{bel27sycap}
  {Beljawsky}, S.\ 1927, Astron.\ Nachr., 230, 349

\bibitem[{Berardi}(2014)]{ber14asassn14jvatel6684}
  {Berardi}, P.\ 2014, Astron.\ Telegram, 6684

\bibitem[{Boyd} et~al.(2007)]{boy07v337cyg}
  {Boyd}, D., {Kracji}, T., {Shears}, J., \& {Poyner}, G.\ 2007, J.\ Br.\
  Astron.\ Assoc., 117, 198

\bibitem[{Breedt} et~al.(2014)]{bre14CRTSCVs}
  {Breedt}, E., {et~al.}\ 2014, MNRAS, 443, 3174

\bibitem[{Brown} et~al.(2015)]{bro15j1915}
  {Brown}, A., {et~al.}\ 2015, AJ, 149, 67

\bibitem[{Brown} et~al.(2010)]{bro10nsv1436}
  {Brown}, S.~J., {Mills}, O.~F., {Osborn}, W., \& {Hoette}, V.\ 2010, J.\
  American\ Assoc.\ Variable\ Star\ Obs., 38, 176

\bibitem[Bruch et~al.(1987)]{BruchCVatlas}
  Bruch, A., Fischer, F.-J., \& Wilmsen, U.\ 1987, A\&AS, 70, 481

\bibitem[Buat-M\'{e}nard, Hameury(2002)]{bua02suumamodel}
  Buat-M\'{e}nard, V., \& Hameury, J.-M.\ 2002, A\&A, 386, 891

\bibitem[{Carter} et~al.(2013a)]{car13SDSSamcvn}
  {Carter}, P.~J., {et~al.}\ 2013a, MNRAS, 429, 2143

\bibitem[{Carter} et~al.(2013b)]{car13sbs1108}
  {Carter}, P.~J., {et~al.}\ 2013b, MNRAS, 431, 372

\bibitem[{Carter} et~al.(2014)]{car14j1730}
  {Carter}, P.~J., {Steeghs}, D., {Marsh}, T.~R., {Kupfer}, T., {Copperwheat},
  C.~M., {Groot}, P.~J., \& {Nelemans}, G.\ 2014, MNRAS, 437, 2894

\bibitem[{Ceraski}(1908)]{cer08suuma}
  {Ceraski}, W.\ 1908, Astron.\ Nachr., 177, 173

\bibitem[{Chochol} et~al.(2012)]{cho12j2138}
  {Chochol}, D., {Katysheva}, N.~A., {Shugarov}, S.~Y., {Zemko}, P.~O., \&
  {Andreev}, M.~V.\ 2012, Contr.\ of\ the\ Astron.\ Obs.\ Skalnat\'e Pleso, 42,
  39

\bibitem[{Cleveland}(1979)]{LOWESS}
  {Cleveland}, W.~S.\ 1979, J. Amer. Statist. Assoc., 74, 829

\bibitem[Cook, Warner(1981)]{coo81zchapdot}
  Cook, M.~C., \& Warner, B.\ 1981, MNRAS, 196, 55P

\bibitem[{Copperwheat} et~al.(2011)]{cop11j0926}
  {Copperwheat}, C.~M., {et~al.}\ 2011, MNRAS, 410, 1113

\bibitem[Cowley et~al.(1984)]{cow84lmcXBID}
  Cowley, A.~P., Crampton, D., Hutchings, J.~B., Helfand, D.~J., Hamilton,
  T.~T., Thorstensen, J.~R., \& Charles, P.~A.\ 1984, ApJ, 286, 196

\bibitem[{Croom} et~al.(2004)]{cro04qz7}
  {Croom}, S.~M., {Smith}, R.~J., {Boyle}, B.~J., {Shanks}, T., {Miller}, L.,
  {Outram}, P.~J., \& {Loaring}, N.~S.\ 2004, MNRAS, 349, 1397

\bibitem[{Dai} et~al.(2009)]{dai09zcha}
  {Dai}, Z., {Qian}, S., \& {Fern{\'a}ndez Laj{\'u}s}, E.\ 2009, ApJ, 703, 109

\bibitem[{Davis} et~al.(2015)]{dav15ASASSNCVAAS}
  {Davis}, A.~B., {Shappee}, B.~J., {Archer Shappee}, B., \& {ASAS-SN}\ 2015,
  American\ Astron.\ Soc.\ Meeting\ Abstracts, 225, \#344.02

\bibitem[{Davis} et~al.(2014)]{dav14j1723atel6455}
  {Davis}, A.~B., {et~al.}\ 2014, Astron.\ Telegram, 6455

\bibitem[{Denisenko}(2011)]{den11dde20dde21}
  {Denisenko}, D.\ 2011, Perem.\ Zvezdy, 31, 3

\bibitem[{Denisenko}(2009)]{den09j2319atel2282}
  {Denisenko}, D.~V.\ 2009, Astron.\ Telegram, 2282

\bibitem[{Denisenko}, {Sokolovsky}(2011)]{den11ROSATCVs}
  {Denisenko}, D.~V., \& {Sokolovsky}, K.~V.\ 2011, Astron.\ Lett., 37, 91

\bibitem[{D'Esterre}(1912)]{des12uvper}
  {D'Esterre}, C.~R.\ 1912, Astron.\ Nachr., 190, 163

\bibitem[{D'Esterre}(1913)]{des13uvper}
  {D'Esterre}, C.~R.\ 1913, Astron.\ Nachr., 193, 281

\bibitem[{Djorgovski} et~al.(2008)]{djo08atel1416}
  {Djorgovski}, S.~G.~., {et~al.}\ 2008, Astron.\ Telegram, 1416

\bibitem[Downes, Shara(1993)]{DownesCVatlas1}
  Downes, R.~A., \& Shara, M.~M.\ 1993, PASP, 105, 127

\bibitem[{Drake} et~al.(2009)]{CRTS}
  {Drake}, A.~J., {et~al.}\ 2009, ApJ, 696, 870

\bibitem[{Drake} et~al.(2014)]{dra14CRTSCVs}
  {Drake}, A.~J., {et~al.}\ 2014, MNRAS, 441, 1186

\bibitem[{Echevarria} et~al.(1983)]{ech83lyhya}
  {Echevarria}, J., {Pocock}, A.~S., {Penston}, M.~V., \& {Blades}, J.~C.\
  1983, MNRAS, 205, 559

\bibitem[{Elvey}, {Babcock}(1943)]{elv43DNspec}
  {Elvey}, C.~T., \& {Babcock}, H.~W.\ 1943, ApJ, 97, 412

\bibitem[{Erastova}(1973)]{era73v701tau}
  {Erastova}, L.~K.\ 1973, Astron.\ Tsirk., 774, 5

\bibitem[Fernie(1989)]{fer89error}
  Fernie, J.~D.\ 1989, PASP, 101, 225

\bibitem[{G{\"a}nsicke} et~al.(2009)]{gan09SDSSCVs}
  {G{\"a}nsicke}, B.~T., {et~al.}\ 2009, MNRAS, 397, 2170

\bibitem[{Garnavich} et~al.(2014)]{gar14v418seratel6287}
  {Garnavich}, P., {Littlefield}, C., {Terndrup}, D., \& {Adams}, S.\ 2014,
  Astron.\ Telegram, 6287

\bibitem[Glasby(1970)]{GlasbyDNbook}
  Glasby, J.~S.\ 1970, The Dwarf Novae (London: Constable)

\bibitem[{Golovin} et~al.(2007)]{gol07j1021}
  {Golovin}, A., {et~al.}\ 2007, IBVS, 5763

\bibitem[{Graham} et~al.(2010)]{gra10j2138cbet2275}
  {Graham}, M.~L., {Broekhoven-Fiene}, H., {Parker}, A.~H., {Sadavoy}, S.,
  {Maxwell}, A.~J., {Hsiao}, E.~Y., \& {Balam}\ 2010, Cent.\ Bur.\ Electron.\
  Telegrams, 2275, 6

\bibitem[Green et~al.(1982)]{gre82PGsurveyCV}
  Green, R.~F., Ferguson, D.~H., Liebert, J., \& Schmidt, M.\ 1982, PASP, 94,
  560

\bibitem[{Greenhill} et~al.(2006)]{gre06oycar}
  {Greenhill}, J.~G., {Hill}, K.~M., {Dieters}, S., {Fienberg}, K., {Howlett},
  M., {Meijers}, A., {Munro}, A., \& {Senkbeil}, C.\ 2006, MNRAS, 372, 1129

\bibitem[{Guthnick}, {Prager}(1933)]{gut33VSnamelist}
  {Guthnick}, P., \& {Prager}, R.\ 1933, Astron.\ Nachr., 249, 253

\bibitem[{Han} et~al.(2015)]{han15oycar}
  {Han}, Z.-T., {Qian}, S.-B., {Fern{\'a}ndez Laj{\'u}s}, E., {Liao}, W.-P., \&
  {Zhang}, J.\ 2015, New\ Astron., 34, 1

\bibitem[{Hartwig}(1915)]{har15uvper}
  {Hartwig}, E.\ 1915, Astron.\ Nachr., 201, 287

\bibitem[{Hartwig}(1917)]{har17uvper}
  {Hartwig}, E.\ 1917, Astron.\ Nachr., 204, 11

\bibitem[{Hartwig}(1920)]{har20uvper}
  {Hartwig}, E.\ 1920, Astron.\ Nachr., 212, 79

\bibitem[Harvey et~al.(1995)]{har95v503cyg}
  Harvey, D., Skillman, D.~R., Patterson, J., \& Ringwald, F.~A.\ 1995, PASP,
  107, 551

\bibitem[Harvey, Patterson(1995)]{har95cyuma}
  Harvey, D.~A., \& Patterson, J.\ 1995, PASP, 107, 1055

\bibitem[{Hirose}, {Osaki}(1990)]{hir90SHexcess}
  {Hirose}, M., \& {Osaki}, Y.\ 1990, PASJ, 42, 135

\bibitem[{Hodgkin} et~al.(2014)]{hod14asassn14fratel6407}
  {Hodgkin}, S.~T., {et~al.}\ 2014, Astron.\ Telegram, 6407

\bibitem[{Hoffmeister}(1963)]{hof63VSS61}
  {Hoffmeister}, C.\ 1963, Ver{\"{o}}ff.\ Sternw.\ Sonneberg, 6, 1

\bibitem[{Hornby}(1975)]{hor75uvper}
  {Hornby}, P.~W.\ 1975, J.\ Br.\ Astron.\ Assoc., 85, 528

\bibitem[Horne(1984)]{hor84superhump}
  Horne, K.\ 1984, Nature, 312, 348

\bibitem[{Howarth}(1978)]{how78uvper}
  {Howarth}, I.~D.\ 1978, J.\ Br.\ Astron.\ Assoc., 89, 47

\bibitem[{Howell} et~al.(1996)]{how96alcom}
  {Howell}, S.~B., {De Young}, J., {Mattei}, J.~A., {Foster}, G., {Szkody}, P.,
  {Cannizzo}, J.~K., {Walker}, G., \& {Fierce}, E.\ 1996, AJ, 111, 2367

\bibitem[{Hudec}(2010)]{hud10j2138atel2619}
  {Hudec}, R.\ 2010, Astron.\ Telegram, 2619

\bibitem[{Imada} et~al.(2006)]{ima06tss0222}
  {Imada}, A., {Kubota}, K., {Kato}, T., {Nogami}, D., {Maehara}, H.,
  {Nakajima}, K., {Uemura}, M., \& {Ishioka}, R.\ 2006, PASJ, 58, L23

\bibitem[{Imada}, {Monard}(2006)]{ima06asas1600}
  {Imada}, A., \& {Monard}, L.~A.~G.~B.\ 2006, PASJ, 58, L19

\bibitem[Ishioka et~al.(2002)]{ish02wzsgeletter}
  Ishioka, R., {et~al.}\ 2002, A\&A, 381, L41

\bibitem[{Isles}(1974)]{isl74suuma}
  {Isles}, J.~E.\ 1974, J.\ Br.\ Astron.\ Assoc., 84, 365

\bibitem[Kato(1990)]{kat90uvper}
  Kato, T.\ 1990, IBVS, 3522

\bibitem[Kato(1995)]{kat95cyuma}
  Kato, T.\ 1995, IBVS, 4236

\bibitem[Kato(1997a)]{kat97tleo}
  Kato, T.\ 1997a, PASJ, 49, 583

\bibitem[Kato(1997b)]{kat97cyuma}
  Kato, T.\ 1997b, VSOLJ\ Variable\ Star\ Bull., 25, 2

\bibitem[Kato(2002)]{kat02wzsgeESH}
  Kato, T.\ 2002, PASJ, 54, L11

\bibitem[{Kato}(2015)]{kat15wzsge}
  {Kato}, T.\ 2015, PASJ, submitted

\bibitem[{Kato} et~al.(2014a)]{Pdot6}
  {Kato}, T., {et~al.}\ 2014a, PASJ, 66, 90

\bibitem[Kato et~al.(1988)]{kat88cyuma}
  Kato, T., Fujino, S., Iida, M., Makiguchi, N., \& Koshiro, S.\ 1988, VSOLJ\
  Variable\ Star\ Bull., 5, 18

\bibitem[{Kato} et~al.(2013a)]{Pdot4}
  {Kato}, T., {et~al.}\ 2013a, PASJ, 65, 23

\bibitem[{Kato} et~al.(2014b)]{Pdot5}
  {Kato}, T., {et~al.}\ 2014b, PASJ, 66, 30

\bibitem[{Kato} et~al.(2015)]{kat15ccscl}
  {Kato}, T., {Hambsch}, F.-J., {Oksanen}, A., {Starr}, P., \& {Henden}, A.\
  2015, PASJ, 67, 3

\bibitem[{Kato} et~al.(2009)]{Pdot}
  {Kato}, T., {et~al.}\ 2009, PASJ, 61, S395

\bibitem[{Kato}, {Maehara}(2013)]{kat13j1924}
  {Kato}, T., \& {Maehara}, H.\ 2013, PASJ, 65, 76

\bibitem[{Kato} et~al.(2012a)]{Pdot3}
  {Kato}, T., {et~al.}\ 2012a, PASJ, 64, 21

\bibitem[{Kato} et~al.(2012b)]{kat12DNSDSS}
  {Kato}, T., {Maehara}, H., \& {Uemura}, M.\ 2012b, PASJ, 64, 62

\bibitem[{Kato} et~al.(2010)]{Pdot2}
  {Kato}, T., {et~al.}\ 2010, PASJ, 62, 1525

\bibitem[Kato, Matsumoto(1999)]{kat99cyuma}
  Kato, T., \& Matsumoto, K.\ 1999, IBVS, 4763

\bibitem[{Kato} et~al.(2013b)]{kat13j1222}
  {Kato}, T., {Monard}, B., {Hambsch}, F.-J., {Kiyota}, S., \& {Maehara}, H.\
  2013b, PASJ, 65, L11

\bibitem[Kato et~al.(2004a)]{kat04nsv10934mmscoabnorcal86}
  Kato, T., {et~al.}\ 2004a, MNRAS, 347, 861

\bibitem[Kato et~al.(1996)]{kat96alcom}
  Kato, T., Nogami, D., Baba, H., Matsumoto, K., Arimoto, J., Tanabe, K., \&
  Ishikawa, K.\ 1996, PASJ, 48, L21

\bibitem[{Kato} et~al.(2014c)]{kat14j0902}
  {Kato}, T., {et~al.}\ 2014c, PASJ, 66, L7

\bibitem[{Kato}, {Osaki}(2013a)]{kat13j1939v585lyrv516lyr}
  {Kato}, T., \& {Osaki}, Y.\ 2013a, PASJ, 65, 97

\bibitem[{Kato}, {Osaki}(2013b)]{kat13qfromstageA}
  {Kato}, T., \& {Osaki}, Y.\ 2013b, PASJ, 65, 115

\bibitem[{Kato}, {Osaki}(2013c)]{kat13j1922}
  {Kato}, T., \& {Osaki}, Y.\ 2013c, PASJ, 65, L13

\bibitem[{Kato}, {Osaki}(2014)]{kat14j1944}
  {Kato}, T., \& {Osaki}, Y.\ 2014, PASJ, 66, L5

\bibitem[Kato et~al.(2001)]{kat01crboo}
  Kato, T., {et~al.}\ 2001, IBVS, 5120, 1

\bibitem[Kato et~al.(2004b)]{kat04v803cen}
  Kato, T., Stubbings, R., Monard, B., Butterworth, N.~D., Bolt, G., \&
  Richards, T.\ 2004b, PASJ, 56, S89

\bibitem[{Kato}, {Uemura}(2012)]{kat12perlasso}
  {Kato}, T., \& {Uemura}, M.\ 2012, PASJ, 64, 122

\bibitem[Kato et~al.(2004c)]{VSNET}
  Kato, T., Uemura, M., Ishioka, R., Nogami, D., Kunjaya, C., Baba, H., \&
  Yamaoka, H.\ 2004c, PASJ, 56, S1

\bibitem[{Kazarovets} et~al.(2006)]{NameList78}
  {Kazarovets}, E.~V., {Samus}, N.~N., {Durlevich}, O.~V., {Kireeva}, N.~N., \&
  {Pastukhova}, E.~N.\ 2006, IBVS, 5721

\bibitem[{Kazarovets} et~al.(2008)]{NameList79}
  {Kazarovets}, E.~V., {Samus}, N.~N., {Durlevich}, O.~V., {Kireeva}, N.~N., \&
  {Pastukhova}, E.~N.\ 2008, IBVS, 5863, 1

\bibitem[{Kazarovets} et~al.(2011)]{NameList80b}
  {Kazarovets}, E.~V., {Samus}, N.~N., {Durlevich}, O.~V., {Kireeva}, N.~N., \&
  {Pastukhova}, E.~N.\ 2011, IBVS, 6008, 1

\bibitem[Kholopov et~al.(1985)]{GCVS}
  Kholopov, P.~N., {et~al.}\ 1985, General Catalogue of Variable Stars, fourth
  edition (Moscow: Nauka Publishing House)

\bibitem[{Khruslov}(2005)]{khr05nsv1485}
  {Khruslov}, A.~V.\ 2005, Perem.\ Zvezdy,\ Prilozh., 5, 4

\bibitem[{Knigge}(2006)]{kni06CVsecondary}
  {Knigge}, C.\ 2006, MNRAS, 373, 484

\bibitem[{Knigge} et~al.(2011)]{kni11CVdonor}
  {Knigge}, C., {Baraffe}, I., \& {Patterson}, J.\ 2011, ApJS, 194, 28

\bibitem[{Krzeminski}, {Vogt}(1985)]{krz85oycarsuper}
  {Krzeminski}, W., \& {Vogt}, N.\ 1985, A\&A, 144, 124

\bibitem[{Kubiak}, {Krzeminski}(1989)]{kun89lyhya}
  {Kubiak}, M., \& {Krzeminski}, W.\ 1989, PASP, 101, 667

\bibitem[{Kubiak}, {Krzeminski}(1992)]{kub92lyhya}
  {Kubiak}, M., \& {Krzeminski}, W.\ 1992, Acta\ Astron., 42, 177

\bibitem[{Levitan} et~al.(2011)]{lev11j0719}
  {Levitan}, D., {et~al.}\ 2011, ApJ, 739, 68

\bibitem[{Littlefield} et~al.(2013)]{lit13sbs1108}
  {Littlefield}, C., {et~al.}\ 2013, AJ, 145, 145

\bibitem[{Lubow}(1991)]{lub91SHa}
  {Lubow}, S.~H.\ 1991, ApJ, 381, 259

\bibitem[{Luyten}(1938)]{luy38propermotion2}
  {Luyten}, W.~J.\ 1938, Publ.\ of\ the\ Astron.\ Obs.\ Univ.\ of\ Minnesota,
  6, 1

\bibitem[{Mayall}(1966)]{may66uvper}
  {Mayall}, M.~W.\ 1966, JRASC, 60, 301

\bibitem[Mennickent, Sterken(1998)]{men98brlup}
  Mennickent, R.~E., \& Sterken, C.\ 1998, PASP, 110, 1032

\bibitem[Misselt, Shafter(1995)]{mis95PGCV}
  Misselt, K.~A., \& Shafter, A.~W.\ 1995, AJ, 109, 1757

\bibitem[{Mitrofanova} et~al.(2014)]{mit14j2138}
  {Mitrofanova}, A.~A., {Borisov}, N.~V., \& {Shimansky}, V.~V.\ 2014,
  Astrophys.\ Bull., 69, 82

\bibitem[{Monard}, {Africa}(2005)]{mon05j1600iauc8540}
  {Monard}, L.~A.~G., \& {Africa}, S.\ 2005, IAU\ Circ., 8540, 3

\bibitem[{Montgomery}(2012)]{mon12negSHSPH}
  {Montgomery}, M.~M.\ 2012, ApJ, 745, L25

\bibitem[{Montgomery}, {Martin}(2010)]{mon10disktilt}
  {Montgomery}, M.~M., \& {Martin}, E.~L.\ 2010, ApJ, 722, 989

\bibitem[{Mroz} et~al.(2013)]{mro13OGLEDN2}
  {Mroz}, P., {et~al.}\ 2013, Acta\ Astron., 63, 135

\bibitem[Munari, Zwitter(1998)]{mun98CVspec5}
  Munari, U., \& Zwitter, T.\ 1998, A\&AS, 128, 277

\bibitem[{Nakano}(2010)]{nak10j2138cbet2275}
  {Nakano}, S.\ 2010, Cent.\ Bur.\ Electron.\ Telegrams, 2275, 2

\bibitem[{Nakata} et~al.(2014)]{nak14j0754j2304}
  {Nakata}, C., {et~al.}\ 2014, PASJ, 66, 116

\bibitem[{Nakata} et~al.(2013)]{nak13j2112j2037}
  {Nakata}, C., {et~al.}\ 2013, PASJ, 65, 117

\bibitem[{Nelemans}(2005)]{nel05amcvnreview}
  {Nelemans}, G.\ 2005, in ASP\ Conf.\ Ser.\ 330, The Astrophysics of
  Cataclysmic Variables and Related Objects, ed. J.-M. {Hameury}, \& J.-P.
  {Lasota} (San Francisco: ASP), p.~27

\bibitem[{Nesci} et~al.(2013)]{nes13j1740ibvs6059}
  {Nesci}, R., {Caravano}, A., {Falasca}, V., \& {Villani}, L.\ 2013, IBVS,
  6059

\bibitem[{Nijland}(1914)]{nij14uvper}
  {Nijland}, A.~A.\ 1914, Astron.\ Nachr., 199, 131

\bibitem[{Nijland}(1915)]{nij15uvper}
  {Nijland}, A.~A.\ 1915, Astron.\ Nachr., 201, 347

\bibitem[{Nijland}(1918)]{nij18uvper}
  {Nijland}, A.~A.\ 1918, Astron.\ Nachr., 206, 223

\bibitem[{Nijland}(1924)]{nij24uvper}
  {Nijland}, A.~A.\ 1924, Astron.\ Nachr., 221, 243

\bibitem[{Nogami} et~al.(1997)]{nog97alcom}
  {Nogami}, D., {Kato}, T., {Baba}, H., {Matsumoto}, K., {Arimoto}, J.,
  {Tanabe}, K., \& {Ishikawa}, K.\ 1997, ApJ, 490, 840

\bibitem[Nogami et~al.(1995)]{nog95rzlmi}
  Nogami, D., Kato, T., Masuda, S., Hirata, R., Matsumoto, K., Tanabe, K., \&
  Yokoo, T.\ 1995, PASJ, 47, 897

\bibitem[{Nogami} et~al.(2004)]{nog04v406hya}
  {Nogami}, D., {Monard}, B., {Retter}, A., {Liu}, A., {Uemura}, M., {Ishioka},
  R., {Imada}, A., \& {Kato}, T.\ 2004, PASJ, 56, L39

\bibitem[Nogami et~al.(2004)]{nog04qwser}
  Nogami, D., {et~al.}\ 2004, PASJ, 56, S99

\bibitem[O'Donoghue(1987)]{odo87brlup}
  O'Donoghue, D.\ 1987, Ap\&SS, 136, 247

\bibitem[{Ohshima} et~al.(2014)]{ohs14eruma}
  {Ohshima}, T., {et~al.}\ 2014, PASJ, 66, 67

\bibitem[{Ohshima} et~al.(2011)]{ohs11qzvir}
  {Ohshima}, T., {et~al.}\ 2011, PASJ, submitted

\bibitem[Olech(1997)]{ole97v485cen}
  Olech, A.\ 1997, Acta\ Astron., 47, 281

\bibitem[{Olech} et~al.(2004)]{ole04ttboo}
  {Olech}, A., {Cook}, L.~M., {Z{\l}oczewski}, K., {Mularczyk}, K.,
  {K{\^e}dzierski}, P., {Udalski}, A., \& {Wisniewski}, M.\ 2004, Acta\
  Astron., 54, 233

\bibitem[Olech et~al.(2003)]{ole03qwser}
  Olech, A., K\c{e}dzierski, P., Z{\l}oczewski, K., Mularczyk, K., \&
  Wi\'{s}niewski, M.\ 2003, A\&A, 411, 483

\bibitem[{Osaki}(1989)]{osa89suuma}
  {Osaki}, Y.\ 1989, PASJ, 41, 1005

\bibitem[{Osaki}(1996)]{osa96review}
  {Osaki}, Y.\ 1996, PASP, 108, 39

\bibitem[{Osaki}, {Kato}(2013a)]{osa13v1504cygKepler}
  {Osaki}, Y., \& {Kato}, T.\ 2013a, PASJ, 65, 50

\bibitem[{Osaki}, {Kato}(2013b)]{osa13v344lyrv1504cyg}
  {Osaki}, Y., \& {Kato}, T.\ 2013b, PASJ, 65, 95

\bibitem[{Osaki}, {Kato}(2014)]{osa14v1504cygv344lyrpaper3}
  {Osaki}, Y., \& {Kato}, T.\ 2014, PASJ, 66, 15

\bibitem[{Osaki}, {Meyer}(2002)]{osa02wzsgehump}
  {Osaki}, Y., \& {Meyer}, F.\ 2002, A\&A, 383, 574

\bibitem[{Osborne} et~al.(2011)]{osb11nsv1436}
  {Osborne}, J.~P., {et~al.}\ 2011, A\&A, 533, A41

\bibitem[{Pagnotta}, {Schaefer}(2014)]{pag14RNcand}
  {Pagnotta}, A., \& {Schaefer}, B.~E.\ 2014, ApJ, 788, 164

\bibitem[{Patterson} et~al.(1996)]{pat96alcom}
  {Patterson}, J., {Augusteijn}, T., {Harvey}, D.~A., {Skillman}, D.~R.,
  {Abbott}, T.~M.~C., \& {Thorstensen}, J.\ 1996, PASP, 108, 748

\bibitem[Patterson et~al.(1993)]{pat93vyaqr}
  Patterson, J., Bond, H.~E., Grauer, A.~D., Shafter, A.~W., \& Mattei, J.~A.\
  1993, PASP, 105, 69

\bibitem[Patterson et~al.(1997)]{pat97crboo}
  Patterson, J., {et~al.}\ 1997, PASP, 109, 1100

\bibitem[Patterson et~al.(2002)]{pat02wzsge}
  Patterson, J., {et~al.}\ 2002, PASP, 114, 721

\bibitem[Patterson et~al.(2003)]{pat03suumas}
  Patterson, J., {et~al.}\ 2003, PASP, 115, 1308

\bibitem[Patterson et~al.(2000)]{pat00v803cen}
  Patterson, J., Walker, S., Kemp, J., O'Donoghue, D., Bos, M., \& Stubbings,
  R.\ 2000, PASP, 112, 625

\bibitem[{Pavlenko} et~al.(2014)]{pav14nyser}
  {Pavlenko}, E.~P., {et~al.}\ 2014, PASJ, 66, 111

\bibitem[{Pearson}(2007)]{pea07amcvnSH}
  {Pearson}, K.~J.\ 2007, MNRAS, 379, 183

\bibitem[{Perez}, {McNaught}(1986)]{per86vyaqriauc4222}
  {Perez}, A., \& {McNaught}, R.~H.\ 1986, IAU\ Circ., 4222, 2

\bibitem[{Petit}(1960)]{pet60suuma}
  {Petit}, M.\ 1960, Journal\ des\ Observateurs, 43, 33

\bibitem[{Petit}, {Brun}(1956)]{pet56uvper}
  {Petit}, M., \& {Brun}, A.\ 1956, Journal\ des\ Observateurs, 39, 37

\bibitem[{Pojma\'nski}(2002)]{ASAS3}
  {Pojma\'nski}, G.\ 2002, Acta\ Astron., 52, 397

\bibitem[{Pojmanski}(2005)]{poj05j1600iauc8539}
  {Pojmanski}, G.\ 2005, IAU\ Circ., 8539, 3

\bibitem[{Prager}, {Shapley}(1941)]{pra41VScatalog}
  {Prager}, R., \& {Shapley}, H.\ 1941, Annals\ of\ the\ Astron.\ Obs.\ of\
  Harvard\ Coll.\, 111, 1

\bibitem[{Price} et~al.(2003)]{pri03uvper}
  {Price}, A., {et~al.}\ 2003, IBVS, 5488

\bibitem[{Prieto} et~al.(2013)]{pri13j1740asassn13adatel4999}
  {Prieto}, J.~L., {et~al.}\ 2013, Astron.\ Telegram, 4999

\bibitem[{Prieto} et~al.(2014a)]{pri14asassn14eiatel6475}
  {Prieto}, J.~L., {et~al.}\ 2014a, Astron.\ Telegram, 6475, 1

\bibitem[{Prieto} et~al.(2014b)]{pri14asassn14doatel6293}
  {Prieto}, J.~L., {et~al.}\ 2014b, Astron.\ Telegram, 6293

\bibitem[{Quimby}, {Mondol}(2006)]{qui06j1202atel787}
  {Quimby}, R., \& {Mondol}, P.\ 2006, Astron.\ Telegram, 787

\bibitem[{Rau} et~al.(2010)]{rau10HeDN}
  {Rau}, A., {Roelofs}, G.~H.~A., {Groot}, P.~J., {Marsh}, T.~R., {Nelemans},
  G., {Steeghs}, D., {Salvato}, M., \& {Kasliwal}, M.~M.\ 2010, ApJ, 708, 456

\bibitem[Ritter(1980)]{rit80oycar}
  Ritter, H.\ 1980, A\&A, 85, 362

\bibitem[{Ritter}, {Kolb}(2003)]{RKCat}
  {Ritter}, H., \& {Kolb}, U.\ 2003, A\&A, 404, 301

\bibitem[Robertson et~al.(1995)]{rob95eruma}
  Robertson, J.~W., Honeycutt, R.~K., \& Turner, G.~W.\ 1995, PASP, 107, 443

\bibitem[{Ross}(1925)]{ros25nsv1436}
  {Ross}, F.~E.\ 1925, AJ, 36, 99

\bibitem[{Ross}(1928)]{ros28sycap}
  {Ross}, F.~E.\ 1928, AJ, 38, 144

\bibitem[{Rykoff} et~al.(2004)]{ryk04j1514j2215}
  {Rykoff}, E.~S., {et~al.}\ 2004, IBVS, 5559

\bibitem[Schmidtke et~al.(2002)]{sch02cal86}
  Schmidtke, P.~C., Cowley, A.~P., Hutchings, J.~B., \& Crampton, D.\ 2002, AJ,
  123, 3210

\bibitem[Schoembs, Hartmann(1983)]{sch83oycar}
  Schoembs, R., \& Hartmann, K.\ 1983, A\&A, 128, 37

\bibitem[Shafter, Hessman(1988)]{sha88yzcnc}
  Shafter, A.~W., \& Hessman, F.~V.\ 1988, AJ, 95, 178

\bibitem[Shafter, Szkody(1984)]{sha84tleo}
  Shafter, A.~W., \& Szkody, P.\ 1984, ApJ, 276, 305

\bibitem[{Shappee} et~al.(2014a)]{ASASSN}
  {Shappee}, B.~J., {et~al.}\ 2014a, ApJ, 788, 48

\bibitem[{Shappee} et~al.(2014b)]{sha14asassn14jvatel6676}
  {Shappee}, B.~J., {et~al.}\ 2014b, Astron.\ Telegram, 6676

\bibitem[{Shears}, {Boyd}(2007)]{she07v701tau}
  {Shears}, J., \& {Boyd}, D.\ 2007, J.\ Br.\ Astron.\ Assoc., 117, 25

\bibitem[{Shears} et~al.(2012)]{she12j0129}
  {Shears}, J., {Brady}, S., {Koff}, R., {Goff}, W., \& {Boyd}, D.\ 2012, J.\
  Br.\ Astron.\ Assoc., 122, 49

\bibitem[{Shears} et~al.(2011a)]{she11nncam}
  {Shears}, J., {et~al.}\ 2011a, J.\ Br.\ Astron.\ Assoc., 121, 355

\bibitem[{Shears} et~al.(2011b)]{she11j0423}
  {Shears}, J.~H., {Gaensicke}, B.~T., {Brady}, S., {Dubovsky}, P., {Miller},
  I., \& {Staels}, B.\ 2011b, New\ Astron., 16, 311

\bibitem[Sherrington et~al.(1982)]{she82oycarIR}
  Sherrington, M.~R., Jameson, R.~F., Bailey, J., \& Giles, A.~B.\ 1982, MNRAS,
  200, 861

\bibitem[{Shumkov} et~al.(2014)]{shu14j0316atel6851}
  {Shumkov}, V., {et~al.}\ 2014, Astron.\ Telegram, 6851

\bibitem[{Simonian} et~al.(2015)]{sim15asassn15bpatel6981}
  {Simonian}, G., {et~al.}\ 2015, Astron.\ Telegram, 6981

\bibitem[{Simonian} et~al.(2014)]{sim14asassn14jfatel6608}
  {Simonian}, G., {et~al.}\ 2014, Astron.\ Telegram, 6608

\bibitem[{Skillman} et~al.(2002)]{ski02j2329}
  {Skillman}, D.~R., {et~al.}\ 2002, PASP, 114, 630

\bibitem[{Smak}(1985)]{sma85vwhyi}
  {Smak}, J.\ 1985, Acta\ Astron., 35, 357

\bibitem[{Smak}(2004)]{sma04EMT}
  {Smak}, J.\ 2004, Acta\ Astron., 54, 221

\bibitem[{Smak}(2008)]{sma08zcha}
  {Smak}, J.\ 2008, Acta\ Astron., 58, 55

\bibitem[{Smak}(1991)]{sma91suumamodel}
  {Smak}, J.~I.\ 1991, Acta\ Astron., 41, 269

\bibitem[{Smart}(2013)]{IGSL}
  {Smart}, R.~L.\ 2013, VizieR\ Online\ Data\ Catalog, 1324

\bibitem[{Smith} et~al.(2002)]{smi02ficet}
  {Smith}, D.~A., {et~al.}\ 2002, IBVS, 5226

\bibitem[{Soejima} et~al.(2009)]{soe09asas1600}
  {Soejima}, Y., {Imada}, A., {Nogami}, D., {Kato}, T., \& {Monard}, L.~A.~G.\
  2009, PASJ, 61, 395

\bibitem[{Solheim}(2010)]{sol10amcvnreview}
  {Solheim}, {J.-E.}\ 2010, PASP, 122, 1133

\bibitem[{Stanek} et~al.(2014a)]{sta14asassn14clatel6233}
  {Stanek}, K.~Z., {et~al.}\ 2014a, Astron.\ Telegram, 6233

\bibitem[{Stanek} et~al.(2014b)]{sta14asassn14hkatel6479}
  {Stanek}, K.~Z., {et~al.}\ 2014b, Astron.\ Telegram, 6479

\bibitem[Stellingwerf(1978)]{PDM}
  Stellingwerf, R.~F.\ 1978, ApJ, 224, 953

\bibitem[Still et~al.(1994)]{sti94lyhya}
  Still, M.~D., Marsh, T.~R., Dhillon, V.~S., \& Horne, K.\ 1994, MNRAS, 267,
  957

\bibitem[Swope, Caldwell(1930)]{swo30abnorbrlup}
  Swope, H.~H., \& Caldwell, I.~W.\ 1930, Harvard\ Coll.\ Obs.\ Bull., 879

\bibitem[Szkody et~al.(2002)]{szk02SDSSCVs}
  Szkody, P., {et~al.}\ 2002, AJ, 123, 430

\bibitem[{Szkody} et~al.(2006)]{szk06SDSSCV5}
  {Szkody}, P., {et~al.}\ 2006, AJ, 131, 973

\bibitem[{Szkody} et~al.(2007)]{szk07SDSSCV6}
  {Szkody}, P., {et~al.}\ 2007, AJ, 134, 185

\bibitem[Szkody, Mattei(1984)]{szk84AAVSO}
  Szkody, P., \& Mattei, J.~A.\ 1984, PASP, 96, 988

\bibitem[{Templeton}(2011)]{tem11nsv1436aan}
  {Templeton}, M.~R.\ 2011, AAVSO\ Alert\ Notice, 434

\bibitem[{Teyssier}(2014)]{tey14asassn14clatel6235}
  {Teyssier}, F.\ 2014, Astron.\ Telegram, 6235

\bibitem[Thorstensen et~al.(2002)]{tho02j2329}
  Thorstensen, J.~R., Fenton, W.~H., Patterson, J.~O., Kemp, J., Krajci, T., \&
  Baraffe, I.\ 2002, ApJ, 567, L49

\bibitem[Thorstensen et~al.(1996)]{tho96Porb}
  Thorstensen, J.~R., Patterson, J.~O., Shambrook, A., \& Thomas, G.\ 1996,
  PASP, 108, 73

\bibitem[{Thorstensen}, {Skinner}(2012)]{tho12CRTSCVs}
  {Thorstensen}, J.~R., \& {Skinner}, J.~N.\ 2012, AJ, 144, 81

\bibitem[Thorstensen, Taylor(1997)]{tho97uvpervyaqrv1504cyg}
  Thorstensen, J.~R., \& Taylor, C.~J.\ 1997, PASP, 109, 1359

\bibitem[Thorstensen et~al.(1986)]{tho86suuma}
  Thorstensen, J.~R., Wade, R.~A., \& Oke, J.~B.\ 1986, ApJ, 309, 721

\bibitem[{Tibshirani}(1996)]{lasso}
  {Tibshirani}, R.\ 1996, J. R. Statistical Soc. Ser. B, 58, 267

\bibitem[{Tovmassian} et~al.(2010)]{tov10j2138cbet2283}
  {Tovmassian}, G., {Clark}, D., \& {Zharikov}, S.\ 2010, Cent.\ Bur.\
  Electron.\ Telegrams, 2283, 1

\bibitem[{Udalski}(1988)]{uda88suuma}
  {Udalski}, A.\ 1988, IBVS, 3239

\bibitem[Udalski(1989)]{uda89uvperiauc}
  Udalski, A.\ 1989, IAU\ Circ., 4885

\bibitem[{Udalski}(1990)]{uda90suuma}
  {Udalski}, A.\ 1990, AJ, 100, 226

\bibitem[Udalski, Pych(1992)]{uda92uvper}
  Udalski, A., \& Pych, W.\ 1992, Acta\ Astron., 42, 285

\bibitem[{Uemura} et~al.(2008)]{uem08j1021}
  {Uemura}, M., {et~al.}\ 2008, PASJ, 60, 227

\bibitem[Uemura et~al.(2002)]{uem02j2329letter}
  Uemura, M., {et~al.}\ 2002, PASJ, 54, L15

\bibitem[{Uemura} et~al.(2010)]{uem10shortPCV}
  {Uemura}, M., {Kato}, T., {Nogami}, D., \& {Ohsugi}, T.\ 2010, PASJ, 62, 613

\bibitem[Uemura et~al.(2001)]{uem01v725aql}
  Uemura, M., Kato, T., Pavlenko, E., Baklanov, A., \& Pietz, J.\ 2001, PASJ,
  53, 539

\bibitem[{Uemura} et~al.(2005)]{uem05tvcrv}
  {Uemura}, M., {et~al.}\ 2005, A\&A, 432, 261

\bibitem[{van Amerongen} et~al.(1987)]{vaname87vwhyi}
  {van Amerongen}, S., {Bovenschen}, H., \& {van Paradijs}, J.\ 1987, MNRAS,
  229, 245

\bibitem[van Amerongen et~al.(1990)]{vaname90zcha}
  van Amerongen, S., Kuulkers, E., \& van Paradijs, J.\ 1990, MNRAS, 242, 522

\bibitem[{Vladimirov} et~al.(2014)]{vla14j1055atel5950}
  {Vladimirov}, V., {et~al.}\ 2014, Astron.\ Telegram, 5950

\bibitem[Vogt(1979)]{vog79ektraiauc}
  Vogt, N.\ 1979, IAU\ Circ., 3375

\bibitem[Vogt(1980)]{vog80suumastars}
  Vogt, N.\ 1980, A\&A, 88, 66

\bibitem[Vogt(1982)]{vog82zcha}
  Vogt, N.\ 1982, ApJ, 252, 653

\bibitem[Vogt(1983)]{vog83oycar}
  Vogt, N.\ 1983, A\&A, 128, 29

\bibitem[{Vogt}(1983)]{vog83lateSH}
  {Vogt}, N.\ 1983, A\&A, 118, 95

\bibitem[Vogt, Bateson(1982)]{vog82atlas}
  Vogt, N., \& Bateson, F.~M.\ 1982, A\&AS, 48, 383

\bibitem[Vogt et~al.(1981)]{vog81oycar}
  Vogt, N., Schoembs, R., Krzeminski, W., \& Pedersen, H.\ 1981, A\&A, 94, L29

\bibitem[{Wagner} et~al.(2014)]{wag14asassn14fvatel6669}
  {Wagner}, R.~M., {et~al.}\ 2014, Astron.\ Telegram, 6669

\bibitem[Walker, Olmsted(1958)]{wal58CVchart}
  Walker, A.~D., \& Olmsted, M.\ 1958, PASP, 70, 495

\bibitem[Warner(1974)]{war74zcha}
  Warner, B.\ 1974, MNRAS, 168, 235

\bibitem[{Warner}(1976)]{war76CVmultipleperiod}
  {Warner}, B.\ 1976, in IAU\ Colloq.\ 29, Multiple Periodic Variable Stars,
  ed. {W.~S.~Fitch} (Dordrecht: D.\ Reidel Publishing Company), p.~247

\bibitem[Warner(1995)]{war95book}
  Warner, B.\ 1995, Cataclysmic Variable Stars (Cambridge: Cambridge University
  Press)

\bibitem[Watanabe et~al.(1989)]{wat89cyuma}
  Watanabe, M., Hirosawa, K., Kato, T., \& Narumi, H.\ 1989, VSOLJ\ Variable\
  Star\ Bull., 10, 40

\bibitem[Whitehurst(1988)]{whi88tidal}
  Whitehurst, R.\ 1988, MNRAS, 232, 35

\bibitem[{Williams} et~al.(2015)]{wil15asassn15bpatel6992}
  {Williams}, S.~C., {Darnley}, M.~J., {Bode}, M.~F., \& {Copperwheat}, C.~M.\
  2015, Astron.\ Telegram, 6992

\bibitem[{Wils} et~al.(2010)]{wil10newCVs}
  {Wils}, P., {G{\"a}nsicke}, B.~T., {Drake}, A.~J., \& {Southworth}, J.\ 2010,
  MNRAS, 402, 436

\bibitem[{Wolf}(1912)]{wol12uvper}
  {Wolf}, M.\ 1912, Astron.\ Nachr., 192, 7

\bibitem[{Wood}, {Burke}(2007)]{woo07negSH}
  {Wood}, M.~A., \& {Burke}, C.~J.\ 2007, ApJ, 661, 1042

\bibitem[Wood et~al.(2002)]{woo02kldra}
  Wood, M.~A., Casey, M.~J., Garnavich, P.~M., \& Haag, B.\ 2002, MNRAS, 334,
  87

\bibitem[{Woudt}, {Warner}(2011)]{wou11j2333atel}
  {Woudt}, P.~A., \& {Warner}, B.\ 2011, Astron.\ Telegram, 3705

\bibitem[{Woudt} et~al.(2013)]{wou13j0450atel4726}
  {Woudt}, P.~A., {Warner}, B., \& {Motsoaledi}, M.\ 2013, Astron.\ Telegram,
  4726

\bibitem[{Yamaoka}(2010)]{yam10j2138cbet2273}
  {Yamaoka}, H.\ 2010, Cent.\ Bur.\ Electron.\ Telegrams, 2273

\bibitem[{Yecheistov} et~al.(2014)]{yec14j0558atel5905}
  {Yecheistov}, V., {et~al.}\ 2014, Astron.\ Telegram, 5905

\bibitem[{Zemko} et~al.(2013)]{zem13eruma}
  {Zemko}, P., {Kato}, T., \& {Shugarov}, S.\ 2013, PASJ, 65, 54

\bibitem[{Zemko}, {Kato}(2013)]{zem13j2138}
  {Zemko}, P.~O., \& {Kato}, T.\ 2013, Astrophysics, 56, 203

\end{thebibliography}
\end{document}